\documentclass[longbib,article,times,12pt]{aastex701} 
\usepackage[utf8]{inputenc}
\usepackage{enumitem}
\usepackage{amsmath}
\usepackage{amssymb,amsmath}  
\usepackage{bm}				
\usepackage[normalem]{ulem}    	
\usepackage{cancel}
\usepackage{mathrsfs}			
\usepackage[mathscr]{euscript}
\usepackage{verbatim}			
\usepackage{graphicx}
\usepackage{natbib}
\usepackage{url}
\usepackage{physics}			

\usepackage{xpatch} 
\xpatchcmd{\thebibliography}{\twocolumngrid}{}{}{}
\makeatletter
\input{aas_macros.sty}
\def\ref@jnl#1{{\jnl@style#1\ }}
\makeatother

\makeatletter
\newcommand{\startappendices}{%
  \setcounter{section}{0}%
  \setcounter{subsection}{0}%
  \renewcommand{\thesection}{\Alph{section}}%
  
  \renewcommand{\section}{\@ifstar{\@appendixstarsection}{\@appendixsection}}%
}

\newcommand{\@appendixstarsection}[1]{%
  \vspace{12pt}%
  \begin{center}%
    \large\bfseries #1%
  \end{center}%
  \vspace{6pt}%
}

\newcommand{\@appendixsection}[1]{%
  \refstepcounter{section}%
  \vspace{12pt}%
  \begin{center}%
    \large\bfseries Appendix~\thesection.\quad #1%
  \end{center}%
  \vspace{6pt}%
  \addcontentsline{toc}{section}{Appendix \thesection. #1}%
}
\makeatother

\usepackage{newcommands}

\numberwithin{equation}{section}

\makeatletter
\newcommand\subsubsubsection{\@startsection{paragraph}{4}{\z@}{-2.5ex\@plus -1ex \@minus -.25ex}{1.25ex \@plus .25ex}{\normalfont\normalsize\bfseries}}
\newcommand\subsubsubsubsection{\@startsection{subparagraph}{5}{\z@}{-2.5ex\@plus -1ex \@minus -.25ex}{1.25ex \@plus .25ex}{\normalfont\normalsize\bfseries}}
\makeatother

\graphicspath{{./}}

\newcommand{\newtext}[1]{{\color{black}{#1}}}   

\newcommand{\newtextR}[1]{{\color{black}{#1}}}		

\begin{document}

\title{\Large Fundamental Noise Processes in Pulsar Timing} 

\author[0000-0002-4049-1882]{James M. Cordes}
\affiliation{Cornell Center for Astrophysics and Planetary Science, and Department of Astronomy, Cornell University, Ithaca, NY 14853, USA}
\email[show]{jmc33@cornell.edu, corresponding author}

\author[0000-0002-4941-5333]{Stella Koch Ocker}
\affiliation{Cahill Center for Astronomy and Astrophysics, California Institute of Technology, Pasadena, CA 91125, USA}
\affiliation{Carnegie Science Observatories, Pasadena, CA 91101, USA}
\email{ }

\author[0000-0002-2878-1502]{Shami Chatterjee}
\affiliation{Cornell Center for Astrophysics and Planetary Science, and Department of Astronomy, Cornell University, Ithaca, NY 14853, USA}
\email{ }


\author[0000-0001-8885-6388]{Timothy Dolch}
\affiliation{Department of Physics and Astronomy, University of New Mexico, Albuquerque, NM 87131, USA}
\affiliation{Eureka Scientific, 2452 Delmer Street, Suite 100, Oakland, CA 94602-3017, USA}
\affiliation{SETI Institute, 339 N Bernardo Ave Suite 200, Mountain View, CA 94043, USA}
\affiliation{Department of Physics, Hillsdale College, 33 E. College Street, Hillsdale, MI 49242, USA}
\email{ }

\author[0000-0003-1082-2342]{Ross J. Jennings}
\affiliation{NANOGrav Physics Frontiers Center Postdoctoral Fellow}
\email{ }

\author[0000-0003-0721-651X]{Michael T. Lam}
\affiliation{SETI Institute, 339 N Bernardo Ave Suite 200, Mountain View, CA 94043, USA}
\affiliation{School of Physics and Astronomy, Rochester Institute of Technology, Rochester, NY 14623, USA}
\affiliation{Laboratory for Multiwavelength Astrophysics, Rochester Institute of Technology, Rochester, NY 14623, USA}
\email{ }

\author[0000-0002-2451-7288]{Jacob E. Turner}
\affiliation{Green Bank Observatory, P.O. Box 2, Green Bank, WV 24944, USA}
\email{ }

\begin{abstract}
{\parskip 10pt \parindent 0pt
\noindent
The stability of spin rates and emission regions of radio-emitting neutron stars (pulsars) has enabled remarkable advances, including the detection of exoplanets, testing  theories of gravity, and 
\newtext{
stimulating the vigorous exploration of a new window in gravitational wave astronomy an nanohertz frequencies.
}

\noindent
An ideal astrophysical clock would emit identical pulses at a constant rate, propagating along single geodesics through a perfect vacuum. In reality, the timing precision achievable with pulsars is limited by spin noise, variability in the beam produced in a complex magnetosphere, and multipath propagation through a turbulent interstellar medium, with additional limitations from  uncertainties in the changing pulsar -- observatory vector and instrumental effects.

\noindent
Here we present the fundamentals of pulsar timing, its current limitations, and prospects for improvements in timing precision. We develop metrics to assess the timing quality of a pulsar and use them to evaluate sources currently included in pulsar timing arrays 
(PTAs)  as a means for characterizing the quality of future pulsar discoveries. With increases in telescope sensitivity, pulsar timing precision will increasingly be limited by propagation effects in the interstellar medium. We identify sources of non-Gaussianity in timing residuals and separate out effects that are chromatic or achromatic, as well as effects that vary on short (seconds to minutes) and longer (days to months) timescales.

\noindent
Modern wide-bandwidth receivers provide simultaneous arrival times across a range of frequencies, enabling the robust mitigation of chromatic interstellar effects and the identification of non-Gaussian events superimposed on random fluctuations. We recommend changes to timing procedures that leverage instrumentation advances to improve timing precision.

}
\vspace {0.75in}   
\end{abstract}

{\parskip -1pt
\setcounter{tocdepth}{1} 
\tableofcontents
}

\clearpage
\providecommand{\Spave}{\langle S_{\nu}\rangle_P}
\providecommand{\Sphiave}{\langle S_{\nu}\rangle_\phi}
\providecommand{\npol}{n_{\rm pol}}
\providecommand{\Tint}{T_{\rm int}}


\section{Introduction}
\label{sec:intro}



\newcommand{\redhighlight}[1]{%
  \colorbox{red!10}{$\displaystyle#1$}}		
  
\newcommand{\bluehighlight}[1]{%
  \colorbox{blue!10}{$\displaystyle#1$}}	
  
\newcommand{\greenhighlight}[1]{%
  \colorbox{green!10}{$\displaystyle#1$}}
  
\newcommand{\yellowhighlight}[1]{%
  \colorbox{yellow!10}{$\displaystyle#1$}}
  
\providecommand{\fspin}{{f_{\rm s}}}
\providecommand{\fsdot}{{\dot f_{\rm s}}}
\providecommand{\fsdotdot}{{\ddot f_{\rm s}}}

\providecommand{\fspinz}{{f_{\rm s_0}}}
\providecommand{\fsdotz}{{\dot f_{\rm s_0}}}
\providecommand{\fsdotdotz}{{\ddot f_{\rm s_0}}}

\providecommand{\FJ}{F_{\rm J}}

The methodology of pulsar timing  exploits the narrow pulses and  
stable spins of neutron stars (NSs) for their use as astrophysical clocks.    
It is now applied across the entire electromagnetic spectrum, primarily at radio, X-ray, 
and $\gamma$-ray frequencies, for a host of unique applications.  
The existence of gravitational waves  was demonstrated by timing studies of binary pulsars  
\newtext{\citep[][]{
Taylor1982,1994RvMP...66..711T,2010ApJ...722.1030W}}, which also provided
fundamental tests of General Relativity and other theories of gravity 
\cite[][]{2021PhRvX..11d1050K},  
along with precise determinations of the orbits and masses of neutron stars 
\newtext{\cite[][]{2016ARA&A..54..401O}} with specific cases reported by 
\citet[][]{2010Natur.467.1081D,2013Sci...340..448A,2020NatAs...4...72C}.  Neutron star masses provide key constraints on the equation of state of supra-nuclear matter, especially in concert with X-ray observations that constrain NS radii
\citep[e.g.][]{2021ARNPS..71..433L, 2021ApJ...918L..28M, 2024ApJ...971L..19R}. Recently,  precision timing of 
 millisecond pulsars has \newtext{provided evidence for the presence of a stochastic background of long-wavelength gravitational waves. }
Propagation delays through intervening plasmas and spacetime are necessarily accounted for in 
pulsar timing and they provide unique information about plasma densities, magnetic fields, and
(upper bounds on) dark matter.  Finally, long timing programs on some pulsars 
now extend to nearly a half century, yielding some of the best upper bounds on variations 
in the gravitational constant $G$ and other fundamental constants \citep[][]{Kaplan_2022}.  

Steady improvement in the precision of time-of-arrival
(TOA) measurements has  underpinned testing theories of gravity with higher-order relativistic \newtext{terms \citep[][]{2021PhRvX..11d1050K}  }
\newtext{
and searching for timing perturbations from low-frequency (nanohertz) gravitational waves (GWs).  The effects of GWs on pulsar timing were first discussed by 
\citet[][]{1978SvA....22...36S,det79} and initial upper limits  were made by \citet[][]{hd83, rt83}  using  data on slowly rotating pulsars (periods $P\sim 1$\,s) from the JPL timing program \citep[][]{1983ApJS...53..169D}. We refer readers to \cite{staylor_book2021,2025arXiv251218822S,mingarelli26} for recent reviews of the state-of-the-art in GW astrophysics with pulsar timing. 
}

Usage of millisecond pulsars (MSPs) for GW detection began with the discovery of 
 the MSP B1937+21 (J1939+2134) \citep[][]{bkh+82}, whose narrow pulse and short 1.56\,ms period  suggested that greater sensitivity to 
cosmological GW backgrounds could be achieved \citep{bcr83}. Initial observations verified this possibility \citep[][]{1984ApJ...277L..53C,1985Natur.315..547D}.   
Comprehensive discussions of pulsar timing by  \citet[][]{1984JApA....5..369B} and \citet[][]{bh86} laid further groundwork for gravity science using pulsars, leading to the notion of a `pulsar timing array' (PTA) by \citet[][]{fb90} for GW detection.

The first MSP-based limits reported by  \citet[e.g.][]{1985Natur.315..547D} were based on \RMS\ timing residuals $\sim 0.9$\,ms over a 2-yr period and 
 expressed as an upper bound on  the GW energy density relative to closure density of the universe, 
$\Omega_{\rm gw} \lesssim 10^{4}$, corresponding to a dimensionless strain $h_{\rm c} \lesssim 3\times 10^{-14}$, more than an order of magnitude larger than the sensitivity of current PTAs.  
Subsequent studies \citep[][]{1990PhRvL..65..285S, ktr94} gave more stringent limits but also identified extra correlated noise in B1937+21 that in current data sets makes this pulsar unsuitable for GW detection compared to other MSPs. 

Discovery of new MSPs combined with improved timing methods   increased the plausibility of detection of nanohertz GWs 
\citep[][]{2001ApJ...562..297L,2004NewAR..48..993K,2004NewAR..48.1413C,jhlm05,jhv+07,hbb+09,2009astro2010S..64D, 2010CQGra..27h4013H}
 and led to the formation of several research groups in pursuit of PTA science. 
In a series of papers published in 2023, evidence for detection of a stochastic GW background was reported 
\citep{2023ApJ...951L...8A,EPTAGW23,2023ApJ...951L...6R,2023RAA....23g5024X} that was corroborated by a more recent analysis using data from the MeerKAT telescope \citep[][]{2025MNRAS.536.1489M}.  

\newtext{
The most demanding timing application is GW detection of $\sim$nanohertz GWs,  which not only  relies on \TOAs\ with small measurement errors but it also requires that {\it all} contributions to the net timing residuals be of order $\sim \mu$s or better. This requires adequate signal-to-noise ratios of measured pulse profiles  but it also requires corrections for chromaticity  to be accurate to the same order or better. 
Chromatic effects are primarily interstellar in origin but are also caused by processes intrinsic to the pulsar emission. 
Evidence for the stochastic GW background emerged from measurements with $\mu$s $\TOA$ precision but further characterization of its spectrum and detection of continuous wave (CW) signals from (e.g.) supermassive black hole binaries (SMBHBs)
requires significant improvements in \TOA\ measurements. 
}
Achieving this  precision brings into play all effects that influence arrival times, from processes interior to the neutron star and its magnetosphere,  to plasma effects along the entire propagation path, and  to the digitized data streams that are analyzed.  Improving timing sensitivity to GWs and for other purposes (e.g. more stringent contraints on theories of gravity)   similarly requires attention to all of these end-to-end effects. 

MSPs provide the timing precision for  GW detection that `canonical pulsars' (CPs)
cannot. CPs,  with surface magnetic fields $\sim 10^{12\pm 1}$~G and periods $P\gtrsim 10$~ms),  show  much larger spin noise
\newtext{\citep[e.g.][]{10.1111/j.1365-2966.2009.15938.x}}
(and pulse shape variations 
\newtext{
\citep[][]{2010Sci...329..408L}
})  
that translate into  pulse phase  variations up to $10^6$ times greater than for MSPs
\newtext{\citep[][]{sc10,2017ApJ...834...35L,2019MNRAS.489.3810P}}.
Nonetheless, CPs are useful for probing the ISM and may present unique opportunities for probing high-value targets,  such as a black-hole pulsar binary (including pulsars orbiting Sgr\,A*). 

\newtextR{Pulsar timing methodology is summarized in {\it The Handbook for Pulsar Astronomy} \citep[][Chapt.~8]{lorimer2005handbook}.
Essential features of pulsar timing are given in 
Sect.~\ref{sec:gist} for the purposes of} this paper, 
which inventories  processes that limit timing efficacy and explores methods for its improvement.   It is therefore both a review and a forward-looking analysis aimed at reducing systematic and random timing errors for individual pulsars,
\newtext{selecting pulsars that yield the lowest astrophysical contributions to timing errors,
and striving for optimization of timing programs, including those for PTAs.
}
We present scaling laws  for
forecasting the measurement precisions of astrophysical parameters.  These provide the means for testing the completeness of our understanding of timing delays as well as for identifying areas of improvements, including the benefits of new telescopes and instrumentation.   The paper builds on a number of works that have characterized arrival time variance on an empirical basis.  References are given in individual sections of the paper that compare  magnetars with 
CPs and MSPs.   Pulsar timing array collaborations have presented the most detailed studies of MSPs.   Though our presentation makes extensive use of data from the NANOGrav collaboration\footnote{\url{https://nanograv.org}},  we 
\newtext{
also use or refer to }
other data sources and the results are intended to be independent of any particular PTA group. 

Optimization of pulsar timing requires assessment of each pulsar's quality as an astrophysical clock, including the intrinsic properties of the neutron star and its magnetosphere, any binary motion, its  immediate environment, and the plasmas encountered along the line of sight (\LoS\ ).    Also relevant is our ability to measure pulse times of arrival (TOAs) precisely and refer them to the solar system barycenter accurately. 

By its focus on timing limitations and optimization,  the paper does not delve into the optimization of model fitting to long arrival-time sequences nor to optimized inference using frequentist or Bayesian methods.   Instead the paper is meant to inform how to construct covariance matrices employed for parameter inference and to minimize the values of covariance matrix elements.

\startlongtable
\begin{deluxetable*}{lcccclll}
\tablecaption{\label{tab:overview} Overview of timing delays and \TOA\ estimation }
\tablehead{ 
\colhead{Contribution} & 
\colhead{Effect\textsuperscript{a}} & 
\multicolumn{2}{c}{\quad\quad Mitigation\textsuperscript{b,c}} &
\colhead{Amplitude\textsuperscript{d}} &
\colhead{RF} & 
\colhead{Fluctuation} &
\colhead{Paper} \\ [-0.7cm] 
\\ 
 & &
\colhead{Type}  &
\colhead{Instr} & 
&
\colhead{Chromaticity\textsuperscript{e}} &
\colhead{Signature\textsuperscript{f}} 
&
\colhead{Section}
}

\startdata
{\bf Source \& Source Environment} & & & & \\
\phantom{0.5in}Spin noise    	& $\dt$ 	& triage 	& $\cdots$ 	& $\sim 100$\,ns -- s		& \quad a                 & R/PL ($f^{-4}$--$f^{-6}$) 	
		& \ref{sec:spin}, \ref{sec:spin_triage} \\
\phantom{0.5in}Pulse shape  	& $\dt$ 	& triage 	& $\cdots$ 	&$\mu$s -- ms          		& \quad  c \ ($\nu^{-0.3}$)& Err 	
		& \ref{sec:pulse_shapes}, \ref{sec:fullstokes} \\
\phantom{0.5in}Pulse jitter 	& $\that$  & NM 	& $T$ 		& $\lesssim \mu$s -- ms 	& \quad  a (c)                & W 		
		& \ref{sec:jitter}\\
\phantom{0.5in}Orbital noise 	& $\dt$	& triage	& $\cdots$ 	&$\lesssim$ ms        		& \quad  a                 & R/PL ($\sim f^{-3}$)	
		& \ref{sec:orbnoise}  \\ 
\hline 
{\bf Interstellar Medium} 		& & & & \\
\phantom{0.5in}Dispersion 	& $\dt$		& C				& $B$		& $\lesssim$ ms 	    &   \quad  C \ ($\nu^{-2}$) & R/PL,LP $r^{-8/3}$ 
		&   \ref{sec:DMt}, \ref{sec:DM_analytics} \\
%
\phantom{0.5in}Multipath [$\DM(\nu)$ 
						& $\that, \dt$	& PC, NM			& $\nu$, $B$	&   $\lesssim 100$\,ns & \quad  C ($\nu ^{-23/6}$)  & R/LP 
		& \ref{sec:DMnu}\\ 
\phantom{0.5in}Faraday rotation  
						& $\dt$		& C				& Pol cal 		& $\lesssim$ ns 	     &   \quad  C \ ($\nu^{-3}$)  & Lin  
		& \ref{sec:prop} \\
%
%
\phantom{0.5in}Pulse broadening 
						& $\that, dt$	& PC, NM, triage	& $\nu$			& ns -- s 		     &  \quad  C \ ( $\sim \nu^{-4}$)  & R/PL,LP 
		& \ref{sec:pbf_gallery} \\
%
\multicolumn{3}{l} {\phantom{0.5in}{Interstellar intensity scintillation (S/N variations)}}\\
\phantom{0.5in}
\hspace{0.3in}Diffractive  		& $\that$		& NM	& $B, T$	 & $\lesssim \mu$s -- ms 	&  \quad  C   & R/LP 
		&  \ref{sec:toahat_eval}, \ref{sec:sclaws}  \\
\phantom{0.5in}
\hspace{0.3in}Refractive   		& $\that$		& NM	& $\nu$	& $\lesssim \mu$s 		& \quad  C  & R/LP  
		& \ref{sec:toahat_eval}, \ref{sec:sclaws}  \\
%
\phantom{0.5in}Angle of Arrival &  $\dt$		& PC 	& $\nu$	& $\lesssim \mu$s 		&  \quad  C \ ($\nu^{-2}$ \& $\nu^{-4}$) & R/PL,LP 
		&  \ref{sec:angular}, \ref{sec:aoatoa} \\
%
%
%
\hline 
{\bf Instrumentation \& Analysis} & & & & \\
\phantom{0.5in}Astrometric \& clock errors
						& $\dt$ 		& C 		&		& $\lesssim 10$\,ns   &  \quad  a & Err 
		& \ref{sec:astrometry} \\
\phantom{0.5in}SSBC uncertainties 
						& $\dt$		& PC, NM	& Ephem	& $<$100\,ns  	&  \quad  a & Osc 
		& \ref{sec:ttransfer}  \\
%
\phantom{0.5in}Polarization 	& $\dt$		& C		& Pol cal	& $\ll \mu$s  		& \quad  C  & Err & \ref{sec:polcal}    \\
\phantom{0.5in}Radiometer noise\textsuperscript{g} 
						& $\that$		& NM	& $B, T$	&   $< \mu$s -- ms 	&  \quad  c \ ($\nu^0 \rightarrow \nu^{-2.7}$) & W 
		& \ref{sec:toa_estimation} \\ 
\hline
{\bf Gravitational Waves} 		& $\dt$ 		& None\!	&		&  $\lesssim 100$ ns & \quad  a & R/PL, Osc 
		& \ref{sec:intro} \\
%
\enddata
\tablecomments{\\
\textsuperscript{a} Effect:  $\that$ signifies an influence on \TOA\ estimation while $\dt$ implies contribution of a systematic   \TOA\  offset that is epoch dependent; \\
\textsuperscript{b} Mitigation type: triage = PSR choice; C = correctable; PC = partially correctable; NM = include in noise model; \\
\textsuperscript{c} Instrumental mitigation:   $\nu$ = radio frequency;  $B$ = bandwidth; $T$ = integration time per epoch; \\
 	\phantom{0.5in}Pol cal = polarization calibration; Ephem = solar system ephemeris \\
\textsuperscript{d} Amplitude of arrival time delay or timing residual either before or after mitigation (as relevant); \\
\textsuperscript{e} Dependence on radio frequency ($=\nu$):  \\
 	\phantom{0.5in}a = achromatic; c $=$ weakly chromatic, C $=$ strongly chromatic  with approximate scaling (if simple); \\
\textsuperscript{f}  Signature of fluctuation spectrum in long time series (yr):  \\ 
	\phantom{0.5in}W $=$ white; R/PL $=$ red with power-law spectrum; R/LP $=$ red with $\sim$\,flat low-pass spectrum;  \\
	\phantom{0.5in}Lin $=$ typically a linear spectrum, \newtext{Osc = oscillatory, Err = erratic or episodic} \\
\textsuperscript{g} The \TOA\ error from radiometer noise depends on the flux density spectrum of the pulsar, which typically scales as $\nu^{-y}$,  with $y\sim 0$\,to 4.
 }
\end{deluxetable*}

Table\,\ref{tab:overview} gives a broad overview of  salient contributions to  arrival times and timing residuals --- from neutron star to digital processing. 
These are grouped into effects that take place in and near the source, those involving propagation through the ISM and other plasmas, and local (solar system) and instrumental delays.   Some of these contribute systematic 
\newtext{
biases
}
\newtext{or random errors}
 (or both) that  are achromatic (or nearly so) while others, especially plasma effects, are highly frequency-dependent.   A wide range of temporal variability is also seen, ranging from uncorrelated perturbations between epochs (white noise) to red noise processes with steep power law spectra $\propto f^{-1}$  to  $f^{-6}$.   Timing residuals include both Gaussian and non-Gaussian 
\newtext{astrophysical} 
processes, the latter from 
\newtext{
processes internal to the neutron star,
} 
magnetospheric effects,  and the ISM.  The table indicates whether an effect contributes to a 
\newtext{
random
} 
error in   a \TOA\ estimate $\that$ or instead introduces a systematic arrival time  offset 
\newtext{
(or bias) 
}
$\delta t$.  The table also indicates how a contribution can be mitigated, where `triage' simply means that only some pulsars warrant inclusion in a timing program.   Representative  amplitudes of timing errors or offsets are given along with their dependence, if any, on radio frequency
$\nu$.   The signature in the power spectrum of timing residuals is given in the penultimate column.  Relevant paper sections are indicated in the last column, which serves as a rough roadmap for some aspects of the review.

\clearpage
The text is organized into seven major parts followed by twelve appendices that contain definitions and derivations:
\begin{list}{Part}{
    \setlength{\topsep}{0pt}
    \setlength{\leftmargin}{1.cm}
    \setlength{\itemindent}{0pt} 
    \setlength{\itemsep}{2pt}}
  \item I.   \textit{Preliminaries} 
  (Sects.~\ref{sec:intro}--\ref{sec:nomenclature}). 
  This includes the Introduction and a short discussion of nomenclature.
  \item II.  \textit{Basics of pulsar timing}  
  (Sects.~\ref{sec:gist}--\ref{sec:instru}). 
  The foundations of pulsar includes descriptions of  pulsar signals and their propagation through the interstellar medium (ISM)  along with a presentation of template matching for \TOA\ estimation. 

  \item III. \textit{Interstellar effects}:
   (Sects.~\ref{sec:ism_scalings}--\ref{sec:chromatic_toas}).
   A summary of what we know about the interstellar plasma is presented following detailed analysis of its effects on individual arrival times.  
  \item IV.  \textit{Chromatic leakage, gaussianity, fluctuation spectra, and red noise}
  (Sects.~\ref{sec:chromatic_leakage}--\ref{sec:red_noise_assessment}).
  The phenomenology of aggregated, multiepoch arrivaltimes is discussed  in the time domain, the radio frequency domain, and the fluctuation-frequency domain. 
  \item V. \textit{Timing equations and mitigations}
  (Sects.~\ref{sec:summary_toa_model}--\ref{sec:toacorrections}).
  Model timing equations and the mitigation of \TOA\ errors are discussed. 
  \item VI. \textit{Prioritizing MSPs for optimization of pulsar timing arrays}
  (Sects.~\ref{sec:spin_triage}--\ref{sec:Q}).
  Previous material is employed for selecting the best MSPs for PTA applications.
  \item VII. \textit{Summary and Recommendations} (Sect.~\ref{sec:summary}). 
  The salient elements of our analysis are summarized by considering how a pulsar timing array might be assembled \textit{ab initio} but taking into account timing phenomenology and methodology covered in the review.    
\end{list}

\section{Nomenclature}
\label{sec:nomenclature}

Discussions of pulsar timing necessarily involve an abundance of symbols and acronyms.   

We use standard nomenclature for electromagnetic frequency $\nu$ and its relationship with vacuum wavelength $\lambda = c\nu^{-1}$ where $c$ is the speed of light.    

\newtext{We use   $f$ to designate ``fluctuation frequencies'' of various quantities, including gravitational waves. 
Fluctuation frequencies are the independent variables for power spectra of time series; for our applications time series extend over multiple years and we refer to their power spectra  as `fluctuation spectra' with frequencies expressed in Hz or cycles\,yr$^{-1}$.

We use  $f_{\rm s}$ for the pulse rates of pulsars, which are closely related to the  spin frequencies of the associated neutron stars (modulo the usually unknown Doppler shift).

We refer to pulsars using standard notation, usually in terms of the J2000 epoch (e.g. J1713+0747) and, for a few well-known objects, the B1950 epoch (e.g. B1937+21).  Where needed we use the PSR, MSP, or CP prefixes (pulsar, millisecond pulsar, and `canonical' pulsar) where CP refers to pulsars that have not undergone accretion-driven spinup (``recycling'').}

Many more definitions are needed for modeling the pulsar signal and for its
propagation through the ISM. 
A comprehensive list is given in Appendix\,\ref{app:nomenclature}.

\section{The gist of pulsar timing and its limitations}
\label{sec:gist}

\newtext{
This section summarizes  the foundations of pulsar timing methodology and the range of factors that contribute to its limitations.   The main intent here is to classify those factors and set the stage for the detailed analysis that follows.
}

Pulsar timing amounts to counting pulses over time spans ranging from minutes to decades, taking into account that they  are only quasi-periodic and pulse frequencies are nonstationary due to spindown of NSs.  
\newtext{
Also, pulses  arrive early or late depending on Earth's location in its orbit and similarly, depending on orbital phase for objects in binary systems.   Most pulsars are isolated but about 2/3 of MSPs are in binary systems with white dwarf companions \citep[ATNF pulsar catalog,][]{ATNFPulsarCat}.}

These deterministic effects are captured in a model for pulse phase, $\Phi(t)$ that parameterizes the NS spin, orbital elements, and astrometric terms.  A perfect model yields integers at measured arrival times but
unmodeled effects 
\newtext{can appear as random and systematic errors in measured \TOAs\ 
and/or to phase residuals,
}
yielding $\Phi(t) = {\rm integer} + \delta\phi(t)$.

Timing methodology  relies on an empirical foundation that has unfolded over the last half century:
\vspace{-0.1in}
\begin{enumerate}
\itemsep -0pt
\item Rotational stability of MSPs is sufficient to allow individual turns of a NS to be predicted with high precision so that only infrequent monitoring is needed to track pulse phase and to look for departures caused by astrophysical effects of interest, including plasma and gravitational perturbations from both static fields and GWs;  
\item Emitted pulses are cross-sectional slices through a beam of radiation whose orientation is locked to the spinning NS crust (in the mean);
\item Relative to a fiducial phase locked to the NS, the phases  (and amplitudes) of single pulses  vary stochastically with stationary statistics; as a result, the average pulse profile obtained from a large number of single pulses \newtext{averaged} synchronously with the rotation converges to a stable form that \emph{largely} does not change with time;
\item 
Average profiles change smoothly as a function of radio frequency; however, these chromatic shapes change very little from epoch to epoch.  
\newtextR{
However, notable exceptions occur, as discussed in Sect.~\ref{sec:shape_shifters} and
Sect.~\ref{sec:intrinsic_chromatic}.
}
\end{enumerate}

\begin{figure}[!htbp]
    \centering
    \includegraphics[width=\textwidth]{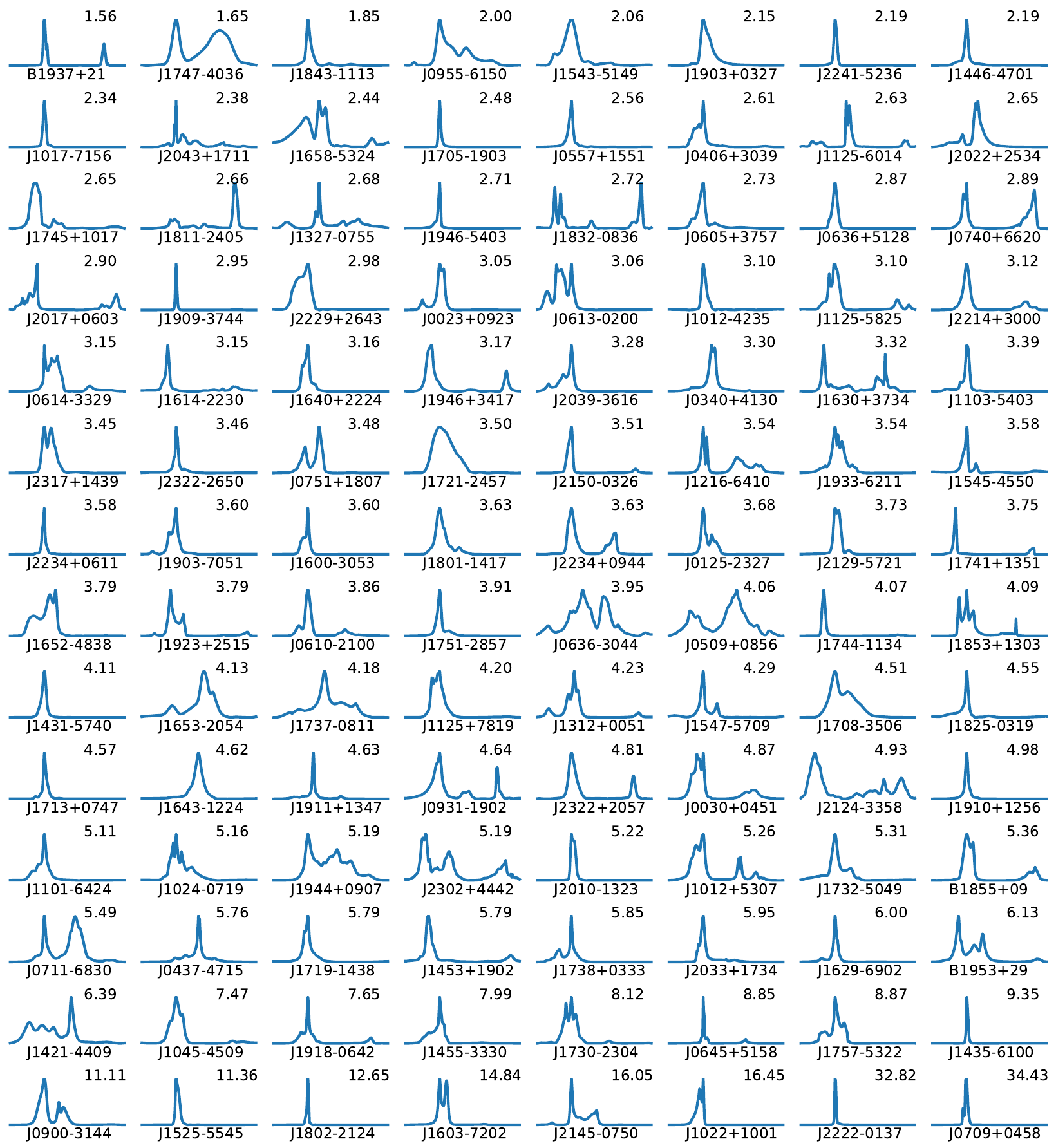}
    \caption{Pulse profiles  (total Stokes intensities) at L-band for \newtext{112 pulsars timed by the IPTA}. Profiles are shown for a full phase cycle and are ordered by increasing pulse period from left to right. The pulse period is shown in milliseconds at the top right of each profile. 
    }
    \label{fig:pulse_gallery}
\end{figure}

As examples, pulse profiles 
\newtext{
for 112 MSPs from multiple PTA programs that contribute to the International Pulsar Timing Array program
}
\newtextR{
\citep[e.g.][]{2019MNRAS.490.4666P}}
are shown in Fig.\,\ref{fig:pulse_gallery}.  Each is the average of more than a million single pulses and represents the shape to which all  averages from a given pulsar are assumed to converge for template fitting.  
\newtextR{While actual profiles closely conform to this assumption, deviations occur from both intrinsic (to the pulsar) and extrinsic processes.}

\begin{figure}[h!]
    \centering 
    \includegraphics[width=0.8\textwidth]{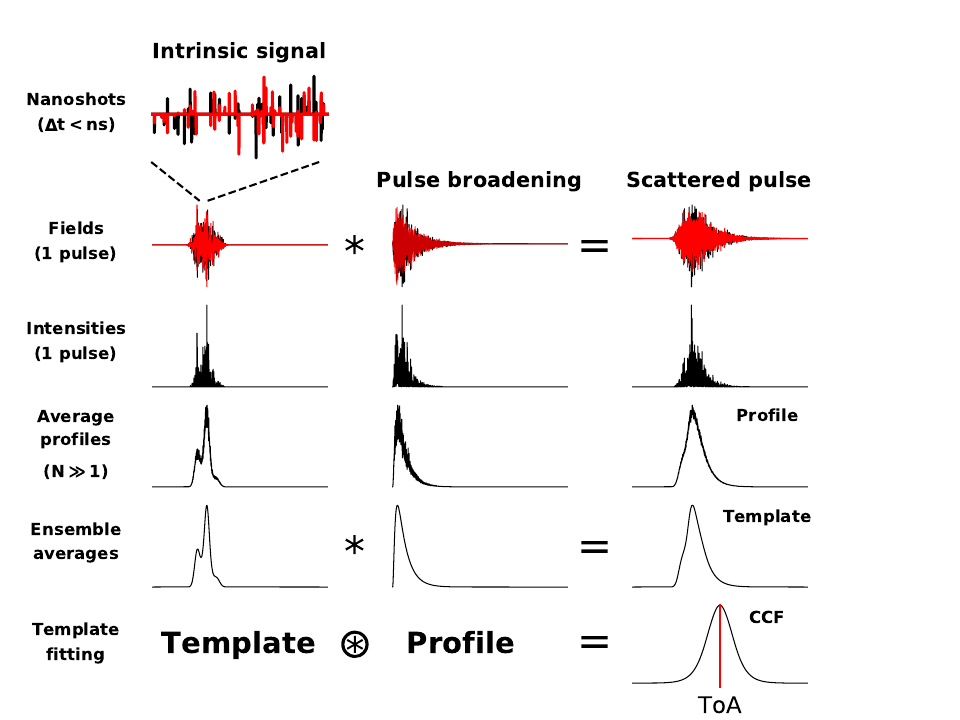}
    \caption{
    Graphical display of the averaging of single pulses to form pulse profiles from which \TOAs\ are obtained 
    using template fitting. The asterisk  and circled asterisk denote convolution \newtext{and cross-correlation} operations, 
    respectively.
\hfil \break
\textit{Columns:}  
    (1) the intrinsic (emitted) signal; 
    (2)  interstellar broadening functions;  and
    (3) the net scattered pulses.  
\hfil \break
\textit{Rows:} 
    (1)  nanoshots,  the fundamental units of coherent emission;
    (2)  field or `voltage' quantities   for a single pulse;
    (3)  intensity quantities equal to the squared magnitudes of row (2);
    (4)  average intensities for a large number of pulses;
    (5)  ensemble averages; and
    (6)  calculation of \TOAs\ by  fitting a template to an individual profile. 
\hfil  \break
 Nanoshots and field quantities are complex if quadrature mixers  are used to obtain baseband voltages; real  and imaginary parts are shown in black and red, respectively.  
 In practice,  average profiles combine two independent polarizations to form the intensity 
 Stokes parameter.   Template fitting is typically implemented by least squares fitting in the Fourier domain. 
 Formally, this is identical to using the maximum  of the cross correlation function (CCF) to determine the \TOA\ but it  has significant practical advantages.   
    \label{fig:shots2pulses}
    }
\end{figure}

Fig.~\ref{fig:shots2pulses} shows the simulated properties  of single pulses and how they are averaged into profiles for calculating arrival times.   Emitted pulses are inherently stochastic.  
\newtext{
As discussed in Sect.~\ref{sec:signal_models},
}
they comprise narrow shot pulses of coherent radio emission.  Shots are grouped incoherently to form single pulses, which can be viewed statistically as an envelope that modulates shots to  give the instantaneous pulse shape.  Pulses incident on  the ISM are broadened by multipath propagation from scattering, which operates as a linear filter.  The convolution of the  broadening function  is with the electromagnetic field components of the pulsar signal and pulse intensities (after squaring field components) are averaged synchronously with the pulse period  following removal of systematic dispersion delays.   The convolution operation does not apply to single pulse intensities  but it does apply to the (hypothetical) ensemble average pulse.  Template fitting is essentially the outcome of cross correlation between a template and a measured profile, as indicated, although it is usually effected in the frequency domain after Fourier transformation  \citep[][and Sect.~\ref{sec:toa_estimation}]{1992RSPTA.341..117T}.

The \TOA\ of a pulse is the net effect of many processes that occur between emission in a pulsar's magnetosphere and reception  at an observatory, with subsequent 
 referencing of the \TOA\ to the solar-system barycenter (SSBC) using a solar-system ephemeris.  GW perturbations contribute to  the overall noise budget as well as to the signal that is the target of PTA observations.

Once \TOAs\ are obtained,   referencing them to the SSBC requires knowledge of astrometric parameters of the pulsar (sky location, proper motion, and parallax), determinable by fitting to a multiyear sequence of \TOAs\ or through very-long-baseline interferometry (VLBI).  An issue with using sky positions from VLBI is that the  International Celestial Reference Frame (ICRF) based on equatorial coordinates differs from the dynamical reference frame based on Solar System ephemerides.  A discussion of reference frame alignment can be found in 
\citet[][]{2026ApJ...999..156S}.

To elucidate the complexity of \TOAs, 
we relate the emission time $\tem$ from  a location $\xem$ in a pulsar's magnetosphere to
the topocentric arrival time $\tA$ at an observatory location $\xobs$ ( 
\newtext{
(with  measurement errors or other perturbations implicit),
}
\be
\tA[\xobs(\tA), \nu)] = \tem[\xem(\tem, \nu) ] + \tprop({\rm ISM, \phi_{grav}}, \nu) + \tgw(\tA),
\ee
where $\tprop({\rm ISM, \phi_{\rm grav}}, \nu)$ lumps together the   plasma and (non-GW) gravitational delays  associated with propagation to the observatory;   $\tgw$ is the gravitational perturbation.  Chromatic contributions to $\tA$  are signified by the  dependence on radio frequency $\nu$.    Referral to the SSBC involves compensation for the R\"omer  and relativistic clock delays by adding a term  $\Delta t_{\rm E, SSBC}$ to $\tA$, yielding
$t_{\rm SSBC} = \tA + \Delta t_{\rm E, SSBC}$.


\subsection{What limits  arrival time precision and  timing predictions?}
\label{sec:limitations}

\TOAs\ are obtained by template matching to 
\newtext{measured}
average profiles \newtext{under the assumptions of matched filtering (see Sect.~\ref{sec:pulse_shapes})}.   \newtext{Average profiles} deviate from the template {\it at minimum} 
\newtext{
due to}
radiometer noise that adds to the pulsar signal.  
If \TOA\ precision were determined solely by the limitations from radiometer noise,  it would be feasible to achieve sub-ns precision for the brightest pulsars\footnote{A notable example is the bright MSP J1939+2134 (B1937+21), with $P = 1.56$\,ms, $S_{0.4\, \rm GHz}  = 240$\,mJy, and $W = 38\,\mu s$
\citep[][]{ATNFPulsarCat}.  Expressions and quantities detailed in later sections 
(e.g. Eq.~\eqref{eq:rmstoaS_gaussian}) are evaluated  with an integration time of $10^3$\,s and bandwidth of $0.1$\,GHz on a large highly optimized telescope with $\Ssys = 1$\, Jy.   For example $\Ssys \sim 1.24$\,Jy at 1.4\,GHz for the FAST telescope
\citep[][]{xi_peng_staveley-smith_for_liu_2022}, though it does not operate currently at 0.4 GHz), the \RMS\ timing error would be $\rmstoaS \simeq 0.04$\,ns.   At  1 to 2 GHz,  the lower flux density $S_{1.4\, \rm GHz}  = 14$\,mJy still yields $\rmstoaS \simeq 0.2$\,ns that is much smaller than actual \TOA\ precisions.  }.
However, the real world intrudes  on predictions based on  idealized matched filtering. 

\textit{Precision} commonly refers to the repeatability of measurements in contrast to  {\it accuracy}, which refers to the fidelity of measurements to the true values of some quantity.  Arrival times are affected by a slew of systematic and random errors but  their true values  are unknown.   We must rely on the self consistency of measurements with underlying physics.   Binary pulsars are the only examples in the pulse timing context where the relativistic descriptions of  their orbits can provide multiple constraints on  measurement  (and model) fidelity.   Essentially all other timing contributions lack any ground truth at the sub-100\,ns level. 

In our analysis, we distinguish between phenomena that affect the time-tagging of pulses --- TOA estimation --- and  the astrophysics of the pulsar clock.   
\newtext{
This distinction is exemplified by considering a delta-function pulse with infinite S/N emitted at a random time by a pulsar with a stochastic spin rate and propagated through the stochastic ISM. The \TOA\  of the impulse at an observatory with a perfect clock would have no uncertainty. However, the timing residuals from a model fitted to a sequence of such \TOAs\ would be affected by the stochastic elements and would have substantial variance.
}

\TOA\ estimation errors  are  often dominated by radiometer noise but in other cases by sometimes  larger effects, namely {\it pulse jitter} intrinsic to pulsars that distorts pulse profiles from the template shape and  
\newtext{{\it diffractive scattering}, which causes intensity scintillation and also causes changes in pulse shape (Sect.~\ref{sec:chromatic_toas})}.

Groupings of terms are
shown schematically  in Eq.~\eqref{eq:toassbc}, including  gravitational wave perturbations comprising two terms associated with the dimensionless strain at the Earth and at the pulsar.
\be
t_{\rm SSBC} &=&
\rm
\bluehighlight{
\text{\parbox{5cm}{\centering{\large\bf Deterministic terms} 
    \\ \hfil spindown polynomial
    \\ astrometric terms 
    \\ pulsar orbit
    \\pulse shape vs . $\nu$
    }}
}
+
\redhighlight{
\text{\parbox{4cm}{\centering{\large\bf Stochastic terms} 
    \\ white noise
    \\ (receiver,  jitter \& ISM)
    \\ red noise
    \\ (spin, ISM
    }}
}
+
\redhighlight{
\text{\parbox{4cm}{\centering{\large\bf Systematic Errors} 
    \\ planetary emphemeris
    \\ time transfer (GPS)
    \\ observatory clocks
    \\ polarization calibration
    }}
}
\nonumber \\
\nonumber \\
&+&
\yellowhighlight{
\text{\parbox{9cm}{\centering{\large\bf Gravitational Wave Perturbations} 
    \\
    Earth term + Pulsar term 
    }}
}
\label{eq:toassbc}
\ee

Several  broad questions arise when assessing the utility of an MSP for precision timing:
\vspace{-0.1in}
\begin{enumerate}
\itemsep -3pt
\item How stable is the spin rate of a NS?
\item How well  are emission regions and directions locked to the spin phase of the NS?
\item How are pulses modified and delayed by the ISM?
and
\item How well can  pulses be time tagged at an observatory and referred to the barycenter of the solar system?
\end{enumerate}
The first two issues relate to the quality of a NS as a clock while the  last two involve determination
of when pulses arrive as  measured against an observatory clock.
Related to these questions is the distinction between pulse phase jitter and red spin noise. Phase jitter
is a zero mean process with respect to a fiducial  point in the pulsar magnetosphere and is physically associated with variations in location and relativistic beaming.  Spin noise, however, represents actual departures in the rotational phase from that of a  smoothly spinning down object.

\providecommand{\Spave}{\langle S_{\nu}\rangle_P}
\providecommand{\Sphiave}{\langle S_{\nu}\rangle_\phi}
\providecommand{\npol}{n_{\rm pol}}
\providecommand{\Tint}{T_{\rm int}}
\providecommand{\Spsr}{S_{\rm psr}}

\providecommand{\fspin}{f_{\rm spin}}
\providecommand{\fdotspin}{\dot {f}_{\rm spin}}

\renewcommand{\arraystretch}{0.8}

\clearpage
\startlongtable
\begin{deluxetable}{l l c l  c}
\tablewidth{450pt}
\tablecaption{\label{tab:keyequations} 
Salient Equations for  arrival time errors 
\newtext{and fluctuations}
\tablenotemark{\footnotesize \!a}}
\tablecolumns{3} 
\tablehead{
    \colhead{Term} & 
    \colhead{Symbol} & 
    \colhead{Nominal Value} & 
    \colhead{Scaling} & 
    \colhead{Equations}  \\[-0.6cm]  \\
   && \colhead{(ns)} \\[-0.6cm] 
}
\decimalcolnumbers
\renewcommand{\arraystretch}{0.5}
\startdata
\\
\multicolumn{2}{l}{
{\bf  Rapid TOA Variations} ($\lesssim$\, hours)
} \\ \\
\quad\quad 
    Radiometer noise &
    \quad\quad
    ${\displaystyle \rmstoaS }$ & 
   $ 15 $ & 
    \quad 
    ${\displaystyle  \frac{\Ssys(\rm Jy)}{\Spsr(\rm mJy)} 
    \frac
        {W_{100}^{3/2}}
        {\sqrt{(\npol/2) B_{\rm GHz}  {\Tint}_{,3}  P_{\rm ms}  }} }$  &
       \eqref{eq:rmstoa_mf_t_tel_phi},  \eqref{eq:rmstoaS_gaussian},  \eqref{eq:rmstoaSQ}
     \\
\\    
\quad\quad 
    Pulse jitter  &
    \quad\quad
	${\displaystyle \rmstoaJ }$ &
	$    33$ & 
      \quad
      ${\displaystyle 
      W_{100} 
     \left(\frac{\FJ}{1/3}\right) 
     \left( \frac{P_{\rm ms}}{{\Tint}_{,3}} \right)^{1/2} }$  &
     \eqref{eq:rmstoaJ}
        \\
\\
\quad\quad 
         Diffractive scintillation &
         \quad\quad
         ${\displaystyle \rmstoaDISS \simeq  \frac{\taud}{\sqrt{\Nscint}} }$ &
         $  10  $ & 
         \quad
         ${\displaystyle 
         \left(\frac{100\,{\rm MHz}}{\dnud}\right) 
         \left(
             \frac{\dtd}{T}\frac{\dnud}{B}
         \right)^{1/2} }$ & 
         \eqref{eq:rmstoaDISS2}
          \\
\\
\multicolumn{2}{l}{
{\bf Slow TOA Variations} ($\gtrsim $\, days) 
} \\  \\
\quad\quad  
	Spin / torque noise &
    \quad\quad
        ${\displaystyle \rmstoaSPIN }$ & 
	$   400 $  &
	\quad
	${\displaystyle 
     P_{\rm ms}^{-0.8} \Pdot_{-20}^{1.1} (T / 10\,{\rm yr})^{2}  }$  &
    \eqref{eq:rmstoaSPIN}
	\\
\\ \quad\quad 
	Scattering variability vs. epoch & 
    \quad\quad
	${\displaystyle \sigma_{\taud(t)}}$ & 
	$ 33$ &
	\quad
	${\displaystyle 
    (\DM / 30\,\DMunits)^{1.5} \nu_{\rm GHz}^{-4.4}}$  & 
	\\
\\ \quad\quad 
         DM(t)  mis-estimation  & 
         \quad\quad
         ${\displaystyle \rmstoaDMt }$ &
         $ 2.2$ &
        \quad
         ${\displaystyle \nu_{\rm GHz}^{-1} \, \left(\frac{3R^2/4}{R^2-1} \right)     \left[ \frac{\Delta T (\rm d)}{\dtd(\rm hr)}  \right]^{5/6} }$ &
         \eqref{eq:dtinfty_async4}
         \\
\\       \quad\quad 
         DM($\nu$) mis-estimation &
         \quad\quad
         ${\displaystyle \rmstoaDMnu}$ &
         various &
         \quad
         ${\displaystyle  \propto 
         \nu_{\rm GHz}^{-23/6} d^{5/6} 
         \Itinfty(R)^{1/2}\SM }$ &  
         \eqref{eq:Var_tinfty} 
         \\
         &&&
         ${\displaystyle 
        \quad\quad \sim \frac{\rm 2.5 \, ns}{\nu_{\rm GHz}^{1/6}} \, \left(\frac{\dnud}{100\, \rm MHz} \right)^{-5/6}}$ 
         \\
\\
\enddata
\tablenotetext{a}  {Definitions:
        \begin{enumerate}
        \item
	\textit{Telescope and observation parameters (default values):} 
	\begin{itemize}
    \vspace{-0.05in}
	\itemsep -1pt
		\item[]
		 $\Ssys$ = system equivalent flux density (SEFD) (1\,Jy)
		 \item[]
		 $\npol$ = number of polarization channels (2)
		 \item[]
		  $\nu$ = center frequency (1 \,GHz)
		 \item[]
		  $B $ = bandwidth (1\,GHz) 
		\item[]
	      		$R > 1$ = frequency ratio of two frequencies used to estimate DM (2)
		 \item[]
		  $\Tint = 10^3\,{\rm s} \times {\Tint}_{,3}$ integration time to obtain a TOA at a single epoch  
        \end{itemize}
	\item
	\textit{Pulsar and line of sight parameters:}
	\begin{itemize}
    \vspace{-0.05in}
	\itemsep -1pt
		\item[]
	         $\Spsr$= period averaged flux density  (1 mJy) 
	    \item[]
	         $P$ = spin period  (1 ms)
	    \item[]
	         $\Pdot_{-20} $ = period derivative in units of $10^{-20}\,$s\,s$^{-1}$
	    \item[]
	         \newtext{$W_{100}$ = pulse width (FWHM)  in units of 100\,$\mu$s}
		\item[]
	         $\DM$ = dispersion measure  in units $\DMunits$
        \item[]
	         $\SM$ = scattering measure  in units $\SMunits$
        \item[]
             $\Itinfty(R)$ = dimensionless factor for calculating frequency-dependent DM
	    \item[]
	         $v_{100}$  = effective transverse velocity  ($\vperp / (100\, {\rm km \, s^{-1}})$)
		\item[]
			$\dnud$ = scintillation bandwidth (100 MHz)
		\item[] 
			$\dtd$ = scintillation time (1 hr)
	    \item[]
            $\Nscint$ is the number of independent scintillation fluctuations (`scintles') included in an observation
    \end{itemize}
	\item
	\textit{Data sets:}
    \vspace{-0.05in}
	\begin{itemize}
    \vspace{-0.05in}
	\itemsep -1pt
		\item[]
	      		$T_{\rm data}$ = length of data set (years)
		\item[]
	      		$\Delta T$ = difference in epochs for two frequency measurements used to estimate DM (days)
	\end{itemize}
	\end{enumerate}
	}
\end{deluxetable}

The relative contributions of  three effects relevant to measuring arrival times can be found in Table\,\ref{tab:keyequations} along with a prominent subset of the many distinct effects comprising  the overall  timing error budget, namely spin noise and ISM propagation effects.   
Nominal values  are given by the coefficients in column (3)  and are based on typical values of observing parameters (see table footnotes) {\it except} for the telescope sensitivity (the system equivalent flux density (SEFD), $\Ssys = 1$\,Jy), which is an order of magnitude better (i.e. smaller) than for a 100-m class telescope and about half the value for the Chinese Five Hundred Meter Aperture Spherical Telescope (FAST).   
 The first three entries in the table correspond to rapidly changing errors  and are listed in order of (current) importance:
radiometer noise (finite signal to noise ratio), phase jitter (intrinsic to pulsars), and diffractive scintillations from turbulence in the ISM.   The remaining, slowly changing processes alter arrival times due to emission and propagation delays but do not alter the precision of individual \TOAs. 
\newtext{
They are not errors, as such, but rather are fluctuations that affect the net timing residuals.
}

 For MSPs used in PTAs,    short-term timing errors are typically dominated by radiometer noise but a bright MSP  like J1939+2134  is significantly affected by jitter and scintillation.   In Sect.~\ref{sec:noise_vs_jitter} we discuss the crossover between noise and jitter-dominated \TOAs\ and the important consequence that improvements in telescope sensitivity do not benefit \TOA\ precision once  it is jitter or scintillation dominated. 

Another conclusion is that 
the slowly changing processes  listed in the table can also exceed, by a large multiple,  the  \TOA\  precision  obtained in decades-long data sets.   J1939+2134 is again illustrative in this regard because strong red noise from spin or orbital noise completely dominates the overall timing error budget.

The inventory and properties of timing effects in Tables\,\ref{tab:overview}, \ref{tab:keyequations}  and Eq.~\eqref{eq:toassbc} 
suggest a schematic model  for arrival times at the SSBC\footnote{
The form given here is a streamlined version of a detailed expression in Eq.~\eqref{eq:toa_with_dm_scattering2} that encompasses the span of processes analyzed in the paper.},
\be
t_\nu = \tinfty &+& 
	\left[\dtSNR +  \dtJ + \dtDISS\right]_{\rm fast} +  \left[\dtspin +  \dtPBF + \dtDMt + \dtDMnu\right]_{\rm slow}
        \nonumber \\
     	&+& \dtOTHERss + \Delta t_{\rm astr} +  \dtOTHERinst + \dtOTHERISM, 
\label{eq:toaeqn1}
\ee
where $\tinfty$ is the arrival time extrapolated to infinite frequency  that would  be measured if there were no contributions from intervening plasmas and if all deterministic terms (e.g. regular spin phase, orbital delays, dispersion terms, astrometric terms, etc.) were modeled perfectly. \newtext{We consider $\tinfty$ to be the arrival time that contains the signal of likely astrophysical interest (e.g. GWs in the case of PTAs).  For this reason we include spin noise as one of the slow TOA fluctuations. }
Included in $\tinfty$ is the perturbation from GWs and any other gravitational effects, such as lensing from intervening objects.   GW perturbations can be stochastic, oscillatory, or burst like.  GWs passing through the solar system generate correlated perturbations while those passing near pulsars are uncorrelated between pulsars. 
The delays that perturb $\tinfty$ are a mixture of achromatic, mildly chromatic, and strongly chromatic terms, as presented in the paper. 
The next seven terms  adding to $\tinfty$ in the first line of Eq.~\eqref{eq:toaeqn1}  correspond to the items in Table\,\ref{tab:keyequations} and are grouped into
`fast'  and `slow' variations with characteristic times of seconds to hours and days to years, respectively. 
 Time offsets between the topocentric arrival time and the SSBC arrival time are contained in 
 $\dtOTHERss$, which includes the R\"omer, Einstein, and Shapiro delays, \newtext{and astrometric terms are contained in $\Delta t_{\rm astr}$}.  
 Instrumental terms are included in
$\dtOTHERinst$ (observatory clock offsets, variable electronic pathlengths, polarization calibration errors) and
$\dtOTHERISM$, which can include angle of arrival effects due to refraction in the ISM, which couples to errors associated with reference of arrival times to the SSBC.
\newtext{
Detailed timing models for binary pulsars 
can be   found in   \citet[][]{bt76,bh86,dd86} 
and  discussions of time transfer to the solar system barycenter 
are given by (e.g.) \citet[][]{,1990SvA....34..496D,2002ASPC..278..251S, 2006MNRAS.372.1549E}.
}

While the tremendous successes of pulsar timing applications pertain to $\tinfty$,
the focus of this review is on the stochastic terms in Eq.~\eqref{eq:toaeqn1} with the goal of allowing quantitative evaluation  timing accuracy on a term by term basis. As well as being pulsar and telescope dependent, the net results are also a strong function of the procedures used in a timing program. 
For an individual pulsar, 
\newtext{
timing precision is ultimately limited by the S/N and shape of the measured pulse profile (the widths of the narrowest components in particular; see Sect.~\ref{sec:toa_estimation}), 
}
on the spin stability of
the NS, and on stochastic variations from the ISM. 
In PTA work,  overall precision can be maximized by pulsar triage, avoiding pulsars that are dim and show  excessive spin or interstellar noise.   Even then,  white and red noise from finite pulse S/N, low-level spin variations, and ISM effects will contend with GWs from the stochastic background and from individual sources.  In order to cover all possible situations involving
radio timing, we discuss a comprehensive set of perturbations that originate everywhere along the line of sight, from inside the NS to the output of a receiver system. 

Our treatment does not cover `noise modeling' in the sense that it is used in
analyses of collections of arrival times.   That term typically refers to the covariance matrix used in frequentist or Bayesian statistical inference of model parameters.  Instead, our aim is to characterize noise contributions and to  differentiate between those effects that can or cannot be corrected for or otherwise mitigated.   Given the slate of options for possible mitigations, it is clear that noise modeling is necessarily methodology dependent.  Of course any effect that cannot be mitigated should be included in the noise covariance matrix. 

\section{Pulsar spin stability and orbital noise}
\label{sec:spin}

The spindown rate $\fsdot \equiv d\fspin/dt$  of nonaccreting  NS is typically modeled with 
 a power law $\fsdot \propto \fspin^n$, where $n$ is the braking index.
Magnetic dipole radiation corresponds to $n=3$ while $n < 3$ for the few  values estimated from timing measurements, presumably due to departures from dipole fields and induced currents in the magnetosphere \citep[][and references therein]{mel97,2024MNRAS.534.3410V}.    This yields a spindown phase written in terms of the phase, spin rate, and its derivative at fiducial epoch $t_0$, 
\be
\phispin(t) 
	= \phi_0 
	+ \fszero  \tau_0\left(\frac{n-1}{n-2} \right)
	 \left\{ 
	   	\left[1 + \frac{t-t_0}{\tau_0 }\right]^{\frac{ n-2}{n-1} } - 1
	 \right\}
	 \simeq
	 \phi_0 + \fszero (t - t_0) + \frac{1}{2} \fsdotzero  (t-t_0)^2 +  \frac{1}{6} \fsdotdotzero (t-t_0)^3 \cdots ,
\ee
where the ``spindown time'' is $\tau_0 = \fszero / (n-1)\fsdotzero$ 
\newtext{
at the start time $t_0$ of a set of measurements
}
and the approximate equality holds
for $t - t_0 \ll \tau_0$.   This is a good approximation since $\tau_0$ is typically millions of years for CPs
and $>$Gyr for many MSPs.   
\newtextR{
Note, however, that it is well known that the spindown time is not a good approximation to the chronological age of a pulsar. The spindown time  depends on the  spin period at its birth and on any evolution  of the magnetic field since birth.   Pulsar kinematics revealed this discrepancy empirically early on
\citep[][]{1975MNRAS.171..579L, 1985MNRAS.213..613L}
for canonical pulsars.  More recent work on the spinup process for MSPs and a comparison ages of  MSPs with their companion white-dwarfs yields a consistent conclusion 
\citep[][]{2012Sci...335..561T}. 
}
Pulse phase models necessarily include a quadratic or,
\newtext{
for canonical pulsars,
}
a cubic polynomial to account for spindown, causing fits to arrival times to remove low frequency power from other contributions, as discussed later. 

Deviations from the smooth spindown phase   result from  processes occurring both inside the neutron star and   in the magnetosphere.   These comprise distinct events (glitches and microjumps) and spin noise, a sustained process distinct from glitches and more relevant to PTA programs. 

{\it Glitches} comprise rapid spinups $ \dffglitch \sim 10^{-9}$ to $10^{-6}$,
often combined with jumps in frequency derivative $\Delta\fsdot$.  
Glitch signatures are largely $\Delta\fs > 0$ and $\Delta\fsdot < 0$ (or +,$-$) in form, with some exceptions. 
Glitches are more common in young pulsars with large spindown rates ($\fsdot$)  and amplitudes of $\Delta\fsdot$ are correlated with those of $\Delta\fs$; the latter has a bimodal distribution with peaks near $\Delta\fs \sim 10^{-8}$\,Hz$^{-1}$ and  $10^{-5}$\,Hz$^{-1}$ \citep[][]{2022MNRAS.510.4049B}.
Glitch recovery corresponds to relaxation of the spin rate $f_s(t)$ to its pre-glitch trajectory.  In some cases this recovery is  incomplete before the next glitch occurs while in others the recovery is slow or indiscernible.  
Evidently,  glitches are primarily due to transfer of angular momentum from vortices in the neutron superfluid to the NS's crust
\newtext{
\citep[][]{AndersonItoh1975, Eya2017}. 
}

Despite the propensity for glitch activity in 
young pulsars,   small, rare glitch-like events have been identified in three  MSPs:
PSR B1821$-$24 in the globular cluster M28 \citep[][]{2004ApJ...612L.125C},  
J0613$-$0200, an MSP included in PTAs \cite[][]{2016MNRAS.461.2809M},
and J0900$-$3144 \citep[][]{2026MNRAS.545f2034B}. 
These are much smaller in amplitude 
\newtext{
than those from canonical pulsars,
}
with $\dffglitch \lesssim 10^{-11}$ and
$\vert \dfdotfdotglitch \vert \lesssim 10^{-4}$.   The J0613$-$0200 event follows the (+,$-$) signature
of large glitches, in contrast to the (+,+) signature for B1821$-$24.
\newtext{
Only the positive $\dffglitch$ has been measured for the third case
(J0900$-$3144). 
It is unclear if these events have the same physical origin as large glitches.    Fortunately,  MSP events are rare, having occurred only three times  
in over a  thousand pulsar-years in MSP timing programs
\citep[][]{2026MNRAS.545f2034B}.  Nonetheless, this rate implies that small glitches in long-duration MSP timing programs are to be expected. 
}




{\it Spin noise}, as used here, refers to sustained stochastic variations in phase distinct from those caused by glitches \citep[][and references therein]{2026MNRAS.545f2034B}.
Some of these may arise from  internal torque variations from crust-superfluid interactions or crustquakes. The interplay of spin noise and glitches while estimating parameters of a two-component crust-superfluid model is recently discussed by \citet[][]{2024MNRAS.530.4648O}, who employ Kalman filters as part of the modeling.  Other variations are clearly caused by changes in the magnetic torque responsible for spindown \citep[e.g.][]{klo+06}.  An early overview of possible effects was given by \citet[][]{cg81}.    

\subsection{Types of spin noise}
\label{sec:types_of_spinnoise}

\begin{figure}[h!]
    \centering
    \includegraphics[width=0.8\textwidth]{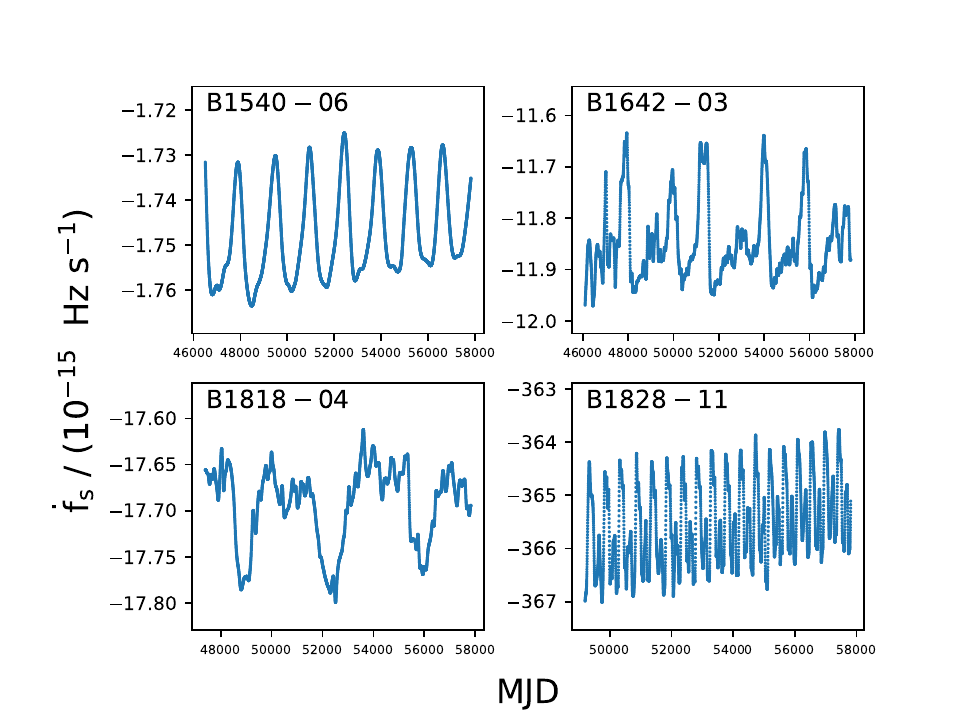}
    \caption{Time series of $\fsdot$ for four pulsars from \citet[][]{2023MNRAS.523.4603K} 
    (data downloaded from zenodo.7664166).
    \label{fig:fsdot_4_pulsars}
    }
\end{figure}

 Empirically, spin noise is manifested as several types of variations in spindown rate $\fsdot$ (Fig.\,\ref{fig:fsdot_4_pulsars}) or in time-correlated timing residuals (Fig.\,\ref{fig:resids_and_sfs}):
\begin{enumerate}
\itemsep -2pt
\item  Slow, smooth variations with no evidence for individual features and a  red spectrum with a steep power law component, $\SR(f) \propto f^{-\xsn}$
with a spectral index $\xsn \sim 4$~to~6 \citep[][]{1997JApA...18....5D}. 
\item  Bandlimited spin noise, similar to the first case, but with a quasi-periodic appearing time sequence in $\fsdot$, signifying a narrower range of frequencies than for a power law spectrum. 
\item Individual events (microjumps) in spin frequency or its derivative that individually have either sign, unlike glitches which largely have $\Delta\nu > 0$ and 
$\Delta\nudot < 0$ 
\citep[CD85;][]{dmh+95}
For the Vela pulsar,  microjump amplitudes in both
$\nu$ and $\nudot$ are about 100 times smaller than glitch amplitudes \citep[][]{1988ApJ...330..847C}. Data spans encompassing many events show consistency with a power law spectrum with slope dependent on whether the spin phase is dominated by microjumps in $\nu$ or in $\nudot$. 
\item Switching between discrete values of spindown rate $\fsdot$ that  persist for weeks to years \citep[e.g.][]{klo+06,2010Sci...329..408L}. 
In some pulsars,  state switching is quasi-periodic
\citep[e.g.][]{2023MNRAS.523.4603K}
and is consistent with stochastic resonance in the CP B1931+24.
Fig.\,\ref{fig:fsdot_4_pulsars} shows quasiperiodicity in $\fsdot$ for four canonical pulsars over time spans of up to 32\,yr.
  Recent work \citep[][]{2025MNRAS.540.2486K} has shown that correlations between profile changes and spindown rate are quite common on time scales of tens to hundreds of days. 
State switching on shorter time scales (seconds to hours), manifested as nulling, profile mode changes, or quantized rates of subpluse drift, appears consistent with Markov processes \citep[][]{2013ApJ...775...47C}.  
State switching so far appears to be a phenomenon occurring primarily 
 in CPs \citep[][]{10.1111/j.1365-2966.2007.11703.x}.
  Nulling, for example,  is either absent in most MSPs \citep[e.g.][]{2014ASInC..13...73R, 10.1093/mnras/stab282} or occurs on too fast a time scale \citep[e.g. MSP B1957+20;][]{Mahajan_2018} to manifest stochastically in timing data.   
  \newtext{
  However, mode changes in the MSP J1909$-$3744 have been reported by \citet[][]{2022MNRAS.510.5908M}.
  }
\end{enumerate}

Red noise in pulsar residuals has
been interpreted in terms of random walk processes \citep[e.g.][]{1972ApJ...175..217B,ch80,1997MNRAS.288..129K}
and analyzed with a variety of time-domain \citep[e.g. CD85,][]{1997JApA...18....5D} and frequency domain methods \citep[e.g.][]{gro75b,db82,chc+11,2013ApJ...766....5S}.  The first two types of spin noise listed above are amenable to spectral methods.  Alternatives to Fourier based methods are necessary because the underlying spectra are steeper than $f^{-2}$, causing spectral leakage and mis-estimation of the spectral index.   Principal component analysis, autoregressive (AR) model fitting (identical to maximum entropy methods), Cholesky decomposition, or Bayesian model fitting avoid leakage effects.   Structure functions are  useful because the spectral index is related to the structure function exponent. 
The third and fourth types of spin noise are best analyzed with time-domain methods if characterization of discrete events is the primary goal. 
Structure functions, for example,  can be employed for testing the reality of apparent events  in residual time series  (CD85).

\subsection{Structure function slopes and spectral indices}
\label{sec:SFs}

\newtext{
Definitions for the variety of spectral representations used in pulsar timing are given in Appendix\,\ref{app:specrep}. 
We make use of structure functions (SFs;  defined and related to power spectra  in Appendix\,\ref{app:SFs}) as tools for inferring the properties of red noise in pulse phase.

We define the spectrum of phase fluctuations between low and high frequencies $\flow$, $\fhigh$ as
\be
S_\phi(f) = S_0 f^{-\xred},  \quad \flow \le f \le \fhigh,
\label{eq:phispectrum}
\ee
where the low-frequency cutoff is assumed to be much smaller than the lowest frequency $\sim 1/T$ sampled in a data span of length $T$.  
The phase variance for the data is then 
the integral of the spectrum over a frequency range $[\sim 1/T, \fhigh]$ 
and phase structure functions 
$\Dphi^{(m)}(\tau)$
of different orders $m$ are similarly related to the spectrum (Sect.~\ref{sec:SFs}).  
}

The best use of SFs requires consideration of the spectral  index $\xred$ in choosing the order $m$.   
The appropriate order  to apply to the phase (or residual) is
$m > (\xred-1)/2$,  
\newtext{
which follows from the integral relation between the spectrum and structure functions defined in Eq.~\eqref{eq:Dmphi} in Appendix\,\ref{app:SFs} and consideration of the interplay between spectral slope $\xred$ and the SF order. 
}
This specification yields an SF  that  is independent of $T$
and has a scaling $\tau^{\xred-1}$ vs. lag $\tau$.  Otherwise the scaling 
is $\tau^{2m} T^{\xred -1 -2m}$, whose $\tau$ dependence is uninformative about the spectral shape. 

Random walk processes are useful models for spin noise because it is physically plausible that discrete jumps in 
\newtext{
spin phase, spin frequency, or spin-frequency derivative (i.e. jumps in torque) 
}
can occur.   Referring to these as  $k^{\rm th}$ order random walks with $k = 1,2,3$ respectively, they correspond to power-law spectra $\propto f^{-2k}$ \citep[][CD85, Table 3]
{1972ApJ...175..217B}.
    
For most pulsar data, the appropriate order of structure function to use for spectral constraints  is $m=3$, for two reasons:  
(a) it is uninfluenced by removal of the quadratic spin phase,   $\phi_{\rm sd}(t) = \phi_0 + \fspin t + \fsdot t^2 / 2$; and
(b) red noise processes are evidently as steep as $\xred = 6$
\newtext{\citep[CD85,][]{1997JApA...18....5D, 2019MNRAS.489.3810P, 2024Univ...10..105Y, 2023ApJ...951L..10A}}
 requiring 
$m > 2.5$.  We note, however, some pulsars show shallower spectra that allow lower-order SFs to be informative.

\subsubsection{Constraints from phase structure functions}
Measured phases  include white noise $\phi_{\rm wn}$ with
variance $\sigma_{\rm wn}^2$ so structure functions of the phase,
$\phi(t) = \phi_{\rm red} + \phi_{\rm wn}(t)$, where the two terms are 
uncorrelated, have the form for $\tau > 0$, 
\be
D^{(m)}_\phi(\tau) = a_m \sigma_{\rm wn}^2  + D^{(m)}_{\phi, \rm red}(\tau),
\label{eq:Dwhite_red}
\ee
where $a_1 = 2$, $a_2 = 6$, and $a_3 = 20$ 
(CD85, Table 3). 
If red noise dominates the white noise at large lags, the scaling index
of the SF $\propto \tau^{\xred-1}$ yields a determination of $\xred$ by fitting the SF in log-log space.  
\newtext{
The  phase variance of the red noise 
}
\be
\rmsred^2(T) = \frac{S_0 T^{\xred-1}}{\xred-1},
\label{eq:varred}
\ee
can be inverted to  
an estimate for the spectral coefficient, 
\be
S_0 = (\xred-1)  T^{-(\xred-1)} \rmsred^2(T),
\label{eq:speccoeff}
\ee
where it is assumed that $f_{\rm Ny}T \gg 1$ and $\xred > 1$. 

\begin{figure}[h!]
    \centering
    \includegraphics[width=0.9\textwidth]{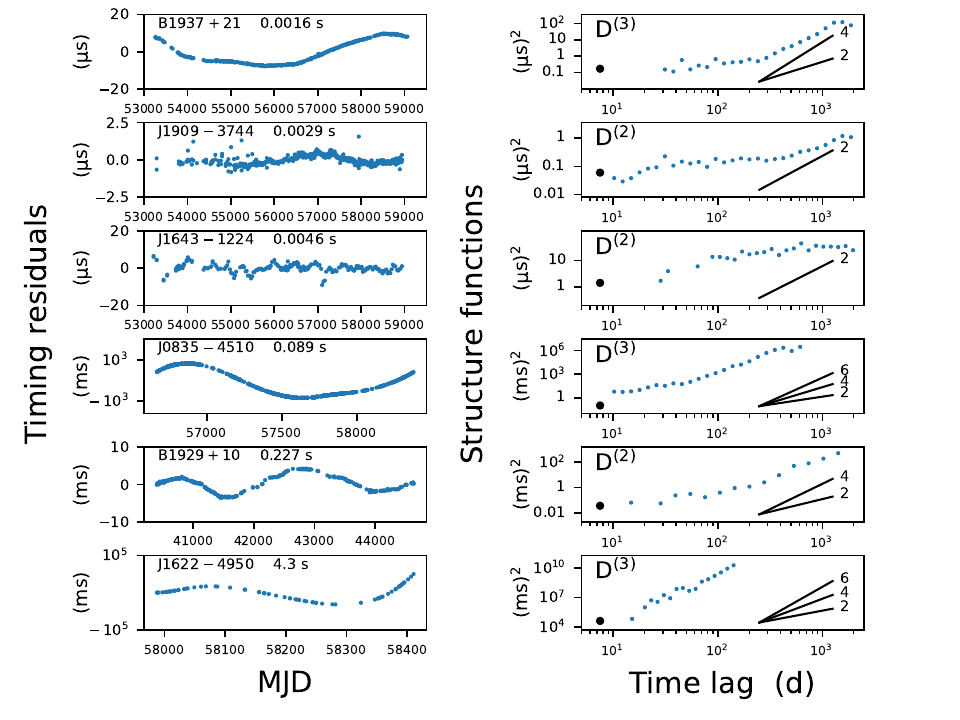}
    \caption{Timing residuals (left) and structure functions (SFs, right) for a sample of pulsars. 
    Top three rows: millisecond pulsar data from NANOGrav's 15 yr data relase \citep[][]{2023ApJ...951L...9A}.  Bottom three rows: residuals for three long period, high magnetic field pulsars. 
    Data for pulsars J0835-1224 (the Vela pulsar)
    and J1622-4950 are from the UTMOST timing program \citep[][]{2020MNRAS.494..228L} and data for B1929+10 are from the JPL timing program 
    (CD85; \citealt[][]{dr83}).
    In each SF panel in the right column, the type of SF  is designated as $D^{(m)}$ with $m = 2$ or 3, blue dotted  points are
    estimates of the binned structure function and the black circle represents the SF value due to white noise only.  
    Slanted lines show scaling laws 
    $D^{(m)}(\tau) \propto \tau^{x_{\rm SF}}$ for 
    $x_{\rm SF} = 2, 4, 6$. 
    \label{fig:resids_and_sfs}
    }
\end{figure}

Timing residuals and structure functions  are shown in Fig.\,\ref{fig:resids_and_sfs} for three MSPs, two CPs, and one magnetar (J1622-4950).   The MSPs shown in the top three panels  have varying degrees of red noise, with B1937+21 having the largest, J1909$-$3744 having  smaller, and J1643$-$1224 with minimal red noise but with distinctly non-white variations.  Accordingly, the structure functions in the right-hand column go from steepest to flattest.     The Vela pulsar (J0835$-$4510) displays strong red noise with a steep SF.  Red noise for the other CP, B1929+10, comprises discrete events (see CD85, Figure\,10q)  that collectively yield a steep SF.  Finally, magnetar
J1622$-$4950 in the bottom row has the strongest, steep spectrum red noise, consistent with spin noise being 
highly correlated with spin frequency derivative and, thus, surface magnetic field strength of the NS. 

\subsubsection{Constraints from scaling of variance with $T$}
The total phase  variance is the quadratic sum of the white and red noise variances,
$\sigma_\phi^2(T) = \rmswhite^2 + \rmsred^2(T)$.   Estimation of $\rmsred^2(T)$ can use the quadratic difference 
$\rmsred^2(T) = \sigma_\phi^2(T) - \rmswhite^2$ for cases where the red noise is strongly dominant and the white noise variance is obtained using the structure function in Eq.~\eqref{eq:Dwhite_red} for small lags where the second term is negligible
(as in CD85, Eq.\,12).  Alternatively, the red noise variance can be obtained from the Bayesian  posterior PDF for the model in Eq.~\eqref{eq:Dwhite_red}.  

The dependence  of the red noise variance on  data span length $T$ can also be used to determine $S_0$ and $\xred$.  This requires that the red and white noise variances are separable for a wide range of sub-spans, which is possible for pulsars with large enough red noise and numerous samples across the total span (e.g. CD85, \S\,V). 

Red spin noise of some pulsars has been demonstrated to be a mixed process, comprising a superposition of events in both $P$ and $\Pdot$.  Individually, each  type of event would produce spectral indices $\xred = 4$ and $6$ and SF scalings
$D_\phi^{(3)} \propto \tau^{3}$ and $\tau^5$, respectively.   The resulting SFs generally comprise at least two red-noise terms, with more than two if events in $P$ and $\Pdot$ occur together.   For several CPs,  the asymptotic scalings of $\rmsred^2(T)$ for large
$T$ and of $D^{(3)}_\phi(\tau)$ for large $\tau$ are dominated by events in $\Pdot$
\citep[CD85;][]{1988ApJ...330..847C,dmh+95},
corresponding to  a strong low frequency component to the spectrum with $\xred \simeq 6$.

\begin{figure}[h!]
    \centering
    \includegraphics[width=0.8\textwidth]{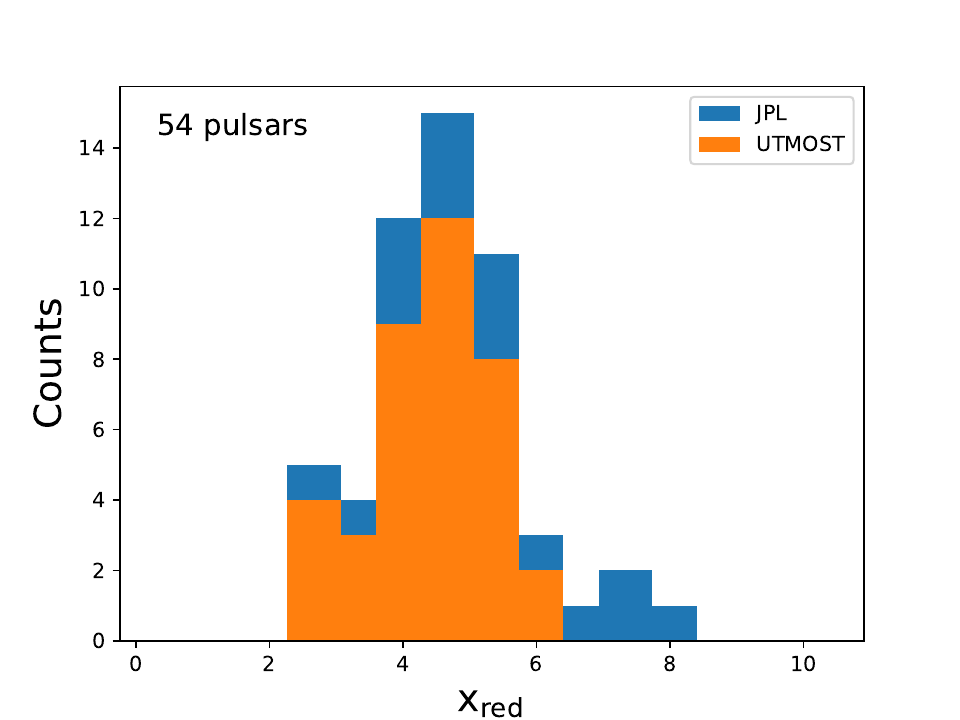}
    \caption{Histogram of inferred spectral indices for red noise assuming it has a power-law spectrum.   Spectral indices are estimated from the asymptotic slope of the third order structure function using the relation $x_{\rm red} = x_{\rm SF} + 1$.
    \label{fig:hist_xred}
    }
\end{figure}

Fig.~\ref{fig:hist_xred} shows spectral index estimates $\xred = \xsf + 1$ derived from SF slopes $\xsf$ evaluated from timing data in the JPL and UTMOST timing programs.   The estimates formally range between 
$\xred = 2$ and $\xred = 8$ but maximize at $\xred \sim 4.5$, which is intermediate between the slope expected for spin noise comprising step functions in $\fs$ and $\fsdot$.  
\newtext{
Individual  sequences of these two types of step functions yield fluctuation spectra $\propto f^{-4}$ and $f^{-6}$, respectively.   Physical contexts likely involve mixtures of both types of step functions, yielding a composite spectrum that can have an intermediate slope index between $-4$ and $-6$ or one or the other extremes, depending on the length of a data set. The Vela pulsar, for example displays both kinds of events
\citep[][]{1988ApJ...330..847C}. 
}

\subsection{Allan variance for red noise}

\begin{figure}[h!]
    \centering
    \includegraphics[width=0.7\textwidth]{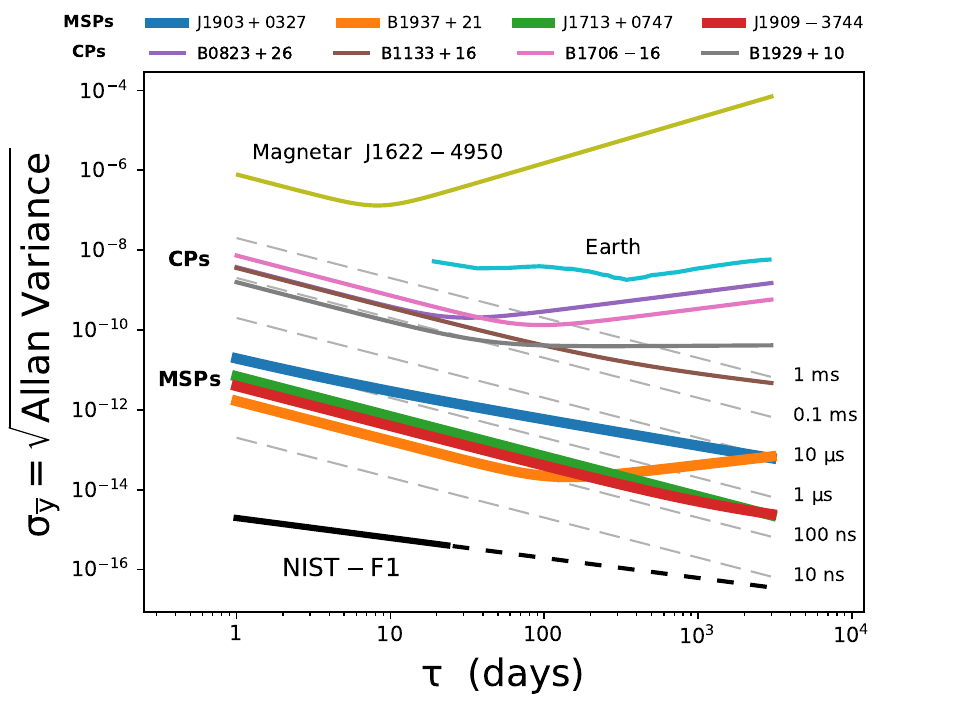}
    \caption{Allan deviation $\sigma_{\ybar}$ for pulsars compared to a cesium clock (NIST-F1) 
\citep[][]{1408302}.
Curves are shown for four MSPs from the NANOGrav 15-yr data set (thick lines labeled 'MSPs') and  four long-period canonical pulsars from the JPL timing program (thin lines labeleld `CPs') 
   ( \citealt[][]{dr83}; CD85)
    Also shown are curves for the magnetar J1622$-$4950 \citep[][]{2020MNRAS.494..228L}  and for length of day variations of the Earth 
    (\url{https://www.iers.org/IERS/EN/DataProducts/data.html}).
    Faint dashed lines correspond to white noise with the labeled 
    RMS values \newtext{while the heavy dashed line for the NIST-F1 has been arbitrarily extrapolated to longer times for visual purposes (we do not have access to measurements for such an extrapolation)}. 
    \label{fig:allanvar}
    }
\end{figure}

Frequency standards and clocks are often characterized using structure functions
\citep[][]{1978IEEEP..66.1048R, rw91}
, including the Allan variance 
\citep[][]{1966IEEEP..54..221A} and the same tools have been applied to pulsar timing data
\citep[CD85;][]{bh86,mte97,2011RvMP...83....1H}.  
Though MSPs once rivaled terrestrial frequency standards in frequency stability,  new technologies have surpassed the stability of the most stable pulsars (Fig.~\ref{fig:allanvar}).    Nonetheless, pulsars are unique tools for  applications in fundamental astrophysics and it is useful to characterize \newtext{and compare} their stabilities using standard methods.  

Let $\fs(t)$ be the time-dependent frequency of a clock or synthesizer
and define the normalized frequency, 
$
y(t) = \fs (t) / \langle \fs \rangle,
$
and its running average  over an interval  $\tau$, 
\be
\ybar(t) = \frac{1}{\tau} \int^{t+\tau}_t d\tp\,y(\tp).
\ee
The Allan variance is the mean-square of  the first increment of $\ybar(t)$,  
$
\Delta_{\ybar}^{(1)}(t, \tau) 
     \equiv \ybar(t+\tau) - \ybar(t)
$,
\be
\sigma^2_y(t, \tau) 
= \frac{1}{2} \left\langle [\ybar (t+\tau) -\ybar (t)]^2 \right\rangle
= \frac{1}{2} \left\langle \left[ \Delta_{\ybar}^{(1)}(t, \tau)\right]^2 \right\rangle.
\ee
 The Allan variance is therefore proportional to the first order structure function of $y$, which is written in terms of the  second order phase SF, 
 \be
 \sigma^2_y(t, \tau) 
 = 
 \frac{1}{2} D^{(1)}_y(t, \tau) 
 \equiv 
 \frac{D_\phi^{(2)}(t, \tau)}{2(\langle \fs \rangle \tau)^2}.
\label{eq:av}
 \ee
 The SF for timing residuals $\resid(t) = \fs \delta\phi(t)$  is 
 $\sigma^2_y(t, \tau) = D^{(2)}_{\resid}(t, \tau) / 2 \tau^2$. 
 
{\em Extension:}
For some processes the Allan variance is independent of $t$ and therefore not dependent on the data set length and  the explicit $t$ dependence is usually dropped
from $\sigma^2_y(t, \tau)$.   For other processes with higher-order nonstationarity,  the $t$ dependence remains.  For example, 
 a random walk  in $\dot \nu$ (i.e. in the second derivative of the phase) requires  a third-order SF to eliminate any $t$ dependence.   This can be used to define an analogous Allan-variance-like quantity (CD85) 
 \be
 \Sigma^2_y(t, \tau) 
 	= \frac{1}{9} D^{(2)}_y(t, \tau) 
 	= \frac{1}{9} 
      \left\langle \left[\Delta_{\ybar}^{(2)}(t, \tau)\right]^2\right\rangle 
    = \frac{D_\phi^{(3)}(t, \tau)}{(3\langle \fs \rangle \tau)^2}
\ee

Figure~\ref{fig:allanvar} compares the Allan deviation (AD) $\sigma_{\ybar}$ of four long period, high magnetic field CPs and four MSPs with a laser-cooled cesium fountain frequency standard 
\citep[][]{1408302}.    
For small lags $\tau$ the AD for  pulsars follows that expected for white noise in pulse phase,  $\sigma_{\ybar} \propto \tau^{-1}$ while the cesium frequency standard shows white noise in frequency, corresponding to $\sigma_{\ybar} \propto \tau^{-1/2}$. 
For $\tau \gtrsim 30$~to~200\,d, the CPs show  upturns in the AD due to achromatic spin noise. 
The MSPs show much smaller ADs.  For the two best MSPs in PTA timing samples, 
J1713+0747 and J1909-3744,  the ADs decline monotonically, with only a slight deviation from a white-noise scaling for J1909-3744, consistent with the NG15 noise modeling \citep[][]{2023ApJ...951L..10A} using alternative methods.   The MSP  J1939+2134 (B1937+21) has the lowest AD at small lags (from white noise) but is the worst MSP in terms of correlated red noise, as seen for  the large upturn in
AD at $\tau \sim 100$\,d.  This red noise is similar, statistically, to that seen from spin noise in CPs but an alternative explanation in terms of a circumpulsar asteriod belt has also been explored \citep[][]{2013ApJ...766....5S}.   The large-DM MSP~J1903+0327 shows a shallow upturn from a white-noise only AD that is apparently
interstellar in origin \citep[][]{2025ApJ...986..191G}.   In particular, this pulsar manifests the effects of incomplete correction for the combined effects of dispersion and scattering (pulse broadening).  For comparison we show an AD curve for length of day variations of the Earth, which shows worse stability than most pulsars but is far more stable than the magnetar J1622$-$4950 also shown  
\citep[data from][]{2020MNRAS.494..228L}.

\subsection{Scaling laws for the variance and spectrum of red spin noise}
\label{sec:spinnoise}

Spin noise is well established for canonical pulsars (CPs),  those with spin periods longer than 
$\gtrsim 10$~ms and with surface magnetic fields $\gtrsim 10^{11}$~G.  However, the underlying physics of torque variations and superfluid dynamics are not well understood.   Spin noise has  a very red power spectrum that causes the variance in timing residuals to have non-stationary statistics dependent on the length of the data span, $T$.
\newtext{
While there is a correlation of the spin-noise variance with a combination of spin frequency and its time derivative 
\citep[][]{1988ApJ...330..847C, dmh+95,sc10, 2019MNRAS.489.3810P}, it is not known how well it extends to MSPs given the significantly lower variance manifested in MSPs.
}

Past analyses 
\newtext{
(e.g. those just cited)}
have used  a scaling law for the RMS spin noise after removal of a quadratic fit, 
\be
\rmsredtwo(T) = C_2 \, \fs^{\alphar} \, \vert\fsdot\vert^{\betar} \, T^{\gammar},
\label{eq:rmstoaSPIN}
\ee
with parameters $C_2, \alphar, \betar$ and $\gammar$.   The  large scatter about this relationship for different objects  is characterized using  a log-normal distribution with standard deviation $\sigma_{\ln \rmsredtwo(T)}$. 

 \citet[][]{sc10} used a large body of measurements on canonical pulsars and a limited amount on MSPs to estimate parameters of the scaling law.    At the time, only two MSPs showed red-noise similar to that of CPs [J1939+2134 (B1937+21) and
J1824$-$2452A], while only upper limits were available on other MSPs.  
 The  maximum likelihood estimates of the spin-noise parameters for  the `CP+MSP' fit in their Table~1 are
\be
\rm
C_2 = 5^{+2.4}_{-1.6}\, \mu s,  \quad\quad \alphar = -1.4\pm 0.1,  \quad\quad \betar = 1.1 \pm 0.1, \quad\quad \gammar=2\pm0.2,
\label{eq:sc10}
\ee
for $\fs$ in Hz,  $\vert\fsdot\vert$ in 
$10^{-15}$~Hz\ s$^{-1}$,  and $T$ in yr.   The empirical spread about the scaling law is 
$\sigma_{ \ln \rmsredtwo} = 1.6\pm0.1$.  Errors on parameters are $\pm 2\sigma$.

Similar scaling laws have been estimated by others.   \citet[][]{2017ApJ...834...35L} included more MSPs than SC10  in their analysis  and obtained for the `CP+MSP$_{\rm 10, PPTA}$ +NANO' fit in their Table 4,
$C_2 =  6^{+1.3}_{-1.6}\,\mu s, \alphar = -1.32\pm0.04, \betar = 1.08\pm0.03, \gammar = 1.96\pm0.08$
and $\sigma_{ \ln \rmsredtwo} = 1.56\pm0.05$.   \citet[][]{2020MNRAS.494..228L} analyzed non-recycled pulsars
(aka CPs) to obtain $\alphar = -0.84^{+0.47}_{-0.49}$ and $\betar = 0.97^{+0.16}_{-0.19}$ and demonstrated consistency with SC10 and other previous results \citet[][]{10.1111/j.1365-2966.2009.15938.x, 2019MNRAS.489.3810P}.

\begin{figure}[t!]
    \centering
    \includegraphics[width=0.7\textwidth]{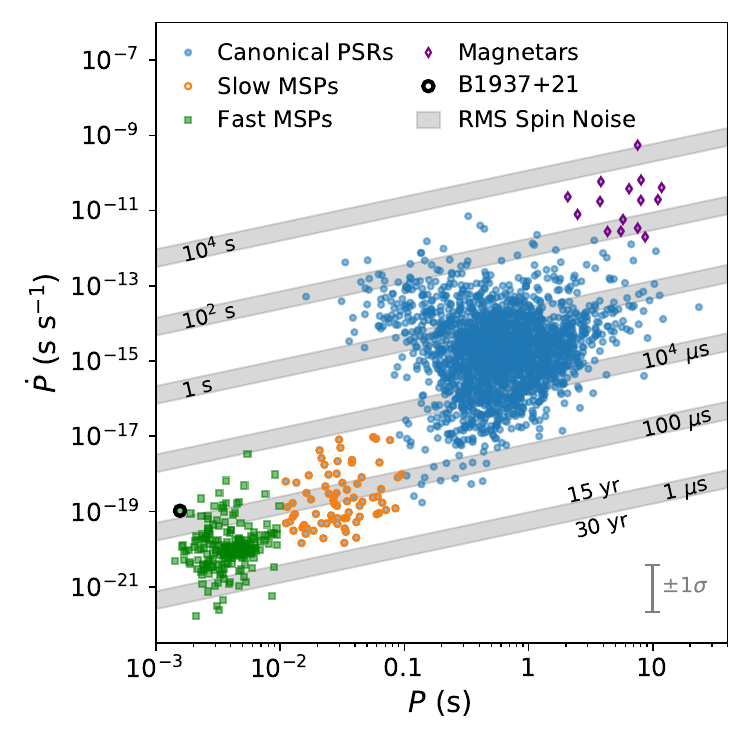}
    \caption{Spin period derivative $\dot{P}$ vs. period $P$ in seconds for all pulsars in the ATNF catalog. $\dot{P}$ has been corrected for the Shklovskii effect where possible.  (The Shklovskii effect is the contribution to 
    $\dot P$ from translational motion of a pulsar.  It is insignificant for
    CPs but can be important for MSPs with small intrinsic $\dot P$.)  Grey bands indicate the range of \RMS\ timing variations due to spin noise for data timespans between 15 and 30 years. The estimated spin noise is based on the scaling relation fit in \cite{2017ApJ...834...35L} across canonical pulsar, MSP, and magnetar populations (Equation~\eqref{eq:rmstoaSPIN}). 
    \label{fig:ppdot_spin}
    }
\end{figure}

Figure~\ref{fig:ppdot_spin} shows the spin noise scaling law projected onto the  period-period derivative ($P$--$\Pdot$) plane with various pulsar types identified.     The gray bands indicate the ranges of spin-parameter values that yield the same \RMS\ \TOA\  for data spans of 15 to 30\,yr.   

{\it Fluctuation spectrum:} 
The  $T$ dependence of $\rmsredtwo(T)\propto T^{2\pm0.2}$ corresponds to a power-law spectrum
$\propto f^{-\xred}$ with $\xred  = 1 + 2\gammar = 5\pm0.4$. 
The spectrum is calculated from $\rmsredtwo(T)$ 
\newtext{
by expressing the coefficient $S_0$ (Eq.~\eqref{eq:speccoeff})  of the phase spectrum (Eq.\eqref{eq:phispectrum}) in terms of the phase variance}
and assuming a lower-frequency cutoff $\eta / T$ 
with $\eta \simeq 1$, 
\be
\Sred(f) =  (\xred-1) \eta^{(\xred-1)}  
    \rmsredtwo^2(T)  T^{1-\xred}  f^{-\xred}. 
\ee
Substituting for $\rmsredtwo$ from Eq.~\eqref{eq:rmstoaSPIN}  yields
\be
\Sred(f) = 2\gamma (C_2 \fs^{\alphar} \vert \fsdot\vert^{\betar})^2 f^{-\xred}.
\label{eq:spectrumSPIN}
\ee
Note there is no net dependence on the length of the time series $T$. 

\newtext{
{\it Utility and complications:}   
The expressions for spin-noise variance (Eq.~\eqref{eq:rmstoaSPIN})
and the corresponding red-noise spectrum (Eq.~\eqref{eq:spectrumSPIN}) are 
statistical descriptions of the pulsar population in the simplest terms.  They are useful for making an initial assessment of the timing quality of a newly discovered pulsar.  Also, as a new pulsar is timed over an initial year or two, any strong departures from the scaling law may suggest that other processes are in play, such as latent orbital motion. 
}

Any specific pulsar can depart from this description in both the amplitude of spin variations and in the scaling law indices,  $\alphar, \betar,$ and $\gammar$.    Moreover, an actual spectrum that has a piecewise power-law form but us analyzed as a single power law will yield an index $\gammar$ and hence  frequency scaling index $\xred = 1 + 2\gammar$ that will show dependence  on the data span $T$.  This arises from the underlying physics of the spin noise, which appears to comprise individual events in the spin torque having the idealized forms of narrow pulses or step functions or combinations of the two.   
These  give fluctuation spectra
$\propto f^{-4}$ and $\propto f^{-6}$, respectively.    The population average  $\gammar \approx 5$ among canonical pulsars suggests that both kinds of events are seen across the population.  The situation is less clear for MSPs but longer timing data sets on more MSPs should provide future clarity. 

Another possible departure from the picture presented here is that  achromatic timing noise may have a different cause than spin fluctuations and the stochastic GW background.   For example, orbital noise from an asteroid belt  can produce a power-law residuals spectrum; asteroids may plausibly  grow out of  debris from the accretion driven spinup proccess   in some cases \citep[e.g.][]{2013ApJ...766....5S, jcc20}.  This is discussed further in Sect.~\ref{sec:orbnoise}.

\subsection{Comparison with the nanohertz GW stochastic background}

It is useful to express the red-noise spectrum in terms that can be compared with those used for
the long-wavelength GW spectrum.   The characteristic dimensionless strain is represented with a dimensionless
amplitude $\Agw$ and a power-law frequency dependence,
\be
\hc(f) = \Agw \left(\frac{f}{\fref}\right)^{-\xh}
\ee
and the corresponding spectrum for  arrival-time residuals is
\newtext{
(Appendix\,\ref{app:specrep})
}
\be
\Sred^{\rm (gw)}(f) 
= \frac{\Agw^2}{12\pi^2 f^3} \left(\frac{f}{\fref}\right)^{-2\xh}
= \frac{\Agw^2 }{12\pi^2 \fref^3} \left(\frac{f}{\fref}\right)^{-(3+2\xh)}.
\label{eq:Sred1}
\ee
Using $\fref = 1\,\cyperyr$ with $\Agw = 10^{-15} \Agwfifteen$ and expressing the spectrum in units
of ${\rm \mu s^2\, yr}$,  the spectrum becomes
\be
\Sred^{\rm (gw)}(f) = \KsR \Agwfifteen^2   \left(\frac{f}{\fref}\right)^{-\xred}
\label{eq:Sred2}
\ee
with $\xred = 3+2\xh$ and the leading coefficient is
\be
\KsR = \frac{10^{-30}}{12\pi^2} ({\rm \mu s\, per\, yr})^2 = 8.41\times 10^{-6}\, {\rm \mu s^2 \, yr}.
\ee
For reference,  the simplest form for $\hc(f)$ results from a population of supermassive black hole binaries 
with orbital decay solely from GW emission, giving $\xh = 2/3$ and $\xred = 13/3$
\newtext{
\citep[][]{rr95a,2001astro.ph..8028P,jb03} This reference spectrum is currently used by all PTA groups as a comparison with empirical spectra.
Actual spectra from merging black hole binaries are affected by the finite number of such systems that contribute significantly to PTA measurements and to other contributions to the spiral-in rate from gas dynamics and stellar interactions \citep[][]{2025arXiv251218822S}.
}

\newtext{Using the same spectral form, we solve for the dimensionless amplitude 
$\Arsnfifteen$ in terms of the residuals spectrum (Eq.~\eqref{eq:Sred2})
and  evaluate it using Eq.~\eqref{eq:spectrumSPIN}},
\be
\Arsnfifteen 
= \left( \frac{2\gammar}{\KsR}  \right)^{1/2} C_2(\mu s) \fs^{\alphar} \vert \fsdot\vert^{\betar}
\approx  3.45\times 10^3\, \fs^{-1.4} \vert \fsdotfifteen\vert^{1.1},
\label{eq:Arsn15}
\ee
 with  $\fs$ in Hz and $\fsdot$ in units of $10^{-15}$\,Hz\,s$^{-1}$ and
where the approximation uses only the  nominal value for $C_2$. 

As an example, the highly stable MSP J1909-3744 with $\nu = 339$\,Hz and 
$\nudot = -1.62\times 10^{-15}$\,Hz\,s$^{-1}$ gives $\Arsnfifteen \simeq 1.7$ \newtext{(comparable to constraints reported for this pulsar by, e.g., \citealt{2024MNRAS.530..287S,2022MNRAS.509.5538C})}, and slightly below
\newtext{the gravitational wave amplitudes $\Agwfifteen \simeq 2.0 - 2.5$ reported by  \citet[][]{2023ApJ...951L...8A,EPTAGW23,2023ApJ...951L...6R}}.

\begin{figure}[t] 
   \centering
   \includegraphics[width=0.6\linewidth]{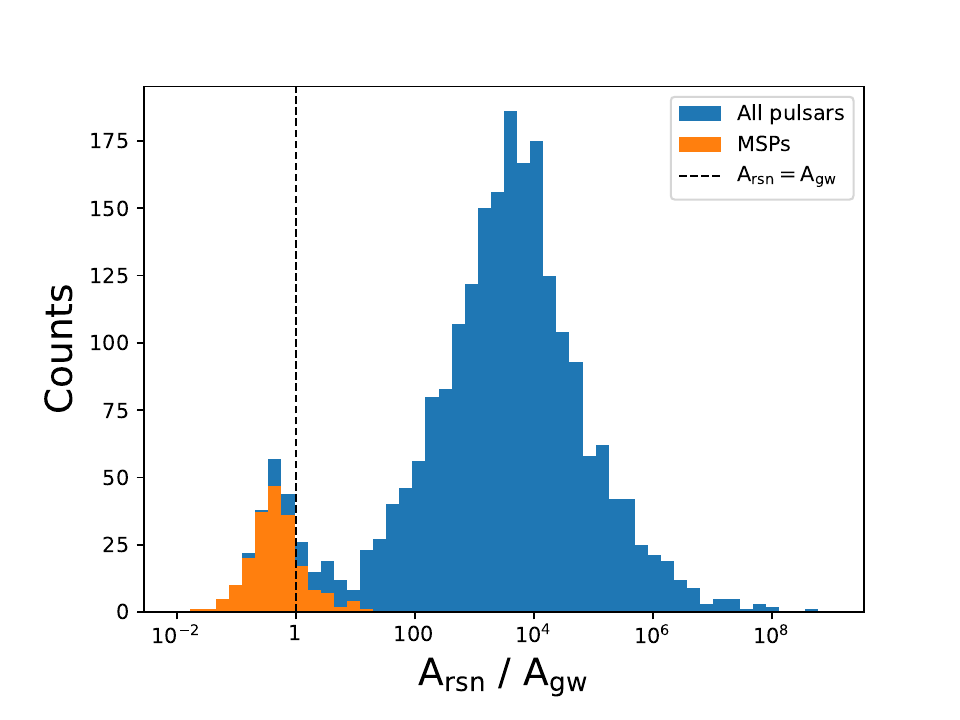} 
   \caption{Histogram of the ratio of \newtext{predicted} dimensionless spectral amplitude $\Arsn$ for red spin noise (using cataloged values for spin frequencies and their derivatives)
   to  $\Agw$ for the nanohertz GW background using the NANOGrav value.   The blue histogram  for all pulsars in
   the ATNF pulsar catalog shows bimodality that reflects the much larger spin noise for
   non-recycled pulsars.   The orange applies to catalog MSPs, defined as $P \le 20$\,ms
   and $\Pdot <  10^{-17}$s\, s$^{-1}$.   The vertical line indicating $\Arsn = \Agw$ demonstrates that the nominal
   red noise for some recycled pulsars is a larger signal than the GW background.}
   \label{fig:histAratio}
\end{figure}

For the pulsar population at large, we compare 
\newtext{
{\it predictions} 
for the dimensionless amplitude $\Arsnfifteen$ for 
red spin noise (using Eq.~\eqref{eq:Arsn15})
}
with $\Agwfifteen$  for  the GW stochastic background using histograms of their ratio $\Arsnfifteen / \Agwfifteen$ (Fig.\,\ref{fig:histAratio}).
The histograms demonstrate that some (but not all) MSPs, which have small spindown rates, have spin stabilities 
yielding $\Arsn \lesssim \Agw$, making them potentially useful for GW detection with PTAs.
We emphasize that additional considerations (e.g. red noise from interstellar propagation and \TOA\ precision vs. S/N) also determine inclusion of any given MSP in a PTA program, as discussed in Sect.~\ref{sec:Q}.
   
\begin{figure}[htbp] 
   \centering
   \includegraphics[width=0.6\linewidth]{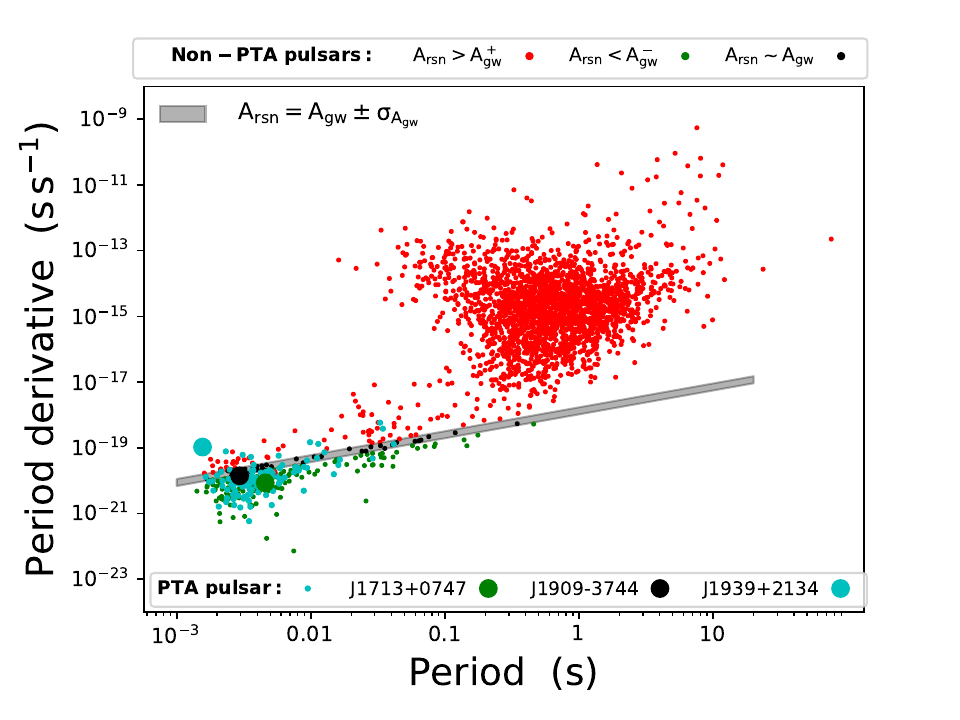}
   \caption{
   Period-Period derivative diagram showing pulsars with
   red-noise spectral amplitudes $\Arsn$ above (red points),  below (green points), or inside (small black points) the gray band where they are approximately equal to the GW background's spectral amplitude $\Agw$. 
   The \newtext{125 pulsars included in PTA programs by the 
   NANOGrav, EPTA, PPTA, InPTA, CPTA, and MPTA groups} are indicated in cyan.  Indicated by large symbols are 
   two MSPs  J1713+0747 and J1909-3744 used in PTA 
   programs because of their high spin stability along with another, J1939+2134 (B1937+21, large black circle), that has poor spin stability.    MSPs in globular clusters have been excluded because their $\Pdot$ values are perturbed by cluster gravitational potentials. 
   }
   \label{fig:ppdotAline}
\end{figure}

For fixed spin period,  pulsars with smaller $\Pdot$ values have greater predicted spin stability. 
\newtext{
Solving for the value of $\Pdot$ that satisfies  the condition $\Arsn \le \Agw$  and using Eq.~\eqref{eq:Arsn15}, we find values that make a pulsar potentially useful for GW detection, 
}
\be
\Pdot \le \Pdot_{\rm max}
&=& 10^{-15} \,{\rm s\, s^{-1}}  
    \left(\frac{\KsR}{2\gammar}\right)^{1/2\betar}     
    \left[\frac{\Agwfifteen}{C_2(\mu s)}\right]^{1/\betar}
    P^{2 + \alphar/\betar} 
\simeq
6.08\times 10^{-19}  \,{\rm s\, s^{-1}}  P^{\,0.73} \Agwfifteen^{0.91}. 
\ee
where $P$ is in s.
Fig.\,\ref{fig:ppdotAline} shows this condition for $\Agwfifteen = 2.4$  plotted as a black line on the $P$-$\Pdot$ diagram along with points color coded for pulsars on above or below the line.  Three MSPs are highlighted, two that easily satisfy the condition (J1713+0747, J1909-3744)  and another, J1939+2134 (B1937+21) that is well known to show significant red noise.   \newtext{Equation 4.19 and Fig.\,\ref{fig:ppdotAline} suggest that for most MSPs with typical periods $P<0.01$ s, sources with $\Pdot_{\rm max} \lesssim 10^{-20}$ s s$^{-1}$ will satisfy the spin stability criterion for expected GW amplitudes.}


\newcommand{\Mbelt}{M_{\rm belt}}
\newcommand{\Mearth}{M_{\oplus}}
\newcommand{\Msun}{M_{\odot}}

\newcommand{\Aast}{A_{\rm ast}}

\subsection{Orbital noise}
\label{sec:orbnoise}

\newtext{
Orbital noise may derive from interactions of a pulsar with its companion, as in `spider' pulsars or from a population of asteroids orbiting a pulsar (and its stellar companion,  if any).   Spider pulsars have low-mass companions that are ablated by pulsar emission and generally show stochastic variations in their orbits that, although of intrinsic interest,  make most of  them unsuitable for precision timing.  \citet[][]{2022MNRAS.509.5538C} discuss timing stability of spider pulsars and identify a few cases where orbital noise is not seen.   

We focus the remainder of this section on asteroids, which may form in the aftermath of the accretion-driven spinup process which may leave residual gas near the neutron star that solidifies into small objects.
The first detected exoplanets around the MSP B1257+12 
\citep[][]{1992Natur.355..145W} exemplify the presence of small bodies around recycled pulsars.  A detailed study of the orbital dynamics of the planets indicate that there are stable orbits in the system 
\citep[][]{2005ApJ...619.1084G}, allowing the presence of dust or larger objects.  Neutron stars with white dwarf companions also have stability regions outside the semi-major axis of  the binary \citep[][]{jcc20}. 

This leads to the possible explanation of red noise seen in timing of the isolated MSP B1937+21 as orbital  rather than spin noise \citep[][]{2013ApJ...766....5S}.  To match the large measured red noise (Sect.~\ref{sec:SFs}) a total mass $\Mbelt \lesssim 0.05\Mearth$ extending to at least 15\,au yields orbital periods longer than the analyzed data span.    This case illustrates that  a small debris mass can, in principle,  produce large timing effects.   Most well studied MSPs 
show substantially less red noise than B1937+21, indicating that any debris disks have very small total masses.    Nonetheless,  as timing data sets grow longer,   the red noise from an asteroid belt will grow along with other red-noise processes, including spin noise and GWs, although with a different dependence on data span.   

An orbital companion  produces a cyclical TOA variation due to recoil motion of the NS.    An ensemble of asteroids contributes a noise process to \TOAs\ with  variance that grows with longer data spans if the maximum orbital period is longer than the data span length \citep[][]{2013ApJ...766....5S}.   Recent work  considering asteroid belts around MSPs with white dwarf companions \citep[][]{jcc20} derives asteroid noise spectra for individual MSPs. 
Asteroids are assumed to be distributed in mass as  $f_m(m) \propto  m^{\alphaa-1}$ and in orbital radius 
as $f_a(a) \propto  a^{\betaa-1}$. For $N$ asteroids with  total mass $\Mbelt$  and NS mass $M$, the spectrum  for
\TOA\ perturbations is  a power law,
  \be
\Sast(f) = 
\left[\frac{G}{(2\pi M)^2} \right]^{2/3} 
\left(\frac{\Mbelt\sin i}{2c \sqrt{N}}\right)^2
\frac{\langle m^2 \rangle}{\langle m \rangle^2} 
\left[\frac{2\betaa f^{-(2\betaa+7)/3}}{3(f_-^{-2\betaa/3} - f_+^{-2\betaa/3})} \right], 
\quad f_- \le f \le f_+, 
\ee
where  the  frequency cutoffs are related to the extent of the asteroid belt,
$f_{\pm} = \sqrt{M} a_{\rm min,max}^{-3/2} \simeq \sqrt{1.4} a_{\rm min,max}^{-3/2}$ for $M$ in solar masses, frequencies in cy~yr$^{-1}$,  and orbital radii in au.  
The spectrum  is expressed in terms of the index $\betaa$ for the distribution of orbital radii $a$, which is related to the index $\beta_f = -2\betaa/3$ used by \citet[][]{jcc20} for the corresponding distribution in orbital frequencies. 
The ratio of moments $\langle m^2\rangle / \langle m \rangle^2$ is determined solely by the mass ratio $r  = m_2 / m_1$ of the largest and smallest asteroid masses, 
\be
\frac{\langle m^2 \rangle}{\langle m \rangle^2} 
 = 
 \frac {(\alphaa+1)^2}{\alphaa(\alphaa+2)} 
 \left[ \frac{(r^\alphaa - 1) (r^{\alphaa+2}-1)} {\left(r^{\alphaa+1} - 1\right)^2} \right]. 
\ee
These expressions
follow from \citep[][]{{jcc20}} except for the use of a distribution in semimajor axis $a$ rather than frequency $f$.   
}

Evaluating for $\alphaa = -5/6$ (the steady-state value for self-similar populations of colliding bodies, as determined by \citet{dohnanyi69}), $\betaa=1$ (corresponding to a uniform distribution in orbital radius), $\Mbelt = 10^{-3} \Mearth$ with $N=10^4$ asteroids, a mass ratio
$r= 10^{-4}$, and a NS mass $M = 1.4~\Msun$, 
\be
\Sast(f) = 
	135\,{\rm ns^2 \ yr}\, 
	\left[
	\frac{\left(\Mbelt\sin i / 10^{-3}\, \Mearth\right)^2}{ N / 10^4}
	\right]
	\left[
	\frac{\langle m^2 \rangle / \langle m \rangle^2}{100}
	\right] 
	\left(\frac{f}{1\, \rm cy\ yr^{-1}} \right)^{-3} .
\ee
Defining  the spectral form $\Sast(f) = \Aast^2 f^{-\gammaa}$, 
again for frequencies in cycles~yr$^{-1}$,   the dimensionless amplitude is
\be
\Aast &= &
	11.6 \ {\rm ns \ yr^{1/2}}
	\left[
	\frac{\left(\Mbelt\sin i / 10^{-3}\, \Mearth\right)}{ \sqrt{N / 10^4}}
	\right]
	\left[
	\frac{\langle m^2 \rangle / \langle m \rangle^2}{100}
	\right]^{1/2} 
\\
\gammaa &=& 
(2\betaa + 7)/3 = 3.
\ee
Later we compare  the spectrum of this achromatic process with other processes.

 \begin{figure}[h!] 
    \centering
    \includegraphics[width=0.6\linewidth]{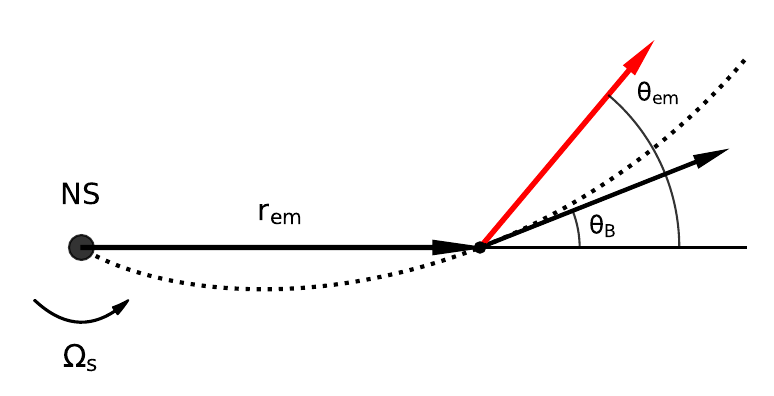}
    \caption{Schematic of emission at radius $\rem$ into a direction angle $\theta_{\rm em}$.   The tangent to the magnetic field line (dotted)   is at an angle $\theta_{\rm B}$ from the radial vector.   The difference between 
    $\theta_{\rm em}$ and $\theta_{\rm B}$ is from rotational aberration and gravitational bending of the ray path.
    A change in  emission radius alters $\theta_{\rm em}$, yielding emission directed toward an observer at an offset  spin phase, altering  the observed pulse phase.  Similarly, a change along a different magnetic field line also changes the observed pulse phase.
    }
    \label{fig:emission_field_line}
 \end{figure}

\section{Elements of pulsar signals and their propagation}
\label{sec:signal_models}

Itemizing the timing error budget requires consideration of emitted wavefields and how they are affected by propagation through intervening plasmas (along with effects from instrumentation). Before introducing the timing model, we must first lay out the key elements of pulsar signals, including their basic emission properties and the plasma propagation effects incurred in the observed signal. Together these elements form a pulsar signal model (e.g. Fig.~\ref{fig:shots2pulses}), which is distinct from and a prerequisite for the timing models discussed later. 

Radio pulsars emit radiation  from their magnetospheres with brightness temperatures as high as 
$10^{42}$~K that entail  coherent emission from highly relativistic particles.
Emission  radii $\rem = \vert \xem \vert$  are  intermediate between the NS  radius $\Rns \sim 10$\,km and the 
light cylinder radius $\rLC = c / \Omspin= c\Pspin/2\pi  \simeq 5\times10^4 P \,{\rm km} \gg \Rns$ for $P$ in s.

Beaming is tangential to the magnetic field  (Fig.~\ref{fig:emission_field_line}) and aberrated by the NS spin into a net emission angle $\thetaem$ relative to the radius vector $\xem$ measured from the NS center
\newtext{
\citep[][]{1978ApJ...222.1006C,2009MNRAS.393.1617K, 2012AA...543A..66H}.
The field direction and aberration angle both depend on the emission location  $\xem$.
Emission is seen when  the beam closely approaches the observer's direction.
Because field lines and thus ray paths are non-radial, gravitational bending must also occur, as analyzed for  the Crab pulsar \citep[][]{2026JPlPh..92E..22M}.
Additionally, polarization-dependent refraction may be significant in some pulsar  magnetospheres
\citep[][]{10.1093/mnras/stu690}.
}

Collective radiation  from large particle numbers is in the form of 
 `nanoshots' i.e. $\lesssim$\, ns-duration impulses  that extend to  $\gtrsim$\,GHz frequencies 
\newtext{
\citep[][see also Sect.~\ref{sec:shotnoise}]{cor76a}.
}
Individual pulses  seen once per spin period  have widths $\gtrsim 30\,\mu$s and up to hundreds of ms  for CPs
 \newtext{\citep[][]{ATNFPulsarCat}.  The MSP observed in multiple PTA programs, J1909$-$3744, in fact has the narrowest average pulse width.} 
 Thus they   generally   comprise incoherent ensembles of many nanoshots\footnote{A special case is the young Crab pulsar from which individual nanoshots have been measured \citep[][]{2003Natur.422..141H}}  with different shapes from one  period to the next owing to variations in both emission radius and  coherence  strength from the dynamic particle flow.    This variability underlies  temporal  jitter with respect to  strictly periodic emission times.  

Evidently particle flows and emission are controlled, in the mean,  by the strong magnetic field of the NS, as evidenced by  the  convergence of average  pulse profiles to distinct shapes, whose stability has been demonstrated  over the decades since pulsars were discovered.   (We note that  magnetars, by contrast, have dynamic magnetic fields that cause their average profiles to vary significantly.)

\newcommand{\filtlp}{B_{\rm lp}}
\newcommand{\hext}{h_{\rm ext}}
\newcommand{\Hext}{H_{\rm ext}}

\subsection{From fields to baseband voltages}
\label{sec:voltages}


Pulsar processing typically involves manipulation of digitized voltages that are proportional to electric field components  selected by the antenna polarization.   The scalar field  in a specific  polarization $\phat$ is
$E(t) = \Evec(t) \cdot \phat$  and the field is typically modified by a transfer function with impulse response 
$\Hext(t)$, which may be polarization specific.   The receiver selects a frequency band according to  an impulse response $R(t)$.   The net narrowband field $\vnarrowband(t)$ is then the convolution of  three real quantities, 
\be
\vnarrowband(t) = E_i(t) * \Hext(t) * R(t),
\ee
where $E_i(t)$ is the initial, emitted field. 
A baseband mixer translates the  field to a complex baseband field through the operation,
\be
\vBB(t) = \left[ e^{2\pi i \nu_0 t} \vnarrowband(t) \right] * \filtlp(t),
\label{eq:vBB1}
\ee
where $\filtlp$ is a low pass filter that enforces limitation of $\vBB$ limited to frequencies $\nu \in [-B/2, B/2]$ with bandwidth $B$.  We simplify $\vBB$ by expanding each of  $E_i, \Hext,$ and $R$  into negative and positive frequency components as (e.g.) $E_i(t) = E_{i,-}(t) + E_{i,+}(t)$.
Hermiticity of the Fourier transform $\Etilde_i(\nu)$ requires $E_{i,-}^*(t) = E_{i,+}(t)$.  
Defining baseband quantities,
$\varepsilon_i(t) =  e^{2\pi \nu_0 t} E_{i,+}(t), 
\hext(t) =  e^{2\pi \nu_0 t} {\Hext}_{+}(t)$
and
$r(t) =  e^{2\pi \nu_0 t} R_{+}(t)$, 
the baseband field is $\vBB(t) = \varepsilon_i(t) * \hext(t) * r(t) * \filtlp(t)$.
By lumping together the last three quantities in the convolution  into an effective
impulse response, $h(t) = \hext(t) * r(t) * \filtlp(t)$,  the baseband field is written succinctly as
\be
\vBB(t) = \varepsilon_i(t) * h(t), 
\label{eq:vBB2}
\ee
where $h(t)$ is understood to include both astrophysical (wave propagation) and receiver contributions.

Generally the impulse response $h(t)$ in Eq.~\eqref{eq:vBB2} could be polarization dependent because it can include  birefringent propagation,  as discussed below, along with scattering. In practice dispersion and Faraday rotation are removed separately, so $h(t)$ then represents only the effects of scattering along with receiver contributions. 

\subsection{Stokes parameters from baseband fields}
\label{sec:Stokes}.

Pulsar emission is generally polarized, requiring separate baseband signals  for two polarization channels $p = x,y$ for measurements with linearly polarized antennas or $p = r, l$ for circular polarization.    The four Stokes parameters ($I, Q, U, V$)  involve auto-and-cross products of $\vBB_{p}$.  Pulsar timing typically  makes use of only the total Stokes intensity $I = \sum_p \vert \varepsilon_{p}\vert^2$ though some studies 
\newtext{
(see Sect.~\ref{sec:fullstokes})
}
have explored the use of sharper features in the linearly polarized intensity $L = \sqrt{Q^2 + U^2}$ or in the circular polarization $V$ (see Sect.~\ref{sec:fullstokes}).  

Adopting either a circular or linear polarization basis, the Stokes vector is expressed in terms of  two baseband fields, $\vBBr, \vBBl$  or $\vBBx, \vBBy$, 
\be
\Svec =  
		\left(
		\begin{array}{c}
		I \\
		Q\\
		U \\ 
		V
		\end{array}
		 \right)	
		 = 	 
		 \left(
		\begin{array}{c}
		\Ir + \Il \\
		2\mathscr{R}\{  \vBBr \vBBl^* \} \\ 
		2\mathscr{I} \{   \vBBr \vBBl^* \} \\
		\Ir - \Il
		\end{array}
		 \right)
		 =
		 \left(
		\begin{array}{c}
		\Ix + \Iy \\
		\Ix - \Iy \\
		2\mathscr{R}\{  \vBBx\vBBy^* \} \\ 
		2\mathscr{I} \{   \vBBx \vBBy^* \} 
		\end{array}
		 \right).	
\label{eq:Stokesvec}
\ee
These are defined without any time averaging, contrary to the usual case, because we wish to consider both the mean and variance of each Stokes parameter, requiring moments $\langle \Svec \rangle$ and 
$\langle \Svec \circ \Svec \rangle$, the latter being the mean of the Hadamard (i.e. element-wise) product. 
In the following we also make use of the complex quantity $L = Q + iU$. 

The variance of the total intensity $I$ is polarization dependent and it also depends on scintillation modulations \citep[][]{cbh+04}.    Instrumental polarization is represented as a transformation  of the true
Stokes vector using a $4\times 4$ Mueller matrix $\Marray$ to give the nominal measured Stokes parameters,
$\Svec_{\rm meas} = \Marray \Svec$.
Without inversion, instrumental polarization yields additional arrival time errors, as discussed later \newtext{(see Sect.~\ref{sec:polcal})}. 

The degrees of polarization are defined in terms of ensemble average Stokes parameters\footnote{For pulsar signals it is useful to think of an ensemble average as an average at fixed pulse phase over an ensemble of pulses for which stationary statistics can apply.  Pulsars are thus said to be cyclostationary, at least in the mean.  However, all quantities are strong functions of pulse phase and thus are non-stationary vs. phase.},
\be
\dL =
 \frac{\sqrt{\langle Q \rangle^2 + \langle U \rangle^2}}{\langle I \rangle} 
\equiv \frac{\vert \langle L \rangle \vert}  {\langle I \rangle},
\quad\quad
\dV = \frac{\langle V \rangle}{\langle I \rangle}
\quad\quad
\dpol = (\dL^2 + \dV^2)^{1/2} \le 1.
\ee

As defined, the variances of the unaveraged Stokes parameters satisfy
\be
\langle \vert L \vert^2 \rangle + \langle V^2 \rangle = \langle I \rangle. 
\ee
This equality follows from the fact that the instantaneous polarization in  Nyquist sampled data is effectively 
100\%.   However, a partially polarized signal varies with time so that
\be
\langle Q \rangle^2  + \langle U \rangle^2 +  \langle V \rangle^2  \le \langle I \rangle^2
\ee

Later we need to refer to Stokes parameters of  a noiselike  polarized signal that is added to unpolarized noise and we  utilize the signal to noise ratio of pulses in each of the Stokes parameters.

The intensity modulation index, 
\be
\mI =
 \frac{\sigma_I}{\langle I \rangle} &=& \left(\frac{1+\dpol^2}{2} \right)^{1/2},
\ee
depends on the total degree of polarization $\dpol$ and, by inspection, is smallest for an unpolarized signal ($\dpol=0$).
\newtext{
This follows because Stokes $I$ for an unpolarized signal has four degrees of freedom while a completely polarized signal has only two. 
}

We need \RMS\ values  for the Stokes parameters for calculations of the
signal to noise ratios of Stokes-parameter pulse profiles.  Expressed as  the ratios of \RMS\ values of $L$, $Q$, $U$, and $V$ to the RMS $I$, we have 
\be
\frac{\sigma_L}{\sigma_I} &=& \left[ \frac{2(1-\dV^2)}{1+\dpol^2} \right]^{1/2} ,
\\
\frac{\sigma_Q}{\sigma_I} &=& 
	\left[ \frac{1 - d^2  +  2\dL^2 \sin^2 2\chi}{1+\dpol^2} \right]^{1/2} ,
\\
\frac{\sigma_U}{\sigma_I} &=& 
	\left[ \frac{1 - \dpol^2  + 2\dL^2 \cos^2 2\chi}{1+\dpol^2} \right]^{1/2} ,
\\
\frac{\sigma_V}{\sigma_I} &=& 
	\left[ \frac{1+\dV^2 - \dL^2}{1+\dpol^2} \right]^{1/2} .
\ee 
These ratios reflect the number of degrees of freedom in each Stokes parameter and depend on polarization-dependent  correlations such as 
$\langle \Ir \Il \rangle = (1/4)(1 + \dL^2 - \dV^2) \langle I \rangle^2$.
A completely polarized signal  ($\dpol = 1$) has perfectly correlated $\Ir$ and $\Il$ and thus fewer degrees of freedom,   so the  Stokes-$I$ modulation index, $ \mI= 1$, is larger than for an unpolarized signal with $\mI= 1 / \sqrt{2}$.

The \RMS\ ratios for a completely polarized signal depend on the type of polarization, 
$\sigma_L  / \sigma_I = \sqrt{1-\dV^2}$, 
$\sigma_Q /\sigma_I  = \dL \vert \sin 2\chi\vert$, 
$\sigma_U / \sigma_I = \dL \vert \cos 2\chi \vert$, 
and 
$\sigma_V / \sigma_I  = \sqrt{1  - \dL^2 }.  $

For the unpolarized case with $\dpol = \dL = \dV = 0$,  the ratios are 
$\sigma_L  / \sigma_I = \sqrt{2}$, 
and
$\sigma_Q /\sigma_I  = \sigma_U / \sigma_I = \sigma_V / \sigma_I =  1$, 
which are all equal to or larger than the fully polarized case.
 
\subsection{Pulsar emission is modulated polarized shot noise}
\label{sec:shotnoise}

\newtext{
Here we investigate the statistical properties of pulsar signals as polarized shot noise that includes a modulation function that shapes either the shot amplitudes or their rates. \citet[][Appendix A]{cw16} show that when shot noise is `dense' (comparable widths and separations of shots),  amplitude-modulated shots and rate-modulated shots are indistinguishable statistically.  In Sect.~\ref{sec:SAMSN} we discuss additional, extrinsic modulations from interstellar scintillation that yields an extended  description of the pulsar signal as \citep[][]{cbh+04} 
`scintillated amplitude-modulated polarized shot noise' or SAMPSN\footnote{A similar portrayal has been made by \citet[][]{2011MNRAS.418.1258O} in what they term `stochastic wide-band impulse modulated self-noise' or SWIMS.   We note that our usage of the terms `self-noise' and `jitter' (see below) differ from that in
\citet[][]{2011MNRAS.418.1258O}}.  
}

Regardless of the radiation physics underlying coherence (e.g. plasma maser vs. charge bunching), 
the fundamental radiation unit is a shot pulse \newtext{having temporal structure} narrower  than the reciprocal, $\nu^{-1}$, of the observation frequency, e.g. sub-nanosecond, 
\newtext{
as follows from Fourier analysis.
}
Large numbers of shots then comprise 
observed pulses with durations and internal structure from microseconds to many milliseconds.  

\newtext{
Let $\shot(t-\tsj)$ be the form of a  single shot
emitted at time $\tsj$ with an impulsive form  determined by the emission process and modified by  the temporal response of a receiver.  Though narrowband shots in the form of oscillatory wave packets are conceivable, there is no empirical evidence in support of their presence. Moreover, the occurrence of broadband shots (i.e. bandwidths $\gtrsim 20$\% of the center frequency) has been established for the Crab pulsar \citep[][]{he07} and for the MSP J0437$-$4715 \citep[][]{2011MNRAS.418.1258O}.  Our assumption of its generality  suffices, statistically, for our analysis  of pulsar timing. 
}

Pulsar signals are often treated as complex baseband voltages  obtained by mixing
the signal selected in a narrowband receiver channel in a single hand of polarization with a complex local oscillator signal, $e^{-2\pi i \nu_0 t}$, and low-pass filtering.  The baseband signal of Eq.~\eqref{eq:vBB2} for shot noise is then
\be
\vBB(t) = \sum_j \shotampj \shot(t-\tsj).  
\label{eq:shots1}
\ee 
\newtext{
where the coefficients $a_j$ are complex amplitudes. 
}
A rectangular receiver bandpass  yields $\shot(t) = \sinc \Br t \equiv (\sin \pi\Br t) / \pi\Br t$. 
The amplitude $\shotampj$ is related to the peak shot flux density as $\vert \shotampj \vert = (2\pi S_{\nu_0, {\rm max}, j} / c \Br)^{1/2}$.  A general statistical model would include a joint distribution
$f_{\rm shots}(\shotampj, \tsj)$; however, empirical and practical considerations imply that it is sufficient to consider 
shot amplitudes and delays to be decoupled\footnote{Examples shown in \citet[][]{he07, 2003Natur.422..141H}] show no strong relation between amplitude and delay. }.

Pulses and their substructures are orders of magnitude wider than shots (e.g. $\mu$s and ms vs. ns for shots) , requiring variable shot amplitudes $\shotampj$ or a variable shot emission rate.  This is  motivated by the 
amplitude modulated polarized  shot-noise (AMPSN) model \citep[][]{ric75, cor76a} with applications to the Crab pulsar that include interstellar scintillation \citep[][]{cbh+04}, and extended to FRB contexts by \citet[][]{cw16,nimmo2022}.  For a single polarization channel the baseband voltage is written as the product,
\be
\varepsilon(t) = a(t) \amnnoise(t),
\label{eq:amn}
\ee
where $a$ is a real modulation and $\amnnoise$ is complex noise. 
In the rest of the paper, we specify the statistical moments of $\amnnoise$ as
\be
\begin{array}{ll}
\langle \amnnoise \rangle  = 0  								& \quad\quad\quad  \text{zero mean} 
\\
\langle \amnnoise(t_1) \amnnoise(t_2) \rangle = 0							& \quad\quad\quad \text{zero correlation}
\\
\langle \amnnoise(t_1) \amnnoise^*(t_2) \rangle = \amnnoisevar \Delta(t_1- t_2) 	& \quad\quad\quad \text{variance $\amnnoisevar$ and auto correlation function $\amnnoiseacf$}.
\end{array}
\ee

The corresponding  single-polarization intensity is
\be
I(t) = a^2(t) \vert \amnnoise(t) \vert^2 \equiv \amnAmp(t) \amnNoise(t)
\ee
where $\amnNoise(t)$ has moments $\langle \amnNoise \rangle = 1$ and $\langle \amnNoise^2 \rangle = 1 + m_{\amnNoise}^2$ written in terms of the
modulation index,  $m_{\amnNoise} \equiv \sigma_{\amnNoise} / \langle \amnNoise \rangle$.   For dense shot noise with separations comparable or less than shot widths, 
 $\amnNoisemod = 1$ whereas sparse shots  $\amnNoisemod > 1$.    Most pulsar investigations yield $\amnNoisemod = 1$ for a few CPs and the MSP B1937+21  \citep[e.g.][]{ch79,jap01} except for high time resolution studies of the Crab pulsar that have revealed individual shots \citep[][]{2003Natur.422..141H, he07}. 

\subsection{Pulse shapes and average profiles}
\label{sec:pulse_shapes}

Pulsar timing methodology is usually based on averages of a large number of single pulses, each  formed by time averaging over $\Delta t \sim 1\,\mu$s intervals for MSPs and up to $\sim1$\,ms for long period objects.   Average profiles are usually formed with 1024 to 4096 time bins per period. This intrapulse averaging reduces  the shot noise modulation index by a factor $(\Delta t B)^{-1/2}$, where
the time-bandwidth product $\Delta t \times B $ is typically $\gg 1$.     The remaining variance from this `self noise'  makes a very small contribution to the overall error in \TOA\ estimates compared to those from additive radiometer noise and pulse jitter, as discussed later.

\begin{figure}[htbp] 
   \centering
   \includegraphics[width=0.7\linewidth]{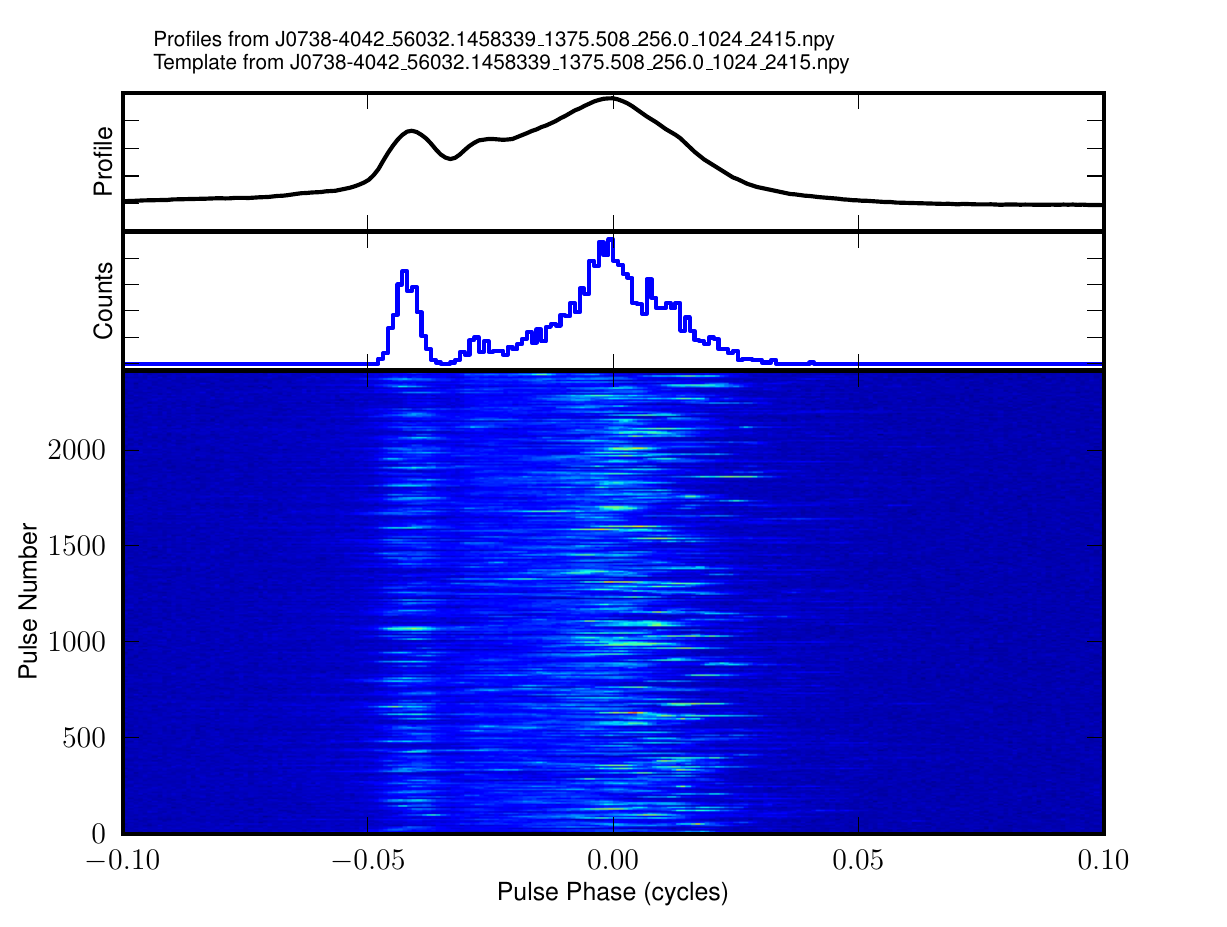}
   \caption{Main panel: a sequence of pulses from PSR J0738-4042 ($P = 0.37$\,s,  DM = $161\, \DMunits$) obtained at 1.38\, GHz with the Parkes telescope (R. Shannon, private communication).  Top panel: average profile from the 2500 pulses shown.  Middle panel:  counts of pulse maxima vs. pulse phase. 
     }
   \label{fig:pulses_J0738-4042}
\end{figure}

Amplitude modulations of the self noise involve multiple pulse components with  varying amplitudes and pulse phases in individual pulses that average into the stable forms shown in Fig.~\ref{fig:pulse_gallery}.   
Pulse components physically correspond to  sub-beam structure in the `lighthouse' beam that rotates through the LoS.    This physical view is consistent with observations showing that jitter is largely independent between components, \newtext{as
can be seen in Fig.\,\ref{fig:pulses_J0738-4042}, which shows a typical example for PSR J0738$-4042$.  This picture is 
substantiated by statistical and Fourier studies of single pulses going back to the early days of pulsar science \citep[e.g.][]{1973ApJ...182..245B}.}
The number of components and their separations are significant factors in the overall timing variance for a given pulsar.
Here we present a suitable model for average profiles that includes  components and their fluctuations.

The envelope of the $j^{\rm th}$ pulse is a sum over $\nc$ pulse components,
\be
A_j(t) = \sum_{i=1}^{\nc} c_{ij} \Ac_{ij} (t - t_{ij}),
\label{eq:ampmod}
\ee
where subpulse amplitudes $\ampij$ and delays $\tij$ vary statistically about mean values from processes intrinsic to the pulsar (e.g. jitter)  but also from propagation through intervening plasmas (next section). 
For pulsars, these quantities generally have well defined mean values that determine the shape of the template profile average  at a specific center radio frequency.   For simplicity we have suppressed the radio frequency dependence of $A_j$ because it is relatively gentle.   A few pulsars deviate from this simple characterization, \newtext{including the Crab pulsar (J0534+2200) that has shown a secular change in component separations \citep[][]{2013Sci...342..598L} and MSP J1713+0747 that has shown a sustained transient evolution of its average pulse shape \citep[][and references therein]{2024ApJ...964..179J}}\footnote{The shapes of fast radio bursts  (FRBs) can be described similarly but with two significant differences.  First, they are not periodic in the same way as pulsars and, second,  the components have different center frequencies and relatively narrow bandwidths.}.

The average profile is the sum over $N$ pulses,  
\be
I_N(t) 
= N^{-1} \sum_{j=1}^N \sum_{i=1}^{\nc} \ampij \ap_i(t-\tij).
\label{eq:IN1}
\ee
A good approximation is that $\ampij$ and $\tij$  are statistically independent of each other and between different pulses (different $j$) \newtext{as seen in Fig.~\ref{fig:pulses_J0738-4042}}.  Pulse sequences from some objects do show correlations in the form of `drifting subpulses' and other effects \newtext{\citep[][]{1975ApJ...197..481B}}, but these do not alter our analysis significantly.   Also, for MSPs, these correlations seem to be largely absent. 

An illustration of this empirical model is shown in Fig.~\ref{fig:pulses_J0738-4042}.   The average profile in the top  panel displays several components.  Single pulses comprise subpulses that occur preferentially in these profile components, as demonstrated in the histogram of the pulse phases in which single-pulse maxima occur.    The main panel shows that subpulse amplitudes and phases  occur independently in different components and with different frequencies of occurrence.    The average profile is determined by both the amplitudes of subpulses and their frequency of occurrence in pulse phase.

The template shape is idealized as the ensemble average  (corresponding to $N\to\infty$), 
\be
\TempAB{}{}(t) 
= \left\langle  I_N(t) \right\rangle 
= \sum_i \cijave \ap_i(t - \tijave).
\label{eq:U}
\ee
In practice, templates are based on finite $N$ averages and application of a denoising procedure to minimize the influence of off-pulse noise on \TOA\ estimates. 

\TOAs\ are calculated through cross correlation or, equivalently, least-squares fitting of the template to a measured profile. The \TOA\  is given by the maximum cross correlation obtained as the solution of $C^{\prime}_{I_N \TempAB{}{}}  (\tau) = 0$ where $C_{I_N \TempAB{}{}} $ is the cross correlation function between the template and a finite-sum profile.    Differences between $U$ and $I_N$ yield \TOA\ errors that add in quadrature to those from additive noise and template fitting.    We refer to these differences as `pulse jitter' in the \TOA\ error budget. 

Implicit in Eq.~\eqref{eq:IN1}-\eqref{eq:U} is that the time $t$ extends over only one period (and thus corresponds to pulse phase $\phi = t / P$ for a period $P$). This requires alignment of pulses according to a previously generated timing ephemeris.  It is also assumed that plasma dispersion delays have been removed with small enough error that pulse shapes are undistorted.   

\begin{figure}[htbp] 
   \centering
   \includegraphics[width=0.495\linewidth]{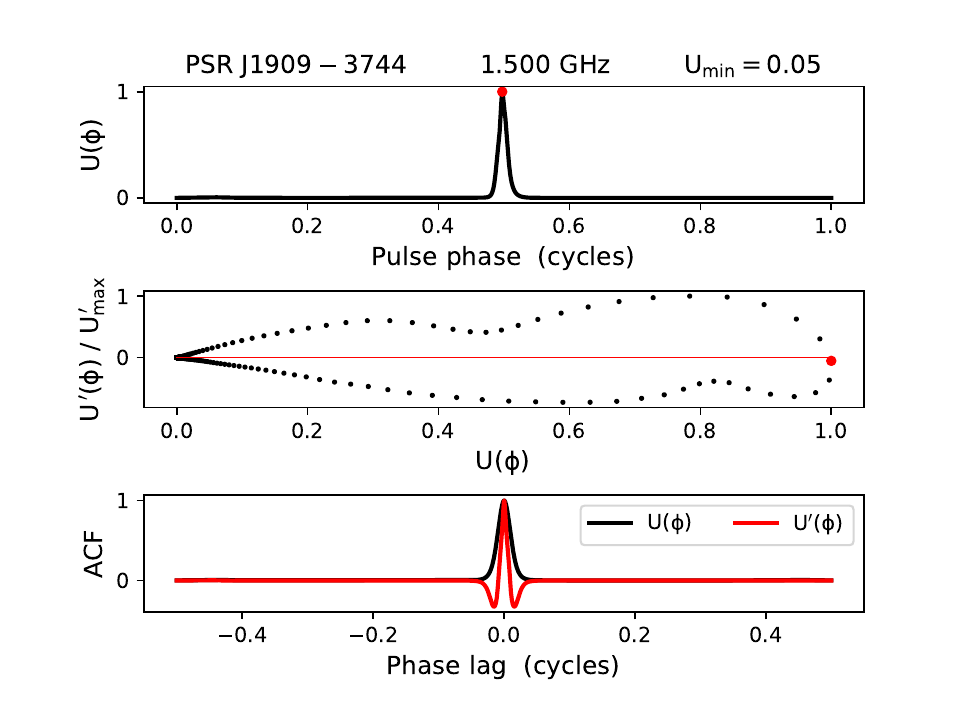} 
   \hoffset -0.5in
   \includegraphics[width=0.495\linewidth]{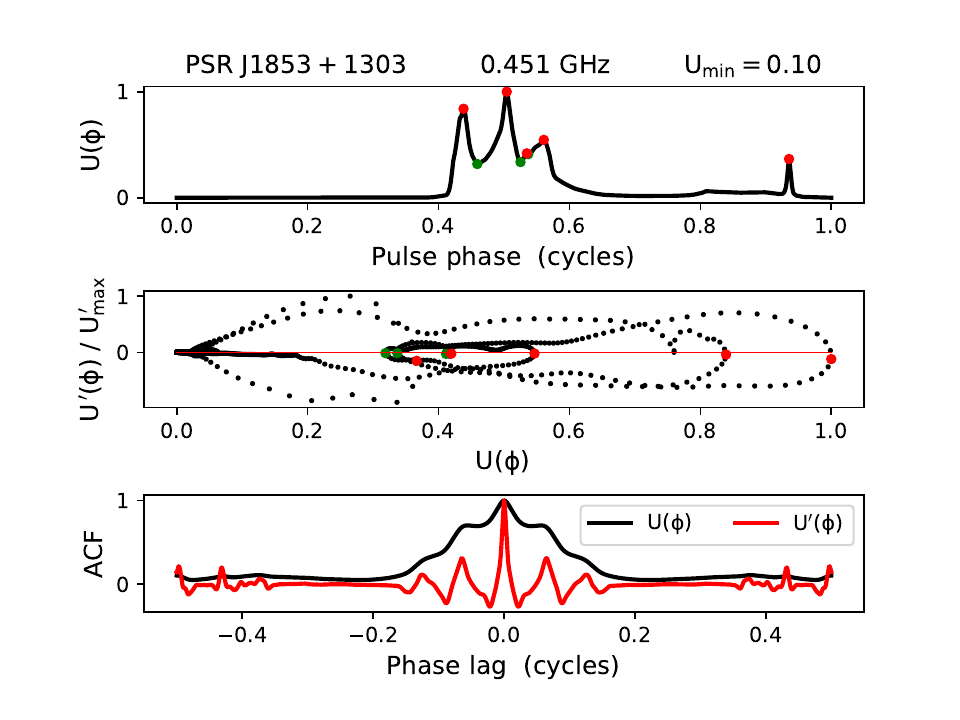}
   \caption{Profile component analysis for J1909-3744 (left) and J1853+1303 (right).  The top panels show the template $U(\phi)$, the middle panels $U^{\prime}(\phi) $vs. $U(\phi)$, and the bottom panels the autocorrelation functions of $U$ and $U^{\prime}$. Data are from NANOGrav's 15-yr data release.}
   \label{fig:uuprime1}
\end{figure}

Identifying and modeling  profile components is another method  for constructing  templates using Gaussian functions.   The motivation is to better understand how the number of components and their separations contribute to the \TOA\ error budget. 
An objective approach to component identification identifies inflection points in the profile $U(\phi)$ using the  first derivative,  $U^{\prime}(\phi)$ (where we now use pulse phase $\phi$ rather than time).    Fig.\,\ref{fig:uuprime1} shows  profiles,  derivatives, and autocorrelation functions for two MSPs.  
\newtext{We have plotted $U^{\prime}(\phi)$ vs. $U(\phi)$ in the middle panels to highlight inflection points in the profile that occur at low as well as high amplitude locations of $U(\phi)$.} 
The red points in the top two panels  indicate zero crossings of $U^{\prime}$ where $U$ exceeds a threshold $U_{\rm min}$, yielding 
$\nc = 1$ for J1909-3744 and $\nc = 5$ components for J1853+1303.  \newtext{The relative  contributions to \TOA\ errors scale as the squares of the component amplitudes because \TOAs\ are determined through template fitting that involves multiplication of a template with a measured average profile. }

\begin{figure}[htbp] 
   \centering
   \includegraphics[width=0.495\linewidth]{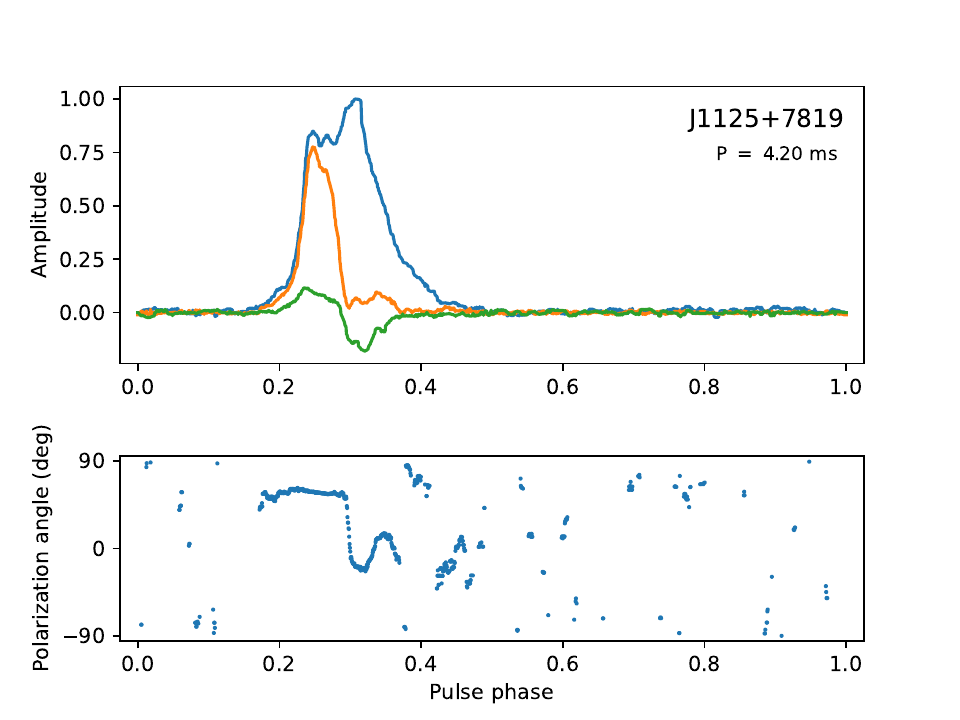} 
   \hoffset -0.5in
   \includegraphics[width=0.495\linewidth]{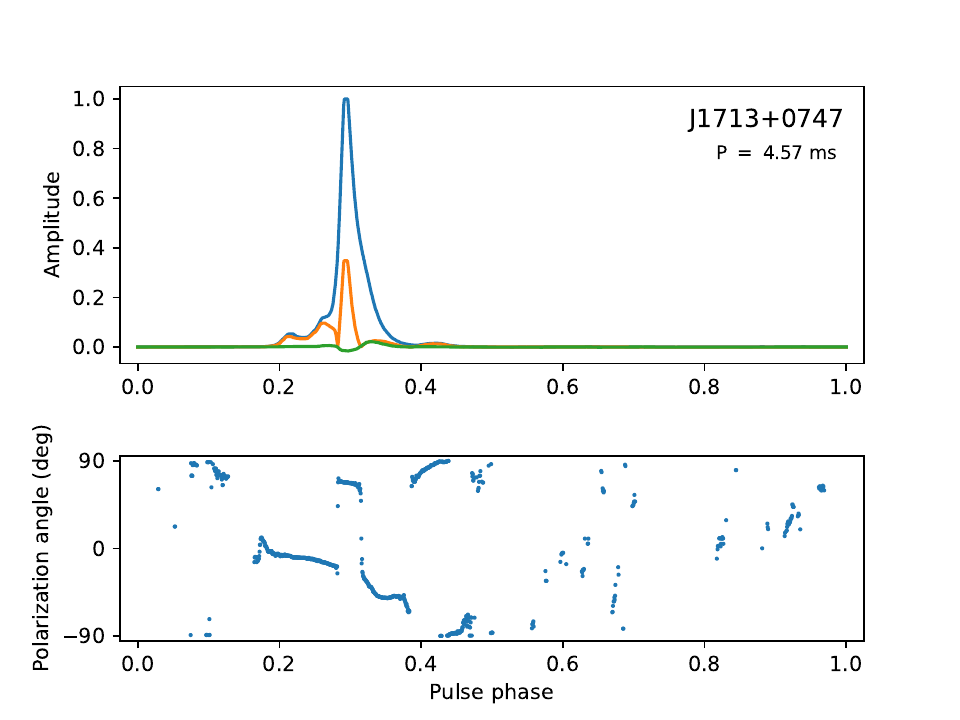}
   \caption{
   Stokes profiles for two millisecond pulsars in NANOGrav's timing program. 
   Profiles are from the analysis of  NANOGrav's 12.5 yr data set \citep[][]{Wahl_2022} and were
   downloaded from \url{https://nanograv.org/science/data/polarization-calibration-data} (Version 2.1).
   For each case the top panel shows Stokes $I$ (blue), $L$ (orange), and $V$ (green) and the bottom 
   shows the polarization angle  for values of $L$ larger than $3\sigma_L$. 
   Left:  J1125+7819, an example of high linear polarization that is narrower than Stokes $I$. 
   Right: J1713+0747, an example complex, narrower structure in $L$ than in $I$ but with 
   less than 40\% linear polarization
    }
   \label{fig:stokes2psrs}
\end{figure}

Examples of polarization profiles (Fig.\,\ref{fig:stokes2psrs}) show how Stokes $L$ and $V$ profiles differ from 
Stokes $I$ and that the degree of linear polarization can be large.    We refer to these profiles in  Sect.~\ref{sec:fullstokes} where \TOA\ estimation using all Stokes parameters is discussed.

\subsection{Plasma propagation effects at radio wavelengths}
\label{sec:prop}

Atomic and molecular gas  have essentially no influence on pulse propagation  except in narrow emission lines that increase the system noise and absorption lines that diminish the pulsar signal \citep[][]{deJager68,dickey81,frail91}.  In extraordinary circumstances the pulsar signal can be enhanced by maser amplification, as seen in the OH line \citep[][]{2005Sci...309..106W}.  

Variations in gravitational potential also affect pulse propagation times and in very rare circumstances may produce multiple images.    Effects from binary companions produce the well known Shapiro delay that has famously been used for precision determinations of NS masses \citep[e.g.][and references therein]{2020NatAs...4...72C}.  That  subject is beyond the purview of this paper. However, Shapiro delays can  be measured from intervening stars, planets, and dark matter clumps \newtext{\citep[e.g.][]{desai18,1999A&A...351..393H}}.   By and large, these instances in the weak field regime are addressed by expressing the gravitational potential
$\Phi$ as an equivalent  index of refraction, 
\newtext{$\nr \simeq  1 - 2\Phi/c^2$ \citep[][\S\,24.5]{1973grav.book.....M}}.

By far the most prominent effects are dispersion and scattering caused by  plasmas encountered along the LoS, including the Earth's ionosphere and the solar wind, but the dominant effects are from partially ionized gas  in  multiscale ISM structures and turbulence extending from  $\sim 100$~km  to kpc scales. \newtext{These media are compared in Table\,\ref{tab:plasmavalues} and dispersion and scattering   are discussed below in this section and in more detail in
Part III `Interstellar Effects.'}

\renewcommand{\Efield}{\vBB}
\renewcommand{\xobs}{\xvec_{\rm o}}

All relevant effects on pulsar signals can be analyzed in terms of  the index of refraction $\nr$ of a cold, tenuous, and magnetized plasma with electron plasma frequency $\nuplasma = \ompe / 2\pi = (\nelec e^2 / \pi \melec)^{1/2}$ and cyclotron frequency $\nuc = \omcy/2\pi = e B / 2\pi\melec c$, where $e$ and $\melec$ are the electron charge and mass, $\nelec$ is the electron number density, and $B$ is the magnetic field.    
Table~\ref{tab:plasmavalues} gives typical values of these quantities and integrated measures for the ISM, which dominates pulsar measurements, along with those for the IPM near the solar system and the ionosphere. 


%

\begin{deluxetable}{lccccc}
\tablewidth{450pt}
\tablecaption{\label{tab:plasmavalues}Plasma parameters for the Ionosphere, interplanetary medium (IPM), and interstellar medium (ISM)}
\tablecolumns{6} 
\tablehead{
\colhead{Quantity} &
\colhead{Units} &
\colhead{Symbol}  &  
\colhead{Ionosphere\tablenotemark{\footnotesize b}} &
\colhead{IPM@1 au\tablenotemark{\footnotesize c}}  &
\colhead{ISM\tablenotemark{\footnotesize d}}  
}
\startdata
Electron density & cm $^{-3}$   & $\nelec$  			& $10^6$ [max] & 5 & 0.03    \\ 
Magnetic field   & $\mu$G  	    & $B$ 				& 3$\times10^5$ & 100 & 3  \\ 
Plasma frequency & kHz  	    & $\nuplasma$			& 9000  & 20  & 2     \\ 
Electron cyclotron frequency & Hz 		& $\nu_{\rm c, e}$		& $10^6$ & 300  & 10  \\
$\delta$(refractive index) @ 1 GHz & ---& $\delta\nr = 1-\nr $		& $10^{-5}$  & $10^{-10}$ &  $10^{-12}$  	  \\
\hline
Integrated measures\tablenotemark{\footnotesize a} \\
\quad Dispersion measure & $\DMunits$ & \DM\ 		& $10^{-6}$ - $10^{-5}$  & $\gtrsim 3\times10^{-5}$   & 3 to 1700  \\
\quad Rotation measure   & $\RMunits$ & $\vert \RM \vert$ 		& 0.5 - 2 & $\gtrsim 4\times10^{-4}$    & few to $> 10^3$ \\
\quad Emission measure   & $\EMunits$ & \EM\ 				& $\sim 3$ & $\gtrsim 10^{-4}$  & $< 1$ to $> 10^5$ 
\enddata
\tablenotetext{a}{Integrated measures are given for lines of sight away from the Sun.   Much larger DM and RM values have been  measured for \LOS\ making close approaches  \citep[e.g.][]{spangler05,Jensen+2018,Kooi+22}. }
\tablenotetext{b}{$\nelec$: \citet[][IRI model]{Bilitza+22}, \citet[][]{2018RaSc...53..724M}; Magnetic field: \citet[][IGRF-12 model]{thebault+15}; RM: \citet[][]{2019MNRAS.483.4100P}}
\tablenotetext{c}{$\nelec$: \citet[][]{ratcliffe72}; B: \citet[][]{2001GMS...125...73R, Borovsky+20}; RM: 
\citet[][]{Kooi+22}}
\tablenotetext{d}{\citet[][]{Draine2011} }
\end{deluxetable}
\

For $\nuc \ll  \nuplasma \ll \nu$ along with quasi-longitudinal propagation at an angle $\theta$ from the magnetic field direction,  and neglecting very small contributions from ions (predominantly protons),  
\be
\nr^2 = \left( \frac{kc}{2\pi\nu}\right)^2
	\approx 1 - \left(\frac{\nuplasma}{\nu} \right)^2
	      \left(\frac{1}{1\mp\nucpar/\nu} \right).
\label{eq:nr2longitudinal}	      
\ee
where  $\nucpar \equiv \nuc \cos\theta$ and the upper and lower signs apply to right-hand and left-hand polarizations, respectively (IEEE and IAU conventions). 

Chromatic effects are time dependent owing to geometrical changes in the propagation path, predominantly caused by  translational motions of the pulsars, which have much higher velocities than main-sequence stars.  In some cases, Earth's motion around the Sun and orbital motion of binary pulsars also contributes significantly to characteristic time scales, especially for MSPs.

Physical effects that need to be accounted for in pulsar timing, either through mitigation or inclusion in the measurement error budget, are as follows: 

\newtext{We make use of standard propagation measures expressed in their standard units (with distances in pc or kpc, $\nelec$ in cm$^{-3}$,
and $\Bpar$ in $\mu$G) and integrations from $0$ at the source to a distance $z$ along the propagation path}, 
\be
\begin{array}{lllll}
\text{Dispersion measure:\quad }   
	&&\DM = \int ds\, \nelec 
	&& \quad \DMunits ,
	\label{eq:DMdefinition}\\
	\\
\text{Faraday rotation  measure:\quad}  
	&&\RM =  \frac{\displaystyle e^3}{\displaystyle 2\pi \me^2 c^4}  \int ds\, \nelec \Bpar 
			= 0.81   \int ds\, \nelec \Bpar
	&&\quad \RMunits,
	\label{eq:RMdefinition}\\
	\\
\text{Emission measure:\quad} 
	&&\EM = \int ds\, \nelec^2 
	&&\quad \EMunits 
	\label{eq:EMdefinition} \\
    \\
\text{Scattering measure:\quad} 
	&&\SM = \int ds\, \Cnsq 
	&&\quad \SMunits .
	\label{eq:SMdefinition}
\end{array}
\label{eq:DMRMEM}
\ee
\newtext{Definitions for these measures can be found in \citep[][]{lorimer2005handbook} with more details in \citep[][]{Draine2011}.}
The first three measures are discussed immediately below while the scattering measure is deferred to Sect.~\ref{sec:plasmaprop}.

{\bf Phase perturbations:} Propagation imposes an extra phase equal to  the integral of $k(\nr-1)$ along the propagation path.   Paraxial optics applies because deflections from vacuum geodesics  are typically less than one arcsecond so the integral can often be taken along an undeflected path.   The phase  at a transverse position $\xvec$  at position $z$ along the path is
\be
\phi(\xvec, z) &\simeq&  
	- \left[ \lambda \relec \DM(\xvec, z) \pm \lambda^2 \RM(\xvec, z) + 
		\left(\frac{\lambda^3 \relec^2}{4\pi} \right) \EM(\xvec, z) \right].
\label{eq:phi2DMetc}
\nonumber \\
&\simeq& 
-\left[
	2.61\times 10^7 \, \left( \frac{\DM}{\nu}  \right)     
	\pm 0.0899\,  \left( \frac{\RM}{\nu^2}  \right)  
	+ 5.25\times10^{-4} \, \left( \frac{\EM}{\nu^3}  \right)
\right] \ {\rm rad} 
\label{eq:phi_dm_rm}
\ee
 where $\relec$ is the classical electron radius and the frequency $\nu$ is in GHz. 
 
\begin{figure}[b] 
   \centering
   \includegraphics[width=0.5\linewidth]{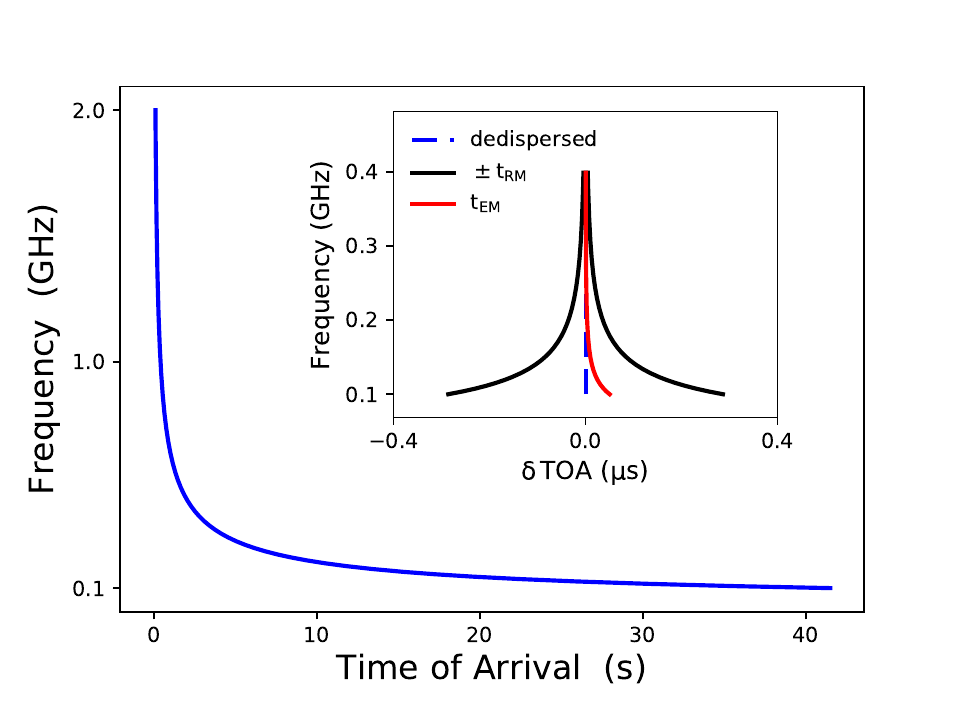}
   \caption{Excess time of arrival vs. frequency for 
   $\DM = 100~\DMunits$, 
   $\RM = 10~\RMunits$ and 
   $\EM = 20~\EMunits$, values chosen to illustrate the effects not to model any particular line of sight.
   Inset:  Excess arrival time after dedispersion of only the DM term.  }
   \label{fig:dm_rm_em_sweep}
\end{figure}
 
{\bf Dispersion:}
 The group delay adding to the vacuum propagation delay $d/c$ is $\td = (1/2\pi) d\phi/d\nu$ or, equivalently,  the integral involving the group velocity 
 $\vgroup= \partial\omega/\partial k =  c (\nr + \nu d\nr / d\nu)^{-1}$
 over the  pulsar-solar system distance\footnote{Note that pulsar timing analyses are referenced to the solar system barycenter, which implies that the Doppler shift from the Earth's motion needs to be included.} $d$,
 $\td=\int_0^d ds\, (\vgroup^{-1} - c^{-1})$.     Evaluating to
 first order in $\nucpar/\nu$ and fourth order in $\nuplasma/\nu$ (\citealt[e.g.][]{2019ApJ...870...29S}; see also
 \citealt[][]{pw91,2012AA...543A..66H}),
 \be
 \td =   t_{\DM} \pm  t_{\RM} + t_{\EM}
 \simeq  
	    \kDM \left(\frac{\DM}{\nuGHz^2} \right)
	\pm \kRM \left(\frac{\RM}{\nuGHz^3} \right)
	+   \kEM \left(\frac{\EM}{\nuGHz^4} \right)  
\label{eq:td}
\ee
where the coefficients are 
\be
\kDM = \frac{e^2}{2\pi \me c}  = \frac{c \re}{2\pi}   = 4.15~{\rm ms} , \quad
\kRM = \frac{c^2}{\pi}  = 28.6~{\rm ps},  \quad
\kEM = \frac{3 e^4}{8\pi^2 \me^2 c}  = 0.251 ~{\rm ps},
\label{eq:tdcoeffs}
\ee

\newtext{for standard units of pc cm$^{-3}$ for DM, rad m$^{-2}$ for  RM, and pc cm$^{-6}$ for EM.} DM fluctuations occur on a wide range of time scales associated with transport of the \LOS\ across multiscale density structures by various motions.   Because DM delays are large, these fluctuations must be tracked vs. epoch, as discussed in the next section.  

Values of plasma quantities in Table~\ref{tab:plasmavalues} indicate that dispersive time delays are dominated by the ISM. Nonetheless, contributions from the ionosphere and IPM make noticeable contributions to the most precise arrival times and their variability on short time scales (hours, days) implies a contribution to the timing error budget if they are not mitigated using multifrequency observations. 
As an example,  a 10\,ns delay at 1\,GHz corresponds to $\delta\DM \sim 2.4\times10^{-6}\,\DMunits$, a value that is reached for all three media characterized in the table.

{\bf Birefringence and Faraday rotation:}
The birefringent RM terms imply that  a narrow pulse splits into  its
differentially arriving circularly-polarized components \citep[][]{2019ApJ...870...29S}.  
The \DM\ term in Eq.~\eqref{eq:td} is by far dominant and it is only in very dense, highly magnetized clouds that the RM and EM terms will matter in low-frequency observations, as demonstrated in Fig.~\ref{fig:dm_rm_em_sweep}.   The inset shows that the RM and EM terms reach $\mu$s  levels only at frequencies $\ll 1$~GHz.    They may be important for burst emitters in dense regions with large magnetic fields (e.g. fast radio burst sources) but for precision timing of  Galactic millisecond pulsars, the EM term is negligible and the RM term is typically less than a few ns.  

 Birefringence is typically measured as rotation of the $E$ vector  vs. wavelength by the angle $\psi = (\phi_+ - \phi_-) /2 = \lambda^2 \RM$.   Ideally it would have no effect on arrival times but instrumental polarization and miscalibration of polarization channels can render it important.  Under those conditions,  imperfect removal of Faraday rotation across the data band  will yield a systematic error that can be epoch dependent.

\begin{figure}[t] 
   \centering
   \includegraphics[width=0.5\linewidth]{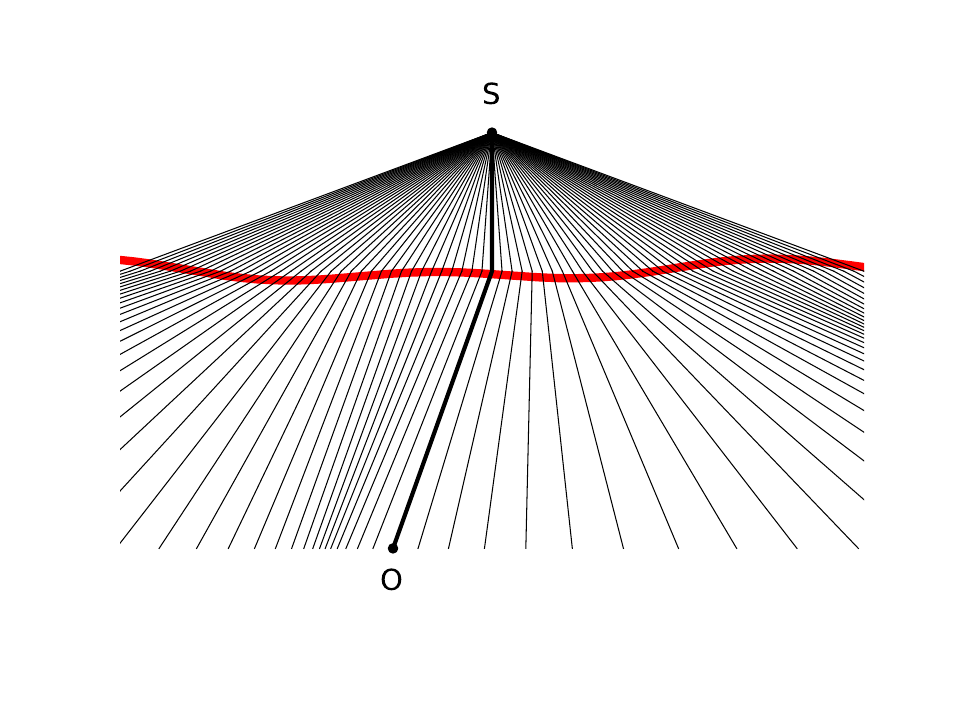}
   \includegraphics[width=0.45\linewidth]{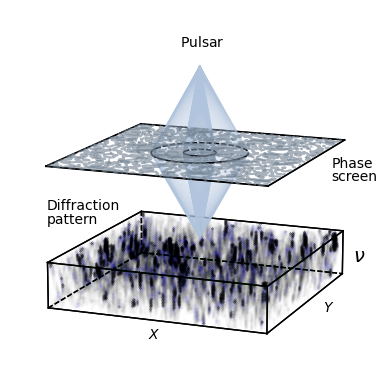}
   \caption{Left: 
   Schematic thin refracting screen that changes  the wavefront (red) and thus the angle of arrival seen by an observer O and the intensity, represented by the variable density of rays in  geometrical optics.  Refraction in the local ISM is sometimes dominated by a single screen-like structure while for distant objects it is a cumulative effect from multiple structures. 
   \label{fig:refracting_screen}
   Right:
   Illustration of scattering from a thin  screen with phase fluctuations produced by Kolmogorov fluctuations in refractive index. 
The cones with apexes at a pulsar
and at an observer's position in the observation plane represent ray path bundles of scattered radiation.  The narrow and wide cones correspond to a high and a low frequency, respectively.  The diffraction pattern is shown as a function of $X, Y$ in the observation plane and as a function of radio frequency $\nu$.
\label{fig:scattering_cone}
   }
\end{figure}


\begin{figure}[b!] 
   \centering
   \includegraphics[width=0.6\linewidth]{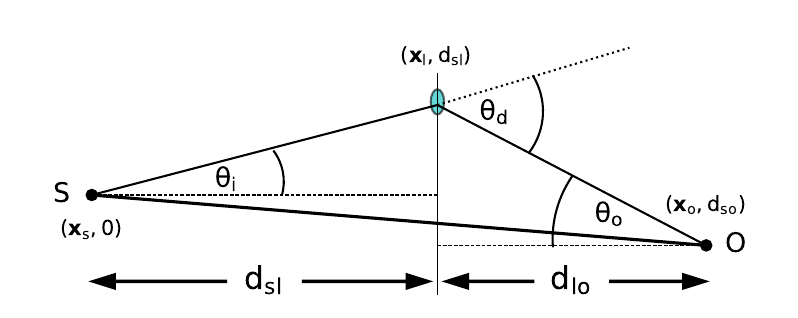}
   \caption{
   Thin screen geometry for diffraction and ray tracing and for  the application of the Fresnel-Kirchhoff diffraction
    integral.   Small angles $\ll 1$\,rad are assumed for diffraction into an angle $\thetad$ in the screen at distance
   $\dso$ from the source S.   Transverse coordinates $\xsvec, \xlvec$ and $\xobs$ define locations in the source, screen layer, and observation planes. 
   } 
   \label{fig:geometry}
\end{figure}

{\bf Refraction:}. Large scale density variations on  $\gtrsim\au$ scales 
  alter the angle of arrival (\AOA)
and cause slow flux density variations
\cite[refractive interstellar scintillations, RISS;][]{rcb84}.  
Fig.~\ref{fig:refracting_screen} (left panel) shows these effects schematically for  a thin screen geometry.
In this case the deflection angle produced by the screen is 
$\thetar = k^{-1} \nabla_{\xvec} \phi(\xvec, z)$ and intensity variations are   related to the second derivative. 
Refractive scintillations alter the signal-to-noise ratio and thus cause episodic changes in achievable 
\RMS\ \TOA\ errors from matched filtering.
Variations in \AOA\ introduce further arrival time  delays to the dispersive delays.      If deflection angles are large enough, bright intensity caustics  and ray crossing   can cause pulse echoes that perturb estimated \TOAs.   Collectively the empirical manifestations of these effects are referred to as `extreme scattering events' (ESEs) even though they are refractive in origin.  Instances of ESEs in pulsar timing   appear to be rare \citep[][]{1993Natur.366..320C, 2018MNRAS.474.4637K}.

{\bf Diffraction (scattering):}.  
Diffraction arises if  the wavelike property of incident radiation plays a role in changing the direction of propagation, in contrast to geometrical optics.  
Fig.\,\ref{fig:geometry} shows the geometry for a thin screen at distances $\dsl$ and $\dso-\dsl$ from the source and an observer, respectively.  A representative diffraction angle $\thetad$ corresponds to an 
initially emitted angle $\thetai$ and an observed angle $\thetao$.  
Density variations on small scales ($\ll \au$) cause forward scattering and  multipath propagation that produce important effects for pulsar timing: 
diffractive intensity scintillations (DISS)  vs. time and frequency and pulse distortion from differential propagation times.  These effects are illustrated in Fig.\,\ref{fig:scattering_cone} (right hand panel).

\begin{figure}[h!] 
   \centering
   \includegraphics[width=0.6\linewidth]{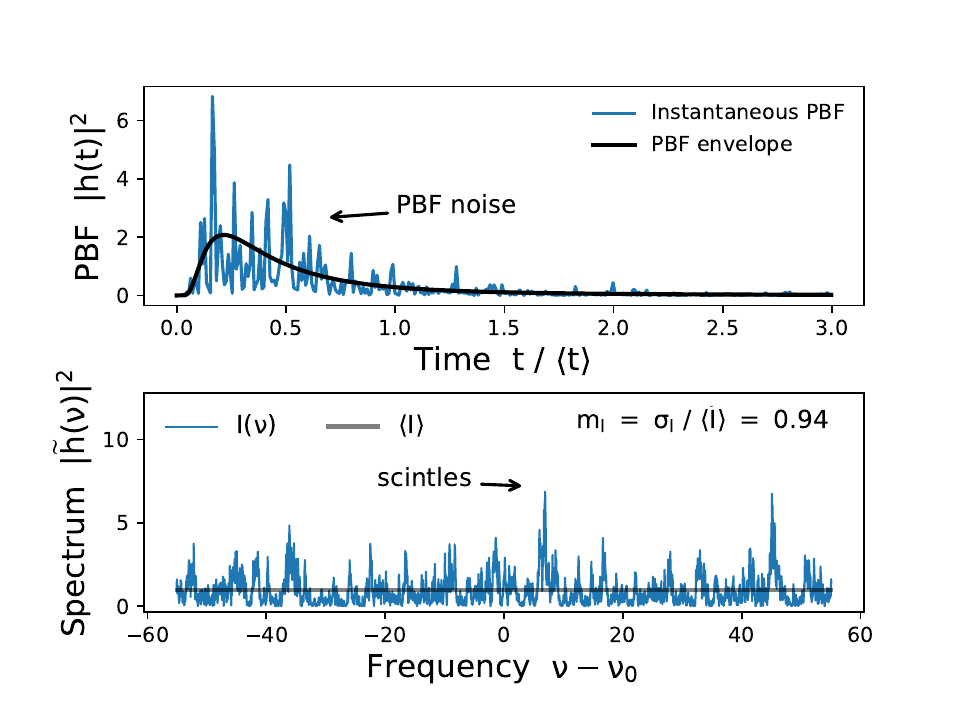}
   \caption{Top: Pulse broadening function for for a pervasive  medium with a Kolmogorov electron density spectrum
   and negligible inner scale.  The single realization shown displays  the noise-like process that is modulated by an       
   envelope function.   The noise process remains constant over a DISS time scale while the envelope is sustained for much longer time spans. 
   Bottom:  Spectrum corresponding to the field PBF normalized to the spectral mean (black line) and plotted 
   against the radio frequency $\nu$ minus a center frequency $\nu_0$. 
   Spectral amplitudes are exponentially distributed so the ensemble
   average modulation of the spectral intensity is unity.   The single realization shown here  slightly departs 
   (statistically) from a modulation index of unity. }
   \label{fig:pbfave}
\end{figure}

 The pulse broadening function (PBF) is typically defined using only geometrical path length differences and represents the scattered pulse shape produced for an emitted delta-function.  In the frequency domain,
 multipath is manifest as constructive and destructive interference that strongly modulates the spectrum.  A spectrum and PBF  are illustrated in Fig.~\ref{fig:pbfave}.  The constructive and destructive interference underlying scintillations   is distinct from caustics that occur in geometrical optics, where deflections are related to phase gradients that occur even in the limit of infinitesimal wavelengths.
 
{\bf Reflection:}   A plasma structure can reflect radiation if the plasma frequency is comparable to the electromagnetic frequency or if there is grazing incidence on an interface between two plasma densities.    Reflections will alter the flux density and polarization. While  currently it is unclear  whether reflections have influenced any pulsar observations to date,
they may be manifested in rare occasions as multiple images and pulse echoes in long timing programs. 

\subsection{Interstellar impulse response functions}
\label{sec:impulseresponse}

The net impulse response of the ISM is a combination of scalar dispersion \newtext{(i.e. not dependent on the wave polarization)}, Faraday rotation, and multipath propagation
(where we  ignore the very small EM term).  
The phase given in Eq.~\eqref{eq:phi_dm_rm} for the two hands of circular polarization indicate that the impulse response due to DM and RM is simply a phase change of Fourier components by a complex factor 
$\hDMRMFT(\nu) = \exp[i\phi(\xvec, z)]$ for the observer's location, which is factorable into separate DM and RM components, $\hDMFT$ and $\hRMFT$.   Coherent dedispersion comprises multiplication of measured Fourier components by the complex conjugate of this factor.   In practice, this operation is applied to baseband data with bandwidth $B$ and, to date, only the DM contribution to the phase is considered in the vast majority of applications. 
Multipath, primarily from diffraction accompanied by refraction is not so simple to mitigate.   Pulse broadening, as already described, involves a `field' PBF denoted as $h_{\rm d}(t)$. The total interstellar impulse response is the
time-domain convolution of the three factors (and a receiver bandwidth factor that we keep implicit), 
\be
h_{\rm ism}(t) = \hDM(t) * \hRM(t) * h_{\rm d}(t).
\label{eq:himpulse}
\ee

Dedispersion of the $\hDM$ term and derotation of the $\hRM$ term are both exercises in matched filtering and can be done exactly if the true DM and RM values are known. 

{\bf Dedispersion:}  The dispersion delay in Eq.~\eqref{eq:td} varies significantly across frequency bands used in pulsar studies and thus must be removed before summing pulse intensities over frequency to improve the signal to noise ratio and to maximize the net time resolution and timing precision.  

An approximate method is `post-detection' or {\it incoherent dedispersion} used for intensities obtained with a multi-channel receiver.  Pulses \newtext{from  each channel} are aligned by shifting each intensity by $\td(\nu_{\rm ref}) -\td(\nu)$,
where $\nu_{\rm ref}$ is a reference frequency.    This method cannot remove the differential delay across a frequency channel and thus distorts the pulse from its emitted form.   The minimum smearing time of the pulse  results when the dispersion delay across a channel $(d\td/d\nu)$ equals the natural channel response time $\Delta\nu_{\rm channel}^{-1}$, so narrower channels can minimize but not remove  this effect. 

The more advanced (and exact) approach is {\it coherent dedispersion} that properly corrects the Fourier phase of the sampled electric field  contained in the baseband voltage \citep[][]{1971ApJ...169..487H, hr75}.     The gist of the method is easily seen by considering the measured electric field $E(t, z) = \int d\nu\, \widetilde{E}(\nu, 0) \exp{i(k(\nu)z-2\pi\nu t)}$ in terms of the Fourier transform of the emitted field $\widetilde{E}(\nu, 0)$ 
  for a medium with constant refractive index $\nr$ given by  Eq.~\eqref{eq:nr2longitudinal}.  The wavenumber
is $k(\nu) = \nr(\nu) k_0$ where $k_0$ is the vacuum wavenumber.   The dedispersed signal is obtained by multiplying the Fourier transform of the measured field $\widetilde{E}(\nu, z)$ by $\exp{-i k(\nu)}$.  
Inspection of Eq.~\eqref{eq:phi_dm_rm} shows that the dominant phase term involves the DM term.    \newtext{In  principle, large values of $\vert \RM \vert$ (or low frequency observations) may require inclusion of the $\nu^{-3}$  \citep[e.g.][]{2021A&A...652A..34B}  term but to date there has not been routine implementation
of coherent Faraday derotation in pulsar timing programs.}

For most cases, DM is  the sole   parameter needed for coherent dedispersion and the required  precision on \DM\  to avoid 
pulse distortion depends on radio frequency and on  the widths of pulse features  \citep[][]{hr75, cm03}.  We discuss the details of DM determination in a later section. 

Implementation of coherent dedispersion with baseband data is discussed in the original papers by 
\citet[][]{1971ApJ...169..487H} and \citet[][]{hr75},
 with applications to full Stokes parameters  \citep[][]{1977ApJ...218..484C,2002ASPC..278..251S}, and
execution of graphics processing units for high data rate applications to pulsars and FRBs
\citep[e.g.][]{2017JAI.....641011R,2018sf2a.conf..301B,2024RAA....24g5011Z}.

In most of the paper, we assume that dedispersion is perfect  across relevant observing bands, requiring that epoch-to-epoch variations in DM are included;    the  primary element of the net impulse response is then
the diffraction term. 

\subsection{Intensity statistics for scintillated amplitude modulated  shot noise}
\label{sec:SAMSN}

\newtext{Emitted signals from pulsars, as discussed previously, are well described as modulated shot noise.  Later, in Sect.~\ref{sec:dtoa_single_pulse_stochasticity}, we discuss contributions to \TOA\ errors from elements of this signal model.   Here we discuss, using a simple approach,  how interstellar scintillation affects the statistics of the pulsar intensity.  A salient point is that in the limit of dense shot noise and strong scintillations,
both the emitted signal and the scintillation modulation conform to Gaussian statistics. The multiplicative aspect of scintillation enhances the net variance of the measured pulsar intensity, which in turn is relevant to the \TOA\ error budget.}

Diffraction causes multipath propagation at  radio frequencies where phase fluctuations exceed about 1\,rad (RMS) over a transverse separation smaller than a Fresnel scale (discussed in detail in Sect.~\ref{sec:diss_analytics}).   Multiple paths replicate individual shots in the  shot noise signal of Eq.~\eqref{eq:shots1}.  Representing the the impulse response for propagation as a weighted sum  of delta functions,
$g(t) \sim \sum_k g_k \delta(t - t_k)$
 the scintillated signal  (with `s' subscript)  becomes  \citep[e.g.][]{cbh+04}
\be
\Efield_{\rm s}(t) = 
\sum_j \shotampj   \sum_k g_k
\Delta(t - \tsj - \tsk).
\label{eq:sampsnnu}
\ee

Intensity statistics in the frequency domain are useful for assessing timing errors.  The voltage 
Fourier transform,  
\be
\EfieldFT_{\rm s}(\nu) 
= 
\widetilde\Delta(\nu) 
\sum_j \shotampj   
e^{-2\pi i \nu \tsj}  
\sum_k g_k
e^{-2\pi i \nu \tsk}  ,
\ee
is the product of two  independent, complex Gaussian noise processes when the number of terms in each sum is large\footnote{The \newtext{amplitude modulated noise (AMN) sum 
$A(\nu)$} has  many terms if the shot noise is dense.   The propagation term has many terms under the conditions of `strong scintillation,'  as defined later.}.  
The   squared magnitude of $\EfieldFT_{\rm s}(\nu)$ is the spectrum having  an overall extent determined by  $\vert \widetilde\Delta(\nu) \vert^2$, 
 \be
 S_{\rm s}(\nu)
=  
\vert \widetilde\Delta(\nu) \vert^2
A(\nu)  G(\nu), 
\quad 
A(\nu) = \left\vert \sum_j a_j   e^{-2\pi i \nu \tsj}  \right\vert^2, 
\quad
G(\nu) = \left\vert \sum_k g_k e^{-2\pi i \nu \tsk} \right\vert^2. 
 \ee
 The random processes $A(\nu)$ and $G(\nu)$ are positive semi-definite with RMS values equal to their mean values in the limit of a large number of terms in their respective sums, which have complex Gaussian statistics.    
 The PDFs for $A$ and $G$  are thus one-sided exponentials because they are   $\chi^2_2$ random variables with two degrees of freedom.    Each quantity  has a characteristic frequency scale determined by the time dependence of the amplitudes $a_j$ and $g_k$.   The shot amplitudes have pulse-like envelopes with duration $\WA$ leading to a characteristic frequency scale $\WA^{-1}$.    The multipath amplitudes $g_k$ follow an impulse response  with width equal to the scattering time $\taud$; its reciprocal is the characteristic scintillation bandwidth.  
 The net modulation index (\RMS/mean) $m_S$ of $S_{\rm s}(\nu)$  is given by 
 \be
1 +  m_S^2(\nu) =  [1 + m_I^2(\nu)] [1 + m_G^2(\nu)].  
 \ee
 For $\chi_2^2$ statistics, $m_I = m_G =1$ and $m_S = \sqrt{3}$. 
 
 \subsection{Gravitational lensing}
 
 Pulses  encounter weak changes in the interstellar gravitational potential which  act as a variable, achromatic refractive  index that adds to small refractive index variations from  magnetized plasmas in the ISM, IPM, and ionosphere.   The \LOS\ integrated gravitational potential from individual stars with masses $m_i$ and locations $\xvec_i$ yields a geodesic noise component to \TOA\ variations  of the simplified form \citep[][]{GOLAT2021136381},
\be
\delta t_{gn} (\xobs, \tA) \simeq  -\frac{2G}{c^3} 
	\sum_i m_i \ln 
		\left [ \frac
			{ \vert \xvec_i \vert - \khat \cdot \xvec_i}  
			{ \vert \xobs - \xvec_i \vert - \khat \cdot (\xobs - \xvec_i)}
		\right]
\ee
based on a thorough treatment by \citet[][]{PhysRevD.60.124002} and
where $\khat$ is the unit vector along the (unperturbed) propagation direction; velocities are assumed small enough to not matter and  retarded times for evaluation of stellar positions are implicit.  The \RMS\ value is 
$\sigma_{\rm gn} \sim 10\,$ns \citep[][]{GOLAT2021136381}, a non-negligible value compared to some of the best \TOA\ precisions.   The characteristic time scale for variations in $\delta t_{\rm gn}$ along any \LOS\ are  very long, decades to centuries \citep[][]{2016MPLA...3150083D}, and any close encounters of the \LOS\ with an individual star have low probability, of order the `optical  depth' $\sim 10^{-6}$ for gravitational lensing.   Over the current $\sim$three  decades of extant timing data,  such variations are removed by fitting for the pulsar spin parameters.  The situation differs for pulsars in globular clusters where the gravitational potential can affect estimates for the spindown rate $\fsdot$. 

Pulsar timing data can be used to place upper limits on dark matter clumps that contribute to $\delta t_{\rm gn}$ \newtext{\citep[][]{2007ApJ...659L..33S, Ramani2020, Chakrabarti2026}. Gravitational lensing from  small objects, though extremely rare, could produce oscillations in intensity vs. time and frequency due to interference between two or more images produced by the lensing.}

\newcommand{\SNRparen}{(S/N)}

\renewcommand{\fp}{f^{\,\prime}}

\section{TOA estimation using template matching}
\label{sec:toa_estimation}

The quality of pulsars as astrophysical clocks is best introduced with a description of how well TOAs are estimated by matched filtering (MF) of  pulse profiles with a template function.    
The fundamental assumption of MF is that the  measured pulse is an exact copy of a template  scaled in amplitude and shifted in time  with noise added. 
These conditions are  largely satisfied but the highest timing precision requires attention to the frequency dependence of pulse shapes and departures in shape   that arise  from emission jitter, from propagation through the ISM, and from instrumental distortions.

Arrival times are most often computed for the pulse profile in Stokes $I$, which we discuss first.  We discuss the  potential  benefits of  full-Stokes template fitting later. 
 Here, we consider  the simplest 1D case of arrival time estimation from a narrow-band profile  in order to illustrate some basic points and in later sections consider 2D estimation that addresses profile chromaticity.  The minimum
 \TOA\ error is achieved when profile and template are exact matches to within additive noise.   All other effects increase the error. 
 \newtext{We discuss template starting with the presentation in \citep[][]{1992RSPTA.341..117T}. }
 Let $U_t$ be the template profile sampled at $\Nphi$ discrete times,  $t \in [0, \Nphi-1]$ corresponding to one cycle of pulse phase.   The data profile  $D_t = \Prof_t + \Noise_t$  thus comprises a scaled and shifted version of the template $\Prof_t = b\, \Temp_{t-t_0}$ added to  noise $\Noise$ having  variance $\sigma_{\Noise}^2$.  Here and below,  $\Prof_t$ is normalized to have unity maximum.   Data are fitted with a numerical model,
 $D_t = a + b\, \Temp_{t-t_0}$,  by minimizing 
the $\chi^2$  cost function to estimate  the \TOA\ $t_0$ along with the baseline $a$ and scale factor $b$, while the \rms\  noise $\sigma_{\Noise}$ is estimated separately.   Least squares fitting is equivalent to finding the \TOA\ from the time lag where  the cross correlation function of the template and profile maximizes. 
With sampled data,  the optimization is far easier in the frequency domain to avoid interpolation issues that arise in the time domain\footnote{Interpolation in the time domain requires use of the appropriate interpolation function,  
which is $\sinc(Bt) = (\sin\pi Bt)/\pi B t$ where $B$ = bandwidth, which has slowly decaying oscillations. \newtext{This interpolation function is discussed in standard derivations of the sampling theorem for uniformly sampled data.}} \citep[][]{1992RSPTA.341..117T}.  
Using tildes to denote discrete Fourier transforms (DFTs) of the data and template, $\FTDataAB{}{f}$ and $\FTTemp_f$, 
with  discrete frequency indices $f \in [0, \Nphi-1]$, 
the TOA estimate is the solution of 
\be
\sum_{f=1}^{\Nphi/2}  
	\Im \{f \FTDataAB{}{f} \FTTempAB{}{f}^* e^{2\pi i f t_0  / \Delta t \Nphi} \} = 0.
\label{eq:tauhat1d}
\ee
The sum excludes the $f=0$ elements, which determine the estimate of the baseline $a$ but are not needed for the TOA; the sum extends only to $\Nphi/2$ by Hermiticity of the DFTs. 
If the template and parameters  yield a good fit, the {\it minimum} TOA error is
\newtext{\citep[see also][]{van06}}
\be
\rmstoaS 
= \frac{ P}  {2\pi \SNRparen} \left(\frac{\Nphi}{2} \right)^{1/2}
	\left[
		\sum_{f=1}^{\Nphi/2}      f^2  \vert \FTTempAB{}{f}\vert^2   
	\right]^{-1/2}.
\label{eq:sigtau1d}
\ee
The $f^2$ factor in the sum causes higher frequency components in the template to be emphasized.   Narrower pulses or those with sharper features yield a larger sum, thus reducing the TOA error. 
The signal to noise ratio  is defined  as  {\rm (pulse \ peak) / (\rms\ \ noise)}, or 
$\SNR \equiv b / \sigma_{\Noise}$.  Eq.~\eqref{eq:sigtau1d}   applies for $\SNR\gg1$ because it is   based on the quadratic expansion of a cost function that assumes small errors.  

The TOA error is also written  in a form that separates
pulsar properties from telescope and analysis parameters,
\be
\rmstoaS
= \frac{\Weff}{\Nphi^{1/2} \SNRparen},
\label{eq:rmstoaS}
\ee
where the effective pulse width involves both the spin period and the pulse shape and is expressed in either the frequency or time domain, 
\be
\Weff 
=
 \left(\frac{\Nphi P}{2\pi \sqrt{2}} \right)
	\left[
		\sum_{f=1}^{\Nphi/2}      f^2  \vert \FTTempAB{}{f}\vert^2   
	\right]^{-1/2}
= 
P \left[\Nphi \sum_t \left( \Temp_{t+1} - \Temp_t  \right)^2 \right]^{-1/2} 
\ \ \xrightarrow {\displaystyle \Delta t \to 0} \ \
\frac{P^{1/2}}    {\displaystyle \left\{ \int_0^P dt\, \left[\Temp^{\prime}(t)\right]^2\right\}^{1/2}}.
\label{eq:weff}
\ee
This expression for $\Weff$ is consistent with the expression given by
\citet[][]{1983ApJS...53..169D} for the \TOA\ error. 
  By inspection of Eq.~\eqref{eq:weff},  narrow pulses or those with sharp features have larger denominators than wide pulses.  This is also evident in the  continuum limit that involves the derivative of the template,  $\Temp^{\prime} \equiv d\Temp/dt$:   larger derivatives correspond to smaller TOA errors.  The continuum limit shows that the effective width involves both the spin period $P$ and the actual pulse shape.   However, it does not depend on the number of samples across the pulse, $\Nphi$, \newtext{as long as all relevant features in the pulse profile are resolved.}  
  
\begin{figure}[h!] 
   \centering
   \includegraphics[width=0.7\linewidth]{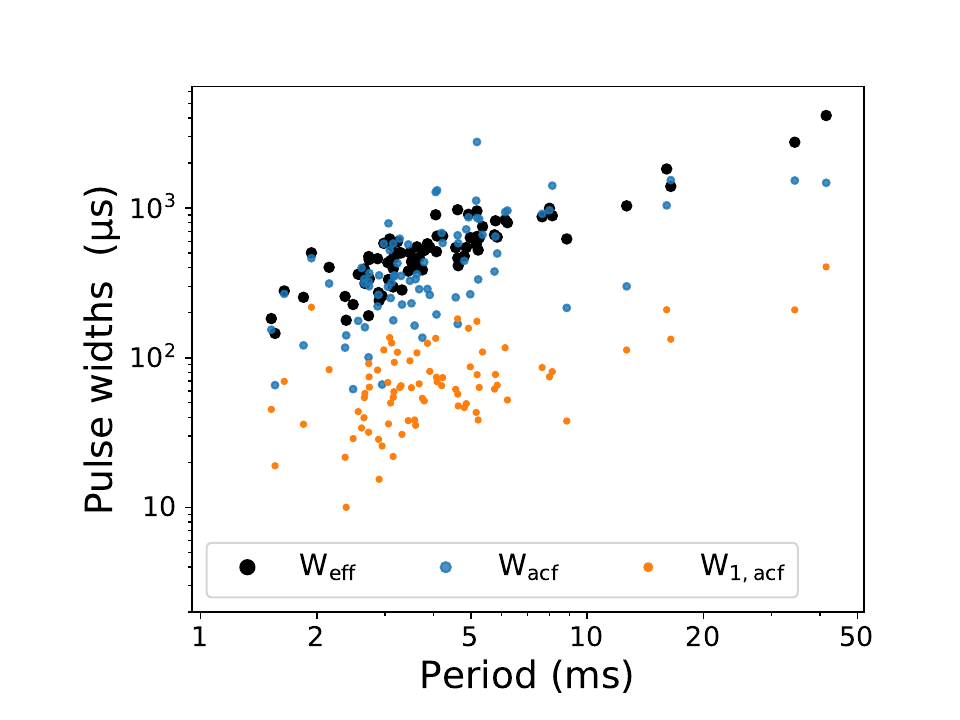}
   \caption{Pulse widths vs. period.   The effective width $\Weff$,
   ACF width $\Wacf$, and single component width $W_{1, \rm acf}$   were calculated from template profiles
    for pulsars in NANOGrav's 15-yr PTA.}
   \label{fig:pulse_widths_Lband_nanograv_pulsars}
\end{figure}

Fig.~\ref{fig:pulse_widths_Lband_nanograv_pulsars} shows $\Weff$ vs. period for pulsars in the 15-yr NANOGrav PTA along with the ACF width of each profile and the widths of single profile components calculated from the ACF of the profile's first derivative (cf.\ Fig.~\ref{fig:uuprime1}).  All three widths show an upward trend with period but with the smallest scatter for $\Weff$ and the largest for individual components. 

\subsection{Evaluation using pulsar and telescope parameters}
\label{sec:toahat_eval}
The \rms\ noise in a pulse profile with $\Nphi$ bins is
$
\sigma_{\Noise} = \Ssys / \sqrt{\npol B \Tint / \Nphi},
$
where $\Ssys = \Tsys / G$ is the system temperature, $\Tsys$, expressed in Jy units using the telescope gain
$G = A_{\rm e} /2k_{\rm B}$ in K~Jy$^{-1}$ (with $A_{\rm e}$ = effective telescope area and $k_{\rm B}$ = Boltzmann constant).  Dominant contributions to $\Tsys$ include receiver noise, ground `spillover' noise, and the sky background that includes free-free and synchrotron emission from the Galaxy, and  the cosmic microwave background.  
Other parameters include  the number of polarization channels used with $\npol = 2$ for Stokes $I$ profiles; $B$ is the bandwidth; and $\Tint$ is the total integration time (typically $\sim$~minutes to an hour). The integration time per profile sample is $\Tint/\Nphi$.

This gives a signal-to-noise 
\be
\SNR = \frac{\Spk}{\sigma_{\Noise}} =
\frac{P\Sphiave}{W} = 
\frac{\Sphiave}{\Ssys} 
\frac{P}{W}
\left(\frac{\npol B \Tint} {\Nphi}\right)^{1/2}
\simeq 3437 \times
	 \frac{\Sphiave(\rm mJy)}{\Ssys({\rm Jy})} 
\frac{P_{\rm ms}}{W_{100}}
\left[\frac{(\npol/2) B_{\rm GHz} {\Tint}_{,3}} {\Nphi / 2048}\right]^{1/2}
\label{eq:SNR_radiometer}
\ee
where  the peak flux density is expressed in terms of the 
catalogued flux density averaged over pulse phase, $\Spk \simeq P\Sphiave/W$, and the pulse duty cycle,  $W / P$.
\footnote{For individual pulses or fast radio bursts we would use the peak flux density directly to calculate $\SNR$ and the \TOA\ error.}

The pulsar's flux density typically scales as a power-law $\Sphiave \propto \nu^{-\alpha_{S}}$ with spectral indices $0 \lesssim \alpha_{S} \lesssim 3$
\newtext{\cite[e.g.][]{Xu2024}}. In some cases the spectrum turns over at low frequencies $\lesssim 0.3$\,GHz and in others the spectrum steepens at high frequencies.   At frequencies $\lesssim 1$\,GHz, the scaling of the synchrotron background,
$T_{\rm sync} \propto \nu^{-2.7}$, is nearly cancelled by that of the  flux density for pulsars with larger spectral indices, yielding S/N that is less dependent on frequency than the individual factors.  
The \TOA\ error is
\be
\rmstoaS  
= 
\frac{\Weff W  \Ssys} {P \Sphiave \sqrt{\Npol B \Tint}}
\label{eq:rmstoa_mf_t_tel_phi}
\ee

{\it Example of a single-component Gaussian pulse of width
$W$} (FWHM): The effective width is  
\be
\Weff = 
\left( \frac{WP} {\sqrt{2\pi\ln 2}}  \right)^{1/2} \simeq 
219~\mu {\rm s} \sqrt{ W_{100\,\mu {\rm s}}P_{\rm ms} },
\label{eq:Weff_gaussian}
\ee
for a nominal 1~ms period and 100~$\mu s$ width. 
The corresponding  TOA error is 
\be
\rmstoaS = 
\frac{1}{(2\pi\ln 2)^{1/4}} \frac{\Ssys}{\Sphiave} \frac{W^{3/2}}{(\Npol B \Tint P)^{1/2}}
\simeq 
    15\,{\rm ns} \, \times
    \frac{\Ssys(\rm Jy)}{\Spsr(\rm mJy)} 
    \frac
        {W_{100}^{3/2}}
        {\sqrt{(\npol/2) B_{\rm GHz}  {\Tint}_{,3}  P_{\rm ms}  }} 
\label{eq:rmstoaS_gaussian}
\ee
where we have used  an integration time
$\Tint = {\Tint}_{,3} \times 10^3 $\,s, a bandwidth in GHz, $\Ssys$ (SEFD) in Jy, and the pulsar period averaged flux density $\Spsr$  in mJy. 
Pulsars show a variety of periods, widths, and strengths, yielding TOA errors ranging from sub-$\mu$s to several ms or more for long period pulsars and magnetars. 

For later use, we express the signal to noise ratio of the average profile 
$\SNR =  \SNRparen_1 / \sqrt{N}$  
in terms of the single pulse value, $\SNRparen_1$, assuming that  amplitudes are statistically independent between 
$N = \Tint / P$ pulses
\!\footnote{Some pulsars show amplitude correlations between neighboring pulses but, for steady pulsars that show no changes in profile state, statistical independence is a reasonable assumption.  State changes are discussed later in Sect.~\ref{sec:shape_shifters}.}
, yielding
\be
\rmstoaS \simeq \frac{4.8~{\rm ns} P_{\rm ms}}{ \SNRparen_1}
	\left(  \frac{W_{100}}  { {\Tint}_{,3}\Nphi / 2048} \right)^{1/2}.
\label{eq:rmstoaS2}
\ee

{\it Multiple component profiles:}
While some pulsars display single-component pulse profiles, many show two or more components with different amplitudes and different degrees of component overlap in pulse phase.   To account for these in a simple way, we let all $\nc$ components have the same width $\Wc$.   Assuming no phase overlap,  the effective width is a factor $\nc^{-1/2}$ smaller (cf. Eq.~\eqref{eq:weff}) and the peak flux density for a component ${\Spk}_{\rm c} = \Sphiave P / \nc \Wc$  is smaller by a factor  $\nc^{-1}$.   The net effect is that the TOA error is larger by a factor $\nc^{1/2}$. 

For profiles comprising   $\nc$ components  with the  same characteristic width $W_{\rm c}$ but different amplitudes $a_j$, the weighted number is
$\ncTwo = \sum_j (a_j / a_{\rm max})^2 \le \nc$, yielding
$\Weff = (  \Wc P / \ncTwo \sqrt{2\pi \ln 2})^{1/2}$.
The period-averaged flux density in Eq.~\eqref{eq:rmstoa_mf_t_tel_phi}  becomes  $\Sphi \to \Sphi / \ncOne$ 
and the net \TOA\ error is
\be
\rmstoaS = 
	\frac{K_U}{\sqrt{\npol B \Tint}}
	\frac{\Ssys}{\Sphi} 
	\left(\frac{\ncOne}{\sqrt{\ncTwo}}\right)
	\left(\frac{\Wc^3 }{P}\right)^{1/2} ,
\label{eq:rmstoaSQ}
\ee
where $K_U \sim 1$ is a pulse-shape dependent factor.  For Gaussian shaped components,
$K_U = [\pi / 32  (\ln 2)^3]^{1/4} = 0.74$. 
\newtext{
Identifying the role of the number of profile components in 
\TOA\ uncertainties helps enable an assessment of the timing quality of, say, a newly discovered pulsar and whether it can contribute significantly to the sensitivity of a PTA program.}

Equations~\eqref{eq:weff}--\eqref{eq:rmstoaSQ}  provide the means for identifying key  factors  and improvements that can be made to minimize the TOA error:
\begin{enumerate}
\itemsep = -1pt
\item The error decreases inversely as  \SNR, as usual for matched filtering, and thus also for bright pulsars; 
\item Interstellar scintillation modulations of the emitted flux density can be large (up to $\sim 100\%$ variations for DISS and $\sim 10$ to 30\% for RISS), making the $\SNR$ and thus the  \RMS\ TOA error  epoch dependent  and stochastic; 
\item Pulsars with smaller effective widths  $\Weff$ yield smaller errors, illustrating how smaller periods  $P$ and
narrower pulses (or narrow pulse structure) are beneficial;
\item Effective widths increased by instrumental or astrophysical broadening degrade timing precision;
\item Larger telescopes and lower $\Tsys$ yield better precision in proportion to smaller $\Ssys = \Tsys/G = 2k\Tsys/A_{\rm e}$; and
\item Larger bandwidths and longer integration times reduce 
$\rmstoaS$\footnote{However,  \newtext{as discussed in Sect.~\ref{sec:DMnu}},  large bandwidths introduce significant complications from scattering that can counteract the improvement of the \TOA\  error from a greater S/N in the matched filtering operation.}.
\end{enumerate}
These features underly our later discussion in Sect.~\ref{sec:Q}  of quality measures for precision timing of MSPs. 

\subsection{Full Stokes template fitting}
\label{sec:fullstokes}

Template fitting using  the full Stokes vector $\Svec$ (Eq.~\eqref{eq:Stokesvec}) has been assessed in
recent work  \citep[e.g.][]{2024ApJ...973...94R, 2024ApJ...977..114D, 2025A&A...699A.104G} that builds upon the formalism developed in \citet[][]{van02, van06, 2013ApJS..204...13V}.   Here we outline the conditions under which improvements over scalar Stokes $I$ template fitting can be expected.  We refer to profiles in $Q, U, L = Q + iU$ and $V$  as polarization profiles distinct from the total intensity profile $I$.   Our treatment here ignores instrumental polarization, a discussion of which is deferred to Sect.~\ref{sec:polcal}. 

Polarization profiles typically show sharper pulse features than in Stokes $I$, as they must given that pulsars generally  show systematic changes in polarization state across pulse phase \newtext{(Fig.~\ref{fig:stokes2psrs})}.    The potential for these sharper features to yield more precise \TOAs\  must be weighed against their lower S/N if  the degree of polarization,
$\dpol \equiv \sqrt{\dL^2 + \dV^2}$, is not large.

To assess this we take into account the properties of polarized signals and unpolarized noise
(Sect.~\ref{sec:Stokes})  and refer to  Eqs.~\eqref{eq:rmstoaS} - \eqref{eq:rmstoaSQ}. 
The \TOA\ error from matched filtering is $\rmstoaS \propto \Weff / \SNR$.   Using a single component profile
as an example, the effective width $\Weff \propto \sqrt{WP}$, where $W$ is the actual component width and $P$
is the spin period.   For an individual pulsar, we can ignore the period dependence  for now.    Defining the 
component width as $W_s$ and the peak amplitude $\vert s\vert_{\rm pk}$  for each Stokes parameter ($s = $\, I, Q, U, V) and taking into account that $Q,U$ and $V$ are signed quantities, 
the MF error for each Stokes parameter is  $\sigma_{t_s} = \sigma_{t, I} \sqrt{W_s} / s_{\rm pk}$. 
Note that the implied signal to noise ratios underlying the \TOA\ error  are based on additive noise having the same \RMS\ value in the Stokes parameters (Sect.~\ref{sec:Stokes}). 
Further, recognizing that peaks do not align in pulse phase, we define $I_{{\rm pk}, s}$ as the Stokes-$I$ value at the phase where the Stokes-$s$ extremum occurs.  This allows us to express \TOA\ errors in terms of the
polarization state using peak values of $\dL$ and $\dV$  and in terms of  the error in Stokes-$I$,
\be
\sigma_{t, I} &\propto& \frac{W_I^{1/2}}{I_{\rm pk}},
\label{eq:sigtI}
\\
\sigma_{t, Q} 
	&\simeq& 
	 \frac{\sigma_{t, I} }{\vert \dL \cos2\chi \vert}_{\rm pk}
				\left(  \frac{W_Q}{W_I}  \right)^{1/2}  
				\frac{I_{\rm pk}}{I_{{\rm pk}, Q}}
\\
\sigma_{t, U} 
	&\simeq& 
	\frac{\sigma_{t, I} }{\vert \dL \sin2\chi \vert}_{\rm pk}
				\left(  \frac{W_U}{W_I}  \right)^{1/2}  
				\frac{I_{\rm pk}}{I_{{\rm pk}, U}}
\\
\sigma_{t, V} 
	&\simeq& 
	\frac{\sigma_{t, I} }{\vert \dV\vert_{\rm pk} }
				\left(  \frac{W_V}{W_I}  \right)^{1/2}  
				\frac{I_{\rm pk}}{I_{{\rm pk}, V}}
\ee
Another expression results by using $L = Q + iU$,
\be
\sigma_{t, L} 
	&\simeq& 
	\frac{\sqrt{2}\sigma_{t, I} }{{\dL}_{\rm pk} }
				\left(  \frac{W_L}{W_I}  \right)^{1/2}  
				\frac{I_{\rm pk}}{I_{{\rm pk}, L}},
\ee
where the $\sqrt{2}$ factor represents the larger \RMS\ noise in $L$ compared to the individual Stokes parameters.
Improved \TOAs\ result if sharper Stokes parameters outweigh the potentially lower S/N .   This holds if
the degrees of polarization are large enough, e.g.
\be
{ \dL}_{\rm pk} > 	\sqrt{2}  \left(  \frac{W_L}{W_I}  \right)^{1/2}   \frac{I_{\rm pk}}{I_{{\rm pk}, L}}
\quad\quad \text{and } \quad\quad
{ \dV}_{\rm pk} >   \left(  \frac{W_V}{W_I}  \right)^{1/2}   \frac{I_{\rm pk}}{I_{{\rm pk}, V}},
\ee
along with similar expressions for $Q$ and $U$ individually.  The intensity ratios 
$I_{\rm pk} / I_{{\rm pk}, L}$ and $I_{\rm pk} / I_{{\rm pk}, V}$ are $\ge 1$ while 
the width ratios $\le 1$, demonstrating the tradeoff between these quantities. 

Joint fitting of all four Stokes parameters is  optimal  because  additive noise
is statistically independent between Stokes parameters.  We illustrate  with a highly simplified approach that estimates  \TOAs\ by minimizing a cost function,
\be
C(a, \tzerohat) = \sum_t \sum_s^{I,Q,U,V} 
	\frac{\left[a U_s(t - \hat t) - s(t)     \right]^2}{\Var(s_{\rm off})},
\label{eq:cost}
\ee
which sums over Stokes parameters $s$ and time $t$ across pulse phase and 
where $U_s$ is a known template shape for Stokes-$s$,  $a$ is a scale factor, and $\that$ is the estimated
\TOA.  Each term is weighted by the off-pulse variance $\Var(s_{\rm off})$. This example assumes that the Stokes profiles $U_s$ are known and the polarization state is epoch independent (after Faraday rotation correction)
so that only a single scale factor is needed.    

From Eqs.~\eqref{eq:sigtI}--\eqref{eq:cost} we estimate the maximum improvement factor for using full Stokes \TOA\ estimates
compared to Stokes-$I$ only.   
\newtext{
Using a representative set of Stokes profiles on 
23 MSPs at 1.5\,GHz from the NANOGrav program\footnote{\url{https://nanograv.org/science/data/polarization-calibration-data}},
we forecast 
maximum improvements in \TOA\ errors  from 5\% to about 70\%. 
Actual results reported by \citet[][]{2024ApJ...973...94R} are based on a combination of enhanced polarization calibration as well as use of full-Stokes template matching for 25 MSPs. They find median improvements in \TOA\ errors of 33\% with no improvement in the worst case and the best (85\%) improvement for bright MSP
J0437$-$4715.   Using similar analyses, \citet[][]{2024ApJ...977..114D} find 
improvements of 14\% to 33\% for the three MSPs they studied in detail.  More recently \citet[][]{2025A&A...699A.104G} found improvements for observations using the Nan\c{c}ay telescope similar to those achieved by \citet[][]{2024ApJ...973...94R}. 
}

While full-Stokes fitting is more complicated than illustrated here \citep[e.g.][and see also Sect.~\ref{sec:polcal}]{2013ApJS..204...13V},  the key points are evident, namely
that the form of the cost function is based on the statistical independence of off-pulse noise and that the net results depend on polarization states and pulse widths that are unique to each pulsar.
PTA measurements with more sensitive telescopes can enable implementation of this benefit, with each  pulsar assessed individually to fully optimize this aspect of \TOA\ estimation as in the work of 
\citet[][]{2024ApJ...973...94R}.

\subsection{Chromatic  changes in profile shape}
\label{sec:intrinsic_chromatic}

\begin{figure}[h!]
    \centering
    \includegraphics[width=0.475\textwidth]{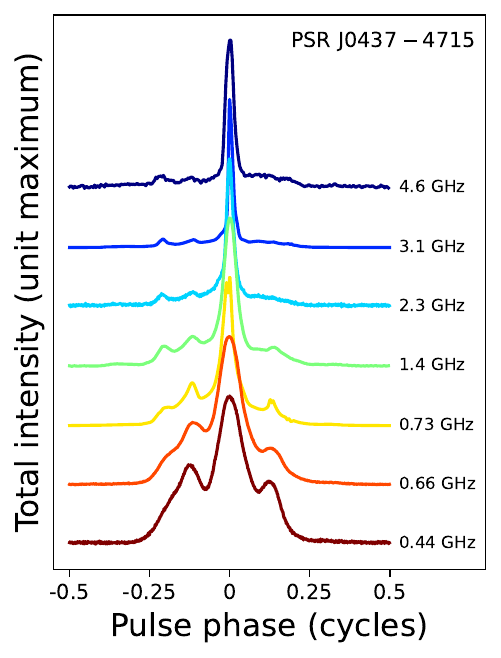}
    \includegraphics[width=0.475\textwidth]{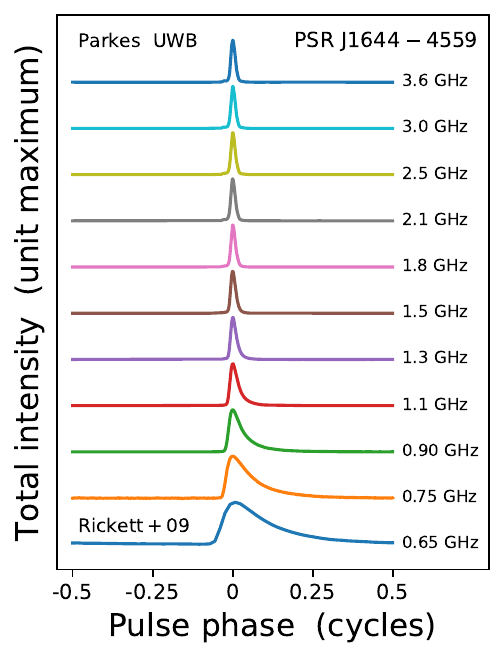}
    \caption{Left: Average profiles for the MSP J0437-4715 showing the change in shape over more than a decade in radio frequency. The profiles have been aligned using the strong central component that appears in all the profiles.   
    Profiles are from the EPN database \citep[][]{mj95,  bbm+97, dhm+15}.
    \label{fig:J0437_profile}
    Right:  Profiles for the heavily scattered
    pulsar J1644-4559 at multiple frequencies obtained  simultaneously using the Parkes Ultra Wideband Feed \citep[][\url{https://doi.org/10.25919/gptm-d012}]{2021MNRAS.504..228S} 
    along with a profile at 0.65\,GHz from 
    \citet[][]{2009MNRAS.395.1391R}. 
    Profiles are aligned by their maxima.   Proper alignment requires fitting for  a contemporaneous value of DM estimated by taking into account actual pulse broadening shapes at each frequency. 
    \label{fig:J1644_profiles}
    } 
\end{figure}

A notable departure from  idealized template matching is the nearly ubiquitous chromaticity of pulse shapes associated with the emission process.  
%
%
Most pulsars show significant variations of average pulse shape vs. frequency, usually as changes in relative amplitudes and spacings of profile features   that can be identified across wide frequency ranges.  All of the MSP profiles shown in Fig.\,\ref{fig:pulse_gallery} for $\sim 1.5$\,GHz evolve with frequency.  An example  of  profile variations over a 10:1 frequency range is shown in Fig.\,\ref{fig:J0437_profile} for the MSP J0437-4715.  Others can be found for CPs \citep[e.g.][]{hr86, Hankins2010, 2012AA...543A..66H} and MSPs \citep[e.g.][]{2016A&A...585A.128K} that show similar but also a diversity of properties. The general trend is for profiles to broaden at lower frequencies  with counterexamples for some CPs by the growing prominence of narrow `core' components relative to broader `conal' components \citep[e.g.][]{ran83}, making the net pulse shape effectively narrower at lower frequencies for these cases. 
Much of the frequency dependence is captured in multicomponent profile models like those described in 
Sect.~\ref{sec:pulse_shapes} with frequency dependent amplitudes, widths, and phases. \newtext{In practice, timing errors induced by chromatic profile evolution are typically modeled by PTAs using a frequency-dependent (FD) model that parameterizes the additional timing errors as a power-law in log-frequency ($\propto C_{\rm FD}({\rm log}\nu)^{-p}$, where $C_{\rm FD}$ and $p$ are the fitted constant and spectral index, respectively) \citep{2015ApJ...813...65N}. This procedure often requires many free parameters to accommodate the chromaticity of different pulse components, as well as to account for differences in data reduction and pulse templates between different observing systems.}

\subsection{Shape shifters: pulsars with epoch-dependent profile shapes}
\label{sec:intrinsic_temporal}
\label{sec:shape_shifters}

%
\begin{figure}[h!]
    \centering
    \includegraphics[width=0.675\textwidth]{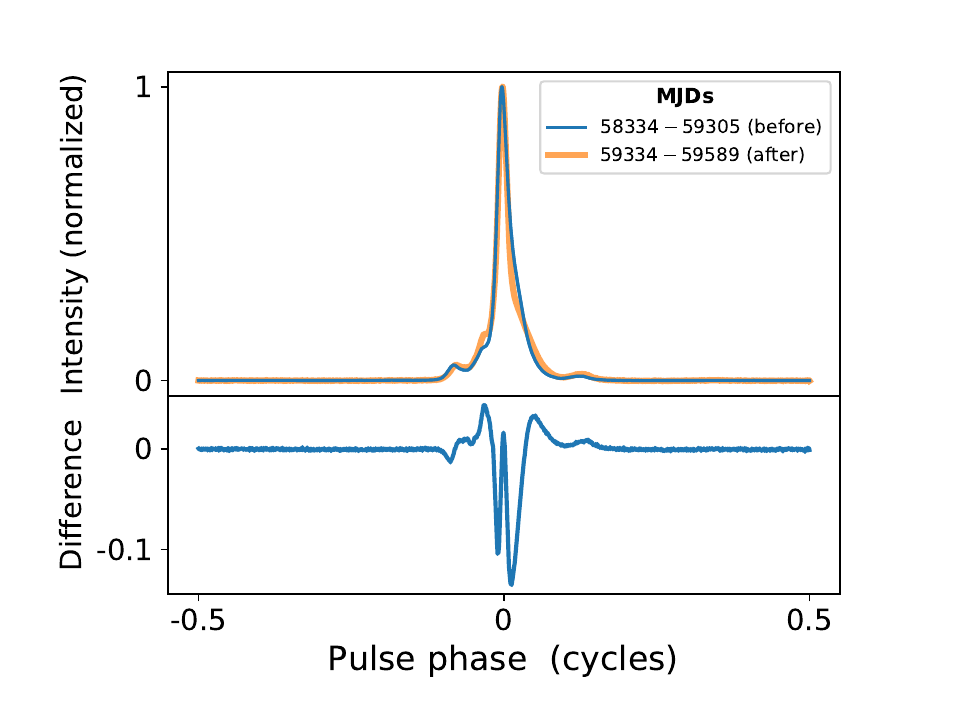}
    \caption{Average profiles for the MSP J1713+0747 before and after a sudden change in shape at
    MJD $\sim 53920$ (top panel).   The profiles were aligned by using the `before' profile as a template for the `after' profile in order that the differences between the two  profiles (bottom panel) reflect the actual change in shape and not include the effects of a time offset. 
    \label{fig:J1713_profile_change}
    } 
\end{figure}

The above considerations are based on the assumption that  pulse variability is  ergodic, i.e.  that pulses are drawn 
from a homogeneous `population' of pulses so that average profiles are 
convergent to a time-invariant (though chromatic)  ensemble average. 
However, even after averaging out single-pulse stochasticity, there are several classes of non-convergent profiles manifested in pulsars.  All of these have significant effects on \TOA\ estimation. 

As mentioned briefly in Sect.~\ref{sec:signal_models},  the shape of an average profile is ultimately linked to the magnetic field topology where radio emission is produced in the magnetosphere of the NS.  This field is a combination of the frozen-in field of the NS and the field produced by magnetospheric currents.  The orientation of the NS's spin vector and its rate then map the emission beam into the observed shape.  The net pulse shape also depends on any variation in emission altitude, which increases the scatter in single-pulse arrival times due to aberration and retardation. 
The emission strength also depends on the relativistic particle flows along the field lines and the degree of coherence.   Empirically, the stability of profiles implies that all of these ingredients are also stable, on average, in most cases.   

 {\it Secular and stochastic profile variations:} The Crab pulsar has been observed for almost 60 yr since its discovery in 1968, representing about 6\% of the time since its apparent birth in 1054.  A secular change in the separation of the primary main-pulse and interpulse components suggests that the magnetic field is evolving on the spindown  time scale \citep[][]{2013Sci...342..598L}.  For the Crab pulsar, this evolution is likely due to migration of the magnetic axis relative to the spin axis.  It is slow enough that it does not limit timing precision; however, substantial glitch activity and spin noise make the Crab pulsar, like most young pulsars, an imprecise pulsar clock. 

 More striking examples can be found among magnetars, which show large  profile changes on time scales of months that are accompanied by changes in the spindown torque e.g. Swift J1818.0$-$1607 
 \citep[][]{2024MNRAS.528.3833F} and PSR J1745-2900 \citep[][]{2018ApJ...866..160P, 2018MNRAS.476.3677Y, 2019ApJ...875..143W, 2021ApJ...921..101S}.  For these objects, stochastic magnetic evolution  drives  shape changes as well as spin variability and X-ray emission.    

{\it Discrete, recurrent states:}  Some objects, primarily CPs, show two or more discrete profile shapes that sustain for multiple spin periods with switching between multistable states occuring rapidly, $\lesssim 1$\,spin period \citep[`mode' changes][]{bac70a, lyn71a, bmsh82,ran86}. Switching appears to be consistent with a Markov process, which gives exponential distributions for the durations of states
\citep[][]{2013ApJ...775...47C}.     Arrival time offsets are substantial between the different modes but can be accounted for in a timing analysis if individual templates are used for each mode.   

Another type of recurrence is `nulling,' where the states are defined by highly disparate pulse intensities,  where they either vanish during a null  or become much smaller for durations of a few to a large number of pulse periods \citep[][]{big92a,Ng_2020}. 
Some pulsars display systematic drifts of pulse centroids over a few to a few dozen pulse periods \newtext{\citep[][and references therein]{2019MNRAS.482.3757B}} and some objects show multiple, repeatable drift rates  that are harmonically related, indicating another class of discrete states \newtext{\citep[][and references therein]{2013ApJ...775...47C}}.

The underlying physics of metastable states and switching between them  is not well understood but appears to involve recurrent states in the particle flows generated by acceleration in pulsar magnetospheres.  One object, the CP B1931+24,  has residence times of weeks in high and low states, where  radio emission during low states has only recently been detected \citep[][]{2025arXiv250704816R}.  State durations are long enough to establish that the spindown torque  is larger in the high state by  $\sim 50$\% \citep[][]{klo+06}.  Torque jumps of this amplitude require changes in current  on the scale of the entire magnetosphere, signifying that radio emission by itself, though a small part of the energy budget of 
a NS, traces large-scale magnetospheric activity. 
Much shorter nulls (tens to hundreds of spin periods)  in other pulsars may also involve changes in torque, but their durations are 
too small to allow changes in $\fsdot$  to be measured.   

 Profile mode changes and nulling appear less common in MSPs and have not yet had a significant impact on timing precision for those MSPs used in PTAs. 

{\it Discrete events:} 
The MSP J1713+0747 has shown a transitory, highly chromatic change in pulse shape 
 \citep[][]{2021ATel14642....1X,2021MNRAS.507L..57S} that recovered to its original form after about two years \citep[][and references therein]{2024ApJ...964..179J}.  Fig.\,\ref{fig:J1713_profile_change} shows profiles averaged over a data span of about $10^3$ days prior to the event and 250 days after. The $\gtrsim  14$\% change is in contrast to $\ll 1$\% changes between high S/N profiles prior to the event. Confinement  of the event to a specific range of epochs is fortunate because this pulsar otherwise shows some of the smallest timing residuals in PTA analysis.  
 This pulsar has shown two similar previous events
 \citep[][]{2018ApJ...861..132L, 2021RNAAS...5..167L}
 with  smaller arrival time shifts  than the event in 2020-2021. Along with sparse  time sampling the weakness of these events has not allowed a detailed study of the pulse shape changes during these two events.  
  It is possible, but as yet unknown, if similar transitory events occur regularly for J1713+0747.  \newtext{Similar events have been seen in the MSPs J1643$-$1224 \citep[][]{Shannon2016} and J0437$-$4715 
  \citep[][]{2024MNRAS.530..287S}}.  The events from all three MSPs correspond to chromatic changes in pulse shape that are not plausibly related to interstellar propagation, although initial reports (see references in above-quoted papers) identified them to changes in \DM\ or to scattering effects. 

  There is ongoing debate as to whether
the J1713+0747 events are intrinsic to the pulsar or a propagation effect \citep[e.g.][]{2021MNRAS.508.1115L} but most indications suggest an intrinsic origin.   Fortunately, reversion to the original shape suggests that its effects are confined to a small range of epochs that can be avoided in PTA analyses. Without correction, they would contribute distinctly non-Gaussian fluctuations into timing residuals.
  
  \citet[][]{2026arXiv260405453J} recently developed a semi-automated method for identifying similar events in \TOA\ data. In a preliminary analysis it reproduced events found in J1713+0747 and J1643$-$1224 (data on J0437$-$4715 were not available) but did not identify events in seven other pulsars (application to a larger sample is ongoing).  

\section{TOA variations from single-pulse stochasticity}
\label{sec:dtoa_single_pulse_stochasticity}

The signal model of Sect.~\ref{sec:signal_models} implies that \TOA\ errors will exceed  those expected   from idealized template fitting discussed in Sect.~\ref{sec:toa_estimation}.    Any departure of an average profile's shape from the template causes an offset in the \TOA.   Here we discuss profile shape changes caused by statistical variations in the $N$ individual pulses used to calculate an average profile.   Departures of profiles  from the template shape  diminish as  $N^{-1/2}$ and thus do not fully vanish.     Even for $N \gtrsim 10^5$ typical for the timing of MSPs, the  $\lesssim 0.3$\% shape variations can cause significant timing offsets. 

\newtext{
In the following we discuss pulsar self noise as a process distinct from pulse jitter.  Self noise was discussed in Sect.~\ref{sec:shotnoise} as the stochastic emission by a pulsar that is modulated by a pulse shape envelope.   \TOA\ variations will arise from this AMN process even if the envelope is identical from one realization to the next.    We define jitter as the change in the pulse envelope that does occur between realizations; it includes amplitude fluctuations along with departures in pulse phase from a strictly periodic sequence of pulses. 
We discuss these two contributions to timing errors separately in what follows. 
}

\subsection{Pulsar self noise}
\label{sec:selfnoise}

The AMPSN model discussed in Sect.~\ref{sec:shotnoise} for the emitted pulsar signal represents `self' noise $Z(t)$ that produces an arrival time error (using  $A(t)$ as a template).   For a single pulse, this  error scales as the pulse width $\WA$ divided by the square root of the number of independent samples,  $n \sim \WA \Br$ where $\Br$ is the receiver bandwidth.   
For a single pulse with $100\,\mu$s width and receiver bandwidth in GHz,  
$\rmstoaAMN  
    = m_Z \sqrt{\WA / \Br} 
    \simeq 316\, {\rm ns} \times  m_Z 
        \sqrt{{\WA}_{, 100} / {B}_{\rm GHz}} $
where the  self-noise modulation index $m_z \simeq 1$ for dense shot noise that is applicable to most pulsar observations  but exceeds unity for sparse noise. 

For a sum of $N = \Tint / P = 10^6 {\Tint}_{,3} / P_{\rm ms}^{-1} $ pulses,  the error is
\be
\rmstoaAMN
    \simeq 0.32~{\rm ns} \times  m_Z
        \left( 
            \frac
                {{\WA}_{, 100}  P_{\rm ms} }
                {{B}_{\rm GHz} {\Tint}_{,3} }
        \right)^{1/2}. 
\label{eq:rmstoaAMN}
\ee

Radio pulsar timing almost always involves the total intensity (i.e. Stokes $I$) equal to the sum of intensities of two polarization channels.  If unpolarized, $Z_p(t)$  (with $p=1,2$) is uncorrelated between the two polarizations and the self noise error is reduced by $2^{-1/2}$. For 100\% polarization, however, there is no reduction because $Z_1 \equiv Z_2$. 

This contribution to the \TOA\ error nominally contributes very little to the timing budget.  However, it is larger than the error expected under the very idealized conditions of template fitting (Sect.~\ref{sec:limitations} , serving as a proof of principle that self-noise alone causes excess \TOA\ errors.   Also, by inspection of Eq.~\eqref{eq:rmstoaAMN},  narrower bandwidths, shorter integration times, wider pulses, and longer periods  can yield  much larger self-noise timing errors.

\subsection{ Timing jitter from stochastic pulse envelope modulations} 
\label{sec:jitter}

Fluctuations in pulsar emission cause pulse envelopes to vary in both amplitude and pulse phase as demonstrated in Fig.\,\ref {fig:pulses_J0738-4042}.    These produce \TOA\ errors that far exceed those caused by self-noise and can be larger than template fitting errors and those caused by interstellar scintillation discussed in the next section. 

We model jitter effects by assuming the single-pulse shapes are the same but their amplitudes and phases vary.
Actual pulses  change shape dramatically, partly due to self noise but also from bona fide  envelope shape variability.  However, envelopes have characteristic widths for a given pulsar and for the purposes of our analysis the extra parameters needed to describe shape variability can be absorbed into a minimal set of parameters.  

\subsubsection{Single component pulses}            

We first consider cases there there is only one pulse component in the sense of Eq.~\eqref{eq:ampmod}. 
The template is equated to an ensemble average ($\Np\to \infty$ assuming no time evolution of the pulse statistics), $U(t) = \langle I_N(t)\rangle$.   This allows expression of 
the template as an integral over a joint distribution for $a_j$ and $\tau_j$,
\be
\Temp(t) = \iint da\, d\tau \, f_{a, \tau}(a, \tau) a A(t-\tau) 
\longrightarrow \langle a \rangle 
    \int d\tau \, f_{\tau}(\tau)  A(t-\tau),
\ee
where the second expression applies when the  joint  PDF $f_{a,\tau}$  for $a$ and $\tau$ is 
separable.
The width $\WJ$ of the jitter distribution $f_\tau$  contributes to  the net width of the template  
$\WU = \sqrt{\WA^2 + \WJ^2}$. \newtext{For a toy model that gives approximate values and a useful scaling relation,}  we assume the width $\WA$  is the same  for all single pulses and  that  both $f_\tau$  and $A(t)$ have Gaussian shapes.

Given these assumptions, the  \RMS\ arrival time  from template fitting \newtext{can be shown to be}
\be
\rmstoaJ = \frac{\rmsJ (1 + m^2)^{1/2}} {\Np^{1/2}}
\equiv \frac{\FJ \WU} {\Np^{1/2}},
\label{eq:rmstoaJ}
\ee
where $\rmsJ$ is the \rms\ time jitter  (the \rms\ of $\tau_j$) and the modulation index is
$m = \sigma_a / \langle a \rangle$.
The second equality
defines  the \rms\ TOA  in terms of a `jitter parameter' defined as
\be
\FJ \equiv  \frac{ (1 + m^2)^{1/2} } {2 \sqrt{2\ln2}} \frac{\WJ}{\WU}.
\ee
For this case (only), amplitude variations alone ($\rmsJ = 0$) do not cause arrival time variations but they enhance
the TOA error  by the factor $(1 + m^2)^{1/2}$ when $\rmsJ > 0$. 
A typical value  for the jitter parameter is $\FJ \simeq 0.3$
although there is substantial variation between pulsars.
\citet[][]{2016ApJ...819..155L} report a median value for an alternative jitter fraction 
defined relative to the spin period,
$k_{\rm J} = \rmsJ / P \simeq 0.010^{+0.023}_{-0.006}$ that corresponds to
$\FJ = k_J P (1 + m^2)^{1/2} / \WU \simeq 0.3^{+0.6}_{-0.1}$ for assumed values $m=1$ and a typical pulse
duty cycle $\WU / P \sim 0.05$.  

Using $N = \Tint / P$ and  nominal values for other parameters, the jitter TOA variation is
\be
\rmstoaJ \simeq 33~{\rm ns} \times W_{100} 
	\left( \frac{\FJ}{1/3} \right)
    \left( \frac{P_{\rm ms}} {{\Tint}_{,3}} \right)^{1/2}.
\label{eq:rmstoaJnom}
\ee

\begin{figure}[h!]
\begin{center}
%
   \includegraphics[width=0.45\linewidth]{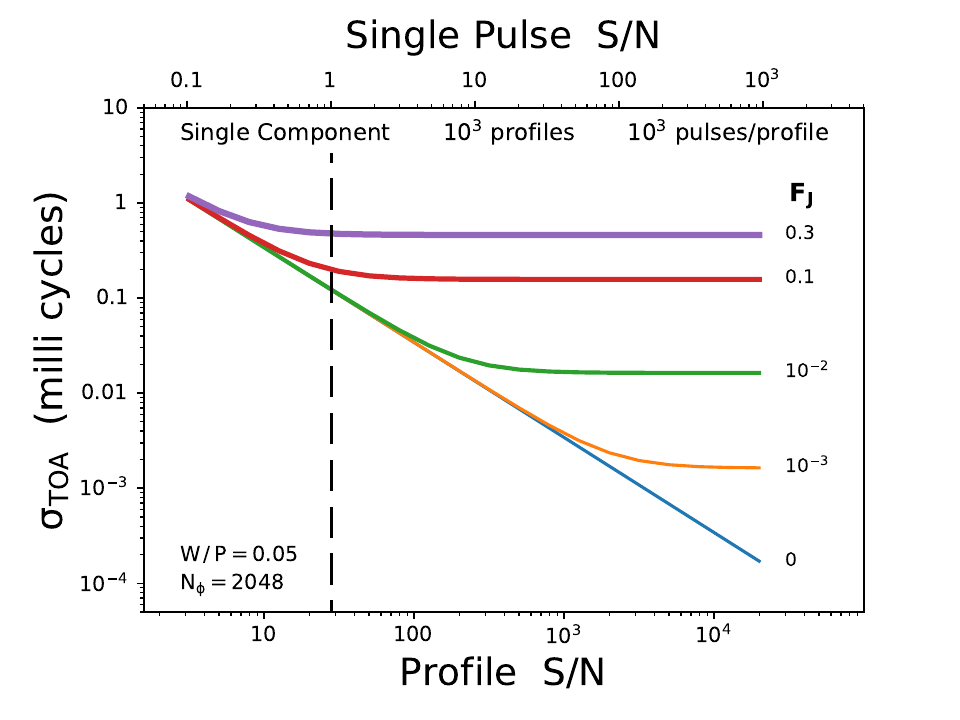}
\includegraphics[scale=0.45]{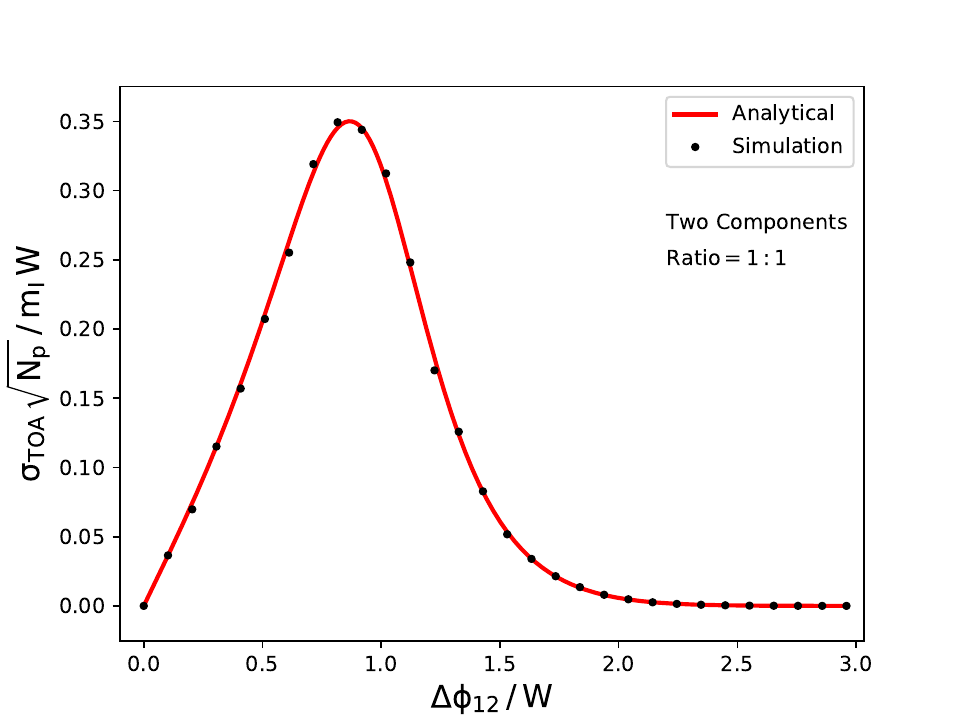}
 \caption{
 Results of simulations of single pulses with jitter and a range of signal-to-noise ratios. 
    Left:
    RMS arrival time  vs signal to noise ratio $\SNR$  for simulated data comprising  pulses with duty cycles $W/P = 0.05$, phase jitter parametrized by $\FJ$ (as labeled),  and additive noise.  TOAs were calculated from averages of $10^3$  pulses  in $\Nphi = 2048$ phase bins to yield a profile \newtext{$\SNR$  $\sim 30$} times larger than the single-pulse $\SNR$.  Plots are based on $10^3$ realizations for each pair of values for $\FJ$ and $\SNR$. 
   \label{fig:rmsjitter_vs_snr}
    Right:
    RMS TOA vs. component separation for two identical Gaussian components with single pulses in each  having amplitude variations but no phase jitter.  The amplitude variations are uncorrelated between components. 
    The points show the RMS TOA vs. component separation,
    where both axes are normalized by the component  width $W$ (FWHM) to make them dimensionless. The plotted line is the analytical expression in Eq.~\eqref{eq:gauss2comp}.
    \label{fig:TCrms_vs_dphi}
 }
\end{center}
\end{figure}

Fig.~\ref{fig:rmsjitter_vs_snr} shows the combined \RMS\ \TOA\ from template fitting and pulse jitter vs.
signal to noise ratio.   For $\FJ = 0$, the \RMS\ error scales as (S/N)$^{-1}$, as expected for matched filtering.  The curves bottom out, however, for  non-zero jitter. 

\subsubsection{Multiple component pulses}

Timing variance changes significantly for pulsars with two or more profile components, especially if they overlap,  as is the case for many pulsars, including those shown in Fig.\,\ref{fig:pulse_gallery}.   In addition to the number of components and their separations,  TOA fluctuations depend on the widths and relative amplitudes of components:
\begin{enumerate}
\itemsep -2pt
\item  {\it Multiple components with no overlap:}
Components with separations larger than their widths  yield  \rms\ TOAs smaller by a factor $\propto \nc^{-1/2}$ compared to  a single component, with a proportionality constant determined by the amplitude ratios of the components.
For the  signal-to-noise dependent TOA error,  $\Weff$  is smaller with $\nc>1$ components because  there are more  terms in the sum with large  template derivative values.   For the jitter contribution,  if $\FJ\WU$ is taken to represent the value for a single component,  independence of jitter between components also implies a reduction by $\nc^{-1/2}$. As with the single component case, there is no TOA variance if there is no phase jitter in any of the components ($\FJ = 0$). 

\item {\it Overlapping components:}   With overlap, amplitude variations in the absence of phase jitter can induce TOA variance.   This is a consequence of amplitude variations in different components affecting the entire profile shape even if some components were to not have amplitude variations (which is not typically the case for radio pulsars). 
The two-component case demonstrates these trends. 
When  two Gaussian components have identical $1/e$ widths $W$,
identical mean and \RMS\  amplitudes of single pulses 
(quantified by a modulation index $m_c$), the 
the \RMS\ \TOA\ variation as a function of mean component 
separation $\Delta_{12}$ is
\be
    \langle \hat\tau^2\rangle^{1/2} =
    \left(\frac{m_c}{W\sqrt{N} }\right)
        \frac{1}{\sqrt 2} \
    \frac
        {(\Delta_{12}/W) \ e^{-(\Delta_{12}  / 2W )^2} }
        {[1 + (1 - (\Delta_{12}/W)^2] e^{-\Delta_{12}  / 2W )^2]}}
\label{eq:gauss2comp}  .
\ee
\end{enumerate}

Figure~\ref{fig:TCrms_vs_dphi} (right panel) shows the RMS TOA error vs. component offset  for the equal amplitude  case when there are only amplitude variations of single pulses.   Plotted points are  based on 1000 realizations of simulated profiles  and the  solid line is an analytical calculation of the TOA error for the assumed Gaussian pulse components with independent single-pulse amplitudes. 
For non-overlapping components, the amplitude fluctuations do not affect the location of the components nor is there any component crosstalk.    For vanishing separation, the two superposed components act as a single component and again  the  amplitude variations produce no TOA variations. 
\citet[][]{2024arXiv241100236J}
give further discussion about \TOA\ errors for multicomponent profiles. 

\subsection{Jitter vs. noise dominated TOA errors}
\label{sec:noise_vs_jitter}

A criterion is needed for assessing whether, on average,  a pulsar has $S/N$  or jitter dominated timing errors.  Improved $S/N$ can be achieved using a larger telescope with more bandwidth but if \TOAs\ are  jitter dominated, the primary means for improvement is with longer integration times.   Jitter is correlated over broad frequency ranges  \citep[$\sim 1$\,GHz, e.g.][]{2014MNRAS.443.1463S}   so jitter is not reduced by frequency averaging over bandwidths  smaller than this correlation bandwidth.  Conceivably, high frequency timing with the ngVLA may incorporate very wide bandwidths that would reduce some of the jitter error in a few cases.   However, this improvement will most likely be negated by other timing errors. 

\begin{figure*}[htb!]
    \centering
    \includegraphics[width=\textwidth]{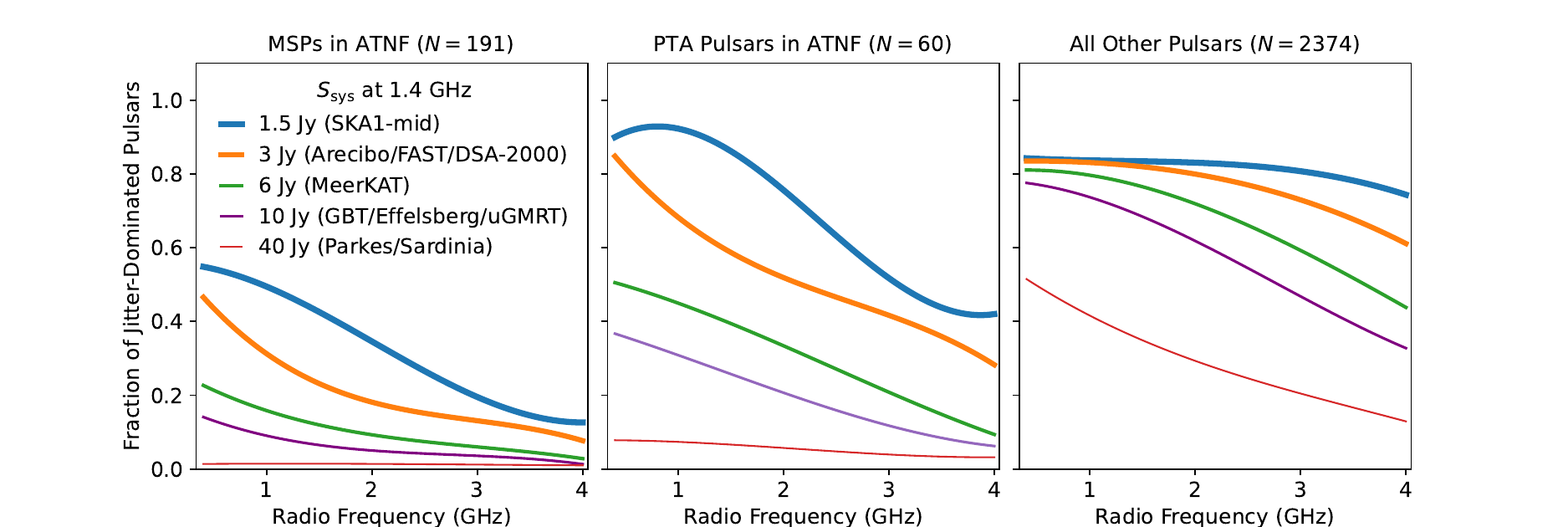}
    \caption{Fraction of pulsars that are jitter-dominated as a function of radio frequency, for MSPs (left; \newtext{191} LOS), PTA MSPs (middle; \newtext{60} LOS), and other pulsars (right; \newtext{2374} LOS) in the ATNF catalog. Fractions are calculated for five values of the system equivalent flux density $S_{\rm sys}$ at 1.4 GHz.
    The curves shown are cubic polynomials fit to the jitter-dominated fraction calculated at $\nu = 0.4$, $0.8$, $1.4$, $2$, $3$, and $4$ GHz, for $10\%$ radio bandwidths $B$ below 1 GHz and $B = 0.4$, $0.8$, $1$, and $1$\, GHz at $\nu = 1.4$, $2$, $3$, and $4$\,GHz. For simplicity, pulse shapes are assumed to be single-component Gaussians, and only pulsars with cataloged widths are used. For $\Ssys = 1.5$, the synchrotron radio background contributes enough to cause a non-monotonic dependence on radio frequency. 
    \label{fig:jitter_fraction}}
\end{figure*}

Fig.~\ref{fig:rmsjitter_vs_snr} (left panel) shows the combined \TOA\ error $\rmstoa = (\rmstoaS^2 + \rmstoaJ^2)^{1/2}$ plotted vs. $\SNR$ for simulated Gaussian pulses  
for several values of the jitter parameter $\FJ$ along with additive noise.  For small $\SNR$ the variation is noise dominated and decreases as $\SNR^{-1}$, as expected from Eq.~\eqref{eq:rmstoaS}, while the variations  are jitter dominated at large $\SNR$.


\begin{figure*}
    \centering
    \includegraphics[width=0.8\textwidth]{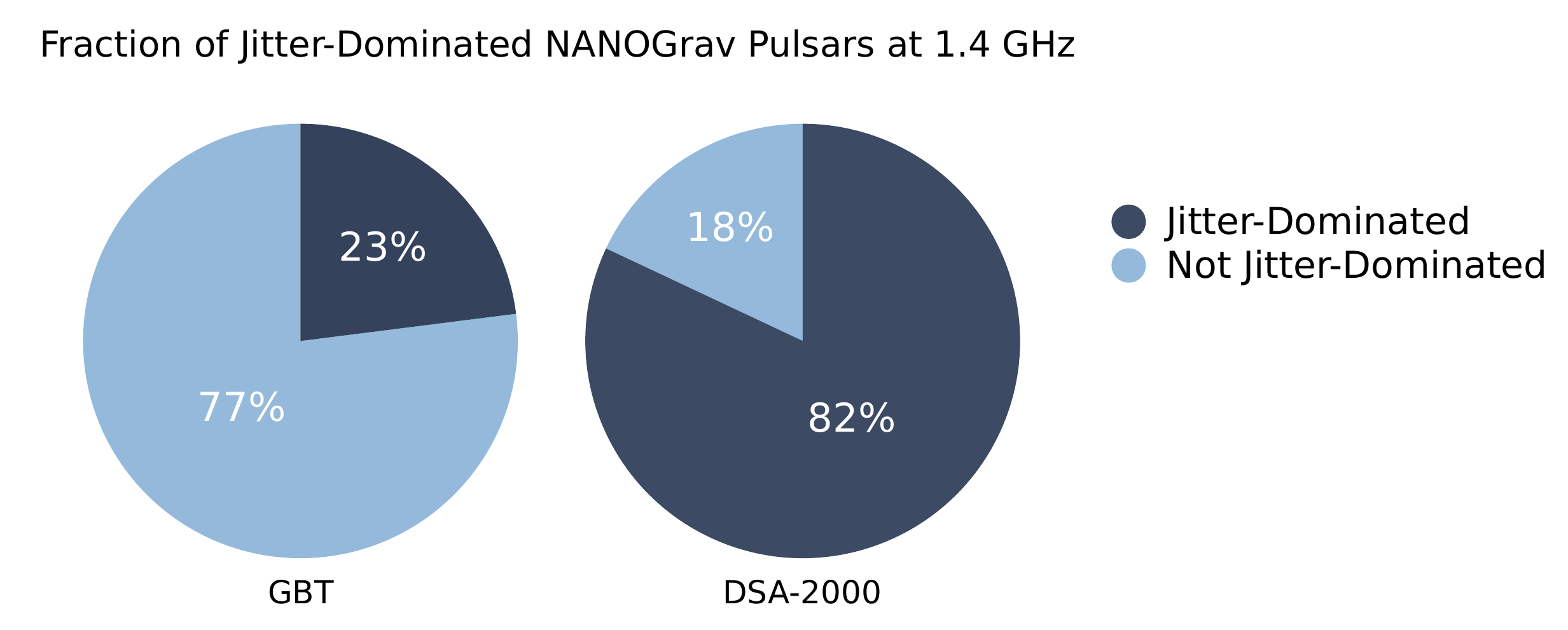}
    \caption{Fraction of NANOGrav pulsars that are jitter-dominated at 1.4 GHz for the GBT and DSA-2000, based on their respective sensitivities and bandwidths. The GBT fraction assumed a bandwidth of 0.5 GHz; increasing the bandwidth to that of the ultrawideband receiver under commissioning would increase the jitter-dominated fraction to about $50\%$.}
    \label{fig:jitter_pie_chart}
\end{figure*}


The transition occurs for a single-pulse signal to noise ratio 
$\SNR_1 = N^{-1/2} \SNR $
obtained by  
equating Eq.~\eqref{eq:rmstoaS} with Eq.~\eqref{eq:rmstoaJ} 
and using Eq.~\eqref{eq:Weff_gaussian} for the effective width
of a Gaussian pulse shape to obtain 
\be
\SNR_{1, \rm trans} 
=
    \frac{1} {(2\pi\ln2)^{1/4} \FJ} 
    \left(\frac{P}{\Nphi W} \right)^{1/2}
\simeq
    0.145 \times 
    \frac{P_{\rm ms}} {(N_\phi W_{100}  / 2048)^{1/2}}
    \left(\frac{1/3}{\FJ} \right).
\ee
\newtext{For many pulsars $P_{\rm ms} / W_{100} \sim 10$,  so a useful rule of thumb is that when single pulses have $\SNR_1  \gtrsim 1$, 
the \TOA\ errors are jitter dominated, regardless of the total integration time.} 

We now assess how many pulsars are expected to have jitter-dominated \TOA\ errors. 
Fig.~\ref{fig:jitter_fraction} shows the fraction of pulsars with jitter-dominated TOA variations as a function of radio frequency, calculated by counting the number of pulsars in the ATNF PSRCAT catalog\footnote{\url{http://www.atnf.csiro.au/research/pulsar/psrcat}} \citep[][]{ATNFPulsarCat}
for which $\rmstoaJ > \rmstoaS$ (or equivalently, $\SNR_1 > 1$). The TOA error due to radiometer noise was calculated assuming a sky background dominated by synchrotron radiation, $T_{\rm sky} \propto \nu^{-2.7}$, and a pulsar spectrum $\propto \nu^{\alpha_p}$. The pulsar spectral index $\alpha_p$ was inferred from flux densities cataloged at 400 and 1400 MHz where possible, otherwise $\alpha_p = -2$ was adopted. Fig.~\ref{fig:jitter_fraction} demonstrates how the fraction of jitter-dominated pulsars increases with telescope sensitivity. Arrival times of pulsars with $\SNR_1 \gtrsim 1$ are not improved by any increases in telescope sensitivity insofar as radiometer noise and jitter are concerned. Larger bandwidths reduce the arrival time variance due to radiometer noise by a factor $B^{-1/2}$. Larger bandwidths can also reduce arrival time variance due to interstellar scintillation
\newtext{(Sect.~\ref{sec:dt_fse})}, which can dominate over radiometer noise and jitter for pulsars with high DMs or observations at low frequencies. 

The fractions shown in Fig.~\ref{fig:jitter_fraction} correspond to three subsets of pulsars in PSRCAT: all MSPs, MSPs currently timed by PTAs, and all  non-MSPs. A significantly larger fraction of PTA pulsars are jitter-dominated because they are chosen to be bright: the mean flux density at 1400 MHz $\langle S_{1400} \rangle = 4$ mJy for PTA pulsars in PSRCAT, whereas $\langle S_{1400} \rangle = 1.6$ \newtext{mJy} for the total MSP population in PSRCAT. TOA variations due to radiometer noise are linearly dependent on flux density, resulting in a factor of $\sim 2$ difference between the jitter-dominated fractions calculated for all MSPs vs. PTA pulsars. The fraction of jitter-dominated pulsars timed by NANOGrav is shown in Fig.~\ref{fig:jitter_pie_chart} for the GBT and DSA-2000, demonstrating the dramatic increase in jitter-dominated pulsars that may be expected for the NANOGrav program on DSA-2000. Regardless of increases in telescope bandwidth and sensitivity, PTAs will always contend with jitter.  

While jitter becomes increasingly dominant at high sensitivities and low radio frequencies for the observed pulsar population, it likely will not be a limiting factor for timing Galactic Center pulsars. Rescaling the flux densities of the entire pulsar population in PSRCAT to their equivalent values at the distance of the Galactic Center yields $50\%$ of pulsars dominated by jitter at 8 GHz for $\Ssys = 0.3$ Jy and $20\%$ at 8 GHz for $\Ssys = 3$ Jy. Optimistically, roughly $50\%$ of Galactic Center pulsars will have jitter-dominated TOA variations for the ngVLA, if observed at high enough radio frequencies. These estimates do not explicitly account for scattering, which is significant near the Galactic Center and further reduces pulsar flux densities.  

For simplicity, we have assumed single-component Gaussian profiles to estimate the fraction of jitter-dominated pulsars shown in Fig.~\ref{fig:jitter_fraction}. However, the presence of multiple pulse components alters the TOA errors due to radiometer noise and jitter, by an amount that depends on the number of components and their phase overlap (see previous section). This effect is compounded by the frequency-dependence of pulse components, making it difficult to quantify across the pulsar population. 
The jitter fractions calculated here nonetheless provide useful benchmarks for understanding the relative roles of jitter and radiometer noise for pulsar timing with current and future telescopes.

\subsection{Estimation of jitter properties}

The above subsections demonstrate how amplitude and phase jitter of single pulses increase \TOA\ errors.   The variety of template shapes seen among pulsars with different numbers and mean amplitudes of components and different component stochasticity implies that the net effect on \TOAs\ is pulsar specific.   For most pulsars it is  difficult to make a multicomponent predictive jitter model unless all parameters for each profile   component can be estimated.   This is easiest if single pulses can be analyzed but in principle they could be estimated using $N$-pulse averages with variable $N$.  However, forward modeling of this kind is not necessary in order to assess the overall timing budget of a pulsar.

Several empirical methods  can be used for identifying and quantifying jitter, some applicable  to cases where single pulses are strong and others when only average profiles can be analyzed:
\begin{enumerate}
\itemsep -2pt
\item 
{\it Analysis of variance of \TOAs}:  Modeling allows separation of  a S/N dependent term (from template fitting) from  S/N-independent contributions, which are often dominated by jitter and sometimes accompanied significantly by interstellar scintillation noise.       \citet[][]{2016ApJ...819..155L} used this approach to characterize the jitter contribution at 1.4\,GHz  in $\sim$60\% of the MSPs  reported in  NANOGrav's 9-yr data set \citep[][]{2015ApJ...813...65N}; the \TOAs\ of the remaining 40\% were dominated by template-fitting errors that prevented jitter from being quantified.   Typically, the \RMS\ jitter is about 1\% of the pulse period.  Similar results were found in an analysis of MSPs observed at Parkes   \citep[][]{2014MNRAS.443.1463S}.
\item {\it Cross correlation analysis of timing residuals  between different frequency bands:}  Jitter is highly correlated across frequency separations $\lesssim 1$\,GHz \citep[][]{2014MNRAS.443.1463S} while additive radiometer noise is uncorrelated and scintillation-caused TOA variations are correlated over  separations $\lesssim$ scintillation bandwidth.   The cross correlation therefore can determine the contribution from jitter in relation to the other contributions.    For single pulses with $\SNR > 1$ and frequency separations  much wider than the scintillation bandwidth, the correlation coefficient is close to unity. 
\item  {\it Autocorrelation function analysis of template and single pulses:}   Pulse jitter always makes the template broader than single pulses.   So too is the ACF of the template vs. time (or phase) lag.   For Gaussian pulse components and a Gaussian phase jitter PDF, the net width of the template  is  the quadratic sum of a typical component width and the \RMS\ jitter multiplied by a factor accounting for multiple components.   Let the widths of the two ACFs be  $\WacfU$ and $\Wacfsp$, then the  width of an {\it effective} phase jitter distribution is $\WJeff \equiv (\WacfU^2 - \Wacfsp^2)^{1/2} / \sqrt{2}$, where  the $\sqrt{2}$ accounts for the ACF width being larger than the widths of pulses by this factor.   This method was used by
\citet[][Fig. 9]{es03a} to show that single pulses are significantly  (40\%) narrower than the template for the CP J1518+4904.   A similar analysis of the MSP J1939+2134 (B1937+21) showed only a small difference in ACF widths (FWHM) for the ACF of template and the ACF of single pulses, $\WacfU - \Wacfsp \sim 2\,\mu s$ with $\WacfU \sim \Wacfsp \simeq 50\,\mu s$, which was incorrectly interpreted as indicating the absence of jitter \citep[][Fig. 4]{jap01}.  Using the quadrature difference in ACF widths
and converting from ACF width to width of jitter distribution 
yields $\WJ \simeq 7\,\mu s$ or $\rmsJ \simeq 3\,\mu s$.   

\item  {\it Crosscorrelation analysis of template and profiles vs. number of pulses averaged}:  The correlation coefficient  between the template and an average profile formed from 
Gaussian  shaped single pulses with amplitude modulation index $\ma$ 
\newtext{can be shown  to be}
\be
\rho(N) = 
	\left[
	\frac{1 + (1+\ma^2) / (N-1)} {1 + (1+\ma^2) S_U / (N-1)} 
	\right]^{1/2}
	\simeq  1 - \frac{(1+\ma^2) (S_U-1)}{2(N-1)},
\label{eq:rhoN}
\ee
where the approximate equality applies for $N\gg1$ and
$S_U = \WU / W_a = \sqrt{1 + (\WJ / W_a)^2}$ is the ratio of the template width to the single-pulse width $W_a$. 
The absence of jitter corresponds to $S_U = 1$ and perfect correlation, $\rho(N) = 1$.
\newtext{Rearranging yields
$S_U \simeq 1 +  [2(N-1) (1-\rho_N) / (1 + \ma^2)]$ which can be used to evaluate  the width of the jitter distribution as
$\WJ  = W_a \sqrt{S_U^2 -1}$. }     

\citet[][]{hmt75} first used this method to demonstrate how average profiles converge to template shapes as $N$ increases.
Application to other CPS showed break points in $1-\rho(N)$ vs. $N$ for some objects,  indicating that pulse to pulse correlations persisted for a few to $\lesssim 100$ pulses \citet[][]{rr95}.    Application to millisecond pulsars by \citet[][]{2012MNRAS.420..361L} showed that  jitter properties of MSPs are very similar to those of  CPs. 
\end{enumerate}

\begin{figure}[htbp] 
   \centering
   \includegraphics[width=0.275\linewidth]{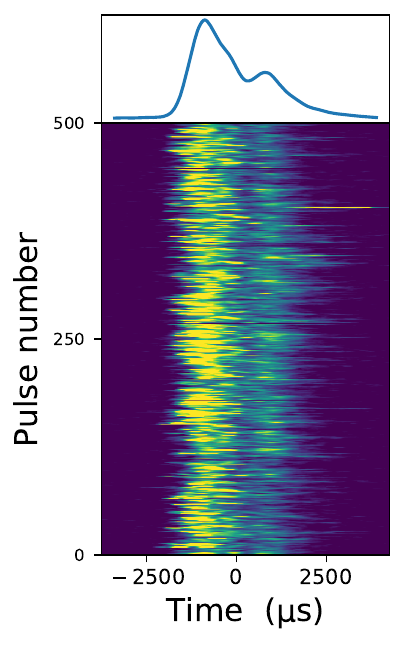}
   \includegraphics[width=0.65\linewidth]{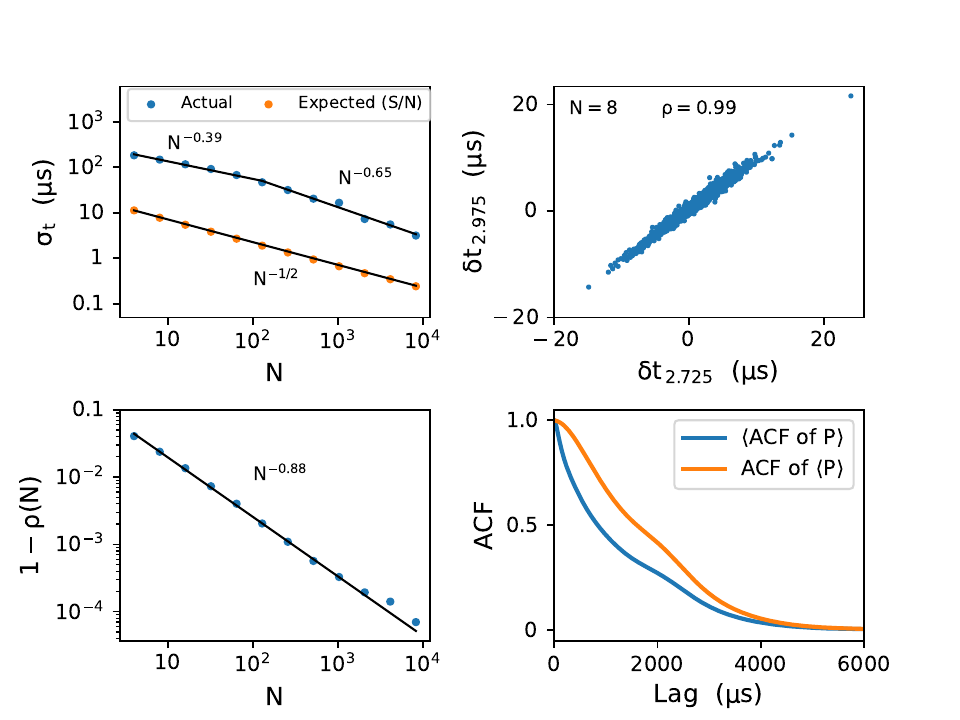} 
   \caption{
   Left panel: sequence of 500 pulses and average profile  from the Vela pulsar J0835-4510 ($P=89$\,ms) 
   	at 2.7 to 3\,GHz.
   Right panels: Analysis of pulse jitter. 
   Top left: RMS arrival time error vs $N$ = number of pulses averaged (blue) and the predicted 
   matched-filtering errors (red).   The scaling with $N$ is shown  for two segments of the actual RMS error and for the expected $N^{-1/2}$ dependence if pulses are statistically independent.
   Top right:  \TOA\ deviations for two frequency bands (2.725 and 2.975 GHz) that show the large correlation.
   Bottom left:  Correlation coefficient $\rho(n)$ between profiles calculated from $N$ pulses 
   and the template shape,  plotted as  $1 - \rho(N)$ vs. $N$. 
   Bottom right: Autocorrelation functions of single pulses ($\langle {\rm ACF \ of \ } P\rangle$) and of the template (ACF of $\langle P \rangle$). 
   \label{fig:Vela}
   }
\end{figure}

We demonstrate  these methods using pulses from  the Vela pulsar obtained with the Parkes telescope at
2 to 3 GHz (Matthew Kerr, private communication; Ross Jennings, private communication).   The  strong single pulses, with average  single-pulse S/N $\sim 12$,  yield   \TOA\ errors dominated by jitter with minimal contributions from either radiometer noise or DISS.    A sequence of 500 pulses is shown in the left panel of 
Fig.~\ref{fig:Vela}. 

The right panel of Fig.~\ref{fig:Vela} shows  diagnostics of jitter in four frames 
\newtext{
similar to results obtained by
\citep[][]{2024arXiv241100236J}. }
The top left shows the actual and predicted (based on matched filtering) \RMS\ \TOA\ vs. the number of pulses $N$ used to compute average profiles.   The actual RMS exceeds the predicted RMS by more than a factor of ten.  The \TOA\ errors are highly correlated between non-overlapping frequency channels separated by 
50\,MHz (top right frame).   The excess \TOA\ variations  are from pulse jitter and not DISS because the latter would be uncorrelated between frequency channels.    Using the values for the smallest $N=4$ value shown,  the RMS jitter for single pulses is $\rmstoaJ(N=1) \simeq 370\,\mu s$. 

Another estimate for RMS jitter results from the template-profile correlation analysis. 
The bottom left frame shows $1-\rho(N)$ vs. $N$ that declines with increasing $N$ as expected for profiles shapes that converge to the template shape.  However, the scaling $1-\rho(N) \propto N^{-0.88}$ is slightly shallower than the $N^{-1}$ dependence expected for pulse jitter that is uncorrelated between pulses.   Using Eq.~\eqref{eq:rhoN} and $1-\rho(4) \sim 0.041$ we obtain $\sigma_J(1)  \simeq 310$ to 390 $\mu s$ for single-pulse modulation indices of 1 to 1.5,  consistent with the jitter value from \TOAs. 

Finally, the ACFs  in the bottom right frame  yield   effective Gaussian widths (FWHM) of  $\WU = 2.3$  and  $W_a = 1.23$\,ms for the template and single pulses, respectively.   These 
yield  a  value, $\rmstoaJ(1) \simeq  830 \mu s$, more than twice the values obtained from the other methods; this is due to the fact that pulses comprise two primary components that cause the template profile width to exceed that expected from single-pulse widths if there were only one component.   Jitter occurs independently   in each of the two components and is thus about half the nominal value, or 415\,$\mu s$, consistent with the other estimates.

It is useful to define a dimensionless {\it jitter parameter} $\fJ$ to allow comparison of pulsars.  Expressing  jitter as an \RMS\ quantity $\rmstoaJ$,  \newtext{we define  $\fJ = 2\sqrt{2\ln 2} / {\rm FWHM}$}.  Using the first two methods applied above to the Vela pulsar yields
$\fJ \simeq (370\,\mus / 2300\,\mus) 2\sqrt{2\ln 2} \simeq 0.37$.   This value is similar to those obtained for other pulsars, including MSPs. 

The chromaticity of average profiles (Sect.~\ref{sec:intrinsic_chromatic}) follows that of  single pulses.   We therefore expect  pulse jitter, though highly correlated between nearby frequencies,  to show only partial correlation for frequency ratios $\gtrsim 2:1$.  Profile component spacing changes slowly with frequency, thus so too will the centroid phases of single pulses.  However,  the amplitudes of single pulses also appear to have limited bandwidths, albeit broad. 
    
\newtext{In practice, PTAs typically account for jitter using the so-called ``ECORR'' parameter, which assumes white noise that is temporally uncorrelated and completely correlated across radio frequency, and adds in quadrature to the template-fitting errors of ToAs. This procedure is known to overestimate the expected jitter noise \citep{2016ApJ...819..155L} and does not account for decorrelation of jitter when observations are performed over very large frequency bandwidths \citep{2024MNRAS.528.3658K}, therefore motivating the need for the alternative methods summarized in this section.}


\providecommand{\JA}{\Jarray_{\rm A}}
\providecommand{\Jg}{\Jarray_{\rm g}}
\providecommand{\thetaM}{\thetavec_{M}}
\providecommand{\thetahatM}{\widehat \thetavec_{M}}

\renewcommand{\fp}{f^{\prime}}
\newcommand{\xt}{x_t}
\newcommand{\xtildef}{\widetilde X_f}
\newcommand{\xtildefp}{\widetilde X_{f^{\prime}}}
\newcommand{\xtildeTf}{\widetilde X_{T,f}}
\newcommand{\Sxf}{S_x(f)}
\newcommand{\Sxfhat}{\widehat{S}_x(f)}
\newcommand{\Cxtau}{C_x(\tau)}
\newcommand{\sigmahat}{\widehat{\sigma}}
\newcommand{\Dxtau}{D_x(\tau)}
\newcommand{\Dytau}{D_y(\tau)}
\newcommand{\Dytauhat}{\hat D_y(\tau)}
\newcommand{\Dxtauhat}{\widehat{D}_x(\tau)}
\newcommand{\dxtau}{\Delta_x(\tau)}
\newcommand{\Pripple}{P_{\rm r}}
\newcommand{\fr}{f_{\rm r}}

\newcommand{\SSB}{\rm SSBC}

\section{Instrumental effects}
\label{sec:instru}

While this paper's primary  focus is on astrophysical noise processes in pulsar timing, understanding and forecasting of timing precision requires consideration of ancillary effects.   Included here are brief summaries of several prominent contributions, including polarization calibration, radio frequency and instrumental interference, digitization issues, time transfer, and the accuracy of solar system ephemerides. 

\subsection{Instrumental polarization and calibration}
\label{sec:polcal}
 
 Radio pulsar timing involves template fitting to Stokes $I$ (total intensity) profiles.  Any alteration of the Stokes $I$ profile from its true shape (or, more specifically, from the template shape) produces a \TOA\ error. The polarized nature of pulsar signals implies that improper calibration will contribute to such errors.     The primary effects to consider are (a) cross coupling between the antenna responses to the two desired polarization modes and
 (b) errors in the relative gain calibration of the the two channels.   The general issues of polarization calibration are well discussed in the literature \newtext{\citep[][]{2002ASPC..278..131H}}, but primarily for continuum and spectral line applications.   Careful calibration of pulsar data is done routinely and \newtext{there is ongoing assessment of its imperfections on arrival times \citep[][]{2024ApJ...973...94R, 2024ApJ...977..114D}}.  
 
 For specificity, we consider two nominal circularly polarized channels.  Generally, these will have responses with some ellipticity and they need not be exactly orthogonal.    The Jones matrix $\Jarray$  encompasses  these effects  as a $2\times 2$ matrix whose complex elements  determine the real elements of the  $4\times4$ Mueller matrix  $\Marray$,  which relates the measured Stokes vector $\Svec^{\,\prime}$ to the true $\Svec$ 
 (defined in Sect.~\ref{sec:Stokes}).  
 
 The Jones matrix for the feed antennas (e.g. \citet[][Appendix II]{ck69};  \citet[][Appendix]{scr+84};  \citet[][]{2002ASPC..278..131H}), 
\be
\JA(\thetavec_{J_{\rm A}}) = 
		\left[
		\begin{array}{cc}
			\displaystyle
			\frac{1}{(1+\epsilon_1^2)^{1/2}}  & \quad
				\displaystyle \frac{\epsilon_1e^{i\psi_1}}{(1+\epsilon_1^2)^{1/2}}  \\
			\\
			\displaystyle 
			\frac{\epsilon_2e^{i\psi_2}}{(1+\epsilon_2^2)^{1/2}}  & \quad
				\displaystyle
				\frac{1}{(1+\epsilon_2^2)^{1/2}}   
				 \\
		\end{array}
		 \right],	
\ee
is characterized with the  parameter vector,  
$\thetavec_{J_{\rm A} } = {\rm col}(\epsilon_1, \epsilon_2, \psi_1, \psi_2)$
and is augmented by a gain matrix.  
A full expression for the corresponding  Mueller matrix $\Marray_{\rm GA}$ that includes ellipticity and non-orthogonality is given by 
\citet[][Appendix]{scr+84}, which builds upon earlier work by \citet[][]{ck69}, and various factorizations
of $\Marray$ are discussed in \citet[][]{PhysRevE.56.6065, 2001PASP..113.1274H, van02, van06, 2013ApJS..204...13V,Gentile_2018}, including treatment of the Stokes vector as a Minkowskian four vector \citep[][]{bri00}.

Gain variations have a strong influence on the accuracy of Stokes parameters. 
Another matrix $\Jg(\thetavec_{J_{\rm g}}= {\rm diag}(g_{\rm R}, g_{\rm L})$ accounts for  departures of the channel gains from unity.  Together the product $\Jg \JA$ transforms the true electric fields into the measured voltages.   The corresponding Mueller matrix is $\Marray_{G}$. 
The  combined Mueller matrix 
 has a total of 6 parameters,
$\thetaM = {\rm col} (\epsilon_1, \epsilon_2, \psi_1, \psi_2, G_{\rm R}, G_{\rm L})$ where
\newtext{$G_{\rm L, R} = \vert g_{\rm L,R} \vert^2$}.  An additional parameter can account for an overall phase but can be absorbed into other quantities. 

The simplest form for $\Marray_{GA}$ is for orthogonal responses (found to be applicable in some cases) with 
$\psi_2 = \psi_1 + \pi$ and small coupling and gain elements, $\epsilon = \epsilon_1 = \epsilon_2 \ll 1$.   
and $\Delta G = G_{\rm R}  - G_{\rm L}  \ll 1$, with $G_{\rm R} G_{\rm L} = 1$, 
\be
\Marray_{\rm GA} \simeq 
		\left[
		\begin{array}{cccc}
			1  			& 	\ \ 0 					& 	\ \ 0 				&	(\Delta G)/2 			\\
			0  			& 	\ \ 1  					& 	\ \ 0				& 	-2\epsilon \cos\psi_1 	\\
			0			& 	\ \ 0 					&	\ \ 1				& 	-2\epsilon \sin\psi_1		\\
			(\Delta G)/2	& 	\ \ 2\epsilon\cos\psi_1	& 	\ \ 2\epsilon \sin\psi_1	& 	1		
		\end{array}
		 \right],	
\ee
This shows that Stokes-$I$ is altered by  Stokes-$V$ and vice versa but the latter is also influenced by the linear polarization, which itself is unaffected by Stokes-$I$.

To illustrate the effects on \TOAs\, we use the simplest case  of  orthogonal polarization responses combined with differential gains for the two hands of polarization, giving intensities $\IR^{\,\prime} =  G_{\rm R} \IR$ and $\IL^{\,\prime} = G_{\rm L} \IL$.  Alteration of  the Stokes $I$ profile depends on  the first row of the Mueller matrix, which gives
\be
I^{\,\prime} = \overline{G} I + \epsilon\Delta G (\cos\psi_1 Q + \sin\psi_1 U) + (\Delta G/2) V,
\ee
where the average  and differential gain are $\overline{G} = (G_{\rm R} + G_{\rm L})/2$ and $\Delta G = (G_{\rm R}  -  G_{\rm L})$ and $\epsilon \ll 1$ is a measure of the cross coupling. 
If all four Stokes parameters  have the same temporal shape,  only the amplitude of Stokes $I$ is altered with no change in \TOA.  However,  polarization profiles of pulsars invariably differ from the Stokes $I$ profile, so pulse distortion and the resulting \TOA\ error are inevitable. 

We use the degrees of linear and circular polarization, $\dL$ and $\dV$, 
to estimate the fractional error in Stokes $I$ 
\be
\frac{\delta I}{I} =   (\Delta G /2) \left [ d_V + 2\epsilon d_L \cos(2\chi - \psi_1)\right],
\label{eq:polerr}
\ee
where $\chi$ is the polarization angle of the incident field,
and $\psi_1$ is the phase of the cross coupling.  
 We have set the average gain $\overline{G} = (G_{\rm R} + G_{\rm L})/2 = 1$ and $\Delta G = (G_{\rm R}  -  G_{\rm L})$.   
For wideband systems now  used routinely in pulsar timing, the frequency dependence  of relevant quantities
 (i.e. $\Delta G, d_V, d_L, \epsilon$ and the argument $2\chi-\psi_1$)  implies 
 frequency-dependent calibration.  The error in Eq.~\eqref{eq:polerr} should then be considered for each subband in a wideband system, as implemented  in some pulsar analysis code packages.
 
Generally,  pulse shapes  in the  polarized components differ from the Stokes $I$ profile, so $\delta I / I$ is a measure of the distortion of the Stokes $I$ profile. \TOA\ errors  scale as $\delta {\rm \TOA} \sim (\delta I / I) W$, where $W$ is the profile width.   In principle, the measured profile can be corrected for cross coupling and the gain calibration difference $\Delta G$ thus  made arbitrarily small.  In practice, however, neither of these is the case and a \TOA\ error is inevitable.     We expect  both $\Delta G$  and $\epsilon$ to be small and often
$d_V \ll d_L$ but the two terms inside the square brackets in Eq.~\eqref{eq:polerr} will typically be comparable unless 
$\epsilon$ is very small.    Using fiducial values, the \TOA\ error (without the variable $\cos$ factor) is
\be
\delta{\rm \TOA} 
\simeq 50\,{\rm ns} 
	\left( \frac{\Delta G}{0.01}\right)
	\left( \frac{W}{100\,\mu s}\right)
	\left( \frac{d_V + \epsilon d_L} {0.1} \right).
\ee
This expression applies  to each subband or to the aggregate effect in a wideband system, where relevant quantities are band averages. 

Ongoing work has demonstrated the correctability of instrumental polarization and gain errors \newtext{\citep[][]{2024ApJ...973...94R,2024ApJ...977..114D}}.
This requires  use of template Stokes pulse profiles for a pulsar combined with  average profiles obtained at different parallactic angles, measurements of unpolarized continuum sources, and noise calibration signals to solve for cross coupling and gain parameters along with arrival times and average flux densities at each epoch \citep[][]{scr+84}; see also  \citep[][]{2002ASPC..278..131H}.

For our purposes here, we include the inference of arrival times along with  Mueller matrix parameters.
Writing the model Stokes vector as $\Svec_{\rm m} = {\rm col}(s), s = (I, Q, U, V)$,
\be
\Svec_{\rm m}(t, \tbar \vert   t_0, \psi_r, \thetaM) = 
	\Marray_{\rm GA}(\thetaM(\tbar) )
	\Marray_{\rm r}(\psi_r)
	\Svec(t - t_0),
\ee
where a rotation matrix $\Marray_{\rm r}$ is included that could account for Faraday rotation of the wavefield before incidence on antennas; 
$t$ and $\tbar$ represent pulse phase and epoch, respectively.
The cost function for template fitting (Eq.~\eqref{eq:cost}) becomes
 \be
C(t_0, \psi_r, \thetaM) = \sum_t \sum_s^{I,Q,U,V} 
	\left \vert 
		\Svec_{\rm m}(t, \tbar \vert   t_0, \psi_r, \thetaM, t_0, \psi_r)  - \Svec(t, \tbar) 
	\right \vert^2 / \Var(s_{\rm off}).
\ee
With seven total parameters in the most general form for $\Marray_{\rm GA}(\thetaM(\tbar) )\Marray_{\rm r}(\psi_r)$ or four parameters for the orthogonal, small coupling case, there are sufficient degrees of freedom in pulsar profiles and ancillary data to estimate parameters and allow inversion of the total 
Mueller matrix to calculate a corrected Stokes vector.

The question is,  how well can instrumental polarization be removed to  benefit  \TOA\ estimation, either from only Stokes-$I$ or from 
full-Stokes estimation?   The  recent studies cited above 
\citep[see also][]{2013MNRAS.430..416O,Gentile_2018,Wahl_2022}
indicate that improvements of order a factor of two or more are obtainable for some but not all cases.   It is beyond the scope of this paper to assess individual pulsars but it is clear that the demand for higher precision \TOAs\  will require close attention to instrumental polarization on a pulsar-by-pulsar basis as well as on a telescope-by-telescope basis.

\subsection{Digitization and interference}
\label{sec:interference}
\label{sec:adc}
\label{sec:ripple}

Further \TOA\ errors arise from aliasing and radio frequency interference .    We describe both of these in general terms by considering terms that add to true profile shapes.   Unbalanced, interleaved analog-to-digital conversion  (ADC) yields an aliased copy of pulses that appear with, effectively, a negative dispersion measure.    NANOGrav data analysis routinely removes this deterministic effect, as described in \cite[][]{Alam_2021}.  It requires determination of the ADC imperfection (time skew with unequal separation of time samples or gain mismatch) and values for relevant parameters.  
\citet[][]{Alam_2021} report correction of the dominant time skew to $\sim10$\% of \TOA\ errors that would otherwise range from a few tens of nanoseconds to $\sim 1\,\mu s$, depending on the pulsar's DM and period.

Radio frequency interference (RFI) has diverse time-frequency signatures, including impulsive and oscillatory forms.   Additional instrumental inteference can originate from leakage of power-line oscillations into the signal path.   In many instances, the resulting pulse shape distortions are dramatic and the data are excised.  In less severe cases, more subtle oscillations may occur across the entire profile as oscillations or ripple in the profile baseline.   Even low-level ripple can skew \TOAs\ but it is potentiallly correctable.   

A derivation of the effect of baseline ripple  on arrival times
\citep[][Appendices A and B]{2021ApJ...915...15J} is summarized and used here.    Let the intensity profile include a sinusoidal ripple term with amplitude $r$ and frequency $\fr$,
\be
I(t) = A\, U(t-t_0) + r \cos [2\pi\fr (t-t_0) + \phi ]. 
\label{eq:Iripple}
\ee
For a small ripple-to-signal ratio, $r / A \ll 1$, the template fitting error is written in terms of the Fourier transform $\widetilde U$, 
\be
\delta t_{\rm r}  = - 
	\left ( \frac{r \fr} {2\pi A} \right)
	\frac
		{{\Im}  \{ e^{-i\phi} \widetilde U(\fr )\} }
		{\int d\fp \,{\fp}^2 \vert \widetilde U(\fp) \vert^2  }, 
\label{eq:dtau_approx}
\ee
where $\Im$   denotes  imaginary part.
For small ripple amplitudes, multiple ripple terms that add linearly to the net TOA error can be easily dealt with, including 
an arbitrarily shaped baseline perturbation represented  as a Fourier sum.

\begin{figure}[h!]
\begin{center}
\includegraphics[scale=0.6]{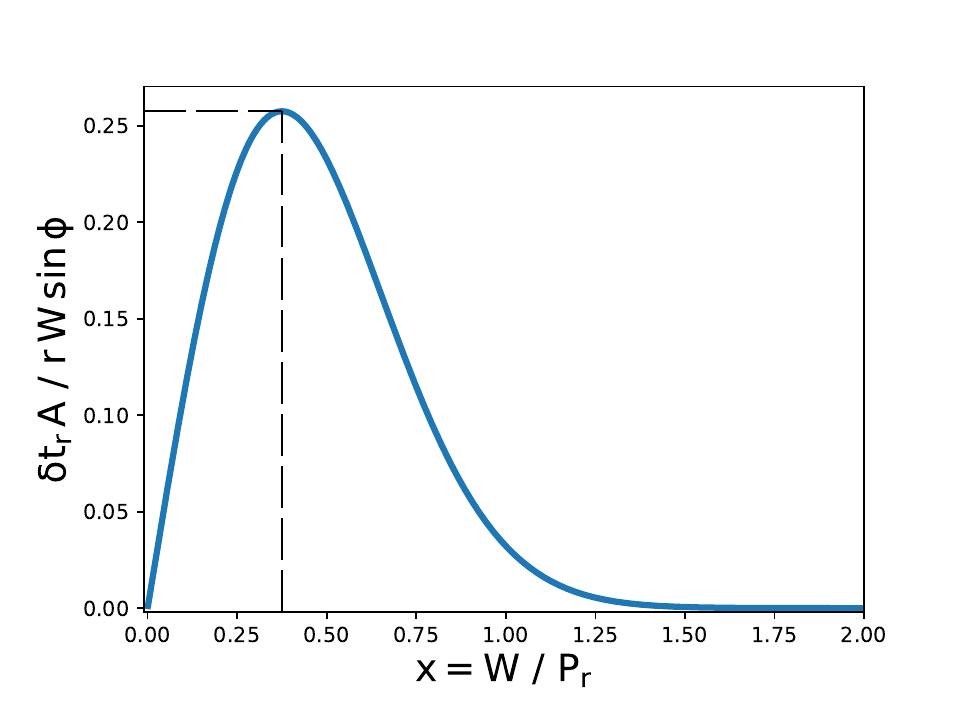}
\caption{ The  normalized \TOA\ perturbation from baseline ripple.  The dashed lines designate $x_{\rm max} = 0.38$ and the maximum of $\delta\tau A / r W\sin\phi =  0.26$. 
\label{fig:curve}
}
\end{center}
\end{figure}

For a Gaussian pulse of width $W$ and sinusoidal ripple, the \TOA\ error is 
\be
\frac{\delta t_{\rm r}}{W} = K \sin\phi
	\left(\frac{r}{A}\right)
	\left( \frac{W}{P_r} \right) 
	e^{\displaystyle -\pi K (W/P_r)^2},
\ee
with $K = \pi / (4 \ln 2)$.
The error vanishes for $\sin\phi = 0$ because  the cosine  maximum aligns with the peak of the Gaussian.  
It also  vanishes as $W/P_r \to \infty$ because the increasing number of integrated ripple cycles integrated  in  template matching leads to cancellation and also when  $W/P_r \ll 1$ because the perturbation is essentially constant over the pulse. 
In terms of $x = W/P_r$, the scaled TOA error follows a universal curve
$\delta t_{\rm r} A / rW\sin\phi  =  K x \exp( -\pi K x^2)$ shown in Fig.~\ref{fig:curve}, which maximizes at 
 $x_{\rm max} =  1 / \sqrt{2\pi K} = 0.375$, corresponding to a ripple period ${\Pripple}_{\rm _{max}} \simeq 2.67W$
and \TOA\ error $\delta t_{\rm r, max} \simeq 0.258 W (r\sin\phi /A)$.   

For  persistent ripple  in a multi-epoch data set, we expect $\phi$ to be  uniformly distributed in $[0,2\pi]$ yielding 
$\sigma_{\sin\phi} = 1/\sqrt{2}$.  Assuming constant $r/A$  at the 1\% level (only for illustration since ripple is episodic) and  other fiducial values, the \RMS\ error  is
\be
\delta t_{\rm r} = 0.80\,\mu s\, \times 
	\left(\frac{r/A } {0.01} \right)
	\left(\frac{W } {100\,\mu s } \right)
	\left(\frac{W} {P_{\rm r} } \right)
	e^{-3.56(W / P_{\rm r})^2}. 
\label{eq:ripple}
\ee
This error is nominally quite large compared to the best timing precision of MSPs $< 100$\,ns.  In most data, the various factors in Eq.~\eqref{eq:ripple} evidently are much less than unity.  However, there are data sets with substantial ripple that limit timing precision 
\citep[e.g.][]{2021ApJ...915...15J}.

\subsection{Time transfer to the Solar System barycenter}
\label{sec:ttransfer}

{\it Topocentric arrival times} are assigned according to  an observatory clock  synchronized to a hydrogen maser 
that, in turn, is synchronized to the  time standard provided by the Global Positioning System  (GPS) or other global navigation satellite system. The maser provides short term stability of the observatory clock while global time standards provide long-term stability.  Commercially-available hydrogen masers  drift by $\lesssim 0.1$~ns per day, and have achieved accuracies of better than 1 part in $10^{16}$ (i.e., $\sim$few ns) over $\sim$year timescales. \newtext{PTAs generally rely on post-corrected timescales from the Bureau International des Poids et Mesures (e.g., TT(BIPM)) for long-term stability.}

\begin{figure}[h!]
\begin{center}
\includegraphics[scale=0.35]{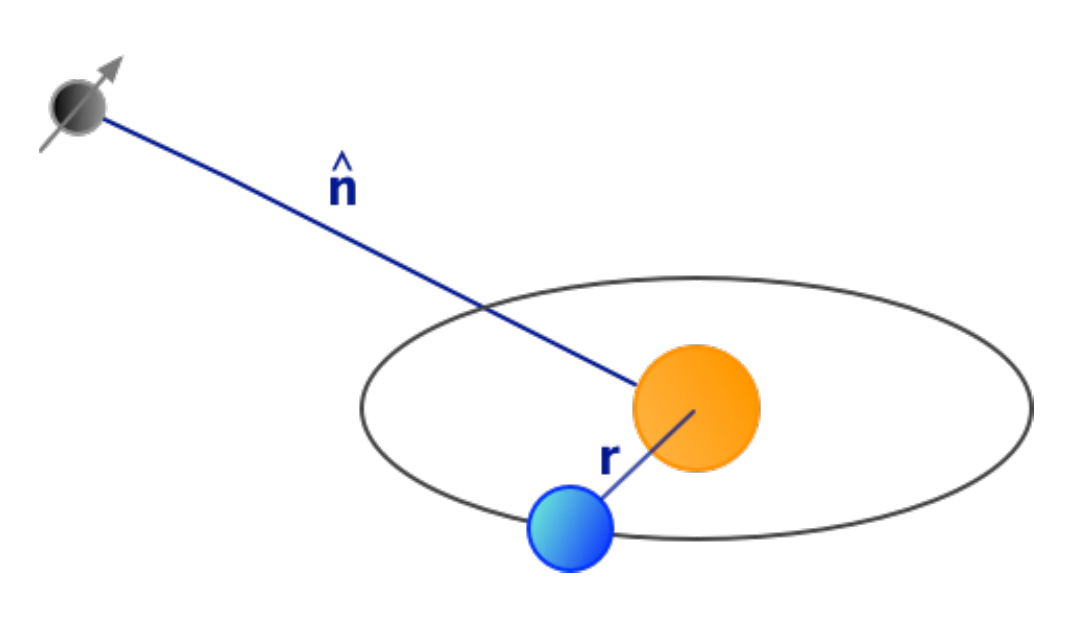}
\includegraphics[scale=0.60]{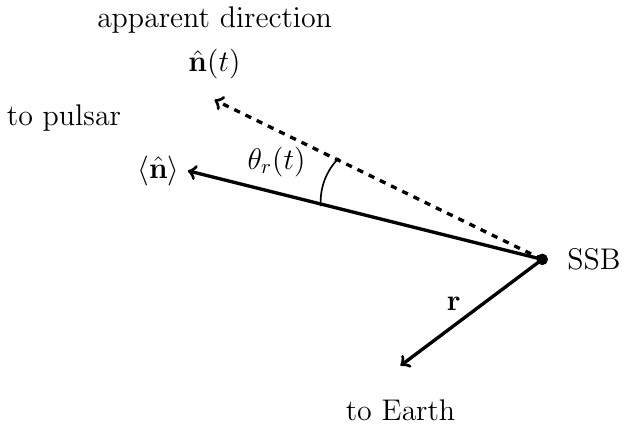}
\caption{Left: The geometry of the observatory-\SSB-pulsar system where $\hat {\mathbf{n}}$ is the unit vector from the 
\SSB\  to the pulsar and ${\mathbf r}$ is the vector from the \SSB\ to the observatory.  Both vectors are  functions of
epoch due to, respectively,  the pulsar's proper motion and the Earth's spin and orbital motions.  Right: In addition,
the pulsar direction is slightly different in the observatory frame from parallax and refraction in the ISM by an epoch dependent angle $\bm{\theta}_{\rm r}$ also can cause the apparent pulsar direction to differ.
\label{fig:sunearth}
}
\end{center}
\end{figure}

{\it Barycentric arrival times:} 
Arrival time analyses are easiest in the quasi-inertial  barycentric frame.   Topocentric \TOAs\ are referenced to
the Solar System barycenter (\SSB) by removing  the light-travel time from the observatory to the SBC, correcting for relativistic time offsets, and correcting for Doppler shifts.   
Given a topocentric \TOA\  $t(\nu)$, the barycentric \TOA\ $t_b(\nu^{\,\prime})$ at the Doppler shifted frequency
$\nu^{\,\prime}$ is
\be
t_b (\nu^{\,\prime}) = t(\nu) + \frac{{\bf r} \cdot \hat{\bf n}}{c} + \frac{({\bf r} \cdot \hat{\bf n})^2 - \abs{{\bf r}}^2}{2cd} - \frac{\KDM\DM}{{\nu^{\,\prime}}^2} + \Delta_{E\odot} + \Delta_{S\odot} + \Delta_{A\odot} ,
\label{eqn:ssb-timing}
\ee
where ${\bf r}$ is the observatory-\SSB\ vector\footnote{The observatory location is strictly the focal point of the telescope.}, $\hat{\bf n}$ is the unit vector from the \SSB\ towards the observed pulsar (see Fig.~\ref{fig:sunearth}), $\KDM\DM/{{\nu^{\,\prime}}^2}$ is the frequency-dependent dispersion delay (introduced in Sect.~\ref{sec:prop} and discussed in detail later sections), and the $\Delta$ terms are, respectively,  the Einstein delay due to gravitational redshift and time dilation, the Shapiro delay due to the curvature of spacetime around the Sun, and the aberration delay due to the Earth's rotation.   Further explanation can be found in \citet{bh86, stairs03} and \citet[][Ch. 3]{staylor_book2021} \newtext{and 
implementation in \citet[][]{2006MNRAS.372.1549E}}. 

Translation of \TOAs\ to the \SSB\  requires a maximum error in ${\bf r}$ tolerated by the astrophysical goals of the timing program.   The most demanding timing application, GW detection,  therefore requires better 
than 50 light-ns precision.    Along with an observatory position on the surface of the Earth \newtext{(updated, e.g., for continental drift for VLBI applications and less critically for pulsar timing)}, an accurate model for the Earth's rotation and orbit around the SSB is required\footnote{We note that the location of the SSB changes with respect to the Sun due to the orbits of the planets (primarily Jupiter and to a lesser extent Saturn) and can be located outside the Sun at certain epochs.}. Solar system ephemerides have been devised primarily for interplanetary spacecraft navigation \citep[e.g.][]{2021AJ....161..105P}
and also rely on  spacecraft missions for measurements of the mass and moment of inertia of the giant planets. Errors in estimating the location of the SSB  impose approximately a yearly sinusoidal variation in pulsar \TOAs 
\newtext{
that have been translated to the SSBC.}
Such a variation would  be {\em correlated} for all pulsars analyzed using the same ephemeris with a dipolar angular signature.  For instance, the loss of the main antenna on the {\em Galileo} spacecraft mission led to reduced tracking accuracy and hence a lower accuracy measurement of Jupiter's orbital elements, introducing correlated timing 
\newtext{residuals}
in the NANOGrav 11-year dataset. We refer the reader to \newtext{\citet[][]{2016MNRAS.455.4339T, 2020ApJ...893..112V}}  for detailed discussion.

As an aside, we note that very long baseline interferometry (VLBI) also involves telescope positions changing by large distances compared to the observation wavelengths during the course of an observation due to the Earth's rotation. VLBI observations are typically referenced to the geocenter as a fixed point, and the reference frame created by VLBI observations of distant quasars (which defines the International celestial Reference Frame; 
\citet[][]{2020A&A...644A.159C}
differs from the timing reference frame defined by solar system ephemerides at the 0.1--1~mas level 
\citep[][]{2013ApJ...777..104M,2019ApJ...875..100D}.  As a result, the positions measured for pulsars via pulse timing (as discussed below) and VLBI imaging can differ at the mas level, and can in turn be used to refine the frame tie between the two systems. 

\subsection{Astrometric errors, proper  motion, and parallax}
\label{sec:astrometry}

Referencing \TOAs\ to the \SSB\ also requires a precise sky position $\hat{\bf n}$ to calculate the delays in the second and third terms of Eq.~\eqref{eqn:ssb-timing}.    An angle error of $\delta\theta =1$\,mas yields an error  in the \SSB\ \TOA\  $\delta t \simeq \delta\theta \times 1\,{\rm \AU} / c \simeq 2.42\,\mu s$.   Errors in sky positions from VLBI are of this order while inclusion of the corresponding terms in the timing analysis allows sky positions  with smaller errors by up to  two orders of magnitude.   

Pulsars are moving targets owing to  their proper motions, which can be  as large as $1$\,arc\,sec\,yr$^{-1}$,  yielding a yearly cyclical variation with growing amplitude if a fixed sky position is used for \SSB\ referencing.  Instead, an appropriate  term is usually included in the timing analysis to solve for the proper motion and remove its contribution to the \TOA.    

Nearby pulsars can also show a sizable parallax (third) term  in Eq.~\eqref{eqn:ssb-timing}, 
$\Delta t_\varpi  = ({\bf r} \cdot \hat{\bf n})^2 - \abs{{\bf r}}^2 / 2cd$, with a 6-month period and  maximum amplitude   $\Delta t_\varpi  \simeq 1.21\,\mu s\,d^{-1}$ for a pulsar in the ecliptic plane at distance $d$ in kpc.   This effect can be measured and removed in the timing analysis of some pulsars and the corresponding 
angular parallax, $\varpi = (1\,\au / 1\,{\rm kpc} ) \times d^{-1} =  1\, {\rm mas}\times d^{-1}$, can be measured with VLBI for some objects.  \newtext{Recent results on timing proper motions and parallaxes are reported by 
\citet[][]{2024MNRAS.530..287S}}. 

Either way, parallax measurements are difficult: timing parallax can be masked by spin  noise and interstellar fluctuations while VLBI is subject to errors from plasma fluctuations in the ionosphere.
 Another issue is that, while VLBI observations can provide independent measurements of the astrometric parameters for pulsars,  incorporating them into timing solutions requires addressing the reference frame ties between the two different systems. In practice, incorporating VLBI measurements can help expedite the refinement of other timing parameters for a newly-timed pulsar \newtext{\citep[e.g.][]{2013ApJ...777..104M,2026ApJ...999..156S}}, but the self-consistency of astrometric parameters measured directly from the timing observations has been prioritized in PTA observations so far.


\section{ISM structure and propagation scaling laws}
\label{sec:ism_scalings}

The ISM is greatly heterogeneous in its density-temperature phase structure and with respect to mixtures of ionized, atomic, and molecular gas.  Radio propagation effects are strongest where the free electron density and magnetic fields are largest.    This then implies that the warm ionized medium (WIM) is the dominant phase contributing to pulse propagation, with a minor contribution from the hot ionized medium (HIM) and minimally by atomic and molecular clouds, where low level ionization occurs primarily from cosmic rays.   Some contributions may involve interfaces between discrete clouds and the WIM or HIM, which are subject to ionizing radiation or ionizing shocks.   

 For this paper, we are concerned with ISM effects on timing, so our focus is on descriptions of the medium insofar as they are needed to quantify timing effects.  Also, since timing is affected only minimally by magnetic fields (cf.\ Eq.~\eqref{eq:td}-\eqref{eq:tdcoeffs}), most of our presentation concerns the electron density and its fluctuations.

Dispersion and scattering both  appear to result from the same regions along a line of sight where the electron density is largest \newtext{\citep[e.g.][]{cwf+91}}.   This need not be the case, of course, because though all ionized regions contribute to DM,  density fluctuations on scales smaller than the Fresnel scale $\rF \lesssim 10^{11}$\,cm ($10^6$\,km) are needed to produce multipath propagation.    This requires the occurrence of processes that create and sustain density fluctuations, such as shocks from supernovae and cloud-cloud collisions, and bow shocks from  super-Alfv\'{e}nic stellar motions. These are local processes whose occurrence is  much higher in the inner Galaxy than at the solar circle and beyond because star formation is more prevalent in the Galactic interior, producing a higher supernova rate.\footnote{\newtext{The topic of Galactic stellar populations is vast.   Recent modeling of the Galactic supernova population 
\citep{2025CoPhC.31109537K}  incorporates astronomical tracers of high-mass star formation.}} 

Dispersion, scattering, and Faraday rotation are used to probe and model ISM structure. 
Galactic electron density models \citep[e.g. NE2001,][hereafter YMW16]{2002astro.ph..7156C, ymw16} are used primarily for estimating pulsar distances from their DMs and for estimating foreground contributions to the DMs of FRBs
\footnote{\newtext{NE2001 has recently been updated to include improvements based on new data obtained over the last quarter century \citep[][]{OckerCordes2026}.}}.   
These models include parameterized large scale structures (thin and thick disks and spiral arms) that are constrained by pulsar DMs. 
 NE2001 also models density fluctuations on large and small scales  using a power-law wavenumber spectrum along with parameters for the strength of the fluctuations in different Galactic locations.    Electron density models are also integral to modeling the Galactic magnetic field \citep[e.g.][]{2024ApJ...970...95U} using large samples of RMs from pulsars and extragalactic sources (AGNs, and FRBs) \citep[e.g. for FRBs][]{2022MNRAS.516.4739P}.

\subsection{Interstellar electron density wavenumber spectrum}
\label{sec:ne_spectrum}

\begin{figure}[t!] 
   \centering
   \includegraphics[width=0.7\linewidth]{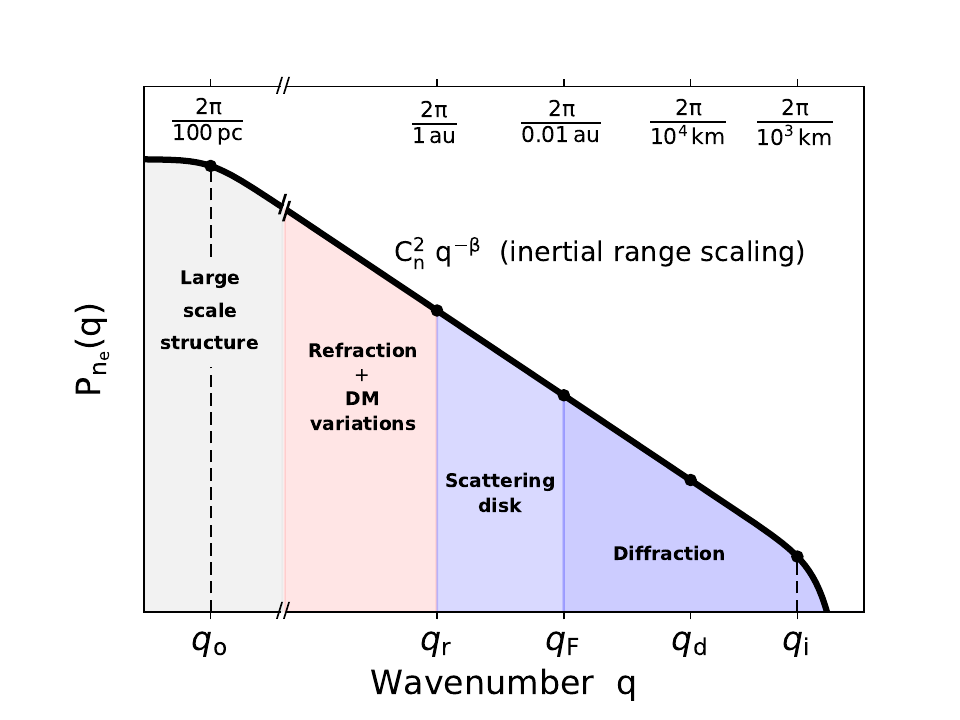}
   \caption{Schematic wavenumber spectrum for electron density.
   \newtext{
   The colored regions indicate wavenumber ranges characterized
   by different  features or phenomena:
   `Large scale structure' (pc to kpc) refers to (e.g.) spiral arms and interstellar clouds; `Refraction + DM variations' are effects caused by scales ($\gg \au$) much larger than those associated with scattering; `Scattering disk' refers to the scale $\sim \au$ spanned by the scattering cone due to  multipath scattering (as in  the cone diameter in the phase screen of Fig.\,\ref{fig:scattering_cone}); and `diffraction' refers to scales $\ll \au$ the cause diffractive effects (angular and pulse broadening).
   }
   Though density fluctuations span more than 14 orders of magnitude in wavenumber, it has not been demonstrated that  a turbulent cascade is responsible nor that any single process underlies all scales.   Additionally, strong evidence exists for the presence of discrete structures on some scales \newtext{(e.g. `extreme scattering events' (ESEs); see text)} that enhance fluctuations that do appear to follow a power-law form. }
   \label{fig:Pq}
\end{figure}

A  power-law wavenumber spectrum is the most frequently assumed form because  fluctuations in electron density are known to exist in the ISM on a large range of scales, 
\newtextR{
as may be seen in theoretical analyses 
\citep[e.g.][]{lj76}, observational constraints on the ionized ISM 
\citep[e.g.][]{lj76, cwb85, ars95, chepurnov2010},
and the very local ISM probed by the Voyager spacecraft
\citep[e.g.][]{lee2019,2021NatAs...5..761O}.
}
In this respect, the ISM parallels the interplanetary medium (IPM), which is much better characterized with {\it in situ} satellite measurements along with ground-based remote sensing.   The IPM 
displays density and magnetic-field fluctuations  with power-law spectra that include an inertial subrange, with  a Kolmogorov slope during some of that range,
\citep[e.g.][]{2009ApJS..182..310S} and dissipation on small scales.  IPM 
turbulence  is anisotropic and combines with other structures generated by solar activity \citep[][]{2021FrASS...7..114S}. 
As with the IPM,  there is evidence for anisotropies in the ISM along some lines of sight, implying that turbulent eddies  have preferred orientations due to magnetic fields  in specific regions in the ISM
\newtext{\citep[][]{smc+01,2018IAUS..337..283R}}.
A general form for the spectrum $\Pne(\qvec)$ with a vector wavenumber  $\qvec$  includes this anisotropy as an argument
$(A q_x^2 + A^{-1} q_y^2)^{1/2} $,
where $A$ is the axial ratio.  
However, to focus on integrated effects along a LoS, which likely average over anisotropies,  
we adopt an  isotropic form for the spectrum,
\be
\Pne(q) = \cnsq q^{-\beta} e^{-(q/\qinner)^2} ,
\label{eq:Pne}
\ee
where  $\cnsq$ is the spectral coefficient and the wavenumber cutoffs or rolloffs $\qinner, \qouter$ correspond to the {\it inner} and {\it outer scales} of the fluctuations,
$\qouter = 2\pi / \louter$ and 
$\qinner = 2\pi / \linner$.
The integral over all wavenumbers gives the  density variance,
$\sigma^2_{\nelec} = \int d\qvec\, \Pne(q)$ with $d\qvec = 4\pi q^2 dq$. 
Examples  of this and similar  spectral forms can be found in  \citet[][]{lj76, cwb85, Coles1987, ars95} and \citet[][]{goodman1985statistical} \newtext{and references therein}.
A Kolmogorov spectrum corresponds to a spectral index $\beta = 11/3$.


\newtext{The  spectrum, shown schematically in  Fig.~\ref{fig:Pq}, 
has an amplitude $C_{\rm n}^2$ that varies by several orders of magnitude across the Galaxy
\citep[][]{2021ApJ...922..233O}, as incorporated in the Galactic electron density models NE2001 and NE2025.}  
Length scales described by the spectrum  span more than ten orders of magnitude
with $\linner \lesssim 10^3$~km and $\louter \sim 100$~pc in the thick disk component of the Galaxy. 
The very large outer scale is essentially irrelevant to timing measurements made over tens or even hundreds of years, so the specific form is not critical. 

Support for the general form of the spectrum includes radio astronomical measurements of  pulsars and other objects \citep[e.g.][]{ars95}; small scales are sampled by scattering measurements while the largest scales for these objects are probed in decades-long time series of dispersion measures (Sect.~\ref{sec:DM_analytics}).  Familiar Galactic structures (atomic and molecular clouds,  HII regions, filaments, etc.), some of which are known to include turbulent motions, comprise the largest scales. 
 
Additional evidence for a broad power-law spectrum comes from studies of the cosmic ray proton energy spectrum, which extends from $\sim 1$ to $10^8$~GeV in a smooth power law (with a change in slope at higher energies).   Protons scatter off of magnetic irregularities on scales comparable to their gyro radii, $r_{\rm g} \simeq 0.074~\au \times E({\rm GeV}) (B / 3~\mu {\rm G})$.   The smoothness of the proton spectrum implies a spectrum of magnetic irregularities spanning a range at least as large as
$\lesssim 0.1$~\au\ to 40~pc.  In theories of compressible turbulence,  these magnetic fluctuations are accompanied by density fluctuations on the same scales
\citep[e.g.][]{1988AIPC..174...48J,1995ApJ...438..763G}.

There are strong variations in the spectral coefficient  $\cnsq$,  the inner,  and the outer scale across the Galaxy. 
The outer scale appears to be smaller in denser regions toward the inner Galaxy and larger in the thick disk component of the electron density distribution.
The notion that a continuous broad spectrum of scales is produced by a single physical process has not been demonstrated in the ionized ISM.    

Though a turbulent cascade plausibly accounts for some of the wavenumber range,  the diversity of density-temperature regimes in the ISM \citep[][]{Draine2011} suggests that multiple processes are at work. 
There appear to be small, discrete structures in the ISM that superpose with power-law fluctuations that distort the wavenumber spectrum from a pure power-law with fixed spectral index.    \newtext{Their presence is manifested in ESEs (`extreme scattering events') manifested in 
AGN light curves \citep[][]{1987Natur.326..675F,  2017ApJ...845...90V,doi:10.1126/science.aac7673}, 
AGN imaging \citep[][]{2023MNRAS.526.5932K}, 
pulsar light curves and timing \citep[][]{1993Natur.366..320C,lrc98,mlc03,2018MNRAS.474.4637K},
refractive scintillations from pulsars \citep[][]{ra82, hwg85, cw86}
 The  statistics
of such structures are not easily modeled because they are both sparse and diverse in nature but they suggest the presence of structures on $\sim \au$ scales that cause plasma lensing \citep[e.g.][]{rbc87,1998ApJ...496..253C}. Their existence in the ISM is subject to a variety of interpretations
\citep[e.g.][]{2007ASPC..365..299W}.}

\newtext{
The importance of the \au\ structures responsible for  ESE-type events is that they induce arrival time variations in a variety of ways, as we elucidate throughout much of the rest of the article. 
}

Fig.~\ref{fig:Pq} also designates regions defined by characteristic propagation scales that are indicated with approximate values along the top horizontal axis.   From smallest to largest, these are the diffraction scale, the Fresnel scale, the refraction scale corresponding to the size of the scattering disk (cf.\ Fig.~\ref{fig:scattering_cone}), and still larger scales that are manifested as variations in DM and refractive scintillations. These scales are discussed in more detail in Sect.~\ref{sec:sclaws}.

\subsection{Propagation through plasma fluctuations  with a power-law wavenumber spectrum}
\label{sec:plasmaprop}

A thin screen suffices for identifying  and analyzing  propagation effects relevant to pulsar timing, which we summarize here.    Thick screens or otherwise extended media yield the same phenomena but with different amplitudes and scaling laws that need consideration for analyzing individual \LoS.   These are described more selectively in what follows.

The {\it phase structure function} appears frequently in analyses of wave propagation because many observables are related to phase differences rather than to the absolute phase. 
Specifying $\phi(\xvec)$ as the electromagnetic phase perturbation induced by a screen and  $\dxvec$ as a transverse spatial separation
across the screen,  the phase SF is
\be
\Dphi(\dxvec) \equiv 
	\left\langle 
	\left[ 
		\phi(\xvec+\dxvec)-\phi(\xvec) 
	\right]^2
	\right\rangle .
\label{eq:dphi}
\ee
where $\xvec$ and $\dxvec$ are vectors in the screen plane transverse to the propagation direction. 
In terms of  the wavenumber spectrum $\Pne(\qvec; z)$, 
\be
\Dphi(\dxvec) =  4\pi (\lambda \relec)^2 
	\int dz\, \int d\qvecperp\, [1 - e^{ i\qvecperp\cdot\dxvec]} \Pne(\qperpvec, q_z=0; z). 
\ee
Application to specific cases must take into account that a source embedded in or near the medium emits diverging, spherical waves, which requires  the replacement  $\dxvec \to (z/\dso) \dxvec$ in the complex exponential, where $\dso$ is the source-observer distance. 

For isotropic density fluctuations, the spectrum  depends only on $\vert \qperpvec \vert$ and integration over azimuthal angle in the $\qperpvec$ plane yields an SF dependent only on the scalar offset $\dx$, 
\be
\Dphi(\dx) = 8\pi^2 (\lambda \relec)^2 
	\int dz\, \int d\qperp [1 - J_0( \qperp\dx)]\Pne(\qperp, q_z=0; z),
\ee
\newtext{where $J_0$ is the zeroth-order Bessel function of the first kind}.

\newtext{
For the spectrum in Eq.~\eqref{eq:Pne}  with isotropic irregularities and plane-wave incidence on a screen, the phase structure function is  \citep[e.g.][]{Coles1987}
\be
\Dphi(\dx) = \frac{8\pi^2 (\lambda \relec)^2 \Gamma(2-\beta/2)} {(\beta-2) \qinner^{\beta-2} }
	\int_0^d ds\, \cnsq(s) 
	\left[
	{}_1F_1\left(1-\beta/2, 1, -\left(  s \qinner\dx / 2d\right)^2 \right) 
	-1
	\right]
\ee
where ${}_1F_1$ is the confluent hypergeometric function of the first kind.
If $\cnsq$ is constant along the line of sight,  the phase SF is 
proportional to  the product  $\lambda^2\SM$, a measure of the scattering strength
where $\SM$ is the \LOS\ integral of $\cnsq$ and  three regimes can be identified,
}
\be
\Dphi(\dx)  
	&=&   (\lambda \relec)^2 \SM \times
	\left\{
		\begin{array}{clcl}
			\pi^2 \Gamma(2-\beta/2)\,  \qinner^{4-\beta} \,\dx^2, 
				&\quad  \qinner\dx \ll 1 
				& \quad \text{I.} & \quad \text{Square-law regime,} \\
		         \\
			f_\beta \,\dx^{\beta-2},
				&\quad  \qinner^{-1} \ll \dx \ll \qouter^{-1}  
				& \quad \text{II.} & \quad  \text{Inertial subrange regime,} \\
			\\ 
			\left(\frac{\displaystyle 8\pi^2}{\displaystyle \beta-2}\right) \,\qouter^{2-\beta} ,
				&\quad  \qouter \dx \gg 1 .
				& \quad \text{III.}  & \quad \text{Asymptotic regime} ,
		\end{array}
	\right.
\label{eq:Dphi_beta}
\ee
where 
\be     
f_\beta = 
    \frac {\displaystyle 8\pi^2 \Gamma\left(2-\beta/2\right)}
    {\displaystyle (\beta-2)2^{\beta-2} \Gamma(\beta/2)}
    \overset{\beta = 11/3} \simeq  88.3. 
\label{eq:fbetadef}
\ee 

The three scaling regimes  are, respectively, for: 
(I)  lags $\dx$ smaller than the inner scale, where  $\Dphi \propto \dx^2 \linner^{\beta-4}$  is a function of the inner scale ($\lesssim 10^3$\,km for the ISM);
(II) intermediate lags between the inner and outer scale (the inertial subrange), where the dependence is shallower, 
$\Dphi\propto \dx^{\beta-2}$ and independent of either scale;
and
(III) lags much larger than the outer scale,  $\dx \gg \qouter ^{-1}$, where the SF asymptotes  to a constant 
$\propto \louter^{\beta-2}$ that can be very large (many square radians) if the outer scale is large. 

 The form of the phase
SF $\Dphi(b)$ is given in Eq.~\eqref{eq:Dphi_beta} for an isotropic medium where only the magnitude
$b = \vert \bvec \vert$ matters. 

The scattered image of a point source is the 2D Fourier transform of the visibility (the van Cittert-Zernike theorem); these are given by
\be
\Gamma(\bvec) = e^{-\Dphi(\bvec)/2}, 
\quad\quad
I(\thetavec) =  \int d\bvec\, e^{i k \bvec\cdot\thetavec} \Gamma(\bvec),
\label{eq:vis_im}
\ee
where $\bvec$ is a vector between two positions in the observation plane. The phase SF in the observation plane is identical to that in the screen plane (with $\dxvec$ replaced by $\bvec$ for scattering through a thin screen). 

For isotropic density irregularties, the expressions depend only on the magnitudes of $\bvec$ and $\thetavec$. 
Many observations of scattered images are in the square-law regime  (I)  in order for scattering diameters to be  resolved with terrestrial baselines.   Observations of Cyg X-3 \citep[][]{1995ApJ...438..708M} spanned the transition between the square-law and inertial subrange regimes and thus provided an estimate of the inner scale
$\sim 300$\,km (50\% error).   A study of angular scattering along eight lines of sight by \citet[][]{sg90} indicates values $\linner \simeq 50$ to 200\,km.  

Radio pulsar timing measurements at $\sim 0.5$ to 1\,GHz are largely in regime II except for distant pulsars observed at low frequencies where the square-law regime applies.   The differences between regimes I and II are relevant to methods for correcting  \TOAs\ for scattering delays.  The asymptotic regime III corresponds to (twice) the total phase variance along a line of sight and is dominated by large scale Galactic structure.

\subsection{Scaling laws for propagation effects from a power-law density spectrum}
\label{sec:sclaws}

Here we define useful quantities for describing dispersion, scattering, and scintillation that we present as scaling laws in Table\,\ref{tab:keyplasmaeq}  along with evaluations for a Kolmogorov spectrum ($\beta = 11/3$).  Where unspecified, the units used are distances in kpc, radio frequencies ($\nu$) in GHz, fluctuation frequencies ($f$) in cycles\,yr$^{-1}$, velocities in $100\, \kms$, and time lags ($\delta t)$ in yr.   Scattering measures are $\SM = 10^{-3.5}\, \SM_{-3.5}\, \SMunits $. 

\renewcommand{\arraystretch}{0.8}

\begin{deluxetable}{c l l l l } 
\tablecaption{\label{tab:keyplasmaeq} 
Dispersion, Scattering, and Scintillation Scaling Laws for Power-law Media  \\
}
\tablecolumns{4} 
\tablehead{
Item  & 
\multicolumn{1}{c}{Quantity} & 
\multicolumn{1}{c}{Expression\tablenotemark{\footnotesize a} }  & 
\multicolumn{2}{c}{
    Evaluation for $\beta = 11/3$\tablenotemark{\footnotesize b}} \\
No. & & & \multicolumn{1}{c}{Coefficient} & \multicolumn{1}{c}{Scaling }    \\[-0.6cm]  
}
\decimalcolnumbers
\startdata
   1 &
   \quad\quad
   $\SM$ & 
    ${\displaystyle
         \frac{(2\pi)^{\beta-2}} {f_\beta \relec^2 c^{(\beta+2)/2}} \  
         \left[ \frac{\taud}{\deff} \right]^{(\beta-2)/2} \nu^{\beta}}$ &
    $10^{-3.26}\,\SMunits$ & 
    ${\displaystyle   \left( \tau_1 / \deff\right)^{5/6}     }$ 
    \\
    \\
    2 &
    \quad\quad
    $\sigma_\DM$ & 
    ${\displaystyle
    \frac{(2\pi)^{(4-\beta)/2}}  {(\beta-2)^{1/2}}\  \louter^{(\beta-2)/2} \SM^{1/2}}$ &
    $0.097~\DMunits$ &  $ \louter^{5/6}(\rm pc)\, \left( \tau_1 / \deff \right)^{5/12}$  
    \\
    \\
    3 &
    \quad\quad
    ${\displaystyle 
    \DDM(\delta t) }$ &
    ${\displaystyle
         \frac{\Dphi(\veff \delta t)}{(\lambda \relec)^2} =
         \frac{(\delta t / \dtd)^{\beta-2}}  {(\lambda \relec)^2} } $ & 
    $\left(2.15\times10^{-4} \, \DMunits\right)^2$ & 
    $\left (\delta t_{\rm yr} / {\dtd}_{100} \right )^{5/3}$ 
    \\
    \\
    4 &
    \quad\quad
    $S_\DM(f)$ &
    $a_\beta \SM \veff^{\beta-2} f^{-(\beta-1)}$ & 
    ${\displaystyle \frac{(7.70\times10^{-5}\, \DMunits)^2} {{\rm cy\,yr^{-1}}}}$ &
    $v_{100}^{5/3} \left(\tau_1 / \deff \right)^{5/6} f^{-8/3}$  
    \\
    \\
    5 &
    \quad\quad
    $\rF$ &
    ${\displaystyle
    \left(\lambda \deff / 2\pi \right)^{1/2} }$ &
    ${\displaystyle 8.1 \times 10^{-3}\ {\rm \AU}}$ &
    ${\displaystyle  (\deff / \nu)^{1/2}}$ 
    \\
    \\
    6 &
    \quad\quad
    ${\displaystyle {\ld}_{_I} = \frac{{\ld}_{_\theta}} {2^{1/(\beta-2)} } } $ & 
    ${\displaystyle 
    \frac{1}{2^{(4-\beta)/2(\beta-2)}}  \frac{1}{2\pi\nu} \left( \frac{c\deff}{\taud} \right)^{1/2} }$ &
    ${\displaystyle
    1.43\times10^4~{\rm km}}$ &  
    ${\displaystyle \nu^{6/5} \left(\deff / \tau_1 \right)^{1/2}}$ 
    \\
    \\
    7 &
    \quad\quad
    $\lr$ &
    $\rF^2 / {\ld}_{_I}$  &
    ${\displaystyle 0.69 \,\AU}$ & 
    ${\displaystyle 
        \nu^{-11/5}\,  \left(\deff \tau_1 \right)^{1/2} }$
        \\
        \\
    8 &
    \quad\quad
    $\thetad$\,(FWHM) &
    ${\displaystyle
    \frac{2\sqrt{\ln 2}}{\pi} \frac{\lambda}{{\ld}_{_\theta}}}$ &
    ${\displaystyle 1.51~{\rm mas}}$ & 
    ${\displaystyle 
        \nu^{-11/5}\,
        \left( \tau_1 / \deff\right)^{1/2} }$ 
    \\
    \\
    9 &
    \quad\quad
    $\thetar$\,(\RMS) &
    ${\displaystyle
    (c / 2\pi\nu) [\Dphi(\lr)]^{1/2} / \lr }$ &
    ${\displaystyle 0.157~{\rm mas}}$ & 
    ${\displaystyle 
        \nu^{-49/30}\,
        {\deff}^{-1/2} 
        \tau_1^{1/3} }$ 
    \\
    \\
    10 &
    \quad\quad
   $\nutrans$ &
   ${\displaystyle 
   \nu (2\pi\taud\nu/C_1)^{(\beta-2)/(\beta+2)} }$ & 
   13.1 GHz &
   $\left(\tau_1 / C_1 \right)^{5/17}$ 
   \\
   \\
    11 &
    \quad\quad
    $\taud$ &
    ${\displaystyle
        \frac{\deff \thetad^2}{8 \ln 2\, c} }$ &
    ${\displaystyle 0.44~\mu{\rm s}}$ &  
    ${\displaystyle
        \nu^{-22/5}
        \deff \theta_1^2({\rm mas}) }$  
        \\
        \\
    12 &
    \quad\quad
    $\dnud$ &
    $C_1 (2\pi\taud)^{-1}$ &
    ${\displaystyle 0.16~{\rm MHz}}$ &
    ${\displaystyle 
        \nu^{22/5} (C_1 / \tau_1) }$ 
    \\
    \\
    13 &
    \quad\quad
    $\dtd$ &
    ${\displaystyle
    \frac{{\ld}_{_I}}{\veff}}$ &
    ${\displaystyle 143~s}$ &
    ${\displaystyle  
     \nu^{6/5}\,
    v_{100}^{-1}\, 
    \left(\deff / \tau_1 \right)^{1/2} }$ 
    \\
    \\
    14 &
    \quad\quad
    $\dtr$ &
    ${\displaystyle \frac{\lr}{\veff} \ {\rm or} \ 
    \frac{1}{\dtd} \left(\frac{\rF}{\veff}\right)^2 }$ &
    ${\displaystyle 11.9~{\rm d}}$ &
    ${\displaystyle  
        \nu^{-11/5}\,
        v_{100}^{-1}\,    
         \left(\deff \tau_1  \right)^{1/2} }$ 
    \\
    \\
    15 &
    \quad\quad
    $\dnusb$ &
    $(2\pi\dtd)^{-1}$ &
    ${\displaystyle 0.97 ~{\rm mHz}}$ & 
    ${\displaystyle 
        \nu^{-6/5}\,
        v_{100}\,
        \left(\tau_1  / \deff \right)^{1/2} }$ 
    \\
    \\
    16 &
    \quad\quad
    $\Nscint$ &
    $\etat\etanu (B/\dnud) (T/\dtd)$  &
    ${\displaystyle 3960}$ & 
    ${\displaystyle 
        \nu^{-28/5}\, B_{\rm GHz}  T_{1000}
        v_{100}\,
        \left(\tau_1^3  / \deff \right)^{1/2} }$ 
    \\
    \\
\enddata
\tablenotetext{a} {
		Expressions are for the strong scintillation regime and  a thin screen with 
                 phase structure function $\Dphi(b) \propto b^{\beta-2}$ .} 
\tablenotetext{b} {
	Numerical results for $\beta = 11/3$ use 
	scattering time $\tau_1 \equiv \taud(1\,{\rm GHz}) = 1\,\mu$s,  
	angular scattering diameter $\theta_1 \equiv \thetad(1\,{\rm GHz}) = 1$\, mas,
	effective distance $\deff = 1$\,kpc.  
	The center frequency  $\nu$ is in GHz. \\}
\vspace{-0.1in}
{Definitions and units:
        \begin{itemize}
        \itemsep -1pt
        \item[]
            All distances are in kpc: ($\dso, \dsl, \dlo, \deff = {\rm effective\ distance} \equiv  \dsl\dlo / \dso $)
        \item[] 
          $f$ = fluctuation frequency (cycles\, yr$^{-1}$)
        \item[]
	         $\SM$ = scattering measure $= \int_0^{\dso} ds\, \cnsq(s) = 10^{-3.5}\,\SM_{-3.5} \,\SMunits $ , 
             $\SMeff = \int_0^{\dso} ds\, \cnsq(s)(1-s/d)^{\beta-2}$
        \item[]  
              $a_\beta = \displaystyle 
                   \left[ \frac{{\pi}^{1/2}}{(2\pi)^{\beta-3}} \frac{\Gamma((\beta-1)/2)}{\Gamma(\beta/2)}\right]
                   \overset{\beta=11/3}{\to} 0.494$, \quad
        	     $f_\beta =  \frac {\displaystyle 8\pi^2 \Gamma\left(2-\beta/2\right)} 
	                               {\displaystyle (\beta-2)2^{\beta-2} \Gamma(\beta/2)}
	      \overset{\beta=11/3}{\to} 88.3$
        \item[]
             $\louter$ = outer scale in pc, 
	     $v_{100}$  = effective transverse velocity  $= \vperp / (100\, {\rm km \, s^{-1}})$
	\item[]
	     $\dnud$ and $\dtd$ = scintillation bandwidth and  scintillation time for DISS, 
	     $\dtr$ = scintillation time for RISS
        \item[]
            $C_1 \simeq 1$ to 2 (dependent on $\beta$ and line of sight distribution of electron density fluctuations)
        \item[]
           $\etat \sim \etanu = 0.3$ (filling factors); $B$ = bandwidth in GHz; $T$ = observation time in 1000 s.
	\end{itemize}
	}
\end{deluxetable}

{\it DM variations:}  
\newtext{The amplitude of the phase structure function can be related to the variance of  spatial variations of pulsar DMs.} 
The asymptotic \newtext{$b \to \infty$ ensemble average phase structure function} implies a total \RMS\ dispersion measure
$\sigma_\DM = (\lambda \relec)^{-1} \left[\Dphi(\infty)\right]^{1/2}$ that is also dominated by the largest scales in the spectrum for $\beta < 4$.    Using $\qouter = 2\pi / \louter$ \newtext{in Eq.~\eqref{eq:Dphi_beta}}, we obtain the first entry in 
Table\,\ref{tab:keyplasmaeq}. 

\providecommand{\vobsperpscalar}{{v_{\rm obs_\perp}}}
\providecommand{\vpperpscalar}{{v_{p_\perp}}}
\providecommand{\vpperpvec}{{\vvec_{\rm p_\perp}}}
\providecommand{\vobsperpvec}{{\vvec_{\rn obs_\perp}}}

{\it DM structure function:}  The DM SF is closely related to  the phase SF,
$\DDM(\dxvec) = (\lambda\relec)^{-2} \Dphi(\dxvec)$, and it is useful for ISM diagnostics 
(cf.\ Sect.~\ref{sec:ismdiagnostics}) to express the spatial offset as a time offset multiplied by an effective velocity. An expression that applies to thin screens and extended media alike \newtext{is obtained by letting 
$b = \veffperp \delta t$,}
\be
\DDM(\delta t) 
	= 
	f_\beta \,
	\int_0^{\dso} ds\,  \cnsq(s) \left\vert \veffperp(s) \delta t \right\vert^{\beta-2},
\label{eq:DDM_pl}
\ee
where $\veffperpvec = (s / \dso) \vpperpvec  + (1 - s / \dso) \vobsperpvec$ 
and the effective scattering measure $\SMeff \le \SM$ is the weighted integral,
\be
\SMeff = \int_0^d d\sbar\, \cnsq(\sbar) \left( 1 - \sbar/d\right)^{\beta-2}.
\label{eq:SMeff}
\ee
For a thin screen at distance $\dsl$ from the source,   $\SMeff = \SM (1-\dsl/\dso)^{\beta-2}$.
If the pulsar velocity dominates the time dependence, the expression simplifies to the form given 
in the second line of the table.   Additional contributions to $\DM(t)$ from discrete ISM structures and from the solar wind augment the SF, as discussed in Sect.~\ref{sec:ismdiagnostics}. 

{\it DM power spectrum:}  The power spectrum is the Fourier transform of the  ACF, $\RDM(\tau)  = \RDM(0) - \DDM(\tau) / 2$.   The result is given in the fourth line of Table\,\ref{tab:keyplasmaeq}. 

{\it RM fluctuations:} Though RM variations are not of great concern for precision timing of low-DM and low-RM pulsars, for completeness, we point out that for a constant magnetic field along the \LOS,  
$\RM =  (e^3 /  2\pi \me^2 c^4) \Bpar \DM = 0.81 \Bpar \DM$
(cf.\ Eq.~\eqref{eq:DMRMEM}), where the numerical value holds for standard units of $\RMunits$ for RM, 
$\DMunits$ for DM, and fields in $\mu$G.     DM structure functions etc. are then easily transformable to RM
structure functions.   However, we note that \citet[][]{ms96b} have shown that electron density fluctuations are accompanied by magnetic field fluctuations, also consistent with a Kolmogorov spectrum, which increase the RM SF over that from density fluctuations alone. 

{\it Fresnel scale ($\rF$):}   The rough division between diffraction and refraction is defined by the Fresnel 
scale.   For a thin screen at $\dsl$ from the source,  we define the Fresnel scale as
$\rF = 2\pi/ \qF= \sqrt{\lambda \deff / 2\pi }\sim 0.01~\au \sim 10^6$~km, where the effective distance
is  $\deff \equiv \dsl \dlo / \dso$ with $\dlo = \dso-\dsl$.  \newtext{Alternative definitions for $\rF$ are found in the literature with $2\pi$ replaced by $\pi$ or unity.  In our case, the Fresnel scale corresponds to a phase difference  between a deflected and a direct ray path of $1/2$ radian. } 

{\it Diffraction scale ($\ld$) and visibility function ($\Gamma(\dxvec)$)}:  
Diffraction is caused by scales $\ld$ smaller than the Fresnel scale. 
  Most pulsar measurements correspond to very small
diffraction scales $\ld = 2\pi/\qd \sim 10^2$\,to $10^4$~km. These appear to straddle the inner scale $\linner$, $\sim 100$~to $10^3$~km  \citep[e.g.][]{2009MNRAS.395.1391R}.

The diffraction scale is related to the size of the scattered image, the pulse broadening time $\taud$,  and the DISS bandwidth $\dnud$.  It is convenient to use two definitions, 
one for which the visibility function = $1/e$ given by 
$\Dphi({\ld}_{_\theta}) = 2~$rad$^2$, and another from     
$\Dphi({\ld}_{_I}) = 1~$rad$^2$, which yields the $1/e$ scale of the intensity rather than the field, as in item 6 of 
Table\,\ref{tab:keyplasmaeq}.

{\it Scattering diameter ($\thetad$):}  
A scattered image represents the distribution of angles into which incident radiation from a point source is scattered.
Scattered images  are generally non-Gaussian but a convenient scattering diameter   is obtained by  matching a Gaussian function to the non-Gaussian image at the half-amplitude value. 
This yields a scattering diameter (FWHM)
$\thetad = 2\sqrt{\ln 2} \lambda / \pi \ld$.   An observer sees a scattering diameter    
$\thetaobs = (\dsl\dlo/\dso) \thetad \equiv \deff \thetad \lesssim \thetad$ that is larger for screens close to the observer.  
The scattering disk is the patch size of the cone of rays  contributing to the scattered radiation  measured  by an observer
(Fig.\,\ref{fig:scattering_cone}).  
The patch or cone size is   $\lcone = \dlo\thetaobs$
for an observed scattering disk size $\thetaobs$ from a screen at distance $\dlo$ from the observer.   

{\it Refraction scale ($\lr$):}   
Refractive scintillations result from scales that focus or defocus incident waves according to geometrical optics.   For a Kolmogorov-like spectrum, refraction is dominated by scales
$\lr \sim \lcone$.    A convenient relation between the refraction and diffraction scales is
$\ld \lr \equiv \rF^2$.
DM variations on time scales up to $\sim 50$\,yr correspond to lengths up to\,$\sim 10^3 \lr \sim 10^3$\,\au. 

The cone of scattered ray paths has a projected diameter $\lcone = \dlo \thetaobs$ at the screen-observer distance $\dlo$.  Motion of the source, screen, or observer causes the scattering cone to move along the screen, acting as a smoothing function of the screen phase.  This manifests as a variation in intensity due to refractive focusing and defocusing, which defines a refraction scale  For  screens with  `shallow' ($\beta < 4$) wavenumber spectra,  intensity variations from refractive focusing and defocusing are dominated by the scale of this smoothing function.   The refraction scale is accordingly  defined as  $\lr \equiv \lcone$.

 {\it Pulse broadening  function $\PBF(t)$and its $1/e$ time scale $\taud$:}  An emitted impulse arrives differentially along multiple scattered ray paths.  For a screen, \TOAs\  are directly related to the square of \AOAs.  The distribution of arrival times is  the pulse broadening function (PBF), usually characterized with a $1/e$ width.    The PBF and $\taud$ typically are measures of only geometrical path length delays and exclude any differential dispersion between individual ray paths.   This and other complications, such as the relation between the mean scattering time and analysis methods are discussed in other sections. 
 
 {\it Diffractive interstellar scintillation (DISS) bandwidth $\dnud$}:  
 Diffraction  yields a spatial  intensity pattern from constructive and destructive interference with a characteristic scale $\ld$ in the observation plane (transverse to the line of sight).   The diffraction 
 pattern also has a characteristic diffraction bandwidth $\dnud$  (also known as the scintillation or decorrelation bandwidth).   Like the pulse broadening time,  it is determined by path length differences and the two are related by an `uncertainty' relation \citep[][]{sut71},
 \be   
 2\pi\dnud \taud = C_1,
\label{eq:uncertainty1}
 \ee
 where $C_1 \simeq 1$~to 2 is a constant dependent on the the index $\beta$ of the wavenumber spectrum.  This relation applies to all cases (not just thin screens) with $C_1$ also depending   on the thickness and location of the scattering region \citep[][]{cr98, lr99}.  At  frequencies $\nu \lesssim 1$\,GHz,  the DISS bandwidth is small, $\dnud \ll \nu$. 
 
 {\it DISS  time scale $\dtd$}:
 Sweep of the diffraction pattern across the \LOS\ by an effective velocity $\veff$ yields the scintillation time scale (or decorrelation time) $\dtd = \ld / \veff$.    In most cases, the effective velocity is dominated by either the pulsar or the observer's motion; bulk or turbulent motions of the screen appear to be small enough so that the diffraction pattern may be considered frozen. 

 {\it Refractive interstellar scintillation (RISS) time scale $\dtr$ and bandwidth $\dnur$}:  Similarly, the refraction time scale is
 $\dtr = \lr / \veff$.   Because RISS does not involve constructive/destructive interference,  it is correlated over wide frequency ranges, $\dnur \sim \nu$. 

%

{\it Diffractive spectral broadening $\dnusb$:} \newtext{While not relevant to broadband pulsar signals,  we include spectral broadening for completeness and because
it is simply the reciprocal of the DISS time scale. 
It  may apply to situations of very strong scattering, such as low-frequency measurements of recombination lines, and may influence narrowband features found in FRBs.}   A narrowband signal is broadened by phase fluctuations that ensue from transport of the diffraction pattern across the \LOS, yielding changes in frequency $\delta\nu \sim (d\phi/dt) \delta t$.  One can also view this as the frequency broadening of a pure sine wave   modulated by DISS.    The broadening bandwidth $\dnusb$ is related to the scintillation time by another uncertainty relation \citep[][]{cl91},
$2\pi \dnusb \dtd \simeq 1.$

 Together, the DISS bandwidth and time scale determine the number of scintles contributing to an observation over a total bandwidth $B$ and time span $T$.  
 
\begin{figure}[h!] 
   \centering
   \includegraphics[width=0.475\linewidth]{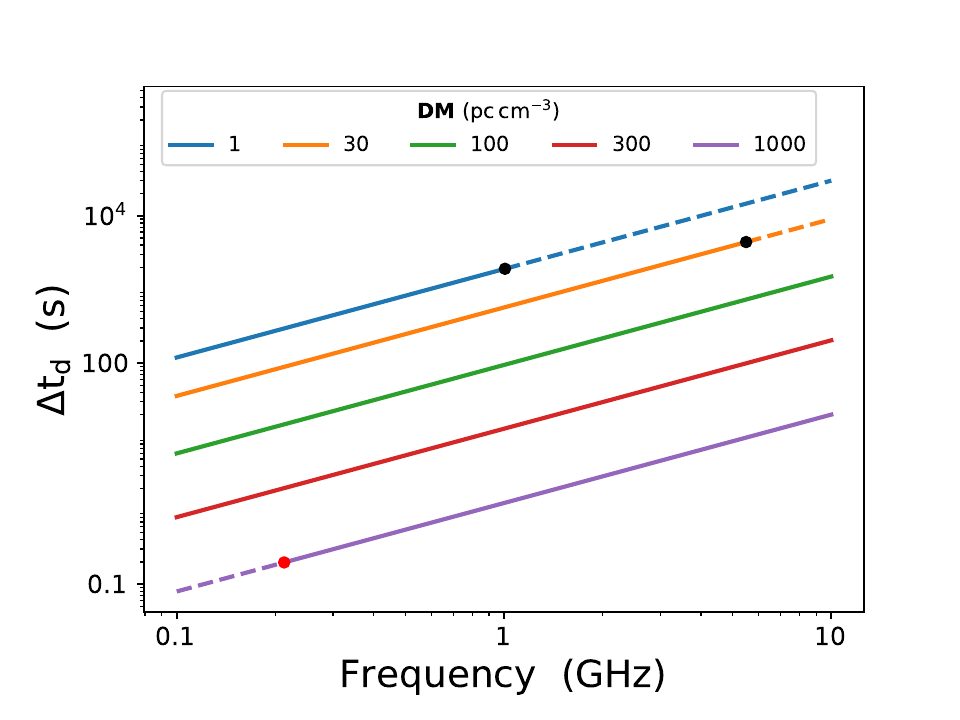}
   \includegraphics[width=0.475\linewidth]{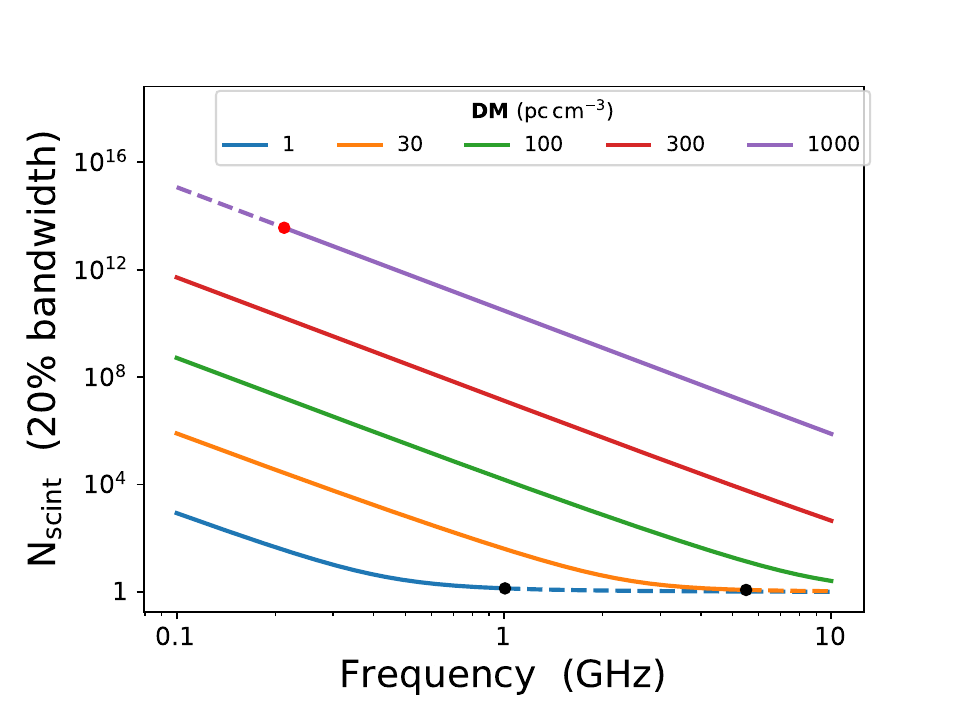}
   \caption{Computed values for the scintillation time (left) and number of scintles (right) for a thin screen with Kolmogorov fluctuations. The number of scintles is calculated for a 20\% bandwidth.  Solid lines indicate where the scintillations are  strong with $\dnud \le \nu$ and where the scintillation time scale exceeds the scattering time. The black dots show the transition to weak scattering with dashed lines indicating the scaling that otherwise would have occurred.  The red dot marks the transition where the scattering time equals the scintillation time scale.   The curves are calculated for a pulsar speed of 100\,km\,s$^{-1}$ and a screen midway to the pulsar,
   giving $\veff = 50$\,km\,s$^{-1}$.    A larger  effective speed shifts the scintillation time  downward  $\propto 1 / \veff$ and the number of scintiles   $\propto \veff$.  }
   \label{fig:scints_vs_nu}
\end{figure}

 Fig.~\ref{fig:scints_vs_nu} shows the scintillation time and number of scintles vs. frequency for five
 DM values.    These scale with frequency as $\dtd\propto\nu^{-6/5}$ and
 $\Nscint \propto \nu^{-28/5}$ for a Kolmogorov spectrum when the scintillations are strong and the inertial subrange of wavenumbers is applicable.   These quantities are calculated from 
 the scattering time $\tau_1$ evaluated using Eq.~\eqref{eq:taudm} and assuming a thin screen midway to the pulsar. They also assume an effective transverse speed of 50\,$\kms$ (corresponding to a pulsar velocity of 100\,$\kms$
 and a screen midway between pulsar and observer ($s=1/2$) and assuming negligible contribution from
 Earth's velocity or from the ISM). 

\begin{figure}[h!] 
   \centering
   \includegraphics[width=0.475\linewidth]{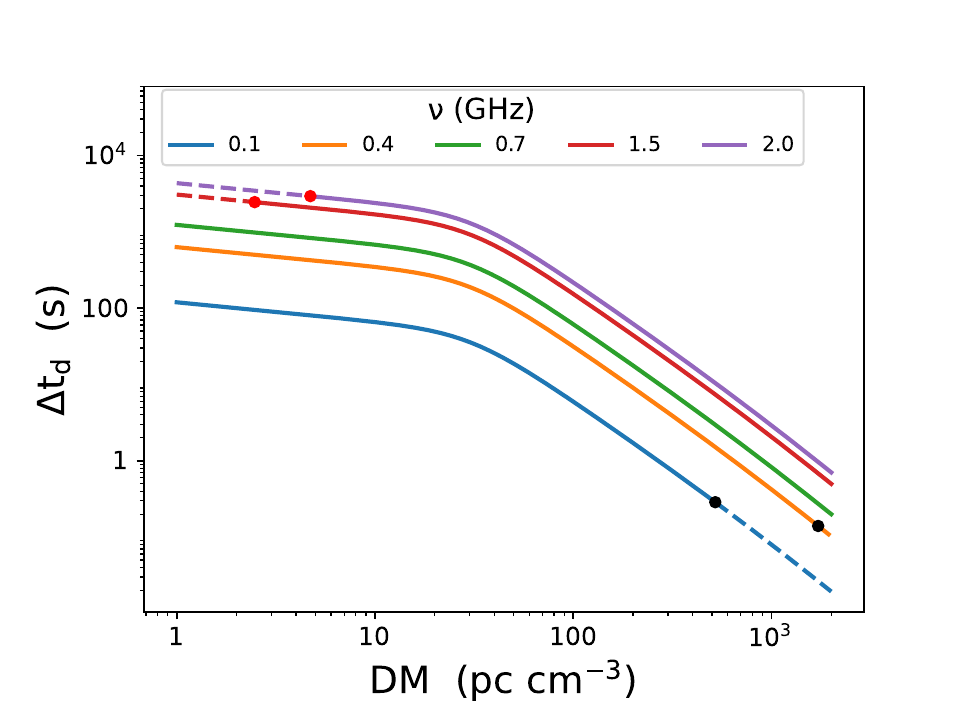}
   \includegraphics[width=0.475\linewidth]{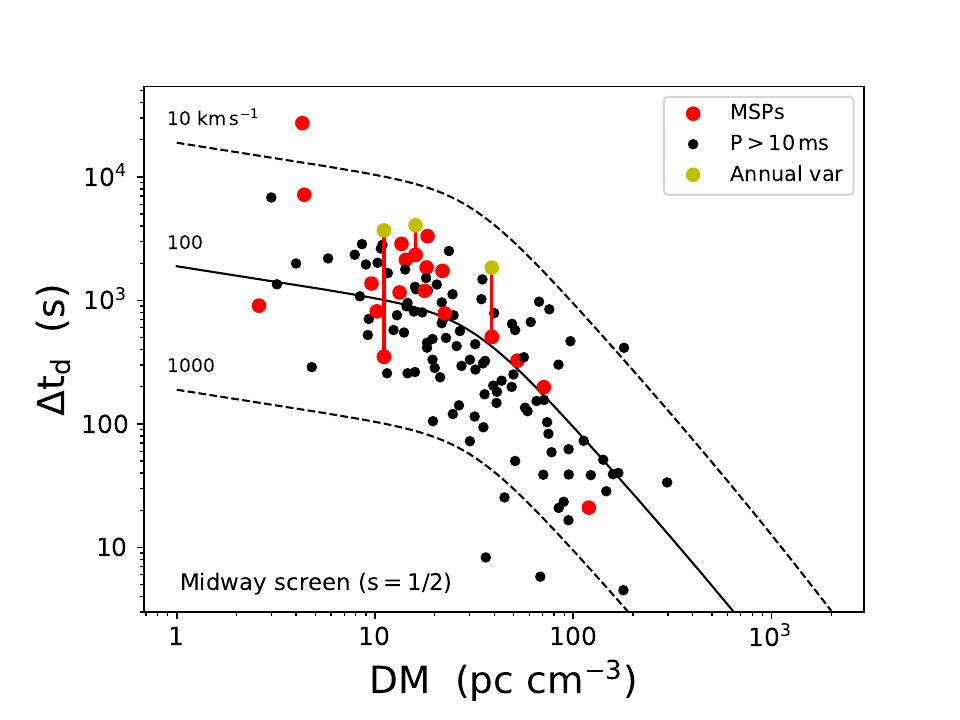}
   \caption{Left: Computed values for the scintillation time vs. DM for different frequencies and a pulsar velocity = 
   100\,km\,s$^{-1}$.  As with Fig.~\ref{fig:scints_vs_nu},
   calculations are  for a thin,  midway  screen ($s=1/2$) with Kolomogorov fluctuations.
   Right: 
   \newtext{Measured values of $\dtd$ are shown for millisecond pulsars (red) and pulsars with $P > 10$\,ms (black).
   The yellow points for three pulsars show maximum values for $\dtd$ that ensue from annual variations.
   The dashed lines show expected values for three different effective, transverse velocities, which bracket the wide velocity distribution of pulsars.} 
   Data are from \citet[][]{cor86,2018A&A...618A.186W,2021MNRAS.507L..41J,Liu_2022}.
    }
   \label{fig:scints_vs_dm} 
\end{figure}

Fig.~\ref{fig:scints_vs_dm} shows scintillation times vs. DM for different frequencies (left) and
different effective velocities (right) at  1\,GHz.   The right-hand panel also shows measured scintillation times
scaled to 1\,GHz for 136 pulsars.   The red points designate MSPs ($P \le 10$\,ms) and the yellow points indicate large scintillation times that occur yearly for three MSPs.      The velocity labels  are the assumed pulsar velocities 
for the three black lines, which apply only for a midway scattering screen.   

\subsection{Scintillation and scattering regimes}
\label{sec:ISSregimes}

\begin{figure}[b!] 
   \centering
   \includegraphics[width=0.6\linewidth]{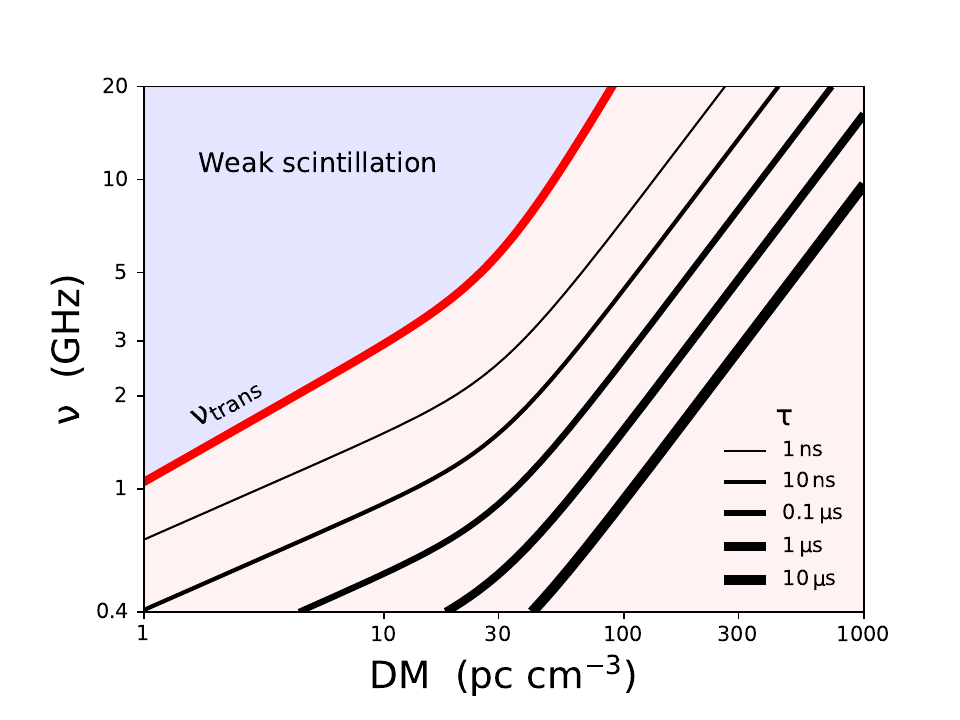}
   \caption{Weak and strong regimes for diffractive interstellar scintillation (DISS). The transition frequency
   $\nutrans(\DM)$ (red line) was calculated by approximating $\nutrans \sim \dnud$ and using the $\taud(\DM)$ `hockey stick'  relation in Eq.~\eqref{eq:taudm}  \newtext{(see Sect.~\ref{sec:pulse_broadening})} with the uncertainty relation Eq.~\eqref{eq:uncertainty1}  to calculate
   $\dnud(\DM)$. The $\nutrans$ line is fuzzy given that there is empirical spread around the nominal $\tau(\DM)$ relation.    Lines of constant scattering time $\taud = (2\pi\dnud)^{-1}$ are also based on this relation.    
   }
   \label{fig:ISSregimes}
\end{figure}

Except for observations at frequencies above $\sim 5$~GHz for the nearest pulsars,   the phase variance 
across one Fresnel scale in the screen  is many square radians.    In this so-called {\it strong scintillation} regime, the various length scales defined above satisfy the inequalities when the diffraction scale is larger than the inner scale ($\ld > \linner$), 
\be
\linner\,  \ll  \, \ld \, \ll \, \rF\, \ll \,  \lr \,  \lll \louter , \quad \text{Strong scintillation}.
\ee
In the  {\it hyperstrong scintillation regime}, the order of $\linner$ and $\ld$ is interchanged when the product $\lambda^2 \SM$ is large:
\be
\ld \, \ll \, \linner \, \ll \, \rF\, \ll \,  \lr \,  \lll \louter ,  \quad \text{Hyperstrong scintillation}.
\ee
If the \RMS\ phase variation on the Fresnel scale is small, i.e. $\Dphi(\rF) \ll 1$~rad$^2$, the diffraction and refraction scales merge and the only characteristic scale manifested in scintillations is the Fresnel scale:
\be
\linner \ll \ld \sim \lr \sim  \rF\,  \lll \louter , \quad \text{Weak  scintillation}.
\ee

The transition from weak to strong scintillation  occurs at a frequency that is \LOS\ dependent and can  be calculated in two ways.  First by finding the frequency where  the phase structure function
$\Dphi(\rF) = 1$~rad$^2$ and second by requiring $\nu = \dnud$.  Either approach gives
$\nutrans \simeq \nu [2\pi\taud(\nu) \nu]^{(\beta-2 / (\beta+2)} = \nu [2\pi\taud(\nu) \nu]^{5/17}$ for a scattering time
$\tau(\nu)$ measured at frequency $\nu$.   Using   Eq.~\eqref{eq:Dphi_beta} and solving for $\nu$ yields the expression for $\nutrans$ in Table\,\ref{tab:keyplasmaeq} in terms of the scattering time $\taud$. 

The boundary between weak and strong scintillation is shown in Fig.~\ref{fig:ISSregimes} for DMs from 1 to 1000\,$\DMunits$, \newtext{based on the empirical $\tau(\rm DM)$ (``hockey-stick'') relation (see Sect.~\ref{sec:pulse_broadening})}.  Strong scintillation corresponds to diffractive scintles being narrower than the radio frequency, $\dnud \ll \nu$.  This becomes $\taud \gg  C_1 / 2\pi\nu > 0.16\,{\rm ns}\,C_1 \nu^{-1}$ ($\nu$ in GHz, as usual) from the uncertainty relation in Eq.~\eqref{eq:uncertainty1}. 

Most pulsar observations  in the strong-scintillation regime 
are below the $\nutrans$ line
where  DISS  is 100\% modulated (if there is no bandwidth smoothing).   In particular, most MSPs observed in PTA programs are in the strong scintillation regime apart from a few low-\DM\ objects observed near the transition line, including J0437$-$4715 (DM $=2.6\,\DMunits$).   Near the transition region $2\pi\taud\nu \sim 1$ there is no bandwidth attenuation of the  $\sim100$\% intensity fluctuations, so arrival time precision will vary markedly from the highly variable S/N.    Conversely, 
large scattering times, e.g. those with $\taud \gtrsim 0.1\,\mu$s, correspond to intensities that are far less modulated  by a factor $\propto  (2\pi\taud B)^{-1}$ for  bandwidths $B\gg 1$\,MHz like those used in current receivers.  

\subsection{Example calculations using Table\,\ref{tab:keyplasmaeq}}

We illustrate usage of Table\,\ref{tab:keyplasmaeq} by considering observations at 1.5\,GHz of a pulsar with dispersion measure $\DM = 10\,\DMunits$, corresponding to a distance\footnote{\newtext{We could also specify Galactic coordinates to obtain a distance using the NE2025 model but it is adequate for our example to use an approximate relation  given by 
$\dso = (\DM/{\nelec}_0) / (1 + \DM/1000) / 1000$ with DM in standard units($\DMunits$) and an electron density ${\nelec}_0 = 0.03$\,cm$^{-3}$.}}$\dso \simeq 0.33$\,kpc.
Using the hockey-stick relation (Eq.~\eqref{eq:taudm}), we obtain a scattering time $\tau_1 = 6.2$\,ns at 1\,GHz that scales to 1.0\,ns at 1.5 GHz. This implies that the strong scintillation regime applies up to $\nutrans \simeq 2.9$\,GHz. 

Adopting a thin screen halfway between the pulsar and Earth, the effective distance,  $\deff = \dso/4 = 0.083$\,kpc, yields
a Fresnel scale $\rF \simeq 1.9\times 10^{-3}$\AU. Electron density fluctuations on scales smaller than $\rF$ yield multipath scattering of the pulsar signal while larger scales cause refraction.   For the assumed Kolmogorov spectrum,  refractive bending of ray paths by angles 
$\thetar \simeq 0.052$\,mas 
is typically smaller than the multipath caused by diffraction, which yields a scattering disk of size $\thetad\simeq 0.26$\,mas. 

Intensity scintillations from refraction and diffraction have estimated characteristic time scales, 
$\dtr \simeq 0.11$\,d and $\dtd \simeq 846$\,s, where 
we have adopted an effective transverse velocity $\veff = 100\,\kms$ (see Sect.~\ref{sec:DM_analytics} for discussion).
The characteristic scintillation bandwidth is
$\dnud \simeq 152$\,MHz which, combined with $\dtr$ and assuming a
0.3\,GHz total bandwidth and time span $T\simeq 1800$\,s for the observations, implies that only a few scintles,
$\Nscint \simeq 3$, contribute to the measurements. 

For a pulsar with $\DM = 100\,\DMunits$ corresponding to a more distant pulsar $\dso \simeq 3.0$\,kpc and $\deff \simeq 0.76$\,kpc, 
the larger scattering time $\tau\simeq 1.16\,\mu s$ is given by 
the scattering disk, $\thetad \simeq 2.8$\,mas which 
refractively wanders by $\thetar \simeq 0.18$\,mas. The refractive scintillation time is now larger,
$\dtr \simeq 11$\,d while the diffraction time scale is smaller,
$\dtd \simeq 77$\,s.  With a scintillation bandwidth 
$\dnud \simeq 0.14$\,MHz, the number of scintles is much larger,
$\Nscint \simeq 5000.$

\newcommand{\fnu}{f_{\nu}}
\newcommand{\ft}{f_{t}}
\newcommand{\fts}{f_{t_s}}
\newcommand{\Itilde}{\widetilde{I}}
\newcommand{\DS}{I}
\renewcommand{\EfieldFT}{\widetilde\varepsilon}

\section{Interstellar  analytics}
\label{sec:ismdiagnostics}

Interstellar propagation effects are significant limiting factors in  timing precision at radio wavelengths with current practices.  To understand these limits and develop  methods for improving arrival times, it is important to characterize and understand the ISM as much as possible.
The phenomena and scaling laws summarized in Sect.~\ref{sec:ism_scalings} and Table\,\ref{tab:keyplasmaeq}
are  diagnosed and tested using various tools applied to observational data for pulsars and other objects.   Our emphasis is on pulse broadening and intensity scintillations, which are more directly intertwined with estimates of timing precision.  
 
\subsection{Dispersion measures}
\label{sec:DM_analytics}

A primary tool  for characterizing DM fluctuations is the DM structure function defined earlier and evaluated
in Eq.~\eqref{eq:DDM_pl} for the power-law electron density spectrum
(Eq.~\eqref{eq:Pne}).  Here we include   further contributions to $\DM(t)$ that are manifested in some pulsar timing data sets.

\DM\ varies because motions  of the Earth, ISM, and  pulsar change the Earth-pulsar distance and direction of the   \LOS.   
For velocities $v \sim 100~\kms$, distances $\sim 1$~kpc, and time scales $t \lesssim$~decades, the fractional change in distance $vt/d \lesssim 10^{-6}$ and it is dominated by velocity components  parallel to the line of sight whereas  the  direction of propagation is due to transverse components.  
Using $\vepar$ and $\vppar$ for the parallel velocities of the Earth and pulsar \newtext{and using an approach similar to that in \citet[][]{2016ApJ...821...66L}}, the time dependence of  \DM\ is 
\be
\DM(t) = \DM(0) 
	+ 
	\left\{
		\vepar \nebar[\xvece(t)] - \vppar \nebar[\xpsrvec(t)] 
	\right\} t
	+
	\int_0^{{\dso_0}} ds\, \Delta \nelec(s, t),
\label{eq:DMt}
\ee
where  the second term is linear in time and involves electron densities 
$\nebar[\xvece(t)]$ and $\nebar[\xpsrvec(t)]$ that
are averaged at the observer's and pulsar's locations over \LOS\ segments of
length $\vepar t$ and $\vppar t$, respectively. 
The integrand in the stochastic  third term, 
\be
\Delta \nelec(s, t) 
	= \nelec[\xvecpzero + \nhat_0 s + \veffperpvec(s) t] 
	- \nelec[\xvecpzero + \nhat_0 s] ,
\label{eq:dne}
\ee
involves the transverse components of the effective velocity,
\be
\veffperpvec(s) =  \veffvec(s) ({\bf I} - \nhat_0 \nhat_0) = (1 - s / \dso) \vpperp  + (s / \dso) \veperp,
\ee
 using the $2\times2$ unit matrix  $ {\bf I}$. 
%
Eq.~\eqref{eq:dne} is general but we now consider only  interstellar contributions.  
The DM structure function then combines the contribution from the power-law density variations
presented in Eq.~\eqref{eq:DDM_pl}   with terms involving the electron density in local interstellar environments of  the pulsar and the solar system.  Breaking these into `linear' and `power-law' terms, 
\be
\DDM(\delta t) 
	&=& \DDMlin(\delta t) + \DDMpl(\delta t) 
	\nonumber \\
	&\simeq&
	\left[
		\vepar \nebar(\xvece) - \vppar \nebar(\xpsrvec) 
	\right]^2 (\dt)^2 
	+ 
	f_\beta \,
	\int_0^{\dso_0} ds\,  \cnsq(s) \left[\veffperp(s) \dt\right]^{\beta-2} .
\label{eq:DDM}
\ee
If the pulsar velocity dominates the time dependence, then 
\be
\DDM(\dt)  
	\simeq	
		 \nebar^2(\xpsrvec) 
	\, 
	(\vppar \dt)^2
	+ 
	f_\beta \,
	\SMeff \,
	(\vpperpscalar\dt)^{\beta-2}.
\label{eq:DDMpsr}
\ee
For nominal values of electron density, scattering measure,  and transverse pulsar velocity,  the
DM structure function terms are 
\be
\DDMlin(\dt) 
	&\simeq& 
		\left[ 
		3.1\times 10^{-6}~\DMunits \times 
		\left(\frac{\nebar}{0.03~\neunits} \right)
		\left(\frac{\vppar}{100~\kms} \right)
	\right]^2
	\times  (\dt_{\rm yr})^2 ,
	\nonumber \\
\DDMpl(\dt) 
	&\simeq& 	\left[ 
		7.8\times 10^{-4}~\DMunits \times 
		\left(\frac{\SMeff}{10^{-3.5}~\SMunits} \right)^{1/2}
		\left(\frac{\vpperp}{100~\kms} \right)^{5/6}
	\right]^2  
	\times (\dt_{\rm yr})^{5/3}.
\label{eq:DDMnom}
\ee
By definition the DISS time scale $\dtd = \ld / \veffperp$ corresponds to a  phase structure function 
value $\Dphi(\ld) = 1$\,rad$^2$.  Using $\DDM = (\lambda \relec)^{-2} \Dphi$, this yields  an empirical scaling law for the power-law component (second term in Eq.~\eqref{eq:DDM}) with $\nu$ in GHz, 
\be
\DDMpl (\dt) = 
    \left(\frac{1}{\lambda \relec}\right)^2
    \left(\frac{\dt}{\dtd} \right)^{\beta-2}
= (3.84\times 10^{-8} \, \DMunits)^2 \times \nu^2
\left(\frac{\dt}{\dtd} \right)^{\beta-2}.
\label{eq:DDM_dtd}
\ee

Measurement errors contribute a constant floor to the structure function that is twice the  noise variance,  $2\sigma_{\DM, \,\rm noise}^2$ for all $\delta t > 0$.   

For some pulsars, published structure functions show features with both the quadratic  $ \tau^2$ term and
the shallower Kolmogorov-like term $\propto \tau^{\beta-2} \to \tau^{5/3}$
\citep[][]{ric90, cwd+90, ktr94,kcs+13, 2016ApJ...821...66L}.    The results are consistent with the overall picture that the ISM includes density structure on a wide range of scales extending from $\lesssim 10^3$~km to $\gtrsim v_{\rm p} T \sim 600~\au$ for 100~$\kms$ and data spans of $T \gtrsim 30$~yr. 

\begin{figure}[htbp] 
   \centering
   \includegraphics[width=0.6\linewidth]{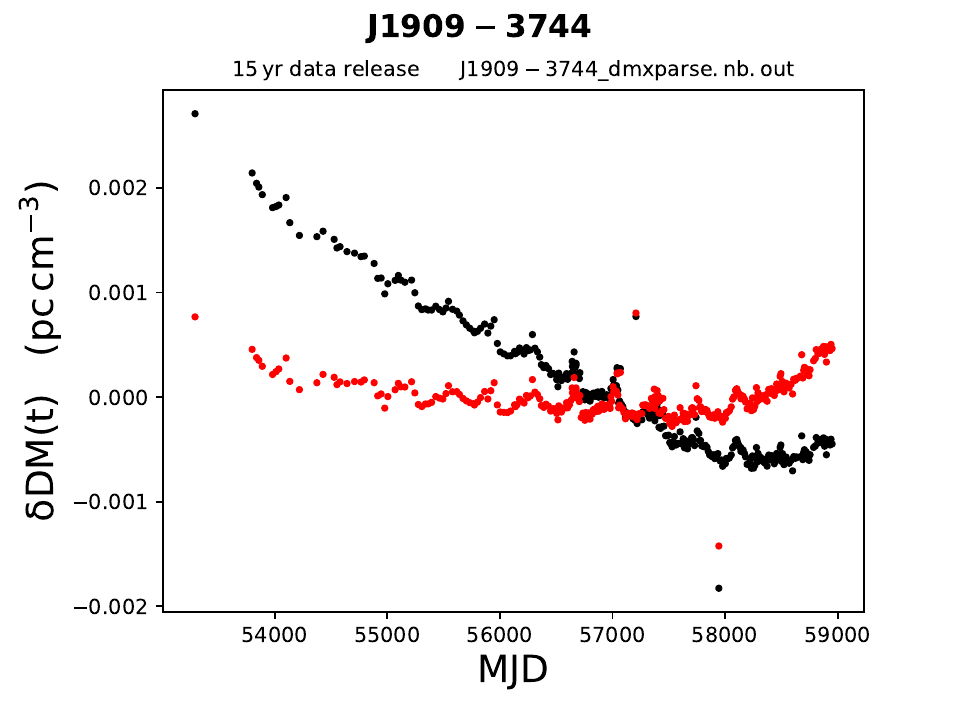}
   \caption{
    Time series of $\delta \DM(t)$ for J1909-3744 from the NANOGrav 15-yr data set \citep[][]{2023ApJ...951L...9A}.  Black points show departure of DM from a constant value; red points show departure from a linear fit to the black points. }
   \label{fig:DM_J1909-3744}
\end{figure}

The DM time series  for MSP J1909-3744 
(Fig.\,\ref{fig:DM_J1909-3744}) shows two additional features not seen for B1937+21: an annual variation and  a linear trend over most of the 15-yr data set.   The annual variation is due to the close approach of the \LOS\ to the solar corona and the linear downward trend that appears to bottom out and reverse suggests that motion of the pulsar through a local region of  variable  electron density is in action. 
In a study of 37\,MSPs in the NANOGrav timing program, 
\citet[][]{2017ApJ...841..125J} found that 19 objects show periodic, yearly variations consistent with contributions from the solar corona and 28 show linear-like trends. \newtext{Similar studies by \cite{you2007,kcs+13,2020A&A...644A.153D,2025ApJ...982..132K} (among others) suggest that many pulsars have DM variations well explained by turbulent ISM density fluctuations, with pulsars at low ecliptic latitudes exhibiting periodic DM variations from the solar wind \citep{2019ApJ...872..150M,2021A&A...647A..84T}, and a general correlation between the magnitude of the DM gradients and the total DM \citep[e.g.][]{2004MNRAS.353.1311H}.}

\begin{figure}[htbp] 
   \centering
   \includegraphics[width=0.6\linewidth]{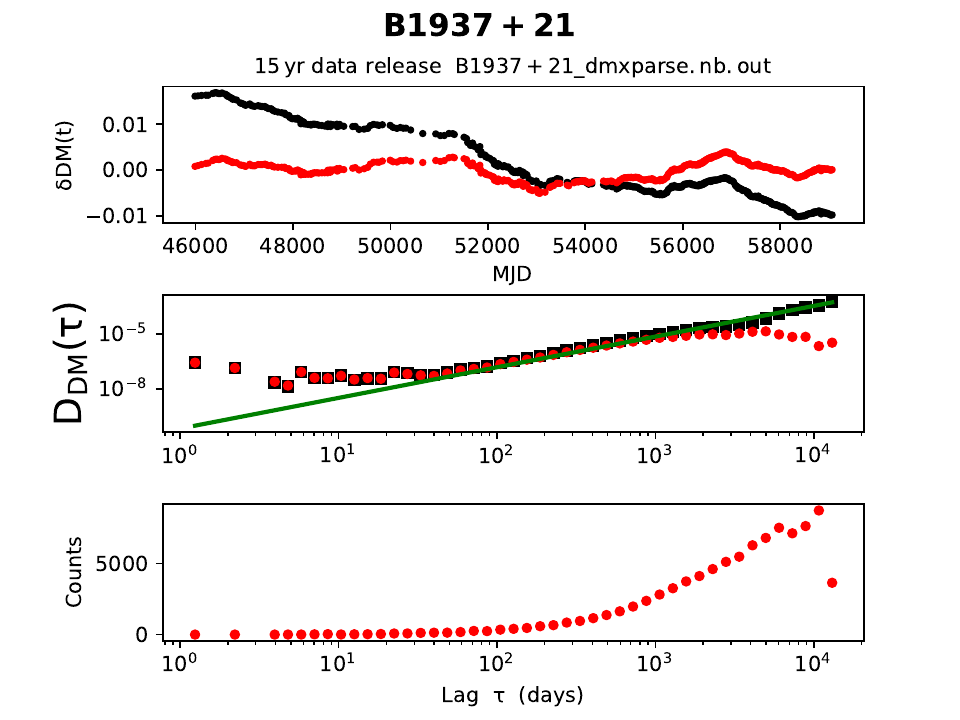}
   \caption{DM Time series and structure function for the MSP J1939+2134 (B1937+21) from the 15-yr NANOGrav data release \citep[][]{2023ApJ...951L...9A}.
   Top:  DM(t) before (black) and after (red) removing a straight line.
   Middle: Structure function before (black) and after (red) removing a straight line. The solid green line is the Kolmogorov inertial subrange scaling $\DDM \propto \tau^{5/3}$.
   Bottom:  The number of counts in each lag bin from the irregularly sampled DM. 
   }
   \label{fig:DMSF_3panels_B1937}
\end{figure}

\begin{figure}[htbp] 
   \centering
   \includegraphics[width=0.6\linewidth]{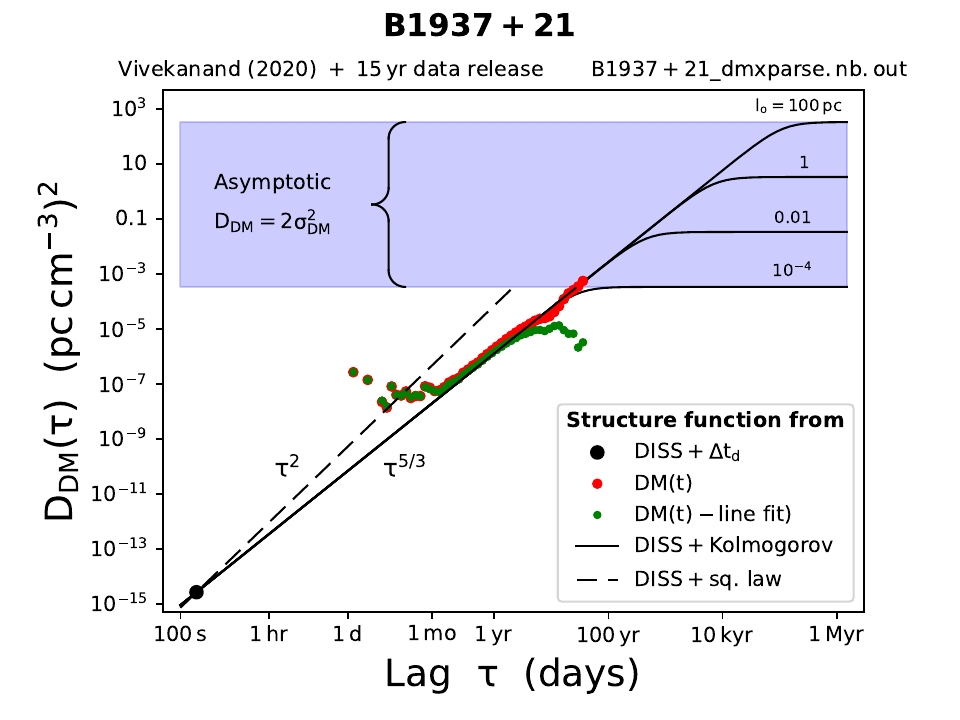} 
   \caption{Structure function of DM for the MSP J1939+2134 (B1937+21) for a 37 yr data set obtained by combining the 15-yr NANOGrav DM time series \citep[][]{2023ApJ...951L...9A} with the 31-yr 
   \newtextR{DM} time series from 
    \citet[][and references therein]{vivekanand2020}.   Where there was overlap,  the NANOGrav time series was used.  The DM SF is shown before (red points) and after (green points) removing a straight line fit to $\DM(t)$.
Points at lags $\lesssim 1$\,month are dominated by additive measurement errors in DM that produce a flat component to the SF. 
Black solid lines show the SF expected from a Kolmogorov medium with different outer scales $\louter$, as indicated.  These lines scale as $\dt^{5/3}$ for intermediate lags between $\sim 1/3$ to  20\,yr and saturate at lags 
$\gtrsim \louter  / \veff$.   The black circle at $\tau \sim 300$\,s indicates  the characteristic DISS timescale 
\newtext{\citep[][]{2024ApJ...972...16T}}
and the implied DM SF calculated from Eq.~\eqref{eq:DDM_dtd}.  The dashed line shows the square-law scaling that is
clearly inconsistent with the measured values. 
    }
   \label{fig:DMSF_asymptotic_B1937}
\end{figure}

Fig.\,\ref{fig:DMSF_3panels_B1937} shows a DM time series  for PSR~B1937$+$21 over a 15\,yr span 
in the NANOGrav data set \citep[][]{2023ApJ...951L...9A}. 
    It is possibly the superposition of a downward secular trend and a stochastic component with a red power spectrum, but, unlike the case of J1909-3744 discussed below, 
it is actually consistent with fluctuations expected from only a power-law process.   The DM structure function shown in the middle panel has a slope consistent with the $\tau^{5/3}$ scaling expected for a Kolmogorov process. 
This result is corroborated by the SF calculated from a longer time series obtained by combining 
earlier measurements of DM$(t)$ \citet[][and references therein]{vivekanand2020} with the NANOGrav
15-yr data.   Fig.\,\ref{fig:DMSF_asymptotic_B1937} shows the resulting SF along with curves indicating 
the $5/3$ scaling law for Kolmogorov fluctuations and a curve for a square-law scaling. 
The noise floor dominates the SF at lags $\lesssim 30$\,days and for larger lags the SF 
appears consistent with the Kolmogorov scaling.   
  The expected scaling using the measured scintillation time $\dtd\simeq 3$ to 5 min at 1.5\,GHz is consistent with the asymptotic part of the SF.  This implies that length scales in the ISM along this \LOS\ span a range of
  at  least $20\,{\rm yr} / {4\, \rm min} = 6.4$ orders of magnitude ($\sim 10^4$\,km to 200\,au using an effective velocity of 50\,$\kms$).

\subsection{Angular deviations and broadening}
\label{sec:angular}

Radio interferometric imaging has provided scattering diameter $\thetad$ measurements useful for establishing the index $\beta$, the inner scale $\linner$, and anisotropies of the wavenumber spectrum \citep[][]{1988ApJ...332..346S,  1990ApJ...348..147M, 1995ApJ...438..708M,2004ApJ...613.1023L} and the outer scale \citep[e.g.][]{ms96b}. \newtext{These interferometric measurements} of the visibility function
in the $\Dphi \propto b^{\beta-2}$ regime have yielded constraints on
$\beta$ for (relatively) weakly scattered objects while heavily scattered objects that also sample the $\Dphi \propto b^2$ regime(cf.\ Eq.~\eqref{eq:Dphi_beta})  have yielded estimates of the inner scale \citep[e.g.][]{sg90}. \newtext{Image ellipticities also seen in the examples reported in these references and toward heavily scattered sources in the Galactic center \citep[e.g.][]{2025ApJ...988..168B} are a direct consequence of anisotropic scattering.} 

Referring to Table\,\ref{tab:keyplasmaeq}, for scattering angles $\simeq 1$\,mas and path lengths $\sim 1$\,kpc at 1\,GHz, the corresponding diffraction scale $\ld \simeq 10^4$\,km and refraction scale $\lr \simeq 1$\,\au, compared to inner scales estimated as $\sim 50$\,to 1000\,km.

\subsection{Scintillations: dynamic spectrum and secondary spectrum}
\label{sec:diss_analytics}

 \begin{figure}[b]
  \centering
  \includegraphics[width=\textwidth]{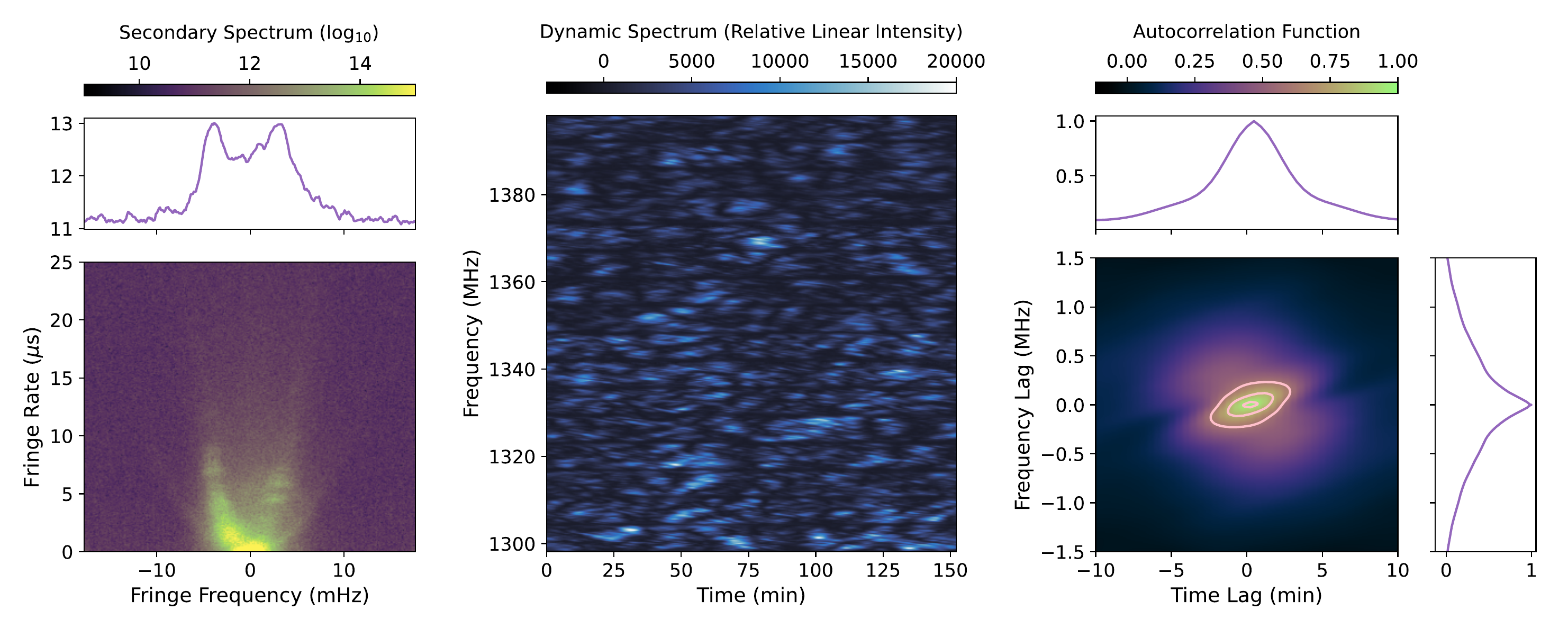}
  \caption{ Scintillation diagnostics for B1937$+$21 from Arecibo observations conducted in 2012 and processed using cyclic spectroscopy. The dynamic spectrum (relative linear intensity as a function of frequency and time) is shown in the middle panel at a time resolution of 28 s and a frequency resolution of 0.61 MHz. The left panel shows the secondary spectrum (2D FFT of the dynamic spectrum), including a 1D average over fringe rates $>10$ $\mu$s. The right panel shows the 2D autocorrelation function (ACF) with three evenly spaced contours between 0.2 and 1, along with 1D slices through the ACF at zero lag along the frequency and time axes. The ACF yields a decorrelation bandwidth and timescale of about 0.4 MHz and 3 minutes, respectively, at 1360 MHz. Data courtesy of Tim Dolch and Jacob Turner (see Turner et al. 2023).}
  \label{fig:DISS_B1937+21}
\end{figure}

Diffractive scintillations  are manifested in
the {\it dynamic spectrum} (DS),  the radio spectrum measured at multiple  times $t$,
\be
\DS(\nu, t) = \langle \vert\EfieldFT(\nu, \tbar)\vert^2\rangle_T,
\label{eq:DSdef}
\ee
where $\EfieldFT(\nu, \tbar)$ is the Fourier transform of the baseband voltage with bandwidth $B$ for a data block of duration $T = N\delta t$ centered on $\tbar$,  where $N$ is the FFT length and  $\delta t = B^{-1}$ is the sample interval\footnote{We use the Nyquist criterion for complex baseband data as opposed to the criterion  $\delta t = (2B)^{-1}$ for real data.}.   
For total intensity data (i.e. Stokes $I$), the DS would be the sum of the individual DS for each of two polarization channels. 
\newtextR{
The time average indicated in Eq.~\eqref{eq:DSdef} typically includes averaging over a narrow window in pulse phase as well as averaging over multiple pulses. 
}

With {\it phase resolved spectroscopy} (PRS), the DS is formed by averaging  (after dedispersion), \newtext{as in Eq.~\eqref{eq:DSdef},} over an on-pulse window (i.e. wide enough to span the duration of the pulse) and  over multiple pulses corresponding to elapsed times  $T$ of order seconds but shorter than the scintillation time $\dtd$. 
\newtextR{This approach has been commonly used to produce pulsar dynamic spectra and has the limitation that spectral resolution is determined by the reciprocal of the pulsar width.}
 A corresponding off-pulse average 
 $\langle \DS_{\rm off}(\nu)\rangle$  is used to remove the   receiver bandpass shape by subtracting and dividing it from the on-pulse spectrum. This approach maximizes the S/N of the scintillation signal before further analysis of the DS.
 An alternative approach for computing the DS, {\it cyclic spectroscopy},  is discussed below. 
 
The two-dimensional ACF of the DS is frequently used to determine scintillation parameters,  
\be
\Gamma_I(\delta \nu, \delta t)
        = \langle I(\nu, t) I(\nu+\delta\nu, t+\delta t) \rangle
        = \langle \vert \EfieldFT(\nu,t)\vert^2
                  \vert \EfieldFT(\nu+\delta\nu, t+\delta t) \vert^2 \rangle
        = \langle I \rangle^2 +
                \vert \Gamma_{\EfieldFT}( \delta \nu, \delta t) \vert^2,
\label{eq:GammaI}
\ee
where the last equality follows for  any stochastic process $\EfieldFT$ having complex Gaussian statistics, including the scintillation modulated  pulsar signal in the strong scintillation  regime \citep[e.g.][]{ric90}.   Such statistics are expected because scintillations typically result from the superposition of a large number of propagation paths (Central Limit Theorem).   \citet[][]{codona1986-twofrequency} have  pointed out that departures from Gaussian statistics can be caused by strong refraction 
from large scales in the medium
along a single or a small number of paths. 
The autocovariance is the ACF of the fluctuating intensity,
$\Gamma_{\delta I}(\delta\nu, \delta t) = \vert \Gamma_{\varepsilon}(\delta \nu, \delta t) \vert^2$.
The scintillation feature in the 2D ACF often is slanted in the $\delta\nu$ - $\delta t$ lag domain, an effect caused by refraction and characterized as a drift rate $d\nu/dt$. 

Assuming $\Gamma_{\delta I}(\delta\nu, \delta t)$ is normalized to unit maximum, the scintillation bandwidth 
$\dnud$ is the HWHM of the 1D slice $\Gamma_{\delta I}(\dnud, 0) = 1/2$ 
and  the scintillation time $\dtd$ is the HWe  of 
$\Gamma_{\delta I}(0, \dtd) = e^{-1}$.  Usage of these particular measures is motivated by a thin screen with a square-law phase structure function, which yields $\Gamma_{\delta I}(\dnud, 0) = [1 + \delta\nu/\dnud)^2]^{-1}$ for the slice along the $\delta\nu$ axis and  $\Gamma_{\delta I}(0, \dtd) = \exp(-\delta t / \dtd)$ along the time-lag axis. 
\newtext{Though a square-law SF is not a viable form for many instances of radio scattering in  the ISM, it is analytically tractable and  provides 
functional forms that are useful for quantifying measurements.    A square-law structure function could 
correpond to  a 
$\beta = 4$ wavenumber spectrum, which corresponds to a medium comprising step functions \citep[][]{lr00} that has mathematically divergent variance.  However, inspection of Eq.~\eqref{eq:Dphi_beta}  indicates the  SF is  square-law in form for spatial lags $\ll$ inner scale, a regime that dominates the measurements of  heavily scattered sources.} 

The Fourier transform of the 2D ACF (equivalent to the
squared magnitude of the 2D Fourier transform of the DS by the Wiener-Khinchin theorem), though containing the same information, has proven to be enormously useful for ISM studies.  Using $\FT_{2D}$ to denote 2D transform,  
this  {\it secondary spectrum} (SS)  is 
\be
S_2(\fnu, \fts) 
= 
\vert \FT_{2D}\{ \DS(\fnu, \ft) \} \vert^2
= 
\vert \FT_{2D}\{\vert\EfieldFT(\nu, t)\vert^2 \} \vert^2,
\label{eq:S2}
\ee
with variables $\fnu$ and $\ft$  that are conjugate to $\nu$ and $t$ with units of time and frequency, respectively.
The second equality in Eq.~\eqref{eq:S2} shows by inspection that the SS is  a fourth moment of
the wavefield $\varepsilon$.    
Often $f_\nu$ is referred to as a ``delay,'' a term we avoid because it may be misinterpreted as the delay of a pulse imposed by ISS rather than as (effectively) the lag of a correlation function.

The SS often displays remarkable {\it scintillation arcs} \citep[][]{smc+01}, parabolic features corresponding to discrete scattering screens along the \LOS\ of interest for ISM studies.   It is also used for assessing  delays from multipath propagation that contribute to \TOA\ stochasticity \citep[e.g.][]{hs08}.     The SS as defined excludes phase information contained in the Fourier transform $\FT_{2D}\{ \DS(\fnu, \ft) \}$; other   authors have exploited phase information  for ISM studies \citep{2010ApJ...708..232B,simard2019,baker2023,zhu2023}.

Fig.\,\ref{fig:DISS_B1937+21} shows examples of these tools for the MSP J1939+2134 (B1937+21), including the DS (center), the 2D ACF and 1D slices through it (right-hand panel), and the SS in the left panel.      The central feature of the 2D ACF is slightly slanted due to refraction from scales larger than those that scatter the radiation. 

\begin{figure}[htbp] 
   \centering
   \includegraphics[width=0.5\linewidth]{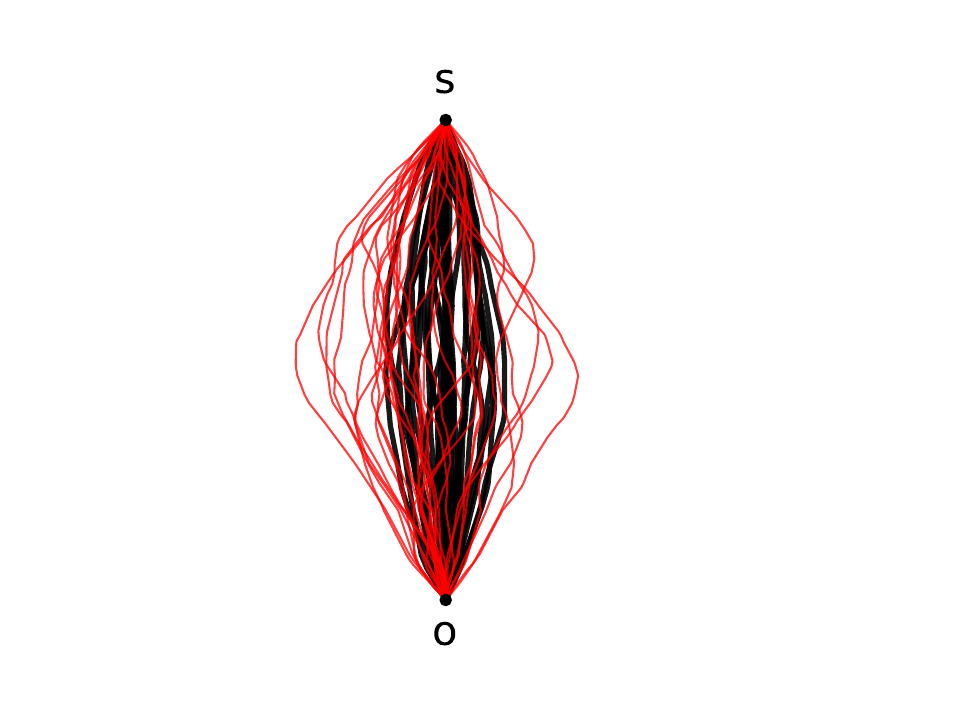}
   \hspace{-0.5in}
    \includegraphics[width=0.5\linewidth]{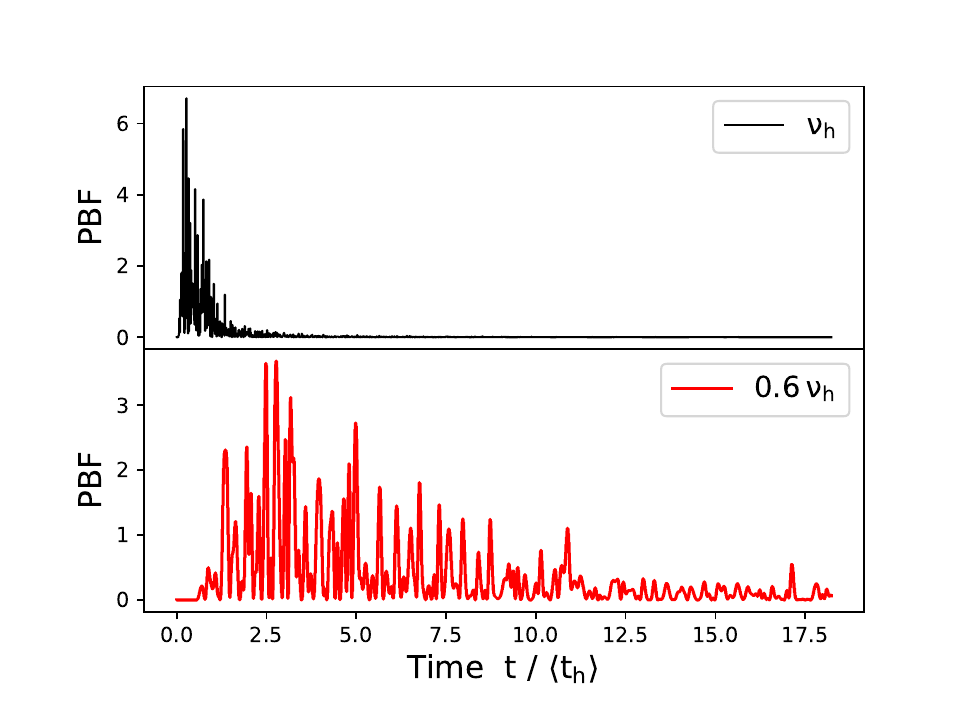}
   \caption{Left: Simulated ray paths for an extended medium with density variations conforming to a Kolmogorov wavenumber spectrum (similar to \citealt[][]{CSS16}), with
  high frequency ray paths in black and 
  low frequency ray paths in red.
    \label{fig:two_freq_raywalks}
   Right: Pulse broadening functions (PBFs) corresponding to the ray paths shown in the left panel.  \newtextR{The horizontal axis is time in units of the mean width of the PBF in the top panel, illustrating the much wider PBF at the lower frequency.} 
   \label{fig:two_freq_pbfs}
   }
\end{figure}

\subsection{Pulse broadening}\label{sec:pulse_broadening}

Scattered pulses from pulsars are the convolution of a pulse broadening function (PBF) with the ``intrinsic'' pulse shape emitted by the pulsar.   This follows from the linearity of propagation through the ISM.  Propagation effects include dispersion and angular deflections from refraction and diffraction.  In the ISM context, the PBF is usually  defined in terms of the geometrical path length differences between the multiple propagation paths in a ray bundle  that contribute   to the measured wave field in the observation plane.  It therefore does not include 
any differential dispersion across the bundle of ray paths because it is a small effect that  can be considered separately (see section on  frequency dependence of \DM).

The PBF quantifies the differences in  arrival times for pulses propagating along different ray paths.
For most pulsar observations (as well as those for fast radio bursts),   the \LOS\ geometry may be considered  invariant over minute to several hour time spans, depending on the path length through the ISM and on the observation frequency.  If scattering occurs primarily in  a thin screen,   the received angles of arrival are directly related to time delays; however distributed scattering that leads to random walk ray paths does not have this simple property.  

When scattering is strong,  the PBF comprises a large number of individual propagation paths, so the net wavefield becomes a complex Gaussian process modulated by an envelope function related to the size of the raypath bundle.  The intensity PBF, calculated as  the squared magnitude,  then comprises a stochastic process with exponential statistics ($\chi^2_2$) modulated by a function that is the ensemble average intensity PBF.   

An impulsive signal incident on a phase changing medium  yields an impulse response in the  measured wavefield
$\varepsilon(t, \tbar) = h(t, \tbar)$ which we call the field  PBF to distinguish it from the 
intensity PBF, $\PBF(t) \equiv \vert h(t) \vert^2$ (for which we drop the $\tbar$ dependence, making it implicit). 
PBFs are calculated most often from the geometrical path length differences along the set of propagation paths that instantaneously reach the observer, which is most easily done for thin screens and for random walks through extended media.   The PBFs can also be calculated using the Fresnel-Kirchhoff integral, which includes dispersion delays along with geometrical delays. 

The field PBF is stochastic,  varying rapidly in time $t$ (e.g. ns to s) but slowly in 
$\tbar$ (s to hr).   The field PBF is constant for times shorter than the DISS time scale,
$\tbar \ll \dtd$.  The  PBF example given earlier  for a thin screen in Fig.\,\ref{fig:pbfave} shows the rapid fluctuations of $\PBF(t)$ with corresponding frequency structure in the spectrum also shown (lower frame).   
An ensemble average (corresponding to averaging the PBF over many scintillation time scales) yields a smooth PBF with a shape determined by properties of the medium (e.g. $\beta$ and inner scale), while the spectrum averages to a constant value.  
To complement the thin screen case shown in  \newtext{Fig.~\ref{fig:pbfave}},  we show in Fig.~\ref{fig:two_freq_pbfs} ray traces (left panel) for a thick medium  and PBFs (right panel) for two frequencies to illustrate their strong frequency dependence along with their  temporal stochasticity.
envelopes that multiply a noise-process.

The  field and intensity PBF are related to  frequency-domain quantities  according to the flow diagram, 
\be
\begin{array}{ccccc}
\displaystyle
\boxed{\text{Field PBF} \ \ h(t)} 
    &&\quad \quad \displaystyle  \mathrm{FT \atop \Longleftrightarrow} \quad\quad 
    &&\boxed{\htilde(\nu)} \\
 \\
{\vert \cdots \vert^2} \ \ \big\downarrow\quad\quad  &&\quad \quad && \quad\quad \big\downarrow\ \  \mathrm{ACF} \\
\\
\boxed{\PBFtext  \ \ \PBF(t) = \vert h(t) \vert^2} &&\quad \displaystyle  \mathrm{FT \atop \Longleftrightarrow}\quad  &&
  \boxed{
   \int d\nu\, \Gamma_{\htilde}(\delta\nu, \nu)      
  }  \  \ ,
\end{array}
\label{eq:pbfflow}
\ee
where the frequency-domain correlation function is
\be
\Gamma_{\htilde}(\delta\nu, \nu) =\htilde(\nuplus)  \htilde^*(\numinus) ,
\quad\quad \nu_{\pm} \equiv \nu\pm\delta\nu/2 .
\label{eq:Gammah}
\ee   
The quantities in each row of Eq.~\eqref{eq:pbfflow} are related by a two-way Fourier transform while vertical operations are irreversible: the squared magnitude in the first column and an autocorrelation function in the second column.    
Recall these are all  bandlimited, baseband quantities  corresponding  to an RF center frequency $\nu_0$ and
an analysis bandwidth $B \ll \nu_0$ (i.e. a wideband system is usually processed in narrower subbands with bandwidths $B$).     

The scintillation spectrum is $H(\nu) \propto \vert \htilde(\nu) \vert^2 =  \Gamma_{\htilde}(0, \nu)$ is the primary quantity of interest for ISM studies and is the essential part of the DS in Eq.~\eqref{eq:DSdef}.  Characterization of
frequency structure in $H(\nu)$ is a key element in \TOA\ corrections for scattering mitigation \newtext{(discussed in Sect.~\ref{sec:mit_fse})}. 

Neither $h(t)$ nor $\PBF(t)$ or their Fourier transforms are known {\it a priori} and they are not directly measurable.  
The field PBF is  manifested through the convolution of $h(t)$ with the emitted pulsar signal and various degrees of time averaging; some of its properties are accessible through  model fitting of pulse shapes  and through frequency-domain analyses (next section). 

The field PBF  has the form $h(t) = \pbfamp(t) \pbfnoise(t)$, similar to the  amplitude modulated noise form for the emitted signal in Eq.~\eqref{eq:amn}. 
The PBF factors $\pbfamp$ and $\pbfnoise$ differ fundamentally from those for emission.   The amplitude $\pbfamp$ is stable over long periods of time ($>$ days),  varying on a refraction time scale,  but is strongly frequency dependent. By contrast,  amplitude variation for emission varies on pulse-period time scales ($\lesssim$ seconds) but varies slowly with frequency.     The PBF noise $\pbfnoise$ is stable over a DISS time scale (seconds to hours) and is correlated over a DISS bandwidth.  By contrast emission noise decorrelates over emission  time scales equal to the inverse radio frequency,  $\nu^{-1} \sim$ ns. 


    The PBF noise process and  the frequency structure of intensity scintillations are two sides of the same coin. Convolution of the PBF with emitted pulses yields small TOA variations for low-DM pulsars at $\sim$GHz frequencies.  Generally, the instantaneous PBF is inaccessible to measurement (with rare exceptions  
    \newtext{involving cyclic spectroscopy discussed 
    in Sect.~\eqref{eq:scattcorr})
    }
    so it is not possible to deconvolve the PBF as a means for reducing \TOA\ errors.  However some degree of correction is feasible by characterizing scintillation frequency structure  and how it changes with time, as discussed in detail in Sect.~\ref{sec:scatt_mit}.

{\it Scattered pulse shape and scintillation frequency structure:}
The measured wavefield  $\varepsilon(t)$ is the convolution (using subscript $e$ to denote emitted wavefield),
\be
\varepsilon(t) = \varepsilon_{\rm e}(t) * h(t) = [a(t) \amnnoise(t)] * [\pbfamp(t) \pbfnoise(t)].
\label{eq:epsscatter}
\ee

Time averaging a large number of pulses with $h(t)$ constant (i.e. over a span $\ll$ scintillation time) yields a measured
pulse
\be
I(t) \simeq \Ie(t) * \vert h(t) \vert^2 \equiv  \Ie(t) * \PBF(t).
\ee
where $\PBF(t)$ has a noisy form as in Fig.\,\ref{fig:pbfave} (top panel)
and Fig.\,\ref{fig:two_freq_pbfs} (right panel). 
Further averaging over many scintillation times then yields a scattered pulse profile  $I(t) = \Ie(t) * \PBF(t)$
like those for pulsar J1644-4559 in Fig.\,\ref{fig:J1644_profiles}.    Determining the scattering time $\taud$ and timing delays from scattering requires model fitting to $I(t)$.   

Accessing scattering information is more straightforward in the frequency domain.  Fourier transforming Eq.~\eqref{eq:epsscatter} yields 
\be
\FTvBB(\nu) = [\amnampFT(\nu) * \amnnoiseFT(\nu)] \times \htilde(\nu).
\label{eq:FTepsscatter}
\ee
The factor in square brackets yields  frequency structure with a characteristic scale $\sim \Wa^{-1}$  equal to the reciprocal of the width  $\Wa$ of $\amnamp(t)$.  This structure is fully modulated (unity modulation index) in the strong scintillation regime that usually applies for $\nu \lesssim 5\,$GHz.  
The average of  the squared magnitude $I(\nu) = \vert \FTvBB(\nu)  \vert^2$ over $\Np \gg 1$  pulses converges to  a constant with reduced modulation index $m_I \simeq \Np^{-1}$.  This yields the scintillation spectrum in isolation, 
$I(\nu) \propto \vert \htilde(\nu)\vert^2$, which can be analyzed to determine the characteristic frequency
of the intensity scintillations, i.e. the scintillation bandwidth $\dnud$. 


\begin{figure}
    \centering
    \includegraphics[width=0.4\textwidth]{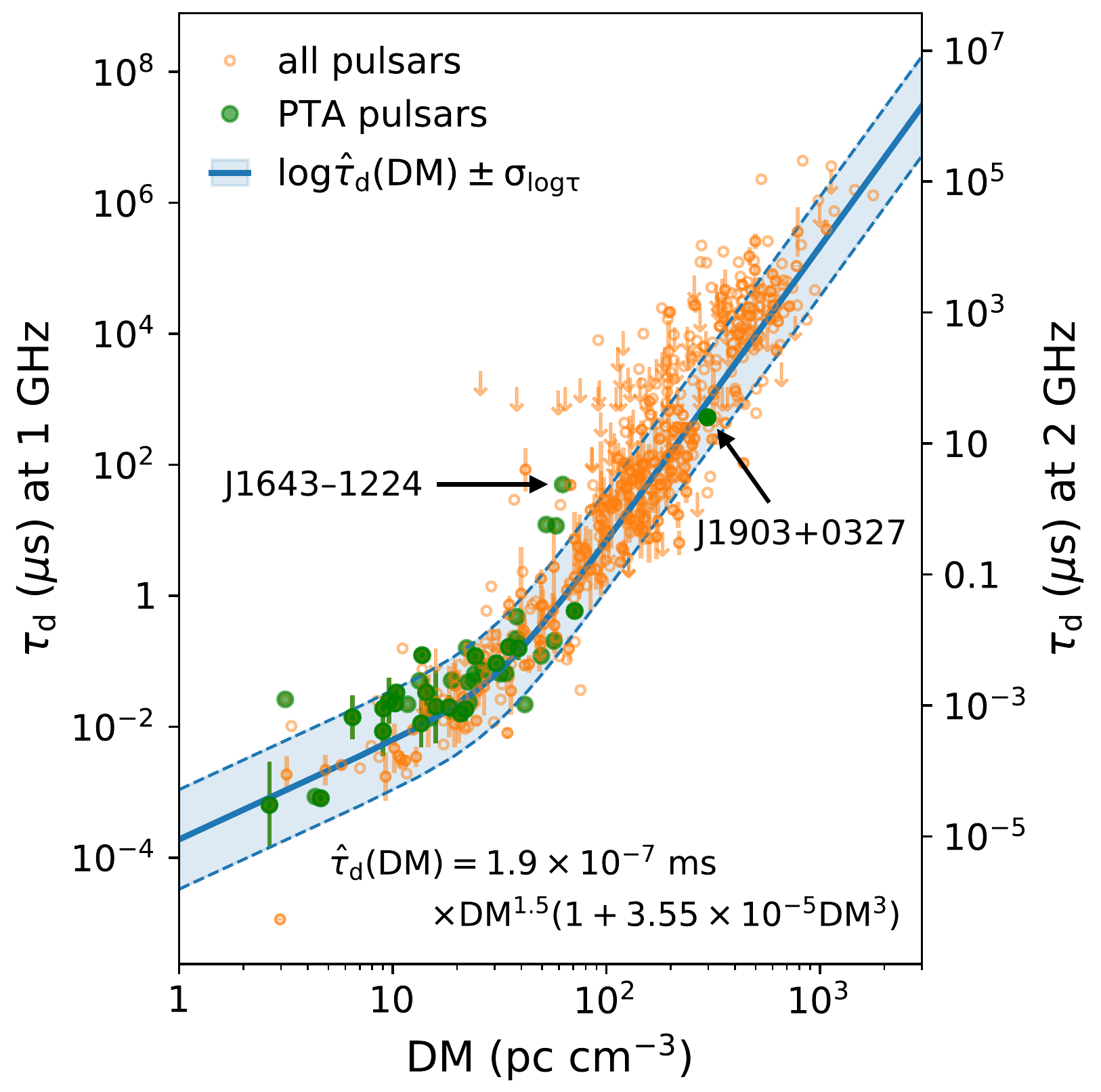}    	
    \includegraphics[width=0.55\linewidth]{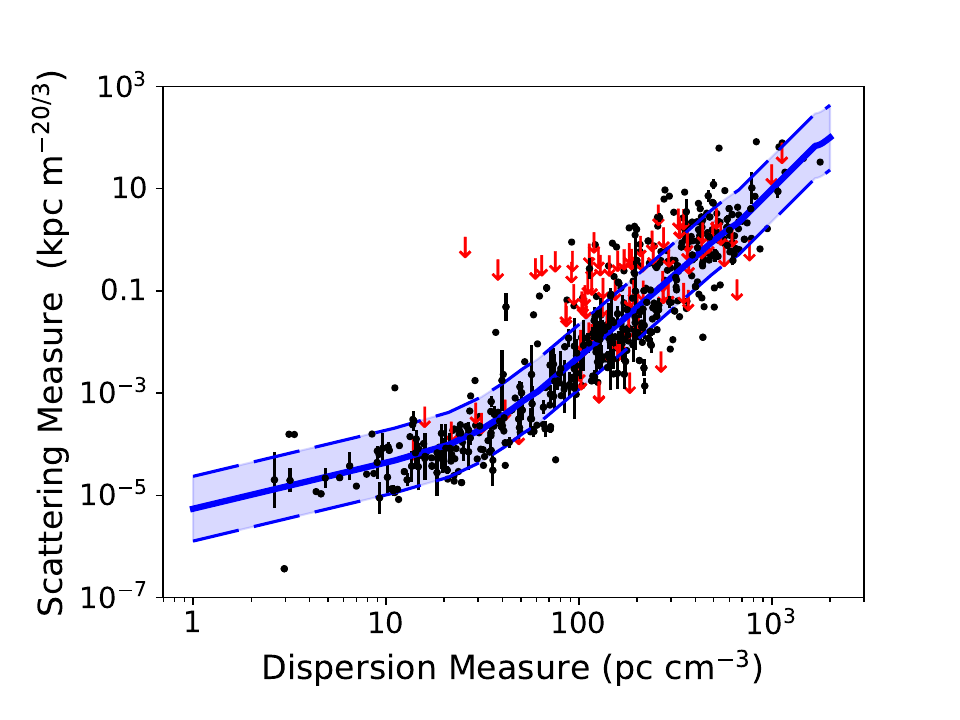}

    \caption{Scattering from known pulsars.  
    Left: scattering time ($\taud$) in $\mu$s at 1 GHz vs. DM for Galactic pulsars (including canonical and MSPs) with scattering reported in the literature. Measurements were scaled to 1 GHz assuming $\taud \propto \nu^{-4.4}$. PTA pulsars are highlighted in green. The solid blue line and shaded blue region show the best-fit empirical relation between $\tau$ and DM, given by the equation shown. For pulsars with multiple published scattering measurements, the scattering times were averaged and the error bars show the standard deviation of the measurements.
    \label{fig:hockeystick}
   Right:  Scattering measures vs. dispersion measures.   SM values are estimated from  pulse broadening times and scintillation bandwidths (see Table~\ref{tab:keyplasmaeq}, assuming $\beta=11/3$). Downward arrows (red) indicate upper limits, black points represent individual measurements, while those with error bars indicate the {\it range} of values for a specific pulsar (not the error on the mean of those values).   The blue lines and band indicate the mean and $\pm \sigma_{\log \tau}$ range of  SM calculated from the $\tau(\DM)$ relation in Eq.~\eqref{eq:taudm}.   The scatter about the mean fit arises from the irregular strength and distribution of scattering regions in the ISM. 
   \label{fig:hockeystick_sm}
   \label{fig:SMvsDM}
   }     
\end{figure}

The pulse broadening time $\taud$ is obtained by either pulse-shape modeling or
via a frequency structure analysis.  
Its importance for \TOA\ correction is discussed later.  Here we briefly present its role  for modeling the Galactic structure of electron density variations and to forecast  pulsar timing precision (and for studies of FRBs).    Several methods  are used to estimate $\taud$:  
\begin{enumerate}
\itemsep -2pt
\item Model fitting of average profiles as an intrinsic shape $\Ie(t)$ convolved with a \PBFtext\  $\PBF(t)$
\citep[e.g.][]{2017MNRAS.470.2659G};
\item Deconvolving average profiles using a template bank of candidate PBFs and the CLEAN algorithm
\citep[e.g.][]{bcc03, 2004ApJ...605..759B,2017PASP..129b4301T, 2019ApJ...874..179K, 2024ApJ...962..131Y};
\item  Inverting scintillation bandwidth ($\dnud$) determinations using Eq.~\eqref{eq:uncertainty1};
and
\item Using cyclic spectroscopy to obtain higher resolution dynamic spectra and estimate the  PBF and characterize its width 
\citep[][]{2011MNRAS.416.2821D, 
2013ApJ...779...99W,
Palliyaguru_2015,
2021ApJ...913...98D,
2023ApJ...944..191T}.
\end{enumerate}
Here we present a brief analysis of a sample of $\taud$ values obtained using the first three methods.  The intent here is to illustrate the variation of the scattering time with DM and show the range of  corresponding  scattering measures.  Later we discuss how constraints on the PBF and $\taud$ can be used to correct 
\TOAs\ for scattering delays. 

It is well known that $\taud$ correlates with \DM\ albeit with significant scatter, \newtext{according to the so-called ``hockey-stick'' relation}.   
Fig.~\ref{fig:hockeystick} shows
measurements of $\taud$ scaled to 1\,GHz 
along with a fit using  the  function 
$\widehat \tau(\DM) = A\times\DM^a (1 + B\times \DM^b)$ \citep[][]{rmd+97}. 
  Fitting to $\log\taud$ yields
\be
\widehat{\tau}(\DM) = 1.9\times10^{-7} \,{\rm ms} \, \times \DM^{1.5} ( 1 + 3.55\times 10^{-5} \DM^3)
\label{eq:taudm}
\ee
with scatter $\sigma_{\log_{10}\taud} = 0.76$  \citep[][]{cordes2022}.
The plotted measurements include forty-three PTA pulsars with published scattering measurements. Most of these MSPs have $\taud < 1$ $\mu$s because they are chosen to have small DMs, which often goes hand in hand with being closer to the observer and  brighter. There are a few notable exceptions, such as  J1643$-$1224 with $\taud = 42$ $\mu$s, an almost $100\%$ excess from the value expected from the $\taud-\DM$ relation ($1$ $\mu$s). This pulsar lies behind the HII region Sh $2-27$ that contributes $90\%$ of its total DM \citep{2020ApJ...897..124O}, and which likely also dominates $\taud$ \citep{mall2022}. J1903$+$0327 has the highest DM and $\taud$ of any pulsar timed by current PTAs, but unlike J1643$-$1224, it is entirely consistent with the $\taud-\DM$ relation.
The pulse broadening times and model fit are used to evaluate scattering measures by inverting
Expression 9 in Table\,\ref{tab:keyplasmaeq}, as shown 
in the right-hand panel of Fig.~\ref{fig:SMvsDM}.

Notable features of the $\taud-\DM$ relation in Fig.\,\ref{fig:hockeystick} include:
(1) the approximate  factor of ten variation in $\taud$ for fixed DM;
and
(2) the large change in slope of the mean relation from $\taud\propto \DM^{1.5}$ to $\DM^{4.5}$ at
$\DM \sim 50\,\DMunits$.    The large scatter in $\taud$ includes estimation errors from pulse-fitting methodology
and from variations in geometrical leveraging: a region midway along the \LOS\ produces 
the largest scattering time.    The increase in slope requires a change in  density variations as quantified   by  a fluctuation parameter that characterizes  small cloudlets having a volume filling factor $f$ and internal density fluctuations with wavenumber spectral index $\beta$ and outer scale $\louter$,  
\be
\Fc \equiv  \frac{\Var(\nelec) / \nelec^2}{\ff\louter^{\beta-3}}   \overset{\beta = 11/3} \simeq  \Fcdefkol
\label{eq:Fckol}
\ee 
where $\zeta$ and $\epsilon$ are dimensionless parameters corresponding to cloud-to-cloud  and internal density variations, respectively \citep[e.g.][]{cwf+91}.    There is a factor of $\sim 20$ difference in $\Fc$ between the local ISM and the inner Galaxy in the NE2001 electron density model that can account for the slope increase
of the $\taud-\DM$  relation.   These results underscore the value of pulsars in the solar vicinity that will show much less scattering-induced \TOA\ variations  than more distant objects.  Though scattering delays can be removed to some extent, methods for doing so are not yet sufficient to make distant pulsars useful for the most demanding timing applications (e.g. nanohertz GW detection). 

\subsection{Cyclic spectrum (field correlation function)}

The cyclic spectrum (CS) is a frequency-domain correlation function that exploits the periodic aspect of a signal \citep{gardner91,antoni07}.  
The primary  utility of the CS is that it  can provide  dynamic spectra  with resolved scintillations in cases where scintles are narrow in frequency. Second is its potential for estimating the  full impulse response $h(t)$ contemporaneously with arrival time data, which, in principle, can aid efforts to correct  \TOAs\ for  multipath propagation delays \citep[e.g.][]{2011MNRAS.416.2821D, 2013ApJ...779...99W, Palliyaguru_2015, 2021ApJ...913...98D, 2023ApJ...944..191T}. Additionally, the fine frequency resolution used to resolved scintles enables fine-frequency RFI mitigation as a consequence.
  
The CS of the pulsar signal,  $\GammaFTE(\delta\nu, \nu)$, is the frequency domain correlation function of the baseband field 
(analogous to $\Gamma_{\htilde}(\delta\nu, \nu)$  for the field PBF), 
\be
\GammaFTE(\delta\nu, \nu) 
\equiv  \EfieldFT(\nup)  \EfieldFT^{\,*}(\num) 
= \Gamma_{\htilde}(\delta\nu, \nu)    \iint dt_1 dt_2\,  a(t_1) a(t_2)  m(t_1) m^*(t_2) 
e^{2\pi i (\nuplus t_1 - \numinus t_2)} .
\ee
The basis for the CS is a  finite FT  calculated over a time block $T$ longer than a pulse period but shorter than the DISS time scale, i.e. $P \ll T \ll \dtd$.   This provides better frequency resolution than PRS for a time block in which the scintillation spectrum $H(\nu)$ is essentially constant.  Note also that because the scintillation bandwidth $\dnud$ is strongly frequency dependent, scaling approximatley as $\nu^{-4}$,  the CS analysis needs to be done over bandwidths narrow enough so that $\dnud$ varies only  by a tolerable amount. 
Including a radiometer noise term,  the CS for a single FT is
\be
\GammaFTE(\Gargs) 
&= & \EfieldFT(\nup)  \EfieldFT^{\,*}(\num)
= \Gamma_{\htilde}(\Gargs)  \Gamma_{\EfieldiFT}(\Gargs)  + \GammaFTn(\Gargs)
+ [\htilde(\nuplus) \EfieldFT_{\rm i}(\nuplus) \ntilde_{\numinus}^* 
+  \htilde(\numinus) \EfieldFT_{\rm i}(\numinus)^* \ntilde(\nuplus)],
\ee
where   cross terms in square brackets between $\widetilde n$ and $\EfieldiFT$ average to zero.  However, they contribute significantly to the variance of  CS estimates considered later. 

\begin{figure}[b!] 
   \centering
   \includegraphics[width=0.475\linewidth]{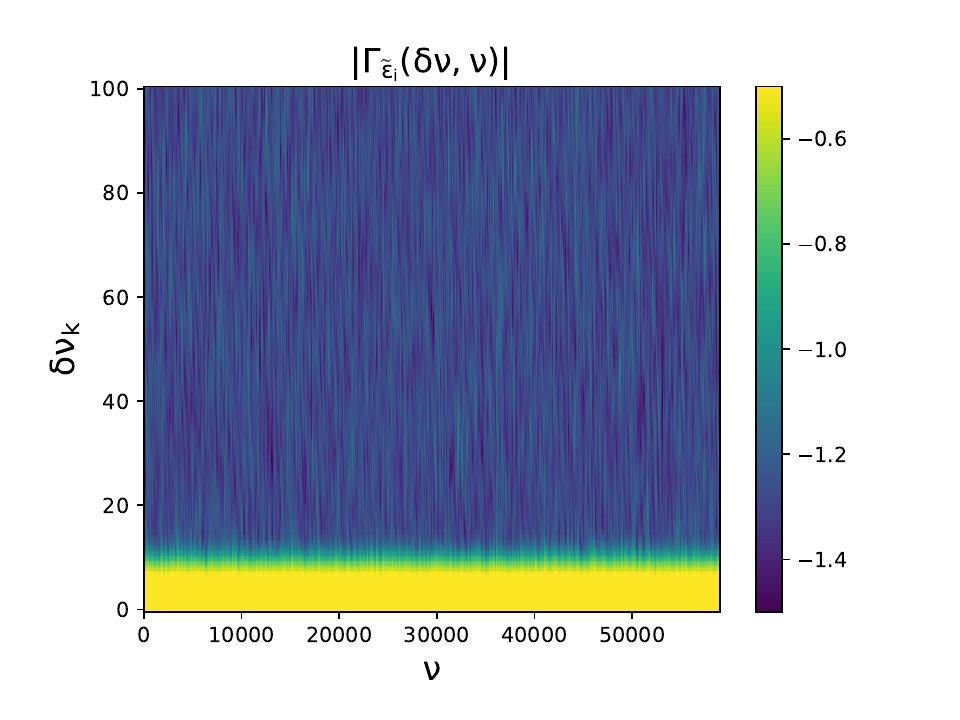}
   \includegraphics[width=0.475\linewidth]{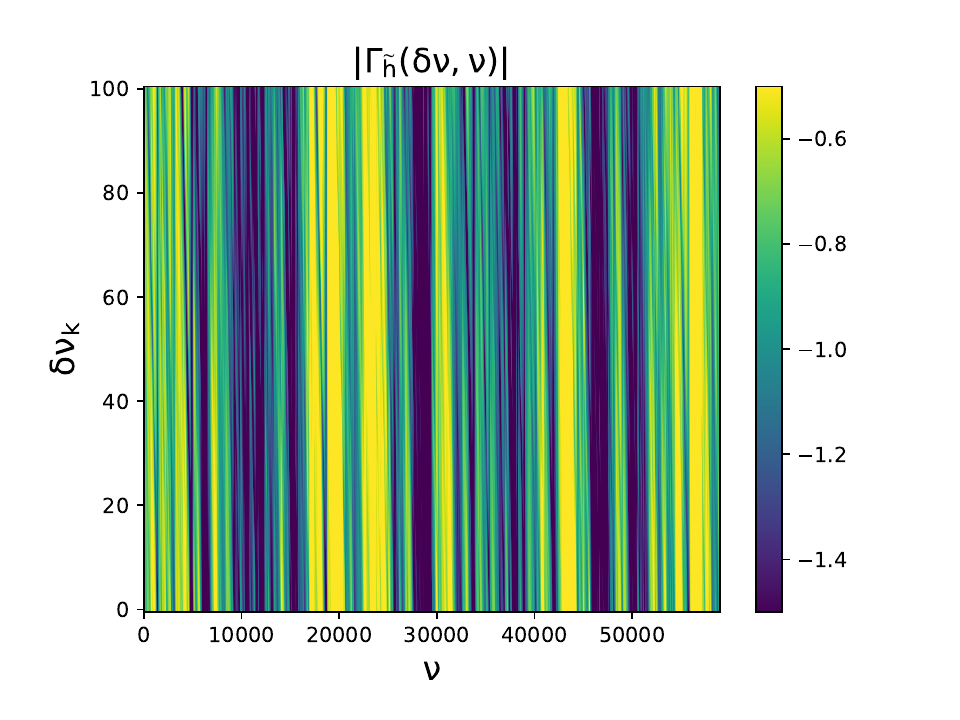}
   \includegraphics[width=0.475\linewidth]{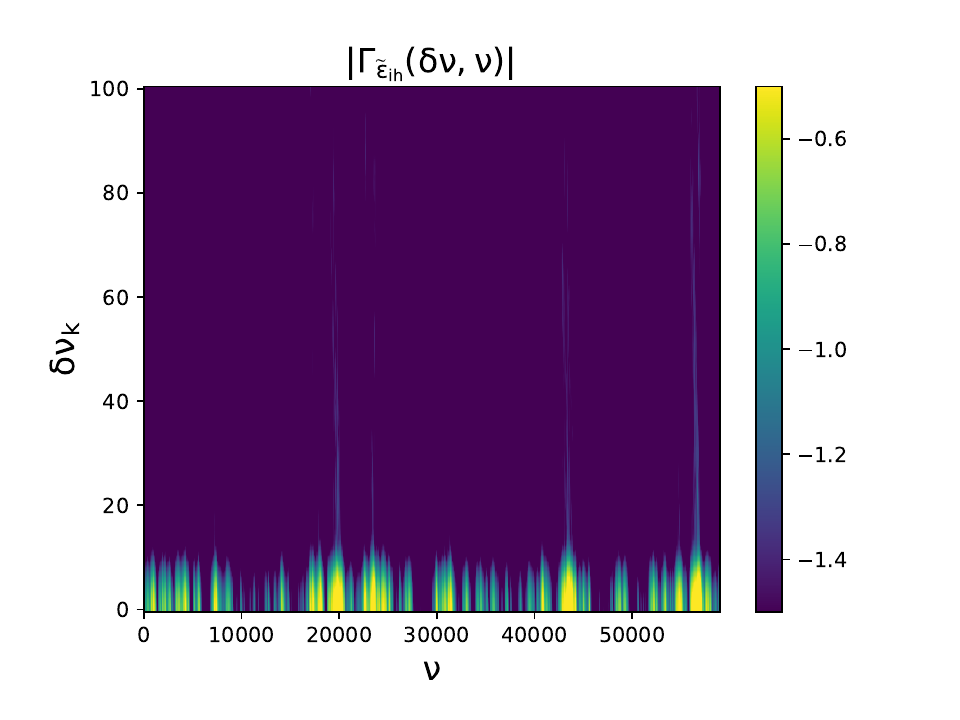}
   \includegraphics[width=0.475\linewidth]{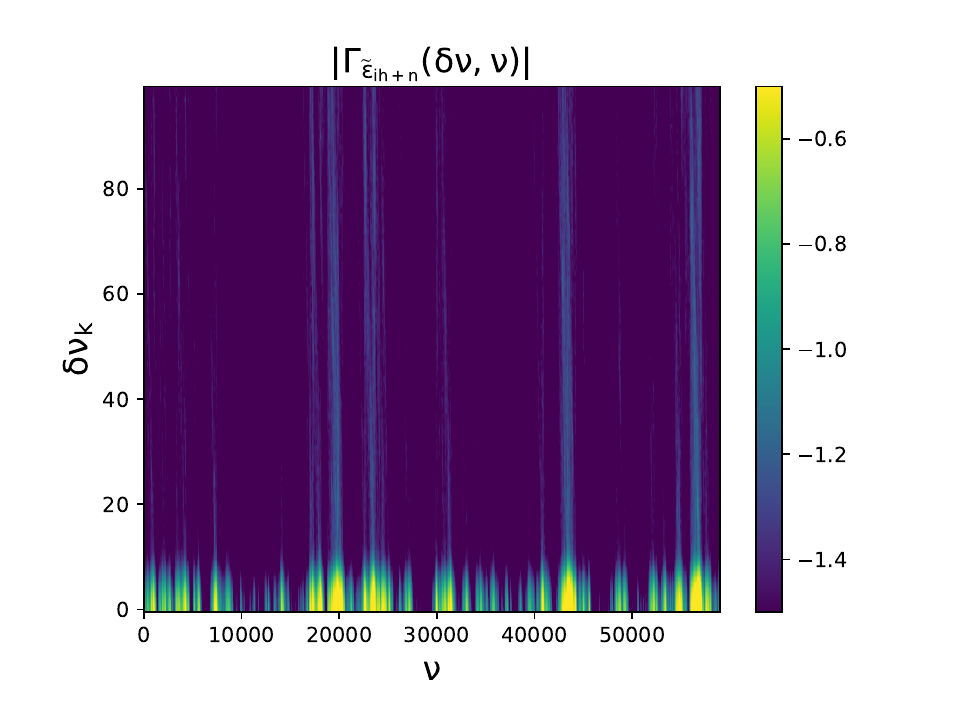}
   \caption{Frequency domain correlation functions (cyclic spectra) of four quantities.  They
   comprise averages over 100 realizations of the pulsar signal while keeping fixed a single 
   realization of  the scintillation modulation  . 
   The plotted quantities are $\log_{10}$ of the absolute values of the correlation function normalized
   to unit maximum. 
   The horizontal axis corresponds to radio frequency (in bins)
   and the vertical axis is the frequency lag $\delta\nu_k = k / P$ corresponding to harmonics in the CS. 
   Each frame is normalized so that the mean across the $\delta\nu_0 = 1$. 
   Top left:  CS for the intrinsic pulsar signal  $\EfieldiFT$ modeled as amplitude modulated noise.
   Top right: CS for the scintillation modulation $\htilde$.
   Bottom left: CS for the modulated pulsar signal $\htilde * \EfieldiFT$.
   Bottom right: CS for the modulated signal + radiometer noise. The case shown here is for
   ${\rm S/N} \equiv \Sphiave / \Ssys = 1$, where $\Sphiave$ is the period averaged flux density and
   $\Ssys = 3$\,Jy is the SEFD for a DSA-2000 class telescope.
    }
   \label{fig:gammas}
\end{figure}

To simplify the correlation function   we exploit the  constancy of $\Gamma_{\htilde}$ over  the time block  and average over the  rapid decorrelation of the  noise process $\amnnoise$.   
We use an ensemble average for the noise while letting  
$\Gamma_{\htilde}$  be constant\footnote{The noise decorrelates on sub-$\mu$s time scales for $>$MHz bandwidths while $h(t)$, if dominated by DISS,  decorrelates over minutes to an hour for nearby pulsars used in PTAs. Integration times of $> 1$\,s yield time-bandwidth products $> 10^6$ for the noise, or a fractional error in the correlation function $\lesssim 1$\%.}.  
The noise correlation $\langle m(t_1) m^*(t_2) \rangle = \amnnoisevar \Delta(t_2-t_1)$ vanishes for 
time differences $\vert t_1 - t_2 \vert$ larger than the inverse bandwidth (e.g.  larger than 0.01 to 1\,$\mu s$) so the amplitude product $a(t_1)a(t_2) \simeq a^2(t_1) \equiv A(t_1)$. 
Using the FTs $\amnnoiseacfFT$ and $\amnAmpFT$ for $\Delta$ and $A$, respectively, yields 
\be
\left\langle 
\GammaFTE(\delta\nu, \nu) 
\right\rangle_{P \ll T \ll \dtd}
\simeq
\amnnoisevar 
 \amnAmpFT(\delta\nu)  \amnnoiseacfFT(\nu) \Gamma_{\htilde}(\delta\nu, \nu) 
 +
\Wn \sigma_n^2  \sinc(\delta\nu T) \amnnoiseacfFT(\nu)
 \label{eq:CSeps}
\ee
where $\sinc(x) \equiv ( \sin \pi x) / \pi x$.   The second (noise) term has a width $1/T$ vs. $\delta\nu$ but extends over the full frequency band vs. $\nu$.

The cyclic aspect of $\GammaFTE$ comes into play when the FT is calculated  over a time 
$T \gg P$ that spans $\Np$ periods with the (assumed) same shape $A_0(t)$,
\be
A(t) = \sum_{j=0}^{\Np-1}  a_j A_0(t-jP) .
\ee
A strictly periodic train of  $\Np\gg 1$ identical pulse shapes $A_0(t)$ with $A_0(0) = 1$ and  an   average amplitude $\Spknu$  has an FT,
\be
\amnAmpFT(\delta\nu) 
\simeq \Np \WA \Spknu
\amnAmpFT_{\rm n}(\delta\nu) \times \vert \sincd(\delta\nu P, \Np)\vert ,
\ee
where $\amnAmpFT_{\rm n}(\delta\nu) = \amnAmpFT_0(\delta\nu) / \amnAmpFT_0(0)$ is  normalized to unit maximum and its effective width is $\WA \equiv \amnAmpFT_0(0)$. The Dirichlet sinc function,
$\sincd(x, N) = \sin(N\pi x) / N\sin(\pi x)$,  yields peaks at discrete frequencies $\dnuk = k / P$ with $k$ extending to
$k_{\rm max }\sim P / \WA$, the reciprocal of the pulse duty cycle.     

The number of  cyclic frequencies where the amplitude of $\GammaFTE$ is significant is of order
the reciprocal of the  pulse duty cycle,   $N_k \sim P / \WA$.   For millisecond periods 
$P_{\rm ms}$ the cyclic frequencies $\delta\nu_k  = (k/P_{\rm ms})$\,kHz are much smaller than the DISS bandwidths for low-DM pulsars observed at 1.5\,GHz.   This means that for
 cyclic frequencies with $k \ll N_k$  the scintillation CS
$\Gamma_{\htilde}(\delta\nu_k, \nu) \simeq \Gamma_{\htilde}(0, \nu) \equiv \vert \htilde(\nu)\vert^2$, which is the standard scintillation spectrum for the time block in question (cf.\ Eq.~\eqref{eq:DSdef}). The frequency resolution is 
$\delta\nu = 1 / T \ll 1/\WA \ll 1/P  \ll \dnud$. 

We rewrite the CS in Eq.~\eqref{eq:CSeps} (after dropping the subscript on the averaging brackets) as
\be
\left\langle \GammaFTE(\Gargsk) \right\rangle
= (\TFT / B)\, \Sphiave\, \bandpass(\nu)
\left[ \amnAmpFT_{\rm n}(\dnuk)  \Gamma_{\htilde}(\Gargsk) + \frac{\sinc\, \dnuk T}{\SNRphi} \right],
\label{eq:Geps2}
\ee
where
the bandpass factor is $\bandpass(\nu) = \widetilde\Delta(\nu) / \widetilde\Delta(0) =  B \widetilde\Delta(\nu)$.
Also,
$\Sphiave \simeq (\WA/P)\Spknu$ is the period averaged flux density (as usually reported for pulsars),
$\bandpass(\nu)$ is the receiver bandpass function with width $B$,
and $\sinc\, x \equiv (\sin \pi x) / \pi x$.   The signal to noise ratio is
$\SNRphi \equiv \Sphiave / \Ssys$ where $\Ssys = \Tsys / G$ is the system effective flux density (SEFD) for a telescope with gain $G$.    The first term in square brackets is from the pulsar signal while the second term is from additive noise reduced by the $\SNRphi^{-1}$ factor.   Though $\SNRphi \ll 1$ for most pulsars,  the second term is important only for $k=0$.

Inspection of Eq.~\eqref{eq:Geps2} shows that the scintillation CS $\Gamma_{\htilde}$ with full phase information can be extracted in cases where the signal-to-noise ratio is large and if the intrinsic pulse shape
$\amnAmpFT_0(t)$ (or equivalently  $\amnAmpFT_{\rm n}$) is known.   The latter condition is not usually the case.

When only the scintillation spectrum is wanted (i.e. without phase information) integration over pulse phases where the pulse is strong can be done as follows.  
The inverse transform of $\GammaFTE(\Gargsk)$ with respect to $\dnuk$ yields the phase resolved CS
or `profile spectrum' 
with a time argument replacing the $\delta\nu$ argument,
\be
\left\langle \GammaFTE(t, \nu) \right\rangle
\propto  \Sphiave\, \bandpass(\nu)  \amnAmp_{\rm n}(t) *  \Gamma_{\htilde}(t, \nu) ,
\ee
where a constant term from the $\sinc$ function has been excluded.
Integration over the time window comprising the pulse then yields the scintillation spectrum. 

Fig.~\ref{fig:gammas} shows cyclic spectra $\Gamma_{x}(\delta\nu, \nu)$ for four quantities:
$x =  $ (pulsar signal,  scintillation modulation, scintillated pulsar signal, and scintillated signal + noise).
The signal to noise ratio of the voltage signal is unity, defined as $\Sphiave / \Ssys$.   The plotted quantities are
sums over 100 realizations and are evaluated at harmonics $\delta\nu_k = k / P$ of the pulsar period $P$.  Features of the four panels include:
\begin{itemize}
\itemsep -1pt
\item  Top left:  The CS of the intrinsic signal is flat across frequency but rolls over vs. $\delta\nu_k$.   The number of significant harmonics is approximately the reciprocal of the duty cycle, $P / W \sim 15$. 
\item  Top right: The CS of the scintillation modulation is nearly constant across harmonics but shows the 100\% modulations expected vs. frequency. 
\item Bottom left:  the CS of the scintillated pulsar signal shows the frequency structure and rolloff vs. harmonic number. 
\item Bottom right: additive noise reduces the significance of the frequency structure.    This can be compensated for by summing over more realizations.   In practice  the number of summed individual CS  is limited by the scintillation time scale.    Most pulsars will have S/N smaller than the unity value  used in the simulation. 
\end{itemize}

%

%


\section{A gallery of pulse broadening functions}
\label{sec:pbf_gallery}

Reasons for considering the variety of PBFs are twofold.  First,   PBFs provide unique information about the ISM, in particular the Galactic distribution of electron density variations  and their  wavenumber spectrum.  Second, optimal timing of pulsars requires removal of chromatic interstellar delays if the primary goal is to obtain achromatic \TOAs\ for the study of pulsar orbits, the neutron star equation of state, and long-wavelength gravitational waves.  The second is of course intertwined with the first because proper removal of chromatic delays requires a sufficient understanding of the ISM, viz. how propagation effects on arrival times scale with frequency. 

In this section we present a gallery of PBFs for selected media and geometries to illustrate their variety and the complexity of \TOA\ correction for chromatic effects.

Most of the issues related to PBF envelope variations can be illustrated using a thin screen geometry that allows an analytical treatment for many quantities.  Quantitative analyses also require consideration of thick   media that  generally depend on  numerical integrations. 

\begin{figure}[b] 
   \centering
   \includegraphics[width=0.7\linewidth]{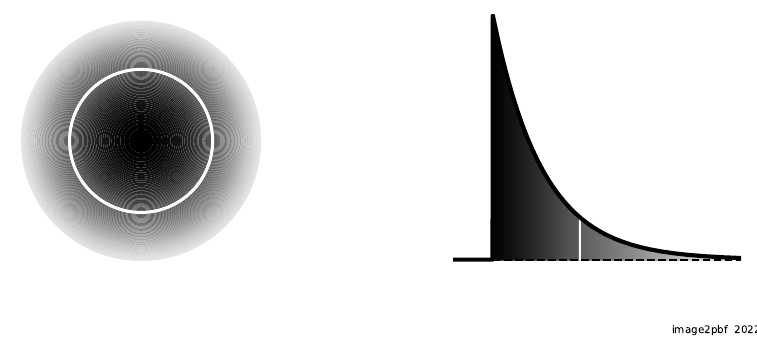}
   \caption{Scattering from a thin screen that produces a circularly symmetric Gaussian image (left) and the corresponding exponential PBF.    Darker shading of the image indicates angles with higher probability.   The white line indicates the 1/e point in the image and in the PBF. }
   \label{fig:image2pbf}
\end{figure}


\subsection {Thin-screen geometry}
\label{sec:thinscreens}

Here we consider scattered images of point sources  $I(\thetavec)$  and PBFs  $\pism(t)$ as  `envelope' quantities   that exclude stochastic noise (speckles and scintles) already discussed.   The PBF is the sum of all {\it geometric} delays over the  scattered image of a point source  \citep[][]{2001ApJ...549..997C},
 \be
\pism(t)  =  \int d\thetavec\,  I(\thetavec) \delta(t - \deff\vert\thetavec\vert^2 / 2c). 
\label{eq:im2pbf}
\ee
No plasma delays are included because the screens are considered infinitesimally thin and the increments in \DM\ corresponding to the phase change at each position along the screen are small. 
Simplification of the integral depends on the shape of the image and on any offset from the direct \LOS\ caused by refraction. 

Most expressions in the literature consider Gaussian images with no offset from the direct propagation path.
For example,   a symmetric Gaussian image with \RMS\ angle $\sigma$ along each axis
gives the often used exponential PBF, $\pism(t) = \taud^{-1} \exp(-t/\taud)\Theta(t)$,  where
$\Theta(t)$ is the Heaviside function and $\taud = \deff \sigma^2 / c$.   Figure~\ref{fig:image2pbf} shows the results from application of Eq.~\eqref{eq:im2pbf} and  how the $e^{-1}$ points in the image and the PBF are related.
A Gaussian image is expected if the phase structure function has a square-law form, $\Dphi(b) \propto b^2$, which is at best only approximated by a screen comprising monoscale density irregularities \citep[e.g.][]{1970Sci...168.1453C, sut71, 1978RQE...20..581O} or a power-law distribution of density scales with 
$\beta \to 4$.    

While the exponential PBF is convenient to use, e.g. for coarse estimation of scattering times $\taud$ by model fitting scattered pulse shapes,  it is overly simplistic  for both ISM and pulsar timing applications.
The reasons for this are that 
(1)  Scattering is generally associated with multiple screens and extended media, rather than single thin screens. While scintillation arcs indicate the prominence of thin screens along some lines of sight, they are often multiple rather than single, even for nearby pulsars \citep{mckee22,ocker24,reardon25}. More distant pulsars encounter even more scattering regions.\footnote{The relevance of extended media to more distant, high-DM pulsars is corroborated by scintillation arc measurements, which indicate that scintillation arcs become less defined (and detectable) at higher DM, suggesting that arcs are ``blurred'' by the presence of multiple screens along the LoS \citep{Stinebring22,main23}. }
Thick media (or an ensemble of thin screens)  yield PBFs with  longer  rise times than the sharp rise of a one-sided exponential function \newtext{\citep{lr99}}.
(2)  Direct imaging of scattered pulsars (and other objects, such as active galactic nuclei and interstellar masers) yield elliptical images in some cases, which cause the PBF to differ from a simple exponential \citep[e.g.][]{desai2001,bower14}. Image anisotropy is also evident in the detection of inverted ``arclets'' in pulsar secondary spectra \citep{2010ApJ...708..232B,sprenger22,Stinebring22}.
(3) PBFs from the ISM appear to deviate from an exponential form \citep[e.g.][]{1975MNRAS.172...97L,2009MNRAS.395.1391R,2017MNRAS.470.2659G}; and
(4) refraction-induced offsets of scattering images alter the shapes of all PBFs, as shown below.
 
Scattering slabs of arbitrary thickness with density variations following a square-law structure function
have PBFs that are the 
 $n$-fold convolution of one-sided exponential functions $\tau_n ^{-1}\exp(-t / \tau_n)\Theta(t)$ where $\Theta(t)$ is the Heaviside function \citep[][]{1975RSPSA.342..131W}.   The $e$-folding times $\tau_n$ are the eigenvalues of the equation that describes diffusion of rays through a slab.   The number of significant eigenvalues depends on the slab's thickness.  For a thin screen, only one eigenvalue is significant.   
 
\begin{figure}[tb] 
   \centering
   \includegraphics[width=0.475\linewidth]{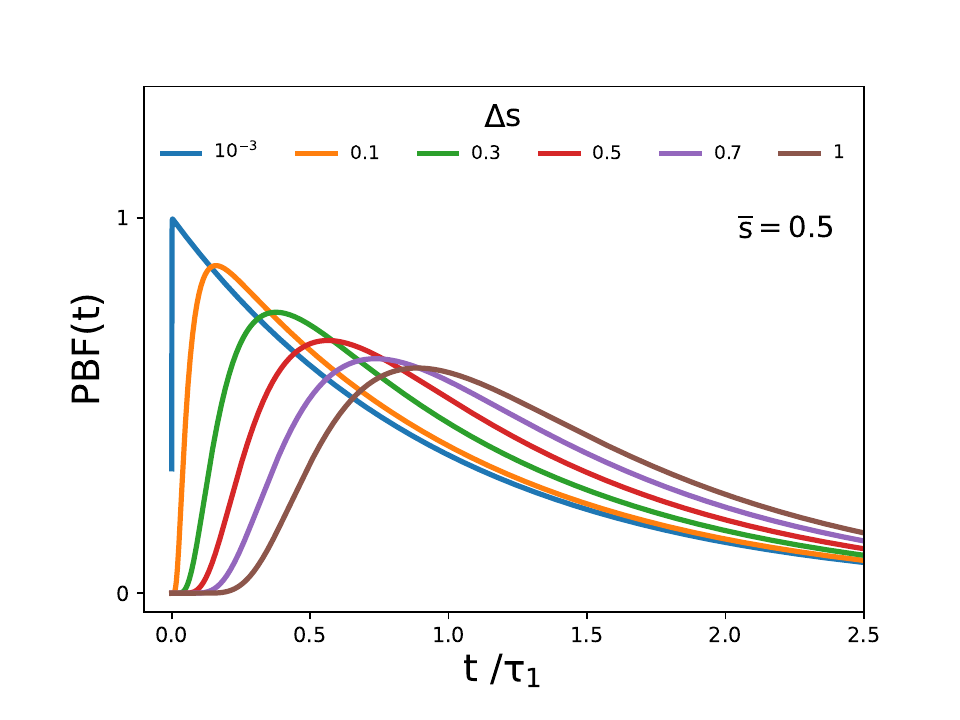} 
   \includegraphics[width=0.475\linewidth]{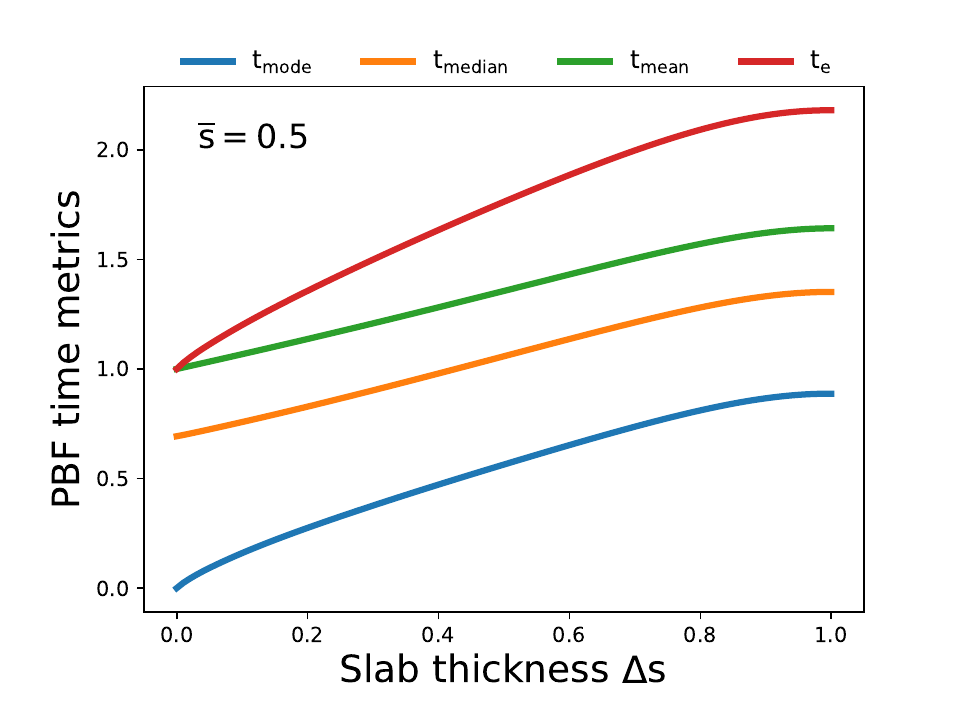}
   \caption{Pulse broadening functions for media with square-law structure functions (Gaussian scattered images).
   Left: PBFs for slabs  midway between source and observer with different  fractional thickness $\Delta s$ for a  square-law structure function.  Time is normalized by $\tau_1$, the $e^{-1}$ time of the PBF for the thin screen case, which has only one eigenvalue.   Finite thickness slabs have multiple eigenvalues that yield PBFs equal to the $n$-fold convolution of one-sided exponential functions.
   Right: Time metrics of PBFs for slabs  midway between source and observer with different  fractional thickness $\Delta s$ for a  square-law structure function. 
    }
   \label{fig:pbfs}
    \label{fig:pbftmetrics}
\end{figure}

 Example PBFs are shown in 
Fig.\,\ref{fig:pbfs} for slabs centered midway along the \LOS\  ($\sbar = 1/2$) with fractional thicknesses $\Delta s$  ranging from $\ll 1$ to 1.  The rise time and width of the PBFs increase with slab thickness, making it clear that model fitting of pulses with the thin screen PBF will yield biased results for pulsars affected by thick-slab scattering.
Indeed \citet[][]{2004ApJ...605..759B} demonstrated with deconvolution of PBFs from measured pulse shapes that  the more rounded rise times of thick-slab PBFs  were required for some pulsars. 

 The mean scattering time  $\tau = \sum_n \tau_n \equiv \int dt\, t \PBF(t)$ 
is equivalent to calculating the mean time for the  PBF  treated as a probability density function.   We compare 
the mean time with the mode, median and $1/e$ time of the PBFs in Fig.\,\ref{fig:pbftmetrics}.   While the mean and 
$1/e$ time are equal for a thin screen with a square-law density structure function, they diverge with increasing slab thickness.   

%
\begin{figure}[t!] 
   \centering
   \includegraphics[width=0.45\linewidth]{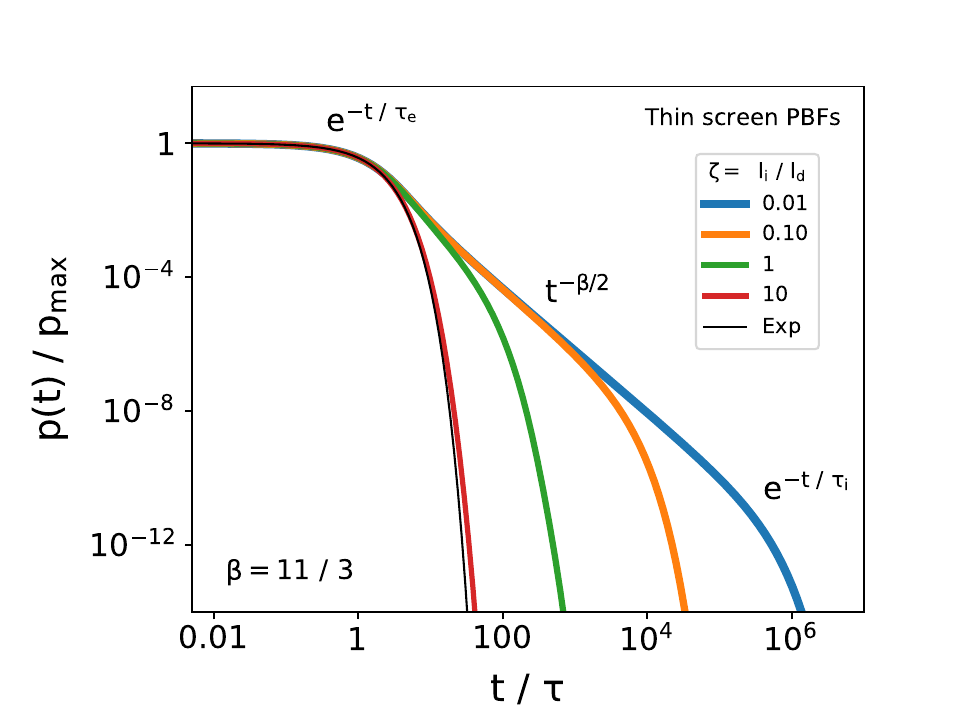}
    \includegraphics[width=0.45\linewidth]{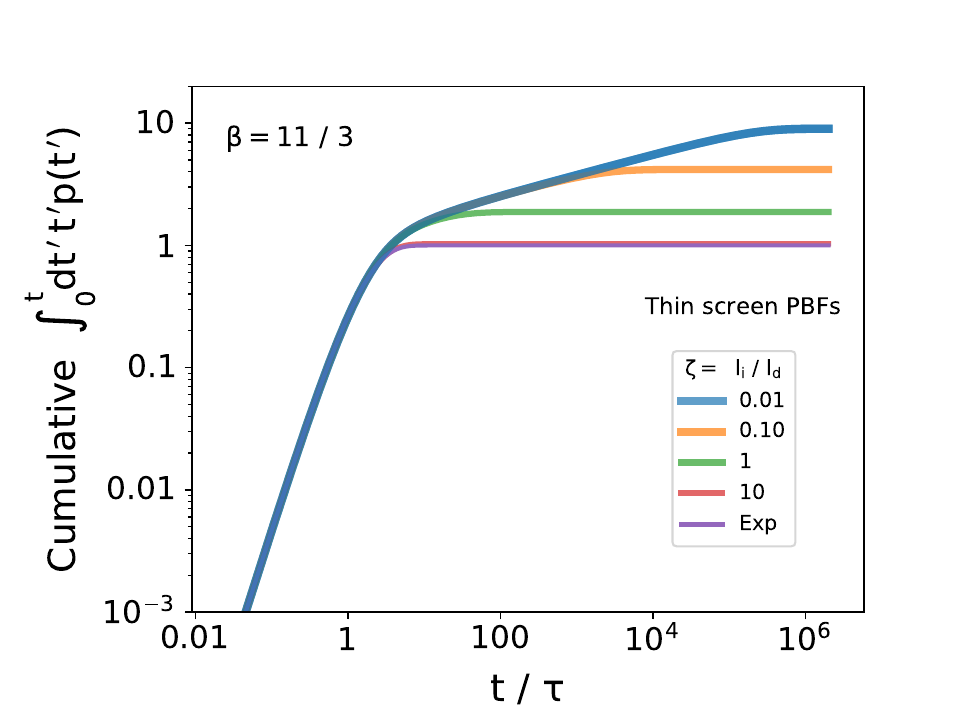}
   \caption{Thin screens with  Kolmogorov density fluctuations ($\beta=11/3$).
   Left: PBFs for an isotropic Kolmogorov spectrum  that illustrate the effects of
   a small or large inner scale $\linner$ relative to the diffraction scale $\ld$.  
   The time axis is normalized by the $e^{-1}$ time scale $\tau_e$. 
   Scaling laws for the smallest inner scale ($\zeta = \linner/\ld = 0.01$)  show three regimes: 
   exponential $\propto e^{-t/\tau_{\rm e}}$,  
   power-law $\propto t^{-\beta/2}$, and 
   exponential $\propto e^{-t / \tau_{\rm i}}$ 
   at small,  intermediate, and large times, respectively. 
   For the largest inner scale, the PBF shows only the exponential scaling that is close to that expected for a Gaussian-shaped scattered image.   
   Right: Cumulative integral  showing the contribution to the mean scattering time from the tail of
      the PBF for different inner scales.   The curve for the exponential PBF is nearly indistinguishable from the 
      $\zeta = 10$ (large inner scale) case. 
      \newtext{
      (Note that the mean scattering time is only indirectly related to the \TOA\ offset caused by scattering (see Sect.~\ref{sec:heavytails}.)}
   }
   \label{fig:pbf_demo1a}
\end{figure}

\subsection{PBFs from thin and thick media with power-law (`Kolmogorov like') wavenumber spectra} 
\label{sec:KlikePBFs}

Media with a power-law wavenumber spectrum spanning a broad range of scales  produce non-Gaussian scattering angle distributions and  PBF envelopes that depart significantly from an exponential form.  In addition, their shapes
as well as their $1/e$ scales are frequency dependent:  thus {\it PBFs are not self similar vs. radio frequency}. 
\citet[][]{1978RQE...20..581O} show for  $2 < \beta < 4$ and zero inner scale that the image scales asymptotically as $I(\theta) \propto (\theta / \thetad)^{-\beta}$  for  $\theta / \thetad \gg 1$,
where $\thetad$ is the diffraction angle, and the corresponding asymptotic PBF is
$\pism(t) \propto (t / \tau)^{-\beta/2}$ for  $t  / \tau \gg 1$ where $\tau = d\thetad^2 / 2c$.
As $\beta \to 4$ from below, the PBF is exponential $\propto e^{-t/ \tau}$ for 
small $t \lesssim 2\tau \ln(2 / (4-\beta))$ 
but includes a power-law component  $\propto (t / \tau)^{-\beta/2}$ for larger $t$. 
A finite inner scale yields a third PBF regime,  an  exponential decay at very long times $t \gtrsim \taui$ where
$\taui \propto \linner^{-2}$ can be many orders of magnitude larger than the $1/e$ PBF scale. 

The three regimes are evident in the PBFs shown in  Fig\,\ref{fig:pbf_demo1a} (left panel). 
For a Kolmogorov spectrum ($\beta = 11/3$) and other cases with $\beta < 4$,  the PBF decays more slowly than an exponential \citep[e.g.][]{1975ApJ...201..532L, lr99} and generally for large $t \gtrsim \tau$, the PBF follows the 
$t^{-\beta / 2}$ scaling until it decays exponentially from  an inner scale. 
For $\beta < 4$,   the mean scattering  time $\langle t \rangle =  \int dt\, t\, \pism(t)$ diverges unless the inner scale is finite. 

\begin{figure}[b!] 
   \centering
   \includegraphics[width=0.5\linewidth]{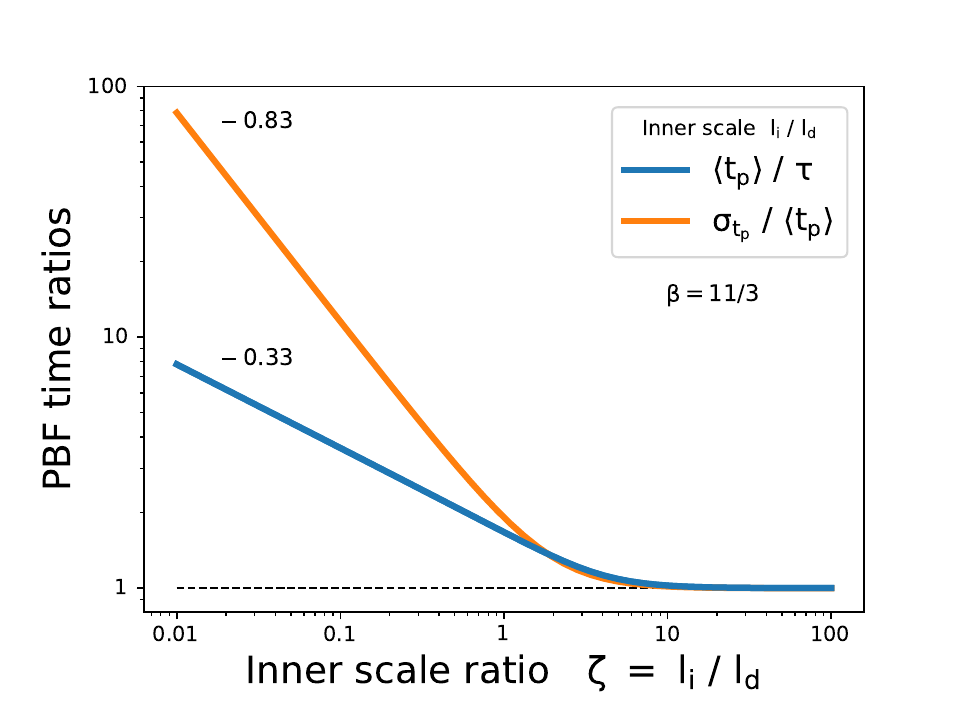}
   \caption{Thin screen scattering time metrics, $\langle t_{\rm p} \rangle / \tau$ and 
   $\sigma_{t_p} / \langle t_{\rm p} \rangle$, vs. the inner scale ratio $\zeta = \linner / \ld$ for a Kolmogorov medium ($\beta = 11/3$) in the
   inertial subrange scaling regime. 
   \newtext{
   These metrics are, respectively, the mean scattering time and the RMS scattering time in units of the $1/e$ scattering time.  They are calculated by treating the PBF as a probability density function. They are much larger than unity for heavy tailed PBFs.} 
   For $\zeta \gg 1$, the PBF asymptotes to a one sided exponential and the two ratios asymptote to unity. 
   The two lines are labeled with their  logarithmic slopes for small values of $\zeta$.
    } 
   \label{fig:Rt_and_Rsigt}
\end{figure}

Fig.~\ref{fig:pbf_demo1a} (right panel) shows the cumulative integral for the  mean PBF delay$\tpbfave = \int dt\, t\, \PBF(t)$ for $\beta = 11/3$.  
With the PBFs  scaled  to  the same $1/e$ time scale, 
$t_e$,   the various curves show that  the contribution can be up to ten times larger than for an exponential form.     
The large delays of long-tailed PBFs are quantified with the ratio  $R_t = \tpbfave \rangle / \tau$
of the average scattering time $\langle t \rangle$ 
to the $1/e$ time $\tau_e$, as  shown in Fig.~\ref{fig:Rt_and_Rsigt} vs. $\zeta = \linner / \ld$. 
This ratio is $R_t \sim 8$ for a small inner scale $\zeta = 0.01$ and declines monotonically to $R_t = 1$, the value for an exponential PBF, as $\zeta$ increases.  Also shown is the ratio $\sigma_{t_p} / \tpbfave$,
which measures the spread in delays relative to the mean; it too is much larger for small $\zeta$.      
For shallower density spectra with $2 < \beta < 11/3$,  these ratios increase and  tend to unity as
$\beta \to 4$. 

As shown above, if the diffraction scale $\ld$ is comparable to or smaller than the inner scale,  the PBF tends toward the monoscale case with only an exponential decay.  \newtext{At $\sim1$ GHz, typical diffraction scales are $\ld \sim 10^4$ km, whereas at $\nu\lesssim100$ MHz they are $\ld \sim 10^3$ km \citep[e.g.][]{2025ApJ...986..191G}.} Empirical determinations of the shape of the PBF or the scaling of the scattering time with frequency   yield  constraints on the inner scale of order 70 to 800\,km \citep[e.g.][]{bcc03, 2009MNRAS.395.1391R}. PBFs at  high radio frequencies  therefore will  show a long power-law tail while at  low frequencies the PBF trend  to the exponential. \newtext{Given that most PTA pulsars are observed near 1 GHz, they largely fall within the $\zeta < 1$ (non-exponential PBF) regime. }  

%

\begin{figure}[t!] 
   \centering
   \includegraphics[width=0.45\linewidth]{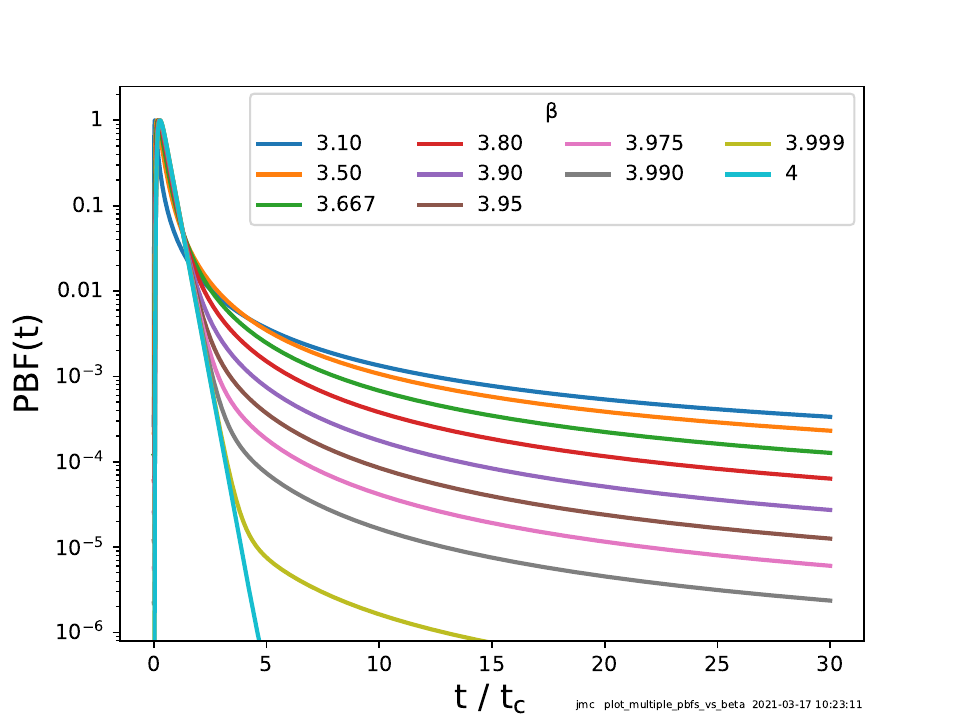}
   \includegraphics[width=0.45\linewidth]{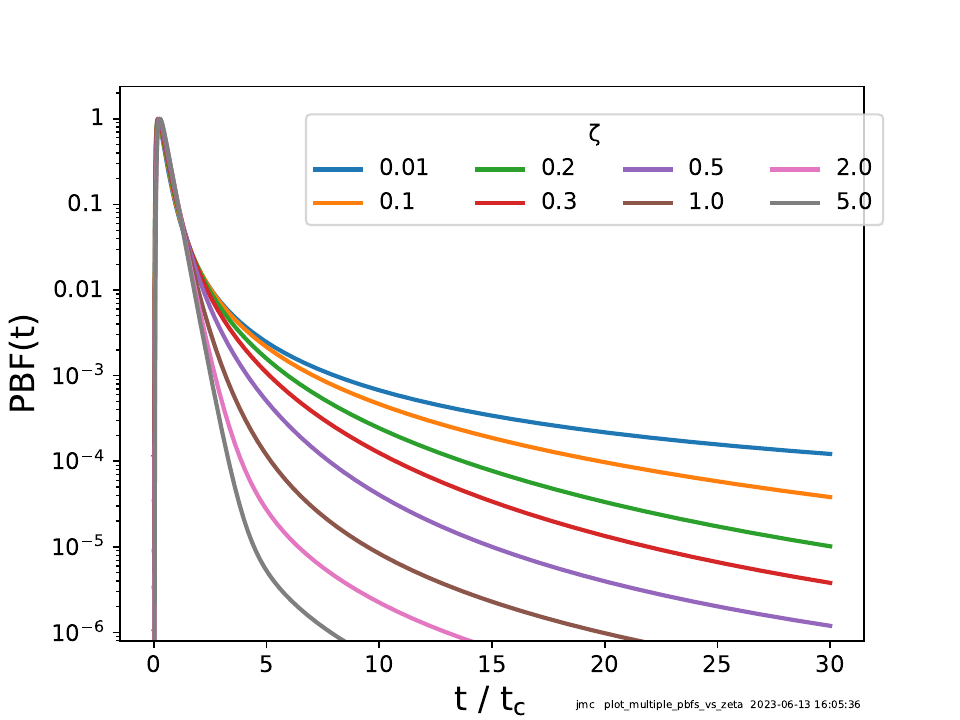}
   \caption{
   PBFs for thick media with power-law wavenumber spectra as a function of $\beta$ for negligible inner scale (left)
   and as a function of  dimensionless inner scale $\zeta = \linner / \ld$  for $\beta = 11/3$ (right).  } 
   \label{fig:KlikeThickMedia}
\end{figure}

As with square-law media, the PBFs for thick Kolmogorov-like  media have finite rise times related to the fact that radiation is  scattered multiple times along each propagation path.   The asymptotic decay of the PBF $\propto t^{-\beta/2}$  is similar to those of  thin screen cases.  Fig.\,\ref{fig:KlikeThickMedia} shows PBFs for different $\beta$ and for different values of inner scale parametrized by $\zeta = l_{\rm i} / \ld$.    The PBFs show a finite rise in contrast to the sharp rises seen from thin screen cases.   Thick media require numerical integration similar to the approaches in  
\citet[][]{1974MNRAS.166..499W,1975ApJ...201..532L,1978RQE...20..581O,ishimaru79} 
and \citet[][]{lr99}; \newtext{our implementation replicates these approaches and will be published elsewhere.}

 
\subsection{Anisotropic scattering and refraction}
\label{sec:aniso_refract}

Long tailed PBFs arise, as we have shown, from screens or extended media with isotropic density fluctuations and small inner scales that produce circularly symmetric scattered images of point sources.   Anisotropic media produce images with elliptical contours and long-tailed PBFs.   Refraction that distorts would-be circular images also yield long-tailed PBFs, as we show here.      These and previous cases indicate, therefore, that long-tailed PBFs are to be expected generally. 

{\it Asymmetric scattered images:}
For some lines of sight, scattering appears to be caused by anisotropic density irregularities  as evidenced in the elliptical shapes of scattered images determined from VLBI and from the properties of pulsar scintillations \newtext{\citep{frail94,2010ApJ...708..232B,bower14,sprenger26}}.   An alternative cause is refractive distortion of scattered images that would otherwise have circular contours.   These two origins for anisotropies result from very different length scales of density variations in the medium, small scales responsible for diffraction and large scales that underly refraction.

Anisotropies are most prominent for  nearby pulsars or for distant pulsars that encounter a local region of enhanced scattering, such as a supernova shock or HII region \newtext{\citep{2010ApJ...708..232B,Stinebring22,mall2022,ocker24,reardon25,sprenger26}}. It is justified for these cases to consider a thin screen and computing  PBFs for  anisotropic screens  involves a simple transform of the PBFs  for  isotropic cases.  

\begin{figure}[t!] 
   \centering
   \includegraphics[width=0.45\linewidth]{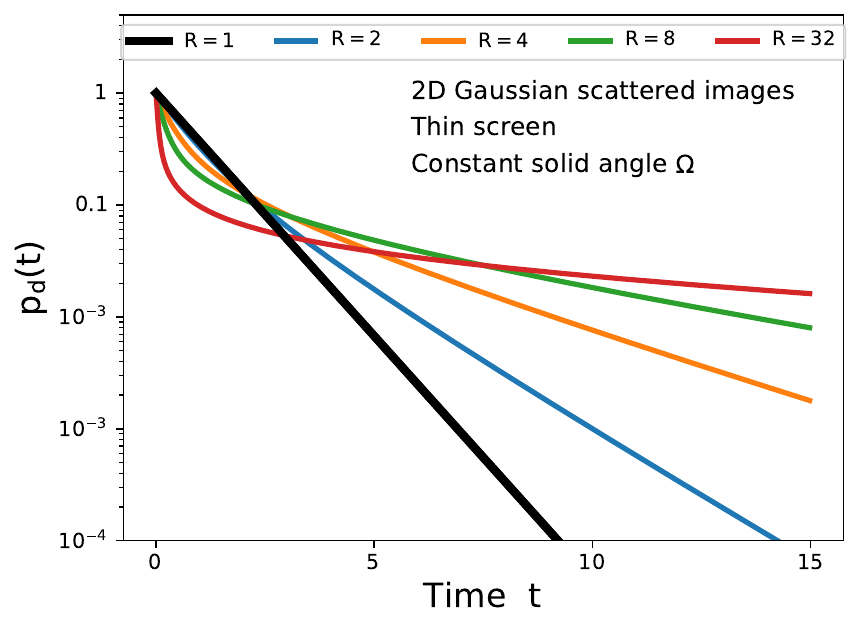}
    \includegraphics[width=0.475\linewidth]{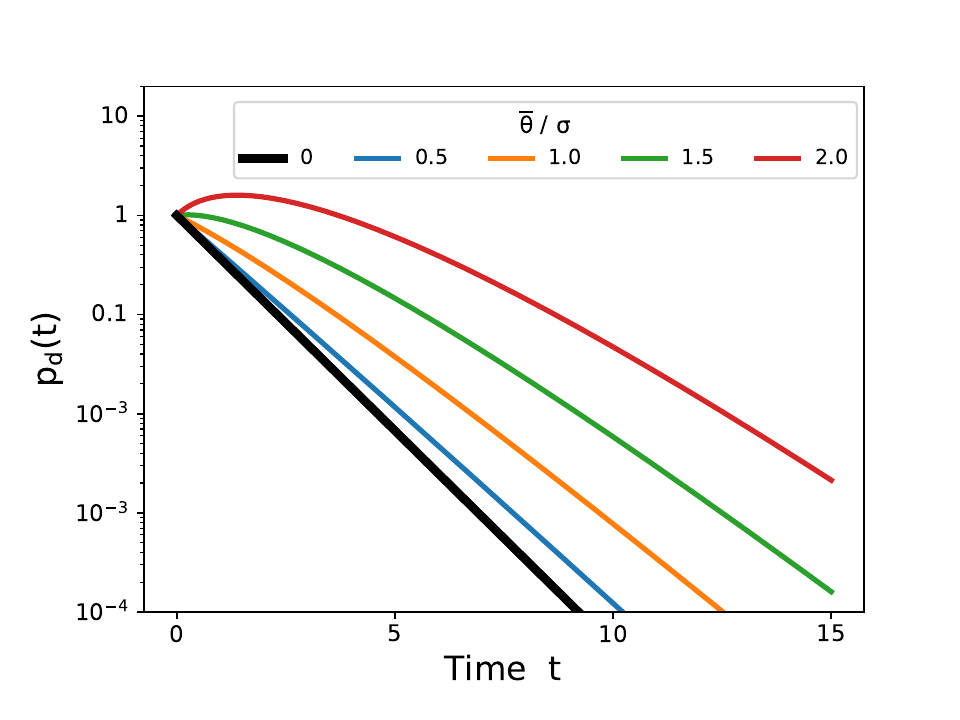}
   \caption{ Left: PBFs for scattered images with different axial ratios, $R$, but calculated from the PBF for a circularly symmetric image, as described in the text. 
   All PBFs are normalized to unit amplitude at $t=0$.
   The time axis is defined so that $t=1$ for the $e^{-1}$ value of the $R=1$ case. 
   \label{fig:2dGaussPDFs_loglin}
   Right: PBFs for a symmetric Gaussian image offset from the direct line of sight by an angle
   $\overline{\theta}$ in units of $\sigma$, the $e^{-1/2}$ width of the Gaussian image. 
    All PBFs are normalized to unit amplitude at $t=0$.
   The time axis is defined so that $t=1$ for the $e^{-1}$ value of the $\overline{\theta} = 0$ case. 
   \label{fig:Gauss_with_offset_PDFs_loglin}
   }
\end{figure}

\newtext{A screen that causes anisotropic scattering  yields a scattered image
that is not circularly symmetric.  Taking this into account when evaluating the delta function in Eq.~\eqref{eq:im2pbf}, we obtain the resulting PBF, }
\be
\pism(t) =  \frac{c}{\deff} \int_0^{2\pi}  d\phi\, I(\sqrt{2ct/\deff} \cos\phi, \sqrt{2ct/\deff} \sin\phi) .
\label{eq:pbf_anisotropic}
\ee
A scattered image with elliptical contours can be written in terms of a symmetric image $I^{\rm iso}$ by replacing
$(\theta_x, \theta_y)$ with $(R^{-1/2}  \theta_x, R^{1/2} \theta_y )$ using an axial ratio $R > 1$ to obtain
$I(\theta_x, \theta_y) =  I^{\rm (iso)}(\theta_x/R^{1/2} , \theta_y R^{1/2})$.
This choice aligns the contours with the coordinate axes, with no loss of generality for the resulting PBF. 
The  PBF in turn becomes an integral over azimuthal angle of the PBF from isotropic  irregularities, 
\be
\pism(t) =  
	\int_0^{2\pi} d\phi\, \pism^{\rm (iso)}\left[ \sqrt{2ct/\deff} \left( R^{-1}\cos^2\phi + R\sin^2\phi\right)\right] .
\label{eq:pbftransform}
\ee
The same transformation applies, in principle, to PBFs from extended media but requires the unphysical assumption that irregularities in the medium have the same axial ratio and orientation along the entire 
\LOS. 

{\it Example of  elliptical Gaussian scattered  images:}  Applying Eq.~\eqref{eq:pbftransform} to the exponential PBF from a circular Gaussian image, we obtain 
\be
\pism(t; R) = \tau_1^{-1}  e^{- [(R^2 + 1) / 2R] (t / \tau_1)} I_0[( (R^2-1) / 2R) (t / \tau_1)],
\label{eq:pismR}
\ee
where $I_0(x)$ is the modified Bessel function and $\tau_1$ is the mean scattering time for $R=1$  (the exponential PBF case).  Note that the PBF is unchanged under a substitution  $R \to R^{-1}$. 
Fig.\,\ref{fig:2dGaussPDFs_loglin}  (left panel) shows examples for five axial ratios and where the image size has been held fixed along one of the axes.    In these cases, the  solid angle of the image scales as $R^{-1}$, causing the PBF to become more concentrated at the origin.   In the limit of 
$R \gg 1$,  the PBF has a characteristic width $\tau = R\tau_1$ and shape, 
$\pism(t) = (\pi \tau t)^{-1/2} e^{-t / \tau}$, 
that differs significantly from the purely exponential form for $R=1$. 
For these PBFs, the ratio of the \RMS\ to mean  time delay is  \be
\frac{\sigma_t}{\langle t \rangle} = \frac{\sqrt{2} (R^2 + R^{-2})^{1/2} }  { R + R^{-1}},
\ee
is unity for a symmetric Gaussian image ($R=1$) but increases to $\sqrt{2}$ for $R \gg 1$ or $R \ll 1$ (by inspection).

{\it PBFs from anisotropies and image offsets:}   Generally, scattered images are asymmetric in shape and offset from the direct line of sight due to refraction.    Asymmetries can arise from asymmetric diffraction combined with refraction, which causes both convergence and shear of the image (described below) or they can be caused be refraction alone or diffraction alone.   

A screen with only a phase gradient  shifts the scattered image and increases propagation delays but does not alter the surface brightness.   For a shift $\thetavecbar = (\thetabar_x, \thetabar_y)$, the PBF is given by Eq.~\eqref{eq:pbf_anisotropic} with the arguments of the image
replaced by $\sqrt{2ct/\deff} \cos\phi - \thetabar_x$ and $\sqrt{2ct/\deff} \sin\phi - \thetabar_y)$. 
A  shifted circularly-symmetric Gaussian image yields, with $\tau \equiv \deff\sigma^2 / c$,
\be
\PBF(t) =
e^{-\thetabar^2 / 2\sigma^2}
I_0(\thetabar \sqrt{t / \tau} /\sigma )
\times
\left[\tau^{-1} e^{-t / \tau} \Theta(t) \right],
\ee
where 
$\thetabar^2 = \thetabar_x^2 + \thetabar_y^2$, 
$I_0$ is the modified Bessel function of the first kind and the square brackets designate the PBF for a centered image ($\thetavecbar = 0$).     Examples are shown in  Fig.~\ref{fig:Gauss_with_offset_PDFs_loglin} (right panel), which demonstrate that  offsets make the PBF wider and non-monotonic.

Next we show results where asymmetries and offsets are both caused by refraction from screens with 
power-law density wavenumber spectra.    For these cases we simulate only refraction from the screen and let it modify  symmetric Gaussian images.   For weak refraction that does not induce any ray crossing, we use a geometric optics approach based on the Fresnel-Kirchhoff diffraction integral.  For the geometry shown in 
Fig.~\ref{fig:geometry}, a given observer position $\xobs$ yields a stationary phase position $\xvec_{\rm l} = \xbarvec$ in the screen.   Expanding the screen phase $\phi_r(\xvec)$ to second order about this point gives
\be
\phir(\xvec) \sim \phir(\xbarvec) +  \bvec \cdot (\xvec-\xbarvec)  + 
	\frac{1}{2} (\xvec-\xbarvec) \cdot  \Carray  \cdot (\xvec-\xbarvec),
\ee
where $\bvec \equiv \grad_{\xvec} \phir(\xbarvec)$  yields the refraction angle (as viewed by the observer), 
$\thetavecbar = -k^{-1} (\dsl/\dso) \bvec$.
The 2$\times$2 matrix  $\Carray$ has diagonal elements
$C_x = \partial^2_x \phir(\xbarvec)$ and $C_y = \partial^2_y \phir(\xbarvec)$  and  off-diagonal elements 
$C_{xy}  = C_{yx} = \partial^2_{xy} \phir(\xbarvec)$, where derivatives are in screen-plane coordinates.   
 A unitary matrix $\Uarray$ transforms $\Carray$ into the diagonal form  $\Carray^{\prime} = \Uarray \Carray \Uarray^{\dagger}$ (where $\dagger$ indicates matrix transpose) with elements 
$C^{\prime} _{x}, C^{\prime}_{y}$.  
The rows of $\Uarray$ are $(\cos\chi, \sin\chi)$ and $(-\sin\chi, \cos\chi)$
where $\tan2\chi = 2 C_{xy} / (C_{x} - C_{y})$ and the angle $\chi$ is the position angle of the distortion. 
As with polarization position angles, the domain of $\chi$ is over $\pi$ radians. 
We define  the diagonal matrix $\gamarray$ with elements $G_x$ and $G_y$ given by  
$G_{x,y} = (1 +  k^{-1} \deff C^{\prime}_{x,y})^{-1}$ where $\deff = \dsl\dlo/\dso$.  

The magnification  of the image is $\mu = \vert \det \Aarray \vert^{-1} = [(1-\kappa)^2 - \gamma^2]^{-1}$,
where 
\be
\Aarray 
=
\left(
	\begin{array}{cc}
	1 - \kappa - \gamma_1  	& -\gamma_2 \\
	\\
	 -\gamma_2 			&  1 - \kappa + \gamma_1
	\end{array}
\right)
\ee
is written in terms of the `convergence'  $\kappa$ and `shear'  $\gamma_{1,2}$ of the refraction.
Defining  $\gamma = \sqrt{\gamma_1^2 + \gamma_2^2}$ in terms of 
 $\gamma_1 = \gamma \cos2\chi$ and $\gamma_2 = \gamma\sin2\chi$, these are
\be
\kappa 	&=& \rF^2 (C_x + C_y)/2 \\
\gamma 	&=& \rF^2 \left[\frac{1}{4} (C_x -C_y)^2 + C_{xy}^2 \right]^{1/2} \\
\ee

If $I_{\rm d}(\thetavec)$ is the scattered (diffracted) image in the absence of refraction,  the refracted image is
\be
I_{\rm dr} (\thetavec) = I_{\rm d}[\Uarray^{\dagger} \gamarray^{-1} \Uarray (\thetavec - \thetavecbar)],
\ee
where $\thetavecbar = -k^{-1} (\dsl/\dso) \bvec$.   Refraction renders the scattered image asymmetric and off center regardless of the shape of the unrefracted (but still scattered) image.   Finally, the PBF follows by substitution of
$I_{\rm dr} (\thetavec)$ into Eq.~\eqref{eq:pbf_anisotropic}.

\begin{figure}[htbp] 
   \centering
   \includegraphics[width=\figwidth]{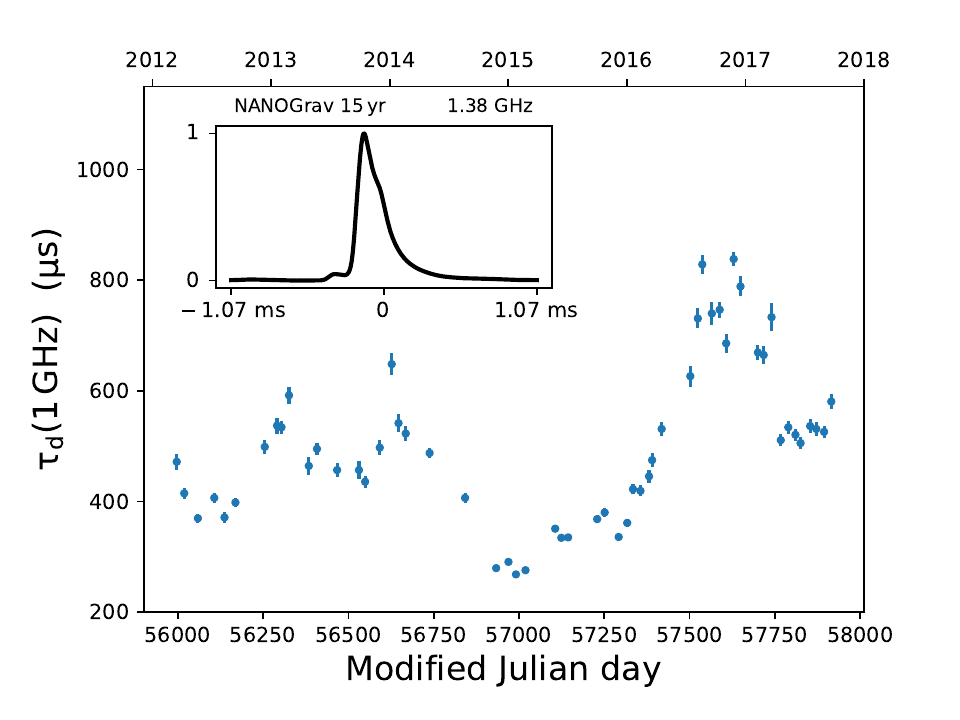} 
   \caption{Scattered pulse shape and variations of the scattering time vs. epoch for the MSP J1903+0327.
   Scattering times were calculated  from pulses in  the NANOGrav
   12.5-yr data set like that shown in the inset \citep[][]{2025ApJ...986..191G}.
   }
   \label{fig:J1903taud_vs_mjd}
\end{figure}

Fig.\,\ref{fig:J1903taud_vs_mjd} shows the $\sim 30$\% variability of the scattering time for the MSP J1903+0327,
with DM = 297\,$\DMunits$ 
\citep[][]{2025ApJ...986..191G}.    The variability time scale $\sim 150$\,d is consistent with the
\DM\ variability and with the  estimated refraction time scale of $\sim 150$\,d. 
 The  similar  $\sim 25$\% variability of $\taud$ for the MSP J1643-1224 with \DM = 62.4\,$\DMunits$
\citep[][]
{2024MNRAS.535.1184S} with $\sim 200$\,d time scale at 0.4\,GHz  is also consistent with the estimated RISS time scale and that for DM variations. \newtext{Measuring scattering time variability for more PTA pulsars will require inclusion of higher time and frequency resolutions in PTA observing configurations, which will become routinely possible with instruments like the cyclic spectroscopy backend on the GBT.}


Figures\,\ref{fig:screenDM10}--\ref{fig:screenDM300a} show the effects of refracting screens on $\delta$DM, intensities, images, and pulse broadening for $\DM = 10$, and 300 $\DMunits$ where the strength of the refraction is consistent with the pulse broadening specified for each DM and assuming Kolmogorov density fluctuations with  negligible inner scale.  [Details of the simulations are available on request.]
The four panels in each of these figures are:
\begin{enumerate}[label=(\alph*), noitemsep]
\item
Slices through the screen for $\delta \DM(t)$,  the scattering diameter and  refraction angle (the centroid of the scattered image),  a parameter $\rho_I$ that is related to the orientation of the refracted image, and the refractive gain modulation $G$ of the flux density and the axial ratio of the refracted image.   The abscissa is in units of the Fresnel scale. 
\item
Wandering of the image centroid in the image plane showing the size of the scattering disk combined with the ensemble average image wandering (gray circle). 
\item
Delays vs. position along the observation plane from diffraction $t_{\rm d}$  (red) and
refraction $t_{\rm r}$ (blue) and their sum (black). 
\item
Images and pulse broadening functions at eight positions along the screen separated by multiple refraction length scales $\lr$.   The pulse broadening time is indicated in each of the PBF frames.
\end{enumerate}

Trends that can be seen with increasing DM are:
\begin{enumerate}[noitemsep]
\item Scattering observables (the scattering disk size $\thetaFWHM$ and the width of the PBF; cf.\ Fig.~\ref{fig:hockeystick_sm}) increase with DM, as expected. 
\item Refraction quantities, including angular wandering and refractive \TOA\ delays, {\it decrease} with increasing DM if refraction is from a wavenumber spectrum consistent with that producing the diffraction.    However this trend can be reversed if there is enhanced power on refraction scales $\gtrsim 1$\, au. 
\item Intensity variations diminish with increasing DM, with $G = G_x G_y$ more closely clustered  around unity.
\item Images become more circularly symmetric (axial ratio $\to 1$). 
\end{enumerate}
\newtext{
These trends are consistent with the large body of scattering measurements cited in Sect.~\ref{sec:ism_scalings}.  In particular, refractive scintillation from a Kolmogorov spectrum is consistent with some of the monitoring observations reported in e.g.\ \citet{1990ApJ...352..207S,1992ApJ...392..530K} for slow pulsars and for MSPS \cite[][]{2023MNRAS.526.3370G}.  These studies also demonstrate that the mean {\it emitted} flux densities of pulsars appear to be nearly constant.   Other cases indicate that discrete structures exist in the ISM that cause enhanced refractive scintillations \citep[e.g.][]{2023MNRAS.520.1311W}.    
}

\newcommand{\fileSa}{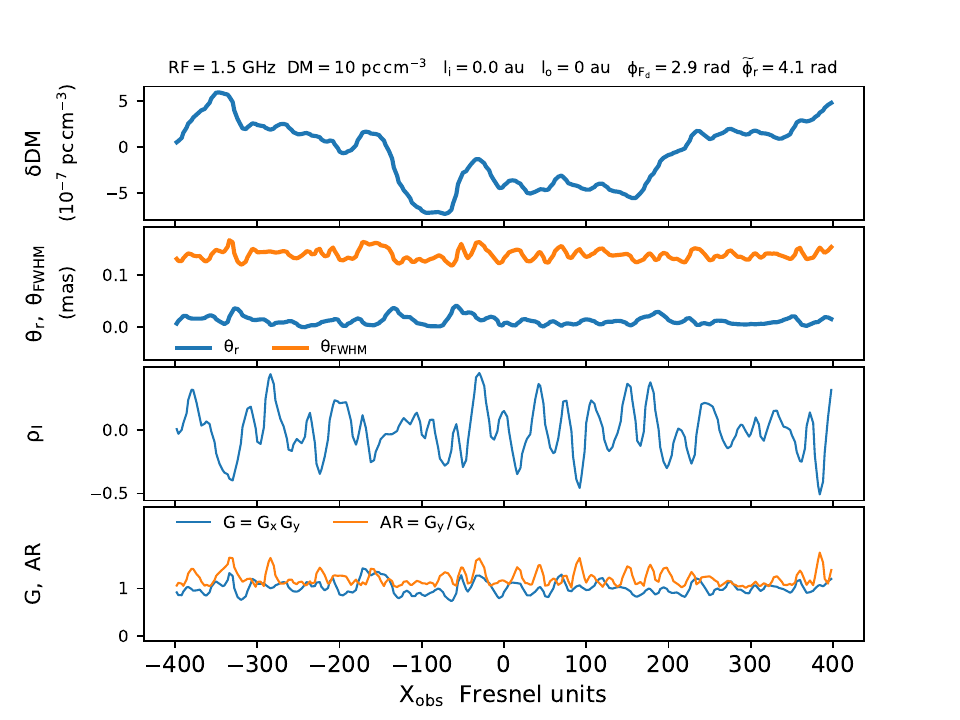}
\newcommand{\fileSb}{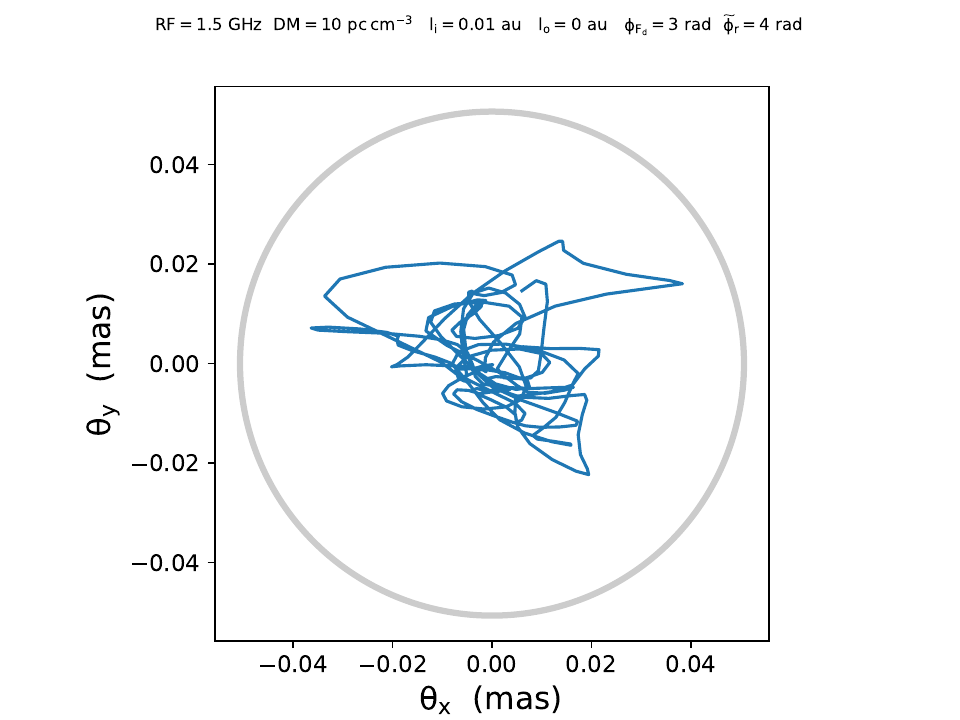}
\newcommand{\fileSc}{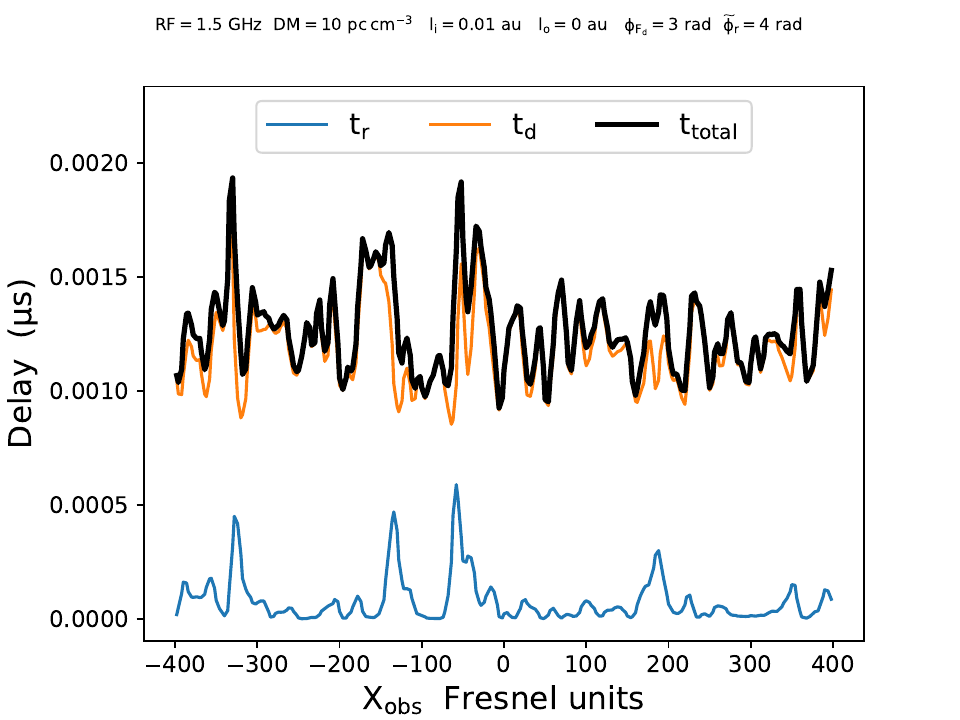}
\newcommand{\fileSd}{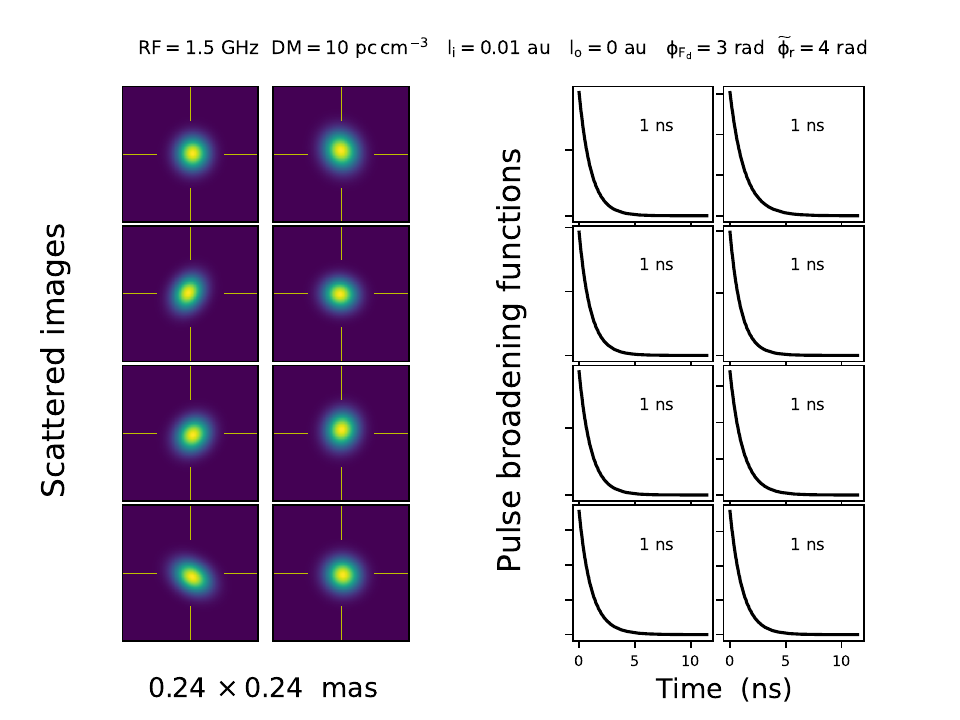}

\begin{figure}[h!] 
   \centering
   \gridline{
   	\fig{\fileSa} {0.45\textwidth}{(a) 1D slices across the phase screen.}
	\fig{\fileSb} {0.45\textwidth}{(b)  Image wandering relative to the direct propagation path.}
	}	
   \gridline{
   	\fig{\fileSc} {0.45\textwidth}{(c)  Arrival time delays (diffractive and refractive).}	
	\fig{\fileSd} {0.45\textwidth}{(d)  Sample images and PBFs at eight positions along the phase screen.} 
	}
   \caption{The effects of refraction on  the scattered  image, pulse broadening function, and \TOA\ delays at 1.5 GHz for a \LOS\ with dispersion measure $\DM = 10\,\DMunits$.  Details are discussed in the text.}
    \label{fig:screenDM10}
\end{figure}

\newcommand{\fileIIIa}{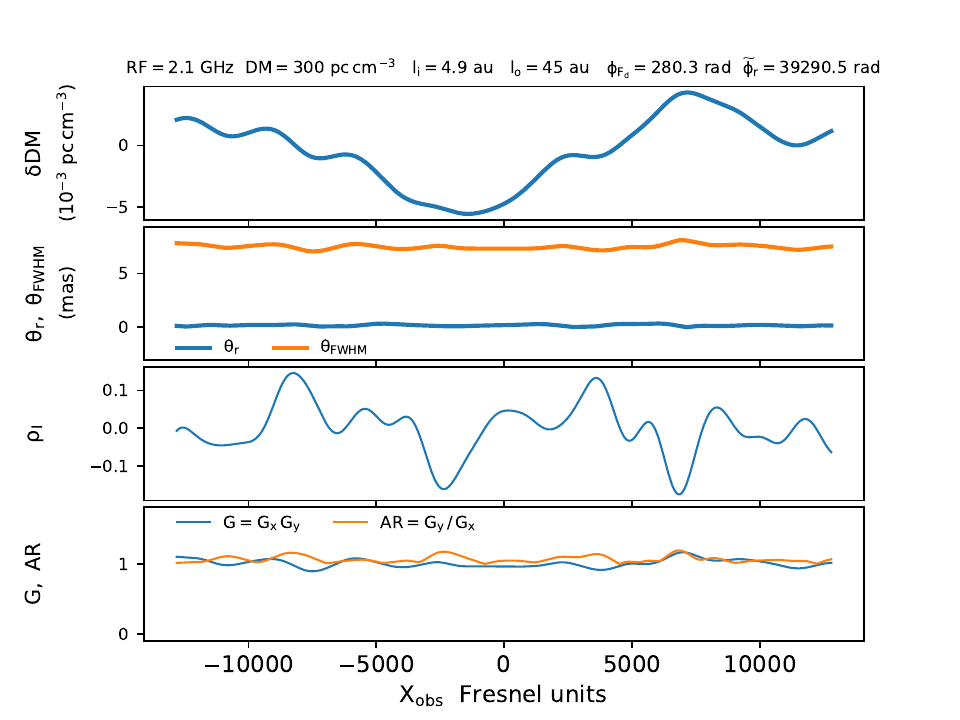}
\newcommand{\fileIIIb}{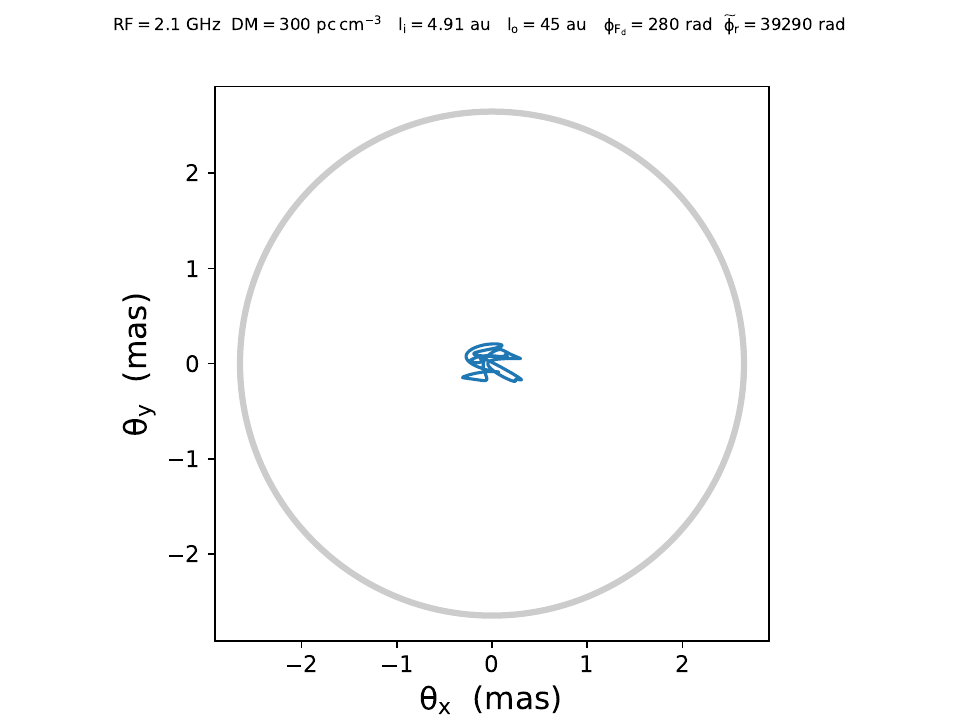}
\newcommand{\fileIIIc}{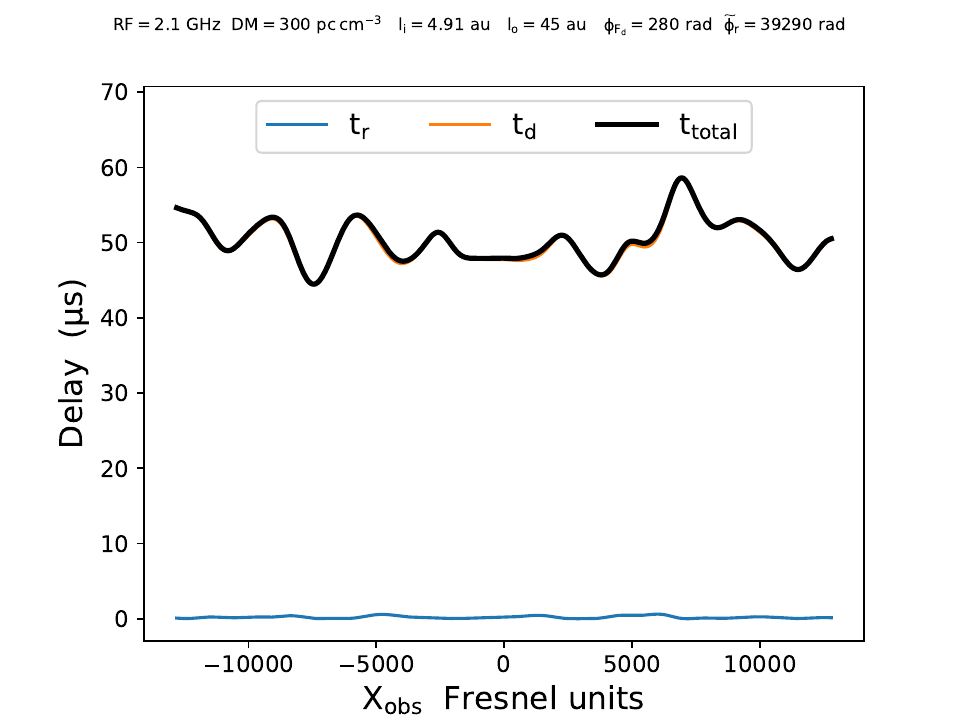}
\newcommand{\fileIIId}{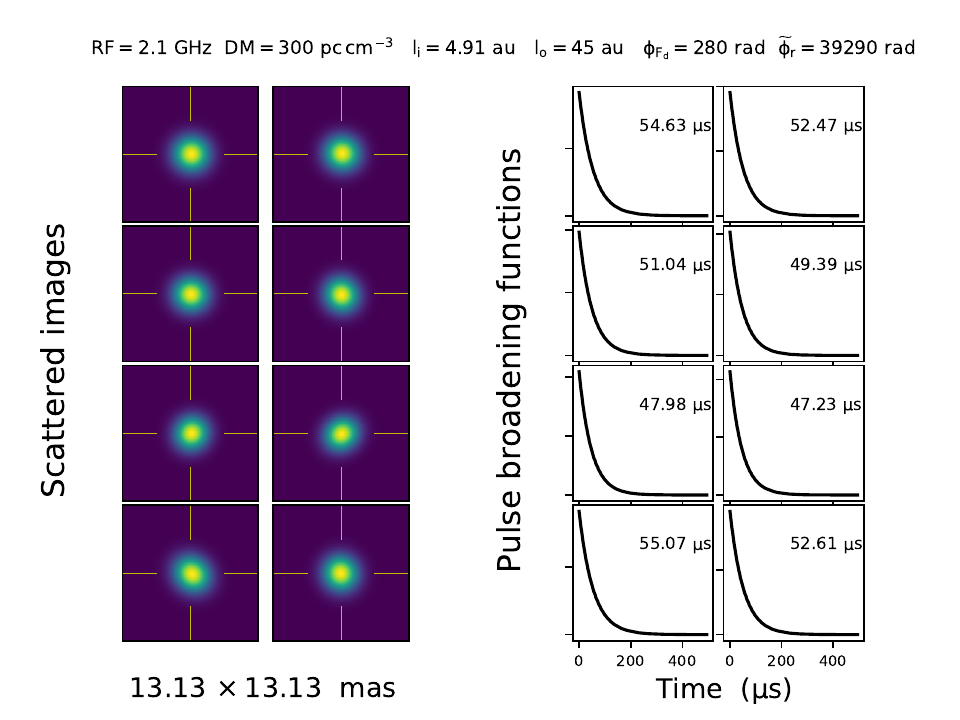}


\begin{figure}[h!] 
   \centering
   \gridline{
   	\fig{\fileIIIa} {0.45\textwidth}{(a) 1D slices across the phase screen.}
	\fig{\fileIIIb} {0.45\textwidth}{(b)  Image wandering relative to the direct propagation path.}
	}	
   \gridline{
   	\fig{\fileIIIc} {0.45\textwidth}{(c)  Arrival time delays (diffractive and refractive).}	
	\fig{\fileIIId} {0.45\textwidth}{(d)  Sample images and PBFs at eight positions along the phase screen.} 
	}
\caption{
	The effects of refraction on  the scattered  image, pulse broadening function, and \TOA\ delays at 2.1 GHz for a \LOS\ with dispersion measure $\DM = 300\,\DMunits$.  Details are discussed in the text.}
    \label{fig:screenDM300a}
\end{figure}

\subsection{Inhomogenous and truncated screens}
\label{sec:truncated}

The `textbook' PBFs often considered in pulsar scattering are based on the (usually) tacit assumption that the strength of scattering (e.g. $\cnsq$) is independent  of position in the screen, leading  to `standard' scaling laws vs. frequency  for scattering and scintillation parameters.    Departures from homogeneity  cause the PBF and scaling laws to differ substantially  \citep{2001ApJ...549..997C}.    In particular, if angles of arrival are bounded by the spatial extent of a scattering region, in contrast to the unbounded \AOAs\ for an infinite screen,   the angular size and pulse broadening time can be only weakly dependent on frequency in some frequency ranges, as demonstrated in \citet{2001ApJ...549..997C} \newtext{(see also \citealt{10.1093/mnras/stad429})}. 

The net frequency dependences depend on the details of the scattering region and will not be discussed further here.  The implication  for arrival time analyses is that  correction of \TOAs\ for scattering delays will likely be imperfect unless it can be demonstrated that scattering regions are well characterized with respect to their spatial distribution as well as for the underlying wavenumber spectrum. 

\clearpage

\subsection{Bandwidth averaging and phase wrapping}
\label{sec:bw_and_phasewrap}

The PBFs applicable to isolated giant pulses from pulsars or one-off FRBs are as described so far. 
For periodic sources, however, the effective PBF is  `phase wrapped,' meaning that the PBF acting on previous pulses affects the shape of any given pulse; in averages of $N$ pulses, the net pulse shape is the wrapped, intrinsic pulse shape convolved with a phase-wrapped PBF.   Inferences from pulsars about the ISM based on the shape of the pulse need to take wrapping into account.   All cases may  furthermore be affected by bandwidth averaging.  The strong frequency dependence of the scattering time,  $\taud \propto \nu^{-\xtau}$ with $\xtau = 2\beta / (\beta-2)\simeq 4.4$ 
implies that averaging of pulses over frequency after dedispersion (e.g. to increase the S/N) 
also alters the net PBF shape.   

\begin{figure}[t!] 
   \centering
   \includegraphics[width=0.7\linewidth]{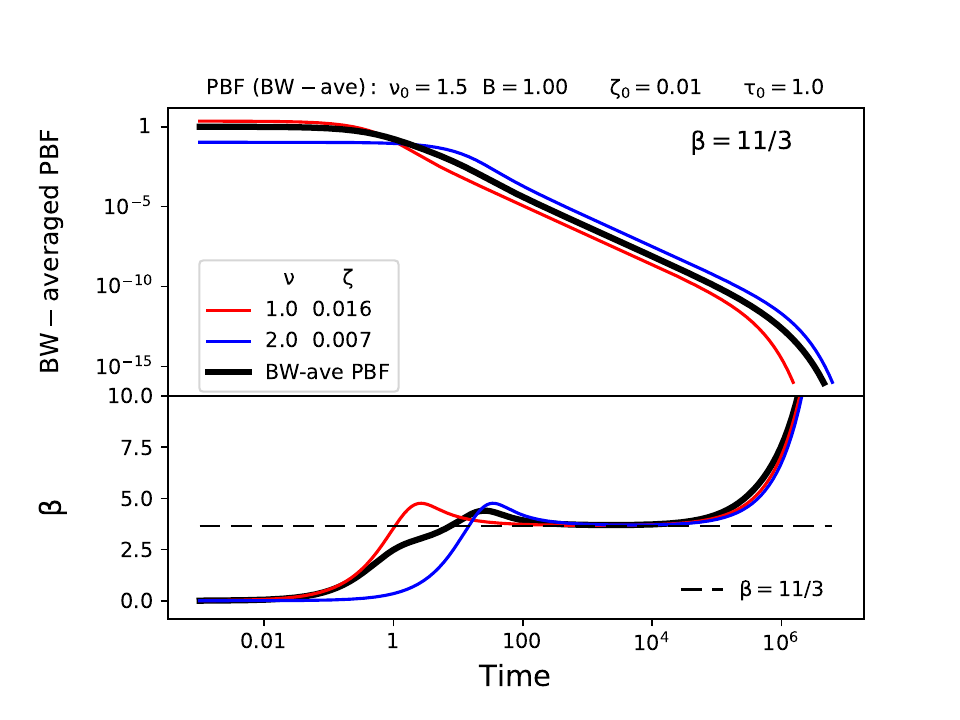}
   \caption{Bandwidth averaging of a power-law PBF for $\beta = 11/3$ for $\zeta_0 = 0.01$ and $\tau_0 = 1$ at  the center frequency $\nu_0 = 1.5$.
   The top panel shows the  PBFs and the bottom panel shows values of  $\beta$ inferred from the bandwidth averaged and individual PBFs at the band edges
   using $\beta = -2 d(\ln\PBF) / d\ln \nu$. 
   }
   \label{fig:pbfPL_bwave}
\end{figure}

Bandwidth averaging of a PL PBF combines PBFs with different $1/e$ scattering times $\taud$ and different power-law spans $\propto t^{-\beta/2}$ before rolling off exponentially at a frequency-dependent time related to the inner scale.   The net shape differs from that applicable to any single frequency and depends on the pulsar's radiation spectrum.   Fig.~\ref{fig:pbfPL_bwave} shows the effects of bandwidth averaging over an octave frequency range.   The top panel shows the PBFs at the lowest and highest frequencies along with the average PBF, where a pulsar spectrum $\propto \nu^{-1.5}$ was assumed.   The bottom panel shows 
$\widehat\beta = (-1/2) d(\ln \PBF)/ d\ln \nu$, which equals the true value, $\beta = 11/3$, over time ranges far beyond the $1/e$ scattering times.  The dependence of the PBF shape on the extent of  bandwidth averaging 
certainly factors into any determinations of $\beta$ from pulse-shape fitting and into arrival time estimation.

\begin{figure}[t!] 
   \centering
   \includegraphics[width=0.7\linewidth]{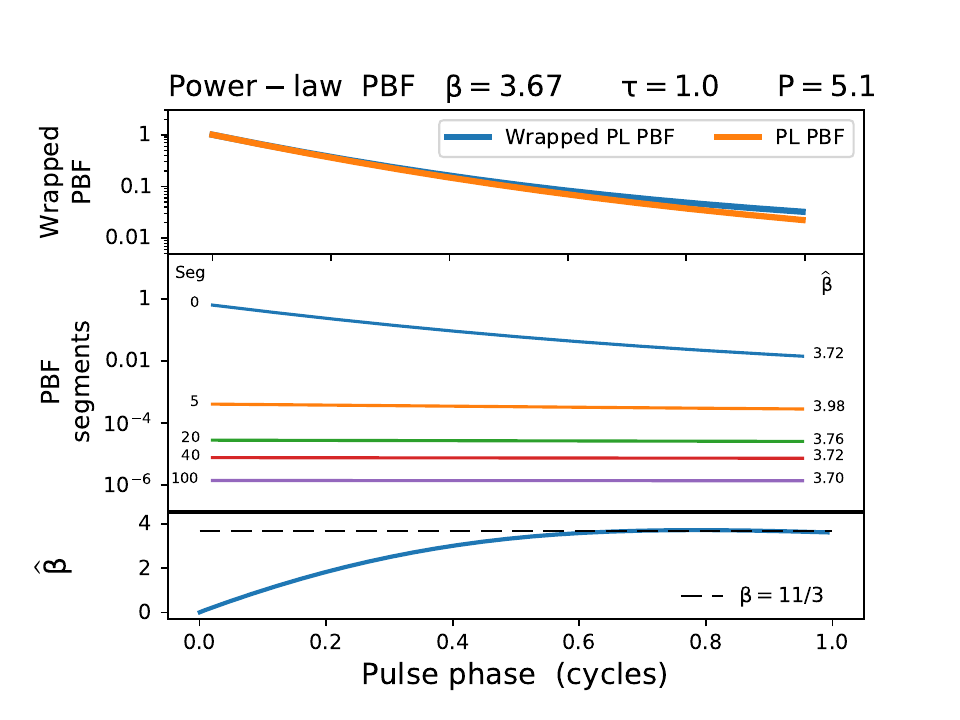}
   \caption{Phase wrapping of a power-law PBF for $\beta = 11/3$, $\tau = 1$, and $P = 5.1$ (in arbitrary time units). 
   (Top panel)  initial and wrapped PBFs. 
   (Middle panel)     segments of length $P$ that are wrapped into one cycle of pulse phase.  The segment
   number is given to the left of each line and the inferred $\widehat\beta = -2d\ln \PBF / d\ln t$ at the midpoint of each segment.    
   (Bottom panel) values of  $\beta$ inferred from the wrapped PBF. 
   }
   \label{fig:pbfPL_wrap}
\end{figure}

Generally, phase wrapping makes heavy-tailed PL PBFs even heavier  (shallower).   The long PBF tails for media with
small inner scales  (small $\zeta$)  can extend over a large number of pulse periods, adding shallower segments to the initial exponential falloff in an average profile comprising many pulses.    Figure~\ref{fig:pbfPL_wrap} shows an example of phase wrapping for $\tau=1$ and a period $P = 5.1$ (arbitrary units, but could be considered ms).   The top panel shows the original and wrapped PBF plotted against pulse phase, while contributions from individual segments from pulse numbers 0 through 100 are shown in the middle panel with inferred $\beta$ values 
given to the right of each trace.   The bottom panel shows the inferred $\beta$ from the apparent power-law portion of the wrapped pulse.

\section{Timing variations from chromatic plasma effects}
\label{sec:chromatic_toas}

 \newcommand{\phipulse}{\phi_{\rm pulse}}
 \newcommand{\dphipulse}{\delta\phi_{\rm pulse}}

\newcommand{\nuspin}{f_{\rm spin}}
\newcommand{\dnuspin}{\delta f_{\rm spin}}
\newcommand{\nupulse}{f_{\rm pulse}}
\newcommand{\dnupulse}{\delta f_{\rm pulse}}

A plethora of systematic and random delays caused by dispersion and scattering contribute to arrival times.  While dispersion and scattering are  are distinct processes, their mitigation is intertwined in several ways, as we discuss in this section. 

A primary chromatic effect on \TOAs\ is simply the change in S/N from intensity scintillations, which can cause large excursions from the mean flux density of a pulsar and affect the matched filtering \TOA\ error discussed earlier. 

The modulation index (fractional intensity variation) for DISS is approximately $\mI \simeq \Nscint^{-1/2}$ where $\Nscint$ is the number of scintles within the time span and bandwidth of an observation.   Low-DM pulsars observed at $\gtrsim 1$\,GHz will yield  $\Nscint \sim 1$\,to\,10, so changes in flux density and $\DtSNR$ are large.    Low frequency observations of higher-DM pulsars will show a much smaller range of flux densities. 

Dispersion measure variations result from the changing direction and distance to the pulsar combined with density fluctuations in the ISM. The solar corona and Earth's ionosphere  contribute  less but still need to be considered in the highest precision timing.
\citep[e.g.][and references therein]{2021A&A...647A..84T}.
Fortunately,  most of the resulting arrival time fluctuations  can be removed by careful estimation of DM at each timing epoch.  However, such removal is based on imperfect  arrival times used to estimate DM, which are  themselves imperfect due to a variety of effects.  These include the achromatic errors already discussed along with pulse shape distortions  from the stochasticity of the PBF, from RFI and instrumental effects (polarization calibration and aliasing), and from the interplay of scintillations and the chromatic aspect of the true pulse shape. 
It is therefore useful to quantify the full extent of DM variations and then assess to what degree they can be removed. 

\subsection{Temporal DM variations $[\DM(t)]$}
\label{sec:DMt}

In this section we concentrate on the effects of DM variations on timing and defer to Section \ref{sec:ismdiagnostics} a discussion of ISM diagnostics and their utility in improving arrival times.   Of all the interstellar effects on timing,  the DM term $\delta t \propto \nu^{-2} \DM$ 
is  the largest.   The largely (but not completely, cf.\ Sect.~\ref{sec:DMnu}) deterministic variation of this term with frequency $\nu$ allows \DM\ to be estimated with multifrequency timing measurements.  One might be tempted to think that this term can be removed completely (to within  random \TOA\ measurement errors).  However, other chromatic effects yield systematic \TOA\ 
\newtext{offsets},
so there is always consequential leakage of DM fluctuations into timing 
\newtext{residuals (see Sect.~\ref{sec:chromatic_leakage})}. 

Figure~\ref{fig:DM_J1909-3744}  shows the DM time series  for PSR~J1909-3744.  It appears to be the superposition of a downward secular trend,  a stochastic ISM component with a red power spectrum, and a yearly oscillation due to  the cyclical intersection of the \LOS\ with the solar corona.  
This and similar time series have been analyzed by a number of authors
 \citep[e.g.][]{
 kcs+13,2015ApJ...801..130L, 
 2016ApJ...821...66L,2017ApJ...841..125J,
 2019ApJ...872..150M,
 2022ApJ...929...39H} 
 using structure function analyses of data from PTA programs on millisecond pulsars.   The motivation of those studies has been to characterize electron density fluctuations of the ISM and the solar corona, of interest in their own right, but  they also contribute to an understanding of the overall noise budget for pulsar timing and to ongoing efforts to improve timing precision.      

We focus on stochastic DM variations from the ISM that have
 red, power-law spectra.  
While early timing analyses did not remove these variations \newtext{\citep[e.g.][]{1983ApJS...53..169D}}, current practice aims to do so.  
The temporal spectrum  of DM is power-law in form for the   inertial range  of wavenumbers in the
stochastic density spectrum of Eq.~\eqref{eq:Pne},  
\be
S_{\DM}(f) = 
	\left[
		\frac{{\pi}^{1/2}}{(2\pi)^{\beta-3}} 
		\frac{\Gamma((\beta-1)/2)}{\Gamma(\beta/2)} 
	\right]
	\SMeff
	v_x ^{\beta-2} 
	f^{-(\beta-1)}.
\label{eq:SDM}
\ee
For Kolmogorov fluctuations ($\beta = 11/3$), the DM spectrum is
\be
S_{\DM}(f)  
	&=& 
	\frac{1}{2^{2/3}\pi^{1/6} }
	\frac{\Gamma(4/3)}{\Gamma(11/6)} 
	\frac{\SMeff}{v_x} 
	\left(\frac{f}{v_x}\right)^{-8/3}
	\nonumber \\
	&\simeq&
	\frac{(5.86 \times10^{-5} \,\DMunits)^2}{\rm cy~yr^{-1}}
	\left( \frac{\SM}{10^{-3.5}\, \SMunits} \right) v_{100}^{5/3}\, f^{-8/3}
	\nonumber \\
	&\simeq&
	\frac{(7.70\times 10^{-5}\,\DMunits)^2}{\rm cy~yr^{-1}}
	 \left( \frac{\tau_1 /  1\,\mus}{\deff / 1\, \rm kpc} \right)^{5/6}  
		v_{100}^{5/3} \, f^{-8/3}
\label{eq:SDMKol}
\ee
where frequencies are in cy~yr$^{-1}$, $v_x$ is in units of 100\, $\kms$; the first evaluation is in terms
of a scattering measure $\SM = 10^{-3.5}\,\SMunits$ and the second evaluation is expressed using  a scattering time
$\tau_1 = 1\,\mu s$ at 1 GHz and a 1\,kpc effective distance. 
The corresponding arrival time fluctuations have 
the same   spectral form,
 \be
S_{t_\DM}(f) 
	= (\kDM/\nu^2)^2 S_{\rm DM}(f)  
	\simeq  
	\frac{(0.319\,\mus)^2}{\rm cy~yr^{-1}}\, 
		 \left( \frac{\tau_1 /  1\,\mus}{\deff / 1\, \rm kpc} \right)^{5/6}  
		 v_{100}^{5/3} \,\nu^{-4}  f^{-8/3}
\label{eq:StDM}
\ee
where $\kDM = c\relec/2\pi \simeq 4149\,\mu s$ for standard DM units. 
Note that these expressions do not include a rolloff at high fluctuation frequencies due to scattering-cone averaging. Rolloff is discussed in Sect.~\ref{sec:red_rolloff}. 

\subsubsection{Pulse frequency variations}   

Though pulse phase and arrival times are the usual focus of timing analyses, 
the effects of $\DM(t)$ on pulse frequency can also be considered  \citep[][]{1984Natur.307..527A,2011PhRvD..83h1301J}.    
The variation in pulse phase,
$\dphipulse = \delta t / P$, corresponds to a change in pulse frequency
$\dnupulse = d\dphipulse/dt$ and a fractional change in pulse frequency
$y \equiv \dnupulse / \nupulse = (c \relec / 2\pi \nu^2) (d\DM/dt)$.   The power spectrum of $y$ is related to the DM spectrum by use of the Fourier derivative theorem,
\be
S_y(y) = (c \relec)^2 \nu^{-4}  f^2 S_{\DM}(f).
\ee
For the inertial subrange of a Kolmogorov spectrum this gives
\be
S_y(y) 
	=  
	\frac{\pi^{1/2} \Gamma(4/3)}  {(2\pi)^{2/3} \Gamma(11/6)} 
	\frac{(c\relec)^2}{\nu^4}   \SMeff \, v_x^{5/3} \, f^{-2/3}
	\simeq   \frac{(3.30 \times 10^{-10})^2}{\rm cy~yr^{-1}}
	\times  \SMeff \, v_x^{5/3}\,   f^{-2/3} .
\ee

\citet[][]{2011PhRvD..83h1301J} discuss the effects of  pulse-frequency variations on the detection of lightyear-wavelength gravitational waves.   To a large extent these variations are removed when 
\TOAs\ are referenced to infinite frequency using DM estimates at each epoch.     However, this procedure is not perfect, yielding residual dispersive delays and thus pulse-frequency fluctuations.   \citet[][]{2011PhRvD..83h1301J}  compare the residual spectrum with other cases where all but 0.1\% or 1\% of the fluctuations in $y$ are removed. 
 
\subsubsection{\TOA\ error  from asynchronous frequency sampling of DM delays}
\label{sec:dm_nonsimultaneous}   

One cause for imperfect removal of dispersion delays is the estimation of $\DM(t)$  from multifrequency \TOAs\  obtained at slightly different epochs \citep[][]{2015ApJ...801..130L}.

 \newtextR{First consider arrival times made at a single frequency (strictly, using a narrowband system at some center frequency $\nu$) that is  corrected for the dispersion delay using a DM value from an unspecified procedure.}
 If a timing precision of $\sigma_t\, \mu$s is needed at frequency $\nu$, the required precision for  DM is  $\delta \DM \lesssim \sigma_t \, \nu^2 / \KDM $
$\simeq 2.41\times 10^{-4}\, \DMunits \times \nu^{2} \sigma_t( {\rm \mu s} )$
\newtextR{with $\nu$ in GHz here and elsewhere in this section.}
The requirement is more stringent at lower frequencies. 
Inspection of Fig.\,\ref{fig:DM_J1909-3744} for J1909-3744  and of the DM time series for another 66 pulsars in 
\citet[][]{2023ApJ...951L...9A} and in papers for other PTA groups cited earlier indicates that DM 
\newtext{
estimation}
errors for the best pulsars satisfy this requirement but high DM pulsars at low frequencies may not, such as MSPs observed with LOFAR \citep[][]{2020A&A...644A.153D}. More widely spaced frequencies yield higher DM precision, all else being equal, but the measurements need to be made at closely spaced epochs.   That has been problematic for some observing programs because telescope logistics can require changes in instrumentation that span several days.   
\newtext{Widely spaced receiver bands used for instantaneous wide frequency  coverage at some observatories and  wideband receiver systems now being deployed with continuous coverage} over  $\gtrsim 3:1$ frequency ranges 
\newtextR{(i.e. highest to lowest frequencies in the receiver band)}
provide the needed simultaneous measurements and obviate the problem.  

Legacy data, however, retain the issues associated with non-simultaneous dual-frequency measurements. The consequences are easily estimated from the DM structure function.  
\citet[][]{2015ApJ...801..130L} analyzed the effects of  dual frequency measurements 
($\nuL, \nuH$)  made
 asynchronously at epochs $t_{1,2}$  spread by a day or more.  The \TOA\ at each frequency and epoch  is modeled as
$
 t_\nu = \tinfty + \nu^{-2} \KDM \DM(t)
 $
 where (for now) we ignore other contributions.  
 Estimating \DM\ as if it were constant yields
 \be
 \DMhat 
 = \frac{\nuL^{-2} \DM(t_1) - \nuH^{-2} \DM(t_2)} {\nuL^{-2} - \nuH^{-2}}
 = \frac{R^2 \DM(t_1) - \DM(t_2)} { R^2-1},
 \ee
 where \newtext{$\DMhat$ is the estimated DM and} $R \equiv \nuH / \nuL$ is the frequency ratio. 
 When used to estimate the DM-corrected \TOA, 
 $\tinftyhat = t_{\nuH} - \KDM \nuH^{-2} \DMhat$, 
  the \DM\ difference 
 $\Delta\DM(t_1, t_2) =  \DM(t_2) - \DM(t_1)$  yields a \TOA\ error, 
 \be
 \delta\tinfty 
 = 
\frac{\KDM}{\nuH^2} 
\left(  \frac{R^2}{R^2-1}  \right)
\Delta\DM(t_1, t_2) .
 \label{eq:dtinfty_async0}
 \ee
 The \RMS\ $\delta\tinfty$ is then written in terms of  the DM structure function,
 \be
 \sigma_{\delta\tinfty}(\Delta t) 
 = 
\frac{\KDM}{\nuH^2} 
\left(  \frac{R^2}{R^2-1}  \right)
\DDM^{1/2}(\Delta t).
\label{eq:dtinfty_async1}
 \ee
 We ignore  the contribution from a  linear variation in DM associated with the change in distance from a pulsar's motion.   This is easier to identify and remove than the stochastic variations.  
 Using previous expressions, we write the \RMS\ \TOA\ error for a Kolmogorov medium (inertial subrange scaling) as
 \be
  \sigma_{\delta\tinfty}(\Delta t) 
  = 
  \left(  \frac{R^2}{R^2-1}  \right) \times
  \left\{
  \begin{array}{l}
  	\displaystyle
  	\frac{23.7\, \rm ns}{\nuH^2} 
		\left(\frac{\SMeff}{10^{-3.5}~\SMunits} \right)^{1/2}
		\left(\frac{\vpperp}{100~\kms} \right)^{5/6}
		\, \Delta t_{\rm days}^{5/6}
	\label{eq:dtinfty_async2}
	\\
	\\
	\quad\quad {\rm or}
	\\
	\\
	\displaystyle
  	 \frac{6.54\, \rm ns}{\nuH} 
	\left[ \frac  {\Delta t_{\rm days}}  {\DtDISS(\nuH) / 10^3\,{\rm s}  }  \right]^{5/6} , 
	\label{eq:dtinfty_async4}
  \end{array}
  \right.
 \ee
 where 
 units  are shown explicitly (except for frequency, which is in GHz.  
 The second form  uses the relation  of $\DDM$ to the phase SF, 
 $\DDM(\tau) = (\lambda\re)^{-2} \Dphi(\vperp\tau)$, 
 and the connection of the scintillation time scale $\DtDISS$  to the spatial scale $\veff \Dtiss$ where $\Dphi(\veff\DtDISS) = 1$\,rad$^2$.
We have used  $\DtDISS$  measured at the higher frequency $\nuH$ and expressed it in units of $10^3$~s, a representative value for a low-DM pulsar.    If an ancillary measurement  of $\DtDISS$ at a frequency $\nu_{\rm ISS}$ is used instead, then $\DtDISS(\nuH)$ in Eq.~\eqref{eq:dtinfty_async4} is replaced by 
$\nuH \nu_{\rm ISS}^{-2/(\beta-2)} \DtDISS(\nu_{\rm ISS})   = \nuH \nu_{\rm ISS}^{-6/5} \DtDISS(\nu_{\rm ISS}) $ for  $\beta = 11/3$. 

\providecommand{\rmstoa}{\sigma_t}
\providecommand{\rmstoaDISS}{{\rmstoa}_{\rm _{DISS}}}
\providecommand{\Beff}{B_{\rm eff}}

\renewcommand{\linner}{l_{\rm i}}


\renewcommand{\nuH}{\nu_{\rm h}}
\renewcommand{\nuL}{\nu_{\rm l}}

\renewcommand{\nuh}{{\nu_{\rm h}}}
\renewcommand{\nul}{{\nu_{\rm l}}}

\subsection{Chromatic DM variations $[\DM(\nu)]$}
\label{sec:DMnu}

Dispersion measures are usually defined as simple \LOS\ integrals of the electron density along a single propagation path, which yields the deterministic $\nu^{-2}$ dependence for the resulting timing delay. 
Measured DM values are  actually chromatic because    
multipath propagation  like that illustrated in Fig.~\ref{fig:two_freq_raywalks} is strongly frequency dependent.
 The DM differs between ray paths and as the geometry changes with epoch,   the collective DM varies.   The time scale for such variations is the time needed for the bundle of ray paths to move by an amount equal to its width, which is the same as the RISS time scale. As a consequence,   the frequency dependence of the ray bundle's transverse extent causes the DM  at any epoch to also   vary with  frequency.  

 The $\DM(\nu)$ effect is intermixed with chromatic variations of the average emitted pulse shape, making it difficult to isolate. However evidence for it has been reported 
 \citep[e.g.][]{cwd+90, rdb+06, 
2019A&A...624A..22D,
2022ApJ...930L..27K}.

The differential DM  between propagation paths results from  variations in electron densities and {\it not} from the different physical path lengths.  The latter dominate the  scattering time 
(pulse broadening time) $\taud$ described earlier, which is  typically much less than a second for most pulsars at radio frequencies $\sim 1$ to 2\, GHz used for precision timing.    The associated DM increment  from  $\lesssim 1$ light second of extra path length is   $\nebar c  \sim  10^9~{\rm cm}^{-2}$ for $\nebar = 0.03~\neunits$, or a negligible $\delta\DM \sim 3\times10^{-10}~\DMunits$. 
    
The frequency dependence of the DM   \citep[][hereafter CSS16]{CSS16} results from 
spatial averaging of ray paths caused by scattering.  At observation  frequency $\nu$ and  location  
$\xvec$, the DM is   the  \LOS\ integral of the convolution of a smoothing function  
$\Anuxz$ with the electron density,   
\be
\DMbar(\nu,\xvec) =
        \langle \DM(\xvec) \rangle
        + 
        \int _0^D  d\zp\int_{-\infty}^{\infty} d\xvecpr\, 
		A_{\nu}(\xvecp, \zp)
		\delta\nelec(\xvecpr, \zp),
\label{eq:DMbar1}
\ee
where the frequency-independent term $\langle \DM(\xvec) \rangle$ is the ensemble average (denoted by angular brackets) of the integrated electron density 
$\nebar$.    The appropriate smoothing function has a  $z$-dependent,   2D area  corresponding to the transverse cross section of the ray bundle.   The scattered image seen by an observer can be used as a starting point but defining $\Anux$ requires knowledge of the strength of scattering vs. $z$.     The image shape is also a function of the type of medium.   In the following we drop the $\xvec$ dependence on $\DMbar$ and consider integrals centered on only a single position for calculating $\DMbar$ at a particular epoch. 

\subsubsection{Two-frequency DM difference}
\label{sec:2freq}

\citet[][]{CSS16} analyzed the consequences for timing errors based on dual-frequency measurements for estimating \DM\ and considered wideband cases using a numerical approach.   After summarizing those results we present a new analysis for measurements over arbitrarily wide bands, which are more in concert with modern timing programs. 

The calculation for the two-frequency case proceeds similarly to the treatment of asynchronous sampling in Sect.~\ref{sec:dm_nonsimultaneous}.
When the \DM\ is estimated from  \TOAs\ obtained at two widely separated frequencies,   
 the measured \TOA\  includes an error $\epsilon_\nu$ (from radiometer noise,
intrinsic pulse jitter, and DISS),
\be
t_\nu = \tinfty + \KDM  \nu^{-2}  \DMbar(\nu) + \epsilon_\nu .
\label{eq:tnu1}
\ee
If \DM\ is assumed to be frequency independent, the  \TOAs\ at 
 two spot frequencies $\nul < \nuh$ measured simultaneously (or nearly so, i.e. within a few hours to a day) yield,
 \be
\DMhat = \frac{t_\nul-t_\nuh}{\KDM (\nul^{-2} - \nuh^{-2})},
\label{eq:dmhat1}
\ee
and an  estimate for $\tinfty$ and its error, 
\be
\tinftyhat = t_\nuh - \frac{\KDM \DMhat}{\nuh^2} = \frac{R^2 t_\nuh - t_\nul}{R^2 -1} ,
\quad\quad
\dtinftyhat \equiv \tinftyhat-\tinfty = \frac{R^2 \epsilon_\nuh - \epsilon_\nul}{R^2 -1},
\label{eq:tinfty1}
\ee
where $R\equiv \nuh / \nul$ (as before).  
The error diverges as $R \to 1$ and the error in the high-frequency \TOA\ dominates the overall error for  a large frequency ratio. 

\begin{figure}[htbp] 
   \centering
   \includegraphics[width=0.6\textwidth]{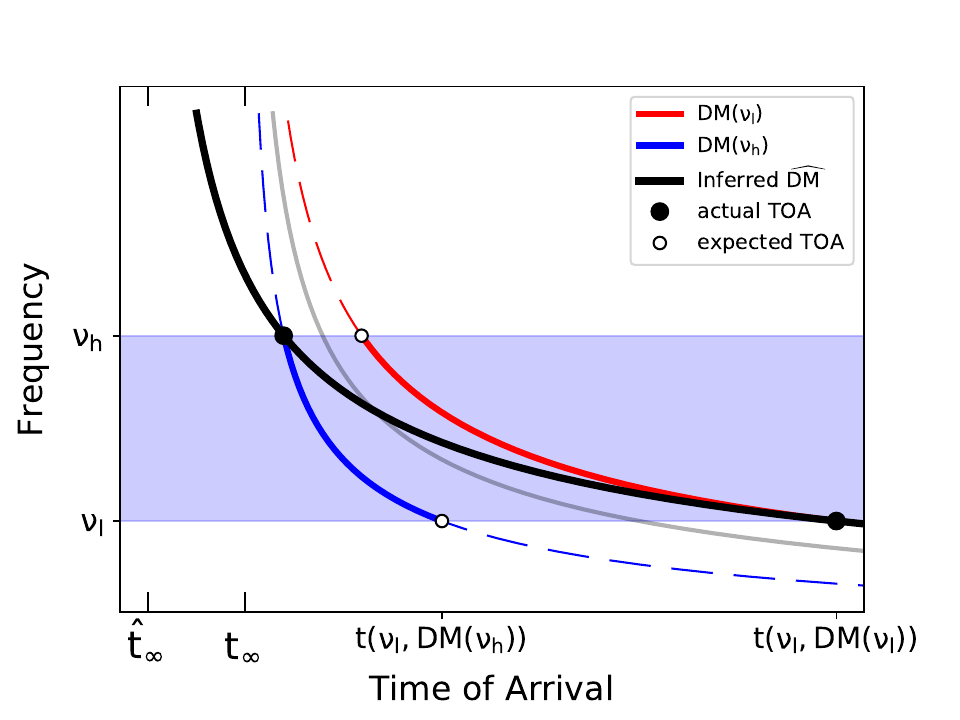} 
   \caption{Illustration of the error in the dedispersed arrival time.   Arrival times at two spot frequencies
   $\nuL,  \nuH$ (filled black circles) are delayed by different dispersion measures, $\DM(\nuL)$ and $\DM(\nuH)$
   (blue and red, respectively).
    Estimation of an assumed frequency-independent  $\DMhat$ (black curve) then leads to an incorrect  infinite frequency  arrival time, $\tinftyhat$, that is earlier than the true value, $\tinfty$. }
   \label{fig:DMnu_effect_demo}
\end{figure}

An illustration of the mis-estimation of arrival times  is shown schematically in Fig.~\ref{fig:DMnu_effect_demo} for \TOA\ measurements at the two spot frequencies $\nuL$ and $\nuH$  on two separate dispersion curves (black points on the two solid lines).  For this figure we ignore the effects from additive noise $\epsilon_\nu$ to focus on the systematic error.
The estimated dispersion measure $\DMhat$ yields a dispersion curve that connects the two black points and extrapolates to an incorrect $\tinftyhat \ne \tinfty$ as shown.

The frequency dependence of \DM\ of course only alters  the error to  
\be
\dtinftyhat = \frac{\KDM }{\nu^2} \left(\frac{R^2}{R^2-1} \right) \Delta\DMbar(\nuh, \nul)
    + \left( \frac{ R^2 \epsilon_\nuh - \epsilon_\nul }{R^2-1}\right),
\label{eq:TOAerror}
\ee
where
$\Delta\DMbar(\nuh, \nul) = \DMbar(\nuh) - \DMbar(\nul)$.

\subsubsection{Wideband timing errors  from chromatic DM variations }

\begin{figure}[htbp] 
   \centering
   \includegraphics[width=0.4\linewidth]{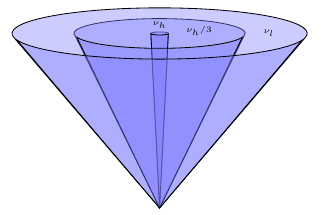} 
   \caption{Scattering cones that illustrate  spatial averaging of electron density variations at the highest 
   and lowest frequencies, $\nuh$ and $\nul$, and an intermediate frequency $\nuh/3$.  The observer is at the apex and the tops of the cones represent cross-sectional areas in  either the plane of a thin screen that scatters incident radiation or midway along the \LOS\ in an extended medium.  The diameters of the conal cross sections are shown to scale $\propto \nu^{-\beta / (\beta-2)} \propto \nu^{-11/5}$  for measurements that sample the inertial range of a Kolmogorov  spectrum ($\beta = 11/3$).}
   \label{fig:Coax_cones}
\end{figure}

The effects of chromatic DMs for wideband systems are obtained by averaging
Eq.~\eqref{eq:DMbar1} over frequency  and calculating the ensemble mean and variance of the result.   This 
averaging takes into account that the ray bundle's  cross section  is a strong function of  location along the line of sight, as illustrated in Fig.~\ref{fig:Coax_cones}.  It can also include the \SNR\ vs. frequency through appropriate weighting.  Here we summarize  results derived in Appendix~\ref{app:DMnu}.

 The analysis yields the error in the {\it estimated} DM for measurements made over a frequency range $\nuL, \nuH$.    Two cases are given: one with uniform coverage over the entire frequency range and another in the limit where two spot frequencies are used to estimate DM (as in the previous section).   The results also can be applied to arbitrary variations of $\cnsq$ along the \LOS, including thin screens, finite-width slabs, or a uniform medium with
$\cnsq = $\, constant.    We obtain
\be
\sigma_{\DMbar}(R, \nuH) 
&=& \left[ \Var(\dDMhat) \right]^{1/2}
= G_\beta Q_\beta \relec  c^{\beta/2}  \dso^{\!\!\!\!(\beta-2)/2}  \nuH^{-\beta/2} \,  \IDMhat^{1/2}(R, \beta) \,\SM,
\\
\sigma_{\tinfty}(R, \nuH) 
&=& \left[ \Var(\tinfty) \right]^{1/2}
= \left( \frac{\KDM}{\nuH^2}\right) \left[ \frac{ \Itinfty(R, \beta)} { \IDMhat(R, \beta)} \right]^{1/2}
\sigma_{\DMbar}(R, \nuH)
\label{eq:Var_tinfty}
\ee
where we note that the values of these \RMS\ quantities are referenced to the upper frequency, $\nuH$. 
The dependence on the frequency range used to estimate DM and $\tinfty$ is contained in the quantities
$\IDMhat(R, \beta)$ and $\Itinfty(R, \beta)$ using the frequency ratio $R = \nuH / \nuL$. 

The dimensionless quantities $\Itinfty$ and $\IDMhat$ are integrals over dimensionless variables $r_{1,2} = \nuH / \nu_{1,2}$ 
are given by
\be
\Itinfty(R, \beta) 
&=& 
		\displaystyle
		\left[  \frac{R(R^3-1)}{(R-1)^4}  \right]^2
		\int_1^R \!\!\!\!\int_1^R
		d\ra d\rb\, 
		\left[ 1 - \frac{3(R-1)\ra^2}{(R^3-1)} \right]
		\left[ 1 - \frac{3(R-1)\rb^2}{(R^3-1)} \right]
		\Phi^2_\beta(\ra, \rb),
\label{eq:Itinftym}
\ee
and
\be
\IDMhat(R, \beta) 
&=& 
		\displaystyle
		\left[  \frac{3R}{(R-1)^3}  \right]^2
		\int_1^R \!\!\!\!\int_1^R
		d\ra d\rb\, 
		\left(\ra^2 / R - 1 \right)
		\left(\rb^2 / R - 1 \right)
		\Phi^2_\beta(\ra, \rb) , 
\label{eq:IDMhatm}
\ee
where equal weighting over frequency has been used and
\be
\Phi_\beta^2(\ra, \rb) 
&=&
2^{-(\beta-2)/2} 
	[ \Vbeta(\ra) + \Vbeta(\rb) - \rb^{\beta} \Vbeta(\ra/\rb)  ] - 1
\ee
with
\be
\Vbeta(x) &\equiv& [1 + x^{2\beta/(\beta-2)}]^{(\beta-2)/2}.
\ee

For two equally weighted spot frequencies these become
\be
\Itinfty(R, \beta)  = \IDMhat(R, \beta)  = 
		\displaystyle
		\left[
		\left( \frac{R^2}{R^2-1}  \right)
		F_\beta(R)
		\right]^2 ,
\label{eq:Itinfty_DM_hat_twofreqm}
\ee
which is obtained by using the identity $\Phi(x, x) = F_\beta(x)$, where $F_\beta$ (first defined in CSS16)
is 
\be
F_\beta(r) &=&
	\left\{
	2^{(4 - \beta)/2} 
		\left[1 + r^{2\beta / (\beta-2)}   \right]^{(\beta-2)/2} - r^{\beta} -1 
	\right\}^{1/2} .
\label{appeq:Fbetam}
\ee

\begin{figure}[!ht] 
   \centering
   \includegraphics[width=0.475\linewidth]{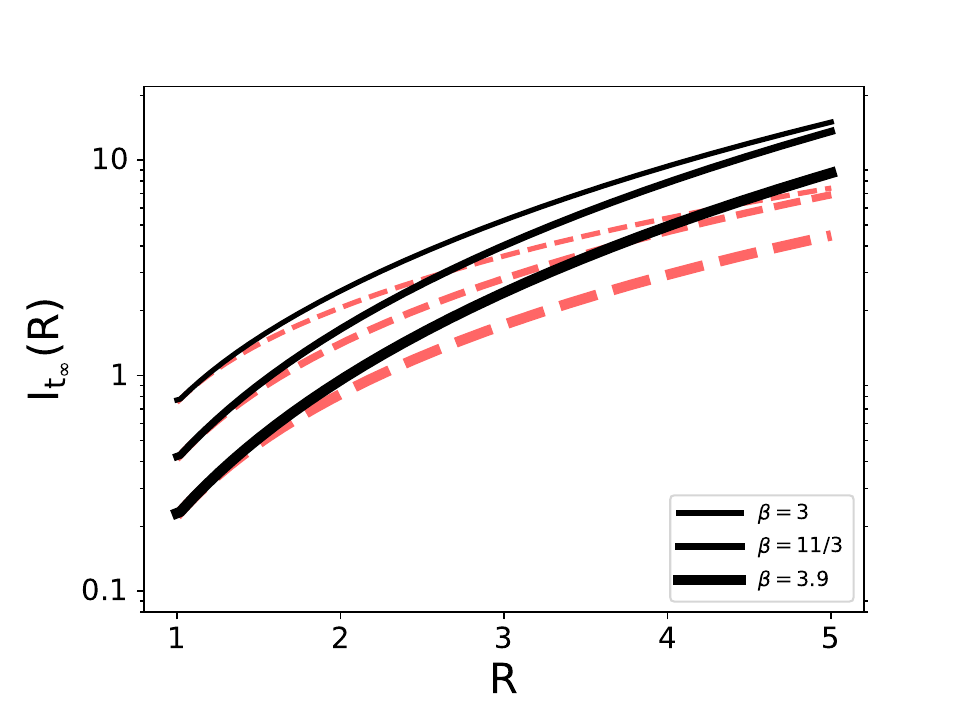}
    \includegraphics[width=0.475\linewidth]{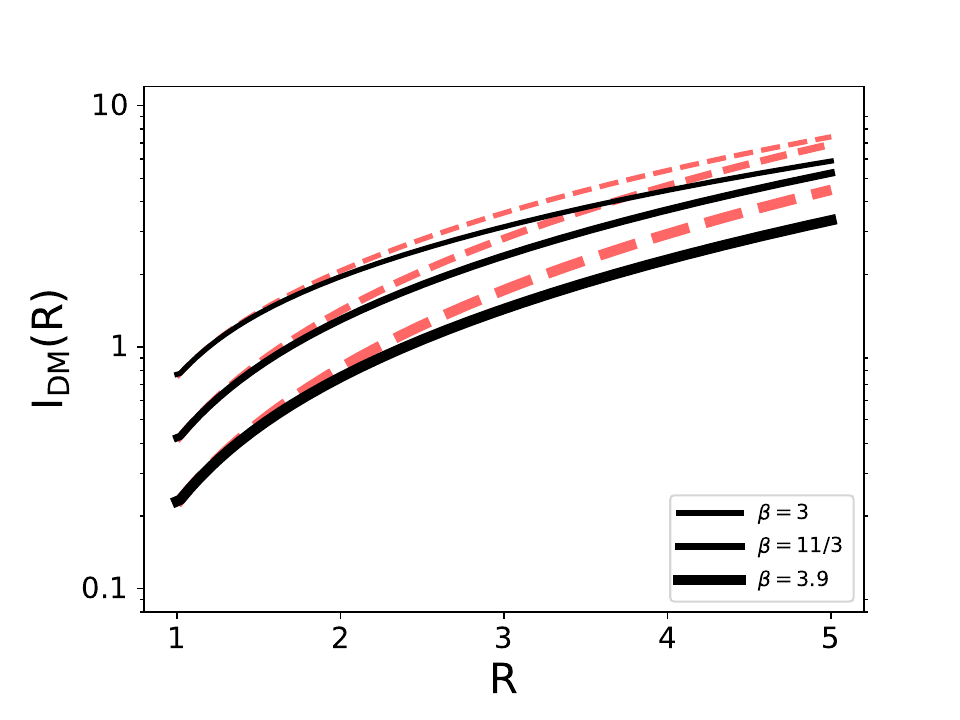}
   \caption{Plot of $\Itinfty(R, \beta)$  and $\IDMhat(R, \beta)$  vs. frequency ratio $R = \nuH / \nuL$ for three values of $\beta$. 
    Solid lines:  wideband case with equal weights using Eq.~\eqref{eq:Itinftym} - \eqref{eq:IDMhatm}.  Dashed lines:  two-frequency cases using Eq.~\eqref{eq:Itinfty_DM_hat_twofreqm}.}
   \label{fig:Iintegrals}
\end{figure}

The $\Itinfty$ and $\IDMhat$ integrals   are shown in Fig.~\ref{fig:Iintegrals} for three values of $\beta$  and for the wideband case (solid black lines) and the two-frequency case (dashed red lines).    While the \RMS\ quantities depend strongly on $R$, the ratio $\Itinfty(R, \beta) /  \IDMhat(R, \beta)$ is less variable. 

The factor  $Q_\beta$ depends on the index of the wavenumber spectrum,  $\beta$,
\be
Q_\beta = \frac{(\sqrt{2}\pi)^{4-\beta} \Gamma(2-\beta/2) f_\beta }{\beta-2},
\ee
where $f_\beta$ is defined in Eq.~\eqref{eq:fbetadef}.    All of the line of sight dependence across the source-observer distance $\dso$ is contained in $G_\beta$,  
\be
G_\beta^2 &=& H_\beta \SMeff / \SM
\ee
where $\SMeff$ and $H_\beta$ involve \LOS\ integrals of $\cnsq$,
\be
\frac{\SMeff}{\SM} =  \frac{\displaystyle  \int ds\, \cnsq(s) (s / \dso)^{\beta-2} }  {\displaystyle  \int ds\, \cnsq(s) } ,
\quad\quad\quad
H_\beta = \frac{\displaystyle  \int ds\, \cnsq(s) [h(s)]^{\beta-2}}  {\displaystyle  \int ds\, \cnsq(s) },
\ee
and $h(s)$ defines the width of the ray-path bundle along the \LOS.   Determination of   
 $h(s)$ generally requires a numerical solution, but for a thin screen it is
\be
h(s) =  
	\left\{
		\begin{array}{ll}
		\displaystyle
			\frac{s}{\dsl} \left(1 - \frac{s}{\dso} \right)		& \quad\quad s \le \dsl, \\
			\\
			\displaystyle
			1 - \frac{s}{\dso}		& \quad\quad s \ge \dsl, 
		\end{array}
	\right.
\ee
which has a value $h(\dsl) = \dlo / \dso$ at the screen itself. 
For a uniform medium it is 
\be
\displaystyle 
h(s) =  \frac{\dsl}{\dso}     \left( 1 - \frac{\dsl}{\dso}  \right). 
\ee

Evaluating for a Kolmogorov index $\beta = 11/3$ with a nominal scattering measure
$\SM = 10^{-3.5}\, \SMunits = 4.53\times10^4\, {\rm cm^{-17/3}}$, we obtain
$Q_\beta \sim 22$ and 
\be
\sigma_{\dtinfty} &=& [\Var(\dtinfty)]^{1/2}   = 156\,{\rm ns}\, \times G_\beta\, \nuH^{-23/6} \dso^{\!\!\!5/6}\, \SM_{-3.5} \Itinfty(R, \beta)
\\
\sigma_{\delta \DMhat} &=& [\Var(\delta \DMhat)]^{1/2}  = 3.76\times 10^{-5}\,\DMunits \times 
	G_\beta \nuH^{-11/6} \dso^{\!\!5/6} \, \SM_{-3.5} \IDMhat(R, \beta) .
\label{eq:dtinfity_DMnueval}
\ee

We note that the dependence on the upper frequency $\nuH$ for the excess DM variation is
$\sigma_{\delta \DMhat} \propto \nuH^{-11/6}$ for a
Kolmogorov spectrum (inertial subrange) while the \ToA\ variation has a much stronger dependence, 
$\sigma_{\dtinfty} \propto \nuH^{-23/6}$.   The actual \ToA\ error also depends on the frequency ratio $R$ and can be much smaller than the nominal values for a narrowband system, particularly if the upper frequency $\nuH$ is well above 1\,GHz.

\subsection{Angle-of-arrival variations}
\label{sec:aoatoa}

Structures in the ISM larger than the Fresnel scale $\sqrt{\lambda \deff /2\pi} \approx 10^{11}$~cm 
 refract radiation, causing the angle of arrival (AOA) to deviate from an assumed fiducial 
direction, which itself may differ from the true direction as defined by the vacuum geodesic.  
Assuming the topocentric and barycentric  \AOAs\ are the same (after accounting for aberration 
and Doppler shift from Earth's motion), pulses will arrive earlier or later than expected for the fiducial
direction.   Translation of topocentric \TOAs\ to the solar system's barycenter (SSBC) involves
accounting for the R\"omer delay (as well as other delays),     
$\Delta t_{\text{ R\"omer}} = \nhat \cdot \rearthvec(t)$, 
where $\nhat$ is the pulsar direction and $\rearthvec$ is the Earth-SSBC vector.

If the actual arrival direction differs from $\nhat$ by a refraction angle
$\thetar$, the associated error in the \TOA\ at the SSBC is
\be
\dtAOA &\approx& \frac{1}{2c}
	\deff
	\thetar^2
	\approx 1.21~\mu s~ \deff \thetar^2
\label{eq:dtoa_thetar}
\ee
for $\deff$ in kpc and $\thetar$ in mas. \newtext{Values for $\thetar$ can be much smaller or much larger than a mas depending on the particular pulsar \LOS\ and on frequency.}
The delay increases at lower frequencies   as $\nu^{-4}$ if refraction is produced by a discrete 
cloud but scales as $\nu^{-49/15}$ for a Kolmogorov medium where ray averaging from diffraction alters the scaling.  The first case scales similarly to the pulse broadening time and would be partially removed by a \TOA\ fitting function designed to account for scattering.   Less removal would occur for the second case. 

The second effect associated with transformation of \TOAs\ to
the SSBC \citep{fc90} is proportional to $\thetar$. Assuming a circular orbital frequency $\Omegaearth$, and defining the SSBC to Earth vector as $\rearthvec$, and source direction $\nhat$, the offset is  
\be
\dtAOABary = c^{-1} \nhat \cdot \rearthvec(t)
	\approx  c^{-1}r_{\oplus} \thetar(t)  \cos \ecliplat \cos \Omegaearth (t - t_p)
	\approx 2.4~\mu s\, \thetar(\rm mas)  \cos \ecliplat  \cos \Omegaearth (t - t_p) ,
\label{eq:dtoa_thetar_ssbc}
\ee
where $\rearth / c \approx 500$~s.
The eccentricity of the Earth's orbit $\sim 0.017$ implies additional harmonics
in the \TOA\ perturbation but these are smaller by powers of the eccentricity. 
This offset scales with frequency as $\nu^{-2}$ if a discrete cloud dominates refraction but  refraction from  a  Kolmogorov medium yields an \RMS\ variation $\thetar({\rm RMS}) \propto \nu^{-49/30}$.  

\newcommand{\ehat}{{\bf \hat e}_{\phi}}

\providecommand{\rmstoa}{\sigma_t}
\providecommand{\rmstoaDISS}{{\rmstoa}_{\rm _{DISS}}}
\providecommand{\Beff}{B_{\rm eff}}

\renewcommand{\linner}{l_{\rm i}}

\providecommand{\ldobs}{\ld^{(obs)}}
\providecommand{\ldscr}{\ld^{(scr)}}
\providecommand{\sigmad}{\sigma_{\rm d}}
\providecommand{\sigmao}{\sigma}

\renewcommand{\gamarray}{\raise0.48ex\hbox{$\bm\gamma$}}   

\renewcommand{\Ii}{U}			
\newcommand{\Id}{I_{\rm d}}
\newcommand{\WIi}{W_{\Ii}}
\newcommand{\Iim}{U_{\rm m}}			
\newcommand{\Wu}{W_{\rm u}}
\newcommand{\Wum}{W_{\rm u, m}}
\providecommand{\Um}{U_{\rm m} }		
\renewcommand{\PBF}{p}
\providecommand{\PBFm}{{p_{\rm m}}}
\providecommand{\PBFme}{{p_{\rm m,e}}}
\providecommand{\xtau}{x_{\tau}}
\newcommand{\xtaum}{x_{\tau,\rm m}}
\newcommand{\xtauhat}{\widehat{x}_\tau}

\providecommand{\tauhat}{\widehat{\tau}}
\newcommand{\taum}{\tau_{\rm m}}
\newcommand{\taue}{\tau_e}

\providecommand{\tpbfave}{\langle t \rangle_{\PBF}}

\renewcommand{\linner}{l_i}
\renewcommand{\louter}{l_o}
\renewcommand{\qinner}{q_i}
\renewcommand{\qouter}{q_o}

\providecommand{\xtaud}{x_{\tau}}
\providecommand{\xdnud}{x_{\dnud}}
\providecommand{\xthetad}{x_{\thetad}}

\subsection{PBF stochasticity from nanoseconds to years}

\begin{figure}[b!] 
   \centering
   \includegraphics[width=0.8\linewidth]{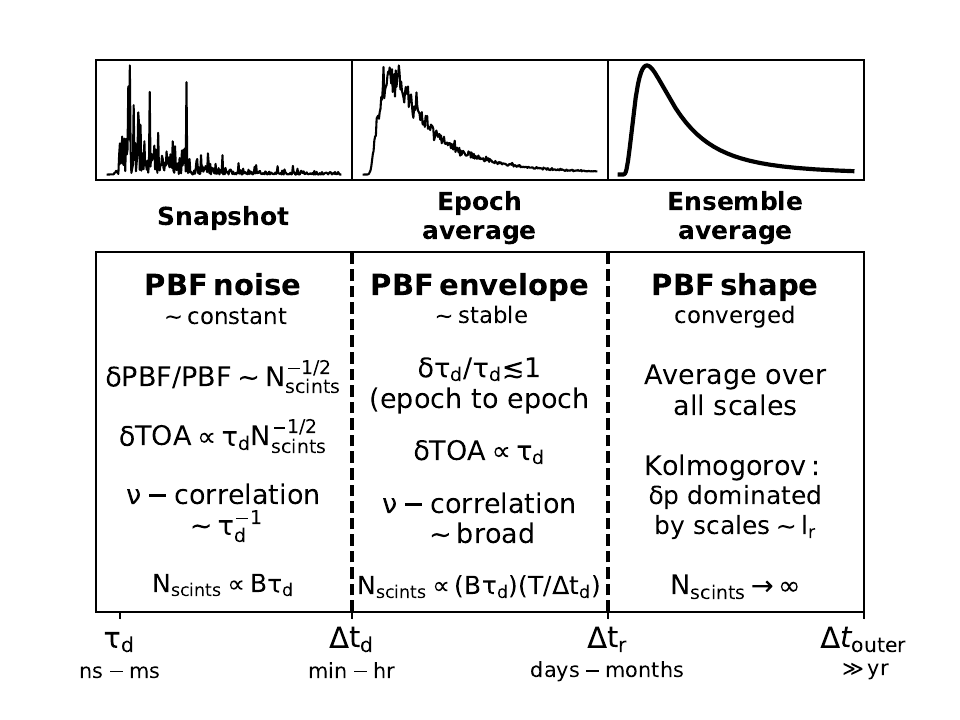}
   \caption{Averaging regimes for PBFs analogous to the description of snapshot and averaging regimes presented by 
   \citet[][]{ng89} for scattered images.
   The snapshot case (left) corresponds to averages over less than 
   a DISS time scale, $\dtd$ that include some number of scintles across frequency.    An epoch average spans many  scintillation times and includes many more scintles, producing a more stable PBF shape than for a snapshot.    However,  such averages are subject to distortions caused by refraction from larger density fluctuations than those that cause diffraction. Only by averaging over all refraction-induced PBF variations is the ensemble average PBF obtained.   For a Kolmogorov spectrum, however, refraction is dominated by scales of order the refraction scale $\lr$ defined earlier. 
   } 
   \label{fig:pbfregimes}
\end{figure}

PBFs vary on time scales of nanoseconds to years, depending on the integration time relative to various scattering and scintillation time scales. Fig.~\ref{fig:pbfregimes} shows three averaging regimes for PBFs defined analogously to scattered images \citep[][]{ng89}.  

An averaging time shorter than the DISS time scale $\dtd \sim$\,minutes to hours yields a noisy PBF like those shown in 
Figs.~\ref{fig:pbfave} and \ref{fig:two_freq_pbfs}.  The noise pattern remains constant for averaging times $\ll \dtd$.  
Radio timing programs generally integrate over multiple  DISS times, yielding a less noisy PBF that better reveals the envelope portion of the PBF.  Even so, an epoch average is still subject to refractive distortion of the PBF by electron density variations much larger than diffraction scales.   Only by averaging over many epochs should a PBF shape approximating an ensemble average be expected.  In principle, this would imply averaging times greater than the crossing time of the outer scale, which could be Myr for an outer scale 
$\louter \sim 100$\,pc
(cf.\ Fig.~\ref{fig:DMSF_asymptotic_B1937}).  

This long a time scale is of low relevance, however.   First, for density fluctuations dominated by a Kolmogorov spectrum,  refractive distortion is dominated by scales $\sim \lr \ll \louter$, the `refraction' scale applicable to RISS, which is of order the typical diameter $\sim \au$ of the ray path bundle along which pulses propagate. Second, other structures exist in the ISM that superpose with Kolmogorov fluctuations and dominate refraction episodically with time scales of days to months \newtext{\citep[e.g.][]{rcb84,Coles1987}}. 

The range of random and systematic errors from scattering discussed in the next sections originate from the effects identified for the three regimes shown in the figure.

\subsection{Rapid stochasticity of PBFs and \TOA\ noise}
\label{sec:rapid_pbf}


\newtext{Multipath effects on \TOAs\ are dominated by changes in the PBF shape, which are both systematic and stochastic, and cause net changes in measured pulse profiles.  These in turn cause mis-estimation of DM that further perturb estimates of DM-corrected \TOAs.}

Scattered  profiles are the convolution (denoted by an asterisk) of a scaled and time-shifted intrinsic pulse shape $I_{\rm i}(t) = a U(t -t_0)$  with the  PBF, 
\be
I(t) = a\, U(t-t_0) * \PBF(t, \tau)
= a \int d\tp\, U(t -  t_0 - \tp ) \PBF(\tp, \tau), 
\label{eq:Iactual}
\ee
where $a$ is a linear scale factor and $t_0$ is the arrival time in the absence of scattering; we include an explicit argument $\tau$ as the characteristic  $1/e$ time scale  of the PBF.

The pulse broadening function is defined as the set of  delays associated with multipath propagation through ionized gas. As such, it comprises geometrical path length differences caused by both diffraction and refraction.  
Stochasticity of the PBF, both short and long term, causes statistical perturbations of \TOAs.   On pulse-to-pulse time scales, the PBF follows the `envelope $\times$ noise' form that yields a systematic delay from the envelope combined with a random error from PBF noise.     For long integrations
at a given epoch ($T \gg \dtd$) and a wide bandwidth ($B\gg\dnud$),  many scintles are averaged over in the \TOA\ estimate, diminishing but not eliminating the random error. 

We discuss two consequences of PBF noise on arrival times.  The first concerns induced variations in the effective center frequency of a measurement if dedispersion is not employed or if  it is executed with an inaccurate DM value. The second  TOA perturbation arises from the short-term  variability of the PBF itself.  Together, they illustrate the utility of using both the time and frequency domains in characterizing these propagation effects as well as mitigating them. 

Both effects arise from the finite number of scintles in the time-frequency plane used to form a pulse profile and calculate an arrival time.   For a total time $T$ and bandwidth $B$,   the number of scintles  is approximately
 \be
 \Nscint  \equiv \Nnu \Nt \simeq (1 + \eta_\nu B / \dnud) ( 1 + \eta_t T / \dtd),
 \label{eq:Nscint}
 \ee
 where $\eta_\nu$ and $\eta_t$ are filling factors for scintles  with values   $\sim 0.2$ to 0.4 that depend slightly on the medium \newtext{and on radio frequency}.  


\newcommand{\Znut}{Z_{\nu t}}
\providecommand{\nubar}{\overline{\nu}}
\newcommand{\nubarmean}{\left\langle \nubar \right\rangle}
\newcommand{\Neff}{N_{\rm eff}}
\newcommand{\Snueff}{S_\nu^{\prime}}

\subsubsection{Scintillation induced changes in effective center frequency} 
\label{sec:dRF}

Dedispersion removes  delays $\propto \DM (\nu^{-2} - \nu_{\rm ref}^{-2})$ with respect to a specified reference frequency.  No error in  arrival time results if the correct value of DM is used.
In some early timing analyses of long period pulsars
\newtext{
\citep[pre-1990s, e.g.][]{1983ApJS...53..169D}
}
differential dispersion delays across narrow receiver bands  were considered  small and no dedispersion was done.   However, without dedispersion, diffractive intensity scintillations across the  band shift the effective center frequency  stochastically on a time scale equal to the scintillation time $\dtd \sim$ (minutes to hours), causing a variable dispersion delay.   
If \newtext{a rectangular bandpass of}  bandwidth $B$  contains a large number of scintles but  is small enough so that the pulsar's intrinsic flux density  and scintillation parameters  are roughly constant, the effective center frequency shifts by
$\Delta \nu_{\rm DISS}$ with an \RMS\ value
 $\sigma_{\nu} = B / \sqrt{12\Nscint}$.   This yields an ensemble
  mean dispersion delay (for DM in $\DMunits$ and $\nu$ in GHz)
 $t_\nu = (c r_e / 2\pi \nu^2) \DM \simeq 4.15\,{\rm ms} \, \nu^{-2} \,\DM$ 
 and RMS fluctuations about the mean delay, 
 \be
 \sigma_{t_{\nu}} 
 \approx 
 \frac{dt_{\nu}}{d\nu} \sigma_{\nu} 
 = 
 \left(\frac{2\sigma_{\nu}}{\nu}\right) t_{\nu}
 \simeq 2.4\,{\rm ms} \left(\frac{B}{\nu} \right) \frac{\DM}{\sqrt{\nu^2 \Nscint}}.
 \label{eq:rmsnu}
 \ee

 Early timing measurements with fractional bandwidths $B/\nu \sim 10^{-3}$ on low-DM pulsars yielded small errors relative to other \TOA\ errors.  
 A counterexample where this scintillation effect dominated other errors  is the bright Vela pulsar \citep[e.g.][]{1986ApJS...62...81D}.
 
Modern dedispersion methods (post-detection dedispersion with channelized intensities or coherent dedispersion), if done with an incorrect value of DM, will also yield (small) \TOA\ errors from this effect.  Post detection methods divide $B$ into narrow channels  with DM delays removed between channels.  Intrachannel dispersion produces a TOA error given by replacing $B$ in Eq.~\eqref{eq:rmsnu} with the channel width $\Delta\nu_{\rm c}$.  If the DM used for dedispersion is in error,   the TOA error is calculated by replacing DM with the DM error $\delta \DM$.   Such errors may arise due to interstellar variance in DM and usage of a non-contemporaneous value of DM in the dedispersion.   DM errors are typically $< 10^{-3}\,\DMunits$, so this will yield a small error $\lesssim 1\,\mu$s that adds to the larger error,
$\delta t_\nu \propto \nu^{-2} \delta\DM$. 

\subsubsection{Rapid changes in PBF shape (the ``finite scintle error'' (FSE))}
\label{sec:dt_fse}

\begin{figure}[t!] 
   \centering
   \includegraphics[width=0.6\linewidth]{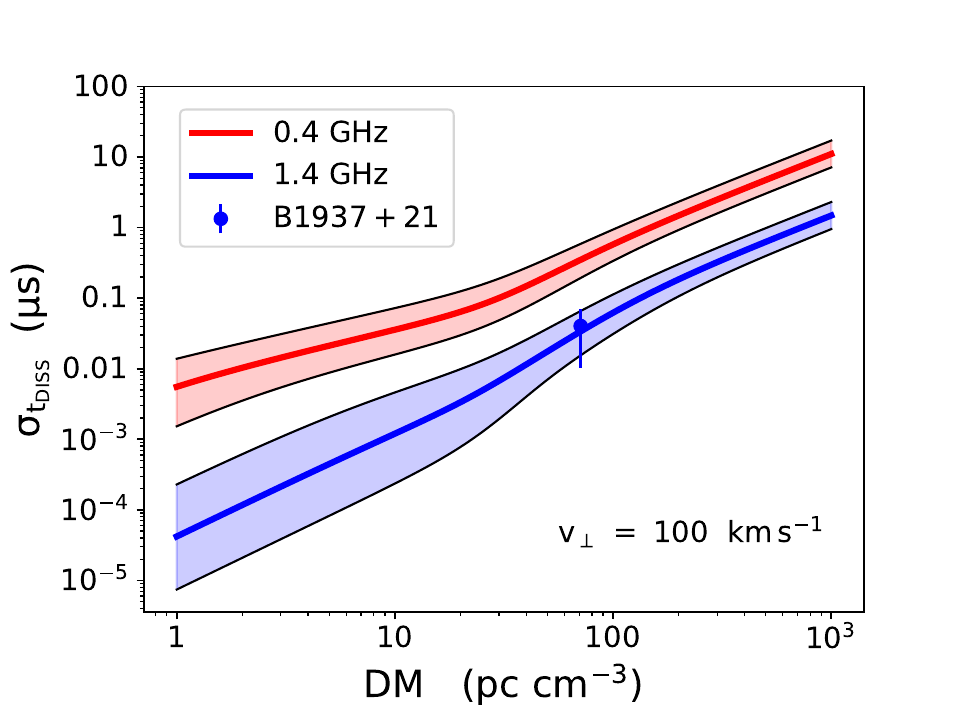}
   \caption{The RMS FSE error $\rmstoaDISS$  plotted vs. DM evaluated using the `hockey stick' relation (Eq.~\eqref{eq:taudm})  to evaluate $\taud$ from  DM and Eq.~\eqref{eq:rmstoaDISS2}. 
   with $B = 0.1$\,GHz.
   The black point (with 10$\times$ actual error) is from a timing analysis of the MSP B1937+21 that determined the  DISS timing error by subtracting (in quadrature) the  S/N and jitter timing errors (JMC, unpublished analysis).
   } 
   \label{fig:dtdiss_vs_dm}
\end{figure}

The  noise process in the PBF (cf.\ Fig.~\ref{fig:two_freq_pbfs}) is due to the specific set of random path delays contributing to the measured wavefield, which also yield    constructive and destructive interference across frequency.   

  For an integration time  \newtext{much less than the diffractive scintillation time,} $T \ll \dtd$  the realization of  
  PBF noise  is approximately constant, so $\Nt = 1$;  $\dtd$ can range from seconds (for heavily scattered pulsars with large transverse velocities) to hours and it introduces an arrival time error that varies on the same time scale. This regime corresponds to the set of inequalities, 
\be
\taud \ll T \ll \dtd \ll \dtr,
\ee
where $\dtr$ is the refraction time scale. 
\newtext{A noisy PBF then typically corresponds to  the presence of multiple  scintles across
the receiver bandwidth, often with $\Nnu \gg 1$.}

The character of the noise process depends on the number of scintles $\Nnu$ in the bandwidth $B$. 
Heuristically, if the band is narrower than a scintle, $B \lesssim \dnud$ \ (e.g. for low-DM pulsars measured at high frequencies), the imposed phase variation across the band $\Delta\phi \lesssim 1$\,rad, yielding a group delay
$\tau \sim \Delta\phi / 2\pi B \lesssim 0.16 \nu^{-1}$\,ns for $B$ in GHz.   When $\Nnu \gg 1$, the $\sim 1$\,rad phase gradient across each scintle gives a larger delay 
$\tau \sim 1 / 2\pi\dnud \sim 0.16 \times (1\, {\rm MHz} / \dnud),\mu$s.  
These delays are statistical with an average $\tau$
consistent with the uncertainty relation in Eq.~\eqref{eq:uncertainty1}. 

For a scattered impulse, the \TOA\ variance\footnote{Note that the \RMS\ FSE effect   is denoted $\rmstoaDISS$ here but in other contexts we will use alternative notation, $\dtDISS$, $\DtFSE$, and $\DtdPBF$, for convenience. This is indicated in Table\,\ref{tab:nomenclature}.} from the finite number of 
scintles is formally given by (Appendix\,\ref{app:FSE}) the integral over the (squared) PBF envelope (i.e. the ensemble average of the PBF, $\langle \pism(t) \rangle$) is, 
  \be
  \rmstoaDISS^2(\nu_0) \simeq  \Beff^{-1} \int dt\, t^2 \langle \pism(t; \nu_0) \rangle^2  
  \label{eq:rmstoaDISS}
  \ee
  where the effective bandwidth is  the integral $ \Beff^{-1} = \int d\nu\, \vert b(\nu) \vert^2$
  over the squared  field (voltage) bandpass function,  $b(\nu)$,  defined to have  bandwidth $B$ and unit area.
  For a rectangular bandpass, $\Beff = B$.   
  
  For long integration times $T \gg \dtd$ (e.g. over an observing session $\sim 1$\,hr) over which the scintillation frequency structure changes  and thus  $\pism$ varies, the net \RMS\ delay involves an integral over the 2D intensity correlation function $\Gamma_I(\delta\nu, \tau)$.    This correlation function can be calculated in closed form only for a square-law structure function but not for realistic media.  The details are not of interest and so we use a heuristic approach in the following.  
  
  Each scintle contributes an \RMS\ delay $\sim \taud$, corresponding to a phase change   $\sim 1$\, rad  across $\dnud$;  the integrated effect over $\Nscint$ scintles in the $B-\Tint$ plane yields an RMS time shift,
  \be
  \rmstoaDISS \sim \taud / \sqrt{\Nscint},
  \label{eq:rmstoaDISS1}
  \ee 
  with $\Nscint$ given approximately by Eq.~\eqref{eq:Nscint}. 
  \newtext{
  For observations with moderate scattering that is determined by electron density fluctuations in the inertial subrange, we derive a simple scaling law by
   calculating $\dnud$ and $\dtd$ in terms of $\taud$ (cf.\ Table\,\ref{tab:keyplasmaeq}).}    From $\taud \sim \deff \thetad^2 / 2c$ we obtain
  $\dtd = \ld / \veff = \lambda / 2\pi\thetad\veff$ and 
  $\dnud = C_1 / 2\pi\taud$.  This gives
  \be
  \rmstoaDISS  
	\simeq  
	\frac
		{[c \deff \taud(\nu)]^{1/4}} 
		{2\pi \sqrt{\eta_\nu \eta_t B T \nu v_{\rm eff} }}
    \simeq 0.052\,\mu s \times 
        \frac
            {(\deff  \taud)^{1/4}}
            {\sqrt{(B_{0.1} T_3 \, v_{100}\, \nu}}
    \propto \nu^{-8/5}, 
  \label{eq:rmstoaDISS2}
  \ee
  where $\taud$ in $\mu$s is evaluated at $\nu$ in GHz,
  the bandwidth $B$ is in units of 0.1\,GHz, 
  $\deff$ is in kpc, and the effective velocity $v_{\rm eff}$ is in units of
  100\,km\,$^{-1}$. 
  
 Fig.\,\ref{fig:dtdiss_vs_dm} shows $\rmstoaDISS$ for
  two radio frequencies using the expression for $\taud(\DM)$ in
  Eq.~\eqref{eq:taudm}.   We have approximated distances as
  $\deff = (\DM / \nelec) / (1 + \DM / 1000\, {\rm pc\,cm^{-3}})$
  with $\nelec = 0.03$\,cm$^{-3}$.    This empirical approach
  implies an asymptotic scaling for the \TOA\ error, $\rmstoaDISS \propto \nu^{-4.9} \DM^{11/8}$.
 
\newtext{We now present scaling laws for the \RMS\ FSE timing variation $\rmstoaDISS$ for  more general conditions.}
The scaling of  $\rmstoaDISS$ with frequency depends on \newtext{the scattering time $\taud$ along with} the number of scintles
included in the time-bandwidth product ($T\times B$) of a timing measurement.   
The scaling law is different for the four cases where
$T$ and/or $B$ are larger or smaller than the characteristic scintle size, i.e. whether  
$\Nt$ and/or $\Nnu$ are unity or much larger than unity.
The scaling laws also depend on whether the scattering is strong
or superstrong, which are defined, respectively,  as whether scintillations are dominated by scattering from
electron density length scales in the inertial subrange $\linner \ll l \ll \louter$, or by 
the inner scale. 
\newtext{
We obtain the scaling laws by using the relations between $\dnud$, $\thetad$, $\dtd$
and $\taud$ given just before Eq.~\eqref{eq:rmstoaDISS2} to calculate $\Nnu$ and $\Nt$ in terms of the scattering time $\taud$.  The frequency scaling of the scattering time is  
$\taud\propto \nu^{-\Xtau}$ with $\Xtau = 2\beta / (\beta-2)$ in the inertial-subrange regime (as in Table\,\ref{tab:keyplasmaeq}) and $\Xtau=4$  (independent of $\beta$) for hyperstrong scattering where the inertial scale of density fluctuations is the dominant scattering scale. 
}

Table~\ref{tab:dtpbf} gives frequency scaling laws  for the different regimes for the
frequency scaling law for $\dtPBF \propto \nu^{-X}$, which shows the dependence on the type of medium ($\beta$),
whether the scattering is strong or superstrong, and on the bandwidth and integration time relative to the DISS bandwidth and time scale.    We do not give a scaling law for the weak scattering regime where  
$\Nt = \Nnu = 1$ and $\taud \sim 1 / 2\pi\nu$ is negligible,  typically  less than a  nanosecond.

\newtext{
In current PTA programs, the scattering strength is typically strong but not superstrong at the $\sim 1$\,to\,2\,GHz frequencies used for observing low to moderate DM pulsars.    Future work that makes more extensive use of sub-GHz frequencies and extends the pulsar sample to higher-DM fluctuations will encounter the superstrong regime.   That will also be true for non-PTA timing studies of high-interest pulsars, such as those in the Galactic center.
}

\begin{deluxetable*}{lcrccccccc}[h!]

\tablecaption{Scaling Laws for Finite Scintle \ToA\ Error ($\rmstoaDISS$)  \label{tab:dtpbf}}
\tablehead{
&&&&& \multicolumn{5}{c}{Frequency index $X$ in $\nu^{-X}$} \\  
&&&&& \cline{1-5} \\
\colhead{Regime} & 
\colhead{$\displaystyle {\rm Scattering\tablenotemark{\footnotesize a} \atop strength}$} &
\colhead{$\displaystyle\frac{T}{\dtd}$} &
\colhead{$\displaystyle\frac{B}{\dnud}$} &  
\colhead{~~~} &
\colhead{$X_{\taud}$} & 
\colhead{$X_{N_t}$} & 
\colhead{$X_{N_{\nu}}$} &
\colhead{$X_{\rmstoaDISS}$} & 
\colhead{$X_{\rmstoaDISS}^{\rm (Kol)}$}
\\
}
\startdata
\\
Low \DM, high $\nu$ & strong & $\ll1$ & $\ll1$ &&
      $\taudscaling$ & 0 & 0 & $\taudscaling$ & $\displaystyle\frac{22}{5}$  \\
\\
Low \DM, high $\nu$, high velocity $v$ & strong & $\gg1$ & $\ll1$ &&
      $\taudscaling$ & $\scalingA$  & 0 & $\scalingB$ & $\displaystyle\frac{19}{5}$  \\
\\
Moderate \DM, low velocity $v$,  & strong & $\ll1$ & $\gg1$ &&
      $\taudscaling$ &  0 & $\taudscaling$ & $\scalingC$ & $\displaystyle\frac{11}{5}$  \\
\\
Moderate-high \DM, typical velocity $v$ & strong & $\gg1$ & $\gg1$ &&
      $\taudscaling$ &  $\scalingA$ & $\taudscaling$ & $\scalingD$ & $\displaystyle\frac{8}{5}$  \\
\\
High \DM, low $\nu$, typical velocity $v$  & superstrong & $\gg1$ & $\gg1$ &&
    4 & 1 & 4 & $\displaystyle\frac{3}{2}$ & $\displaystyle\frac{3}{2}$  \\
\\
\\
\enddata
\tablenotetext{a}{{\it Strong scattering} corresponds to scaling laws applicable for 
the inertial subrange of the electron density spectrum.\\
{\it Superstrong scattering} corresponds to scintillations dominated by the inner scale of
the density spectrum.}
\end{deluxetable*}

\subsection{Slow stochasticity of PBF envelopes and \TOA\ bias}

\TOAs\ are often obtained using integration times and bandwidths encompassing   a large number of scintles, which make the  finite-scintle error subsidiary to other errors.    Epoch to epoch, however, the PBF envelope could change through any inhomogeneity of the scattering strength,  e.g. variations in scattering measure $\SM$; this  is unlikely on time scales of years given that $\SM$ is plausibly  determined by  ISM structures on large scales that might take centuries or more for lines of sight to cross. More likely is  variable refraction from more rapidly crossed scales that alter the ray bundle  as the  \LOS\ 
traverses the medium.  Refraction associated with the density spectrum will cause changes on time scales similar to intensity variations from
 RISS, which varies on time scales of hours to days for  pulsars that are weakly scattered (low \DM\ and high $\nu$  measurements)  and weeks to years for larger scattering (high \DM\,, low $\nu$).   These time scales differ substantially between different pulsars  for the same $\nu$-\DM\ combinations due to differing effective velocities.
Also, refraction from larger-scale structures ($\gtrsim$\,AU) that superpose with stochastic density variations can yield much larger changes in PBFs. 

Temporal variations in the PBF shape are also frequency dependent.  Coupled with the chromaticity of intrinsic pulse shapes, the cumulative, non-dispersive arrival time variations have dependences on frequency that require a careful  mitigation approach unique to each pulsar.

\subsubsection{Inapplicability of the `mean-shift' scattering regime} 
\label{sec:mean_shift_not}

The stark difference between exponential and PL PBFs is   demonstrated by considering the 
{\it mean-shift regime}.   If the PBF is much narrower than the intrinsic pulse (width $\Wu$)  for {\it all} times where the PBF contributes significantly to the convolution in Eq.~\eqref{eq:Iactual}, a first-order Taylor  expansion of the intrinsic shape,  $U(t - \tp - t_0) \simeq U(t - t_0) + (dU(t-t_0) / dt) \tp$  yields a scattered profile  that is simply  a shift of the intrinsic shape, $\Id(t) \simeq a U(t - t_0 - \tpbfave)$,
where $\tpbfave = \int dt\, t\, \PBF(t)$.
For an exponential PBF with $\tau \ll \Wu$, the mean-shift regime applies. 
 However, this deceptively simple result masks two issues.  First, the mean  shift $\tpbfave$ is not the same as the characteristic scattering time $\tau$, as demonstrated in Sect.~\ref{sec:heavytails}.  
 Second,   the linear expansion underlying the mean-shift breaks down  for  heavy-tailed PBFs\footnote{`Heavy tailed' means that the PBF is not exponentially bounded for a significant  time range, though we emphasize that PBFs are exponentially bounded at very large times due to the inner scale of the medium.}  that extend well beyond the intrinsic pulse width. 
 
\begin{figure}[htbp] 
   \centering
   \includegraphics[width=0.45\linewidth]{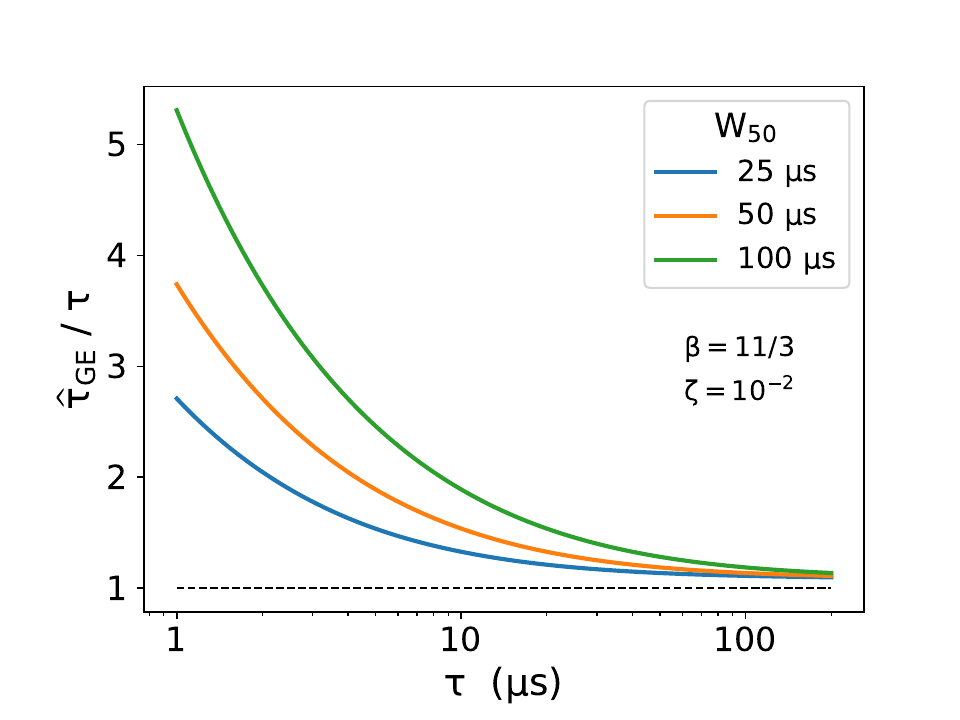} 
   \includegraphics[width=0.45\linewidth]{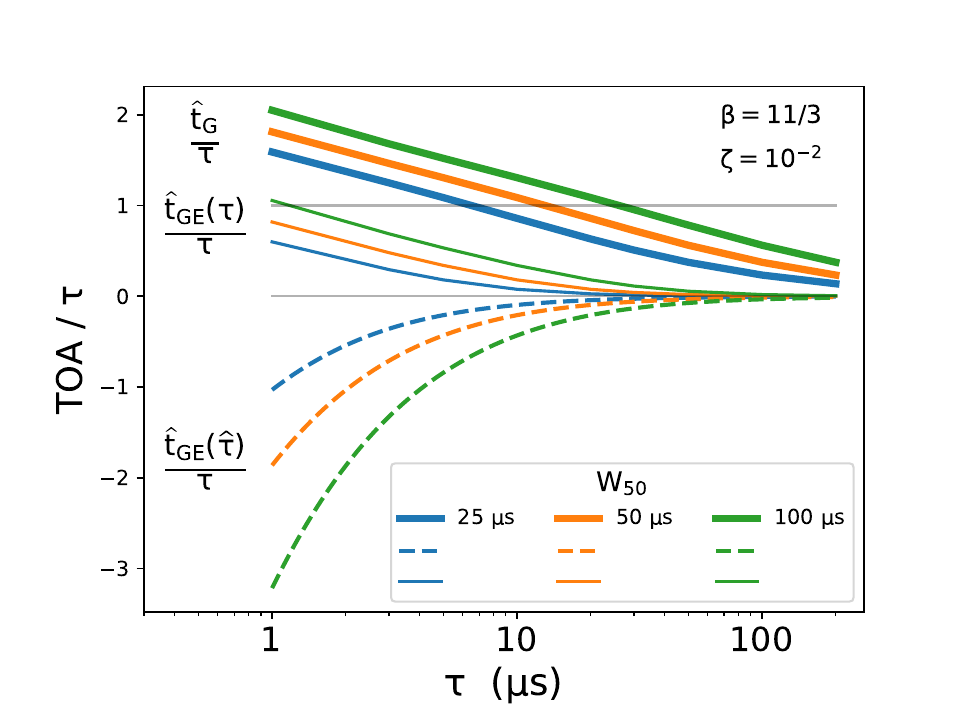}
   \caption{Bias of scattering time and arrival time estimates when exponential PBFs are used to
   fit Gaussian pulses scattered by a  Kolmogorov medium  ($\beta = 11/3$) with a small inner scale,
   $\zeta = 0.01$.  The three  Gaussian widths used are designated in the legend.
   Left: The ratio of estimated  $\widehat{\tau}_{\rm G*E}$ to true scattering time $\tau$  for a G*E template applied to a G*PL pulse.   
%
The ratio  decreases as the scattering time becomes a larger fraction of the Gaussian width
 $W_{50}$. 
 Right: The ratio of \TOA\ estimate to scattering time $\tau$ for three different pulse-fitting approaches
 and three different Gaussian widths; 
 $\widehat{t}_{\rm G}$ is the \TOA\ obtained with a Gaussian-only  (G) template.
 $\widehat{t}_{\rm G*E}(\tau)$ is the \TOA\ obtained with a G*E template with a scattering time fixed to the true
 scattering time;
 and
 $\widehat{t}_{\rm G*E}(\tauhat)$ is the \TOA\ obtained with a G*E template with an estimated scattering time
 $\tauhat$ that optimizes the pulse fitting (minimum mean-squared difference). 
    }
   \label{fig:demo_toa_bias}
\end{figure}

 Fig.\,\ref{fig:demo_toa_bias} exhibits the biases that result from fitting with  an exponential PBF when a power-law PDF actually applies.  
 \newtext{We  simulated measured average profiles by  numerically scattering an
 assumed Gaussian pulse as the emitted pulse using a heavy-tailed PBF like those shown in Sect.~\ref{sec:pbf_gallery} for a Kolmogorov spectrum ($\beta = 11/3$) and a small inner scale (small $\zeta$).
 Template fitting was then done using  incorrect templates comprising   a Gaussian pulse convolved with an exponential PBF. We show results for three different widths for the `emitted' pulse to illustrate that the degree of bias depends on the widths of both the emitted pulse and the PBF.}
 
 The left panel shows the scattering time estimate $\widehat{\tau}_{\rm GE}$ for a  Gaussian pulse  scattered by a PBF with $\beta = 11/3$ and
 $\zeta = 0.01$.   The arrival-time bias  (right panel) depends on the kind of template used for estimating arrival times:
 \begin{enumerate}
 \item
 If scattering is ignored by usage of a Gaussian template (wide solid lines), the ratio of \TOA\ to $\tau$ exceeds unity for small $\tau$ and declines to $< 1$ for larger $\tau$.  The \toa\  bias is larger for
 wider pulses. 
 \item If a G*E template (i.e. a Gaussian convolved with an exponential PBF) is used with scattering time fixed to the true value of $\tau$ (thin solid lines), the bias is smaller but shows the same trends as with a G-only template;
 \item If $\tau$ is optimized with a G*E template,  the \TOA\ bias is {\it negative} (dashed lines).   
 \end{enumerate}
 The excess of the scattering time estimate over the true scattering time  declines with larger
 $\tau$ and  wider Gaussian pulses. 
 While the  ratio of \toa\ to $\tau$ asymptotes to zero for large $\tau$,   the \toa\ bias is still significant
 for values of $\taud$ encountered in many pulsar timing contexts. 
 \newtext{
 Lastly, the bias is expected to decrease for PBFs from media with 
 $\beta \to 4$ or those for which scattering is dominated by a `large' inner scale.
 We note that PTA measurements in the 1 to 2\,GHz range on  typically low to moderate DMs will involve heavy tailed PBFs like that used in the simulations.
 }
 
\begin{figure}[ht!] 
   \centering
\includegraphics[width=0.45\linewidth]{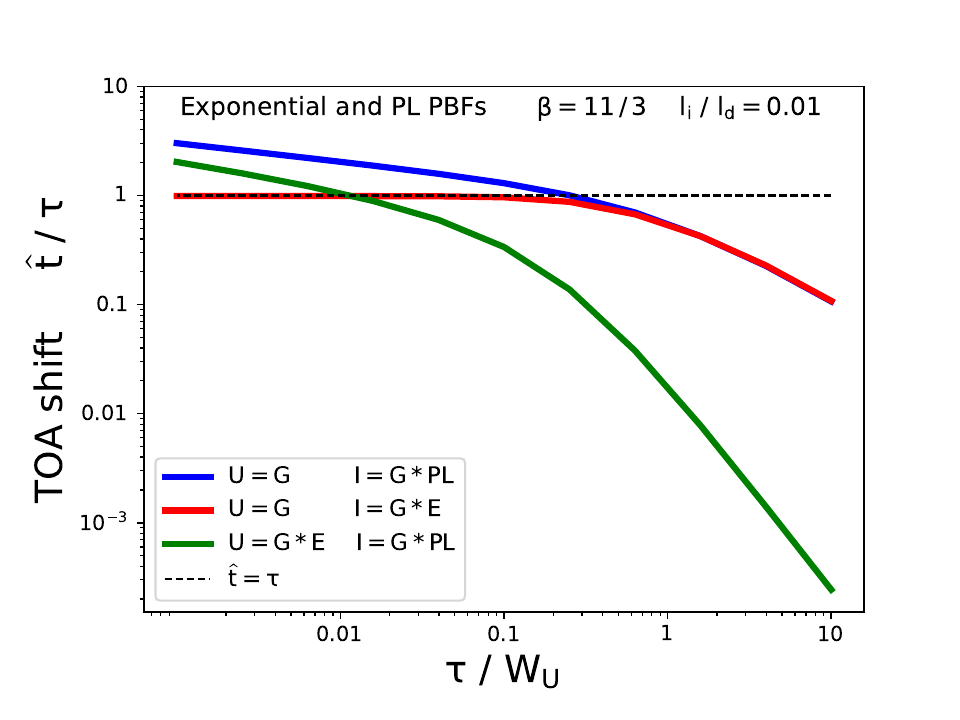}
\includegraphics[width=0.45\linewidth]{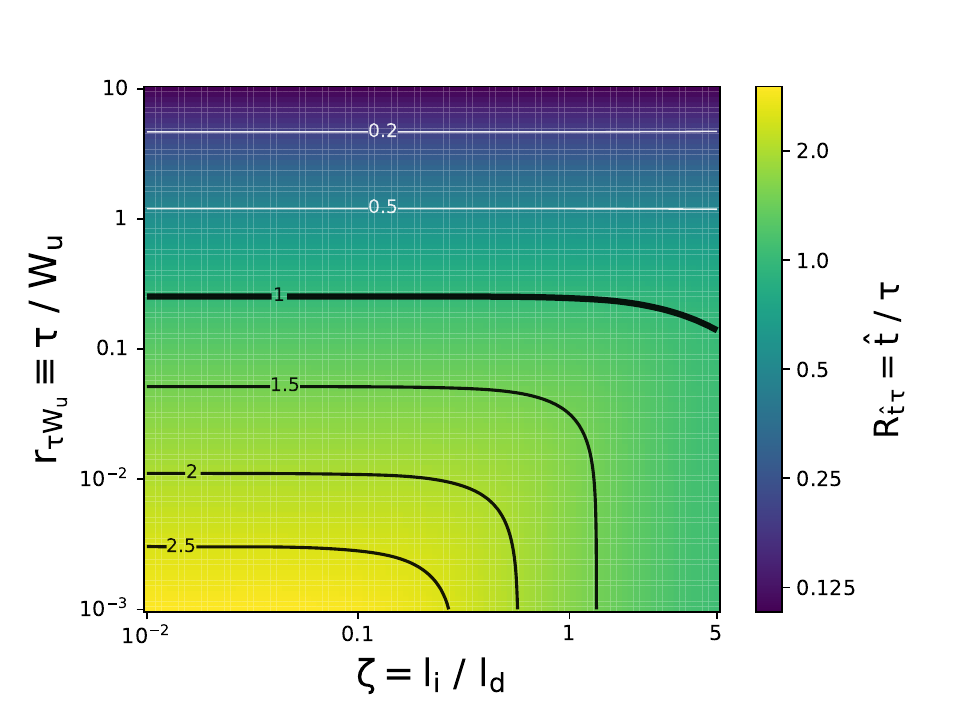}
%
   \caption{Simulated arrival time shifts resulting from template fitting of Gaussian pulses scattered by thin phase screens.
   Left:
   The normalized shift $\widehat t / \tau$ is plotted vs. $\tau / \Wu$, where $\Wu$ is the width (FWHM)  of the assumed Gaussian pulse scattered by either an exponential PBF or a PBF from a power-law medium
   with $\beta = 11/3$  and a small inner scale, $\zeta = \linner / \ld = 0.01$. 
   The scattering time $\tau$ is the $1/e$ width for both kinds of PBF. 
   The red curve is for a Gaussian-exponential pulse with a Gaussian template.   
   The blue curve is for a power-law PBF convolved with a Gaussian that is also  template fitted with the same Gaussian.
   The green curve results from the same pulse shape as for the blue curve but with a template comprising the Gaussian convolved with an exponential PBF having the same $\tau$ as the power-law PBF.  For larger values of $\linner / \ld$, the 
   blue lowers to values asymptotically closer to unity for $\tau / \Wu \lesssim 0.1$. 
   \label{fig:ttau_vs_tauWu}
   Right:  
   Shift ratio $R_{\widehat{t} / \tau} = \widehat{t} / \tau$  vs. $\zeta = \linner / \ld$ and 
  $r_{\tau\Wu} = \widehat\tau  / \Wu$, shown as pseudo color and contours with levels of  
  $R_{\widehat{t} / \tau} $ marked. 
   \label{fig:ttau_vs_zeta}
       } 
\end{figure}

\newtext{We conducted additional simulations to quantify  how  the mean-shift regime applies to (pseudo) data profiles that involve an exponential PBF but fails for those with  PL PBFs.
For one set of cases, we generated Gaussian emitted pulses scattered with an exponential PBF (G*E) but template fitted using only a Gaussian template (G).   Other cases involve data profiles comprising Gaussian pulses convolved with a power-law PBF (G*PL) that are template fitted with either a G or a G*E template.  We do not show results for a G*PL template because they would show zero bias. 
}

Fig.~\ref{fig:ttau_vs_tauWu} (left panel) shows the ratio $\that / \tau$ plotted against $\tau / \Wu$ to indicate the degree of \TOA\ bias associated with a particular template-fitting procedure.  
For the exponential PBF, the mean-shift regime can be seen  to apply for $\tau / \Wu \ll 1$ where  $\that / \tau = 1$ (red curve in the left-hand panel).    However, the mean-shift result is violated by the Kolmogorov cases (blue and green curves) which show
$\that > \tau$ for small $\tau/\Wu$.   This discordance follows from the long tail of the Kolmogorov PBF
when the inner scale is small (Sect.~\ref{sec:pbf_gallery}).  For both kinds of PBF, $\that < \tau$ by increasing amounts
for $\tau / \Wu > 1$.      This indicates that in most cases,  the \TOA\  shift differs from $\tau$.  
The ISM yields epoch-dependent $\tau$ values, so any correction for variable PBF envelopes must take into account the ratio of $\tau$ to the intrinsic width.   The trends shown here for a Gaussian pulse extend to arbitrary pulse shapes with finite widths. 
\newtext{
The right panel of Fig.~\ref{fig:ttau_vs_tauWu} shows the bias in $\that / \tau$ as a function of  the normalized inner scale $\zeta$ along with  $\tau / \Wu$. 
}

In practice, \TOAs\ are obtained over a wide frequency range for estimation and removal of chromatic propagation delays.  These estimates are affected by the well known chromaticity of the scattering time, $\tau \propto \nu^{-\xtau}$ with
$\xtau \sim 4$ but they are also affected by the non self-similar
PBF shapes from the strong variation of the diffraction scale $\ld$ with frequency, which alters $\zeta = \linner / \ld$ and thus the PBF shape.  

For a thin screen and strong scattering and using
${\ld}_{_\theta}$ 
(cf.\ Table\,\ref{tab:keyplasmaeq}, item 6), 
\be
{\ld}_{_\theta}  = \frac{1}{2\pi \nu} \left[\frac{c\deff}{\taud(\nu)} \right]^{1/2} 
\simeq 	1.53\times 10^4\,\rm km\ 
		\times
		\frac{1}{\nu}
		\left[\frac{\deff} {{\taud(\nu)}} \right]^{1/2}
\label{eq:ld_for_zeta}
\ee
for  $\nu$  in GHz, $\deff$ in kpc, and ${\taud}$  in microseconds. 
Using Eq.~\eqref{eq:ld_for_zeta} and $\linner = 10^3 \, {\rm km}\, {\linner}_{3}$, 
\be
\zeta \equiv \frac{\linner}{{\ld}_{_\theta} } 
\simeq 
0.0653 \times (\nu{\linner}_3)
\left ( \frac{{\taud}} {\deff} \right)^{1/2},
\label{eq:zetaeval}
\ee
so $\zeta \ll 1$ and  a heavy tailed PBF is expected for nominal values,
which are relevant to most MSPs observed in PTA programs.  For heavily scattered pulsars,
$\zeta \gtrsim 1$ and the PBF can satisfy the requirements for the mean-shift regime because it tends toward an exponential form. 

Table\,\ref{tab:scattering_correction_regimes} shows the phase space for scattering corrections in terms of $\zeta$ and the ratio $\taud / \Wu$. 
For the mean-shift regime to apply (upper left quadrant in the table),  the PBF must be exponentially bounded with   the ratio  $r_{\tau\Wu} = \taud / \Wu \ll 1$.  
A power-law medium  with an asymptotically exponential PBF also requires $\zeta \gg 1$.  
Together, these two conditions constrain the scattering time to the range, 
\be
 234\,{\rm \mu s} \times   (\nu {\linner}_3)^{-2}  \deff  \  \ll  {\taud}   \ll   \ 10^3\, {\rm \mu s}\,  \times  \Wu. 
\label{eq:inequalities1}
\ee
The mean-shift regime  applies to heavy scattering of  pulsars with wide pulses.    The other three quadrants require full pulse-shape modeling\footnote{Note that a template that includes scattering, albeit with a fixed amount, will yield \TOA\ errors due to epoch dependence of the scattering.}
  or deconvolution of the PBF from measured pulses  in timing analyses. 
One of these is the  lower left
quadrant applicable  to the precision timing of  MSPs for PTAs.    

\providecommand{\Wu}{W_U}
\renewcommand{\arraystretch}{0.8}

\begin{deluxetable}{l c l l }[ht!]
\tablewidth{600pt}
\tablecaption{\label{tab:scattering_correction_regimes}
Scattering Correction Phase Space }      
\tablecolumns{4} 
\tablehead{\\
 \multicolumn{2}{c}{\large Inner scale  regime}   & \multicolumn{2}{c} {\large $ r_{\tau W} = \tau / \Wu$}   \\		 
 \\
 \multicolumn{2}{c}{-----------------------------------------------------------}  &  \multicolumn{2}{c}{-----------------------------------------------------------} \\
     \multicolumn{2}{c}{$\zeta  = \linner / \ld $}             &  \multicolumn{1}{l} {\quad\quad $\ll 1$}        
                 & \multicolumn{1}{l} {\quad\quad $\gg 1$}
 }
\renewcommand{\arraystretch}{0.7}
\startdata
\\
&  
	&  \multicolumn{1}{l} {\bf  Mean shift regime}  
	&  \multicolumn{1}{l} {\bf  Modeling required}  \\
	&  \ \ &
	$I(t) = a U(t-\langle t \rangle_p$  &   \ \ $I(t) = a U * p$ 
	\\
	& \ \  &
	PBF $\to$ Exp  & \ \ PBF $\longrightarrow$ Exp 
	\\
	& \ \  $\zeta \gg 1$  \quad\quad\quad  &
	Large $\DM$, low $\nu$      & \ \  Large $\DM$, low $\nu$ 
	\\
	&  ($\ld$ small  $\to  \tau$ large) &  Large $\Wu$                       & \ \ Small $\Wu$ 
	\\ 
	& \ \ &
	Large $P$ pulsars, magnetars & \ \  Many pulsars 
	\\ \\
\cline{1-4} &
\\ 
\\
&
	&  \multicolumn{1}{l} {\bf  Modeling required}  
	&  \multicolumn{1}{l} {\bf  Modeling required}  \\
	&  \ \ &
	$I(t) = a U * p$   & \ \  $I(t) = a U * p$ 
	\\
	& \ \  &
	Heavy-tailed PL PBF   & \ \   Heavy-tailed PBF 
	\\
	& \ \ $\zeta  \ll 1$ \quad\quad\quad  &
	Small $\DM$, high  $\nu$      & \ \  Small  $\DM$, high $\nu$ 
	\\
	&  ($\ld$ large $\to \tau$ small)  &
        Small to large $\Wu$                      & \ \ Very small $\Wu$ 
	\\
	& \ \ &
	MSPs to magnetars                       &  \ \ Microstructure, nanoshots
	\\
	& \ \  &
	Most PTA pulsar data  
	& \\
\\
\enddata
\end{deluxetable}

\subsubsection{Time delays from heavy tailed  PBF envelopes}
\label{sec:heavytails}

The variety of PBF envelopes has implications for diagnosing and correcting arrival times for scattering delays. 
Here we make the important distinction between the {\it PBF delay}, the mean delay calculated by treating the PBF as a probability density function,  and the {\it excess \TOA} induced by the PBF
through convolution with the intrinsic pulse shape.  Generally they are not the same.     We first discuss  PBF delays  and  then analyze measureable \TOA\ delays that are of greater importance to pulsar timing applications. 

Arrival time delays caused by PBFs are generally smaller than the mean PBF.   

Template fitting to obtain an arrival time estimate is equivalent to maximizing the cross correlation
$\CIU(t)$ between the profile and template $U$, 
\be
\partial_{t} \CIU(\that) = 0 
\quad\quad \text{where} \quad\quad
\CIU(t) = \int d\tp\, U(\tp)  I(\tp+ t -t_0), 
\label{eq:CIU}
\ee
Inspection of $\CIU$ indicates that a template identical to the profile shape,  $U \propto I $, yields an unbiased estimate $\that = t_0$.    However, the variability of interstellar effects and the also chromatic  and sometimes variable intrinsic pulse shapes guarantee that this will rarely  apply\footnote{X-and $\gamma$-ray timing are not affected by  interstellar  plasma propagation but profiles are energy dependent.   Low energy X-ray pulses can be affected by scattering from interstellar grains as well as being absorbed.
These effects are not expected to be epoch dependent, at least on time scales of years.
}.  
\newtext{
Whether  departures of profile shapes from template shapes are important depends on the emitted pulse profile, on the ISM along the line of sight, and on the radio frequency. Low DM pulsars observed at high frequencies are less likely to  manifest the consequences of pulse shape discrepancies.  Given that timing programs cover increasingly wider frequency spans (especially to lower frequencies) and are likely to extend to higher-DM pulsars with more sensitive telescopes,  shape mismatches are highly likely to be important in order to obtain the best timing precision. }

Power-law media yield heavy-tailed PBFs and resulting   delays  that can be substantially  larger than those expected from an exponential PBF, as shown in Sect.~\ref{sec:KlikePBFs}.

\begin{figure}[t!] 
   \centering
   \includegraphics[width=0.6\linewidth]{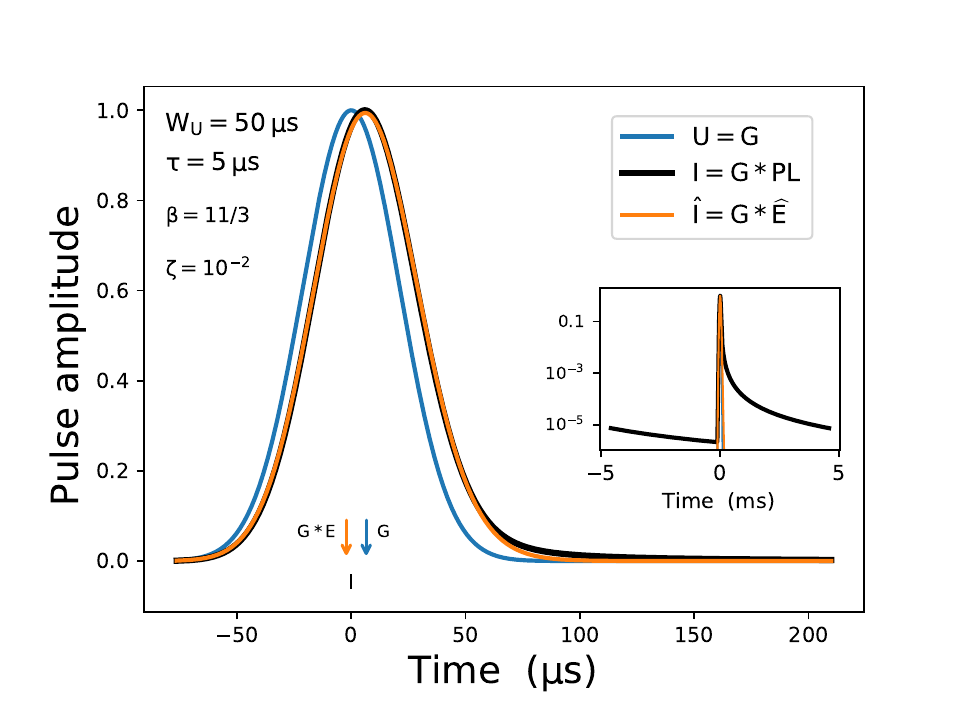} 
   \caption{Fitting of scattered pulses and the resulting \TOA\ 
   \newtext{offsets}.  The Gaussian intrinsic profile (blue) is scattered with a Kolmogorov PBF with $\zeta = 0.01$ (corresponding to an inner scale that is 1\% of the diffraction scale,
   $\ld$).   The scattered profile (black) is fitted with a Gaussian (G)  template that has the same width $W_U$ as the intrinsic profile and it is also fitted with a template comprising a Gaussian convolved with an exponential PBF (G*E, orange);  the exponential PBF has an $e^{-1}$ scale that yields a best fit to the scattered profile.  The true \TOA\ in this case is zero but both templates lead to an error, as shown with the downward arrows.   The G template yields a positive \TOA\ error while the G*E template yields a negative \TOA\ error. Inset:  a zoom-in of the pulse on a log vertical scale that shows the long tail of the actual PBF,   which would extend further than  the 10\,ms period of the simulated pulse but is truncated at one period. }  
   \label{fig:profiles_GPLE}
\end{figure}

Fig.\,\ref{fig:profiles_GPLE} shows an example Gaussian pulse scattered with a Kolmogorov PBF with a finite inner scale ($\zeta = 0.01)$.  The Gaussian width (FWHM) is 50\,$\mus$ and the scattering time is $\taud = 5\,\mu s$. A perfect template (equal to the Gaussian convolved with the true PL PBF)  would yield zero \TOA\ for this example (black tick mark shown at the bottom of the figure).   A Gaussian-only (G) template yields a positive \TOA\ error of 
$6.53\,\mu s$ while a best fit using an exponential PBF convolved with a Gaussian  (G*E) yields a negative \TOA\ error of $-2.16\,\mu s$.   The inset shows the wraparound of the long tail of the PBF.     The scattered profile can be evaluated as the convolution of the intrinsic pulse  over a single pulse period $P$ with a synchronously averaged (`folded') PBF envelope given by
\be
{\PBF}_{\rm wrap}(t) = \sum_{m=0}^{M} \PBF(t + mP).
\ee
Heavy tailed PBFs can extend beyond the $1/e$ time scale $\taud$ by many orders of magnitude.  The multiplicative  factor is $(2\pi / \zeta)^{2} = (2\pi\ld  / \linner)^2 \gg 1$ for $\zeta \ll 1$.   Even for low-DM pulsars with small $\taud = 10$\,ns, the tail contributes significantly across the entire period of most MSPs, as demonstrated in the inset of  Fig.\,\ref{fig:profiles_GPLE}.

\begin{figure}[h!] 
   \centering
\includegraphics[width=0.475\linewidth]{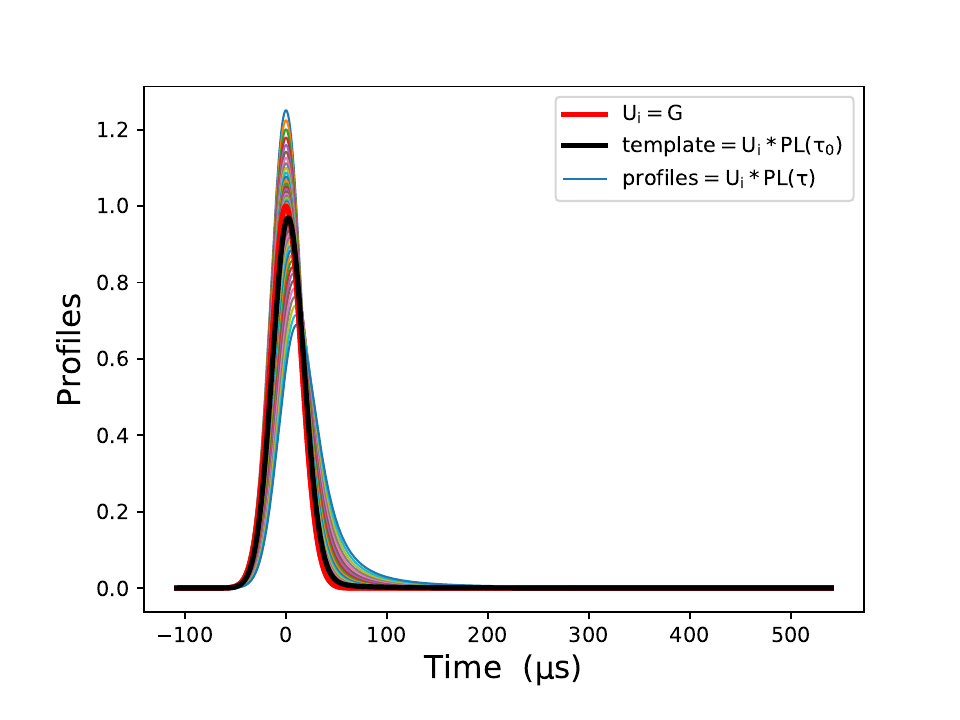}
\includegraphics[width=0.475\linewidth]{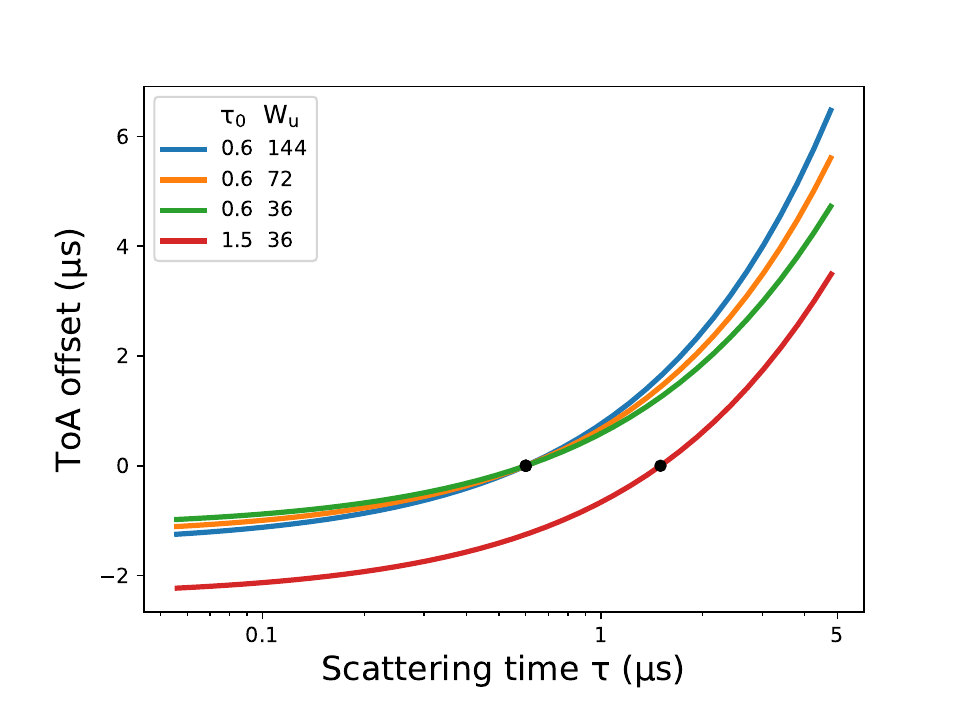}
\caption{Arrival time analysis of scattered pulses using a fixed template. 
   Left: Simulated scattered pulse profiles.   The intrinsic shape is a Gaussian function with 
   width (FWHM)  $\WU = 36\,\mu$s.   The template equals the intrinsic shape  convolved with a PBF 
   with fixed ${\taud}_0 = 1.5\,\mu$s.    The heavy-tailed PBF shape is for a thin screen having Kolmogorov fluctuations 
   and small inner scale. 
   Individual profiles are calculated with scattering times ranging from 
   0.1 to 10$\times {\taud}_0$. 
 Right: 
 \TOA\ offset vs scattering time when a  template with fixed scattering time ${\taud}_0$ is applied to 
 profiles having a range of scattering times $\taud$.   The curves apply to different combinations of template 
 width $\WU$ and template scattering times, as indicated in the legend in $\mu$s.  Black circles show the expected
 zero \TOA\ offsets that result from identical   profile and template scattering times. 
   \label{fig:profiles_toas_pl}
   }
\end{figure}

 Results for power-law PBFs are shown in 
 Fig.~\ref{fig:profiles_toas_pl}  (left panel) for a Gaussian 
 intrinsic pulse shape and  a variable (e.g. with epoch)  $\taud(\tbar)$, giving
 $I =  {\rm G*PL}(\taud)$ .
 A template comprising  a Gaussian pulse and  PBF  with a fixed  ${\taud}_0$ 
 $U = {\rm G*PL}({\taud}_0)$
 yields  systematic   \TOA\ offsets   when scattering differs from that of the template   (right panel). 
 The different cases show that the \TOA\ error depends on both the scattering time and the widths of the unscattered  pulses and is positive or negative depending on the difference  $\Delta\taud = \taud(\tbar) - {\taud}_0$. 

\subsubsection{\TOA\ shifts from scattering islands}
\label{sec:islands}

\newtext{
The PBF envelopes  utilized in the analysis of the previous section and that commonly appear in the literature \citep[e.g.][]{1975ApJ...201..532L, 1978RQE...20..581O, lr99}
}
are highly idealized because they are based on simple scattering geometries and media.   Refraction, truncated screens, and  inhomogeneous scattering across screens 
(Sect.~\ref{sec:truncated}) present numerous types of departures from the simplest geometries, too many to make
generalized conclusions.  Instead we consider 
an alternate setup where multipath takes the form of multiple  ($\Ni$) distinct  images, each yielding a pulse that is  a copy of the template.   

The relevance of multiple image  islands comes from studies of scintillation secondary spectra, which sometimes indicate the presence of two or more separate image patches \citep[e.g.][]{hs08, 2010ApJ...708..232B}.    These typically comprise a dominant image combined  with subimages with substantially less flux density.  However, the corresponding pulses can arrive significantly later than the primary pulse.   Here we assess how the strength of a subimage and its delay combine to influence the net arrival time.

The  measured pulse is the superposition of a strong,  primary pulse with unit amplitude
and $\Ni$ weaker pulses with amplitudes $\epsilon_j \ll 1$ and delays $\Delta_j$, 
\be
I(t) = U(t-t_0)  + \sum_{j=0}^{\Ni-1}  \epsilon_j U(t-t_0 -\Delta_j). 
\label{eq:tmultiple_mages}
\ee
Template fitting with $U(t)$ gives a \TOA\ offset from $t_0$ written in terms of the template's ACF, $R_U(\tau) \propto \int dt\,U(t) U(t+\tau)$,
\be
\delta t_{\rm mult} 
\simeq 
	\frac{ \sum_j \epsilon_j R_U^{\,\prime}(\Delta_j)} 
	       { R_U^{\,\prime\prime}(0) + \sum_j \epsilon_j R_U^{\,\prime\prime}(\Delta_j)} 
\ \ \xrightarrow[{W_{\rm a} =  {\rm width} @}\ 1/e\  ]{\rm Gaussian}\ \ \sum_j \epsilon_j  \Delta_j e^{-(\Delta_j / W_{\rm a})^2}, 
\ee
where the final expression assumes a Gaussian pulse with width $W = W_{\rm a} / \sqrt{2}$ and that
$\sum_j \epsilon_j R_U(\Delta_j) \ll 1$. 

For a single extra image, the  maximum offset is  $\delta t_{\rm dual} \simeq \epsilon\Delta / e$ for $\Delta = W$ and the offset  is zero if the two pulses coincide ($\Delta = 0$)  or are far apart  ($\Delta\gg W$). 
An example indicates the sensitivity to the presence of a weak secondary image.  For $\Delta = W = 100\,\mu$s,  $\delta t_{\rm dual} < 100\,{\rm ns}$ requires $\epsilon < 2.7\times10^{-3}$.     On the other hand, a smaller
separation $\Delta = 1\,\mu$s yields the same limit on $\delta t_{\rm dual}$ for a brighter secondary pulse
with $\epsilon < 0.27$.   

Extra subimages render additional stochasticity to PBF shapes that certainly require empirical characterization rather than textbook modeling.   However, the stability of the PBF envelope over multiple epochs can help.   Monitoring of pulsars indicates that scattering patches can persist for months or years 
\citep[][]{hsa+05} with smooth evolution of their structure. 

\newcommand{\xthattau}{x_{\that\tau}}
\newcommand{\xthat}{x_{\that}}

\subsubsection{Implications for the frequency scaling of timing offsets from scattering}
\label{sec:nu_toaoffsets}

Of interest here is  the frequency scaling of systematic timing offsets  for low-DM MSPs used in PTAs.  
Recall that if the actual scattering were from an exponential PBF and $\taud \ll \Wu$, the offset would be $\that  = \taud$.   
The left hand panel of Fig.~\ref{fig:ttau_vs_zeta} shows that the \TOA\ offset $\that / \tau > 1$ if scattering is from  
a heavy-tailed PBF and  template fitting is done with  an unscattered pulse shape\footnote{There is also an offset if the  template includes scattering with an exponential PBF.}.  The deviation from unity increases  for smaller values of  $r_{\tau W}$.    For a specific pulsar, this corresponds to higher frequency measurements.     

The \TOA\ shifts shown in Fig.~\ref{fig:ttau_vs_zeta} (right panel)
for different $\zeta$ values indicate that
smaller values of $\tau / \Wu$ yield larger fractional \TOA\ shifts
$\that / \tau$ and the largest \TOA\ shifts are for $\zeta \lesssim 0.1$.   This regime applies to nearly all of the
MSPs used in PTAs.   We also see for $\zeta \ll 1$ that $R_{\that\tau} = \that / \tau$ is nearly independent of $\zeta$ but depends systematically on $r_{\tau W}$ (e.g. blue line in the left panel of Fig.~\ref{fig:ttau_vs_zeta}).   

{\it Frequency scaling:}
The net scaling law with frequency combines the scaling $\taud \propto \nu^{-\xtau}$ 
with the scaling of $\that / \taud$ with $\taud / \Wu$ when a Gaussian template is used. 
Numerically we find for $\taud/\Wu \ll 1$ that $\that/ \taud \propto  (\taud / \Wu)^{-\xthattau}$ with $\xthattau \simeq 0.168$ (blue curve in left panel of Fig.~\ref{fig:ttau_vs_zeta} ).    Then 
$\that \propto \nu^{-\xthat}$ with $\xthat = \xtau(1 - \xthattau) \simeq {3.66}$ for $\xtau = 2\beta / (\beta-2) = 4.4$ for $\beta = 11/3$.    For larger ratios $\taud / \Wu \gtrsim 1$, 
$\xthattau$ increases as the blue curve steepens, yielding smaller $\xthat$ and a shallower variation of $\that$ with frequency.
{\it The essential point is that the  dependence of the \TOA\ delay on frequency is weaker  than that of   the scattering time $\tau$  and it gets progressively weaker at lower frequencies.}  

{\it Scattering corrections:} Improvement of arrival times  can be implemented in two alternate ways.  First is pulse-shape modeling (template fitting) that includes PBF effects in the templates used to obtain  \TOAs.     The second approach obtains \TOAs\ using a template that incorporates scattering incorrectly (via an incorrect $\tau$ or incorrect shape)  but makes post-facto  scattering corrections  by using using appropriate terms in a \TOA\ model.   The latter approach is made problematic by chromatic  PBFs  that lead to non-trivial scalings of scattering-induced delays with frequency.     

An example case is for an MSP  with $\DM = 10\,\DMunits$ and scattering time $\taud \simeq 6.2$\,ns at 1\,GHz estimated from the hockey-stick relation (Eq.~\eqref{eq:taudm}).  This gives $\zeta \simeq 0.018\, {\linner}_3 \nu^{-6/5}$ for $\beta = 11/3$.   
\newtext{
Applying this scaling law to multifrequency observations spanning 0.4 to 2\,GHz, we obtain ranges of   7:1  and 1200:1 for   $\zeta$ and $\tau$, respectively, which imply that profiles and \TOAs\ would  span two or more of the quadrants identified in  Table\,\ref{tab:scattering_correction_regimes}. }  The degree to which dispersion and scattering delays are mitigated depends on  how carefully systematic PBF envelope effects are dealt with
(along with finite-scintle effects as well). 

\citet[][]{2024MNRAS.535.1184S} explored scattering corrections on simulated data using an exponential PBF and two alternative frequency scalings for the scattering time,
$\taud \propto \nu^{-22/5}$ (the Kolmogorov inertial subrange value) and $\taud \propto \nu^{-4}$, which applies to a Kolmogorov medium with a large inner scale or to a monoscale medium with scales $a \gg \ld$.
In their study, 
scattered profiles generated over a wide frequency band  yielded biased DM estimates when a template devoid of scattering was applied, as  expected,  given the positive time shifts imposed by the asymmetric PBF. 
Bias was eliminated when the template included scattering, also as expected.  We note two issues with the approach used.  First, it is inconsistent to use an exponential PBF and any frequency scaling other than $\taud \propto \nu^{-4}$.   Second, while  simulations showed that scattering imposed with an exponential PBF could be recovered by model fitting with an exponential PBF, application to actual pulses that include scattering by a heavy-tailed PBF would not recover the scattering, \newtext{as we have shown in Sect.~\ref{sec:heavytails}}. The bias and enhanced random errors due to mismatch between model and actual PBFs is discussed further in
the next section.

\subsection{Systematic timing errors from mismatched templates and PBFs}
\label{sec:systematic_errors_tau_toa}

Removal of PBF-shape effects in \TOA\ estimates raises several thorny issues.   
First,  the epoch dependence of PBFs   requires a template  that  has the  PBF for each epoch `baked in'  if scattering delays are to be removed perfectly.   
Second,  both the width and the shape of the  PBF are  chromatic.  
 The PBF's characteristic width scales as the pulse broadening time $\taud \propto \nu^{-\xtau}$ with $\xtau \sim 4$. More importantly,  its actual shape depends on frequency if the inner scale is finite and comparable to the diffraction scale $\ld$ at any of the observation frequencies.  This is more likely at lower frequencies given the scaling  $\ld \propto \nu^{2\beta / (\beta -2)}  \propto \nu^{22/5}$ for the Kolmogorov inertial subrange.    An optimal template therefore needs to be frequency dependent for this reason as well as to account for the chromaticity of the intrinsic pulse shape (Sect.~\ref{sec:intrinsic_chromatic}).
Third,  while significant prior information is known about scaling laws of pulse broadening from the ISM,  uncertainties in those scaling laws 
and the effects of variable refraction yield systematic errors in \TOA\ estimates.   
%
The last conundrum is the desireability of using a template that is constant vs. epoch, which guarantees that mismatches with the true pulse shape at any epoch require post-fitting mitigation. 

To assess systematic timing errors, we  consider application of Eqs.~\eqref{eq:Iactual}-\eqref{eq:CIU} using   
 model  shapes $\Iim$ and $\PBFm$ to form a profile model,
\be
I_{\rm m}(t) = a_{\rm m} \Iim(t-t_{\rm m} ) * \PBFm(t, \tau_{\rm m}),
\label{eq:Imodel}
\ee
that is fitted to a measured  data profile, where frequency dependences of all quantities are implicit. 
%

Optimal TOA estimation requires fitting with $\Iim = \Ii$ and $\PBFm = \PBF$, to obtain the correct \TOA\ $t_0$, pulse broadening time $\tau$, and pulse amplitude $a$ to within uncertainties determined by additive noise\footnote{Actual uncertainties also include pulse jitter and the finite scintle effect, but for the sake of discussion in this section we ignore those distinct contributions.}. However, both  $\Iim$ and $\PBFm$ will differ from the true shapes.  In practice, this optimal situation is never realized due to imprecision in the knowledge of either $\Ii$ or $\PBF$, largely caused by the  chromaticity of intrinsic pulse shapes and the epoch dependence of PBFs.  

Fitting with  $\Iim \ne U$ or $\PBFm \ne \PBF$ yields systematic errors in arrival time and also in the scattering time $\taum$.   These errors propagate into the \TOA\ estimate for $\tinfty$ (cf.\ Eqs.~\eqref{eq:toaeqn1} and \eqref{eq:tnu1})
because multifrequency \TOAs\ are used to estimate \DM\ for extrapolation to infinite frequency and, ideally, there should be a similar extrapolation for  scattering delays.  As for characterization of the ISM using the scattering time, its scaling with frequency, $\taum \propto \nu^{-\xtauhat}$ is also highly affected by mismatches. 
 
 Various methods have been used to model intrinsic profile shapes on a pulsar-by-pulsar basis; 
they  reduce but do not eliminate the difference between the true and modeled  profile shapes at a specific frequency.    One method  adopts a model profile equal to a measured high frequency profile  (minimally affected by scattering)  convolved with a fixed-shape PBF (such as the one-sided exponential) with a scale parameter
$\tau(\nu) \propto \nu^{-\xtau}$ where $\xtau$ is either fixed at some value or also included as a fitting parameter.
\newtext{This approach originated shortly after the discovery of pulsars and is summarized briefly in 
\citep[][and references therein]{1972MNRAS.157...55W}
and analyzed in \citep[][]{2016MNRAS.462.2587G}.}
Given that  the exponential PBF is unlikely to be an accurate representation of the true PBF in most cases, systematic errors should be expected not only in \TOAs\ but also in $\tau$ and $\xtau$, \newtext{as also discussed by 
\citep[][]{2016MNRAS.462.2587G}}.

\begin{figure}
    \centering     
    \includegraphics[width=0.45\linewidth]{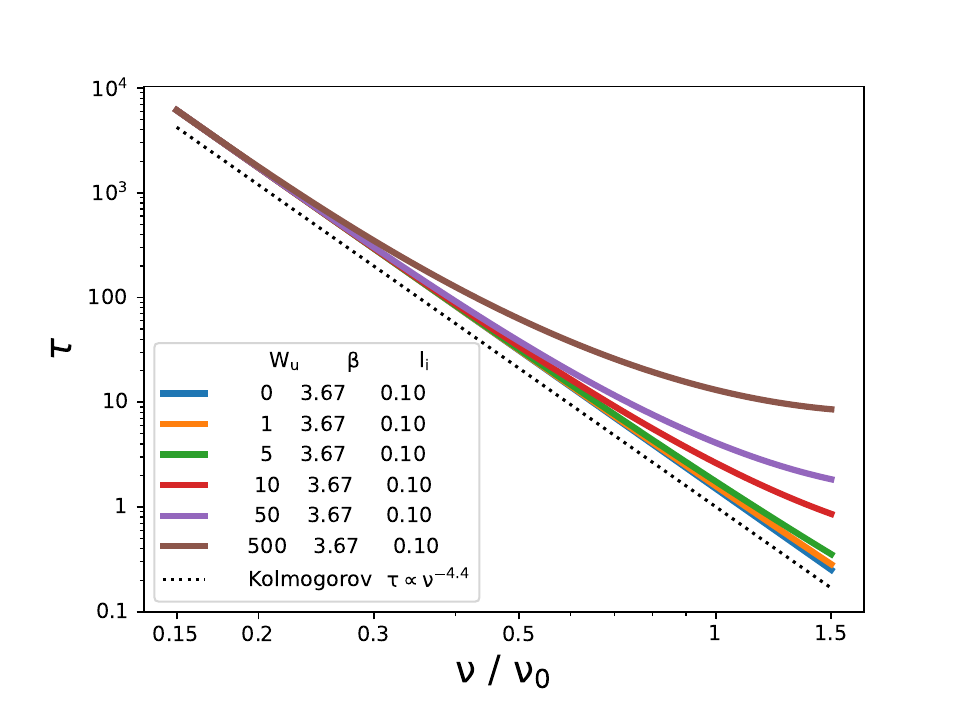}
    \includegraphics[width=0.45\linewidth]{plot_xtau_vs_nu_tau_vs_nu_Wu_500_siU_3.67_liU_0.10_2023-08-0.pdf}
    \caption{Effects of an incorrect PBF: Scaling of the estimated scattering time $\tau$ vs. frequency
    for power-law scattering with  $\beta=11/3$ and a small inner scale $\linner = 0.1$ in units 
    where the diffraction scale ${\ld}_{_0} = 1$ at a reference frequency $\nu_0$.   
    The quantity $\zeta(\nu) = (\linner / {\ld}_{_0}) (\nu / \nu_0)^{-2 / (\beta-2)} \to \zeta_0 (\nu / \nu_0)^{-6/5}$
    increases with decreasing frequency, shortening the tail of the PBF. 
    Left:  $\tau(\nu)$ vs. $\nu$ showing curves for different intrinsic (unscattered) pulse widths $\Wu$ 
    where $\tau$ \newtext{(arbitrary units)} is determined by model fitting the intrinsic shape scattered with an exponential PBF 
    instead of the true power-law PBF.  The dotted line shows the Kolmogorov scaling 
    $\propto \nu^{-4.4}$ for the inertial subrange. 
    Right: Exponent of the apparent scaling law $\tau(\nu) \propto \nu^{-\xtau}. $}
    \label{fig:xtaud}
\end{figure}


Fig.~\ref{fig:xtaud} shows the effects of mismatches on the scaling of scattering times with frequency.    Simulated Gaussian pulses with different widths $\Wu$ are scattered with a heavy tailed PBF for  a  value $\zeta = 0.1$ at a reference frequency $\nu_0$ and a scaling $\zeta(\nu) \propto \zeta \propto (\nu / \nu_0)^{-2 / (\beta-2)}$.
  Fitting scattered pulses using an exponential PBF yields values of $\tau$ that deviate from true values by amounts that are 
  frequency dependent.   High frequency values for $\tau$ are biased high because the exponential PBF needs to be broader to accommodate  the heavy tail of the `true' PBF (left panel).   The effect is exacerbated for larger $\Wu$.    The implied scaling laws $\tau \propto \nu^{-\xtau}$  have exponents (right panel) that are always smaller than the index of 4.4 if the correct PBF had been used.    

Examples of shallow frequency scalings for $\tau$ (i.e. $\xtau < 4$) include 
an average index $\xtau \simeq 3.44\pm0.13$ \citep[][]{2001ApJ...562L.157L}
from an analysis of nine strongly scattered  pulsars and 
$\xtau \simeq 3.4 \pm 0.2$ for B1933+16 \citep[][]{2004A&A...425..569L}. 
A wide range of index values above, below, and consistent with $\xtau=4.4$
were reported by \citet[][]{10.1093/mnras/stt989} 
while low-frequency observations at 118-180\,MHz \citep{10.1093/mnras/stad429,10.1093/mnras/stae399} show shallow scalings.

The overall tendency is for shallow scalings to be more prevalent for distant, high-DM pulsars observed at higher frequencies and for somewhat lower DMs observed at lower frequncies.  This might suggest greater departures from a Kolmogorov spectrum for high-DM pulsars in the inner Galaxy but that cannot account for the shallow scalings for low-frequency observations.   We are  inclined to interpret many of these results  as due to fitting bias ensuing  from mismatches of adopted and actual PBFs and intrinsic profile shapes.   Exponential PBFs used in all of these studies combined with chromaticity of intrinsic shapes will tend to yield  underestimates of the true $\xtau$.   In some cases, however, the underestimates may be real and caused by confinement of the dominant scattering region transverse to the \LOS\ \citep[][]{2001ApJ...549..997C} rather than being due to a genuine non-Kolmogorov scaling.  This latter interpretation was applied to the scattering seen from J0826+2637 
\citep{10.1093/mnras/stae399,10.1093/mnras/stad429}.   
These authors also found that measurements of scintillation structure  at higher frequencies yielded $\xtau = 4.58\pm 0.16$ (consistent  with the Kolmogorov index $\xtau = 4.4$) while pulse shape modeling for lower-frequency data using an exponential PBF and a Gasussian pulse shape yielded $\xtau = 2.7\pm 0.1$.  This discordance  could be due to transverse scattering confinement, as the authors concluded or it instead may simply be the result of fitting bias.   A thorough discussion of scaling laws,  fitting bias, and conclusions about the scattering properties of the ISM is deferred to another paper (in preparation). 


\begingroup

\renewcommand{\nul}{{\nu_{\rm l}}}
\renewcommand{\nuh}{{\nu_{\rm h}}}
\providecommand{\nulh}{{\nu_{\rm l,h}}}
\providecommand{\tbarnul}{{\tbar_{\rm l}}}
\providecommand{\tbarnuh}{{\tbar_{\rm h}}}
\providecommand{\tbarnulh}{{\tbar_{\rm l,h}}}

\section{Chromatic leakage into achromatic arrival times and residuals}\label{sec:chromatic_leakage}

The goals of most, though not all, timing analysis concern achromatic terms in arrival times, such as  those due to gravitational waves or post-Newtonian terms in orbital motions of pulsars in binaries. For these applications chromatic plasma effects are a nuisance; however,  for studies of Galactic structure of ionized gas and for probing  very compact, discrete structures that underlie ESEs, they are a unique source of information.   

Chromatic leakage into hoped-for achromatic timing residuals is inevitable.   The overriding question is what is  its amplitude relative to  achromatic terms  and whether they are mitigable.
In this section, the removal of chromatic terms from arrival times is discussed with special attention paid to particular procedures that cause
 chromatic leakage.        

To emphasize leakage effects, we re-visit some of the previous results. 
The barycentric \TOA\ at frequency $\nu$ (where the delay and Doppler corrections to the topocentric TOA have been made) is expressed similarly to Eq.~\eqref{eq:tnu1}, but with the scattering-averaged DM delay combined with  a lump chromatic term and excluding the error term $\epsilon_\nu$:  
\be
\tnu = \tinfty + \KDM\DMbar(\tbar, \nu) / \nu^2  + \tC(\tbar, \nu),
\ee
where the  dependence  on epoch $\tbar$ is shown explicitly.

We again consider  \TOAs\ obtained at two frequencies  $\nulh$, with  ratio $R\equiv \nuh / \nul > 1$,  and  at two  epochs $\tbarnulh$, respectively.   These are the center frequencies of two well separated (non-overlapping)
bands.  Then $\DMhat$  is calculated under the false assumption 
that \DM\ is  constant and that the $\tC$ terms are absent.  The achromatic \TOA\ is estimated as $\tinftyhat = t_{\nuh} - \nuh^{-2} \KDM \DMhat$,  which is in error by an amount,
\be
\dtinftyhat = \tinfty-\tinftyhat 
=  
+  \frac{\KDM}{\nuh^2} \left(\frac{R^2}{R^2 - 1}  \right)   \Delta \DMbar_{1,2} 
+ \frac{ {\tC}_1 - R^2 {\tC}_2 }{R^2 -1} ,
\label{eq:dtinfty_leakage}
\ee
where the simplified notation uses  $\DMbar_1 \equiv \DMbar(\tbarnul, \nul)$, ${\tC}_1 = \tC(\tbarnul, \nul)$ etc. 
and defines $\Delta\DMbar_{1,2} = \DMbar_1 - \DMbar_2$.  
%
Ideally  $\dtinftyhat = 0$ but the assumptions underlying  Eq.~\eqref{eq:dtinfty_leakage} lead to non-zero residuals  with superposed terms having different fluctuation statistics, including several lumped together into $\tC$. 

{\it Contributions from DM stochasticity:}  Variations vs.   both epoch and frequency (cf.\ Sect.~\ref{sec:DMt}-\ref{sec:DMnu})  yield chromatic errors in $\tinftyhat$:    
 \begin{enumerate}
\item 
{\it Asynchronous arrival times:}  The epoch dependence of \DM\ alone (without frequency dependence or
 contributions from $\tC$)  yields a chromatic error when two epochs are used to calculate $\DM$.
 Using the  DM structure function $\DDM(\Dt) \propto (\Dt)^{\beta-2}$ (in the inertial subrange for the ISM), 
 \be
 {\rmstinfty}_{,\DM(t)} = 
 \frac{\KDM}{\nuh} 
 \left( \frac{R^2}{R^2 - 1}\right)
 \DDM^{1/2}(\tbarnuh-\tbarnul). 
 \ee
 This contribution vanishes of course when observations are at the same epoch\footnote{Interstellar 
 Kolmogorov-like DM variations of significance are not expected  for time separations less than about one day.   Small ionospheric contributions vary diurnally  and stochastically. Interplanetary contributions will vary with solar elongation and the solar cycle.  Only for lines of sight near the Sun might there be significant variations on sub-day variations.}.
 \item
 {\it Frequency dependent DM:}  Now including DM chromaticity, same epoch observations ($\tbarnul = \tbarnuh = \tbar$) give a contribution  
  \be
 {\rmstinfty}_{,\DM(\nu)} = \frac{\KDM}{\nuh}  \left( \frac{R^2}{R^2 - 1}\right)  \DDMbar^{1/2}(0, \nul, \nuh). 
 \ee
 \end{enumerate}
 
 {\it Contributions from interstellar refraction and diffraction:}  
 The summed  contributions to 
 $\tC(\tbar, \nu) = \sum_j {\tC}_{j} (\tbar, \nu)$ 
 include random, zero-mean and systematic biases that arise from interstellar  refraction and diffraction. The chromaticity of emitted pulses (`profile evolution') also contributes but is not explicitly quantified here.  The different terms have a variety of dependences on RF $\nu$ and fluctuation frequency $f$ that make their removal problematic.   \newtext{Attempts to do so have mostly relied on subtracting a single term in the \TOA\ expression with an assumed RF dependence \citep[e.g.][]{hs08}
 or have made use of Gaussian process modeling 
 \citep[e.g.][]{2024MNRAS.528.3658K},
 like that used for DM fluctuations \citep[e.g.][]{2024MNRAS.528.3658K} with an assumed or fitted
 RF dependence.} Chromatic fluctuations inevitably  remain  in timing residuals with these approaches and that may remain true regardless of the approach.  The contributions of these undesired terms are pulsar and \LOS\ dependent, so they must be managed on a  case by case basis. 
\newcommand{\tCtaud}{{\tC}_{\taud}}
\newcommand{\tCtaur}{{\tC}_{\taur}}
\newcommand{\tCthetar}{{\tC}_{\thetar}}
\newcommand{\tCssbc}{{\tC}_{ssbc}}

 \begin{enumerate}
 \item  {\it Diffractive pulse broadening ($\tCtaud$):} As demonstrated \newtext{in Sect.~\ref{sec:heavytails}}, the time delay  $\tCtaud$ from pulse broadening  is typically less than  the $1/e$ broadening time $\taud$.   There is a complicated dependence of the ensuing 
 error ${\tinfty}_{,\taud}$ on the particular forms of the emitted pulse shape and PBF as well as on the template used for  \TOA\ estimation.   While the amplitude and \RMS\ value of  ${\tinfty}_{,\taud}$ cannot be expressed generally, the fluctuations of $\taud(\tbar, \nu)$ with epoch will be `low-pass' in form with a characteristic time scale comparable to the RISS  time scale $\dtr$ of order days or more. 
 \item {\it Refractive delays ($\tCtaur$):}   The refractive delay $\taur \simeq \deff \thetar^2 / 2c$ is a consequence
 of image or ray-path wandering, differing from  the effects of multipath propagation that underly the diffractive delay.  
 The refraction angle $\thetar < \thetad$ for Kolmogorov like media with $\beta < 4$ and small inner scales
 (cf.\ Fig.\,\ref{fig:screenDM10}-\ref{fig:screenDM300a}) and varies on RISS time scales.  The delay
 $\taur \propto \thetar^2$ will vary more rapidly by about a factor of $\sqrt{2}$ but both will have a low-pass spectrum. 
 \item {\it Refractive error in barycentric \TOAs\  ($\tCssbc$):}   The SSBC transformation error discussed previously scales linearly with $\thetar$ and varies on RISS time scales. 
 \end{enumerate}

This summary  of prominent chromatic effects illustrates that the net chromatic \TOA\ error involves a variety of RF dependences that are accompanied by a mixture of fluctuation spectra, discussed further in the next section. Mitigating the effects is unlikely to be optimal in operations on flawed \TOAs, at least not with just one or two model terms.   The alternative or complementary approach is to minimize chromatic effects on
\TOAs\ to begin with. 

\endgroup
\section{Gaussian and non-Gaussian  timing fluctuations}
\label{sec:nongauss} 

Timing fluctuations generally  comprise  Gaussian-like  processes (via the central limit theorem) intermixed with those with non-Gaussian statistics.   Contributions to these categories come from effects intrinsic to the pulsar and from  extrinsic propagation effects.  
{\it Gaussian elements:} 
Spin stochasticity takes on several forms.  Large-amplitude glitches  occur sparsely and with preferred signs, i.e. spinups with
$\Delta \fspin > 0$.  Spin noise (aka `timing noise') is a sustained process that tends to be random-walk like (cf.\ Sect.~\ref{sec:types_of_spinnoise}) and, for some cases, is made up of identifiable,  discrete jumps in spin frequency or its derivative.  Unlike glitches, these events are much smaller and show both signs.  In other objects,  individual events cannot be identified due to lack of statistical sensitivity 
and possibly due to a high event rate (e.g. many events per month).  Generally, spin noise does not appear to be a Gaussian process for cases with a low event rate while high-rate objects can be modeled as such. 
The prominent red noise in J1939+2134 is consistent with a dense random walk process that is consistent with a Gaussian process. 

Rapidly changing \TOA\ errors from pulse jitter, finite scintle numbers from DISS, and radiometer noise are consistent with Gaussian, white-noise processes. Jitter is largely statistically independent between pulses\footnote{Correlations between single pulses are seen in CPs in the form of drifting subpulses and mode changes, but these typically decorrelate after $\sim$\,ten or so spin periods \newtext{\citep[][and references therein]{2019MNRAS.482.3757B}}.}, leading to a Gaussian error distribution when a large number of pulses is averaged, as is usually the case.  The same holds for additive noise and for errors from template fitting.   A \TOA\ is usually estimated 
over a time-frequency block spanning many scintles, so the CLT applies for the finite scintle effect as  well. 

Propagation through the ISM will produce some fluctuations that are Gaussian.  Those that do will be associated with Kolmogorov density variations and those that do not arise from discrete structures that superpose with Kolmogorov variations. 
Gaussianity  likely holds for DM variations produced by Kolmogorov density fluctuations because DM is contributed to along most of the path length and there is also averaging over the bundle of ray paths reaching an observer (Sect.~\ref{sec:DMnu}).  
Variations in the scattering time $\taud$ on long time scales will also  be Gaussian if refraction is induced by Kolmogorov-like density fluctuations.   
 
{\it Non-Gaussian elements:}  
Spin variations are non-Gaussian for pulsars that manifest discrete events as glitches or as the elements of  the red noise process.   Most MSPs do not show such events except for a few that show glitches.   However, a sizable fraction of MSPs does show red noise consistent with spin noise  \newtext{as discussed below in Sect.~\ref{sec:red_noise_assessment}}.  

Emission in pulsar magnetospheres can produce pulse variations other  than single-pulse phase jitter.  Changes between pulse-shape modes are consistent with stationary  Markov processes in some cases
\citep[][]{2013ApJ...775...47C}, involving  a small number of discrete states that persist for several to a large number of spin periods\footnote{State changes discussed here are changes in pulse shape or in pulse fluctuations.  They do not refer to objects that show transitions between radio and X-ray emitting phases
\citep[e.g.][]{2014ApJ...790...39S,2023A&A...677A..30B}.}. Arrival time variations will be clustered in multiple peaks in histograms of \TOA\ residuals if only a small number of state changes (e.g. profile mode changes with a small set of preferred shapes) occur during an integration time (typically $\lesssim 1$\,hr).   If many occur, \TOA\ variations will tend toward a Gaussian process.   Fortunately, MSPs show very little such state changes though there are a few examples in the literature. 

Refractive caustics from propagation through discrete plasma structures produce several timing effects that are event-like in nature.   These include 
(1) strong intensity variations that influence template fitting errors;
(2) discrete DM events accompanied by refraction angle variations;
and
(3) refractive distortion of pulse broadening functions with attendant  timing offsets, as discussed earlier.  
Collectively, these  effects are  associated with the `extreme scattering event' (ESE) phenomenon\footnote{These events are more refractive  than scattering (diffractive) in nature, so the term is a misnomer.}. 

\begin{figure}[h!] 
   \centering

  \includegraphics[width=0.45\linewidth]{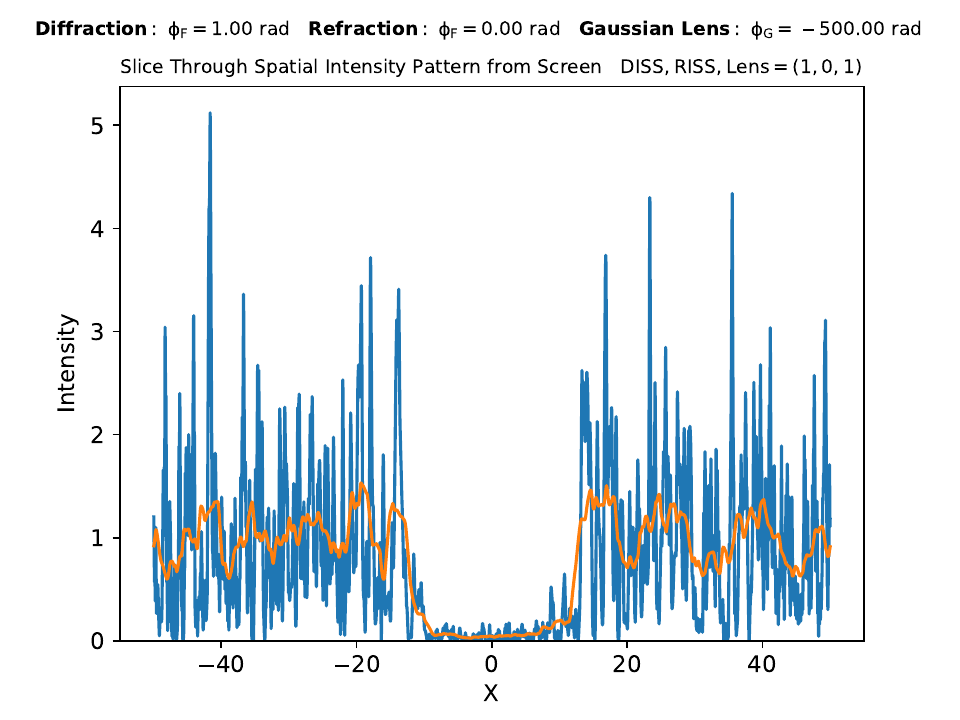}
   \includegraphics[width=0.45\linewidth]{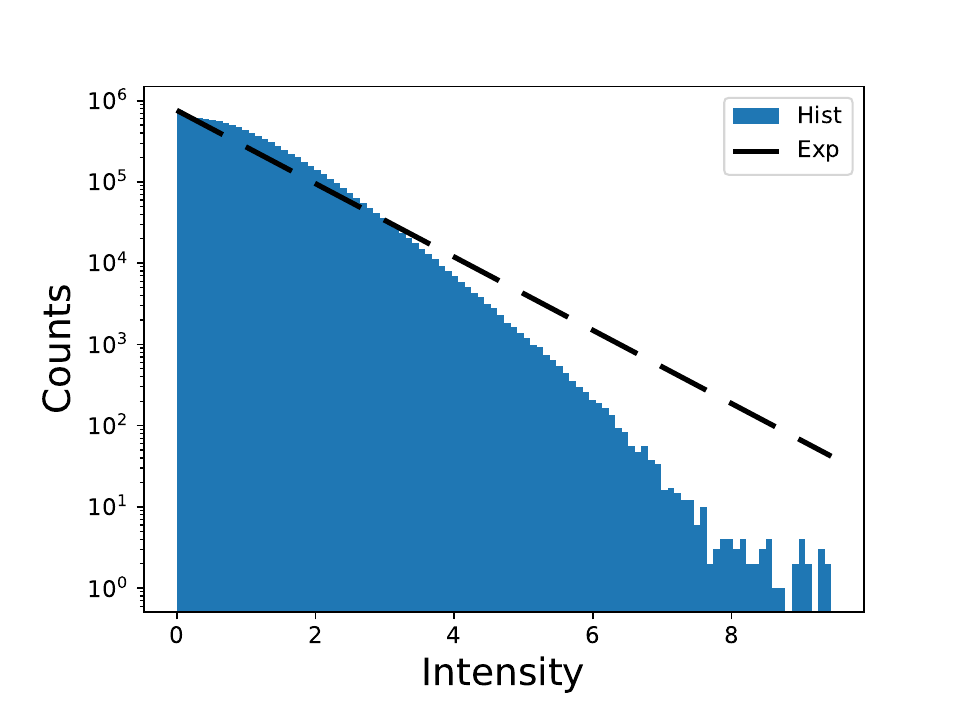}
   
 \includegraphics[width=0.45\linewidth]{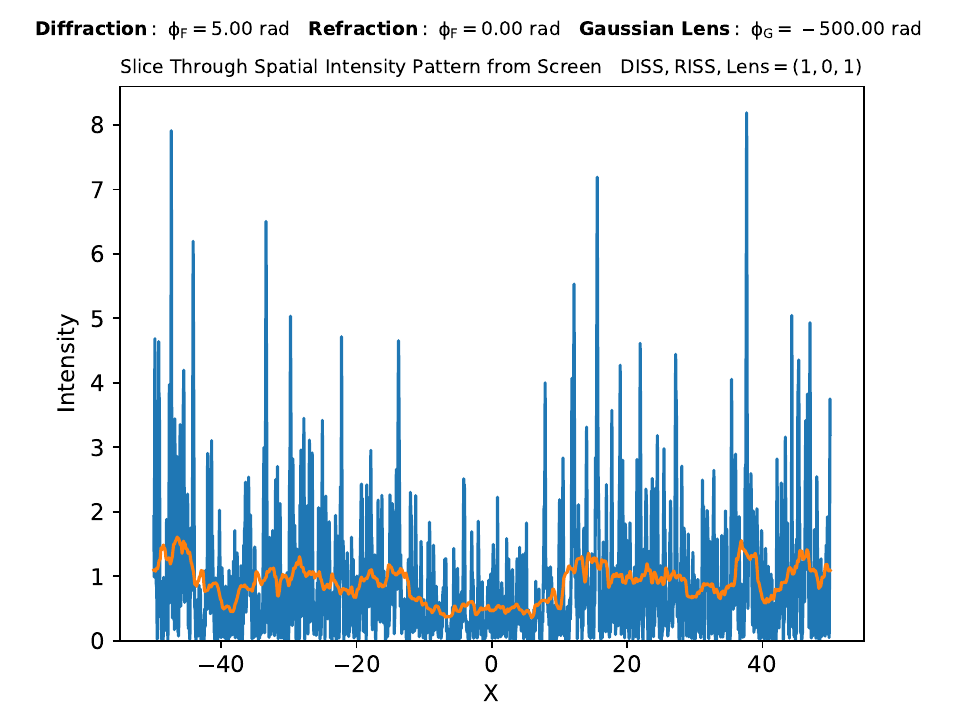}
  \includegraphics[width=0.45\linewidth]{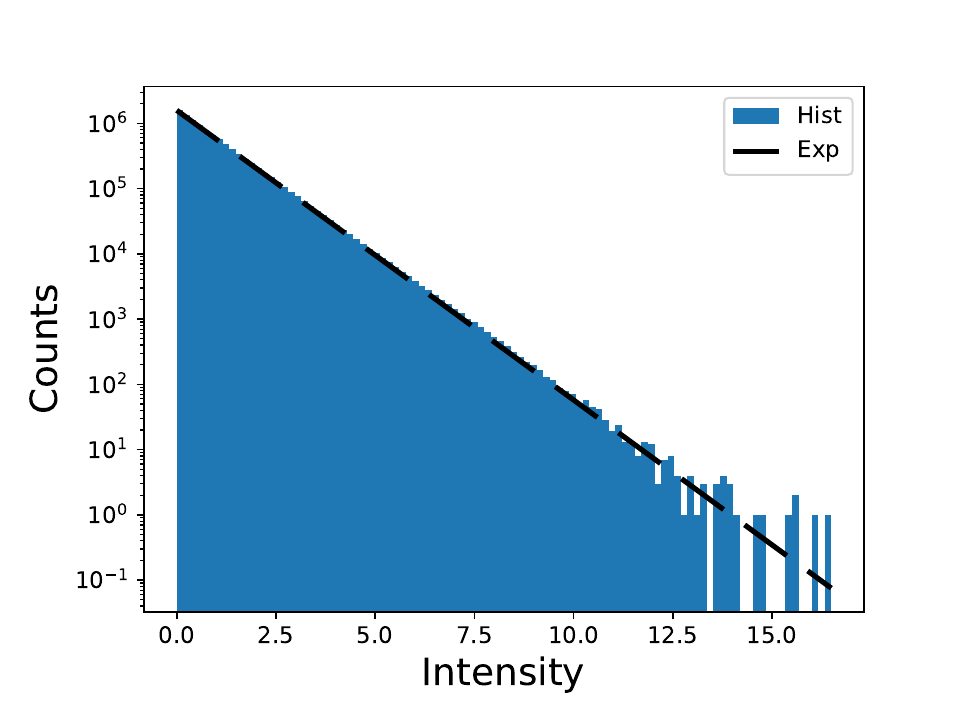}
 \caption{Simulated effects of combined scattering and  plasma lensing on  a pulsar's intensity.  
 Left panels: the  apparent intensity plotted against a slice across the observation plane. 
 Right panels; histograms of intensities across the entire 2D observation plane. 
 Top row:  for a weak scattering screen with ${\phiF}_{\rm d} = 1$\,rad.
 Bottom row: for a stronger screen with ${\phiF}_{\rm d} = 5$\,rad. 
 Intensities are normalized to a mean of unity. 
 The simulated data were produced by evaluating  the Fresnel-Kirchhoff diffraction integral for a phase
 screen that includes diffraction and a Gaussian plasma lens.
 The simulation method is described in 
 Sect.~\ref{app:sims}.
 }
   \label{fig:kdiplasmalens}
\end{figure}

\begin{figure}[htbp] 
   \centering
  \includegraphics[width=0.7\linewidth]{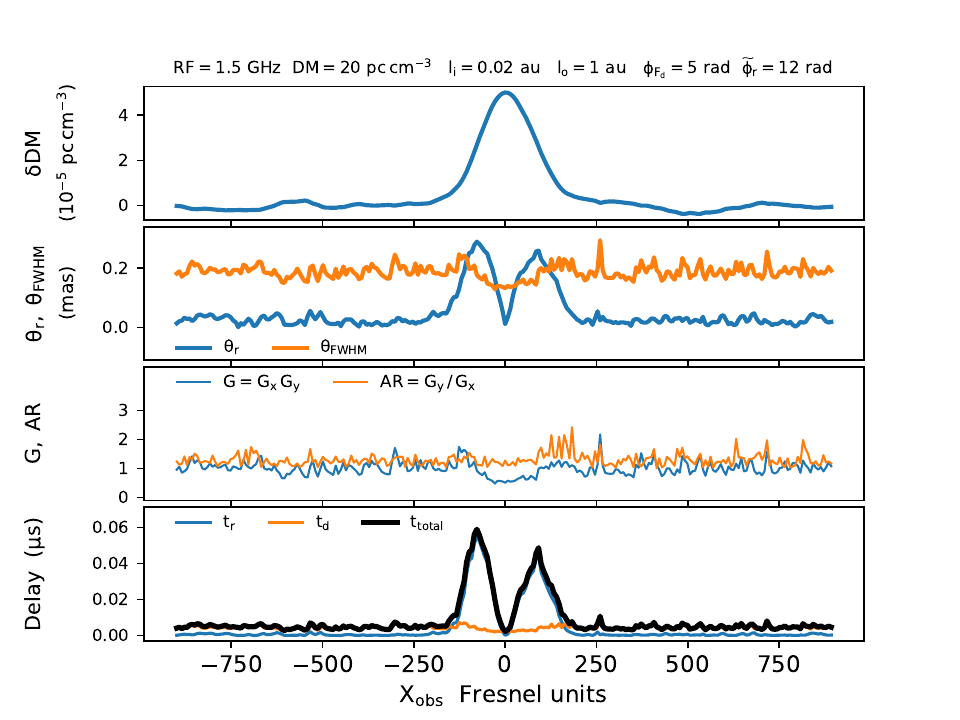}
\caption{Results for a screen with Kolmogorov refractive variations combined with a plasma lens.  Quantities 
are plotted vs.  position along the screen plane that passes through the plasma lens at $X=0$.   
Top frame: DM plotted vs. position in the observation plane.
Second from top: angular size $\theta_{\rm FWHM}$ and centroid refraction angle $\theta_{\rm r}$.
Third from top: Geometrical optics gain $G$  and image axial ratio $G_y / G_x$.  
Bottom:  Time delays: the scattering time $t_{\rm d}$ is determined by the width of the image while the refraction time $t_{\rm r}$ includes the offset of the image from the direct path, which is substantial for positions on either side of the direct path ($X=0$). Lengths are measured in Fresnel scale units, $\rF = (\lambda D / 2\pi)^{1/2} \sim 10^{11}$\,cm.
 }
   \label{fig:ese_slices}
\end{figure}

ESEs involve 
plasma lensing  effects that have been seen in radio light curves of AGNs, including the original discovery by \citet[][]{1987Natur.326..675F}  and other notable cases \citep[e.g.][]{doi:10.1126/science.aac7673} and in pulsar timing data \citep[][]{2015ApJ...808..113C,2018MNRAS.474.4637K}, including the MSP J1939+2134 (B1937+21) 
\citep[][]{1993Natur.366..320C} and the MSP J1643-1224 that is behind the HII region Sh 2-27 \citep[][]{mlc03,2020ApJ...897..124O,2023MNRAS.519.4982D}.   As with the simulations shown in Figs.~\ref{fig:kdiplasmalens} and \ref{fig:ese_slices}, plasma lensing can be more prominent in timing signatures than in  flux densities.  This is particularly the case for low-DM MSPs monitored in PTA programs because DISS variations can easily mask the lensing's effect on the intensity.  The main point,  however is that the episodic nature of plasma events represent departures from Gaussian statistics that might otherwise prevail. 

ESE events seen in intensity time series are accompanied by changes in DM and  angle of arrival, as  discussed in Sect.~\ref{sec:aoatoa}, and by alterations of PBFs discussed in Sect.~\ref{sec:aniso_refract}. 
Refraction angles $\thetar$  from a discrete plasma structure scale natively  as $\nu^{-2}$ and the related timing delays scale as $\nu^{-2}$ and $\nu^{-4}$, respectively,  
and thus match or nearly match those from DM delays and scattering.
However, time delays along propagation paths where  diffraction and refraction both occur will depart from these scalings. 

The effects of a plasma lens are demonstrated in a simulation that also includes Kolmogorov fluctuations.    
Fig.~\ref{fig:kdiplasmalens} shows simulated intensity slices and  histograms. In the top row the intensity trough is easily seen amid the scattering from a screen with ${\phiF}_{\rm d} = 1$\,rad but is not obvious in the bottom panels where lensing is combined with
a stronger screen having ${\phiF}_{\rm d} = 5$\,rad.  The intensity histogram for the first case shows a strong departure from the exponential form expected without the lens.   For the stronger diffraction screen, the histogram is consistent with an exponential except for depletion at intensities about 13 times the mean. 

Fig.~\ref{fig:ese_slices} shows the resulting  images, PBFs, and arrival times obtaind by ray tracing.  
The simulation parameters are approximately match  a pulsar  with DM = 20$\,\DMunits$ combined with a plasma
lens with a small DM depth of $5\times10^{-5}\,\DMunits$.   The first panel shows $\delta \DM (t)$ and the other three panels show the resulting changes in (1) angular size and location of the scattered image (second panel from top);
(2) the intensity modulation `gain' and image axial ratio in the third panel;  and (3) the timing perturbations (bottom panel). Note that the timing perturbations as shown exclude DM delays and thus represent extra variations that are a consequence of angular broadening and angular wandering of the image. 

\begin{figure}[htbp] 
   \centering
 \includegraphics[width=0.7\linewidth]{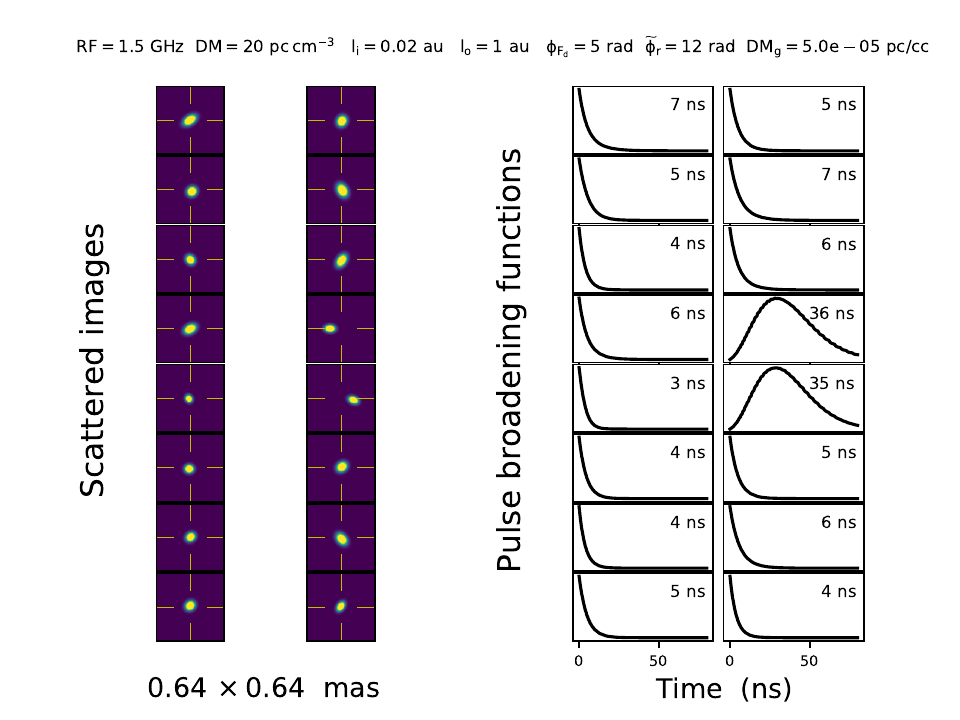}
 \caption{Simulation results  for a refracting screen with Kolmogorov fluctuations combined with a plasma lens.
 Left: images at 16 locations along the observation plane, showing their variable shapes and offsets.
 Right: the corresponding PBFs and mean delays. 
 }
   \label{fig:ese_images_pbfs}
\end{figure}

Fig.~\ref{fig:ese_images_pbfs}  shows how the image shape and location and the corresponding   PBF varies between  16 locations across the observation plane. Images are well centered on the direct line of sight except at locations near the plasma lens.  PBFs maximize at $t=0$ for  most cases but dramatically shift to larger times when the lens is intersected.  

\clearpage

\begingroup
\renewcommand{\nul}{\nu_{\rm l}}
\renewcommand{\nuh}{\nu_{\rm h}}

\section{Fluctuation spectra of timing delays and residuals}
\label{sec:flucspec}

The numerous processes already discussed contribute  to the net fluctuation spectrum of timing residuals.   Some have  white noise signatures while others  have red spectra with power concentrated at lower frequencies.   We distinguish between a red power-law spectrum $\propto f^{-X}$ that extends to frequencies less than $T^{-1}$ (where $T$ is the length of a time series) and a 
`low-pass' spectral form that is flat and then declines for $f$ greater than a bandwidth $B$.   Red power-law processes, such as pulsar spin noise and the GW stochastic background, yield increasing timing residual variance as $T$ increases given the rise in spectral amplitude at lower frequencies. By contrast, processes with low-pass spectra have constant variance  for $T\times B > 1$, examples of which are  the `white' noise processes\footnote{Like all real-world processes, these do not have spectra that are  strictly white but are simply flat over a wide bandwidth.  
\newtext{
In sampled data, values of these processes are assumed to be statistically independent.}} discussed earlier: template fitting errors, jitter, and the DISS finite scintle effect.   Other interstellar effects also approximately show the
low-pass form. 
Here we summarize the spectral models for the salient  noise processes discussed in previous sections.  

\begingroup
\centering 
\newtext{\textit{Nomenclature: We remind the reader that in the following subsections, $\nu$ refers to radio frequency, $f_s$ refers to spin frequency, and $f$ refers to fluctuation frequency.}}
\endgroup

\subsection{White noise processes}  

{\it Radiometer noise and jitter:}
 Signal-to-noise dependent  timing errors from radiometer noise ($\DtSNR$) are  uncorrelated for \TOAs\ obtained from  profiles computed  from independent sets of pulses.  The same is true for jitter errors ($\DtJ$).   Across radio frequency, they differ 
 \newtext{
 because radiometer-noise
 }
 errors are uncorrelated between non-overlapping frequency channels while jitter is highly correlated. 
In the time and RF domains, the cross correlations are then
$
\langle \dtRN(t_i, \nu_k) \dtRN(t_j, \nu_l)\rangle =  \rmstoaS^2  \delta_{ij} \delta_{kl}
$
and
$
\langle \dtJ(t_i, \nu_k) \dtJ(t_j, \nu_l)\rangle =  \rmstoaJ^2  \delta_{ij}
$.
The  ensemble average fluctuation spectra  are thus  constant in fluctuation  frequency  but chromatic in radio frequency $\nu$ via the noise temperature, receiver characteristics, and  pulse chromaticity, 
\be
\Srn(f, \nu) = {\Srn}_{0}(\nu) \quad \text{and} \quad 
\SJ(f, \nu) = S_{\rm J, 0}(\nu).
\ee
The spectral coefficients equal the time-domain variance divided by the Nyquist frequency,
$\fNy = \fcadence / 2$, where $\fcadence$ is the rate of TOA sampling, e.g. $\sim 10$\,cy\,yr$^{-1}$. 
The S/N  contribution is  uncorrelated in both epoch and frequency\footnote{Radiometer noise is uncorrelated between different time samples and frequency channels of a multichannel system, for example.} so the \RMS\ error for thet two-frequency measurement is of the form
${\rmstinfty}_{\rm ,W} = \sqrt{R^4 \sigma_2^2 + \sigma_1^2} / (R^2 - 1)$,
where $\sigma_{1,2}$  are the \RMS\ errors at the two frequencies, which generally differ.
The  spectral level for $\DtSNR$  is affected by the variable  signal strength of the pulsar caused by  DISS, RISS, and caustics from plasma lensing.
Jitter is highly correlated across frequency, giving a different form,
 $\rmstinfty = R^2 \sigma_J  / (R^2 - 1)$ where $\sigma_J$ is the error in an individual measured \TOA.

{\it Finite scintle effect:} 
The rapid shape changes of PBFs from finite numbers of scintles yield a \TOA\ spectrum that is technically of low-pass form, 
\be
\Sfse(f, \nu) = \Slowpass(f, \bandwidth{fse}, \nu),
\ee
where the bandwidth $\bandwidth{fse} \simeq \dtd^{-1}$ is roughly the inverse of the DISS time scale,
i.e. $\bandwidth{fse} \simeq 1\, {\rm mHz} \, (10^3\,{\rm s} / \dtd)$.   On a practical basis, however, \TOAs\ are usually obtained from profiles with integrations $T \gtrsim \dtd$ so the \TOA\ error from this effect is effectively 
independent between \TOAs.   Consequently, the spectrum for the residual time series from this effect is essentially white,
\be
\Sfse(f, \nu) = S_{\rm fse, 0}(\nu),
\ee
but it is strongly dependent  on RF and $T$ (Sect.~\ref{sec:dt_fse}).   The spectral coefficient is
$S_{\rm fse, 0}(\nu) = \rmstoaDISS^2 / \fNy$.

\subsection{Achromatic red noise with power-law spectral components}

{\it Spin noise:}   Stochastic spin variations (as opposed to glitches, which are large discontinuous events; \newtext{Sect.~\ref{sec:spin}}), display red spectra with a range of spectral indices among pulsars \newtext{(see Sect.~\ref{sec:spinnoise})}
\be
\Sspin(f)  = S_{\rm spin,0}  f^{-\Xspin},   \quad 4\lesssim \Xspin \lesssim 6.
\ee
Referring back to Sect.~\ref{sec:spinnoise}, some pulsars reveal that the spin noise comprises the superposition of individual events in the spin frequency $\fs$ or its derivative $\fsdot$.   Step functions in either of these yield  corresponding spectra scaling as $f^{-4}$ or $f^{-6}$, respectively. 

\newtext{
{\it GW stochastic background} (GWSBG):  The  GWSBG that is apparently manifested in PTA data
(Sect.~\ref{sec:intro})
may originate from}
 the superposition of GWs from a population of supermassive black-hole binaries, that appears as a power-law like spectrum for arrival-time perturbations with the possibility of flattening at the lowest frequencies sampled by the $T\simeq 15$-20 yr data sets.
The angular correlation of  timing residuals  between pulsars results from GWs affecting spacetime near the Earth \citep[][]{hd83}.  However, GWs passing through each pulsar's location contribute  timing variance equal to that of the Earth-term GWs that is uncorrelated between pulsars.    The pulsar-term GWs therefore contribute power-law noise to timing residuals  that we represent with the same form as spin noise,
\be
\SGW(f) = S_{\rm GW,0}  f^{-\Xgw},   \quad  \Xgw \simeq 4 \ {\rm to} \  4.3.
\ee 

{\it Asteroid noise:}  The spectrum of timing noise produced by an asteroid belt around a pulsar (Sect.~\ref{sec:orbnoise}) is typically shallower than that produced by GWs or spin noise, but steeper than that produced by interstellar processes \newtext{(Sect.~\ref{sec:orbnoise})}:
\be
\Sast(f) = S_{\rm ast, 0}  f^{-\Xast}, \quad \Xast \simeq 3 \ {\rm to} \ 3.67.
\ee
The spectral index $\Xast = (2\betaa + 7)/3$,  where $\betaa$ is the index in the radial distribution of asteroids, $f_a(a) \propto a^{\betaa-1}$. \newtext{Example} values of $\betaa$ include $\betaa=1$, corresponding to a uniform distribution in orbital radius, and $\betaa=2$, corresponding to a disk with a uniform surface density of asteroids. These correspond to $\Xast=3$ and $\Xast=11/3$, respectively.

\subsection{Chromatic red noise with low-pass cutoffs}
\label{sec:red_rolloff}

Most of the interstellar effects  comprise  time-correlated noise with stationary statistics, yielding a variance contribution 
with little or  no dependence on the length of time series $T$ if $T$ is greater than the correlation time.   The exception is DM variations, which are sensitive to the full range of scales in electron density and for which an outer scale 
$\sim 1$\,pc would require $T\gtrsim 10^4$\,yr.   In all cases, however, the spatial averaging from multipath propagation,  which underlies the chromaticity of DM,  yields smoothed variations in DM and other quantities with a time scale  of order the refractive scintillation (RISS)  time scale  $\dtr$ (Sect.~\ref{sec:sclaws}).  This truncates fluctuation spectra for frequencies $f \gtrsim \dtr^{-1}$.    We account for this low-pass filtering with a factor $\Slowpass(f, \bandwidth{})$. It has an unspecified form but is approximately flat and rolls off  for frequencies $f \gtrsim  B$ with $B \propto \dtr^{-1}$.

{\it Dispersion measures:}   A single propagation path through the ISM  yields $\DM(t)$  with structure function 
$\DDM(\Dt) \propto (\Dt)^{\beta-2}$ (in the inertial subrange) for a medium with a power-law wavenumber spectrum. 
This corresponds to a red fluctuation spectrum $\SDM(f) \propto f^{-(\beta-1)} \overset{\beta=11/3} {\to} f^{-8/3}$.
Ideally, $\DM(t)$ is estimated at each epoch and dispersion delays are removed from arrival times.  
 However, the multifrequency  methodology used for DM estimation combined with  multipath propagation introduces 
 the low-pass factor into the spectrum, 
\be
\SDMbar(f, \nu) \propto  f^{-\XDM} \Slowpass(f, \bandwidth{\DMbar}, \nul, \nuh), \quad\quad \XDM = \beta-1
\ee
with a fluctuation bandwidth dependent on the frequency range $[\nul, \nuh]$  used, over which $\dtr$ varies as 
$\nu^{-\beta/ (\beta-2)} \overset{\beta=11/3} {\to} f^{-11/5} $.  The highest frequency $\nuh$ yields the fastest RISS so the spectrum will extend to $\bandwidth{\DMbar} \sim \dtr^{-1}(\nuh)$.
 An example (simulated) spectrum is shown in  \citet[][Fig. 5]{CSS16}.

{\it DMs from asynchronous arrival times:}   The TOA error from asynchronous dual-frequency measurements scales as
$\Delta \DM = \DMbar(\tbar_1, \nul) - \DMbar(\tbar_2, \nuh)$ 
\newtext{
and its variance is calculated from the DM structure function (Sect.~\ref{sec:dm_nonsimultaneous}; see also 
\citet[][]{2015ApJ...801..130L}). 
}
For small $\vert \tbar_2 - \tbar_1\vert$ this difference acts as  a derivative.  From the Fourier derivative theorem this  yields a factor of $f^2$ that multiplies the spectrum of $\DM$.  Including the low-pass factor, the spectrum is 
\be
\SDMasync(f, \tbar_1, \tbar_2 ) \propto  f^{-(\beta-3)} \Slowpass(f, \bandwidth{\Delta\DM}).
\ee
For $\beta = 11/3$ the spectrum is shallow, $\SDMasync \propto f^{-2/3}$ for $f \ll \bandwidth{\Delta\DM} \simeq 1 / \dtr(\nuh)$.  
 
 {\it Frequency dependent DMs:}    Scattering cone averaging is responsible for the radio frequency dependence
 of $\DMbar(\tbar, \nu)$, so the \DM\ increment $\Delta\DMbar$ has 
  a low-pass fluctuation spectrum,
\be
\SDMbar(f, \nul, \nuh) \propto  \Slowpass(f, \bandwidth{\DMbar}, \nul, \nuh),
\ee
with $\bandwidth{\DMbar} \simeq \dtr(\nuh)^{-1}$, as \newtext{can be seen by inspection of} the  simulated spectrum  shown
in  \citet[][Fig. 5]{CSS16}, which demonstrates  that the spectral amplitude decreases as $\nul \to \nuh$
and the bandwidth increases with larger $\nuh$ because $\dtr$ is smaller at higher radio frequencies. 

{\it  Variations in scattering and refraction parameters:} Relevant quantities here are the scattering time $\taud$,
the scintillation bandwidth $\dnud \propto \taud^{-1}$, the DISS time scale  $\dtd$, and the RISS time scale $\dtr$. 
The refraction angle  $\thetavec_r = k^{-1} \grad_{\perp} \phi(\xvec)$ has  
a wavenumber spectrum  $\Pthetar(\qperpvec) = \qperp^2 \Pphi(\qperpvec, 0)$ 
related to but less steep (by virtue of the gradient)  than either the phase spectrum $\Pphi$ or the spectrum for the electron density $\Pne$.    Sampling the  refraction angle $\theta_r$ across the observation plane gives a temporal spectrum   $S_\thetar(f) \propto f^{-(\beta-3)}$ multiplied by the low-pass factor.  
\AOA\ variations thus have a shallow spectrum for a Kolmogorov medium, $S_\thetar(f) \propto f^{-2/3}$. 

Refraction from power-law media varies on the refraction time scale $\dtr$; refraction alters the  shape of the ray-path bundle and thus affects all multipath effects (Sect.~\ref{sec:aniso_refract}).  Stationary fluctuations with a characteristic  bandwidth $\bandwidth{r}  \sim  1 / \dtr$ (i.e. lowpass in form) are expected where  $\bandwidth{r} $  is RF dependent because $\dtr \propto \nu^{-11/5}$ so $B_r \propto \nu^{11/5}$.   

For a Kolmogorov-type medium,  refraction delays are smaller than $\taud$, so we discuss the latter first. 
As indicated in Sect.~\ref{sec:heavytails}, \TOA\ delays  (and thus timing residuals) are related to but not equal to the scattering time $\taud$ and the RF dependence of \TOA\ delays is nontrivially related to that of $\taud$ because it depends on the methodology used to estimate \TOAs\ while addressing the scattering delay. We consequently write the spectrum for timing perturbations   caused by epoch dependence of $\taud$ in the low-pass form,   
\be
\Staud(f) = S_{\rm \taud, 0} \Slowpass(f, \bandwidth{\taud}, \nu). 
\ee

{\it Angle of arrival (\AOA) variations:} Refraction of the ray-path bundle causes a refraction delay
$\dtAOA \simeq \deff \thetar^2 / 2c$ that has a shallow spectrum.   A detailed treatment  for a thin screen  and Kolmogorov spectrum  ($\beta = 11/3$)  yields a refraction-angle  spectrum (Appendix\,\ref{app:AOA_spectrum}),
extending to $\bandwidth{\AOA} \simeq \dtr^{-1}$, 
\be
\Sthetar(f) = 
\frac{1 } {\sqrt{2}(2\pi)^{1/6}} 
\frac{\Gamma(1/3)}  {\Gamma(5/6)} 
\frac{(\lambda^2 \relec)^2 \,\SM }{v_x} 
\left(\frac{f}{v_x}\right)^{-2/3}, \quad f \lesssim \bandwidth{\AOA} \sim \dtr^{-1}.
\ee

Note that low-DM pulsars observed at high frequencies have small refraction times, $\dtr \lesssim$\,few days,  so the spectrum extends to $\gtrsim$cy\,yr$^{-1}$.   With a typical observing cadence  $\Delta t_{\rm obs} \sim 1$\,month,  the spectrum is aliased to form essentially a flat spectrum.   In that situation, a low-pass spectral form
with bandwidth $\sim \Delta t_{\rm obs}^{-1}$ would be appropriate, 
\be
\SdtAOA(f) = S_{\rm \delta t, AOA, 0} \Slowpass(f,\Delta t_{\rm obs}^{-1}, \nu).
\ee

The spectrum of $\thetar^2$ is the convolution of $\Sthetar(f)$ with itself and scales as $f^{-1/3}$ for a Kolmogorov spectrum 
and generally as $f^{-(2\beta -7)}$ for $3  < \beta < 4$ with a maximum at a small wavenumber determined by a combination of the inner and outer scales. 
Thus we expect this contribution to \TOA\ fluctuations to scale as 
\be
\SdtAOA(f) \propto  f^{-1/3} \quad \text{for} \ f \lesssim \bandwidth{\AOA}.
\ee

\subsection{Chromatic bandpass noise}

\AOA\ variations introduce another timing error related to the translation of topocentric \TOAs\ to the solar system's barcyenter. If the  assumed pulsar direction $\nhat$  differs from the actual, refracted direction
\citep[][]{fc90},   the \TOA\ error has the form (assuming for simplicity a circular orbit),
$\dtAOABary \simeq \thetar (r_E/c) \cos[\Omega_{\rm E}(t-t_p)]$,
where $r_{\rm E}$ and $\Omega_{\rm E}$ are  the Earth's orbital radius and orbital frequency
and $t_p$ is an offset related to the pulsar's direction.       

The ensemble average spectrum is  a low-pass form 
(for $\thetar$) convolved with a delta function from the sinusoidal variation, yielding a shift of the low-pass spectrum
\be
\SdtAOAssbc(f) = S_{\rm AOA, SSBC, 0} \Slowpass(f - \Omega_{\rm E}/2\pi, \Br, \nu).
\ee
The bandwidth of the low-pass factor is determined by the refraction time scale,   $\Br \simeq (\dtr)^{-1}$.   For low-DM pulsars $\dtr \ll 1$\,yr so $\Br \gg 1$\,cy\,yr$^{-1}$ and the orbital shift is barely discernible.    Stronger scattering from high-DM pulsars or even 
low-DM pulsars observed at low frequencies $< 1$\, GHz will have longer RISS time scales;  the resulting  narrower bandwidths make the orbital shift more manifest and thus the spectrum departs even more from a power-law form. 

The sampling cadence $\Tcadence$ of a PTA program combined with the refraction time scale $\dtr$ and total time span $T$ determine whether $\dtAOABary$ is manifested as  
\newtext{
white noise in a specific radio band (and which will differ between bands; we refer to this as `chromatic white noise')}, a low-pass spectrum, or a bandpass spectrum. 
\newtext{The refraction time $\dtr$ is strongly frequency and \LoS\ dependent but can range from hours for high-frequency observations of low-DM pulsars to months or years for high-DM pulsars at low frequencies:}
\begin{enumerate}
\itemsep -1pt
\item {\it White noise:}  $\dtr < \Tcadence \ll T$:  When refraction angle variations are undersampled,  they are statistically independent, corresponding to data with a white-noise representation.  This regime applies
to low-DM pulsars or high frequency observations.  
\item {\it Correlated noise:}  $\Tcadence < \dtr  < \,1\, {\rm yr} \ll T$: A longer refraction time  that is  still small
 compared to one year yields  $\dtAOABary$ that  appears as red noise with a low-pass form. 
 (Moderate DM pulsars observed at $\sim 1$\,to\,2\,GHz frequencies.)
 \item {\it Bandpass noise:}  $\Tcadence < 1\,{\rm yr} \, < \dtr \ll T$:   Refraction times longer than about one year 
 allow the coherence of the cosine term in $\dtAOABary$ to manifest.  (Large DM pulsars or low frequencies.)
\end{enumerate}

\begin{figure}[b!] 
   \centering
   \includegraphics[width=0.475\linewidth]{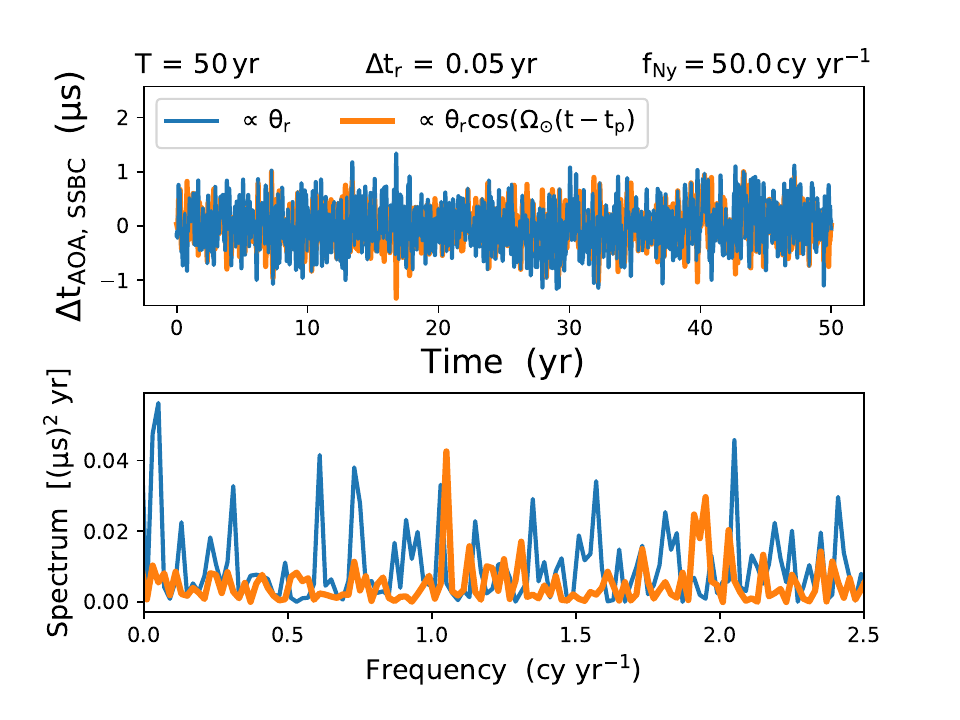}
   \includegraphics[width=0.475\linewidth]{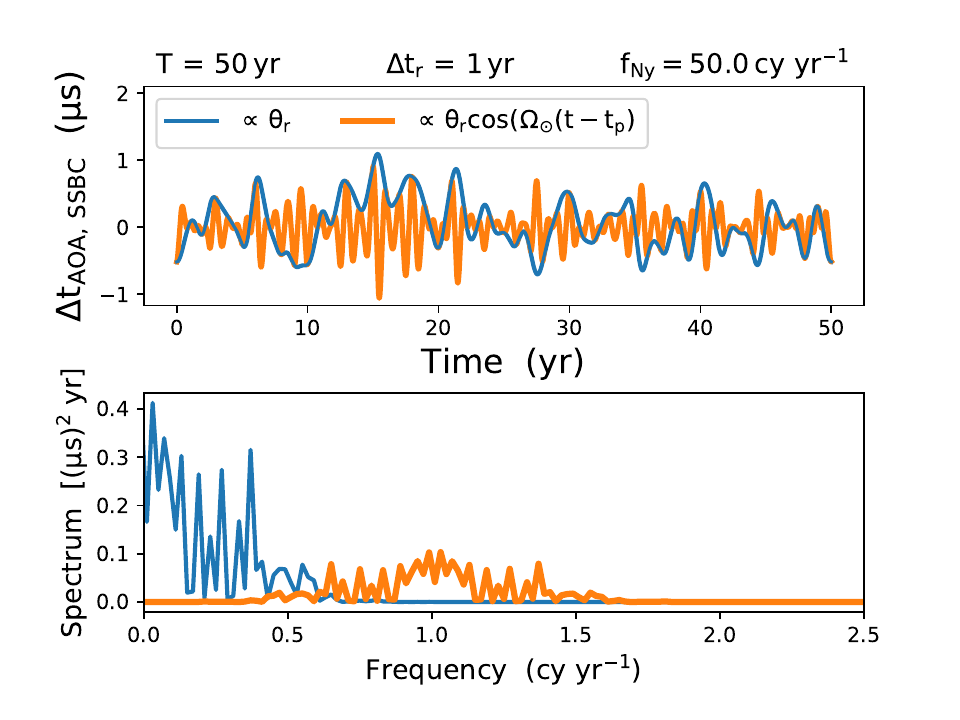}
   \caption{Time series and spectra for the refraction angle $\thetar(t)$ and the barycentric delay $\dtAOABary$.  
   Simulated refraction angles with 1\,mas RMS  comprise 5000 samples over 50 years that yield
  \RMS$(\dtAOABary) \sim 0.27\,\mu s$ for a direction in the ecliptic plane.   Refraction angles were simulated
   as a process with a Gaussian correlation function that captures the essential features but is only a coarse
   approximation to a Kolmogorov medium. 
   Left: undersampled variations in refraction angle with $\dtr  = 0.05\, {\rm yr} \, \ll \Tcadence$ yield uncorrelated samples in  $\dtAOABary$ and a white noise spectrum.
   Right:  a longer refraction time $\dtr = 1\,$yr   allows the coherence of the cosine term in $\dtAOABary$ 
   to be manifested in the bandpass spectrum. 
   }
   \label{fig:aoassbc_spectrum1}
\end{figure}

Figure~\ref{fig:aoassbc_spectrum1} shows results from a toy model for refraction angles that illustrate the white noise and bandpass noise regimes.   Refraction angles were generated by convolving  white noise with a Gaussian kernel with width equal to $\dtr$.   Time series for $\dtAOABary$ are shown in the top panels with and without the cosine factor to demonstrate its effect.    Spectra are shown in the bottom panels.   

Timing programs require measurements over wide frequency ranges to enable estimation of plasma propagation delays.  These ranges can cut across the regime boundaries shown in the figure.    The lowest-DM pulsars are largely immune to crossover because the barycentric delay will appear as `chromatic white noise' at all relevant frequencies.   Moderate-DM pulsars up to $\DM = 100\,\DMunits$ will show temporal correlation at low frequencies and white noise at high frequencies. 

Figure~\ref{fig:aoassbc_regimes} shows the noise regimes in the DM-$\nu$ plane where the lines correspond to
values of \DM\ where $\dtr = T_{\rm test}$ with values for $T_{\rm test}$ equal to representative observing cadences of 7 and 30\,d and the yearly oscillation period of the barycentric delay.  

Figures~\ref{fig:aoassbc_spectrum1} and \ref{fig:aoassbc_regimes}  illustrate temporal aspects of the barycentric delay and their dependence on DM and radio frequency.   The \RMS\ refraction angle scales
as $\nu^{-49/30} {\taud}_1$ (where ${\taud}_1$ is the scattering time at 1\,GHz, cf.\ Table\,\ref{tab:keyplasmaeq}).
For a specific pulsar and epoch, the  barycentric delay (and refraction angle)   will vary differently from the 
$\nu^{-2}$ frequency dependence of the DM delay and thus  will not be removed without an additional chromatic term in the fitting function at each epoch. 

\begin{figure}[t!] 
   \centering
   \includegraphics[width=0.5\linewidth]{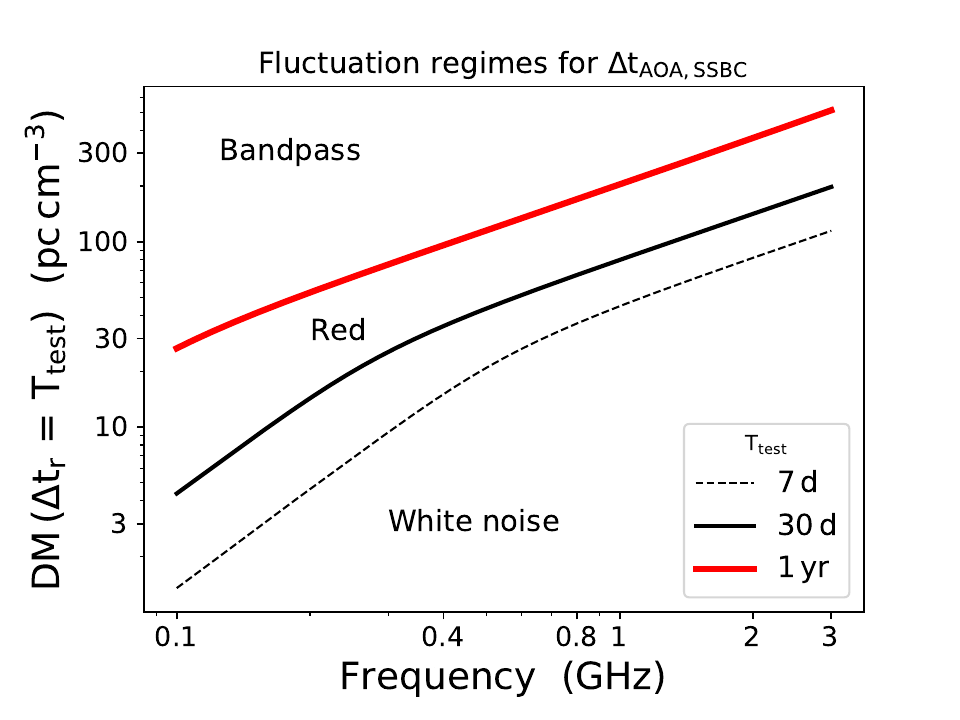} 
   \caption{Plot showing regimes for the noise character of the stochastic barycentric correction delay.
\   As a function of observation frequency,  the DM corresponding to equality of the refraction time scale $\dtr$ 
   with a specified test time $T_{\rm test}$ is shown.    Two  test values correspond to observation cadence   (7\,d
   and 30\,d) and another corresponds to the yearly period of the cosine in the delay expression.
   }
   \label{fig:aoassbc_regimes}
\end{figure}

\subsection{Intrinsic pulse shape variations}
\label{sec:intrinsic}

The simplest summary of intrinsic  pulse profiles  is that large-$N$ pulse averages are epoch independent  but chromatic. Pulsars that show state changes between profile shapes (`mode changes') require $N$ to be large enough to average over mode changes.  The average profiles of  pulse sequences for an individual mode are consistent with having  epoch independent shapes.   A notable exception
to epoch independence is the MSP J1713+0747 that showed a sudden change in shape that was both chromatic and time dependent as it relaxed back to its original shape after about two years 
\citep[][]{2021ATel14642....1X,2021MNRAS.507L..57S,2024ApJ...964..179J}.
These changes lend themselves  to a  description in the time domain  rather than in  Fourier-space.    Timing residuals nonetheless show variations on time scales comparable to those of GW perturbations in the nanohertz band that must be carefully mitigated.   It is possible that smaller events like those in J1713+0747 data are also present.  Collectively, if frequent enough, they could produce variations not unlike those associated with spin noise.  Indeed, profile changes associated with changes in orientation or structure of the magnetosphere might reflect torque events that would produce spin noise. 

\subsection{Instrumental discontinuities}

It is inevitable that changes in instrumentation occur during multiyear timing programs.  These can involve any elements along the signal path from the feed antennas to the analog to digital converters.   Discontinuities or jumps also occur in the synchronization of observatory clocks to UTC.    Pulsar timing analyses take these (mostly) known jumps into account but there are certain to be residual effects.  Discontinuities in  arrival times occurring at random epochs correspond to random walks in the residuals that have an $f^{-2}$ spectral signature (Sect.~\ref{sec:SFs}).    If these are due to clock offsets, the spectrum is achromatic.   But jumps in imperfect polarization calibration are chromatic \cite[e.g.][]{Rogers2024} and would have an $f^{-2}$ fluctuation spectrum. 

\subsection{Aliasing of fluctuation spectra}
\label{sec:aliasing}

Figure~\ref{fig:sampling} (left panel) shows schematically  the integration time and cadence for typical PTA data sets where an integration time $\Tint$ is used to obtain a TOA at epochs separated by an  interval $T_{\rm cadence}$.
The Nyquist frequency is $\fNy = 1/2\Tcadence$. 
\newtext{
Matching this to the bandwidth of the integrated signal $\sim 1 / \Tint$  corresponds to $T_{\rm cadence} = \Tint / 2$.  
}

PTA data are  grossly undersampled because  $\Tcadence \gg \Tint$ due  to telescope time constraints 
combined with choices about how target signals (e.g. GW perturbations) can be sampled. 
The corresponding schematic spectrum (right panel)  shows  the unaliased and aliased frequencies for this situation
and indicates which frequencies are aliased into the Nyquist band.
The integration time for typical PTA data sets $T_{\rm int} \sim 0.5$~hr corresponds to a bandwidth for
TOA variations $f_{\rm rolloff} = 1 / T_{\rm int} \sim 0.24$~mHz (HWHM), which happens to be much larger than the frequency range of GW signals thought to be strong enough to detect with PTAs.

\begin{figure}[h!]
\begin{center}
\includegraphics[scale=0.4]{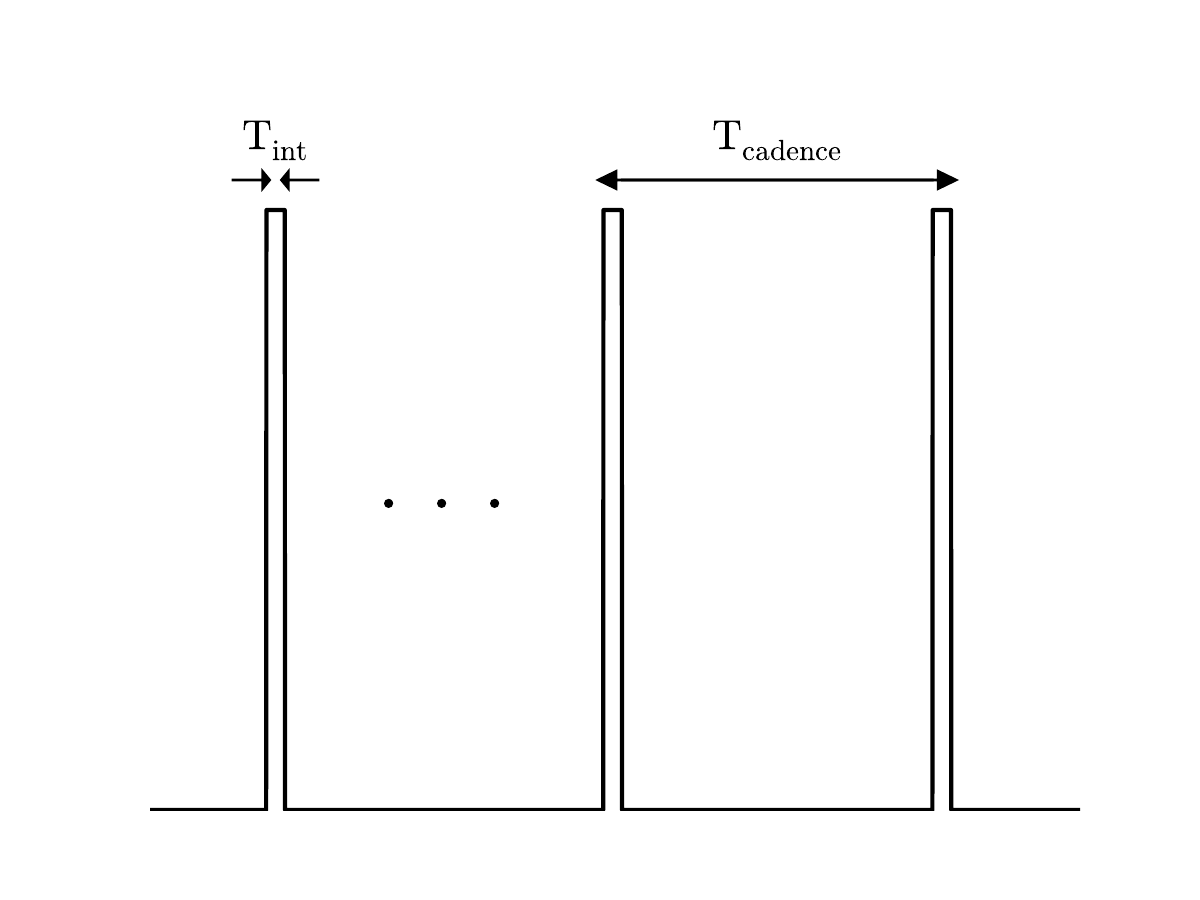}
\includegraphics[scale=0.4]{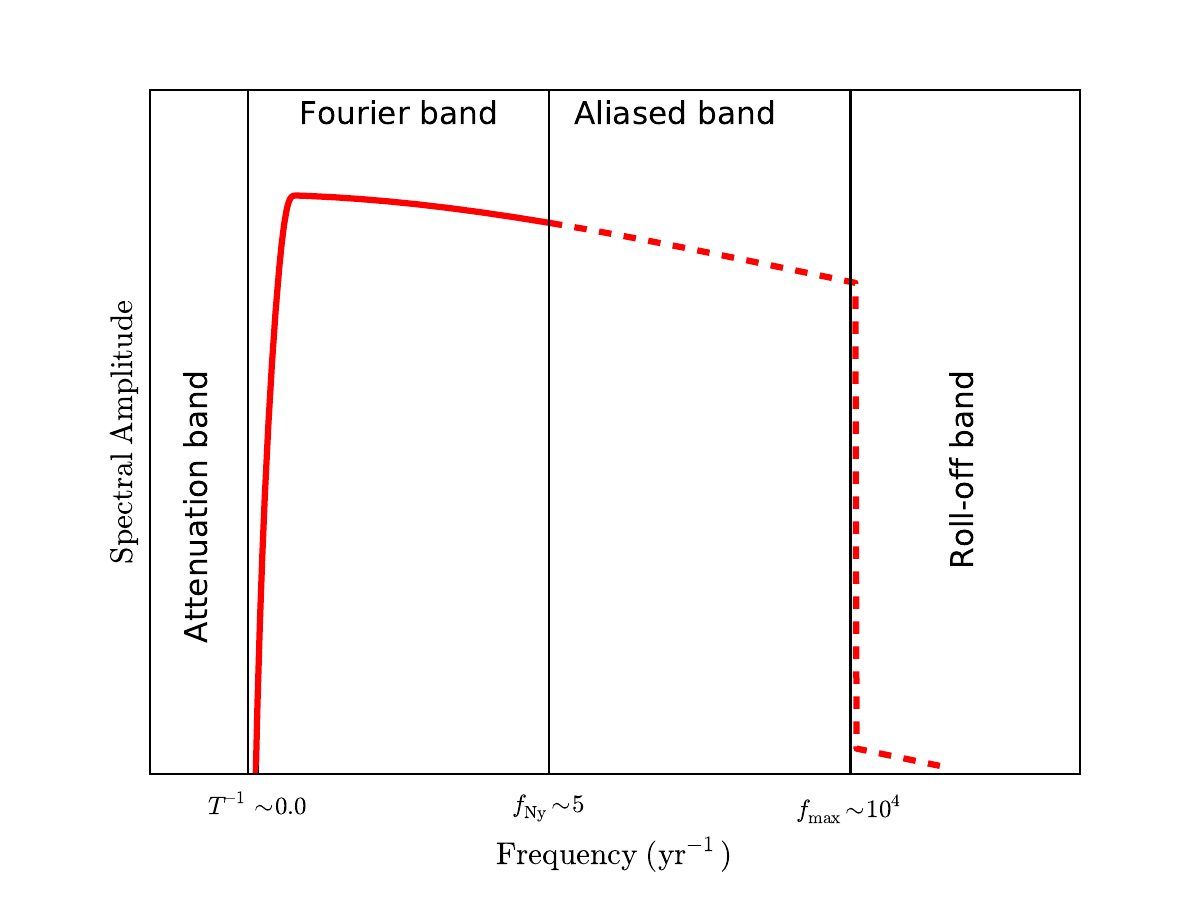}
\caption{\footnotesize 
Left:
Sampling of pulsar timing data showing the integration-time interval  $T_{\rm int}$
at each epoch  and the  separation $T_{\rm cadence}$ between epochs (`cadence') . 
Right:
Schematic spectrum showing the \newtext{attenuation} band where low frequencies are attenuated by the finite data span $T$, the Fourier band of unaliased frequencies below the Nyquist frequency, the aliased band, and the rolloff band related to the bandwidth of each sampled TOA for frequencies
$f \gtrsim f_{\rm max} \sim T_{\rm int}^{-1}$.
\label{fig:sampling}
}
\end{center}
\end{figure}

For typical PTA data, the aliasing is severe.   We show  that it is inconsequential for steep power laws \newtext{(steeper than $f^{-3}$)} but can be important for chromatic leakage into timing residuals by interstellar delays. 
Let $[0, \fNy]$ be the range of frequencies into which all spectral amplitudes are mapped.
True frequencies in this interval are unaliased but the estimated spectrum $S(f)$ for  $f \in [0, \fNy]$  receives contributions from 
\be
f_{\rm true}  = 
\left\{
\begin{array}{ll}
  f  & 
  \\
  2j \fNy - f  	& 	 j = 1, 2, \ldots, j_{\max}
  \\
 2j \fNy + f 	&	 j = 1, 2, \ldots, j_{\max}
 \end{array}
 \right.
 \ee
 with $j_{\rm max} \sim (\fmax - \fNy) / 2\fNy$. 
The  aliased spectrum is \citep[e.g.][]{2005PhRvE..71f6110K}
\be
S_a(f) = S(f) + \sum_{j=1}^{j_{\max}} \left[ S(2j \fNy - f ) + S( 2j \fNy + f) \right].
\ee
%
Figure~\ref{fig:pl_aliases} shows examples of power-law spectra $S(f) \propto f^{-X}$
 with spectral indices $X$ relevant to timing fluctuations from interstellar  processes, spin noise, and GWs.  For most of these cases, distortion by aliasing is minimal, becoming significant only for shallow spectra with
 $X  < 2$.  The apparent spectral index $X_{\rm a}$ (also shown in the legend) departs significantly from $X$ for $X= 5/3$ and especially for $X=2/3$ where the aliased spectrum is nearly flat with $X_{\rm a} = 0.04$. This last case applies to chromatic leakage of DMs obtained from asynchronous arrival times, which are relevant to much of the PTA used to report discovery of the stochastic GW background.  The
$X=2/3$ case also applies to refraction and the transfer of \TOAs\ to the SSBC. \newtext{Note that when white noise is aliased, its spectrum remains flat.}

\begin{figure}[h!]
\begin{center}
\includegraphics[scale=0.6]{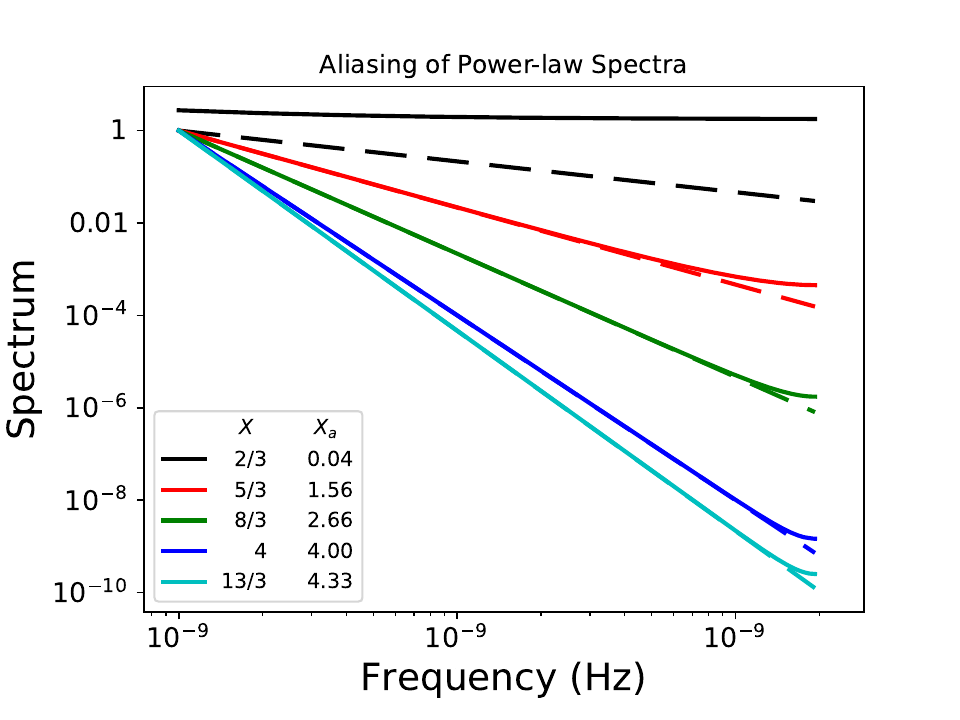}
\caption{\footnotesize 
Examples of power law spectra $S(f) \propto f^{-X}$  that extend to $1.7\times10^{-3}$~Hz (dashed lines)  but are aliased into  the Nyquist band, $f \le \fNy = 2\times10^{-7}$\,Hz and distorted (solid lines).   The aliased spectra
have apparent spectral indices $X_{\rm a}$ evaluated at $f =1$\,cy\,yr$^{-1}$,   as indicated in the legend.
\label{fig:pl_aliases}
}
\end{center}
\end{figure}

\clearpage
\subsection{Example  spectra for four MSPs}
\label{sec:exampleMSPs}

To illustrate fluctuation spectral shapes and the differences between pulsars,  we show schematic spectra for four exemplar 
MSPs.    They are chosen to represent the span of conditions encountered across the MSP population, including those with predicted low and high noise variance from  the neutron star spin  or  the ISM along the \LOS.   

Spectra display a subset of the effects described in the paper in order to highlight the more prominent effects in comparison to the red noise from the stochastic GW background.  The included effects are 
red spin noise, residual DM variations, the arrival time variation resulting from the $\dtAOABary$ term, and from the stochastic GW background. 
Spectra are shown for 30-yr data sets, about 50\% longer than typical data sets for MSPs that dominate PTA sensitivity to GWs.  

{\bf J0509+1856} ($P = 4.06$\,ms, $\DM = 38.3\,\DMunits$): This pulsar shows large white noise due to its relatively wide pulse and pulse jitter.   However, it is predicted to have low spin noise and will contribute significantly to GW detection sensitivity in multi-decade long data sets.

{\bf J1903+0327} ($P = 2.15$\,ms, $\DM = 297.5\,\DMunits$): This relatively weak, large-DM pulsar shows large scattering broadening of its pulse.  Consequently it will never contribute to GW sensitivity but it is an important object for demonstrating ISM effects on timing precision
\citep[e.g.][]{2025ApJ...986..191G}.  The oscillations at frequencies near $10^{-7}$\,Hz are due to the $\dtAOABary$ term, which are prominent for this object and not the lower-DM cases because of the large scattering. 

{\bf J1909$-$3944} ($P = 2.95$\,ms, $\DM = 10.4\,\DMunits$):  This is one of the best MSPs used in PTAs owing to its narrow pulse.  It contributes significantly to the \newtext{evidence for}  detection of the GWSBG reported in 2023. 

{\bf J1939+2134 (B1937+21)} ($P = 1.56$\,ms, $\DM = 71.02\,\DMunits$):  This bright MSP has a narrow width, like J1909$-$3944, and would be the best MSP for PTA work were it not for its very strong, achromatic red noise from spin fluctuations (or possibly from asteroid noise).
The oscillations at frequencies near $10^{-7}$\,Hz are due to the $\dtAOABary$ term, as with J1903+0327.  

\begin{figure}[h!]
\begin{center}
\includegraphics[width=0.475\linewidth]{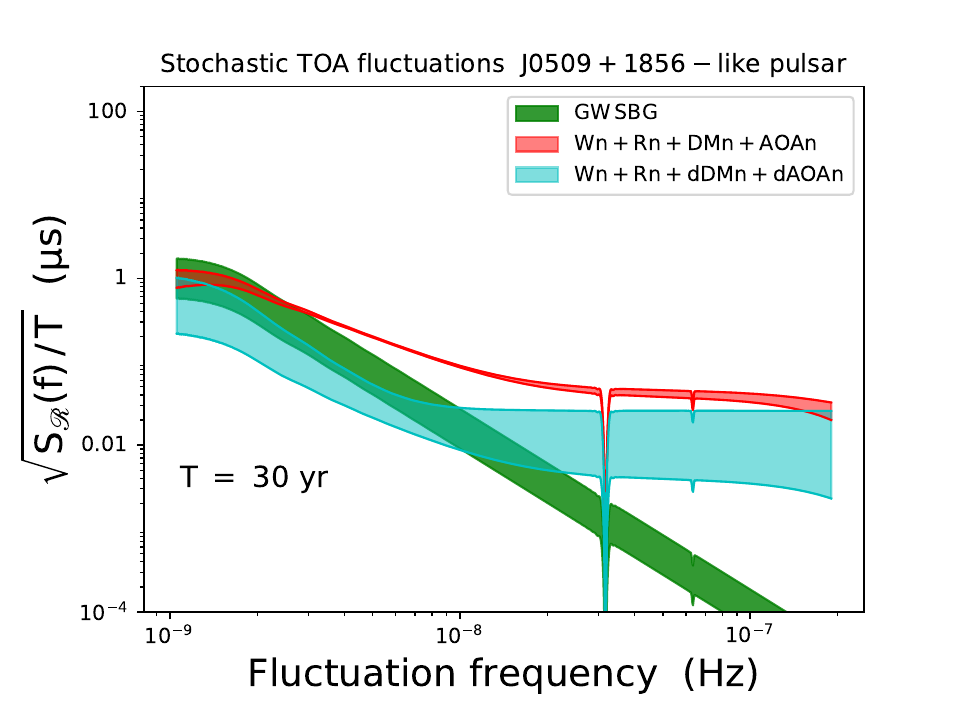}
\includegraphics[width=0.475\linewidth]{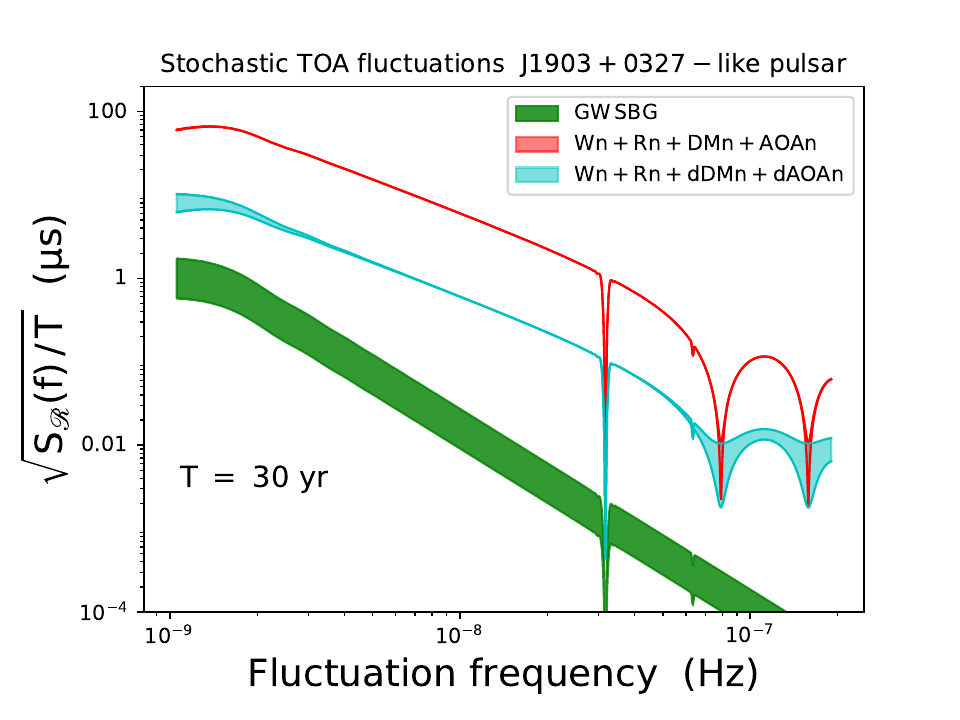}
\includegraphics[width=0.475\linewidth]{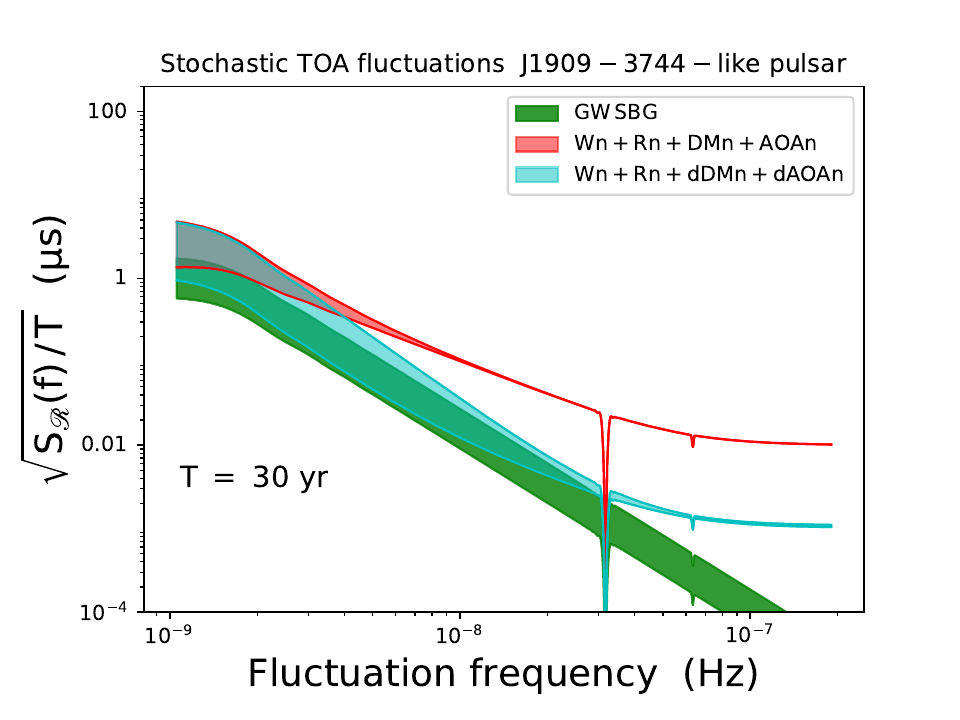}
\includegraphics[width=0.475\linewidth]{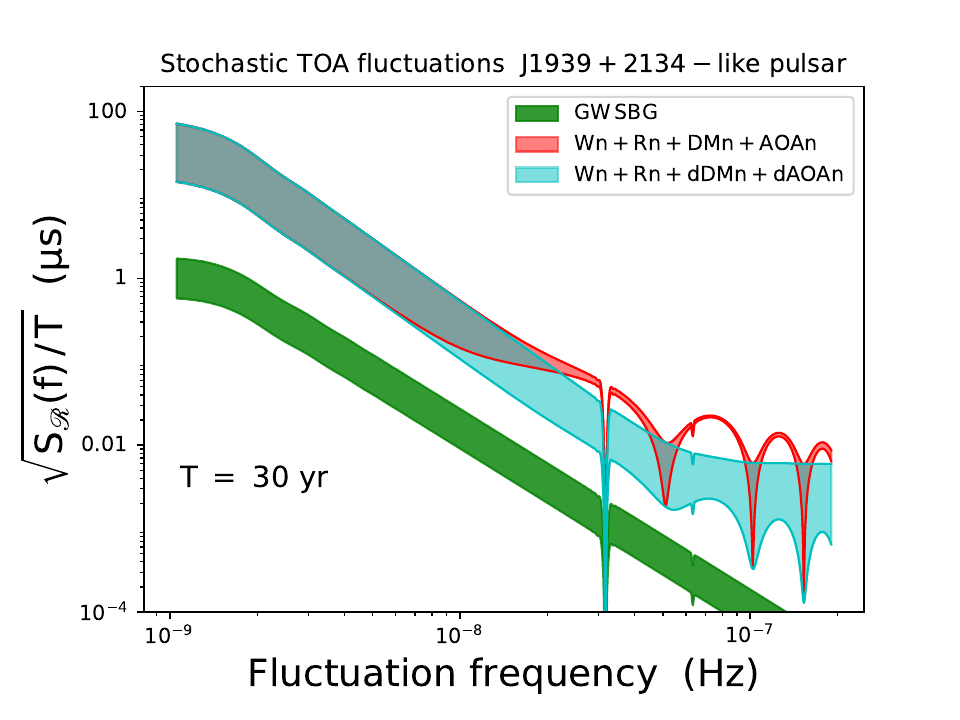}
\caption{
Example noise spectra based on {\it predicted} contributions from white noise, spin noise, and ISM fluctuations for four pulsars.
The spectra are based on time series spanning $T = 30$\,yr and are shown  as $\sqrt{S_{\mathscr R}(f) / T}$, where
$S_{\mathscr R}(f) $ is the spectrum of timing residuals, obtained by multiplying the prefit spectrum by the transmission function.
The GW stochastic background (green) is compared with the total noise spectrum both before (red) and after (cyan) mitigation of 
interstellar effects.  It is assumed that mitigation reduces the spectral contribution by 10\%.
Top left: J0509+1856 has a wider pulse than the other three MSPs, so its template fitting error is the largest by a factor $\sim 15$;
Top right: J1903+0327 is a high DM pulsar with a discernible scattering tail on its pulse shape, so its noise spectrum even after 10\% mitigation is dominated by interstellar noise. 
Bottom left: J1909-3744 is a low-DM pulsar with a narrow pulse; it has a relative high $\fsdot$ so its predicted spin noise contributes significantly to its overall noise spectrum. 
Bottom right: J1939+2134 (B1937+21) is dominated by achromatic red noise from spin variations (or possibly orbital  noise from an asteroid belt around the pulsar.) \newtext{Here and in the top right panel, oscillations at frequencies near $10^{-7}$\,Hz are due to the $\dtAOABary$ term.}
\label{fig:exemplar_spectra}
}
\end{center}
\end{figure}
\endgroup

\clearpage

\section{Red noise assessment of MSPs in PTAs}
\label{sec:red_noise_assessment}

Among the set of papers announcing the identification of the nanohertz GW background is a description of the
properties of the NANOGrav PTA as a GW detector 
\citep[][]{2023ApJ...951L..10A}. Similar detector properties were
presented for 
the Australian PPTA 
\citep[][]{2023ApJ...951L...7R},
the Chinese PTA 
\citep[][]{2023RAA....23g5024X}, 
the European PTA 
\citep[][]{2023A&A...678A..49E},
the Indian PTA 
\citep[][]{2023PhRvD.108b3008S},
and subsequently for the MeerKAT PTA
\citep[][]{2025MNRAS.536.1489M}.

Table\,2 of \citet[][]{2023ApJ...951L..10A} presents characterizations of  red noise in the timing residuals of  the 25 MSPs for which it could be identified  out of the 67 MSPs comprising the NANOGrav PTA.   Each MSP has an attributed power spectrum of the form $\ARN (f / \fref)^{-\gammaRN}$ with
$\ARN$ in dimensionless units and  $\fref = 1$\,cy\,yr$^{-1}$
(Appendix\,\ref{app:specrep}). Mean values for $\ARN$  range from 
\newtext{$\sim 1.3\times 10^{-15}$ to $1.3 \times 10^{-12}$}.
In the same dimensionless quantity, the strength of the background is
$\Agw = 2.4^{+0.7}_{-0.6} \times 10^{-15}$ if the frequency index is held fixed at $\gammaGW = 13/3$
or $\Agw = 6.4^{+4.2}_{-2.7} \times 10^{-15}$ for a best-fit index $\gammaGW = 3.2\pm0.6$. 
We list the results from \citet[][]{2023ApJ...951L..10A} in Table\,\ref{tab:rednoiseanalysis}.

\newtext{
Similar analyses by other PTA groups   also separate timing residuals into chromatic (ISM) and achromatic
(spin and GW) terms.    The ranges of $\ARN$ and $\gammaRN$ for the 31 MSPs  in the PPTA program \citep[][Table 1]{2023ApJ...951L...7R}, six in the EPTA program \citep[][Table 5]{2022MNRAS.509.5538C}, and eight in the InPTA program \citep[][Table IV]{2023PhRvD.108b3008S}  are similar to those in Table\,\ref{tab:rednoiseanalysis}, and characterization of the high-quality MSP J1909$-$3744 is in good agreement, except for the InPTA analysis, which gives a much higher amplitude and shallower index than the other PTAs.  The predicted red-noise amplitude $\ARN$ for J1909$-$3744 also agrees with the NANOGrav, PPTA, and EPTA measurements.  However, other MSPs show measurements from these three PTAs that are  only in modest agreement, generally,  that likely  reflect the different time spans, data quality,  and analysis methods.  A detailed comparison of all PTA data sets is clearly needed but is beyond the scope of this article.
}

\begin{figure}[h!] 
   \centering
   \includegraphics[width=0.65\linewidth]{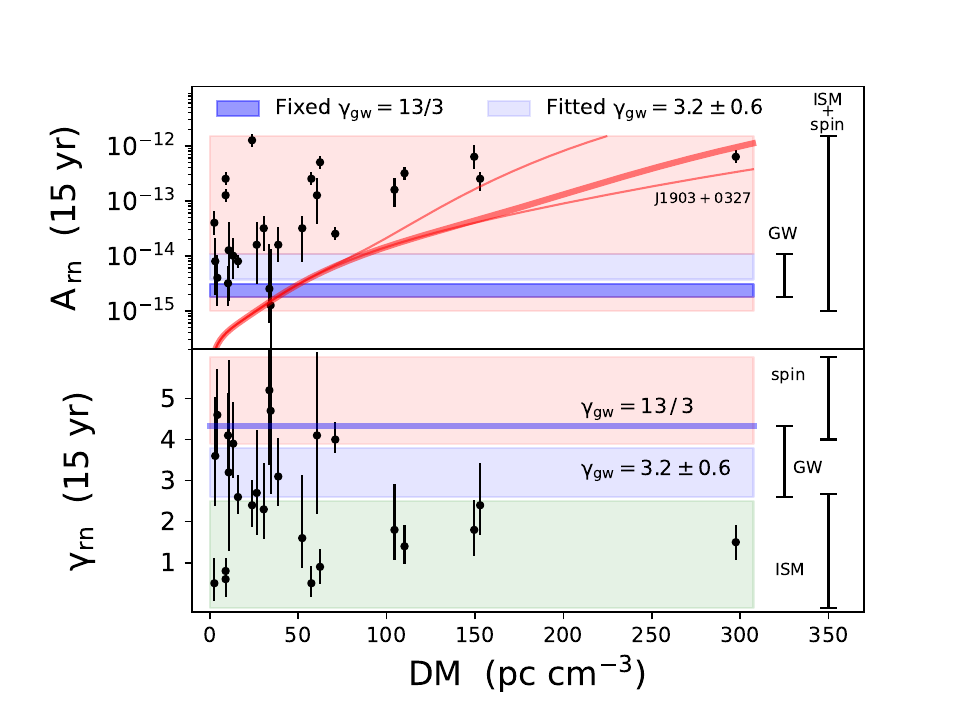} 
   \caption{Red noise analysis of 25 pulsars  presented in Table 2 of \citet[][]{2023ApJ...951L..10A} for NANOGrav's 15 yr dataset. 
Top panel:  the dimensionless spectral amplitude $A_{\rm rn}$.  Amplitude ranges for the GW stochastic background 
are indicated for two cases: (dark blue)  the GW frequency index was held fixed at $\gamma_{\rm gw} = 13/3$  and (light blue) the index was fitted for along with the amplitude.  
The wide range of amplitudes for ISM and spin noise is indicated in red shading.
The red lines indicate a preliminary analysis of the ISM fluctuations from frequency-dependent DMs,  errors in the translation of \TOAs\ to the SSBC, and  variations in scattering time $\taud$.  The heavy red line is based on the mean
$\taud(\DM)$ relation and the thin lines are the $\pm \sigma_{\log_{10}\taud}$ departures from the mean. 
The high DM MSP J1903+0327 is indicated (rightmost black circle with error bar on the top and bottom frames) as an example  where ISM fluctuations clearly dominate its red noise, consistent with the small $\gamma_{\rm rn} < 4$ spectral index. 
Bottom panel: frequency index $\gamma_{\rm rn}$ with ranges for spin noise, ISM effects, and GWs indicated. 
   }
   \label{fig:rn_assess}
\end{figure}

Fig.\,\ref{fig:rn_assess} shows the amplitudes and indices for the 25 pulsars plotted against their dispersion measures. 
In the top panel, the range of values  for  $\Agw$  quoted in \citet[][]{2023ApJ...951L..10A} are shown as blue bands.   Also shown  as a red band is the wide range estimated for red noise from interstellar propagation effects and spin noise.   Red lines show the estimated range for several combined interstellar effects  that produce red noise with a low-pass spectral form as discussed in Sect.~\ref{sec:flucspec}.  The three lines are based on  the $\taud(\DM)$ relation of Eq.~\eqref{eq:taudm} including the uncertainty
$\sigma_{\log_{10}\taud} = 0.76$ for $\taud$.

To calculate the range of interstellar contributions, we include the variability with epoch of \TOA\ variations associated with:
(1) the scattering time $\taud$;
(2) the frequency dependence of DM; and 
(3) the effects of refraction on the translation of \TOAs\ to the SSBC.  
All three quantities vary roughly on the same time scale $\dtr$ as RISS.
The first two effects dominate at large DMs but are subdominant at low DMs where refraction produces a bigger effect on \TOAs.

Variations in the  scattering time $\taud$ are represented as $\delta\taud \sim \epsilon_{\taud} \taud$ with 
$\epsilon_{\taud} = 0.1$, in accordance
with the fluctuations seen in the high-DM MSP J1903+0327 (e.g. time series in Fig.\,\ref{fig:J1903taud_vs_mjd} and  labeled in Fig.\,\ref{fig:rn_assess}).  
We also assume that  only a fraction $m_{\taud} = 0.2$ of $\delta\taud$  is manifested in arrival times (cf.\ discusssion in Sect.~\ref{sec:heavytails}), yielding  
$\sigma_{t_{\taud}} = m_{\taud}  \epsilon_{\taud} =  0.02 \taud$.  
Variations in $\taud$ at the lower frequency(ies) used in the timing analysis dominate the net error in the DM-corrected \TOA. 
We conservatively evaluate $\taud$ at  $\nu_0 = 1.5$\, GHz because we emphasize the range at larger values of DM, such 
as $\DM = 297\,\DMunits$ for J1903+0327, for which this frequency would be appropriate.   Lower frequencies are used for
most MSPs monitored in PTAs, so we underestimate the role of this effect at low DMs, but it is subdominant to the refraction effect anyway. 

For the contribution from  frequency dependent DMs, we use Eq.~\eqref{eq:dtinfity_DMnueval} with the scattering measure
calculated form $\taud$ (cf.\ Table\,\ref{tab:keyplasmaeq}, item 1).   We assume a uniform medium and $\beta=11/3$  that gives
$G_\beta = 1 / \sqrt{\beta-1} \to \sqrt{3/8}$ and use $\nuh = 2$\,GHz with $R=1$ (a 2:1 frequency range), giving
$\Itinfty(R, \beta) \simeq 1.5$.  

Finally, the \RMS\ refraction angle $\thetar$ is given  in terms of $\taud$ by item 8 of Table\,\ref{tab:keyplasmaeq}, which is the refraction angle produced by length scales equal to the width of the ray-path bundle  (i.e. scattering cone) of multiple propagation paths.  Geometrical optics for a fixed plasma cloud would yield $\thetar \propto \nu^{-2}$ but the scattering cone gets larger at lower frequencies, yielding a net Kolmogorov scaling $\thetar \propto \nu^{-49/30}$.   This differs from the native refraction scaling and also from the dispersion scaling $\propto \nu^{-2} \DM$.   The \TOA\ variation is given by Eq.~\eqref{eq:dtoa_thetar_ssbc}.
We ignore the sinusoidal dependence (see discussion in Sect.~\ref{sec:aoatoa}) but assume that 
 DM fitting removes 50\% of the \TOA\ variation and we use a nominal 1.5\, GHz frequency for this term. 

Calculation of the net dimensionless parameter for ISM effects,  $A_{\rm ism}$,  uses Eq.~\eqref{eq:ALP} after combining the three effects in quadrature to yield $\sigma_{t, \rm ism}$.  The RISS time scale (Item 13 in   Table\,\ref{tab:keyplasmaeq}) is used to calculate the bandwidth of the combined low-pass process as $B_{\rm L} = 1 / \dtr$, which is a function of radio frequency $\nu$ and epoch.  For the shape of the low-pass spectrum we use a Gaussian function, $S_{\rm L}(f) = \exp[-(f / 2 B_{\rm L})^2]$.  
The ISM $A$ parameter is necessarily a function of fluctuation frequency and for slow RISS with $\dtr \gg 1$\,yr,
the low-pass factor $S_{\rm L}(f = \fref =  \rm 1\, cy\,yr^{-1}) \ll 1$.  This becomes important for large-DM pulsars but is not a large effect for the DM range shown in Fig.\ref{fig:rn_assess}.

Armed with the analysis of the red noise properties of the general pulsar population in Sect.~\ref{sec:spin}  and the fluctuation spectra of chromatic processes in Sect.~\ref{sec:chromatic_toas}, we assess the dominant cause for red noise in each of  the 25 objects.   

Table\,\ref{tab:rednoiseanalysis} presents  assessments for the 25 MSPs analyzed in \citet[][]{2023ApJ...951L..10A}. Columns 2-4  give the DM and spin parameters, which pertain to interstellar and spin noise effects, respectively.
The NANOGrav 15\,yr values for $\ARN$ and $\gammaRN$ are repeated here  in columns 5-6 from Table\,2 of \citep[][]{2023ApJ...951L..10A} and the spin noise analysis of Sect.~\ref{sec:spin} is presented
as $A_{\rm rsn}$ in column 7 and its ratio to $A_{\rm}$ from the NANOGrav analysis in column 8.   The red-noise classification based on both analyses is in column 9.   The overall assessment in column 10 combines factors based on interstellar processes combined with the red noise  and spin analyses.   This also involves inspection of
time series for DM and timing residuals given in \citet[][]{2023ApJ...951L..10A}.  The assessments in the final column Table\,\ref{tab:rednoiseanalysis}, though based in part on these quantitative criteria, are ultimately subjective in nature. 

We use the following criteria to assign a pulsar to one of three classes:
\begin{enumerate}
\item {\bf GW background:}  Dominance by the stochastic background is indicated  if  the timing residuals yield $\ARN \sim \Agw$ and a steep power is identified with $\gammaRN \sim 4$.  \newtext{We use estimates for $\Agw$ based on the common red noise process identified by multiple PTA groups. } ISM effects can be ruled out on this basis and also if the pulsar's DM is small (e.g. $\lesssim 20\,\DMunits$), which implies chromatic leakage is small.  Further corroboration follows if the predicted red spin noise is smaller than observed. 
\item {\bf Red spin noise:}  If a steep spectrum $\gammaRN \gtrsim 4$ is indicated and the measured red noise amplitude is significantly larger than $\Agw$,   dominance by spin noise is likely (or possibly orbital noise from asteroids, though there is little precedence for conclusive evidence for asteroid belts, only plausibility).  
Low-DM lines of sight will produce only weak or negligible contributions and they will not yield steep spectra. 
\item {\bf Interstellar noise leakage:}  Interstellar red noise from refraction  is indicated  for large-DM pulsars if the frequency index  $\gammaRN \lesssim 2$ and if $\ARN > \Agw$.   ISM effects give native frequency indices  $\gammaISM$ ranging from $2/3$ to $5/3$ for a Kolmogorov spectrum, the smallest of these for non-contemporaneous frequency observations for DM estimation or for the effect of  angular wandering on barycentric \TOAs.   Aliasing flattens the noise process for $\gammaISM < 1$ (cf.\ Fig.\,\ref{fig:pl_aliases}). 
\end{enumerate} 

%
%


\begin{deluxetable}{l r c c l l h  c r c c}[h!]
\tablewidth{5in}
\tablecaption{\label{tab:rednoiseanalysis}
Individual RN model parameter values and 68\% credible intervals for pulsars with significant 
detections of RN. 
Note that $\ARN$ is expressed in strain amplitude to match  $A_{\rm gw}$ from the GW analysis; both are unitless.}
\tablecolumns{11}
\tablehead{
& & & & \multicolumn{2}{c}{NANOGrav 15yr\tablenotemark{\footnotesize a}} && \multicolumn{3}{c}{Spin noise analysis}\\ 
\cline{5-6}  
\cline{8-10} 
    \colhead{Pulsar} &
    \colhead{DM} &
    \colhead{P} &
    \colhead{$\rm \dot P$} &
    \colhead{$\log_{10}\ARN^{\rm 15\,yr}$} &
    \colhead{$\gammaRN^{\rm 15\,yr}$} & 
    \colhead{} &
    \colhead{$\log_{10}A_{\rm rsn}$} &
    \colhead{$\displaystyle\frac{\strut A_{\rm rsn}}{ A_{\rm gw}}$} &
    \colhead{$\displaystyle\frac{\rm \lower.48ex\hbox{RN}}{\rm type}$} &
    \colhead{$\displaystyle {\rm Overall} \atop {\displaystyle\rm assessment}$}
    \\ 
    & $(\DMunits)$ & (s) & ($10^{-20}$\,s\,s$^{-1}$) 
    \\
    \colhead{(1)} &
    \colhead{(2)} &
    \colhead{(3)} &
    \colhead{(4)} &
    \colhead{(5)} &
    \colhead{(6)} &&
    \colhead{(7)} &
    \colhead{(8)} &
    \colhead{(9)} &
    \colhead{(10)} 
} 
\startdata
  B1855$+$09   &    13.3 &   0.0054 & 1.78   & ${-14.0_{-0.4}^{+0.3}}$ & ${3.9_{-0.8}^{+1.0}}$ & &    -14.9 &     0.56 & gw  & gw   \\
  B1937$+$21   &    71.0 &   0.0016 & 10.50  & ${-13.6_{-0.1}^{+0.1}}$ & ${4.0_{-0.3}^{+0.4}}$ & &    -13.6 &    10.65 & rsn & rsn  \\
  B1953$+$29   &   104.5 &   0.0061 & 2.97   & ${-12.8_{-0.3}^{+0.2}}$ & ${1.8_{-0.7}^{+1.1}}$ & &    -14.7 &     0.89 & gw  & ism  \\
  J0030$+$0451 &     4.3 &   0.0049 & 1.02   & ${-14.4_{-0.5}^{+0.4}}$ & ${4.6_{-0.9}^{+1.1}}$ & &    -15.1 &     0.33 & gw  & gw   \\
  J0437$-$4715 &     2.6 &   0.0058 & 5.73   & ${-13.4_{-0.2}^{+0.2}}$ & ${0.5_{-0.4}^{+0.6}}$ & &    -14.3 &     1.92 & rsn & ism? \\
  J0610$-$2100 &    60.7 &   0.0039 & 1.23   & ${-12.9_{-0.5}^{+0.3}}$ & ${4.1_{-1.9}^{+2.0}}$ & &    -14.9 &     0.49 & gw  & rsn  \\
  J0613$-$0200 &    38.8 &   0.0031 & 0.96   & ${-13.8_{-0.3}^{+0.3}}$ & ${3.1_{-0.7}^{+0.9}}$ & &    -15.0 &     0.45 & gw  & gw?  \\
  J1012$+$5307 &     9.0 &   0.0053 & 1.71   & ${-12.6_{-0.1}^{+0.1}}$ & ${0.8_{-0.3}^{+0.3}}$ & &    -14.9 &     0.55 & gw  & ism  \\
  J1600$-$3053 &    52.3 &   0.0036 & 0.95   & ${-13.5_{-0.6}^{+0.2}}$ & ${1.6_{-0.7}^{+1.5}}$ & &    -15.0 &     0.39 & gw  & ism? \\
  J1614$-$2230 &    34.5 &   0.0032 & 0.96   & ${-14.9_{-0.8}^{+1.0}}$ & ${4.7_{-2.0}^{+1.6}}$ & &    -15.0 &     0.44 & gw  & gw   \\
  J1643$-$1224 &    62.4 &   0.0046 & 1.85   & ${-12.3_{-0.1}^{+0.1}}$ & ${0.9_{-0.4}^{+0.4}}$ & &    -14.8 &     0.66 & gw  & ism  \\
  J1705$-$1903 &    57.5 &   0.0025 & 2.15   & ${-12.6_{-0.1}^{+0.1}}$ & ${0.5_{-0.3}^{+0.4}}$ & &    -14.5 &     1.28 & rsn & ism  \\
  J1713$+$0747 &    16.0 &   0.0046 & 0.85   & ${-14.1_{-0.1}^{+0.1}}$ & ${2.6_{-0.4}^{+0.5}}$ & &    -15.2 &     0.28 & gw  & gw   \\
  J1738$+$0333 &    33.8 &   0.0059 & 2.41   & ${-14.6_{-0.6}^{+0.8}}$ & ${5.2_{-1.8}^{+1.3}}$ & &    -14.8 &     0.73 & gw  & gw?  \\
  J1744$-$1134 &     3.1 &   0.0041 & 0.89   & ${-14.1_{-0.6}^{+0.4}}$ & ${3.6_{-1.2}^{+1.4}}$ & &    -15.1 &     0.33 & gw  & gw   \\
  J1745$+$1017 &    24.0 &   0.0027 & 0.27   & ${-11.9_{-0.1}^{+0.1}}$ & ${2.4_{-0.5}^{+0.6}}$ & &    -15.5 &     0.13 & gw  & ?    \\
  J1747$-$4036 &   152.9 &   0.0016 & 1.31   & ${-12.6_{-0.2}^{+0.1}}$ & ${2.4_{-0.7}^{+1.0}}$ & &    -14.6 &     1.03 & rsn & ism  \\
  J1802$-$2124 &   149.6 &   0.0126 & 7.26   & ${-12.2_{-0.2}^{+0.2}}$ & ${1.8_{-0.6}^{+0.7}}$ & &    -14.5 &     1.33 & rsn & ism  \\
  J1853$+$1303 &    30.6 &   0.0041 & 0.87   & ${-13.5_{-0.4}^{+0.2}}$ & ${2.3_{-0.7}^{+1.1}}$ & &    -15.1 &     0.32 & gw  & gw?  \\
  J1903$+$0327 &   297.5 &   0.0022 & 1.88   & ${-12.2_{-0.1}^{+0.1}}$ & ${1.5_{-0.4}^{+0.4}}$ & &    -14.5 &     1.24 & rsn & ism  \\
  J1909$-$3744 &    10.4 &   0.0029 & 1.40   & ${-14.5_{-0.4}^{+0.3}}$ & ${4.1_{-0.9}^{+1.0}}$ & &    -14.8 &     0.70 & gw  & gw   \\
  J1918$-$0642 &    26.6 &   0.0076 & 2.57   & ${-13.8_{-0.7}^{+0.4}}$ & ${2.7_{-1.0}^{+1.5}}$ & &    -14.8 &     0.63 & gw  & ism? \\
  J1946$+$3417 &   110.2 &   0.0032 & 0.32   & ${-12.5_{-0.1}^{+0.1}}$ & ${1.4_{-0.4}^{+0.5}}$ & &    -15.5 &     0.13 & gw  & ism  \\
  J2145$-$0750 &     9.0 &   0.0161 & 2.98   & ${-12.9_{-0.1}^{+0.1}}$ & ${0.6_{-0.4}^{+0.5}}$ & &    -15.0 &     0.41 & gw  & ism  \\
  J2234$+$0611 &    10.8 &   0.0036 & 1.20   & ${-13.9_{-0.9}^{+0.5}}$ & ${3.2_{-1.9}^{+2.7}}$ & &    -14.9 &     0.50 & gw  & gw?  \\
\enddata
\tablenotetext{a} {From \citet[][]{2023ApJ...951L..10A}.}
\end{deluxetable}

\newpage


\section{Profile modeling summary and arrival time models}
\label{sec:summary_toa_model}

We assert the point of view that optimal 
estimation of {\it achromatic} \TOAs\ (i.e. after mitigation of chromatic effects)  should be done \newtext{epoch-by-epoch.} 
\newtext{This requires contemporaneous estimation of DM and scattering delays (through \textit{profile modeling})} 
rather than analyzing chromatic \TOAs\ in long, multiepoch time series to remove offsets peculiar to any given epoch. This echoes the conclusion in 
\citet[][]{2024A&A...692A.170I} that an 
``epoch-wise''' approach yields greater accuracy for removing DM variations than methods that utilize  piecewise linear fitting, the DMX method, \citep[e.g.][]{2015ApJ...813...65N,2021ApJ...909..219S}), or Gaussian process (GP) modeling \citep[e.g.][]{2014PhRvD..90j4012V}\footnote{We note that alternate criteria 
discussed by 
\citet[][]{2024A&A...692A.170I} imply that GP modeling and DMX fitting can be superior to epoch-wise fitting in some cases. However, epoch-wise fitting is more heavily influenced by pulse signal-to-noise ratios 
so improved telescope sensitivity can alter this comparison.}.
Moreover, our statement applies to {\it all} chromatic terms that vary with epoch\footnote{By and large this implies interstellar refraction, scattering, and dispersion but also applies to any systematic shape changes in emitted pulse profiles, which in some cases are also chromatic.}, which require more intricate fitting of \TOAs\ than usage of only a  $\nu^{-2}$ dispersion term, as is well known 
\citep[][]{cwd+90, 
rdb+06, 
cs10, 
CSS16, 
2018ApJ...861...12L, 2022ApJ...930L..27K,
2024MNRAS.535.1184S,
2019A&A...624A..22D,
2020ApJ...892...89L}.   
Inclusion of non-dispersive chromatic terms  has been explored with 
GP noise modeling 
\citep[e.g.][]{2023A&A...678A..49E,2024ApJ...972...49L}, but with scattering expressed in timing equations with explicit $\nu^{-4}$ frequency scalings.   We have shown that timing delays differ from scattering times and that delays originate from multiple processes that are epoch dependent with differing frequency dependences.  This leads to mismatch between templates and profiles 
\newtext{
resulting in systematic \TOA\ offset errors.}
Moreover, these errors generally do not scale as $\nu^{-4}$. 

The recourse, at the very least, is to weight  contemporaneous measurements more strongly than  those from adjacent epochs.   This might entail a more sophisticated GP modeling approach.  For example, GP modeling of {\it profiles} rather than of individual terms in measured \TOAs\ could better account  for slowly varying chromatic effects.  However, it would not capture the more rapidly varying effects (e.g. finite scintle effect, refraction).

Solely using single-epoch data for achromatic \TOA\ estimation is therefore the optimal approach. 
This is much more feasible now that wideband (3:1 or larger) receivers are available to provide simultaneous frequency coverage and more sensitive telescopes are available.   Previously,   DM estimation required separate dual or multiple band observations spread over multiple days  or longer, necessitating multiepoch fitting. With determinations from single-epoch measurements,  the quality of the results will be less dependent on the cadence of timing observations.

Myriad effects contribute to arrival times, many  of which are  distractions from the goals of some applications  but are nonetheless  diagnostic  of instrumental parameters important for establishing the veracity of an analysis and may be  of interest for  ancillary astrophysics. 
This section consolidates the dominant effects into  empirical timing model(s) suitable for quantitative analysis, taking into account that \TOA\ estimates  are methodology dependent. 
 Addressing  biases requires consideration of how the effects of interstellar plasma processes are intertwined with the chromaticity of emitted pulse shapes.  A detailed analysis  in Appendix\,\ref{app:notation_and_toa_estimation_errors}   is summarized here.

\subsection{Profile modeling}
\label{sec:profile_model}

Profiles are formed by synchronously averaging $M\gg1$ single pulses  (equivalent to an integration time  $M\times P$, where $P$ = pulse period) in a bandwidth $B$  at center  frequency $\nubar$ and epoch $\tbar$.  In the following, $M$  and $B$ are implicit and we ignore any distortions from instrumentation (including polarization miscalibration) and RFI. These distortions are of course unavoidable in actual measurements and could dominate if astrophysical delays are mitigated to a large degree. 

Define $\Ui$ as the intrinsic average profile (i.e. upon emission)   that is scattered by the PBF $\PBF$ to produce $\Us = \Ui * \PBF$.   
The three quantities $U, \PBF$, and $\Us$ are all stochastic. 
We make use of two kinds of averages: an ensemble average denoted by angular brackets $\langle \cdots \rangle$ and an average over epochs $\langle \cdots \rangle_{\tbar}$  indicated with a subscript $\tbar$.  We assume ergodicity so that  in the limit $\tbar\to\infty$ the epoch average converges to the ensemble average. 

 The intrinsic and scattered profiles and scattering function  are written in terms of their ensemble averages and deviations, 
 \be 
\Ui =  \Uiave + \dUi, \quad\quad
\Us = \Usave + \dUs, \quad \text{and} \quad
\PBF =  \PBFave{} + \dPBF.
\ee
The  scattered profile is then
\be
\Usave &=& \Uiave * \PBFave{} 
\\
\dUs &=&\dUi * \PBFave{} + \Uiave * \dPBF + \dUi * \dPBF.
\ee
With complete generality,  all terms in these expressions are functions of $t, \nubar$, and $\tbar$ but the  mean intrinsic pulse shape $\Uiave$ is  now assumed to be epoch independent (with exceptions, as previously mentioned).   The deviation $\dUs$  includes intrinsic pulse jitter in $\dUi$ and fast and slow variations of the PBF in $\dPBF$, all of which are epoch dependent and chromatic.

The  measured profile  in Stokes $I$ is 
\be
I(t, \nubar, \tbar) 
= A(\nubar, \tbar)  \times \Us(t - t_0(\nubar, \tbar),  \nubar, \tbar) + N(t, \nubar, \tbar). 
\label{eq:Imod}
\ee
The pulse amplitude  $A(\nubar, \tbar)$ includes scintillation modulations  (DISS and RISS) of the emitted flux density that are epoch dependent and chromatic.  DISS modulations are attenuated by bandwidth and temporal averaging for high-DM pulsars or low-frequency observations.    Modulations can be large for low-DM pulsars at 1 to 2\,GHz, thus affecting the \SNR-related \TOA\ error 
(cf.\ \Eqs\,\eqref{eq:rmstoaS}, \eqref{eq:rmstoaS_gaussian}, and \eqref{eq:rmstoaS2}).

The arrival time $t_0(\nu, \tbar)$ includes contributions to the net propagation time from pulsar to observatory but {\it  not} those effects that  alter the pulse shape.  It therefore includes interstellar dispersion and its temporal variability along with single-path refraction but it excludes shape-changing effects from  multipath propagation.  
As such, this \TOA\ term is intended to correspond to propagation from a specific location  in a pulsar's magnetosphere  that corotates with the neutron star. 
The additive noise term $N(t, \nubar, \tbar)$ is from radiometer noise, which is uncorrelated between different 
$\nubar$ and $\tbar$. 

Without any  shape changes ($\dUs = 0$),  the pulse model  in Eq.~\eqref{eq:Imod} would strictly satisfy the conditions for matched filtering (Sect.~\ref{sec:toa_estimation}) and the only \TOA\ error would be from additive noise, $N(t, \nubar, \tbar)$.   That shape changes do occur implies that \TOA\ errors will always be larger than those predicted for matched filtering.

\subsection{Template modeling \newtextR{and matching}}
\label{sec:templates}

Errors in \TOAs\  depend on the methodology for implementing template matching and in particular the choice of template.  A matched filter by definition maximizes the \SNR\  of the filter output.  However, 
even if a template nearly yields the maximum theoretical  \SNR, it does not eliminate \TOA\ {\it bias}.   The chromaticity and epoch dependence  of measured pulse profiles essentially guarantees that this bias will always be present. 
 Existing practices make use of different kinds of templates in the matched filtering process described in Sect.~\ref{sec:toa_estimation}.  
%

\begingroup

\renewcommand{\SNR}{{\rm S/N}}

\renewcommand{\arraystretch}{0.8}

\begin{deluxetable}{l l l l l }[h!]
\tablewidth{600pt}
\tablecaption{\label{tab:template_types} Templates and timing errors induced by scattering}      
\tablecolumns{5} 
\tablehead{\\
\colhead{Template type} &
\colhead{Intrinsic Pulse Shape} &
\colhead{PBF} & 
\colhead{Chromatic \TOA\ cause} &
\colhead{TOA correction}  \\
 & \colhead{($U$)} & \colhead{($\PBF$)} & \\
}
\renewcommand{\arraystretch}{1.0}
\startdata
\\
\multicolumn{3}{l}{{\bf True (hypothetical)} \quad  (for center frequency $\nubar$ and epoch $\tbar$)} \\
	    						&   $\Ui(t, \nubar, \tbar)$ 			& $\PBF(t, \nubar, \tbar)$ 			&  \multicolumn{1}{l}{None \tablenotemark{\footnotesize a}}  				&  \multicolumn{1}{c}{0} \\
	    						&   $\Ui(t, \nuref, \tbar)$ 			& $\PBF(t, \nuref, \tbar)$ 				&  \multicolumn{1}{l}{Chromatic $\Ui$}  								&  \multicolumn{1}{c}{$ \DtUs(\nu, \nuref, \tbar)$}
\\ \\
\multicolumn{3}{l}{{\bf Empirical}  (where $\langle \cdots \rangle_{\tbar}$ indicates average over available epochs, $\tbar$)} \\
\quad\quad 1. 2D  \ \ ($U_2$)		& ${\Uave{(t, \nu, \tbar)}}_{\tbar}$ 	& $\PBFave{(t, \nu, \tbar)}_{\tbar}$  	& \multicolumn{1}{l}{Epoch dependences\tablenotemark{\footnotesize b}}  	&  \multicolumn{1}{c}{Customized} \\
\quad\quad 2. 1D \ \ ($U_1$)		& ${\Uave{(t, \nuref, \tbar)}}_{\tbar}$ 	& $\PBFave{(t, \nuref, \tbar)}_{\tbar}$ & \multicolumn{1}{l}{Epoch dependences and} 						& \\ 
  \quad\quad\quad (fixed at $\nuref$)	&							&							&  \multicolumn{1}{c}{chromaticity  \tablenotemark{\footnotesize b}}  			& \\
\\
{\bf Synthetic model} 					& Multi-Gaussian: \tablenotemark{\footnotesize c}  	& Exponential: \tablenotemark{\footnotesize d}  \\
							& \quad \quad $U_{\rm g}(t, \nubar, \tbar)$  		&   \quad\quad $ e^{-t / \tau(\nubar, \tbar)} \Theta(t)$   &   $U, \,\PBF$ mismatches &  $ \DtGE(\nubar,\tbar)$	
\\ 
		&															& Power law: \tablenotemark{\footnotesize e} \\
		&															& \quad\quad $  \PL(t, \nubar, \taud(\tbar), \Thetavec_{\PL}) $  & $U$ mismatch & $\DtGPL(\nubar, \tauhat_{\rm d}(\tbar), \delta \widehat{\Thetavec}_{\PL})$ \\
		& 															& \quad\quad $\Thetavec_{\PL}  = (\beta, \linner, \thetavec_{\xvec}) $  & Parameter errors 	
\\ \\
{\bf Coherent}  & 	Estimated $\widehat U$  					& Estimated $\widehat{\PBF}$ 		&  Estimation errors 	& $\DtCS(\nubar, \that)$  \\
{\bf deconvolution} & \\
\\
\hline
\\
\multicolumn{3}{l}{Estimates for scattering times,  $\tauhat_d(\tbar)$} \\
	 \multicolumn{3}{l}{\quad\quad 1. Dynamic spectrum analysis (ACF, secondary spectrum)}  & Estimation error		& $\Dttaud(\tauhat_d(\nubar, \tbar \vert U, \PBF)$		\\
	 \multicolumn{3}{l}{\quad\quad 2. Cyclic spectrum analysis} 						& Estimation error		& $\Dttaud(\tauhat_d(\nubar, \tbar \vert U, \PBF)$ 		\\ 
\\
\enddata
\tablenotetext{a}  {No \TOA\ offset from template-profile mismatch.   \TOA\ errors from white noise effects are implicit.}
\tablenotetext{b}  {Epoch dependence of $\PBF$ due to  $\taud(\nu, \tbar)$ and (rarely) epoch dependence of $U$. Chromaticity of  intrinsic pulse and PBF shapes.}
\tablenotetext{c}  {Multicomponent Gaussian model: $U_{\rm g}(t, \nu, \tbar) = \sum_{j} a_j \exp[-(2\sqrt{\ln 2} (t - t_j(\nu, \tbar)) / W_j(\nu, \tbar))^2]$}
\tablenotetext{d}  {$\Theta(t)$ = Heaviside function}
\tablenotetext{e}  {$\Thetavec_{\PL}$ = parameter vector for power-law PBF that includes $\thetavec_{\xvec}$, a vector of parameters for the spatial distribution of scattering regions (e.g. thin screen, thick slab, transversely bounded clouds, etc; $\delta\Thetavec_{\PL}$ denotes 
errors in these parameters. 
}
\end{deluxetable}
\endgroup

Briefly,  several alternatives have been employed, as summarized in 
Table\,\ref{tab:template_types} and elaborated on here:
 \begin{list}{}
\item  {\bf \newtext{Analytic} template model:}  A template comprising one or more analytic components (e.g. Gaussian functions) typically  does not explicitly include scattering; however, scattering in data used to form synthetic templates may have affected the fitting of these components, albeit at low levels for low-DM pulsars.   It will therefore yield  both systematic and random  \TOA\ errors. 
To match actual profile chromaticity, particularly with multicomponent profiles,  synthetic templates need to include frequency dependent amplitudes, widths, and pulse phases. 

A common approach  models the intrinsic pulse shape $\Uiave$ as a superposition of Gaussian components, 
\be
\Um(t, \nu, \tbar) = \sum_{j} g_j(\nu, \tbar) \exp[-(2\sqrt{\ln 2} (t - t_j(\nu, \tbar)) / W_j(\nu, \tbar))^2],
\label{eq:synthetic_template}
\ee
with individual amplitudes $g_j$, locations $t_j$ and widths $W_j$ (FWHM).  Generally, all three quantities vary with
frequency and also, in principle, with epoch.    However, pulsars providing the best timing results evidently have very little intrinsic  variability with epoch.  The Crab pulsar is a counterexample: its spindown time $\sim 10^3$\,yr is short enough that secular changes in pulse shape have been discerned since its discovery more than half a century ago \citep[][]{2013Sci...342..598L}.   Another is the MSP J1713+0747 that shows changes in profile in the form of episodic events (cf.\ Sect.~\ref{sec:shape_shifters}. 
\newtextR{
Recent work \citep[][]{2026MNRAS.552g1715B} suggests that a significant fraction of MSPs displays subtle epoch-to-epoch profile changes that are achromatic while others show chromatic changes mostly due to scattering.
}
\item {\bf Empirical template at each frequency:}  Multifrequency templates built from measured  pulses necessarily include scattering, albeit from an epoch or range of epochs that differ from those of profile data used to obtain \TOAs.  This  removes intrinsic profile chromaticity (to the extent that it is epoch independent) but will only  partially remove systematic \TOA\  errors from scattering. 

 \item {\bf Empirical template at a single frequency and epoch} ($\nu_{\rm template}$  and $\tbar_{\rm template}$): Application of a single-frequency template to multifrequency profiles yields chromatic \ToA\ errors if, as is common, profiles evolve with frequency.   Correction for profile chromaticity requires a lookup table of timing offsets  (`FD' parameters).   Scattering baked into the template generally differs from that in 
the profile even after scaling a representative scattering time to each frequency (epoch dependence!).   Template fitting thus removes some but not all scattering contributions.  
\end{list}

Template matching can be applied in two ways:
 simultaneously over a wide band using a 2D template or on a narrowband,  channel-by-channel  basis covering the same total band. 
 \newtextR{In the literature, the phrase `wideband timing' is often applied to the first of these (template matching in two dimensions,  time and frequency) while 'narrowband timing' applies to the channel-by-channel approach.  There are two ways to accomplish the channel-by-channel approach: one is to use a single one-dimensional template that typically yields systematic frequency-dependent biases requiring  subsequent correction; the other is to use multichannel templates, one for each frequency channel,  that ideally removes frequency dependent biases.  We consider the `wideband' vs. `narrowband' nomenclature to be misleading given that the same data are used in both methods.   Our main purpose in the following is to describe the ensuing errors due to departures from the underlying assumptions of these methods.   }
 
{\newtextR{\it Application of two-dimensional empirical templates} }:
The 2D template is the profile shape averaged over many epochs  for each frequency $\nubar$ in a multichannel system spanning a total bandwidth $B$ and denoised to minimize \TOA\ estimation errors.   Ignoring the multiplier $A$ in Eq.~\eqref{eq:Imod}, the 2D template,
\be
\Uhat_2(t, \nubar, \tbar) = \langle I(t, \nubar, \tbar) \rangle_{\tbar},
\ee
is the basis for  `pulse portrait'  fitting pioneered by       
\citet[][]{liu14} and \citet[][]{2014ApJ...790...93P,2016ascl.soft06013P}. 
It is related to the ensemble average shape as 
$\Uhat_2 = \Usave + \langle \dUs\rangle_{\tbar}$.   Averaged over a large number of epochs, the difference 
$\langle \dUs\rangle_{\tbar}$ reduces to a small value that we ignore  in the following. 

Construction of 2D templates is done empirically by integrating a large number of pulses from multiple epochs.  In \citet[e.g.][]{2014ApJ...790...93P}, measured pulses are further modeled as multiple Gaussian components as is also done for 
\newtextR{1D templates} discussed above. 
\newtext{With 2D timing (the so-called `wideband timing' method}) the 2D template can be applied as a 2D cross correlation (or by least squares fitting in the 2D Fourier domain) with 2D profiles to yield $\DM(\tbar)$ and a \TOA\ at a specified reference frequency
\citep[e.g.][]{2022MNRAS.512.1234N}.

 At each frequency $\nubar$, the cross correlation of the template and profile, $\Uhat_2(t, \nubar, \tbar) * I(t, \nubar, \tbar)$  includes two terms.  The  leading term is the delayed autocorrelation of $\Usave$,  yielding a  \TOA\ estimate equal to the true value, $t_0(\nubar, \tbar)$.   The second term is the  cross correlation
 $\Usave * \dUs$, which  adds a perturbation  $\Delta t_{\dUs, \dPBF}(\nubar, \tbar)$ from pulse jitter and \PBFtext\ fluctuations, both fast (finite number of scintles) and slow (e.g. refractive modulation of the PBF).

\newtextR{{\it Application of one-dimensional templates}}: 
Alternative to 2D template fitting, individual \TOAs\ can be obtained for each subband frequency $\nubar$ across the total bandwidth $B$, which are then fitted to give $\DM(\tbar)$ and a \TOA\, at a reference frequency.  
\newtextR{One version of this method employs a single 1D template at a reference frequency $\nuref$}, 
 \be
\Uhat_1(t, \nuref, \tbar) = \langle I(t, \nuref, \tbar) \rangle_{\tbar} \equiv  \Uhat_2(t, \nuref, \tbar),
\ee
that is applied to individual profiles at each frequency $\nubar$. 
In this case, the first term of the CCF of $\Uhat_1$  with a profile adds a bias to the true \TOA\ due to profile chromaticity,  $\Delta t_{\Uref,U}(\nubar)$.
Profile chromaticity manifests as  changes  of   widths and pulse phases of profile components
with frequency.   The second term yields a stochastic delay from pulse jitter and \PBFtext\ variations like that for the 2D template, $\Delta t_{\dUs, \dPBF}(\nubar, \tbar)$, but it will differ quantitatively due to the different template. 
The net \TOA\ is written as
\be
\widehat{t} (\nubar, \tbar) = t_0(\nubar, \tbar) + \Delta t_{\Uoneref,U}(\nubar) + \Delta t_{\dUs, \dPBF}(\nubar, \tbar).
\ee

\newcommand{\DtPE}{\Delta t^{(\rm pe)}}

The first perturbation, $\Delta t_{\Uoneref,U}(\nubar)$, is the systematic error from profile chromaticity (`profile evolution').  It is  addressed in NANOGrav timing analyses by modeling it as a power series in $(\ln\nubar)$, 
\be
\widehat{\Delta t}_{\Uoneref,U_1}(\nubar) = \sum_{l=1}^{N_l} \DtPE_l (\ln\nubar)^l.
\ee
The coefficients $\DtPE_l$ and their optimal number $N_l$ are determined from a fit to profiles over many epochs 
for each pulsar.    Random errors in the coefficients are small if profiles have large-enough  
S/N but systematic errors arise if profiles show secular evolution.   Note that the  coefficients include contributions from scattering as well as intrinsic profile chromaticity. 

\newtextR{
A second version of channel-by-channel template fitting uses different
templates for each frequency channel that are essentially 1D cuts through the 2D template described above.   This approach avoids the bias that drives the need for FD parameters and is very similar to a full 2D fitting approach (indeed covering the same overall frequency band).  However, it may yield slightly different responses to frequency-dependent scattering and RFI that affect profiles chromatically. 
}

\subsection{PBF models}
\label{sec:pbfs} 

A  model for the pulse broadening function $\PBFm$ enters into \TOA\ estimation either directly through inclusion in  template modeling as in Eq.~\eqref{eq:synthetic_template} or indirectly with empirical templates. 
For any model, the $1/e$ time $\taud$ is strongly frequency dependent and also epoch dependent from refraction (and possibly other causes). 

{\it Exponential PBF:}
A common choice is the single-parameter
 exponential form, $\PBF_{\rm E}(t, \taud) = \taud^{-1} \exp(-t / \taud)\Theta(t)$ ($\Theta$ = Heaviside function) described earlier (Sect.~\ref{sec:thinscreens}).  The exponential PBF is almost always  a poor choice because it can apply only under special conditions that are not applicable to precision timing of low-DM MSPs.   First, it requires a thin screen scattering geometry but more importantly it applies to either a monoscale medium or one with a large inner scale, $\linner$, compared to the diffraction scale, $\ld$    
 \citep[e.g.][]{lr00}.
 Such a medium is inconsistent with observations of low-DM pulsars.  Nonetheless it has seen extensive, but problematic, use in pulse modeling.

{\it Power-law PBFs:}
A better choice is  $\PBF_{\rm PL}$, a `power-law'  PBF from  a medium with a power-law electron density spectrum, such as the Kolmogorov spectrum with a small inner scale (Sect.~\ref{sec:KlikePBFs}).  The parameters of PL models are the $1/e$ scattering time $\taud$ and  $\Thetavec_{\rm PL} = (\beta, \linner, \Theta_{\xvec})$, 
where $\beta$ is the spectral index of the density spectrum, $\linner$ is the inner scale, and
$\Theta_{\xvec}$ is a vector of parameters for the spatial distribution of scattering regions along and transverse to the \LOS.

{\it Consequences of mismatched profile elements:}
A template model  inevitably  differs  from a measured  profile shape  due to mismatches between   the modeled and actual intrinsic shapes or  PBF shapes.   
Template fitting accordingly yields systematic \TOA\ 
\newtext{
offsets}
that accompany random errors.

As with empirical templates, we express the model intrinsic profile and model PBF in terms of their average forms or shapes  
and departures from them, 
\be
\Um &=&  \Uiave + (\Um - \Uiave) \equiv \Uiave + \dUm
\\
\PBFm &=& \PBFave{} + (\PBFm - \PBFave{}) \equiv \PBFave{} + \dPBFm .
\ee
Together these give a model scattered profile,
\be
\Ums &=& \Um * \PBFm = \Usave + \dUms , 
\ee
that deviates  from the actual scattered pulse $\Usave$ by 
\be
\dUms =  \dUm * \PBFave{} + \Uiave * \dPBFm + \dUm * \dPBFm. 
\ee

\subsection{Arrival time equations}

There is no unique choice of equation for modeling long sequences of
multiepoch \TOAs. The adequacy of a \TOA\ expression is illustrated by considering a sequence of models ranging from simplistic to comprehensive. 

\subsubsection{No ISM terms}
 
First, suppose  ISM effects are completely absent   but pulse profiles are chromatic and epoch independent. This is  hypothetical  in the radio context but would apply to 
very high frequency ($\gg 10$\,GHz) measurements, including non-radio observations from the millimeter band to gamma-rays. Though propagation effects are largely achromatic in these bands\footnote{Note however that scattering off interstellar grains can induce multipath propagation even in X-rays \citep[e.g.][]{1986ApJ...302..371M}.}
 emitted pulses are frequency or energy dependent, typically. 

The \TOA\ model is written compactly 
(with the center frequency  $\nubar$ and epoch $\tbar$ implicit) as, 
\be
\tnuhat 	&=& \tinfty +  \Dtmismatch  + \DtSNR + \DtJ ,
\label{eq:toa_no_ism}  	
\ee
where $\tinfty$ is the target \TOA\ that would include all achromatic effects related to the spin  and orbit of the neutron star combined with astrometric terms and gravitational wave perturbations\footnote{We associate $\tinfty$ with the rotational phase of a fiducial point on the neutron star's surface or, equivalently, the average location in the magnetosphere for emission at a particular frequency,  which is achromatic and epoch independent (at least over time spans much less than the spindown time). Differential emission times vs. frequency in the pulsar's magnetosphere (cf.\ Fig.\,\ref{fig:emission_field_line}) are  then included in the chromaticity of the pulse profile.}.
The $\Dtmismatch$ term is nonzero if a reference-frequency template is applied to profiles at all frequencies (or energies in the high-energy context). 
Errors from radiometer noise and pulse jitter have temporal white-noise statistics.  We keep them separate here because they differ statistically across frequency, as discussed previously. 
If epoch  independent,  $\Dtmismatch$ is determinable from measurements built up over many epochs on each pulsar.
This is the basis for the FD (`frequency dependent') parameters used in NANOGrav's chromatic mismatch model,
\be
\Dtmismatchhat(\nu) = \displaystyle \sum_{k=1}^n c_k (\ln\nu)^k,  
\label{eq:FD}
\ee  
where the number of terms and coefficients $c_k$  are determined on a pulsar-by-pulsar basis.
This appears adequate for many pulsar timing applications \citep[][]{2014ApJ...794...21D,2015ApJ...813...65N,2015ApJ...809...41Z}.

\begin{figure}[h!]
    \centering
    \includegraphics[width=0.475\textwidth]
    {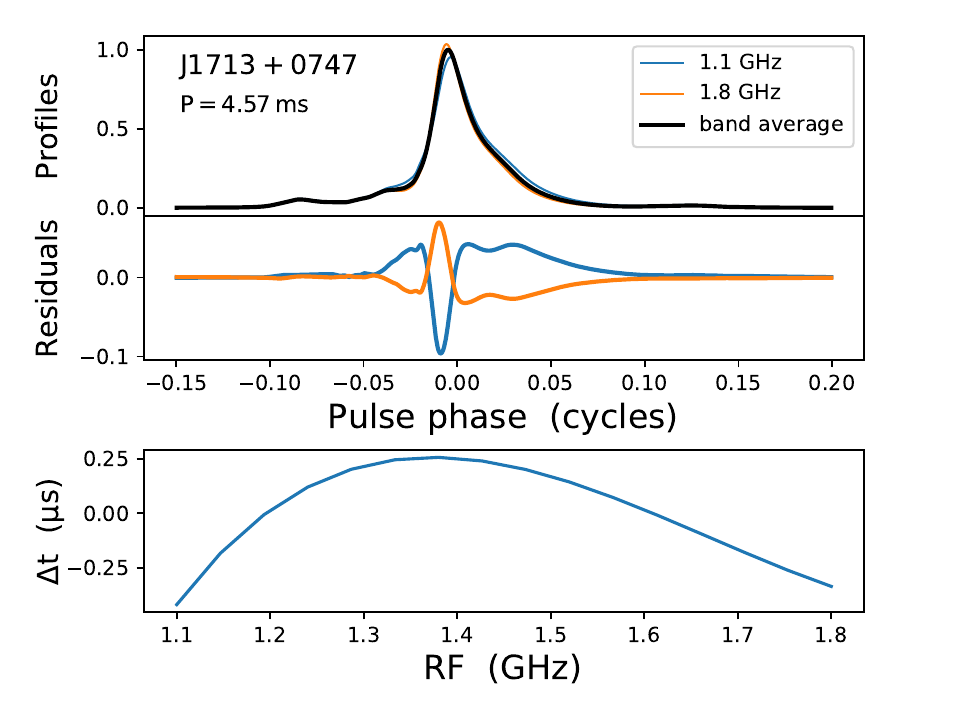}
    \includegraphics[width=0.475\textwidth]{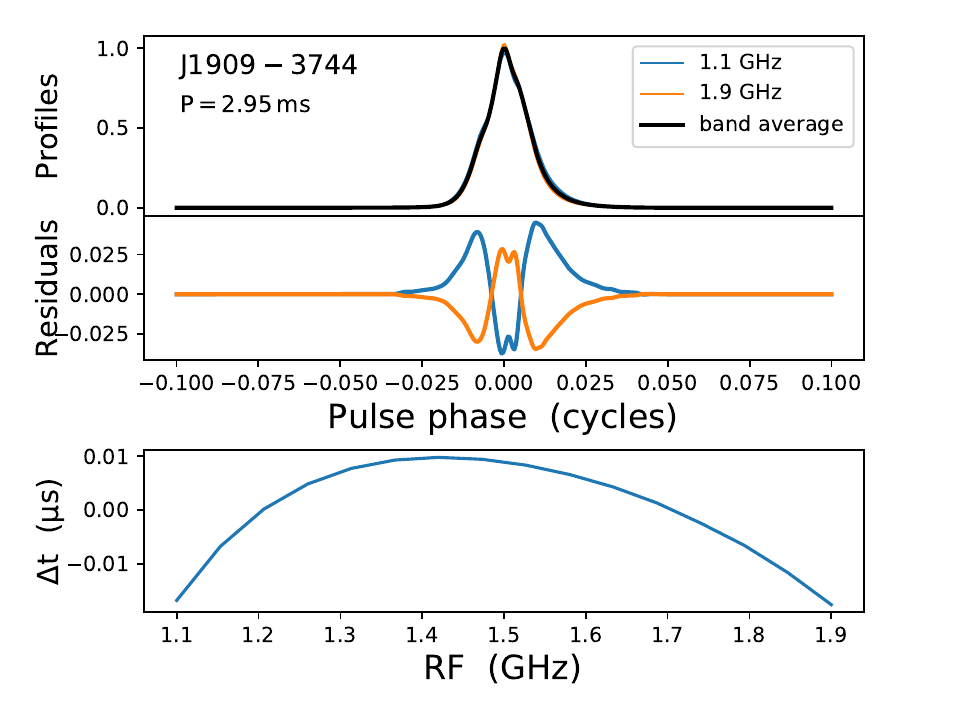}
    \caption{Timing offsets caused by profile chromaticity for two MSPs.  Top panels: profiles at band edges and a full-band average profile. 
    Middle panels: profile differences from the band average.
    Bottom panels: relative timing offset by using the band average as a template and removing a fit for a $\nu^{-2}$ DM term. 
    Profiles are generated using the interpolation method
    developed by \citet[][]{Pennucci_2019} as applied in the 
    15-yr NANOGrav data set  
    (\url{https://zenodo.org/records/16051178}) and using software 
    obtained at \url{https://github.com/pennucci/PulsePortraiture} and implemented by M. Lam. 
    }
    \label{fig:profile_chromaticity_and_timing}
\end{figure}

Fig.~\ref{fig:profile_chromaticity_and_timing} shows chromatic  template profiles, difference profiles,  and  timing offsets  across $\sim 1$ to 2 GHz.  These are based on the 2D profile modeling by
\citet[][]{Pennucci_2019}. 
Though the profiles differ very little by eye, \TOA\ differences of a sizable fraction of a microsecond occur for MSP J1713+0747 though they are only $\sim 10$\,ns for J1909-3744 owing to its narrow pulse.  The offset curve for J1713+0747 is consistent with that based on measured profiles shown by \citet[][Figure 4]{2014ApJ...794...21D}. 

The shape of $\Dtmismatch$ vs. $\nu$ is  methodology dependent, contingent on the template used to determine \TOAs.  As such it generally crosses zero with an arbitrary   zero point.  This implies that  a power-law model such as $\Dtmismatch(\nu) \propto \nu^{-\Xmismatch}$ cannot account for this term.   On the other hand, the chromatic mismatch model
(Eq.~\eqref{eq:FD}) can handle zero crossings of $\Dtmismatch$. 
 
Similarly the two dimensional template method for \TOA\ estimation over wide frequency bands
\citep[][]{2014ApJ...790...93P} also accounts for profile chromaticity. 
  With strict epoch independence, the $\Dtmismatch$ term can be removed to  essentially arbitrary precision under the conditions outlined here.   In the radio band  chromatic, epoch dependent ISM effects (and RFI) \newtext{may} obviate this possibility because they require attention to epoch dependent scattering and  the chromaticity of 
  PBF shapes for  power-law media. 

\subsubsection{Inclusion of an epoch independent  dispersion term}

Inclusion of a dispersion term with  DM that varies with  epoch but is achromatic (no scattering)  yields the next simplest case,
\be
\tnuhat 	&=& \tinfty +  \frac{\KDM\DM(\tbar)}{\nu^2} + \Dtmismatch  + \DtSNR + \DtJ    ,
\label{eq:toa_with_constant_DM}
\ee
In principle,  multifrequency observations and prior determination of $\Dtmismatch$ can remove 
 the DM term to a precision limited only by the white noise terms.   
 If applied to radio pulsars, this model would lead to chromatic leakage into estimates of $\tinfty$ caused by scattering, which causes
 \DM\ to be chromatic.

\subsubsection{Full expression for \TOAs\ at the SSBC}

Motions of the pulsar, ISM,  and observer introduce epoch dependence in all propagation effects, including  refraction and multipath propagation from scattering that also require differing chromatic terms in  the timing equation.  Pulse distortion from the convolution with the PBF induces short term and long term timing offsets.   Refraction, which varies on day to several month time scales, causes image wander and distortion of the PBF (Sect.~\ref{sec:aniso_refract}).

We account for the epoch and frequency dependent DM term
$\DMbar(\nu, \tbar)$ by expanding $\DMbar(\nu, \tbar) = \langle \DMbar(\nubar, \tbar)\rangle_{B} + \delta\DMbar(\nu, \tbar)$ as an average over the receiver bandwidth and  a fluctuating part.
Translation of \TOAs\ to the SSBC requires clock corrections and accounting for Doppler shifts and   astrometric errors.    Details  of these and other corrections are well discussed in (e.g.) \citet[][]{stairs03}, \newtext{\citet[][]{2006MNRAS.372.1549E}}, and \citet[][]{staylor_book2021}.    Here we include an additional error term in the translation to the barycenter related to the R\"omer term  for the difference in propagation times to a terrestrial observatory and to the SSBC;
 it is nominally achromatic, involving  the geometric path length difference with parallax and proper motion of the pulsar included.    Inclusion of the  refraction (AOA) term $\dtAOABary(\tbar, \nubar)$ renders the R\"omer term chromatic \citep[][]{fc90} in expressions for the arrival time  and radio frequencies referenced to the SSBC\footnote{Both the solar wind and the ISM contribute to refraction but our focus is on the interstellar contribution.  \newtext{For impact parameters  of a line of sight greater than  20\,deg $\ll 1$\,\au, refraction in the solar wind 
 \citep[][]{Fomalont2009}
 is smaller than  from the ISM.
 Some pulsars are observed close to the Sun, so refraction from the Sun should be investigated in more detail for those cases.}
 },
\be
\tnuhat   =  
\tinfty &+& \Dtmismatch(\nubar, \nuref) +   \frac{\KDM \langle \DMbar(\nubar, \tbar)\rangle_B}{\nu^2}   
\nonumber \\
&+&  \DtSNR + \DtJ + \DtdPBFall  
\nonumber \\
&+&   \DtdPBFbarall  + \frac{\KDM \delta\DMbar(\nu, \tbar) }{\nu^2} +  \dtAOABary(\tbar, \nubar).
\label{eq:toa_with_dm_scattering2}
\ee
The terms are ordered so that the 
first line of Eq.~\eqref{eq:toa_with_dm_scattering2} includes 
the usual target of interest, the achromatic $\tinfty$ term,  the epoch-independent  mismatch term from chromatic pulse shapes, and the $\DMbar$ term with an achromatic DM at epoch $\tbar$.  
The second line includes the well-recognized white-noise terms from the finite S/N, jitter, and DISS/FSE that change rapidly (i.e. statistically independent between non-overlapping data sets),  while the 
third line includes three \TOA\ offsets that vary slowly from changes in PBF shape, the chromatic DM term, and  the AOA/SSBC term.

The physical origin of the terms and their fluctuation properties  with respect to epoch $\tbar$ and fluctuation frequency $f$ are 
summarized in more detail: 
\begin{enumerate}
\item $\tinfty$ includes all achromatic contributions from a deterministic spin model, including spin noise,  emission location and direction, and orbital terms (all added to the mean propagation time related to the secularly changing pulsar distance,) and GW perturbations.   Spin noise and GW perturbations  have  red, power-law like spectra and any continuouse-wave GW signals add spectral lines. 
\item $\Dtmismatch(\nubar, \nuref)$ 
 is the timing offset due to profile changes with frequency when the template at $\nuref$ is used for template fitting at other frequencies. It is defined to be epoch independent;   pulsars that show epoch dependent profiles do so on a variety of time scales; those with discontinuous `mode' changes would show $f^{-4}$ spectra while variations like that of J1713+0747 would have amorphous spectra (note however that this kind of profile change would be mitigated through modeling in the time domain, not the spectral domain). 
\item $\KDM \nu^{-2}  \DMbar(\tbar, \nubar)_B$  is the dispersion term associated with the average DM across the bandwidth $B$ for center frequency $\nubar$ and scales strictly as $\nu^{-2}$. Its variations with epoch conform to  a power law spectrum determined by the wavenumber spectrum of the electron density with enhancements  by  discrete interstellar structures.  Additional contributions from the solar wind and Earth's ionosphere are typically much smaller but are variable over a wide range of time scales.
\item $\DtSNR + \DtJ + \DtdPBFall$ from radiometer noise, pulse jitter, and the DISS/FSE effect have  different radio frequency dependences but combine as white temporal noise  for \TOAs\ obtained from data that do not overlap in epoch or, for $\DtdPBFall$,  epochs separated by more than a DISS time scale (minutes to hours). 
Across frequency, $\DtSNR$ is uncorrelated between frequency channels (subbands) while $\DtJ$ is highly correlated over (at least) an octave in frequency.   The $\DtdPBFall$ term is intermediate with a correlation bandwidth $\dnud$ much smaller than an octave but often spanning multiple frequency channels.
\item $\DtdPBFbarall(\tbar, \nubar)$ is the timing offset from long term temporal changes in PBF shape; it  has a flat spectrum that rolls off at approximately  the inverse of the RISS time scale, $f \sim \DtRISS^{-1}$. 
\item $\KDM \delta\DMbar(\nu, \tbar) / \nu^2$ is the chromatic error from the scattering-induced frequency dependence of \DM.  It also has a flat spectrum that rolls off at $f \sim \DtRISS^{-1}$
(\citet[][Figure 5]{CSS16} and Sect.~\ref{sec:DMnu}).
\item $ \dtAOABary(\tbar, \nubar)$ from the barycentric translation of \TOAs\  caused by interstellar refraction  that varies both stochastically and cyclically with a yearly period. The spectrum is a shallow power law with rolloff at  $f \sim \DtRISS^{-1}$  shifted to a centroid of 1\,cy\,yr$^{-1}$. 
\end{enumerate}


\newcommand{\SP}{S_{\rm p}}
\newcommand{\SPave}{ \langle \SP \rangle}
\newcommand{\dSP}{\delta S}

\renewcommand{\tP}{ t_{\rm p}}
\newcommand{\tPave}{\langle  t_{\rm p}\rangle}
\renewcommand{\dtP}{\delta t_{\rm p}}

\newcommand{\tc}{t_{\rm c}}

\newcommand{\sigSP}{\sigma_{\SP}}
\newcommand{\sigtP}{\sigma_{\tP}}

\newcommand{\rhoSt}{\rho_{\rm S,t}}
\newcommand{\sigtc}{\sigma_{\tc}}

\section{\TOA\  error corrections}
\label{sec:toacorrections}

%

Raw arrival times from template fitting include all delays and pulse distortions imposed at the pulsar and during propagation (and instrumental effects peculiar to each observing system.). 

Any departure of a measured pulse profile's shape from  that of the template  yields an arrival time offset.  Shape differences can be imposed by propagation, instrumentation, or RFI.   They also occur statistically from the finite number of pulses used to form a profile.   Systematic changes in shape occur as `mode changes'
(predominantly in CPs) where single pulses are drawn from two or more pulse populations whose averages converge to  different template shapes.  

\subsection{Mitigation of intrinsic (emitted)  pulse shape variations (jitter etc.)}

Correlations between profile shape and timing offset can be exploited as the basis for \TOA\ error correction.   These methods have been explored using a variety of techniques, including principal component analysis (PCA), \newtext{as we will discuss here.}  

Suppose a pulse shape parameter $\SP$ is  determined for a sequence of average profiles, each comprising the sum of $N$ pulses.  Profiles differ for the reasons cited above (along with additive noise) and the shape parameter could be as simple as the amplitude ratio of two components or it could be the projection of a profile onto one or more of the basis functions from PCA.   Departures  $\dSP = \SP - \SPave$
from the shape parameter 
$\SPave$ of the template may correlate with  corresponding timing offsets $\dt = \tP-\tPave$.  
A linear  \TOA\ corrector  that minimizes the mean-square difference between  the corrected \TOA\ $\tc$ and the true \TOA\ is
\be
\tc = t  +  (\sigtP / \sigSP )\rhoSt   \dSP,
\label{eq:toacorr_jitter}
\ee
where  $\rhoSt$ is the  correlation coefficient, 
  $t$ is the nominal \TOA\ and the correction involves the \RMS\ values of $\tP$ and $\SP$.
If the shape parameter, $\SP$, depends on the absolute phase alignment of the profile, this procedure can lead to overcorrection, reducing \TOA\ differences from all sources rather than only those caused by differences in profile shape. For this reason, only shape parameters which are independent of profile alignment (such as the amplitude ratio of two components, or the overall skewness of the profile, but not, e.g., the central phase of a component) should be used for correction.
The efficacy of the error correction, as measured by the \RMS\ corrected \TOA, 
$
\sigtc = \sigtP (1 - \rhoSt^2)^{1/2}
$
requires a large correlation coefficient to make significant improvements.   To reduce the \RMS\ error by 50\% requires $\vert \rhoSt\vert = \sqrt{3}/2 \simeq 0.87$.

The earliest attempt of this type of \TOA\ correction was by \citet[][Appendix]{1986ApJS...62...81D} in their analysis of timing variations based on a profile shape analysis method. 
Trying several types of shape parameters, they found no significant correlation for the Vela pulsar and only $\sim 30$\% correlation for two pulsars (B1818$-$04 and B1911$-$04) out of the other 21 pulsars in the JPL timing program \citep[][]{dr83}.   A retrospective analysis \newtext{using results in Sect.~\ref{sec:dRF}} implies that the lack of correlation for the JPL sample is due to the absence of dedispersion of the S-band data\footnote{The JPL program used a narrow bandwidth at S-band that yielded small dispersion smearing of pulses, so no dedispersion was deemed necessary.}.   As shown in Sect.~\ref{sec:dRF}, interstellar scintillation causes the effective center frequency to vary on a diffractive scintillation time scale, producing timing variations (cf.\ Eq.~\eqref{eq:rmsnu}) without significant changes in pulse shape.  For the bright Vela pulsar, this effect dominates the jitter error, which in turn is larger than the radiometer noise contribution.  In this situation, no correlation is expected, because changes in effective center frequency do not modify the pulse shape. 

Usage of PCA for pulse shape and related  studies was done by \citet[][]{Blaskiewicz_1991PhDT.........1B}, 
\citet[][]{Demorest_2007PhDT........14D},  and \citet[][]{2011MNRAS.418.1258O}.  Correction of \TOAs\ from the very bright MSP J0437$-$4715 using PCA yielded  a 20\% reduction in \TOA\ error   
\citep[][]{2011MNRAS.418.1258O} \newtext{and a similar reduction on a long-period pulsar by \citet[][]{2018MNRAS.475.1323L}}. Recent work further developed PCA-based tools for characterizing pulse shape variations of simulated data and pulses from the Vela pulsar \citep[][]{2024arXiv241100236J}  and transitory shape variations from the MSP J1713+0747 \citep[][]{2024ApJ...964..179J}.   Application to \TOA\ correction to the Vela pulsar yielded a 30\% reduction in \TOA\ error \citep[][]{Jennings_2021PhDT........00J}.  Shape parameters derived from principal components are not necessarily independent of profile alignment, and so, in general, \TOA\ estimators based on them can overcorrect. For small \TOA\ differences, this overcorrection can be largely eliminated by ensuring the principal components used are orthogonal to the derivative of the average profile shape \citep[][]{2011MNRAS.418.1258O}. Alternatively, the \TOA\ can be estimated directly using a maximum likelihood or maximum \emph{a posteriori} method (Jennnings et al., in prep.), which eliminates overcorrection even when \TOA\ differences are large.

Simulations of average profiles indicate that intrinsic phase jitter for  pulsars with single-component profiles yields negligible correlation between pulse shapes and arrival times.  However, there are significant correlations (and thus \TOA\ improvements) for multicomponent profiles with strong amplitude variability of individual pulses with some degree of overlap in pulse phase  \citep[][]{Jennings_2021PhDT........00J}.   More dramatic changes from shape shifter pulsars, like those that display  profile `mode changes' (mostly seen in canonical pulsars) produce large correlations and thus significant corrections to \TOAs\ \cite[e.g.][]{2023MNRAS.523.4405N}.

\subsection{Removing DM variations}
\label{sec:toa_dm_removal}


While DM delays dominate all other chromatic perturbations, completely removing them is confounded by refraction, scattering, and intrinsic pulse profile chromaticity.   
In Sect.~\ref{sec:summary_toa_model} we briefly summarized two main approaches, one that operates on single-epoch pulse profiles and 
\TOAs\ and an alternative that separates chromatic terms in long sequences of many-epoch \TOAs\ (e.g. using DMX and Gaussian process models).  Comparisons of methods have been presented by 
\citet[][]{2024A&A...692A.170I}.
Contributions to \DM\ from the solar wind (SW) are absorbed in DM estimates so, to within the precision of those estimates, no special modeling of the SW is required.  However, close approaches of lines of sight to the Sun yield rapidly changing $\DM(t)$ that requires special attention, including the possibility of excision of those epochs from timing analyses.  

In the following we first consider pulsar-frequency combinations where scattering can be ignored even if $\ll 100$\,ns precision timing is striven for.  Then we consider alternative approaches for
making single-epoch corrections for dispersion and scattering. 

\subsection{Conditions where \TOA\ scattering bias is negligible}
\label{sec:noscatt}

To motivate the selection of pulsars based on allowable  tolerance levels  for scattering,  we use
a simplified \TOA\ expression,  $t_\nu = \tinfty + \nu^{-2} \KDM \DM + \epsilon_\nu$,
where the \TOA\ at frequency $\nu$ is the sum of the true \TOA\ $\tinfty$, a dispersion term, and an error term 
$\epsilon_\nu$. 
Though timing measurements are now made with broadband receivers, the essential issues are illustrated using \TOAs\ at two frequencies $\nul < \nuh$ with ratio $R = \nuh / \nul > 1$.  These are used to estimate and remove the DM term, yielding an   error for $\tinfty$, 
\be
\dtinftyhat = \tinfty - \tinftyhat 
= \frac{\epsnul - R^2 \epsnuh} {R^2 -1}  
= \epsnuh  \times \frac{(R^{\Xeps}  - R^2 )} {R^2 -1},
\label{eq:dtinfty}
\ee 
The high frequency error  is enhanced by the $R^2$ factor  but if  \TOA\ errors scale as 
$\epsilon _\nu \propto \nu^{-\Xeps}$ with $\Xeps > 2$, the low frequency  error dominates the net error, as in the second equality. 
We showed earlier that the chromatic error is generally smaller than the scattering time and the scaling index 
$\Xeps$ is itself chromatic and satisfies $\Xeps \lesssim \Xtau$ where $\Xtau = 22/5$ for the Kolmogorov inertial subrange.   
It is therefore conservative to require the scattering time $\taud_{\nul}$  to be less than some specified amount.  Assuming $\Xeps > 2$ for \TOA\ errors, Eq.~\eqref{eq:dtinfty} implies that the lowest frequency used by a PTA is the primary determinant of the DM-corrected \TOA.
Scattering delays can be ignored if they are below a chosen maximum and we  
invert the empirical expression for scattering times in Eq.~\eqref{eq:taudm} to derive the corresponding maximum
DM. 

No correction is needed if $\taud(\nu) < {\taud}_{\rm max}$ for a fixed value of ${\taud}_{\rm max}$, 
\footnote{An alternative is to require that only one scintle  fills the  bandwidth $B$, equivalent to having no frequency-dependent scintillation modulation, but since bandwidths differ between receivers, we prefer a constant maximum value.}, 
\newtext{
which we choose to be 
$\taudmax = 10$\,ns as a best-case value, 
}
yielding   a constraint on frequency,
\be
\nu \ge (\tau_1 / \taudmax)^{1 / \Xtau} \simeq  2.8 \,{\rm GHz}\, \times [(10\,{\rm ns} / \taudmax)\tau_1]^{5/22}.
\ee

\begin{figure}[h!] 
   \centering
   \includegraphics[width=0.7\linewidth]{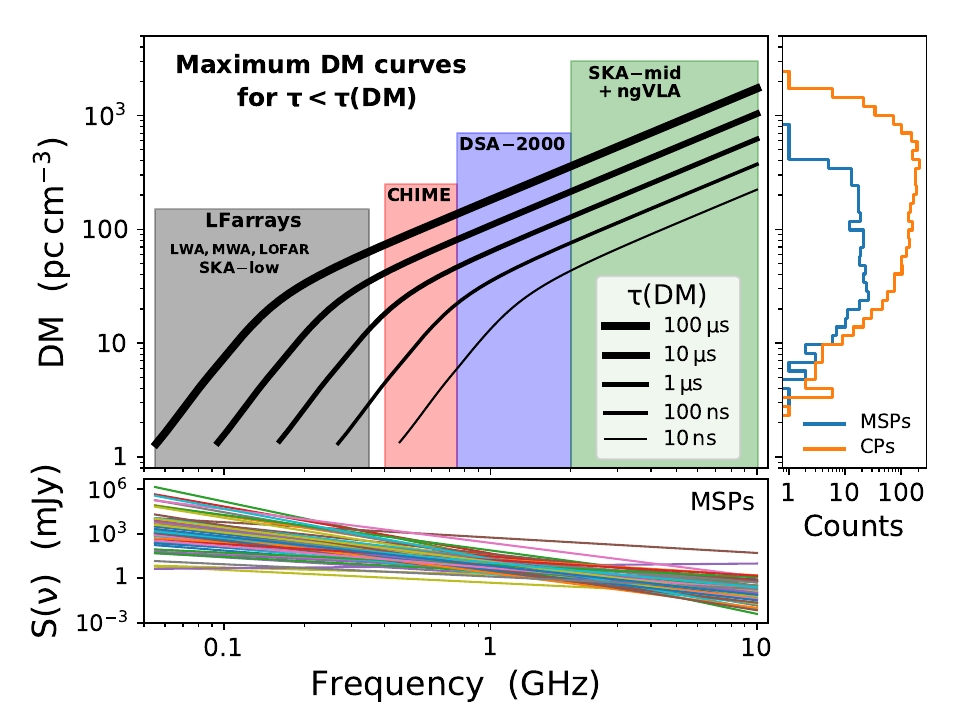}
   \caption{Scattering times  vs. \DM\ and radio frequency along with flux density spectra for MSPs and 
   DM number counts for MSPs and CPs. 
   Frequency bands of particular telescopes are indicated. 
   Largest frame: lines of constant $\taud$ vs   DM and frequency (black curves) \newtext{calculated using the mean of the $\tau(\DM)$ relation (Eq.~\eqref{eq:taudm}) shown in 
   Fig.~\ref{fig:hockeystick}. There is substantial variation about these curves.}
   Bottom frame: flux density vs. frequency for MSPs that have catalogued  flux densities at 0.4 and 1.4 GHz
    \citep[][]{ATNFPulsarCat}. Extrapolations to $\lesssim 0.1$ GHz are uncertain given that some pulsars show 
    turnovers in their spectra.
   Right panel: DM histograms for MSPs and CPs in the Milky Way disk; globular cluster pulsars are excluded.    }
   \label{fig:DMmax_vs_tau}
\end{figure}

Fig.~\ref{fig:DMmax_vs_tau} shows lines of constant $\taud$ as a function of \DM\ and frequency for scattering times from 10\,ns to $100\,\mu s$ along with frequency ranges for several low-frequency arrays and CHIME along with bands covered by forthcoming arrays (DSA-2000, SKA-mid, and the ngVLA). Current PTA programs by 
worldwide collaborations (CPTA, EPTA, InPTA, \newtext{MPTA}, NANOGrav, and PPTA)  use a large number of telescopes  approximately coincident with the band indicated for DSA-2000.  

  For a telescope covering  a specific frequency range,  like those indicated in the figure,  the lowest frequency determines whether the timing analysis can avoid mitigation of scattering.  
 For example,  pulsars with $\DM \lesssim 4\,\DMunits$ satisfy the condition $\tau(\DM,\nu) \le 100$\,ns  for the CHIME frequency band (0.4 to 0.8\ GHz), $\DM \lesssim 20\,\DMunits$ for the 0.7 to 2\,GHz band for DSA-2000, and 
$\DM \lesssim 75\,\DMunits$ for timing programs with 2\,GHz as the lowest frequency.   Low-frequency arrays will require scattering mitigation to achieve  100\,ns precision.  Conducting observations at high frequencies is limited by the generally steep decline in flux densities, as shown in the bottom panel of Fig.~\ref{fig:DMmax_vs_tau}.  Given the typically steep spectra of radio pulsars, observations from $\sim 1$ to 2 GHz generally  offer a good compromise between  minimizing  interstellar delays and maximizing  signal-to-noise ratios.  Magnetars have flat radio spectra, allowing observations at much higher frequencies where scattering is minimal,  but even at low frequencies, interstellar scattering delays are  completely dominated by intrinsic  spin noise in those objects. 

While restricting PTA programs to low-DM pulsars is one strategy, the DM histograms in the right-hand panel of 
Fig.~\ref{fig:DMmax_vs_tau} show that scattering mitigation for   MSPs   with  $\DM > 20\,\DMunits$  could increase the sensitivity to GWs, perhaps significantly, by virtue of the sheer increase in pulsar numbers. 

Timing programs with other goals, such as orbit monitoring of relativistic double neutron star binaries, need not have such demanding requirements on scattering times because other factors can  determine the achievable precision  in those cases, such as the  longer periods with their correspondingly wider pulses and their larger attendant pulse jitter.   

\subsection{Mitigation of  rapid PBF variations related to scintillation  fluctuations }
\label{sec:mit_fse}

\TOA\ errors from scattering include rapid
 variations from the finite scintle effect (FSE, Sect.~\ref{sec:dt_fse}) combined with an epoch-dependent bias related to the scattering time $\taud$. 
We consider the FSE here and \TOA\ bias in the next sections. 
The FSE error is a zero mean error resulting from the variable number of scintles contributing to a TOA or, equivalently,  from the stochasticity of the intensity PBF.  

\begin{figure}[h!] 
   \centering
\includegraphics[width=0.6\linewidth]{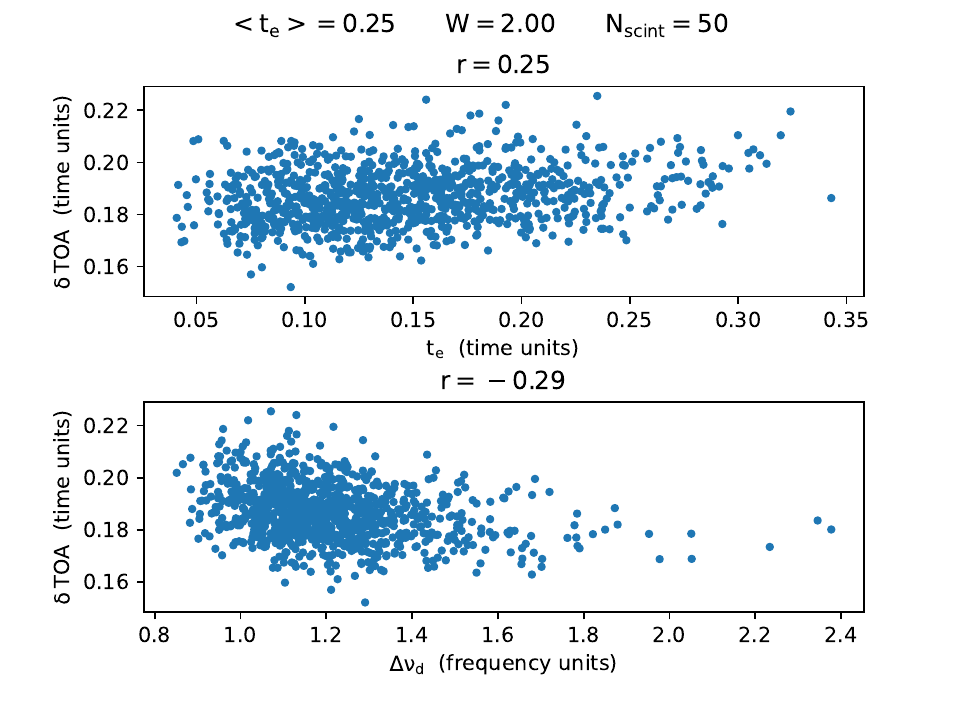}
   \caption{Simulations for a thick medium with a square-law structure function (1000 realizations). 
   Top:  $\delta$\TOA\ vs $1/e$ scattering time estimated directly from the noisy  PBF.
   Bottom: $\delta$\TOA\ vs. scintillation bandwidth estimated from an ACF analysis of the scintillation spectrum $H(\nu)$. 
	 }
   \label{fig:toas_vs_dnud}
\end{figure}

Fig.~\ref{fig:toas_vs_dnud} shows simulated \TOA\ variations and scintillation bandwidth estimates
for 1000 realizations of a PBF for a thick medium (the results do not depend strongly on the type of medium). 
The left panels show $\delta$\TOA\ plotted against the $1/e$ scattering time estimated directly from PBF realizations and plotted against the scintillation bandwidth estimated from ACFs of the scintillation spectrum.
  The right hand panel shows a scatter plot of $\delta$\TOA\ and
  $\dnud$ along with a regression line.  
  The \TOA\ and $\dnud$ are typically (in these simulations) anticorrelated at $\rdtdnud = -39$ to $-25$\%. 

Anticorrelation of $\delta$\TOA\ with $\dnud$ implies a positive correlation with the scattering time,
$\rdttau = - \rdtdnud$.
This correlation  allows correction of the TOAs using an expression analogous to 
Eq.~\eqref{eq:toacorr_jitter}
$t_{c} = t - \rdttau (\sigma_{\TOA} / \sigma_{\taud}) (\taud - \langle\taud\rangle),$
yielding a reduction in \TOA\ error by a factor  $R_{\rm c} = (1 - \rdttau^2)^{1/2}$.
For $\rdttau \simeq 0.35$,   the reduction factor is $R_{\rm c}  \simeq 0.94$, i.e. only a 6\% reduction.

\providecommand{\Um}{U_{\rm m} }	
\providecommand{\FTI} {\widetilde{I}}
\providecommand{\FTPBF}{\widetilde{\PBF}}
\providecommand{\DC}{{\rm f_{\rm DC}}}

\newcommand{\phiCS}{\phi_{\rm CS}}
\newcommand{\phiHT}{\phi_{\rm HT}}

\newcommand{\HT}{{\rm HT}}

\subsection{Mitigation of slow epoch-to-epoch scattering time ($\taud$) variations}
\label{sec:scatt_mit} 

To tackle the epoch-dependent variability of scattering,   mitigation of scattering  for timing purposes is best done using at-least contemporaneous and preferably simultaneous data for DM and scattering timing determinations.  \newtext{These quantities vary on a refractive time scale ($\dtr$, Table\,\ref{tab:keyplasmaeq}; see also \citet[][]{2025ApJ...986..191G}) which can range from hours to months or longer, depending on radio frequency and the particular line of sight.} 
The alternative use of scattering/scintillation data from separated epochs can only  partially mitigate scattering at a particular epoch.  We focus here on the simultaneous case. 

Use cases for scattering mitigation  divide into two categories defined by  the degree of scattering:
\begin{enumerate}
\itemsep -1pt
 \item  {\it Incoherent deconvolution of scattering:}
 This applies to strongly scattered pulsars with unresolvable scintillation structure.    
 Distant pulsars observed at low frequencies  will manifest scattering  solely as pulse broadening because  the scintillation bandwidth
 $\dnud$ is too small to resolve due to a combination of   FT resolution limits and low \SNR.   

   A notable use case for incoherent deconvolution  is  its application to  any special  pulsar of interest  at a large distance in the Galactic plane owing to its rarity.    Large scattering times $\gtrsim 1$\,ms, for example, would correspond to scintles with widths $\dnud \lesssim 160$\,Hz and rapid DISS time scales too small to be isolated in most data sets.  This implies contributions to  pulse broadening  in a typical data set  from a large number of scintles (cf.\ Eq.~\eqref{eq:Nscint}) and an uncorrectable \TOA\ error from the finite scintle effect (Sect.~\ref{sec:dt_fse}).  However, the much larger systematic \TOA\ offset can be mitigated through deconvolution. 
 
 Arrival times can be a byproduct of deconvolution of an intensity PBF or pulse-shape modeling, with accuracy dependent on the fidelity of the modeled  to actual shapes of the intrinsic pulse  and PBF. 
 
 Incoherent deconvolution of  an intensity pulse broadening function from measured pulse shapes would be followed by  template fitting at each frequency to obtain \TOAs; these would be  fitted for DM and  dispersion delays then removed, as usual.
 
Alternatively,   the measured pulse can be forward modeled  by convolving a model intrinsic pulse shape with candidate PBFs to identify the best PBF \citep[e.g.][]{bcc03,2025ApJ...986..191G}.   As discussed previously, a mis-modeled intrinsic shape will introduce biased values for the scattering time and \TOA\  as will a mis-matched PBF.  This procedure can be repeated for all frequency channels and then a DM-corrected \TOA\ can be estimated. 

The general flow of these methods is
\be
\begin{array}{lll}
\text{profile} &  I(\phi, \nu; \tbar) 	&   \\
							&+&  \quad \longrightarrow \text{model pulse and \TOA }\  \that_{\nu} \\
\text{PBF} &  \PBF(t, \nu; \tbar)
\end{array}
\ee

\item
\textit{Coherent deconvolution of scattering and high-resolution spectroscopy:}.  
Pulsars with resolvable scintillation structure allow alternative procedures.   MSPs observed  in current PTAs and other low-DM pulsars fall into this category.    Timing error corrections can be based on detailed analysis of scintillation structure using several approaches, including phase-resolved spectroscopy (PRS) and cyclic spectroscopy (CS). 

To correct biased arrival times, first estimate the pulse broadening time $\taud$ or the DISS bandwidth $\dnud = C_1 / 2\pi\taud$ and then use the known pulse template to determine the \TOA\ offset caused by $\taud$.  Recall that the \TOA\ offset is  generally not a simple function of $\taud$ (cf.\ Sect.~\ref{sec:heavytails}).
 Calculate biased multifrequency arrival times using  a fixed template followed by fitting  and removal of  multiple chromatic terms from the \TOAs, one of which is for scattering, which is epoch dependent,  and another for intrinsic profile chromaticity, which is largely epoch independent\footnote{The transitory pulse shape changes for J1713+0747 represent a counterexample 
 \citep[][]{2018ApJ...861..132L,2021ATel14642....1X,2021RNAAS...5..167L,
 2021MNRAS.508.1115L,2021MNRAS.507L..57S,2024ApJ...964..179J}}.
 
 The flow of this method involves two parallel steps
\begin{enumerate}
\item
\TOAs\ at frequency $\nu$ are obtained through template fitting of a profile at frequency $\nu$ and epoch $\tbar$:
\be
\text{profile} \ I(\phi, \nu, \tbar) \longrightarrow \text{template fit} \longrightarrow \that_{\nu}.
\ee
\item
In parallel, the dynamic spectrum (DS)  is produced  by phase resolved spectroscopy (PRS)  or by cyclic spectroscopy (CS) and  an estimate for the \TOA\ correction $\dtoacorr(\taudhat)$ is   derived  from the scintillation bandwidth $\dnudhat$,
\be
\text{PRS or CS} \longrightarrow
\text{dynamic spectrum (DS)} \ I(\nu, t, \tbar) \longrightarrow \dnudhat \longrightarrow \taudhat \longrightarrow  \delta\that_{\nu} (\taudhat).
\ee
\end{enumerate}
For high signal to noise cases,  {\it coherent} deconvolution is a possibility.  It makes  use of the phase information in the CS to obtain the field PBF $\htilde$ and the intrinsic pulse shape.  This effectively   deconvolves the field PBF
from the  measured voltage signal: 
\be
\text{CS} \longrightarrow \text{unscattered pulse shape and PBF} \longrightarrow \text{corrected \TOA}.
\ee
\end{enumerate} 

  \subsubsection{Intensity (incoherent) deconvolution of pulse broadening}
 
 Inversion of scattering through deconvolution with noisy data is almost always problematic.   With high enough S/N, an intensity PBF, if known, can be deconvolved from profiles.   
  Two   methods for doing so make use of  Fourier transform division or application of the CLEAN algorithm.  
 The  first is based on  the convolution  theorem applied to the measured profile and an assumed  PBF shape.   With no additive noise,  the FT of the intrinsic shape is simply  the ratio of FTs,
 $\widetilde \Ii(\nu) \propto \FTI(\nu) / \FTPBF(\nu)$  \citep[][]{1993MNRAS.260..724K}. However, additive noise makes this approach essentially infeasible because noise causes large excursions of the ratio.    
 
 The CLEAN algorithm, developed for deconvolution of the point spread function (PSF)  from radio images obtained with aperture synthesis, has been used for a broader range of applications than interferometic imaging.  It operates by iteratively subtracting a scaled and angle-shifted PSF from an image.  For pulse profiles, an assumed PBF plays the role of the PSF and the pulse profile for the image.   
 Without {\it a priori} knowledge of  the PBF shape,  iteration over a bank of PBF shapes can be applied to both identify  the best PBF and determine the accompanying  intrinsic pulse shape \citep[][]{bcc03,2004ApJ...605..759B,2017PASP..129b4301T,2024ApJ...962..131Y}. 
 A boundary condition that aids identification of the optimal PBF is that the resulting intrinsic intensity profile must be non-negative\footnote{Deconvolution with CLEAN has not been applied to other Stokes profiles, e.g. $Q, U, V$ or $L = \sqrt{Q^2 + U^2}$.   Particular objects may provide an advantage for discerning the PBF if the Stokes profile is narrower than the total intensity. However, $Q, U$ and $V$ are not constrained to be non-negative while $L$, though non-negative, is biased by off-pulse noise  that needs subtraction.}.
 Application to heavily scattered pulsars  \citep[][]{2004ApJ...605..759B} shows that some \LOS\  favor a thin screen PBF having a rapid rise time while  others  favor a thick medium with a slower rise.  
 
 Intensity deconvolution  has the advantage that it can be applied to  pulse profiles with long integration times $\gg \Dtiss$ and potentially  high signal-to-noise ratios.  However, to be useful, the pulse broadening  must be ``discernible,'' which we define  as $\taud  \gtrsim \epsilon W = \epsilon \DC P$ ($W$ = unscattered pulse width, $\DC$ = pulse duty cycle) with $\epsilon \sim 0.1$.    

\begin{figure}[htbp]
   \centering
   \includegraphics[scale=0.7]{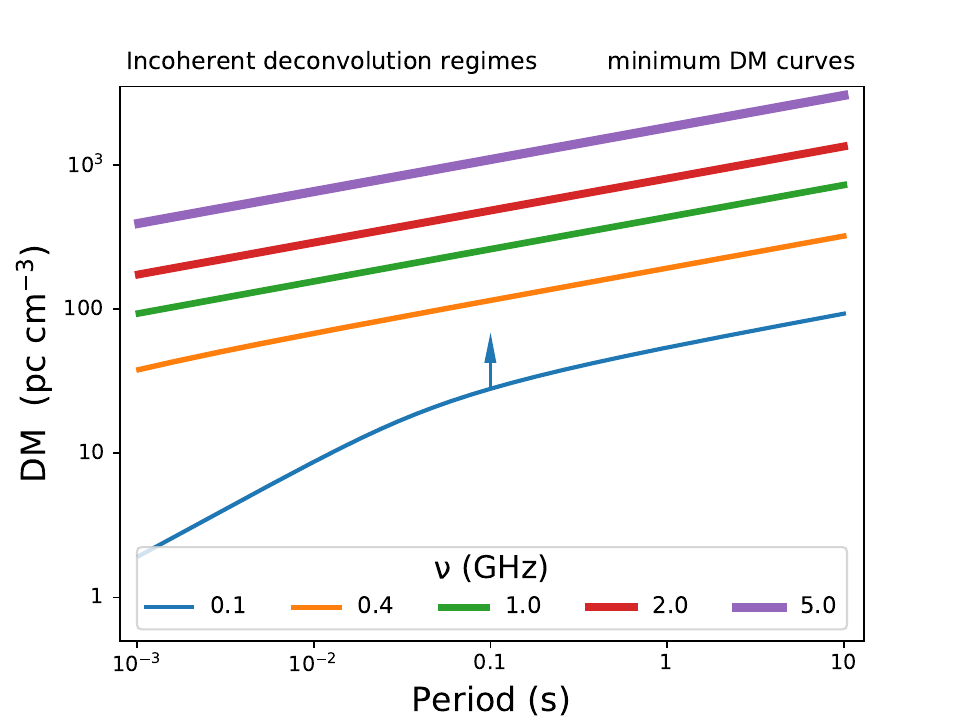}
   \caption{\footnotesize 
   Regions in the DM-Period plane where it is efficacious to incoherently
   deconvolve the pulse-broadening function.    The area above each line 
   indicates  the appropriate  DM and  pulse-period  range for each frequency as indicated by the arrow.
   \label{fig:incoherent_regime}
   }
\end{figure}

 Figure~\ref{fig:incoherent_regime} shows  regions in DM-period space where incoherent deconvolution is feasible for frequencies from 0.1 to 5\,GHz assuming a constant duty cycle $\DC = 0.05$.
If a period-dependent duty cycle is used, e.g. $\DC \propto P^{-1/2}$, the curves are not dramatically different except for small periods where the duty cycle is large, requiring a larger broadening time to satisfy the condition.   Since the MSPs used in precision timing have pulse widths much narrower than predicted by the $P^{-1/2}$ scaling that applies to long-period pulsars
\newtext{\citep[][]{1990ApJ...352..247R,2012MNRAS.424.1762M}},
the fixed duty-cycle curves are appropriate.

 \subsubsection{Forward modeling of pulse shapes}
 
Forward modeling  involves convolution of an intrinsic profile  model with candidate PBFs to identify the best PBF among a family of PBFs.   As discussed in Sect.~\ref{sec:pbfs} , the best  model PBFs are those drawn from electron densities with  power-law wavenumber spectra, which can
account for the variety of shapes encountered across all degrees of scattering.   However, these (and any other analytical shapes) do not account for PBF distortions from 
 refraction (Sect.~\ref{sec:aniso_refract}), leading to mismatches between the model and actual PBF.  
As discussed previously, mis-modelling of the  intrinsic shape and/or  the PBF cause   bias in the estimated  scattering time and \TOA\ \newtext{\citep[e.g., see also][]{lentati2017}}.    

This procedure can be repeated for each frequency channel, making use of the expected frequency dependence of both $\taud$ and the PBF shape to aid estimation of  the PBF parameter vector
$\Thetavec_{\rm PL}$.    The resulting \TOAs\ can then be analyzed to estimate DM and $\tinfty$. 

 {\it Customized profile modeling:}   As argued in Sect.~\ref{sec:nu_toaoffsets}, profile modeling with the correct shapes for the intrinsic pulse and PBF can completely remove the \TOA\ error associated with the PBF envelope applicable at a given frequency and epoch.   

A potential modeling scheme can proceed as follows. 
The PBF envelope $\PBF(t, \nu; \tbar)$ depends on the inner scale of the electron density wavenumber spectrum and is not  generally self-similar vs. frequency. In contrast,  an exponential PBF scales only  by the single parameter, $\taud(\nu)$.    It is most likely that the inner scale $\linner$ is   epoch independent for most pulsars, at least over  the decades long time spans of PTA data.   The key determinants of  the PBF shape are $\beta$, the spectral index of the wavenumber spectrum, and the ratio $\zeta = \linner / \ld(\nu)$, where the diffraction scale $\ld(\nu)$ depends on the strength of scattering as well as on $\beta$.   The method would be applied to each pulsar, requiring  a preliminary analysis to determine basic parameters, and then an epoch-by-epoch  analysis to obtain \TOAs:
	\begin{enumerate}
	\item Use multiepoch, multi-frequency observations to determine $\beta$ and $\linner$ for the line of sight.   At each epoch,  simultaneous fitting over a bank of functions for both the intrinsic pulse and the PBF  ideally will yield a determination of $\linner$ and $\beta$.  An analysis similar to this has been done by 
\citet[][]{2025ApJ...986..191G}.   In the few cases where $\linner$  has  been determined,  values range from $\sim 35$ to $\sim 1000$\,km \citep[e.g.][]{1990ApJ...348..147M,sg90,1995ApJ...438..708M, 2004ApJ...605..759B,2009MNRAS.395.1391R}. The inner scale may be  associated with the thermal  proton gyroradius   or the proton inertial length, which involve the electron temperature, magnetic field, and electron density.  Since all of these vary across the ISM, the inner scale almost certainly varies along an \LoS\ as well as between different 
\LoS.   
	\item The applicable value of $\beta$ from pulse shape fitting may show much more regularity than exists in the literature \newtext{\citep[e.g.][]{10.1093/mnras/stt989,lewandowski2015,2017MNRAS.470.2659G}} owing to the bias effects discussed in Sect.~\ref{sec:systematic_errors_tau_toa}.  A working conjecture is that  use of a fixed value for $\beta=11/3$  in pulse modeling will suffice for the estimation of
the inner scale in step 1.
	\item  For a sustained timing program,  the intrinsic shape model  $\Um(t, \nu)$ and values for 
$\beta$ and $\linner$ from preliminary stages are used to model the pulse shape at each epoch and subband, yielding both the $e^{-1}$ scattering time and the \TOA$(\nu)$ as well as additional corroboration for the  input values for $\beta$ and $\linner$ based on subband to subband results from the modeling.
	\end{enumerate}
Complications in pulse modeling result from interstellar refraction, which causes distortion of the applicable PBF that varies  on refraction times of days for low-DM pulsars and weeks,  months,  or longer for high-DM pulsars.     The severity of refractive distortion needs to be assessed on a case by case basis.  At worst, corrections of \TOAs\ for scattering will be limited by un-modeled  refraction effects so that 
 mere subtraction of a delay estimate based on the $e^{-1}$ broadening time of method 1 is the best approach.

\subsubsection{Correction of \TOAs\ using  extant or contemporaneous estimates of the pulse broadening time $\taud$.} 

This approach  uses a bulk,  value for $\tau$ based on multiepoch measurements, which cannot  correct for rapid FSE errors that are statistically independent between epochs.   It can partially correct for slow, epochal changes in the shape of the PBF envelope (e.g. from refraction) if multi-epoch measurements are employed.  The method involves estimation  
in subbands narrow enough so that $\taud(\nu)$ is close to constant over the subband and then use 
$\taud(\nu)$ from all subbands to correct \TOAs\ for  the scattering-induced error  $\DtdPBF$. 
  This error is a function of the shapes of the intrinsic pulse and the PBF with frequency dependence unique to each pulsar.   Simulations show that the systematic error  is generally  negative for some frequencies and positive for others.   This cannot be captured as a Gaussian process but is better dealt with (for wideband systems) by model fitting \TOAs\ obtained at each individual  epoch.

\subsubsection{Scattering corrections using template fitting + dynamic spectrum analysis}
\label{eq:scattcorr}

 When scintillation structure is resolved in frequency, scattering corrections to \TOAs\  can be derived from the scintillation bandwidth  obtained from the DS.   The DS itself is obtained from  phase-resolved spectroscopy (PRS) or cyclic spectroscopy (CS), the choice depending on the ratio of scattering time to pulse width.   
 Using a DS based on the same data (same time span and frequency band) as the \TOA\ estimate, 
 an autocorrelation analysis of the DS vs. frequency lag  yields the scintillation bandwidth.   The scattering time then follows from
 the uncertainty relation (Eq.~\eqref{eq:uncertainty1}), 
$\taudhat = C_1 / 2\pi \dnudhat$. 
The \TOA\ correction $\dtoacorr(\taud)$ is  generally a nonlinear function of  $\taud$ dependent on the intrinsic pulse shape and PBF (Sect.~\ref{sec:systematic_errors_tau_toa}).   It therefore needs to be  customized for each pulsar as  a predetermined lookup table.  One type of error associated with this procedure derives from not knowing the exact value of $C_1$ for real-world PBFs.  

\begin{figure}[b!]
   \centering
   \includegraphics[scale=0.8]{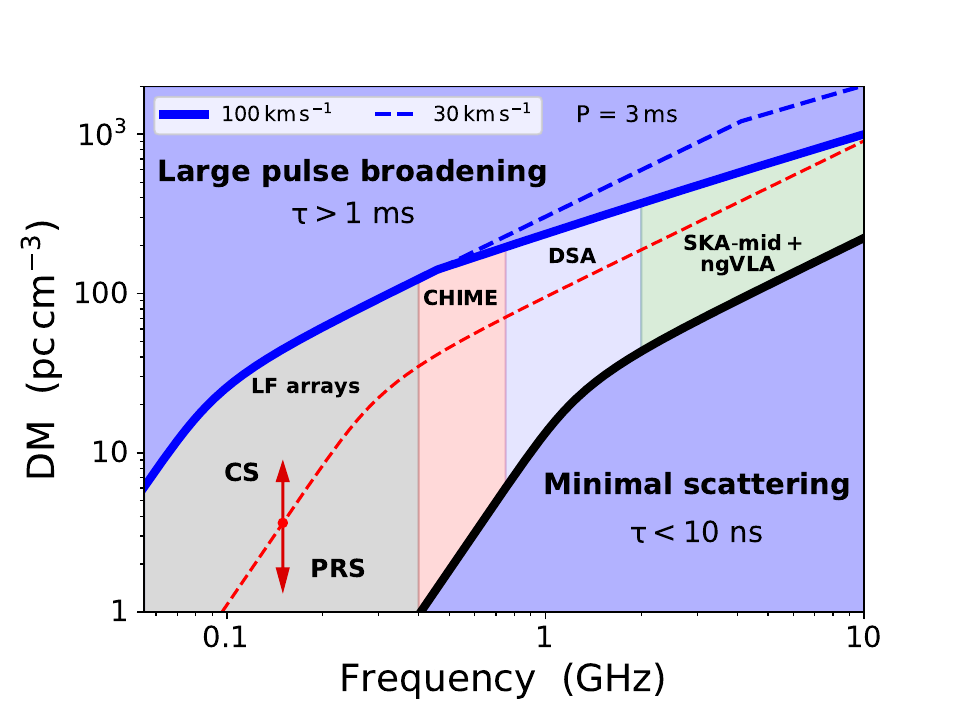}
   \caption{\footnotesize 
   The DM-frequency  plane for scattering mitigation using DS analyses. The region between the heavy blue and black lines  is amenable to estimation of scattering corrections with DS obtained either using cyclic spectroscopy (CS) or phase-resolved spectroscopy (PRS) with the boundary indicated by the red dashed line.  
   DMs below the black line correspond to negligible scattering in no need of mitigation while 
   above the heavy blue line  scintles are too small to allow  a DS analysis,   The blue dashed line shows the case where the effective speed is 30\,$\kms$ rather than 100\,$\kms$. 
   \label{fig:DM_RF_cs_prs_regimes}
   }
\end{figure}

Fig.~\ref{fig:DM_RF_cs_prs_regimes} identifies areas in the DM-frequency plane where a DS analysis is useful 
vs. those  where scattering is too large for precision timing or too small to require mitigation.   CS is required for large DMs and/or
lower frequencies because $\dnud$ cannot be resolved with PRS, as indicated.   Also shown are the relevant frequency ranges for the same telescopes indicated in Fig.\,\ref{fig:DMmax_vs_tau}.   
  
The potential for coherent deconvolution is if the field PBF can be determined contemporaneously from the same data used to calculate \TOAs. Doing so makes use of the phase information in the CS to obtain
the field PBF $\htilde$ and the intrinsic pulse shape.  This effectively   deconvolves the field PBF
from the  measured voltage signal  \citep[e.g.][]{2011MNRAS.416.2821D, 2013ApJ...779...99W, Palliyaguru_2015, 2021ApJ...913...98D, 2023ApJ...944..191T,2024ApJ...972...16T}.
Coherent deconvolution  strives to use  CS to obtain the complex $\hism(t)$, which is essentially a task of {\it phase retrieval} of its Fourier transform,  $\phiCS(\nu) = \arg\{\htilde(\nu)\}$.  
This has been demonstrated for the MSP J1939+2134 (B1937+21) \newtext{in the afore cited papers} but it has not yet been used to obtain arrival times with greater precision than standard methods.   \newtext{Further work is needed in technique development and real-time access and processing of voltage data.}
   
One approach uses the phase of $\left\langle  \Gamma_{\htilde}(\delta\nu, \nu) \right\rangle_{P \ll T \ll \dtd}$,
\be
\Phi_{\Gamma}(\delta\nu, \nu) = \phi_{\htilde}(\nu+\delta\nu/2) - \phi_{\htilde}(\nu - \delta\nu/2), 
\label{eq:PhiGamma}
\ee
(where additive noise is ignored) to estimate  the derivative for small $\delta\nu$,
\be
\frac{d\phi_{\htilde}(\nu) }{d\nu} \simeq  \frac{\Phi_{\Gamma}(\delta\nu, \nu)}{\delta\nu}.
\label{eq:dPhiGammadnu}
\ee
Upon integration, this yields an estimated  phase $\widehat{\phi}_{\,\htilde} \equiv \phiCS$ to within a constant.   Unfortunately, unless  statistical errors in $\Phi_{\Gamma}(\delta\nu, \nu)$ are very small,  the derived phase will random walk away from the true solution.   Also, numerical evaluation of the phase will inevitably introduce multiple $2\pi$ phase wraps that need to be dealt with.

An alternative approach avoids direct use of the phase by seeking an iterative solution for $\htilde$ that yields convergence to the measured $\left\langle  \Gamma_{\htilde}(\delta\nu, \nu) \right\rangle_{P \ll T \ll \dtd}$. 
Simulations demonstrate the viability of this approach on idealized data and for the bright MSP J1939+2134
   \citep[e.g.][]{2013ApJ...779...99W, Palliyaguru_2015, 2021ApJ...913...98D}.   As of yet,   the method has not yielded improvements in \TOA\ estimates. 
   
%
\begin{figure}[t!] 
   \centering
   \includegraphics[width=\figwidth]{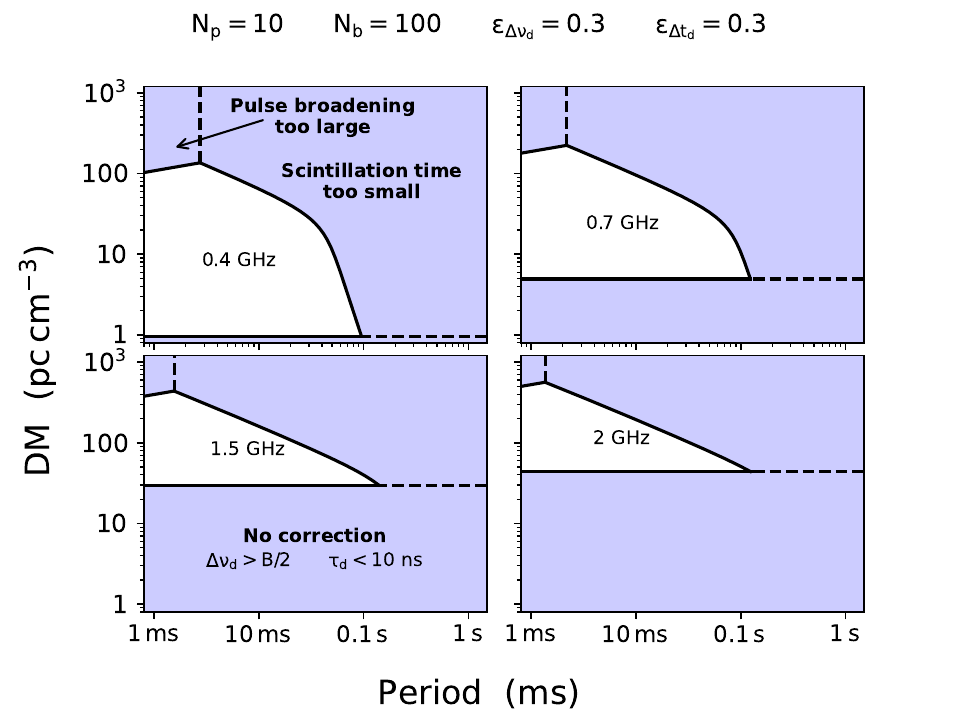} 
   \caption{Regions in the DM vs. period plane where CS is potentially viable at four different frequencies (white).  Excluded regions are based on Eq.~\eqref{eq:nucond1}, \eqref{eq:nucond2}, and \eqref{eq:nucond3} and described in the text.}
   \label{fig:DMP4pan}
\end{figure}

It is possible that coherent deconvolution will be fruitful in future, high S/N data.   
Though determination of $\hism(t)$ from the cyclic spectrum is not unique,  it may be sufficient to at least partially correct TOAs and deconvolve the pulse shape.   The regime in which this approach may work is  expressed with the inequalities,
\be
\WA \sim 2\pi\tau_d \ll P \le T \ll  \frac{\dtd}{\Nb},
\label{eq:reqcohdecon}
\ee
based on requiring the intrinsic pulse width $W_A$ and scattering time to be much less than the pulse period, which in turn is much less than the time $T$ used to calculate an individual  FT.  A large number of separate FT blocks need to be averaged to reduce errors.  The aggregate time of  $\Nb \times T$ data blocks must be much less than scintle durations ($\dtd$).   The pulsar signal must have high \SNR, which is a separate requirement from
Eq.~\eqref{eq:reqcohdecon}. 
Using the hockey-stick expression for $\taud$ in Eq.~\eqref{eq:taudm} and $\dtd$ expressed in terms of $\taud$, we produce the allowed ranges in DM and $P$ shown in 
Fig.\,\ref{fig:DMP4pan} that satisfy the inequalities
of Eq.~\eqref{eq:reqcohdecon}.   It is clear that coherent deconvolution is viable only for pulsars with  short periods and small DMs for frequencies  $\sim$\,0.4 to 1.4\,GHz.   The cases shown are for $\Nb = 100$. 

In principle, the phase space for coherent deconvolution can potentially be applied  to MSPs with $\DM \gtrsim 100\,\DMunits$ for timing programs covering 0.4 to 2 GHz (or higher).   However, signal to noise considerations imply that only a few MSPs may satisfy the conditions using 100-$m$ class telescopes.  However $\sim$ 300-$m$ class telescopes (Arecibo,  FAST, DSA-2000, and SKA) can  enlargen this number.   At present, coherent deconvolution requires further experimentation.  Nonetheless, even when a full coherent deconvolution is not possible, the cyclic spectrum a remains a useful tool for RFI mitigation given the fine frequency resolution it enables. Whether this should be done routinely is a resource-specific question, but should become increasingly feasible with growing compute power. 
\subsubsection{Phase retrieval with the Hilbert Transform (HT) }
\label{sec:toa_errcorr}

The scintillation spectrum $H(\nu, \tbar) \equiv \vert \htilde(\nu, \tbar)\vert^2$, though possessing  no phase information, can be used as the basis for deriving a {\it minimum delay field PBF}.    As with CS, implementation of an Hilber transform (HT)-based method requires high S/N data.   Though not yet used in scattering mitigation, it may find application in niche   circumstances.  Here we summarize the utility of the HT.   A more detailed,  exploratory analysis is given in NANOGrav Memorandum 9  (2012-2025)\footnote{J.M. Cordes, {\it Correction Methods for Interstellar Pulse Broadening}, \url{https://nanograv.org/science/memos-white-papers}}.

Generally a function cannot be determined uniquely from the magnitude of its FT  or from its autocorrelation
function \citep[e.g.][]{sca81}.
Additional conditions may yield a unique result,  modulo S/N uncertainties as with the CS approach.
A thorough discussion of conditions required for phase retrieval is given in \citet[][]{Klibanov_1995}.   
For a causal function,  the real and imaginary parts of its FT are related by an HT, as with 
 the Kramers-Kronig relations for dispersion relations in dielectrics.
%
The Hilbert transform  yields a phase $\phiHT(\nu)$ corresponding to  the {\it minimum delay} solution for the time-domain function,
\be
\phiHT(\nu) = \frac{1}{\pi}\, 
	{\cal P} 
		\int_{-\infty}^{\infty} d\nu^{\prime}\, 
			\frac{\ln \vert \tilde h(\nu) \vert} {\left( \nu -\nu^{\prime}\right) },	
\ee
where ${\cal P}$ denotes Cauchy principal component and  $\vert \htilde(\nu) \vert = \sqrt{H(\nu, \tbar)}$ is the  
FT magnitude at  epoch $\tbar$ for a snapshot shorter than the scintillation time $\dtd$.  That is, $H(\nu, \tbar)$ is well approximated by the DS at a fixed time if the DS has sufficiently averaged out fluctuations from pulsar self noise. 
The implied  signal flow going from the DS to an HT-based field PBF is  
\be
\left\langle \vert \EfieldFT(\nu)\vert^2 \right\rangle_{P \ll T \ll \dtd}  \
	\longrightarrow \ \vert \htilde(\nu) \vert^2 \
	\longrightarrow \ \phiHT(\nu)  \
	\longrightarrow \ \htilde_{\rm HT}(\nu) \
	\longrightarrow  \ h_{\rm HT}(t).
\ee
If the field PBF decreases monotonically, such as the one-sided exponential function,  it can be determined uniquely using the HT.    
Empirical studies \newtext{based on secondary spectra show
\citep[][]{2010ApJ...708..232B, Stinebring22}}, however,  that PBFs are not monotonic and can be multimodal, albeit with peaks that generally are substantially smaller than the PBF's maximum.      In this case the width of the HT solution is biased low.  For quasi-monotonic PBFs\footnote{I.e. those for which there is a prevailing decay superposed with small bumps.},   there is a systematic relationship between the width of the HT and actual PBFs.   An example is given in 
Fig.\,15 of Memo\,9 cited above, showing  that the  mean delay of the HT-PBF is  on average $\sim$70\%  of the mean of the true PBF.    This delay is, of course, not the same as the \TOA\ correction, so further analysis like that described earlier is needed to 
relate the PBF width to the correction. 

Cyclic spectroscopy may play a role   in  providing high-resolution intensity dynamic spectra for cases where scintles are too small in frequency extent to be resolved with standard setups for pulsar timing (i.e. where frequency resolution is determined by dedispersion contraints).    However, data from  low-DM pulsars with resolved scintles  can be treated without recourse to CS.

\subsection{Assessment of  MSPs for scattering mitigation}
\label{sec:scattassess}

\TOA\ correctability for a specific MSP depends on its pulse properties, the DM (and scattering) depth along its line of sight, and especially on observation frequency.  Here we evaluate individual MSPs  at four frequencies (0.4, 0.7, 1.5, and 2\,GHz) for 20\% fractional bandwidths using conditions derived in Appendix\,\ref{app:ds_errors_prs_cs}.  The primary constraint is that the net  arrival time have an error below a specified ceiling of 1\,$\mu s$, \newtext{which we choose as a useful fiducial value}.    Some MSPs require no correction at the higher frequencies while others cannot be plausibly corrected to this precision. 

In the following we use the following units:
 Frequency $\nu$  and bandwidth $B$ are in GHz; 
period $P$ and pulse width $\WA$ are in ms. 
The system equivalent flux density (SEFD) $\Ssys$ is in Jy and the period-average flux density $\Sphi$ is in mJy.  
Scattering times $\tau_1$  at 1\,GHz are in $\mu s$ and $\dnud$ is the scintillation bandwidth in MHz. 
The effective scattering screen distance $\deff$ is in kpc and the effective velocity $\veff$ is referenced to 
100\,$\kms$.

Expressions used to assess each MSP are as follows:

1. {\it Scattering times:}
We evaluate scattering times using the empirical $\tau(\DM)$ relation (Eq.~\eqref{eq:taudm} and a  frequency scaling $\tau\propto\nu^{\Xtau}$ with index $\Xtau=22/5$, appropriate for scattering dominated by scales in
the inertial subrange of a Kolmogorov spectrum.   High DM pulsars observed at low frequencies will manifest the effects of the inner scale and reduce the index to $\Xtau = 4$. 
We invert Eq.~\eqref{eq:taudm} to obtain a DM value or frequency  corresponding to a specified scattering time. 

2. {\it No scattering correction needed:}  
Requiring  $\taud \le \taudmax$ for a fixed maximum $\taudmax$ implies a lower bound on frequency.  For  
$\taudmax = 10$\,ns (but  using $\tau_1$ in $\mus$, as above), 
\be
\nu >  (\tau_1 / \taudmax)^{1 / \Xtau} \simeq  2.8 \,{\rm GHz}\, \times [(10\,{\rm ns} / \taudmax)\tau_1]^{5/22}.
\ee

Fig.~\ref{fig:DMmax_vs_tau} shows curves of  constant $\taud$ in the DM -$\nu$ plane.  These can be used to identify where the scattering time is too small to warrant correction.

3. {\it  Scattering too large:}  Scattering is deemed too large to correct if  a $100\epstaud$\% accuracy
 determination of  $\taud$ exceeds $\delta\taudmax$.  Using $\tau_1$ and $\delta\taudmax$ with the same units, 
\be
\nu <  \left( \epstaud \tau_1 / \delta\taudmax \right)^{1/\Xtau} = 0.35\,{\rm GHz} \times 
	[ (\epstaud / 0.01) (\tau_1/ \delta\taudmax) ]^{5/22}. 
\ee
The nominal 1\% accuracy in determining $\taud$ is generous in the sense that it is difficult to achieve,  causing fewer  pulsars to be ruled out than might be more realistic.

4.  {\it Cyclic spectroscopy required:}  This condition occurs when resolving scintles in frequency requires FFTs longer than the pulse width. 
 If $\taud(\nu)$ is small enough,  standard phase-resolved spectroscopy (PRS) 
is sufficient to provide a dynamic spectrum (DS) from which the scintillation bandwidth $\dnud$ and $\taud$ can be estimated.  For PRS, the maximum FT length is restricted to the pulse width $\WA$, yielding  frequency resolution
$\Delta\nu \gtrsim \WA^{-1}$.    We require that it be a fraction $\epsdnud = 0.1$ of the scintillation bandwidth, $\WA^{-1} <   \epsdnud \dnud$.   With $\WA$  in ms units,
\be
\nu <  (2\pi\tau_1 / \epsdnud C_1 \WA)^{1/\Xtau} \simeq 0.53 \,  {\rm GHz} \times (\tau_1 /C_1 \WA)^{5/22}.
\label{eq:PRScriterion}
\ee

5. {\it Errors on estimates of scintillation bandwidth from a dynamic spectrum:}  An ACF analysis  of the DS yields
$\dnud$ with errors from radiometer noise (N) and from the finite number of scintles (FSE) spanned by the DS.   The fractional errors 
\newtext{
are derived in Appendix\,\ref{app:ds_errors_prs_cs}:
}
(a) Fractional error in $\dnud$ from  radiometer noise (based on PRS with frequency resolution $\dnu = \WA^{-1}$):
\be
\fracerrordnudN 
&\simeq& 
\frac{2 \Kdnud\Ssys}{\Sphiave} \left( \frac{\WA / P} {B \Ttotal \WA \dnud}  \right)^{1/2}
\nonumber \\
&\simeq& 10^{-4} \, \times
		\left(\frac{\Ssys / 3\, {\rm Jy}} {\Sphiave /1 \, {\rm mJy}} \right)
		\left ( \frac{\WA } {P } \right)^{1/2}
		\frac{1}{\sqrt{(B  (\Ttotal / 1\,{\rm h}) \WA  \dnud ]}}.
\label{eq:epsdnud_N}
\ee

(b) Fractional error in $\dnud$ from the finite number of scintles:
\be
\fracerrordnudFSE 
\simeq \Nscint^{-1/2}
\simeq  8.4 \times 10^{-3}  \times  \nu^{14/5}\,  
	\left(\frac{\deff}{\tau_1^3}  \right)^{1/4} 
	\frac{1}{\sqrt{B  (\Ttotal / 1\, {\rm h}) (v / 100\,{\rm km\,s^{-1}})}}   
\label{eq:epsdnud_fse}
\ee
where the number of scintles contributing to the DS is assumed large, 
\be
 \Nscint  \simeq 
 	\etat\etanu (B/\dnud) (T/\dtd)  
	\simeq {\displaystyle 3960}  {\displaystyle 
        \nu^{-28/5}\, B_{\rm GHz}  T_{1000}
        v_{100}\,
        \left(\tau_1^3  / \deff \right)^{1/2} } \gg 1. 
\ee 

\begin{figure}[h!]
   \centering
   \includegraphics[scale=0.55]{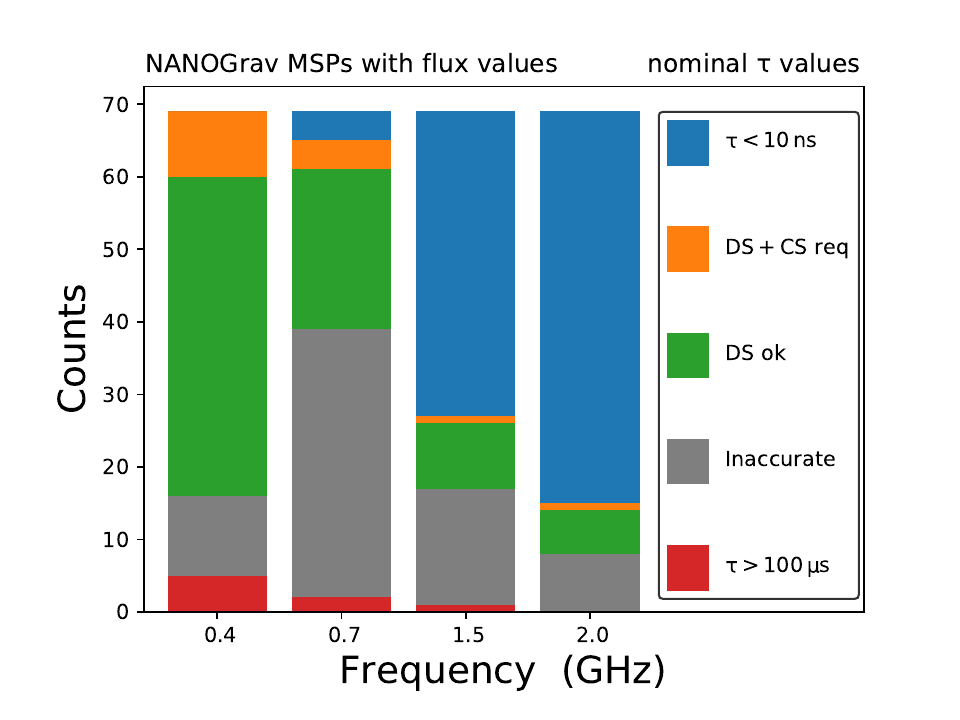}
   \includegraphics[scale=0.55]{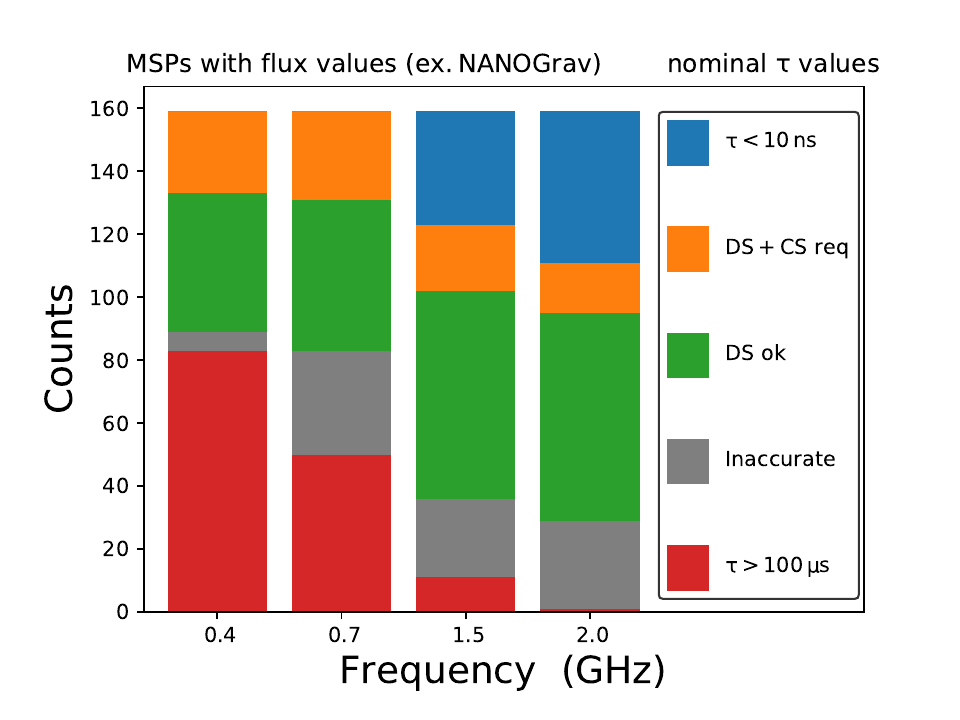}
   \caption{\footnotesize 
   Bar charts showing counts in the scattering categories.
   Left: NANOGrav pulsars (only those with flux density values included)
   Right: Other MSPs with flux density values. 
   \label{fig:cs_prs_regimes}
   }
\end{figure}

Fig.~\ref{fig:cs_prs_regimes} shows the number of MSPs  that fall into the different correctability categories for four frequency bands.   Assessments are made only if flux density measurements are available in the ATNF pulsar catalog
\citep[][]{ATNFPulsarCat}.   Results for the NANOGrav sample (69 objects) are shown in the left panel 
and for  a larger, non-overlapping  sample of 159 MSPs in the right panel.  
\newtext{We emphasize that the NANOGrav sample overlaps significantly with those of other Northern-hemisphere PTA programs (those by the CPTA and EPTA), so we consider it to be representative of the MSPs included in PTA programs. 
The non-NANOGrav sample includes some of the MSPs observed in other PTA programs but the key feature of the non-NANOGrav sample
for the discussion here is that it extends to significantly higher DMs, for which interstellar effects are much larger.
}

\newtext{
A summary of trends that can be inferred from the figure include:
}
\begin{enumerate}
\itemsep -1pt
\item The fraction of scattering influenced \TOAs\ (those with
scattering times $> 10$\,ns 
\newtext{
shown in colors other than blue)
}
 is sizable at 0.4 GHz in the NANOGrav sample 
 \newtext{
 (left panel)
 }
 but declines rapidly with increasing frequency
(DISS = diffractive interstellar scintillations).
\item The fraction of scattering influenced \TOAs\ is significantly larger for the 159 MSP sample that excludes 
NANOGrav MSPs 
\newtext{
(right panel)
}
than for the 69 NANOGrav-only  MSPs.  This reflects  the smaller  mean DM for the NANOGrav sample, $\langle \DM_{\rm NANOGrav}\rangle  = 29\,\DMunits$ compared to
$\langle \DM_{\rm ex-NANOGrav}\rangle  = 105\,\DMunits$.
\item The corresponding mean scattering times are a factor of ten larger for the 159 MSP sample compared to the NANOGrav sample, \newtext{based on the mean $\tau-$DM relation (Eq.~\eqref{eq:taudm})}. 
\item The fraction of pulsars that can benefit from  scattering determinations from DS analysis  depends on whether rapid or slow distortions of profiles are considered:
	\begin{enumerate}
	\item Rapid variations from the finite number of scintles in a DS: 
	about 17\% at 0.4\,GHz, dropping to 5\% at 0.7\,GHz and zero at  1.5\,GHz and above.
	\item Slow variations from changes in the scattering time and PBF shape (from refraction):
	about 34\% at 0.4\,GHz and 10\% at 1.5\,GHz. 
	\end{enumerate}
\item For MSP-frequency combinations that can  benefit from an analysis of dynamic spectra,  phase-resoloved spectroscopy (PRS) suffices 
for the majority of the cases 
\newtext{
(green)
}
while CS is needed for $\sim$10 to 30\% 
\newtext{
(red), 
}
depending on frequency and whether the NANOGrav or non-NANOGrav sample is considered. 
\end{enumerate}

\clearpage
\section{Spin noise triage}
\label{sec:spin_triage}

MSPs are used in PTAs because they are much more spin stable  than canonical pulsars and, per unit telescope time,  they provide a much larger number of pulses  for averaging down emission self white noise (e.g. pulse jitter).   However, red spin noise (or possibly orbital asteroid noise that mimics spin noise) is manifested in some MSPs at levels that can mask the red GW background signal.   Since, by definition, red spin noise is correlated over long time spans, its effects in a PTA can be averaged down only by summing over a large number of MSPs.  This mandates that MSPs should be culled for PTA science on the basis of displaying minimal spin noise. 

To select MSPs for PTAs, red spin noise for a newly discovered pulsar initially can  use  measurements of the spin period and its time derivative to evaluate the empirical spin noise scaling law  from SC10 (Eq.~\eqref{eq:rmstoaSPIN},\eqref{eq:sc10}).  Actual measurements of spin noise require a year or more of timing data to distinguish
systematic astrometric errors from red stochastic spin noise.   Auxiliary VLBI-based astrometry can aid this process \citep[][]{2013ApJ...777..104M,2017MNRAS.469..425W},
but it too requires at least a year of data to measure proper motion and parallax.
In addition, linking the extragalactic and solar-system reference frames poses additional uncertainties at the early stages of a VLBI program 
\citep[][]{2023A&A...674A.187L,2024arXiv240113373E}.


 The current pulsar sample comprises $\sim 3750$ pulsars, including
$\sim 545$ MSPs with periods $P \le 10$ ms, of which 356 are not in globular clusters \citep[][]{ATNFPulsarCat}. 
These numbers are expected to grow  dramatically from ongoing pulsar surveys given that the total number of active pulsars in the Galaxy may be as large as 100$k$ \citep[e.g.][]{2014ApJ...787..137S}. 
  A model dependent estimate is that only about 20\% of MSPs are  sufficiently spin stable over a ten-year span to be useful for inclusion in a PTA \citep[][]{sc10}.   Other factors of course also limit inclusion, including all of the effects discussed previously. 

A reasonable approach is to place a cap on the predicted \RMS\ spin-noise over a specified time block for an MSP to warrant inclusion in a PTA.  We estimate the data span $T_{\rm r}  \propto \sigma_{\rm spin \, noise}^{1/2}$ needed to reach a nominal maximum level of red spin noise
$\sigma_{\rm spin \, noise} = 0.1\,\mu$s; it utilizes the SC10 scaling law
(Eq.~\eqref{eq:rmstoaSPIN}) along with the assumption  that spindown rates $\fsdot$ follow that for magnetic dipole radiation. 
This yields the time needed for the \RMS\ spin noise to reach  $\sigma_{\rm spin \, noise}$, 
\be
T_{\rm r} = 
	\left(
		\frac{\sigma_{\rm spin\, noise}}  {C_2 \fs^{\alphar} \, \vert\fsdot\vert^{\betar}}
	\right)^{1/\gamma_{\rm r}}
	\simeq 
	0.4\,{\rm yr} \left(\frac{P_{\rm ms}^{0.95}} {B_9^{1.1}} \right)
	\left(\frac{\sigma_{\rm spin\, noise}}{0.1\, \mu s} \right)^{1/2} ,
\ee
for spin periods in ms and surface fields $B = 10^9B_9$\,G.
We have expressed $\fsdot$ in terms of the surface magnetic field 
assuming magnetic dipole radiation dominates spindown of the neutron star. 
The surface magnetic field is related to
$\Pdot$ by $B_{\rm surface}  = 10^9 B_9\,G$ where 
$B_9 \sim 3.2 (P \Pdot_{-20})^{1/2}$  
with $\Pdot_{-20} = \Pdot / (10^{-20}\, s\,s^{-1})$. 

\begin{figure}[htbp] 
   \centering
   \includegraphics[width=0.6\linewidth]{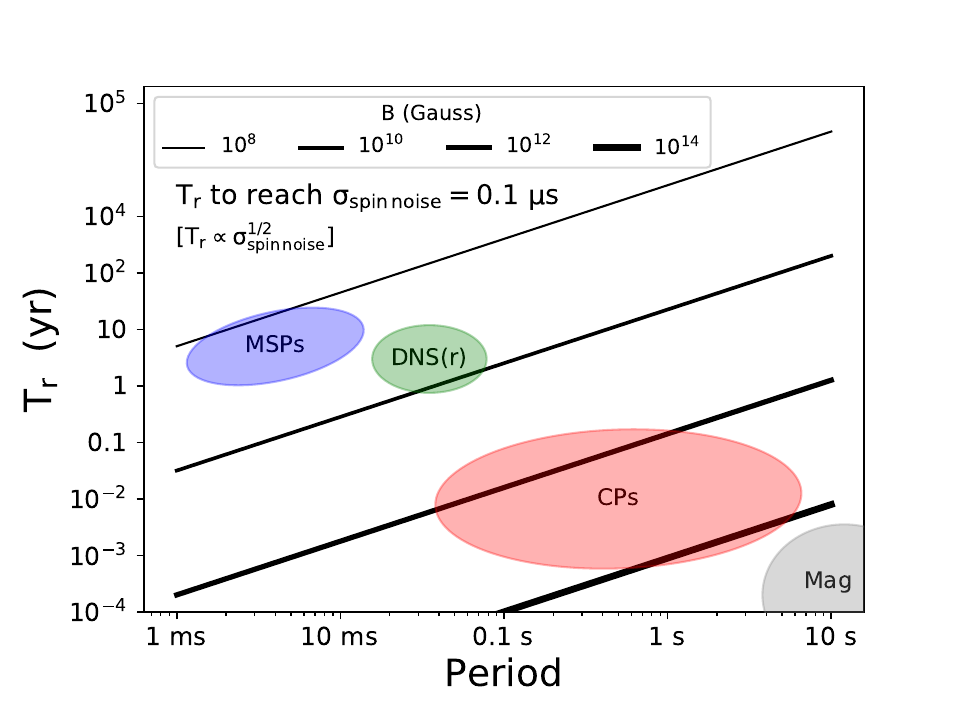} 
   \caption{Time $T_{\rm r}$  needed for a timing data set to yield red spin noise at an \RMS\ level 
   $\sigma_{\rm spin\, noise} = 0.1\,\mu$s vs spin period.  
   It scales as $T_{\rm r} \propto \sigma_{\rm spin\, noise}^{1/2}$. 
   Lines are shown for  surface magnetic field strengths from $10^8$ to $10^{14}$\,G.
   Ellipses indicate the approximate locations of distinct pulsar populations:  MSPs,  double neutron star binaries (recycled component), canonical pulsars, and magnetars. 
   The curves are only indicative of approximate times because there is substantial variance in the
   red spin noise scaling law and they are based on a simplistic relation between field strength and spin frequency derivative based on the magnetic dipole scaling (i.e. a braking index $n=3$).
   }
   \label{fig:time_to_reach}
\end{figure}

Fig.\,\ref{fig:time_to_reach} shows $T_{\rm r}$ vs. spin period for surface magnetic field strengths that span 
MSPs to magnetars, with recycled pulsars in double NS binaries  and canonical pulsars in between. 
Most MSPs have estimated surface fields between $10^8$ and $10^9$\,G. 
The figure indicates that the lowest field objects will take several years for spin noise to be manifested in timing residuals whereas those with the shortest periods and larger fields $\gtrsim 10^9$\, G will show red spin noise in less than a year.  
A similar approach can be used to prioritize MSPs with negligible spin noise on time scales of interest for GW detection (tens of years)  and those with spin noise too large  to be useful. 

Timing stability better than 0.1\,$\mu s$  over a span  $T = 10$\,yr, say, requires short period MSPs with the smallest magnetic fields.     Objects with larger spin noise can be included in PTAs if their number 
$N_{\rm pulsar}$ is large enough to average spin variance down as  $N_{\rm pulsar}^{-1/2}$ detection statistics for   correlated Earth-term GW effects. 

\section{Evaluating timing quality of individual pulsars}
\label{sec:Q}

  As pulsar timing methodology has developed over the last few decades, several  factors that influence the utility of a pulsar  for precision timing applications  have been identified.   So far the achievable 
  S/N  
of an average pulse profile has been the dominant factor for many objects,  though  pulse distortions from interstellar propagation are also prominent.  However, arrival time precision is only part of the story because other physical effects influence  the true arrival times. The regularity of  pulse periodicities is affected by stochastic effects occurring in or near the NS such as spin fluctuations, emission fluctuations, and orbital noise  that alter  arrival times.   Interstellar delays, if imperfectly removed, also contribute to the overall timing noise budget for a pulsar. 

Here we define quality measures that enable a thorough comparison of MSPs with regard to {\it  achievable \TOA\ precision} and {\it timing stability}.    In aggregate the measures can be used for  assessing whether a given  MSP  would
be competitive among other MSPs in its contribution to the overall sensitivity  of a pulsar timing array (PTA).  
For other applications  where a pulsar is monitored for purposes unique to it, such as orbit and mass determinations, the quality measures provide the basis for \TOA\ optimization, such as using longer integration times or alternative radio bands.  

We define quality measures (QMs) $Q_{\rm a}$ such that larger values mean better \TOA\ precision or better timing stability, where ${\rm a} = {\rm SNR,} {\rm J},  {\rm DISS}, {\rm spin}, {\rm PBF},  \DM(t), \DM(\nu)$.   The first three of these correspond, respectively, to \TOA\ errors from the finite SNR of a pulse profile, from pulse jitter, and diffractive scintillations that perturb \TOAs\ on short time scales (seconds to hours).   The others  produce slowly changing  perturbations from spin variations and changes in DM and   scattering as the line of sight traverses different ISM electron density fluctuations.   \newtext{These effects are major subjects of this paper, with reference to 
Sect.~\ref{sec:spin} (spin and orbital noise),
Sect.~\ref{sec:toa_estimation} (template fitting), 
and 
Sect.~\ref{sec:chromatic_toas} (chromatic effects, primarily the ISM).}

Table\,\ref{tab:QM} summarizes the QMs with partial scaling laws and pointers to equations developed later in this section.  Mitigation possibilities are also listed. 

Several phenomena are  not included in this brief  inventory of  quality measures. 
First are those that represent {\it departures from the assumed  average pulse shape stability vs. epoch}.  By and large,   average profiles and their chromaticity appear to be epoch-independent once instrumental factors are accounted for \citep[e.g.][]{2018ApJ...868..122B}.   A few notable exceptions
are discussed in Sect.~\ref{sec:shape_shifters}. 

Second are {\it telescope site dependent factors}, especially  the level of radio frequency interference (RFI) that differs greatly between sites and is episodic (e.g. diurnal variations) and worsening secularly.    Another is calibration methodology that mitigates instrumental polarization.  The generally large linear (and sometimes significant circular)  polarization fraction among pulsars requires accurate calibration to avoid  pulse shape changes that affect \TOA\ estimates.    The efficacy of polarization calibration is intertwined with contamination from RFI  and is therefore equally difficult to quantify. 

Third is {\it orbital noise} from asteroids or from spin-orbit interactions in ``black widow'' binaries where
white dwarf companions undergo ablation from the pulsar wind, producing variable amounts of orbital debris \citep[][]{2015ApJ...813L...4B}
Black widow pulsars are generally excluded from PTAs for this reason, but the presence of an asteroid belt is
plausible, though it has not been firmly established for any MSP.

After defining quality measures, we apply them to  MSPs in the current NANOGrav PTA \newtext{as a representative sample of MSPs to exemplify their utility.  Other PTA groups in the northern hemisphere  monitor many of the same pulsars and future studies can compare the results and incorporate MSPs from the southern hemisphere in the PPTA and MPTA programs.} 
After showing results that put high weight on template fitting errors, we then recalculate the quality measures for anticipated improvements in telescope sensitivity and greater flexibility in frequency band usage and time allocations on a pulsar by-pulsar basis. \newtext{The Python code used to calculate these quality measures is described in Appendix\,\ref{app:QMcode} and is available on GitHub (\url{https://github.com/jmcordes/PTQ}).} 


\begin{deluxetable}{l l c  l}
\tabletypesize{\footnotesize}
\tablewidth{450pt}
\tablecaption{\label{tab:QM} Pulsar Timing Quality Measures \tablenotemark{\footnotesize a}}
\tablecolumns{4} 
\tablehead{
	\colhead{\hspace {-55pt}Measure} &
	\colhead{Expression \tablenotemark{\footnotesize b, c} }  &  
	\colhead{Eq. Number }  &
	\colhead{Mitigation/Comments \tablenotemark{\footnotesize d}}  
}
\startdata
{\bf Rapid TOA Variations} ($\lesssim$\, hours) \\
\quad\quad  
	$\Qsnr$ 	  	& $6.8 (P / \nc W^3)^{1/2} \Sphi$  & \eqref{eq:Qsnr} & SEFD,  $B$, $\Tint$ \\
\quad\quad 
	$\Qj$ 	 	& $30 / (3 \Wc \sqrt{P})$ & \eqref{eq:Qj} &  $\Tint$    \\
\quad\quad 
         $\Qdiss$    	& $21.2 ( \nu v_{100})^{1/2}  / (D \taud)^{1/4}$ & \eqref{eq:Qdiss}  &  \DM, $\nu$, $B$, $\Tint$  \\
\\
{\bf Slow TOA Variations} ($\gtrsim $\, days) \\  
\quad\quad  
	$\Qspin$ 		& $252\, P^{0.8} \Pdot_{-20}^{-1.1} T_{\rm yr}^{-2}$  & \eqref{eq:Qspin}  & None except pulsar choice  \\
\quad\quad 
	$\Qtaud$  		&$ \taud^{-1}$    & \eqref{eq:Qtaud}  & Fitting vs. $\nu$, $\DM, \nu$ \\
\quad\quad 
         $\Qdmt$    	& $32 \, \nu^{1/6} (1-R^{-2}) [(v_{100} \Delta t_{\rm days})^2 \taud / D]^{-5/12}$  &\eqref{eq:Qdmt2}  & \DM, Simultaneous $\nu$ coverage \\
\quad\quad 
         $\Qdmnu$ 	& $1.9 \, \nu^{1/6} \taud^{-5/6} \Itinfty^{-1}(R, \beta=11/3)$   
         &\eqref{eq:dmnu2}   & \DM, Fitting function vs. $\nu$. \\
                                  &                                                  &  & Coefficient  is for $R \sim 2$ .
\\
\hline  
{\bf Total QM }  & $\sum_j w_j Q_j$       &   \\
\enddata
\tablenotetext{a}  {{\bf Units: }
        \\
	All quality measures have units of $\mu s^{-1}$. 
	\\ 
	{\it Telescope/observation parameters:}
		 $\Ssys = 3$\,Jy, $\npol=2$, $B = 0.1$\,GHz, $\Tint = 10^3$\,s, $\nu=1$\,GHz, $R = \nu / (\nu-B)$.
	\\ \quad
	{\it Pulsar parameters:}
	        $\Sphi$ in mJy, $D$ in kpc, $P$ in ms;  $\Pdot_{-20} = \Pdot / (10^{-20} \,$s\,s$^{-1})$; \\
	        $W, \Wc, \taud$ in $\mu s$; $\nc$ dimensionless,  
	        $\DM$ in $\DMunits$, $v_{100} = \vperp / (100\, {\rm km \, s^{-1}})$.
	}
	
\tablenotetext{b} { Scaling laws show pulsar-dependent parameters but   telescope/observation  parameters are implicit; see indicated equation numbers for dependences.  }

\tablenotetext{c} { $\taud$ can be measured  directly or from the scintillation bandwidth, $\taud \simeq (2\pi\dnud)^{-1}$. It can also be estimated from the $\taud(\DM)$ `hockey stick' relation, 
$\taud(\mu s) \simeq  \nu^{-4.4}  (\DM / 303)^{3/2} [1 + (\DM / 30.4)^3]$.  }

\tablenotetext{d} {Mitigations:  choosing pulsars (and telescope parameters)  with large QMs  optimizes a PTA. 
	\\ \quad
	 \DM: smaller DM, $\Ssys$:  smaller $\Ssys$ = system equivalent flux density,  
	 $\nu$: higher frequency, $B$: larger bandwidth, $\Tint$:  longer integration time.}
\end{deluxetable}

\subsection{Quality measures for fast \TOA\ fluctuations}

Additive noise in template fitting, pulse jitter, and diffractive scintillations perturb arrival times  that change during intraday observing sessions \citep[e.g.][]{2016ApJ...819..155L}.      The template fitting error depends on properties of both the pulsar and the telescope.  Jitter is intrinsic to the pulsar while diffractive scintillations are caused by the ISM along the line of sight.    The template fitting error is a white noise process vs. both time and frequency.  Jitter is also a white noise process vs. time but is highly correlated across frequency.   DISS differs by being correlated over finite ranges of time and frequency:  seconds to hours and kHz to 100s of MHz, respectively.  

We assess these three processes separately and later combine them as a weighted sum.

\subsubsection{Template fitting}

The quality measure is the reciprocal of the \TOA\ error   $\rmstoaS^{-1}$ from Eq.~\eqref{eq:rmstoa_mf_t_tel_phi} - \eqref{eq:rmstoaSQ}, 
\be
\Qsnr 
= \rmstoaS^{-1} \propto \Sphi  \left(\frac{\ncTwo P}{\ncOne^2W_{\rm c}^3} \right)^{1/2} .
\label{eq:Qsnr}
\ee
An expression for $\SNR$ is given in Eq.~\eqref{eq:SNR_radiometer}. 
Pulsars rank higher  if they  are brighter, have longer periods, and  have profiles with narrower (and more)  components\footnote{Note $\ncOne \le \ncTwo^{1/2}$ by the Cauchy-Schwarz inequality.  If all $\nc$ components have the same amplitude, then $\Qsnr \propto \sqrt{\nc}$. }. 

A more sensitive telescope or longer  net integration time can improve  a weaker pulsar to the same timing precision as a stronger pulsar  with  identical period and pulse-shape factors.  
 A quality measure that excludes the flux density is
\be
\Qsnrnoflux = \frac{\Qsnr}{\langle S_\nu \rangle} \propto \left(\frac{\ncTwo P}{\ncOne^2W_{\rm c}^3} \right)^{1/2}.
\label{eq:Qsnrnoflux}
\ee
\newtext{This measure is useful, for example, when considering a pulsar that is too weak for current telecopes but becomes viable with a larger telescope.
We use this measure in Sect.~\ref{sec:initcomp}.} 
Among MSPs in the NANOGrav 15-yr data set,   the largest $\Qsnr$ are for J1939+2134 (B1937+21),  
J0437$-$4715, and  J1909$-$3744 in rank order  while for $\Qsnrnoflux$, which excludes flux densities,  J1909$-$3744 has the top rank owing to its narrow pulse followed by B1937+21.     Dropping to 54-th is J0437$-$4715 due to its wide pulse and \newtext{in spite of} its  longer period. 

\subsubsection{Pulse jitter}

The \rms\  \TOA\ error from pulse jitter is given by Eq.~\eqref{eq:rmstoaJ} - \eqref{eq:rmstoaJnom}.
If we account for $\nc$ separate components with the same width $\Wc$ and same jitter parameter,   the  \TOA\ error is smaller by a factor $\sqrt{\ncFour} / \ncTwo$ where $\ncFour  = \sum_j (a_j / a_{\rm max})^4 \le \ncTwo$.
The jitter quality measure is then
\be
\Qj = \rmstoaJ^{-1} = 
	\frac{1}{\FJ \Wc}
	\left(\frac{\ncTwo}  {\ncFour^{1/2}} \right)
	\left( \frac{\Tint}{ P}\right)^{1/2} .
\label{eq:Qj}
\ee
The jitter measure favors pulsars with more narrow components and shorter periods.

\subsubsection{Diffractive ISS PBF variations (`finite scintle effect')}

For pulsars that show strong scintillations, 
\RMS\ \TOA\ errors associated with stochastic variations in the PBF are $\sim \taud / \Nscint^{1/2}$, where
the number of scintles $\Nscint$ is given by Eq.~\eqref{eq:Nscint}. 
For nearby pulsars and high-frequency observations, the number of scintles is  small, but no less than unity as long as strong scintillation applies.   Above a transition frequency of
about 2\,GHz for the nearest pulsars,  scintillations become weak and PBF variations become unimportant.   However, PTA programs use widely spread frequencies  that invariably include those where scintillations are strong  even if higher frequencies are not.

Using $\taud \simeq \dso \thetad^2 / c$  along with 
$\dnud = C_1 (2\pi \taud)^{-1}$ 
and $\dtd \sim \ld / v_{\rm eff}$
where $\ld \sim \lambda / 2\pi \thetad$,  we obtain
\be
\rmstoaDISS = \frac{\taud}{\Nscint^{1/2}}  
	\simeq  
	\frac
		{[c \deff \taud(\nu)]^{1/4}} 
		{2\pi \sqrt{\eta_\nu \eta_t B T \nu v_{\rm eff} }}
\label{eq:rmstoaDISSQ}
\ee
where  the broadening time $\taud$ is measured at  frequency  $\nu$.


The DISS quality measure is the inverse of $\rmstoaDISS$ (Eq.~\eqref{eq:rmstoaDISS1} -  \eqref{eq:rmstoaDISS2}), 
\be
\Qdiss = \rmstoaDISS^{-1}
	= \frac	
		{2\pi \sqrt{\eta_\nu \eta_t B T \nu v_{\rm eff} }}
		{[c \deff \taud(\nu)]^{1/4}} 
\label{eq:Qdiss}
\ee
This measure favors pulsars that are less scattered, closer, and have higher effective velocities.

\subsection{Quality measures for slow timing variations} 

\subsubsection{Spin fluctuations}

We use the results given in  Sect.~\ref{sec:spinnoise} and  express $\rmstoaSPIN$  in terms of 
period $P = \nu^{-1}$ and period derivative $\dot P = - \dot\nu P^2$ to obtain a quality measure
\be
\Qspin 
=  \rmstoaSPIN^{-1}
= C_2  P^{\alpha + 2\beta} \dot P^{-\beta}  T^{-2} 
\simeq 
 252\,{\rm \mu s}^{-1} P_{\rm ms}^{0.8\pm0.22} \dot P_{-20}^{-1.1\pm0.1},
 \label{eq:Qspin}
\ee
where $\dot P_{-20}$ is $\dot P$ in units of $10^{-20}$\,s\,s$^{-1}$ and the spin period is in ms. 

This measure appears to favor longer periods and smaller period derivatives.    However,
$P$ and $\dot P$ are related  using the relation for the surface magnetic field $B \propto \sqrt{P\dot P}$,
yielding $\Qspin \propto B^{-2\beta} P^{\alpha+3\beta}$.   The net effect is that MSPs with low magnetic fields
have larger $\Qspin$ than canonical pulsars  in spite of the period dependence that favors longer-period pulsars. 

\subsubsection{Long term variations of the pulse broadening time}

To develop this quality measure,  we employ a simplified timing model for  two-frequency observations. 
A more detailed account would address the use of broadband receivers covering 2:1 or larger frequency ranges.
Denoting  arrival times  as $t_\nu \equiv t(\nu)$ and $t_{1,2} = t_{\nu_{1,2}}$ at two spot frequencies with $\nu_2 > \nu_1$ , we naively  model them
with $t_\nu = \tinfty + \tDM + \varepsilon(t)$  using a dispersion delay $\tDM = \nu^{-2} \KDM \, \DM$ and 
a (for now unspecified)  chromatic error term $\varepsilon(t)$ adding to the desired `infinite frequency' \TOA, 
$\tinfty$. The errors in the estimated DM and $\tinfty$ 
are similar to the analysis yielding Eq.~\eqref{eq:tinfty1} - \eqref{eq:TOAerror} in  Sect.~\ref{sec:2freq},  
\be
\dDMhat \equiv \DMhat - \DM = \frac{\nu_2^2 (\varepsilon_1 - \varepsilon_2)}{\KDM (R^2-1)}, 
\quad\quad \text{and} \quad\quad 
\dtinftyhat \equiv \tinftyhat - \tinfty  = \frac{R^2 \varepsilon_2 - \varepsilon_1}{R^2-1} ,
 %
\ee
where, as before, the frequency ratio $R = \nu_2 / \nu_1 > 1$. 

If exact templates are used at the two frequencies, the systematic errors $\varepsilon_{1,2} = 0$.  (Recall we are analyzing the no radiometer-noise case here and we furthermore ignore any intrinsic chromaticity of the pulse shapes).   This applies if the PBFs and intrinsic pulse shapes are known exactly.  This is never the case, as we have demonstrated, so systematic \TOA\ errors arise from epochal variability in the pulse shape whether intrinsic or, more likely, from changes in the PBF. 

As seen in Sect.~\ref{sec:mean_shift_not}  - \ref{sec:nu_toaoffsets},  \TOA\ errors due to  mismatched templates from  scattering  can exceed the scattering time $\tau$, especially for small $\tau$ values, and are a different proportion of the scattering time at different frequencies.  Also, depending on template choice, the \TOA\ error can be of either sign.   Given the variety of conditions and choices, we take a simplified approach to develop a metric. 


We proceed by associating  the systematic error with a multiple of the scattering time,
$\epsilon_{1,2} = m_{1,2} \tau_{1,2}$ and we adopt a frequency scaling 
$\tau \propto \nu^{-\xtau}$ with $\xtau = 22/5$ and $\zeta = \linner / {\ld}_{_\theta} \propto \nu^{-6/5}$
 for the Kolmogorov inertial subrange regime.    
 
 For low-DM MSPs,  $\tau / \Wu \ll 1$ and $\zeta$ is also small at L-band frequencies.   For example, scaling
 $\tau = 0.1\,\mu s$ at 1\,GHz implies $\tau / \Wu \sim 10^{-3}$ for a 100\,$\mu s$ pulse width and using 
 Eq.~\eqref{eq:zetaeval},  $\zeta \lesssim 0.1$ even at 0.4 GHz.    Inspection of Fig.\,\ref{fig:ttau_vs_zeta}
 then shows that 
 $\dtinftyhat / \tau \simeq m_1 \sim m_2 \sim 2.5$ for the entire frequency range $\gtrsim 0.4$\,GHz. 
For $R^2 \gg 1$ (wide frequency separation),  the error is approximated as 
\be
\dtinftyhat \simeq  - m_2 \sqrt{\tau_1 \tau_2} = -m_2 \tau_0 (\nu_1 \nu_2 / \nu_0^2)^{-\xtau/2}.
\label{eq:Qdtinftyhat}
\ee

If $\tau_\nu$ were constant vs. epoch  this error would simply be an unimportant bias in \TOAs\ that would be absorbed in the timing analysis if observing frequencies were the same at all epochs.  
However, as shown in Sect.~\ref{sec:aniso_refract}, 
refraction from large-scale electron density fluctuations in the ISM will cause  
$\tau_\nu$ to vary on time scales from days to years for weakly scattered and strongly scattered pulsars, respectively.  
In principle, these variations can be partially mitigated by measuring the scintillation bandwidth $\dnud \propto \taud^{-1}$ at each epoch vs. frequency  and correcting \TOAs\ accordingly.  Actual \RMS\ variations in $\taud$ 
\newtext{
on month to year time scales can be as large as 100\% (\S~\ref{sec:chromatic_toas}; \citealt[][]{hs08})
}
and correction of \TOAs\ will not be perfect.    

We define a quality measure  from Eq.~\eqref{eq:Qdtinftyhat}  by simply using the reciprocal of
$\tau(\nu)$, the scattering time at a fiducial frequency $\nu$ and using  the 
$\taud(\DM)$ expression of Eq.~\eqref{eq:taudm}, 
\be
\Qtaud =  \tau(\nu)^{-1}   \simeq  1\,\mu s^{-1} \times \nu^{4.4}  (\DM / 303)^{-3/2} [1 + (\DM / 30.4)^3]^{-1}.
\label{eq:Qtaud}
\ee
This measure favors less scattered pulsars and/or \TOAs\ obtained at higher frequencies.
While the \TOA\ error is dominated by the lower (or lowest)  frequency used in a multi-frequency timing program,
a comparison of pulsars can use a single representative frequency for evaluating $\Qtaud$.   However, any detailed comparison needs to consider specific radio frequencies used for observations of different MSPs. 


\subsubsection{Temporal DM variations and asynchronous multifrequency observations}

Most extant timing data have relied on multiple frequency measurements obtained at different epochs. 
\citet[][]{2015ApJ...801..130L} evaluated the timing error incurred from measurements made at slightly different epochs in terms of the DM structure function.   For  epoch separations of
$\Delta t$ (a few days or less), the \TOA\ error is given by Eq.~\eqref{eq:dtinfty_async0} - \eqref{eq:dtinfty_async2}
and the quality measure is evaluated for $\beta=11/3$ as the reciprocal, 
\be
Q_{\DM(t)} = {\rmstoa}_{_\infty}^{-1}  
=
2\pi \nuH
(1 -R^{-2}) 
\left[ \frac{\Delta t}{\Dtiss(\nuH)} \right]^{-5/6}  
\label{eq:Qdmt1}
\ee

An alternative form  follows by expressing $\Dtiss$ in terms of the pulse-broadening time and effective 
velocity,
\be
Q_{\DM(t)}   
&=&
(2\pi \nuH)^{(4-\beta)/2} (1 -R^{-2}) 
(v_{\rm eff} \Tint)^{-(\beta-2)/2} 
\left[ \frac{\taud(\nuH)}{c\deff} \right]^{-(\beta-2)/4}  
\nonumber \\
&\simeq& 
(2\pi \nuH)^{1/6} (1 -R^{-2}) 
(v_{\rm eff} \Tint)^{-5/6} 
\left[ \frac{\taud(\nuH)}{c\deff} \right]^{-5/12}  .
\label{eq:Qdmt2}
\ee

\subsubsection{Frequency-dependent DMs}

 RMS timing errors  from the frequency dependence of \DM\   are given by Eq.~\eqref{eq:dtinfity_DMnueval}.
 We evaluate it using a constant  geometric factor $G_\beta=1$. 
The resulting QM is 
\be
Q_{\DM(\nu) }  
= \rmstoaDMnu^{-1} 
=  \frac{ 6.41\, \mu s^{-1} \, \nu^{23/6}} {\dso^{5/6} \SM_{-3.5} \Itinfty(R, \beta=11/3)}.
\label{eq:dmnu1}
\ee

Casting $\SM$ in terms of the pulse broadening time, we have an alternate form 
\be
Q_{\DM(\nu) } 
=  
  \frac{1.88  \mu s^{-1} \,\nu^{1/6}} {\taud^{5/6}  \Itinfty(R, \beta=11/3)}.
\label{eq:dmnu2}
\ee
This measure also favors less-scattered pulsars but is agnostic to effective velocity.

For our analysis below, we use  a nominal 2:1 ratio that gives $\Itinfty(R=2, \beta=11/3) \sim 1$.   A PTA comprising pulsars with a wide range of DMs and different frequency ranges would likely be used for different objects; in that case $\Itinfty$ would be included in the evaluation of $\sigma_{\dtinfty} $ and the quality measure.  


\begin{figure}[ht!] 
   \centering
   \includegraphics[width=0.8\linewidth]{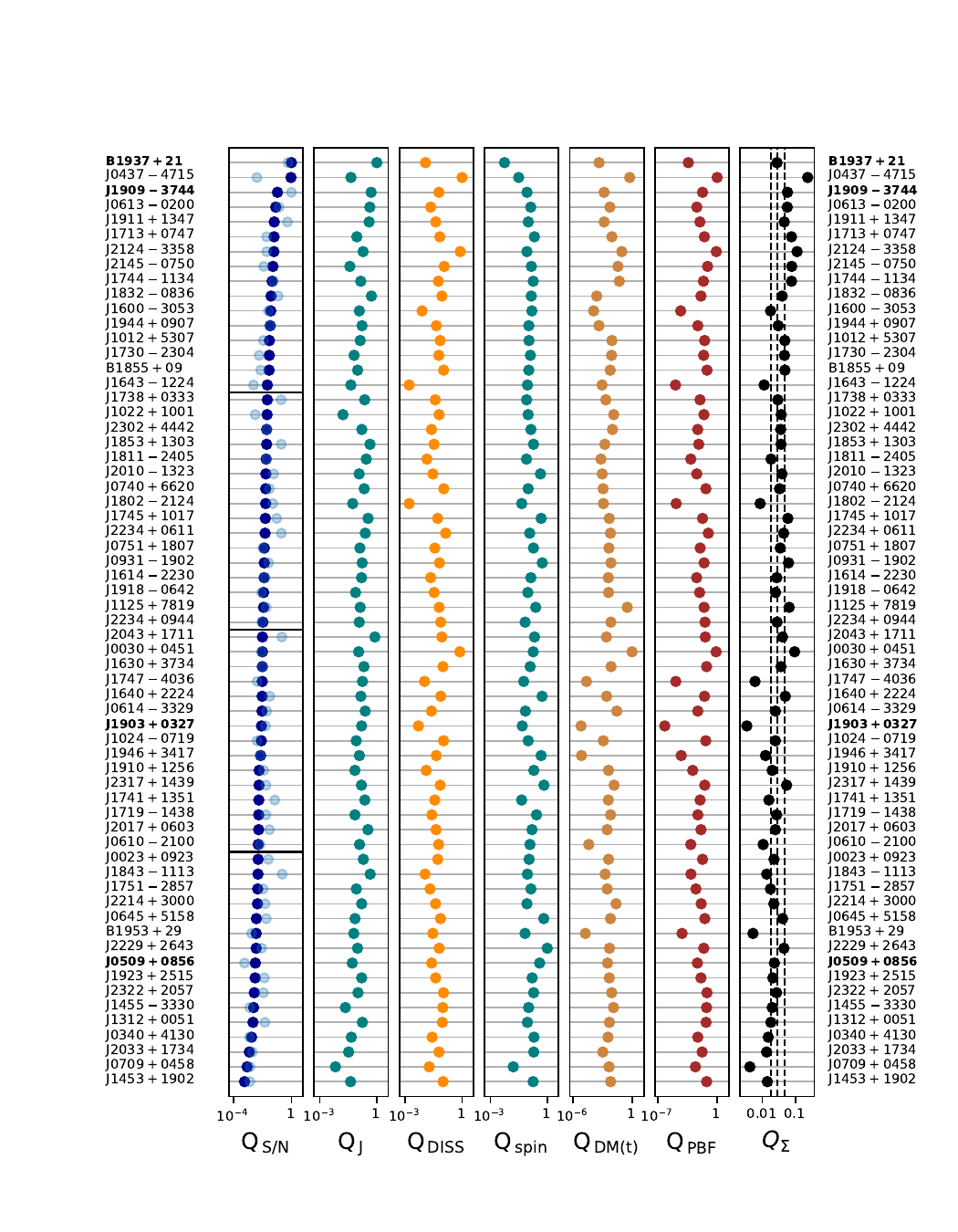} 
    \caption{
    Quality measures for 63 MSPs ordered using $\Qsnr$ in the first column.  
    Darker horizontal lines indicate quartile boundaries  in $\Qsnr$.   
    Also shown in the first column with fainter symbols is $\Qsnrnoflux$, which measures the relative potential for improved \TOAs\ using more sensitive telescopes; it has been rescaled to cover the same range as $\Qsnr$.   
    The next five columns are for $\Qj$, $\Qdiss$, $\Qspin$, $\Qdmt$, and $\Qpbf$ as defined in the text.   
    The last column gives the weighted sum $\Qsum$ of the logarithms of the other $Q$ values.
    Vertical dashed lines indicate quartile boundary values for $\Qsum$. 
    Pulsars in bold font are discussed as exemplars of timing quality in NANOGrav's detector characterization paper \citep[][]{2023ApJ...951L..10A}. 
   \label{fig:Qcolumns6}
   }
\end{figure}

\subsection{Meta-analysis of quality measures}

Fig.~\ref{fig:Qcolumns6} shows  $Q$ values for 63 pulsars in NANOGrav's timing program.
Pulsars are ranked according to $\Qsnr$ in the first column, which also shows $\Qsnrnoflux$ values as fainter symbols.   The $Q$ values for each of the first five columns are normalized to unit maximum and the last column gives the weighted sum 
\be
 \Qsum =  \sum_j  w_j \log Q_j, \quad j = {\rm (S/N, \, J,\, DISS, \, spin, \,\DM(t), \,PBF)}.
\ee
Weights sum to unity, $\sum_j w_j = 1$.
We use logarithms  (base 10)  to mitigate the dominance of any one quality measure given the several to many orders of magnitude that they span.  

\newtext{
In the following we choose weights to illustrate trends in the ranking of MSPs; they are not to be taken as optimal in any sense. However, we expect that individual and aggregated quality measures like the ones we suggest will be helpful for incorporating additional MSPs into PTAs.  Eventually, it is conceivable that weights can be based on quantitative measures of a pulsar's contribution to the overall PTA sensitivity to GWs.  However, the type of GW (i.e. stochastic background vs. GWs
produced by individual sources of CW signals or memory bursts) influences sensitivity calculations and thus will affect the choice of weights.
}

\subsubsection{Initial MSP comparisons}
\label{sec:initcomp}

For the initial comparison shown in Fig.~\ref{fig:Qcolumns6}, 
we use   weights $w_j = (0.35, 0.05, 0.05, 0.35, 0.1, 0.1)$ 
 that reflect  current 
 \newtext{PTA}
 observations where most MSPs have \TOA\ estimates  limited by  template fitting errors with some contribution from jitter and DISS and where overall timing residuals receive contributions from red spin noise,   unmodeled epoch-dependent
DM variations, and changes in the pulse broadening from scattering.   Our subjective choice of weights 
Vertical dashed lines in the last column indicate divisions between quartiles of $\Qsum$. 
Four `exemplar' pulsars indicated in bold font, J0509+0856, J1903+0327, J1909$-$3744, and J1939+2134 (B1937+21), are discussed in detail in NANOGrav's 15-yr detector characterization paper 
\citep[][]{2023ApJ...951L..10A} and we therefore refer to them here. 

 Inspection of the first column shows that B1937+21 has the best quality with respect to \TOA\ precision obtained from template fitting but it is in only the second to best quartile of $\Qsum$ because it manifests significant red  spin noise (or possibly stochastic asteroid noise \citep[][]{2013ApJ...766....5S} and scintillations also contribute to short-time scale TOA perturbations \citep[][]{2016ApJ...819..155L}. 

Comparison of the values shown in the first two columns indicate that \TOAs\ can be improved for a significant number of pulsars. If all pulsars are timed  with the same telescope, receiver, and integration time, $\Qsnr$ is the relevant quantity for comparing MSPs.   However, a significant number of pulsars show values for  $\Qsnrnoflux$ (faint symbols) that substantially exceed  $\Qsnr$.  This indicates that they possess  intrinsic properties (period, pulse width, and number of pulse components) that can be exploited to  optimize  \TOAs\ by using longer integrations or a larger telescope (for example).        By contrast, J0437$-$4715 ranks second in $\Qsnr$  because it is bright but is intrinsically worse than many MSPs because it has a wide pulse.   Nonetheless with weights that favor $\Qsnr$ and the proximity of the pulsar to the solar system (hence small interstellar effects), this pulsar has a large $\Qsum$. 
In the next section we use an alternate  scheme that weights spin noise more strongly and puts
J0437$-3744$ at much lower rank.
The third ranked pulsar in $\Qsnr$, J1909$-$3744 (another exemplar) ranks in the first quartile of $\Qsum$  because it has a narrow pulse and has small interstellar contributions.   

The two remaining exemplars,  J0509+0856 and J1903+0327, rank in the second lowest and lowest quartiles
of $\Qsum$, respectively.  The first of these has poor values of $\Qsnr$ and $\Qsnrnoflux$.
The  large interstellar contributions to the timing of  the large-DM object J1903+0327  make it the lowest ranked object of the entire set of pulsars in $\Qdmt$, $\Qpbf$, and subsequently $\Qsum$.   In principle,  the timing of this pulsar could  be improved significantly  using observations at higher frequencies, as discussed in \citet[][]{2018ApJ...861...12L}, especially with a future large collecting area telescope such as the Next Generation Very Large Array (ngVLA).  


\begin{deluxetable}{l    c   l   l    c}
\label{tab:top2quantiles}
\tabletypesize{\footnotesize}
\tablewidth{450pt}
\tablecaption{\label{tab:quants} Top two pulsar quartiles ranked by $\Lambda_{\Sigma}$. }
\tablecolumns{5} 
\tablehead{
	\colhead{ MSP} &
	\colhead{Quartile }  & 
	\colhead{$\Lambda_{\Sigma}$ }  & 
	\colhead{$\Qsum$ }  &
	\colhead{$\log_{10} \Qsnr$ }  
}
\startdata
     J1745+1017 &        1 &         0.68 &        0.57 &         0.66 \\ 
     J0931-1902 &        1 &         0.66 &        0.54 &         0.66 \\ 
     J1640+2224 &        1 &         0.64 &        0.48 &         0.62 \\ 
     J2317+1439 &        1 &         0.63 &        0.39 &         0.62 \\ 
     J2043+1711 &        1 &         0.63 &        0.49 &         0.58 \\ 
     J2124-3358 &        1 &         0.63 &        0.79 &         0.64 \\ 
     J1713+0747 &        1 &         0.63 &        0.79 &         0.67 \\ 
     J1832-0836 &        1 &         0.63 &        0.72 &         0.58 \\ 
     J1744-1134 &        1 &         0.62 &        0.75 &         0.66 \\ 
     J0030+0451 &        1 &         0.61 &        0.49 &         0.62 \\ 
     J1909-3744 &        1 &         0.61 &        0.85 &         0.59 \\ 
     J2229+2643 &        1 &         0.60 &        0.31 &         0.59 \\ 
     J2010-1323 &        1 &         0.60 &        0.59 &         0.60 \\ 
     J0613-0200 &        1 &         0.60 &        0.82 &         0.62 \\ 
     J1853+1303 &        1 &         0.59 &        0.61 &         0.57 \\ 
     J1125+7819 &        1 &         0.59 &        0.52 &         0.62 \\ 
     J0645+5158 &        2 &         0.59 &        0.32 &         0.58 \\ 
     J1911+1347 &        2 &         0.58 &        0.79 &         0.57 \\ 
     J2234+0611 &        2 &         0.58 &        0.57 &         0.56 \\ 
     J2145-0750 &        2 &         0.58 &        0.77 &         0.64 \\ 
     J1012+5307 &        2 &         0.55 &        0.68 &         0.57 \\ 
       B1855+09 &        2 &         0.55 &        0.68 &         0.57 \\ 
     J1630+3734 &        2 &         0.55 &        0.48 &         0.53 \\ 
     J0740+6620 &        2 &         0.55 &        0.58 &         0.51 \\ 
     J0437-4715 &        2 &         0.54 &        0.96 &         0.62 \\ 
     J0751+1807 &        2 &         0.54 &        0.54 &         0.55 \\ 
     J1730-2304 &        2 &         0.53 &        0.68 &         0.57 \\ 
     J1944+0907 &        2 &         0.53 &        0.70 &         0.52 \\ 
     J2017+0603 &        2 &         0.53 &        0.38 &         0.49 \\ 
     J2302+4442 &        2 &         0.52 &        0.61 &         0.54 \\ 
     J1946+3417 &        2 &         0.51 &        0.43 &         0.49 \\ 
\enddata
\end{deluxetable}
 
 \subsubsection{Forward looking MSP comparison}
 
 
 Future timing observations generally will be made with more sensitive telescopes having greater collecting areas and improved receivers.   For these,  errors associated with $\SNR$  become secondary for many objects.    Wideband systems deployed on current and future telescopes will 
 \newtext{likely decrease}
  timing errors associated with temporal DM fluctuations, though frequency-dependent DMs would be important for low frequency observations of  higher DM objects  \citep[][]{CSS16}.   
 For these reasons we compute a second sum that  weights jitter, spin, and DISS more strongly,     
 $w_j = (0.25, 0.15, 0.15, 0.35, 0, 0.1)$.
 
\begin{figure}[htbp] 
   \centering
   \includegraphics[width=0.8\linewidth]{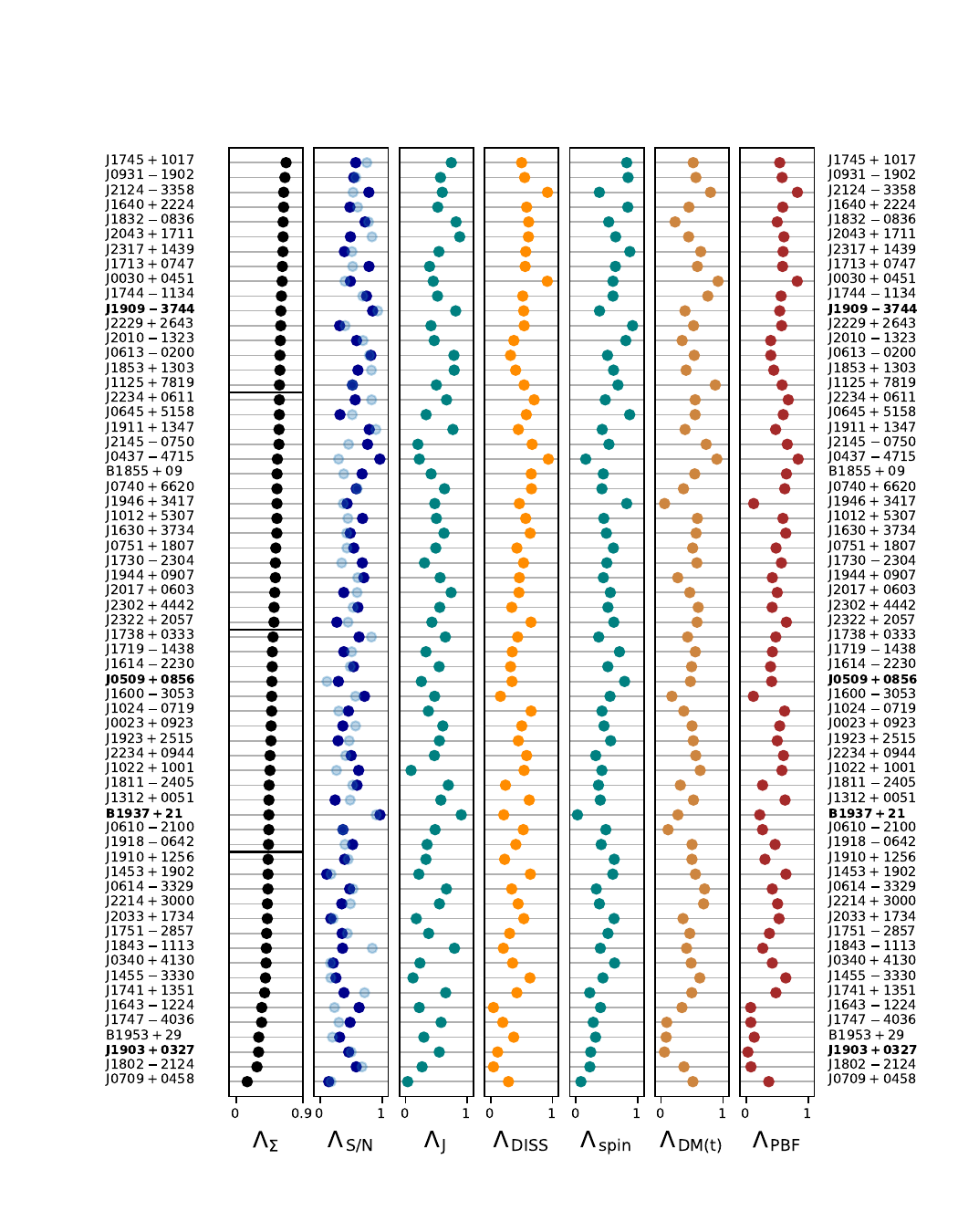}  
    \caption{Quality measures for 63 MSPs ordered using $\Qpsum$ in the first column.  
    Darker horizontal lines indicate quartile boundaries. 
    Other columns show individual transformed $Q$ values, including both 
    $\Qpsnr$ and $\Qpsnrnoflux$ in the second column. 
    Pulsars in bold font are discussed as exemplars of timing quality in NANOGrav's detector characterization paper \citep[][]{2023ApJ...951L..10A}.   
   \label{fig:Qlogistics_columns}
   }
\end{figure}

 We also employ a sigmoid transformation  that normalizes all of the (log) quality measures  into the same range, $[0,1]$, and further limits the domination of an aggregate quality measure by any single measure.  Using a logistic function  $\Sigmoid(x) = (1 + e^{-x})^{-1}$ and letting $l_j \equiv \log Q_j$ we  transform the measures  using  the median $l_{\rm m}$ and \RMS\ $\sigma_l$ to calculate a weighted sum $\Lsum$ of $\Lambda_j \in [-1, 1]$, 
 where
  \be
\Lambda_j  = \Sigmoid \left(   \frac{l_j - l_{\rm m}} {\sigma_l}  \right) ,
 \label{eq:Lambda}
\quad\quad 
\Lsum  = \sum_j w_j \Lambda_j.
\label{eq:Qpsum}
\ee
Results are shown in Fig.~\ref{fig:Qlogistics_columns}, which demonstrates how the new weights affect the placement of pulsars  and especially the exemplar MSPS.  Table\,\ref{tab:top2quantiles} lists the MSPs in the top two quartiles.

\begin{figure}[h!] 
   \centering
   \includegraphics[width=0.9\linewidth]{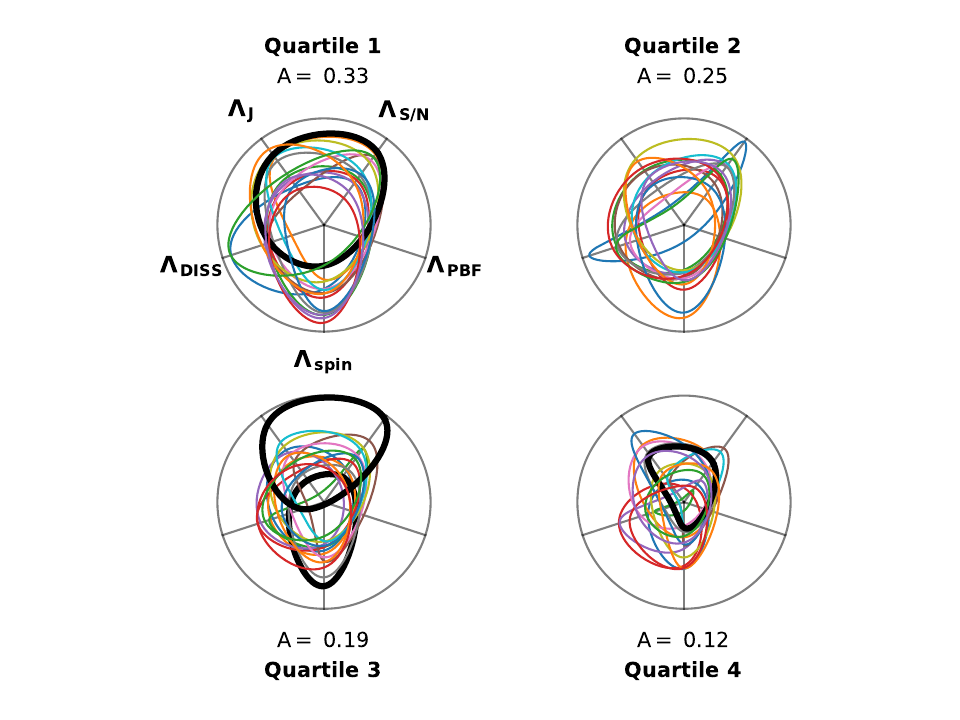} 
    \caption{Logistic quality measures for 63 MSPs split into four quartiles in $\Lambda_{\Sigma}$.
For each pulsar a closed curve is plotted  that connects $\Lambda$ values  on axes corresponding to each of the individual  measures.  The  average enclosed  area of  the curves  is given for each quartile, normalized to unit maximum for the  ideal pulsar.
The four thick black curves correspond to the four exemplar objects discussed in the text.   The high quality
MSP J1909$-$3744 is in quartile 1 while  J1939$+$2134 (B1937+21)  and J0509$+$0856  are in quartile 3 as the upper and lower black curves, respectively,  and the large-DM MSP J1903$+$0347 is in the fourth quartile due to its large  consequential scattering.
   \label{fig:Qcurves_noQLSsum}
   }
\end{figure}

To enable MSP comparisons more visually, 
Fig.~\ref{fig:Qcurves_noQLSsum} shows $\Lambda_j$  for each pulsar plotted as a closed curve calculated from a spline fit to the individual values.   Objects are divided into four quartiles in $\Lsum$.    The enclosed area of each curve is normalized by the area ($\pi$) of the unit circle and the average normalized area is given for each quartile.
An ideal pulsar would appear as the  circle  shown with unit radius and would have unit normalized area.     

The heavy black lines in the figure correspond to the four exemplar objects discussed above.  One of the best pulsars in PTA observations  is  J1909$-$3744  and it correspondingly falls in quartile 1.
The  large DM (297\,$\DMunits$)   MSP  J1903$+$0327  is in the fourth quartile because its large scattering is not compensated by other factors.     The other  two objects are in the third quartile for differing reasons:  B1937$+$21 shows large achromatic spin noise and large interstellar contributions that outweigh its high quality in terms of its jitter and S/N, while  J0509$+$0856 has a large template fitting error given its wide pulse, which is about 15 times wider than those from J1909$-$3744  and B1937$+$21 and three times that of J1713$+$0747.

%
\section{Summary and Recommendations}
\label{sec:summary}

This review presented an end-to-end study, from the neutron star to the radio telescope, of effects and processes that influence the utility of radio pulsars in timing metrology. 

We emphasize again that the best timing precision is obtained from millisecond pulsars  due to their much lower spin noise compared to other pulsars and because,  in a fixed measurement time, they provide a much larger number of pulses for averaging down pulse jitter. 
However, not all MSPs are suitable for the most demanding application: as detectors of long wavelength (light year) gravitational waves. A combination of spin noise, diverse pulse widths and jitter, gravitational potential noise (in globular clusters), and measurement noise combine with stochastic interstellar propagation to render only a fraction of MSPs suitable for PTA work. 
\newtext{\citet[][]{sc10}, for example, estimate 20\% of MSPs are suitable based on spin noise considerations alone.}
Some, but not all, of these factors can be mitigated.  We summarize key findings below, placing emphasis on considerations for inclusion of MSPs in PTAs and on the opportunities and limits for TOA error mitigation. 

\subsection{Building a Pulsar Timing Array}

A useful way to summarize our results is to  consider building a PTA from scratch.
This first involves  selecting MSPs  based on criteria like those already used for existing PTAs and using elements from our study.  Next is to optimize \TOA\ estimation and identify constituents of the covariance matrix used for \TOA\ modeling.

We describe the main steps as follows, where we take a forward looking approach that assumes broad frequency coverage during observations at a single epoch. 
Simultaneous measurements over a 3:1 frequency range, for example, allow \TOAs\ to be optimized  for each epoch at the `front end' of an overall pipeline rather than through covaiance matrix modeling that is applied to multiepoch \TOAs.   We also eschew usage of dimensionless fudge factors (e.g. EFAC and ECORR) to alter error
contributions from template matching and pulse jitter, which actually can be calculated using known information about observations for each pulsar.

\begin{enumerate}
\itemsep -1pt
\item 
\textbf{Select/cull pulsars for the array:} Some pulsars are simply unsuitable for high-precision timing, due to effects that are both intrinsic (primarily, red spin noise) and extrinsic (e.g., ISM-induced timing offsets). To aid the selection of MSPs for PTAs and identify methods for optimizing  timing programs, we developed a set of  metrics for assessing pulsar quality.  These metrics take into account timing errors related to spin noise,  pulse jitter,  the signal-to-noise ratio of pulse measurements,  and interstellar scattering, each with its own quality factor that are combined into a composite quality factor.  These were applied to and calibrated against the current MSP sample in NANOGrav's PTA.  Appendix\,\ref{app:QMcode} describes Python code that calculates  quality factors for any MSP either in absolute terms or in comparison with the current best MSP observed in multiple PTA programs, MSP J1909-3744\footnote{The code is available 
on GitHub (\url{https://github.com/jmcordes/PTQ)})}. 

Spin noise, in particular, cannot be mitigated when it is too large; it can only be avoided by excluding pulsars from a PTA. Spin noise is achromatic (not dependent on radio frequency) and is to some extent degenerate with power-law noise in the stochastic GW background.  Fortunately they can be distinguished by the fact that GW perturbations are partially correlated between pulsars, while spin noise is of course not. In principle, GW detection involves combining signals from all pulsars in the PTA to build up the GW signal relative to other red noise contributions, which are statistically independent between MSPs.   However, there are insufficient numbers of MSPs in current PTAs to  allow inclusion of those with large spin noise
because they will  inhibit GW detection and characterization. In  special cases, such as a CP orbiting a black hole,  the scientific value of the target pulsar  may override concerns about  red noise. 

\item
\textbf{Design an optimized observing program:} Once pulsars have been selected for a PTA, an optimal combination of observing frequency, dwell time, and cadence must be determined. These considerations should take into account the following factors: 

Large observing bandwidths are critical for precise estimation of all chromatic noise effects, most of which are induced by propagation through the ISM. Wideband systems provide greater bandwidth and thus larger pulse signal-to-noise 
ratios, and they can provide better determination of plasma propagation terms in the arrival time equation. However, wideband systems are also subject to departures of the simplest dispersion law ($\Dt \propto \DM / \nu^2$) caused by multipath propagation. The choice of observing frequency thus involves a trade-off between pulsars’ steep intrinsic spectra, which yield higher flux densities at lower frequencies, and ISM-induced timing offsets, which also increase inversely with frequency. 

The optimal dwell time should beat down pulse jitter, but will also depend on the ISM characteristics along the LoS. In general, the scaling of jitter-induced timing errors with integration time means that shorter integrations can be accommodated for pulsars with shorter periods and narrower pulse components. However, when the scintillation timescale is long compared to the dwell time, one must consider TOA variance due to diffractive scintillation (the finite scintle effect). Large observing bandwidths mitigate the finite scintle effect, and thus provide another means of optimizing the dwell time per pulsar. 

The cadence of observing epochs is most relevant when combining datasets that cover disparate frequency bands in order to determine DM (as well as scattering). 
The TOA error induced by asynchronous DM estimation scales almost linearly with the temporal separation between observing epochs (e.g. Eq.~\eqref{eq:dtinfty_async4}).
Here again, wideband systems reduce the need for  observations at different frequencies made at slightly different epochs, because the DM can be determined more precisely from a single observing epoch with a single receiver.

\item
\textbf{Estimate \TOAs\ at each epoch, including corrections:}
Individual TOAs should  be optimized  empirically at each epoch.  
Even with suitable choices of template,  \TOA\ errors in excess of those expected from idealized matched filtering  are expected. 
Equation~\eqref{eq:toa_with_dm_scattering2} illustrates the full arrival time equation, including timing delays from all relevant effects.  
Terms that are most amenable to correction are $\Dtmismatch$  and the average $\DMbar$ term (epoch dependent, achromatic); this latter term is addressed by all current \TOAs\ but may not have been optimized with respect to the influence of scattering. 
Terms that may be partially correctible include $\DtJ$ and $\DtdPBFall$.   A jitter correction may emerge from PCA applied to pulse profile sequences; attempts to date have given only modest improvements.  Short term PBF variations (finite number of scintles) may be correctible if scintillations are carefully monitored through DS with sufficient time and frequency resolutions.  
Uncorrectable terms or residual errors include $\DtSNR$, $\DtJ$, and  $\DtdPBFall$.   Optimized observations and processing can minimize these but they are mostly  irreducible.  We now discuss possible corrections.

{\it Rapid profile variations and  resulting \TOA\ errors:}  Systematic template mismatch, pulse jitter, and rapid PBF variability (finite scintle effect) are potentially correctible, to varying degrees.  Template mismatch produces a systematic error while jitter and the FSE effect yield random errors.   Jitter is potentially correctible if the net profile change in an $N$-pulse average is systemtically related to the TOA offset.  The FSE effect is correctible if CS can be used to determine the instantaneous PBF on time scales $\lesssim \dtd$, the DISS time scale $\sim$\,minutes.  

{\it Slow changes correlated over multiple epochs:}  Intrinsic profile changes, like the transitory events seen in MSP J1713+0747, which eventually decay to the pre-event profile shape,  can be mitigated through epoch-dependent profile modeling.  This can be done using PCA, for example, but whether it is better than simple excision of \TOAs\ from epochs during an event needs to be evaluated on a case-by-case basis.  \newtext{The same statement holds for other pulsars that show episodes of profile shape changes \citep[e.g.][]{Shannon2016}.}

Several ISM perturbations have  time scales on the order of $\dtr$, the characteristic time scale for RISS (refractive scintillations).  In general, this is determined by the transverse extent of the bundle of scattered rays and the speed with which the \LOS\ changes.  For a single, thin screen along the \LOS\, the width of the ray bundle $\sim \lr$ and net time scale are simply related and the characteristic refraction timeis $\dtr \sim \lr / \veff$ (cf.\ Table\,\ref{tab:keyplasmaeq}).

ISM perturbations include slow departures of PBFs from the ensemble shape due to variable refraction, which shifts the ray bundle,  slow AOA variations that alter the translation of \TOAs\ to the SSBC, and the frequency dependence of DM.  Analysis of dynamic spectra using PRS or CS  can provide `at epoch' estimates of the scattering time or possibly the PBF shape, which can track refractive distortions of the PBF.   We emphasize that determination of the PBF shape and its effect on \TOAs\  need to be based  on the same data employed for  initial
\TOA\ estimates (e.g. using template matching with a fixed template).  We also underscore that usage of an exponential PBF is unsuitable in PTA contexts because it introduces biases in \TOA\ corrections. 

Once scattering time estimates are available, 
\TOA\ corrections for slow scattering variability can be based on a look-up table for \TOA\ offsets based on the actual intrinsic profile shape and relevant PBF. This is described in more detail below.  AOA variations are harder to track but might be quantified using drift rates in dynamic spectra, $d\nu/dt$ of scintles.    

Scattering-corrected \TOAs\ will lead to better (unbiased) estimates for \DM.  Whether the chromaticity of DM can be corrected is unclear.      The sustained form for $\DM(\nu)$ over a refraction time may provide the basis for its determination.
Further exploration of this is rqeuired.   On the other hand, the consequences of
DM chromaticity are very small for low-DM (e.g. $\lesssim 50\,\DMunits$) pulsars observed at $\gtrsim 0.8$\,GHz frequencies.   Utlizing higher-DM pulsars and lower frequencies may require special attention.

Discrete interstellar refraction and delay events, seen as `extreme scattering events' are sustained over periods of weeks to months.   Delays from ESEs are generally expected, through modeling, to include a mixture of dispersion delays, AOA delays, the combined effects from multiple images that may be produced.   These are not easily modelable and, as discussed earlier, will produce non-Gaussian timing residuals. 

These considerations will be increasingly important as PTAs expand the number of included MSPs, which will likely require adding  higher-DM pulsars than those in current PTAs.    
As shown in Sect.~\ref{sec:scattassess},  scattering corrections for higher-DM pulsars (or observations at lower frequencies than typically used in PTAs even for the current sample) require resolution of scintillation frequency structure.  This is more challenging for high-DM/low-frequency combinations and in some cases requires incorporation of cyclic spectroscopy methods.

\item
\textbf{Model remaining errors and variances:}  
Not all terms in the arrival time equation can be fully corrected and some are simply uncorrectable. The remaining \TOA\ errors require inclusion  in   the covariance matrix  used to  optimize parameter estimation of a comprehensive timing model and  for GW detection.

 Three terms are well known to always  require inclusion in the covariance matrix: (1) Template fitting errors from radiometer noise, which are dependent on epoch, telescope, receiver, and the degree of refractive and diffractive scintillations; (2) Uncorrectable jitter, whose variance is largely epoch-dependent and is broadband,  decorrelating only  over frequency ranges larger than about an octave; and (3) Red noise, including spin and orbital noise, as well as residual ISM noise that is only partially corrected.

 Additional terms arise from uncorrected PBF variations on both rapid and slow time scales.  Quantifying their contributions to the covariance matrix requires consideration of the specific methodology used for template matching.  The same holds for intrinsic profile variations and whether TOAs are corrected or excised. 

Attention is also required for the stationarity and gaussianity of residual errors 
and fluctuations.  The simplest form of nonstationarity is the variation in radiometer noise with epoch due to changes in receivers (and telescopes).  Also, RISS alters the effective flux density of a pulsar, causing  modulations of 
 the template-fitting variance.  Terms that are well approximated as Gaussian processes include \TOA\ errors from radiometer noise, jitter, and rapid PBF changes.   Spin variations (cf.\ discussion in Sect.~\ref{sec:spinnoise}), yields mixtures of (likely) Gaussian variations and discrete events.    DM and multipath effects are also Gaussian processed intermixed with strong modulations of observables due to caustic events. 

\end{enumerate}

\subsection{Customized scattering mitigation for individual MSPs }
\label{sec:msp_customized}

We have demonstrated with simulations how \TOA\ offsets depend on astrophysical aspects of scattering as well as methodological details of template matching.  
The most general conclusion is that \TOA\ offsets are smaller than  than the scattering time $\taud$ at a given frequency and also  scale with frequency differently than the scattering time. Furthermore, the  \TOA\ offset is function of both the  intrinsic pulse shape as well as the PBF shape.  

Scattering corrections of \TOAs\  thus require a customized analysis for each pulsar.  
These would comprise simulations that convolve the adopted template profile (assumed to be devoid of any scattering\footnote{Of course templates based on measured pulses will likely include some level of scattering but this is small for low-DM pulsars.  In practice, some templates are based on Gaussian component modeling of measured profiles, which excludes scattering to a large degree.}) with a family of assumed PBFs.  These PBFs would be parameterized with the underlying 
density spectrum index $\beta$ and inner scale $\linner$.  Measured scattering times $\taud$ provide an estimate for the diffraction scale $\ld(\nuref)$ at a reference frequency $\nuref$ which yields
the parameter $\zeta(\nu) = \linner / \ld(\nu)$ for power-law PBFs. Recall that an exponential PBF with an achromatic shape is generally a poor choice for constructing templates for precision timing.  Asymptotically, however, very strong scattering yields a small
$\ld$ and a large value fir $\zeta$ for which the PBF tends toward an exponential if scattering is confined to a small fraction of the line of sight. 

The steps of a multifrequency analysis would be:
\begin{enumerate}
\itemsep -2pt
\item Choose a template to use as a starting point. Ideally this would be a `de-scattered' 
profile or one obtained at a frequency where scattering has a minimal effect on the shape. 
Intrinsic profile chromaticity is also an issue that can be addressed by multifrequency profile 
modeling  that excludes scattering but includes frequency dependent amplitudes, widths, and positions of Gaussian components \citep[e.g.][]{kwj+94, 2025ApJ...986..191G}.
\item Choose a PBF shape according to $\beta$ and inner scale $\linner$.
\item Use a measurement of scattering time at a reference frequency $\nuref$ to evaluate
$\zeta(\nu) = \linner / \ld(\nu)$ where $\ld(\nu)$ is related to $\taud(\nu)$.
\item Generate frequency dependent PBFs using $\beta$, $\zeta(\nu)$, and $\tau(\nu)$. 
\item Convolve the intrinsic shape $U(t, \nu)$ with each PBF $\PBF(t, \nu, \tau)$ using a grid of $\tau$ values.
\item Using a so-generated  template (which may differ from that used in step 1), determine
the \TOA\ offset as a function of $\tau$  and frequency.  
\item Use the results as a lookup table for a \TOA\ correction that is a function of $\nu, \tau(\nu)$ and implicitly depends on the intrinsic profile shape
and on ISM turbulence parameters ($\beta$ and $\linner$).  The correction is most robust for a contemporaneous measurement of  $\tau(\nu)$  based on a large number of scintles (cf.\ Sect.~\ref{sec:toa_errcorr}).
\end{enumerate}

{\it Acknowledgements:}
We thank numerous colleagues, including those  in the NANOGrav, the EPTA, the PPTA, the MPTA, the InPTA, and the CPTA collaborations along with others for discussions about pulsar timing over the last several decades.  
We thank the NANOGrav collaboration for providing data from its
15-yr data release \citep[][]{2023ApJ...951L...9A} used to make Figs. \ref{fig:pulse_gallery}, \ref{fig:resids_and_sfs}, 
\ref{fig:allanvar}, \ref{fig:uuprime1},
\ref{fig:pulse_widths_Lband_nanograv_pulsars},
\ref{fig:J1713_profile_change}, \ref{fig:DM_J1909-3744},
\ref{fig:DMSF_3panels_B1937},  and
\ref{fig:DMSF_asymptotic_B1937}.
We thank M. Vivekanand for providing data from his
paper \citep[][Fig 5]{vivekanand2020} also used in Fig.~\ref{fig:DMSF_asymptotic_B1937}.
We thank Ryan Shannon for unpublished data used for 
Fig.~\ref{fig:pulses_J0738-4042}. 
We thank Matthew Kerr for providing data on the Vela pulsar used in
Fig.~\ref{fig:Vela}. 
This research has made use of the EPN Database of Pulsar Profiles maintained by the University of Manchester, available at: \url{http://www.jodrellbank.manchester.ac.uk/research/pulsar/Resources/epn/}.
It also made extensive use of the ATNF Pulsar Caralogue
\citep[][]{ATNFPulsarCat}
\url{ http://www.atnf.csiro.au/research/pulsar/psrcat}.
This work was supported by NSF Physics Frontiers Center award numbers 1430284 and 2020265,  which support the NANOGrav project.

\section*{Statements and Declarations}

{\bf Competing interests:} There are no competing interests for any of the authors.

{\bf Funding:} All authors are members of the NANOGrav Physics Frontiers Center, which has been supported by the U.S. National Science Foundation  Physics Frontiers Center under award numbers 1430284 and 2020265.

\clearpage

\startappendices
\section*{\centerline{Appendices}}
\section{Nomenclature}
\label{app:nomenclature}

\startlongtable
\begin{deluxetable*}{ l l c }
\tabletypesize{\footnotesize}
\centering 
\tablecolumns{3}
\tablecaption{Acronyms, terms,  and symbols \label{tab:nomenclature}}
\tablehead{ 
    \colhead{\large Item } & \colhead{\large Definition} & \colhead{\large Typical Units}    
}
\startdata
\\
\multicolumn{2}{l}{\bf \large Acronyms and terms:}\\
\multicolumn{2}{l} {\it \large Astrophysical}\\
AGN         & Active galactic nucleus & \\
CP          & Canonical pulsar \\
DGR		    & Differential galactic rotation &  --- \\
DNS         & Double neutron star & \\
ESE         & Extreme scattering event & \\
GC          & Galactic center & \\
IDV         & Intraday varaible (AGN) \\
IHV         & Intrahour variable (AGN) \\
MSP         & Millisecond pulsar & --- \\
MW          & Milky Way & --- \\
\\
\multicolumn{2}{l} {\it \large Methods and statistical quantities}\\
ACF, ACV, CCF    & Autocorrelation, autocovariance,  and crosscorrelation  functions & \\
AD, AV      & Allan deviation $\sigma_{\ybar}$ and Allan variance
 $\sigma^2_{\ybar}$ of dimensionless frequency $y$ & --- \\
AMN         & Amplitude modulated noise & --- \\
AMPSN       & Amplitude modulated shot noise (SAMPSN = Scintillated AMPSN) & --- \\
CS          & Cyclic spectrum or spectroscopy & --- \\
DS          & Dynamic spectrum  (intensity vs. time and frequency)& \\
FWHM        & Full width at half maximum & \\
HWHM        & Half width at half maximum & \\
MF          & Matched filter & \\
PCA         & Principal component analysis & \\
PRS         & Phase resolved spectrum or phase resolved spectroscopy  & --- \\
SEFD $\equiv \Ssys$        & System equivalent flux density = $\Tsys / G$ & Jy \\
SS          & Secondary spectrum (power spectrum of dynamic spectrum)  \\
SWIMS       & Stochastic wideband impulse modulated self-noise & --- \\
VLBI        & Very long baseline interferometry & \\
\\
\multicolumn{2}{l} {\it \large Propagation}\\
AOA         & Angle of arrival & various  \\
DISS		& Diffractive interstellar scintillation & --- \\
Line of sight Measures: \\
\quad DM		    & Dispersion Measure 			& $\DMunits$    \\
\quad $\DM(\nu)$    & Chromatic DM due to multipath averaging & $\DMunits$ \\
\quad EM              & Emission measure 			& pc\,cm$^{-6}$ \\
\quad SM              & Scattering measure                	& kpc\,m$^{-20/3}$ \\
\quad RM              & Faraday rotation measure          & rad\,m$^{-2}$ \\
PBF         & Pulse broadening function & --- \\
RISS        & Refractive interstellar scintillation & \\
Scintles    & Intensity features in the time-frequency plane caused by diffractive scattering & intensity like\\
TOA         & Time of arrival & various \\
\\ 
\multicolumn{2}{l} {\it \large Units}\\
$\AU$         & Astronomical unit & \\
Jy          & Jansky (unit of flux density) & \\
mas         & milliarc second \\
\\
\multicolumn{2}{l}{\bf \large Symbols:}\\
$\langle X \rangle$ & Ensemble average of a quantity $X$  & \\
$\langle X \rangle_{x} $ & Average of $X$ over another  quantity $x$ (e.g. time, frequency, or phase) & \\
$\alpha$		& Exponent in phase structure function		& \\
$\beta$		& Exponent in wavenumber spectrum for $\nelec$ (11/3 for Kolmogorov)& \\
$B$         & Receiver bandwidth & MHz, GHz \\
$\cnsq$		& Coefficient in electron-density wavenumber spectrum & m$^{-20/3}$ \\
$C_1$		& Constant in uncertainty relation, $2\pi\dnud\tau_d = C_1$ & --- \\
$c$		& Speed of light			& cm\,s$^{-1}$ \\
%
$D_{\DM}$	& Structure function for dispersion measures    & $\DMunits^2$ \\ 
$D_{n_e}$	& Structure function for electron density   	& cm$^{-6}$ \\
$D_{N_e}$	& Structure function for electron column density   	& cm$^{-4}$  \\ 
$\Dphi$		& Phase structure function & radians$^2$	\\
$d_{\rm p}$ & Total degree of polarization & --- \\
$\Delta(\tau)$  	& Normalized ACF of noise & \\
$\dnud$		& Decorrelation bandwidth for diffractive scintillation (DISS); & kHz or MHz \\
            & (also scintillation or diffraction bandwidth) \\
$\dnur$		& Decorrelation bandwidth for refractive scintillation (RISS); & MHz or GHz \\
$\dnusb$    & Spectral broadening bandwidth & Hz \\
$\dt$		& Time Lag	&			\\
$\delta I$	& Intensity fluctuation &		\\
$\dne$		& Fluctuation in free electron density   	&  cm$^{-3}$\\
$\dxvec$	& Spatial offset in observation plane & \\
$\Delta\omega$ 	& Difference in galactic rotation rate & \\
$\dtd$      & Diffractive scintillation (DISS) time scale                      & various \\
$\dtr$      & Refractive scintillation (RISS) time scale                      & various \\
$\dlo$      & Distance from Lens (or layer or screen) to observer  & kpc	\\
$\dsl$      & Source-lens (or layer or screen) distance & kpc	\\
$\dso$      & Source-observer  distance & kpc	\\
$E$ and $\varepsilon$ 	&  Electric field and baseband electric field & \\
$\epsilon$ &  fractional electron density variation = $\RMS(\dne) / \nelec$  \\
$\zeta$ &  Dimensionless inner scale or  fractional density variance \\
$\eta$		& Mean-square scattering angle per unit distance & mas kpc$^{-1}$ \\
$\eta_{\rm t}, \eta_{\nu}$ & Dimensionless coefficients for calculating numbers of scintles & --- \\
$F_{\rm c} = \zeta \epsilon^2/f_{\rm f} \louter^{\beta-3}$ & Cloudlet model parameter for relating SM to DM & pc$^{-2/3}$ (Kolmogorov) \\
$\Ftilde = F_{\rm c} / \linner^{4-\beta}$   &  Cloudlet model parameter for calculating scattering time & ${\rm (pc^2\, km)^{-1/3}}$ (Kolmogorov)  \\
$\FJ$       & Jitter parameter & ---  \\
$f_{\rm f}$ & Volume filling factor in cloudlet model & \\
$f_{\rm Nyq}$ & Nyquist frequency & \\
$\fs, \fsdot, \fsdotdot$ & Pulsar spin frequency and its first two time derivatives & Hz, Hz\,s$^{-1}$, Hz\,s$^{-2}$\\
$f_{\rm yr}$ & Frequency of 1~yr$^{-1}$ & yr$^{-1}$ \\
$f_{\alpha}$	& Numerical factor in structure function & \\		
$\fJ$       & Dimensionless jitter factor for calculating TOA jitter & \\
$\Gamma_I$	& Intensity correlation function &	\\
$\Gamma_{\delta I}$	& Intensity covariance function &	\\
$\Gamma_{\varepsilon}$ &  Spatial correlation function of wavefield & \\  
$G$         & Telescope gain & K\,Jy$^{-1}$  \\
$I, Q, U, V, L=Q+iU$		    & Stokes parameters 	&	Jy or mJy 		\\
$\Itinfty$  & Factor in the calculation of TOA error from frequency-dependent DMs & ---   \\  
$\IDMhat$   & Factor in calculation of frequency-dependent DMs & ---  \\
$\KDM = c \relec / 2\pi  \times {\rm pc / GHz^2}$      & Constant in the expression for dispersion time delay
		$ t = \KDM \nu^{-2} \DM$ & s \\
$\lambda,\nu$	& Electromagnetic wavelength and frequency	& cm, Hz \\
$\ld, {\ld}_{_\theta}, {\ld}_{_I}$		& Spatial scale in diffraction pattern (alternative definitions)  	& various \\
$\ell, b$	& Galactic coordinates			& deg \\
$\linner, \louter$   & Inner and outer scales of electron-density variations 
    ($\equiv 2\pi/\qinner, 2\pi/\qouter$) & various \\ 
$\miss$         & Scintillation modulation index (= rms / mean) & --- \\
$m_M$           & Modulation index of M &  --- \\
$\nu$           & Radio frequency & GHz \\
$\ndof$         & Number of degrees of freedom in scintillations & --- \\
$\Nscint = \Nnu\Nt$         & Number of independent scintillation features (scintles) averaged & ---   \\
$\Nnu$ & Number of independent scintillation features averaged over frequency --- \\
$\Nt$  & Number of independent scintillation feasures averaged over time --- \\
$\npol$         & Number of polarization channels & ---  \\
$n_e$		& Free electron density		& cm$^{-3}$ 	  \\ 
$\Pne(\qvec)$	& Wavenumber spectrum for the electron density & (length)$^{-3}$  \\
$P$         & Pulse period & s  \\
$\Pdot$     & Period derivative = $dP/dt$ &  s\,s$^{-1}$ \\
$\qvec$		     & Wavenumber (3D vector) 			& (length)$^{-1}$  \\
$\qouter, \qinner$	& Lower and upper wavenumber cutoffs of $\Pne$ & various \\ 
$\Qsnr$,  $\Qj$,  $\Qdiss$, $\Qtaud$,   &  
        Quality measures for TOA effects & $\mu s^{-1}$  \\ 
\quad  $\Qspin$,, $\Qdmt$,  $\Qdmnu$,     \\
\quad $\Qpbf$,  $\Qsum$,  $\QLSsum$ & &      \\
%
$r_e$		& Classical electron radius &	cm	\\
R           & Ratio of a pair of frequencies ($\ge 1$) & --- \\
$\sigma_I, \sigma_Q, \sigma_U, \sigma_V, \sigma_L$ & RMS values of Stokes parameters & \\
$\rmstoaS$  & RMS TOA error from radiometer noise (S/N dependent) & ns or $\mu$s  \\
$\rmstoaJ$  & RMS TOA error from pulse jitter  & ns or $\mu$s  \\
$\rmstoaDISS \equiv \dtDISS \equiv \DtdPBF$ & RMS TOA error from diffractive scintillations (finite scintle effect) & ns or $\mu$s \\
$\rmstoaSPIN$ & RMS residual from spin or torque noise of a neutron star & ns or $\mu$s \\
$\rmstoaDMt$ & RMS residual due to asynchronous multifrequency observations  
        & ns or $\mu$s \\
$\rmstoaDMnu$ & RMS residual due to chromaticity of DM from multipath propagation  
        & ns or $\mu$s \\
$\sigtheta$	& RMS angle of arrival 		& \\
$\Ssys$     & SEFD = $\Tsys / G$            & Jy \\
$\Spsr$     & Pulsar period averaged flux density & mJy, Jy \\
$\mathrm{S/N}$ & Signal-to-noise ratio & --- \\ 
$\sinc(x)$ & sinc function = $(\sin \pi x)/ \pi x$ & --- \\
$s$		& Location along line of sight, $s=0$ at source & \\
$\tau$        & lag time in correlation function or characteristic pulse scattering time & various  \\
$\tau_{\rm d}$	& Characteristic pulse  scattering time	& various  \\
$t$		& Time 		&	various		\\
$\ts$   & Slow time (e.g. $\gg$\, s) & various \\
t, TOA     & Time of arrival & various \\
$\tinfty$  & TOA with chromatic delays removed (`infinite frequency') & ns or $\mu$s \\
$t_{\rm e}$ & Emission time at source & various \\
$t_C$ & Non-dispersive chromatic delays & $\mu$s \\
$\tau$ or $\taud$  & Scattering time constant (1/e) & various \\
$t_{\rm sep}$ & Separation time between multi-frequency observations & \\
$\Tint$ & Integration time or data span at a single epoch  & various \\
$T_{\rm data}$ & Total time span data set for an individual pulsar & yrs \\
$\Tsys$     & System (radiometer) temperature & K \\
$U$         & Template shape normalized to unit peak amplitude & \\
$\phi$		& Phase perturbation from refractive index perturbations &  rad \\
$\veffvec$	& Vector effective velocity & km\, s$^{-1}$	\\
$\vobsvec$		& Observer's velocity	&		\\
$\vpvec$		& Pulsar velocity	&		\\
%
$W$ or $\WU$  & Pulse width (FWHM) & various \\
$\xtau$      	& Exponent in scaling law  for pulse broadening time 
        $\tau(\nu) \propto \nu^{-\xtau}$ & \\
\enddata
\tablecomments{\newtext{Wherever a symbol appears in the text with a caret ($\widehat{\ }$) it refers to an estimate of that quantity.}}
\end{deluxetable*}


\begingroup
\newcommand{\Sr}{S_{\mathscr R}}
\newcommand{\Srbar}{S_{\,\overline{{\!\mathscr R}}}}

\renewcommand{\hc}{h_{\rm c}}
\renewcommand{\fref}{f_{\rm ref}}
\providecommand{\Agw}{A_{\rm gw}}
\providecommand{\Xgw}{X_{\rm gw}}
\providecommand{\fNy}{f_{\rm Ny}}

\clearpage
\section{Spectral representations}
\label{app:specrep}

Spectra are defined as one-sided (positive frequencies only) so that the variance $\sigma_X^2$ of a quantity $X$ is related to its spectrum $S_X(f)$ by
\be
\sigma_X^2 = \int_0^\infty df\, S_X(f). 
\ee

A variety of spectral representations are used in the pulsar, PTA, and  GW  literature, which are summarized  here, though not all are utilized in the paper:
\begin{center}
\begin{tabular}{lll}
Power spectrum of TOA perturbations $\Delta t(t)$:  && $\Sdt(f)$ \\
Power spectrum of timing residuals $\mathscr{R}(t)$: &&  $\Sr(f)$ \\
Power spectrum of dimensionless strain $h(t)$:  && $S_h(f)$ \\
Amplitude spectrum for dimensionless strain: && $\hc(f)$ \\
Energy spectrum of GWs: && $S_E(f)$ \\
Energy density of GWs per unit log frequency: & & $\rho_{\rm gw}(f)$ \\
Power spectrum as fraction of closure density: && $\Omega_{gw}(f) = \rho_{\rm gw}(f) / \rho_c  $ \\
\end{tabular}
\end{center}

These are related according to the following relations \citep[e.g.][]{2015CQGra..32a5014M}:
\begin{eqnarray}
h_c^2(f)
&=& fS_h(f)  
=\left(\frac{3H_0^2}{2\pi^2} \right) \frac{\Omega_{\rm gw}(f)}{f^2}
= \frac{4G}{\pi c^2} \frac{S_E(f)}{f}
= 12\pi^2f^3 \Sdt(f),
\label{eq:specrelations}
\end{eqnarray}
where $H_0$ is the Hubble constant and $G$ is the gravitational constant. 
The coefficient $12\pi^2 = 3(2\pi)^2$ in the last equality involving $\Sdt(f)$ includes  $(2\pi)^2$ from the Fourier derivative theorem and a factor of 3  from  all-sky averaging of the TOA perturbation. 

The residuals spectrum $\Sr(f)$ is the power spectrum of the 
 timing residual ${\mathscr R}(t)$,  the difference between the timing perturbation $\Delta t(t)$  and a model fit  $\widehat{\Delta t}(t)$,
\be
{\mathscr R}(t) = \Delta t (t)- \widehat{\Delta t}(t). 
\ee


Empirical spectral models often use a power law form with an implicit frequency range,
\be
S(f) = S_0 \left(\frac{f}{\fref}\right)^{-X},
\label{eq:plspec}
\ee
where $\fref$ is a reference frequency (often specified as 1\,cy\,yr$^{-1}$ in PTA contexts).
Motivated by representations for the GW amplitude spectrum,  
\be
\hc(f) = A \left(\frac{f}{\fref}\right)^{\alpha}
\ee
 using a dimensonless amplitude,  $A$, the spectrum for   
\TOA\ perturbations from GWs  (before any model fitting that removes fluctuation power),  is
\be
\Sdt(f) = \frac{A_{\rm gw}^2}{12\pi^2} \left(\frac{f}{\fref}\right)^{-\Xgw},   \quad \quad  \Xgw = 3 +  2\alpha.
\label{eq:SdtAsq}
\ee
\newtext{
The residuals spectrum $\Sr(f)$ has the same form as $\Sdt(r)$ but with an implied lower frequency cutoff related to the data span length $T$, $f_{\rm min} \propto 1 / T$.  More generally,  the residuals spectrum is the \TOA\ perturbation spectrum scaled by a {\it transmission function} that accounts for variance removed by the multi-parameter fit to the arrival times.
}

For example,   the simplest model for the nanohertz stochastic background gives $\alpha = -2/3$ and
$\Xgw = 13/3$ \citep[][]{2001astro.ph..8028P,jb03}.

Generalizing the relation between Eq.~\eqref{eq:plspec} and Eq.~\eqref{eq:SdtAsq}, we have
\be
A = 2\pi\sqrt{3} \left[ \fref^3 (f / \fref)^X S(f)\right]^{1/2}.
\label{eq:Ageneral}
\ee

Application to several classes of stochastic noise includes the following:
\begin{enumerate}
\item {\it Power-law noise:}  
The dimensionless amplitude, $A_{\rm PL}  = 2\pi \sqrt{3}  \left[ \fref^3 S_0\right]^{1/2}$
is independent of  fluctuation frequency, 
\item {\it White noise:}  With spectral index   $X = 0$,  noise 
with  time-domain \RMS\  $\sigma_{\rm w}$ in a bandwidth $B_{\rm w}$ has  amplitudes
$S_0 = \sigma_{\rm w}^2 / B_{\rm w}$ and 
$A_{\rm w}  = 2\pi \sqrt{3}  (\fref / B_{\rm w})^{1/2} \fref \sigma_{\rm w} $.
A time series of sampled data with uniform spacing $\delta t$ has a bandwidth $B_{\rm w}$ 
equal to the Nyquist frequency,  $\fNy = 1 / 2\delta t$.
%
%
For instance, white noise \TOA\ errors with $\sigma_{\rm w} = 0.1\,\mu$s in data obtained roughly monthly with
$\delta t = 0.1$\,yr  has Nyquist frequency 
$\fNy = 5$\,cy\,yr$^{-1}$ and  amplitudes
yields a spectral amplitude
\be
S_0 &=& 2\times 10^{-3}\, {\rm \mu s^2\, yr} 
	\left(\frac{\delta t}{0.1\, \rm yr}\right)
	\left(\frac{\sigma_{\rm w}}{0.1 \, {\rm \mu s}} \right)^2,
\ee
and a dimensionless strain spectrum with $\alpha = 3/2$ and 
\be
A_{\rm w} &=& 1.54\times 10^{-14} 
	\left(\frac{\delta t}{0.1\, \rm yr}\right)^{1/2}
	\left(\frac{\sigma_{\rm w}}{0.1 \, {\rm \mu s}} \right).
\ee

\item {\it Low-pass spectra:}
A spectral form  $S(f) = S_0 L(f)$ with low-pass   shape normalized to $L(0) = 1$ has bandwidth
$B_L \sim 1 / W_L$ and  $W_L$ is the correlation time for the process.  
If the variance is $\sigma_L^2$, then
$S_0 = \sigma_L^2 / B_L = \sigma_L^2 W_L$.    This relation is largely unaffected by the removal of a second-order polynomial, as is done in pulsar timing analysis to fit for a pulsar's spin rate and its derivative, if the length of the time series $T \gg W_L$, which we assume. 

While a power-law form is not a good model for a low-pass spectrum,  PTA studies characterize the GW background with a power law that is then compared with contributions from other processes, whether or not they have power-law spectra. 
Non power-law spectra  have equivalent $A$ values that are frequency dependent, as seen by inspection of Eq.~\eqref{eq:Ageneral}.  Estimated values for the corresponding $X$ values are also frequency and data dependent  in other ways.   
For a low-pass spectrum 
the dimensionless amplitude is
\be 
A_L(f) = 2\pi\sqrt{3} \left[ \fref^3 (f / \fref)^X  S_L(f) \right]^{1/2}.
\label{eq:ALP}
\ee 


Evaluating at $f = \fref$ and assuming $B _L  > \fref$ (or $W_L < 1\,$yr) we have  $S_L(\fref) = S_0$ and 
$A_L(\fref) = 2\pi\sqrt{3} \left(\fref^3  \sigma_L^2 W_L \right)^{1/2}$.
For $\sigma_L = 0.1\,\mu$s, $W_L$ in yr, and $\fref = 1$\,cy\,yr$^{-1}$ the spectral amplitudes are
\be
S_0 &=&  10^{-2}\, {\rm \mu s^2\, yr} 
	\left(\frac{W_L}{\rm 1\ yr}\right)
	\left(\frac{\sigma_{L}}{0.1 \, {\rm \mu s}} \right)^2,
\\
A_L(\fref) &=& 2\pi\sqrt{3} \left(\fref^3  \sigma_L^2 W_L \right)^{1/2} 
	= 3.45\times10^{-14}
	\left(\frac{W_L}{\rm 1\ yr}\right)^{1/2}
	\left(\frac{\sigma_{L}}{0.1 \, {\rm \mu s}} \right)^2,
\ee

Processes with low-pass spectra will yield data-dependent values for $A$ and $X$ for a power-law model.
Using simulations, we find that frequency indices $X$  for low-pass processes are smaller than those for
the GW background and pulsar spin noise, i.e. $X_L \lesssim 3$. 
Simulated time series comprising low-pass noise with \RMS\ $\sigma_{ L}$ added to  white noise with
\RMS\ $\sigma_{\rm w}$ were analyzed by first subtracting a second-order polynomial to mimic the removal of a spin-down polynomial from arrival time data.   The resulting spectra $S(f)$ were  fitted with a  model of the form $a + X\log(S) +c$ for $f \le f_{\rm c}$ where $f_{\rm c}$ is a cutoff frequency and a constant amplitude $c$ for $f > f_{\rm c}$.    Least squares fitting for all four parameters yields frequency indices $X$ that depend on the bandwidth of the low-pass noise and the `signal'-to-noise ratio, ${\rm S/N} = \sigma_{L} / \sigma_{\rm w}$.
 Indices averaged over 1000 realizations and over S/N   from 0.5 to 5  were $\langle X\rangle  \simeq -1.8 \pm 0.8$.
 For different S/N values, $\langle X\rangle$ and $\sigma_X$ were $-1.2$ and 0.9 for S/N = 0.5
 and $-2.3$ and 0.6 for S/N = 5.

\end{enumerate}

\endgroup

\clearpage
\section{Structure functions}
\label{app:SFs}

Structure functions are a useful tool for parameterizing red spin
noise that complement frequency-domain methods. 
Following \citet[][]{1978IEEEP..66.1048R}, CD85,
\citet[][]{dmh+95}, and \citet[][]{2015ApJ...801..130L},
the $m^{\rm th}$ order structure function of the spin
phase\footnote{We  freely refer to the phase in either units of cycles
or temporal units (usually microseconds), depending on the context.} is the mean-square expectation,  
\be
D_\phi^{(m)} (t,\tau) \equiv \langle [\Delta_\phi^{(m)}(t,\tau)]^2\rangle ,
\ee
where the 
$m^{\rm th}$ phase increment is
\be
\Delta^{(m)}_\phi (t,\tau) \equiv
\displaystyle{\sum^m_{\ell=0}} (-1)^\ell 
\left ({{m} \atop {\ell}}\right) \,\phi\,[t+(m-\ell)\tau] .
\ee
Phase increments are useful for removing  trends in time series. A polynomial of order
$m-1$ is removed by the $m^{\rm th}$ order increment, so the mean value of a time series is removed by the first order increment $\Delta_\phi^{(1)} = \phi(t+\tau)-\phi(t)$ while a linear trend is removed by the second order increment.  Removal of  a quadratic by the third order increment implies that the structure function for
timing residuals formed by fitting and removing a quadratic spin phase, $\widehat\phi_2(t)$, is identical to the \newtext{third} order structure function of the original phase.  
I.e. if $\resid(t) = \phi(t) - \widehat\phi_2(t)$,
then $D^{(3)}_{\resid}(\tau) = D^{(3)}_\phi(\tau)$.
In practice, fitting functions include more terms than just the spindown polynomial, but these  remove  much less variance  of the red noise. 

Phase increments vs. $t$ are useful for event detection in the phase or one of its derivatives.   A step function in the $k^{\rm th}$ derivative appears as a pulse in the $(k+1)^{\rm th}$ increment and the corresponding structure function is the variance for a comparison  Gaussian process that serves as a null hypothesis for the presence of individual events, as discussed in 
CD85 (\S\,IVa).

Structure functions are good probes of non-stationary, red noise processes  because they are less vulnerable to spectral `leakage,' by virtue of the fact that they act similarly to derivatives, whose spectra are less steep than the original  process\footnote{The Fourier derivative theorem states that $dx(t)/dt$ of a process with spectrum $S_x(f)$ has a spectrum $(2\pi f)^2 S_x(f)$.}. 
The power spectrum $S_\phi(f)$ is defined so that the phase variance is the integral over non-negative frequencies $\langle \phi^2(t) \rangle \equiv \int_0^{\infty} df\, S(f)$.   To keep notation simple, the  spectrum can correspond to  any or all of the stochastic phase terms  discussed earlier and it  can stand for either the pre-fit phase or post-fit residuals.

The phase autocorrelation function (ACF) $R_\phi(\tau)$ is the Fourier transform of the spectrum (the Wiener-Khinchin theorem) using a Fourier kernel $e^{2\pi i f \tau}$. 
Any time (or long term epoch)  dependence of the ACF  is implicit but we recognize that phase fluctuations may have nonstationary statistics.   Red noise combined with a finite data span $T$ has this property and the phase variance $\rmsred^2(T) = R_\phi(0)$ generally depends on $T$ raised to some positive power. 

Structure functions of any order can be written in terms of the ACF
and thus related to the spectrum 
\citep[][]{1978IEEEP..66.1048R} by
\be
D^{(m)}_\phi(\tau) = \int df\, S_\phi(f) \sin^{2m}(\pi f\tau).
\label{eq:Dmphi}
\ee
Depending on the process, one or more of the SFs can be independent of time and depend only on $\tau$.   This is more likely  for higher order SFs.  
The relevant frequency range for the ensemble average implied in 
Eq.~\eqref{eq:Dmphi} is determined by the spectrum itself, but for finite data sets of length $T$,  the lower frequency limit 
for the spectrum of {\it phase residuals}
is $\flow \equiv \eta /T$ where $\eta \simeq 1$ depends on the fitting function used to form the residuals.   The upper frequency limit $\fhigh$  becomes the Nyquist frequency
$f_{\rm Ny} = 1 / 2\Delta t$ for data sampled at time intervals $\Delta t$.

\clearpage
\section{Modulated complex noise}
\label{app:modnoise}
\label{app:noiseproducts}

Time-domain processes comprising modulations of a complex noise process appear in two contexts: the emitted pulsar signal and the impulse response from multipath propagation through the interstellar medium.    The statistical properties of each of these processes are similar, so we present here a generic description that applies to both 
cases.

Define $z(t) = x(t) m(t)$ as the product of a real pulse-like amplitude $x(t)$ and complex, 
Gaussian noise $m(t)$ with bandwidth $B$. The noise has stationary statistics with 
zero mean $\langle m \rangle = 0$,  
variance $\langle \vert m \vert^2\rangle = \sigma_m^2$ 
and autocorrelation function
\be
\langle m(t) m^*(\tp) \rangle = \sigma_m^2 \Delta(t-\tp). 
\ee 
The quantity $\Delta(t)$ has a width equal to the reciprocal of the signal bandwidth $\sim B^{-1}$ and is assumed to be much narrower than $x(t)$.  The noise $m$ is therefore  flat (i.e. `white') over the bandwidth. The `intensity' is the squared magnitude $Z(t) = \vert z \vert^2 = x^2(t) \vert m(t) \vert^2$. In the following we consider $x(t)$ to be deterministic (not stochastic). 

Properties of the squared modulated noise  $M = \vert m \vert^2$ are
\begin{enumerate}
\itemsep -2pt
\item
Modulation index, $m_M$:  $1 + m_M^2 \equiv  \langle M^2 \rangle / \langle M \rangle^2 = 2$.   This value is a property of 
complex Gaussian noise.    If $m$ were real, the value would be three. 
\item 
Spectrum: the spectrum of $z(t)$,  $S_z(\nu) \equiv \vert {\widetilde z}(\nu) \vert^2$,  has mean
$\langle S_z(\nu) \rangle = \sigma_m^2 \widetilde\Delta(\nu) R_x(0)$
where $R_x(0) = \int dt\, x^2(t)$. 
The mean square of the spectrum is $\langle S_z^2(\nu) \rangle = 2 \langle S_z(\nu) \rangle$.
\item
Spectral autocorrelation function:
\be
R_{S_z}(\delta \nu) 
= \sigma_n^4 
\left\{
	R_x^2(0) \widetilde\Delta(\nu) \widetilde\Delta(\nu+\delta\nu)
	+
	[\widetilde\Delta(\nu+\delta\nu)]^2 S_{X}(\delta\nu) 
\right\},
\label{eq:Sacf}
\ee
where $X \equiv x^2$ and
\be
S_{X}(\nu)
= \left\vert   \widetilde {X}(\nu) \right\vert^2 
= \left\vert  \int dt\, X(t) e^{2\pi i \nu t} \right\vert^2 .
\ee
\end{enumerate}
The  two equal-amplitude terms of the frequency correlation function 
in Eq.~\eqref{eq:Sacf} 
comprise a broad component determined by the receiver bandpass function $\widetilde \Delta$ and a narrow spike associated with the modulation $x$.

{\it Product of two complex Gaussian noise processes:}
Let $z = xy$ be the product of two complex noise processes $x$ and $y$ like that described in Appendix\,\ref{app:modnoise}. 
Defining  $X = \vert x \vert^2$, $Y = \vert y \vert^2$, and $Z = \vert z \vert^2$, 
the squared  modulation indices $m_X^2 \equiv \langle X^2 \rangle / \langle X \rangle^2 - 1$ etc. are related as
\be
1 + m_Z^2 = (1 + m_X^2) (1+m_Y^2). 
\ee
If $x$ and $y$ have (complex) Gaussian statistics, then $m_X = m_Y = 1$ and $m_Z = \sqrt{3}$. 

If $x$ and $y$ are Fourier transforms of time-domain processes, they will generally have complex Gaussian statistics by the central limit theorem.  The resulting spectrum $Z$ will manifest two characteristic frequency scales determined by those of $X$ and $Y$ individually. 


\newcommand{\nua}{\nu_1}
\newcommand{\nub}{\nu_2}
\renewcommand{\nuc}{\nu_3}

\clearpage
\section{\TOA\ variations from frequency-dependent dispersion deasures}
\label{app:DMnu}

The chromaticity of dispersion measures $\DMbar(\nu)$ caused by multipath propagation was initially identified by
\citet[][]{cwd+90,rdb+06} and also by 
\citet[][]{2019A&A...624A..22D,2022ApJ...930L..27K,2025ApJ...982..132K}
subsequent to  the theory developed by 
\citet[][hereafter CSS16]{CSS16}. 
CSS16 presented  the consequences for pulsar timing for both  dual-frequency estimates of \DM\  and  wideband measurements using analytical and simulation methods, respectively.  

Here we present a more general analytical  treatment for arbitrary frequency coverage that extends the analytical approach in CSS16 for two spot frequencies. 
A multifrequency data set 
${\cal D} = \{t_\nu,  \nuL \le \nu \le \nuH\}$ comprising $\Nnu \gg 1$ frequency channels is analyzed here in the continuum approximation.  The total frequency ratio 
$R \equiv \nuH / \nuL$ can be much larger than unity for recently developed wideband receivers.  

To calculate the effects of $\DMbar(\nu)$  in isolation of other chromatic propagation effects, we adopt a minimal  model for \TOAs\ that includes  the infinite frequency (`true') \TOA\ , a dispersion term, and measurement error, 
\be
t_\nu = \tinfty + \KDM\, \nu^{-2} \DMbar(\nu) + \epsilon_\nu,
\ee
where the overbar denotes that DM is averaged over scattered propagation paths, 
$\KDM = c \relec / 2\pi$,  and additive noise from measurement errors  is  statistically independent between frequencies    and has variance $\sigma_\nu^2$.   

We estimate $\tinfty$ and \DM\ as parameters of  a fitting function,
\be
t_\nu = \tinftyhat  + \KDM\, \nu^{-2} \DMhat ,
\ee
using a cost function $C = \int d\nu\, \wnu (t_\nu - \tnuhat)^2$ with weights\footnote{For simplicity we assume a diagonal covariance matrix for the errors $\epsilon_\nu$.  It is trivial, but less useful analytically, to extend the analysis to an arbitrary covariance matrix.}  $\wnu \propto \sigma_\nu^{-2}$. 
Unless stated otherwise, integrals are over the entire frequency interval $[\nuL, \nuH]$. 
The resulting least squares solutions are
\be
\tinftyhat &=& 
	t_\infty + \frac{\KDM}{\Delta} (j_4 l_2 - j_2 l_4) 
\nonumber \\
\DMhat &=&
	\Delta^{-1} (j_0 l_4 - j_2 l_2), 
\ee
where we have defined
\be
j_n &=& \int d\nu\, \wnu \, \nu^{-n} ,
\\
l_n &=& \int d\nu\, \wnu \, \nu^{-n} \DMbar(\nu) ,
\\
\Delta &=& j_0 j_4 - j_2^2.
\ee
To assess the extra variance that results from DM chromaticity  
we reference  $\DMbar(\nu)$ to the highest frequency, $\nuH$,
\be
\DMbar(\nu) = \DMbar(\nuH) 
		    + \left[ \DMbar(\nu) - \DMbar(\nuH) \right] 
		    \equiv \DMbar(\nuH) + \Delta\DMbar(\nu, \nuH). 
\ee
Parameter estimates are then
\be
\tinftyhat &=& 
	t_\infty + \frac{\KDM}{\Delta} 
	\int d\nu\, \wnu\, \nu^{-2} \left[j_4  - j_2 \nu^{-2}\right]
	\Delta \DM(\nu, \nuH) , 
\\
\DMhat &=& 
	\DM(\nuH) +  \Delta^{-1} 
	\int d\nu\, \wnu\, \nu^{-2} \left[ j_0 \nu^{-2}- j_2\right]
	\Delta \DM(\nu, \nuH).
\ee

\subsection{Errors in \TOA\ and DM Estimates}

To express results in terms of frequency ratios, we make a change of variable to $r = \nuH / \nu$ to obtain errors in $\tinfty$ and $\DMhat$ due solely to the frequency dependence of $\DMbar$, 
\be
\delta\tinftyhat &=&
\frac{\KDM}{\nuH \Delta} 
\int_1^R dr\, W_r  \left[ j_4 - j_2 (r / \nuH)^2 \right]   
\Delta \DM(\nuH / r, \nuH)
\\
\delta\DMhat &=& 
(\nuH \Delta)^{-1} 
\int_1^R dr\, W_r  \left[ j_0 (r / \nuH)^2 - j_2\right]   
\Delta \DM(\nuH / r, \nuH).
\ee

We apply these results to  two cases, one with constant weights $w_\nu = 1$ over the frequency range $[\nuL, \nuH]$ with ratio $R = \nuH / \nuL$  and another where $\nuL$ and $\nuH$ are  two spot frequencies.    The weights for the two-frequency case are
\be
w_\nu = (1/2) [ \delta(\nu-\nuL) + \delta(\nu-\nuH)], 
\ee
which, for dimensionless frequencies,  corresponds to 
 \be
 w_r = (r / 2) [ \nuH^{-1} \delta(r-1) + \nuL^{-1} \delta(r-R)] 
 = 
 \frac{r}{2\nuH} [ R \delta(r-R) + \delta(r-1) ].
 \ee
 
 The  errors in $\tinftyhat$ from the frequency dependence of $\DMbar$ for the two cases are
 \be
\delta\tinftyhat &=&
\frac{\KDM}{\nuH^2} \times
\left\{
\begin{array}{ll}
	\displaystyle	 
	\frac{R(R^3-1)}{(R-1)^4} 
	\int_1^R dr\, \left[1 - \frac{3(R-1) r^2}{(R^3-1)}  \right]   
	\Delta \DM(\nuH / r, \nuH) ,
	& \quad\quad w_\nu = \text{constant} 
	\\
	\\
	\displaystyle
	- \left( \frac{R^2}{R^2-1}  \right)
	\Delta \DM(\nuL, \nuH) ,
	& \quad\quad \text{two frequencies}.
\end{array}
\right.
\ee

The corresponding errors in the estimated DM are
\be
\delta\DMhat &=& 
\left\{
\begin{array}{ll}
	\displaystyle	 
	\frac{3R}{(R-1)^3} 
	\int_1^R dr\, \left(\frac{r^2}{R} - 1   \right)   
	\Delta \DM(\nuH / r, \nuH) ,
	& \quad\quad w_\nu = \text{constant} 
	\\
	\\
	\displaystyle
	\left(   \frac{R^2}{R^2-1}  \right)
	\Delta \DM(\nuL, \nuH) ,
	& \quad\quad \text{two frequencies}.
\end{array}
\right.
\ee

\subsection{Variances}

The variances of $\dtinftyhat$ and $\delta \DMhat$ are given by the ensemble averages of 
$(\delta\tinftyhat)^2$ and $(\delta\DMhat)^2$.   These involve the cross correlation function
\be
\GammaDDM(\nua, \nub, \nuc; \beta)  
	\equiv  
	\left\langle \Delta \DM(\nua, \nuc) \Delta \DM(\nub, \nuc) \right\rangle, 
\ee
which is evaluated in \S \ref{app:derivationGamma} along with the quantity
$\Phi_\beta(\ra, \rb)$,
\be
\GammaDDM(\nua, \nub, \nuc; \beta) =
	\Phi^2_\beta(r_1, r_2) 
	\JDM(\nuH, \beta)
\ee
where $\JDM$, evaluated later in this section, integrates the scattering medium over the bundle of ray paths, 
\be
\JDM(\nuH, \beta) = 
\frac{(2\pi)^2 \Gamma(2-\beta/2) }{\beta-2}
	\int d\zbar\, \cnsq(\zbar) \,\sigma_X^{\beta-2}\zbar, \nuH). 
\label{eq:JDM}
\ee

The variances of the estimates for arrival time and DM are then
\be
\Var(\dtinftyhat) 
&=&
\left( \frac{\KDM}{\nuH^2}\right)^2 
\JDM(\nuH, \beta ) \,\Itinfty(R, \beta)
\label{appeq:Var_tinfty}
\\
\nonumber \\
\Var(\dDMhat) 
&=&
\JDM(\nuH, \beta ) \, \IDMhat(R, \beta).
\label{appeq:Var_DMhat}
\ee
The dimensionless integrals $\Itinfty$ and $\IDMhat$ over dimensionless frequency ratios  $r_{1,2} = \nuH / \nu_{1,2}$ for the wideband case with $w_\nu$ = constant are given by
\be
\Itinfty(R, \beta) 
&=& 
		\displaystyle
		\left[  \frac{R(R^3-1)}{(R-1)^4}  \right]^2
		\int_1^R \!\!\!\!\int_1^R
		d\ra d\rb\, 
		\left[ 1 - \frac{3(R-1)\ra^2}{(R^3-1)} \right]
		\left[ 1 - \frac{3(R-1)\rb^2}{(R^3-1)} \right]
		\Phi^2_\beta(\ra, \rb),
\label{eq:Itinfty}
\ee
and
\be
\IDMhat(R, \beta) 
&=& 
		\displaystyle
		\left[  \frac{3R}{(R-1)^3}  \right]^2
		\int_1^R \!\!\!\!\int_1^R
		d\ra d\rb\, 
		\left(\ra^2 / R - 1 \right)
		\left(\rb^2 / R - 1 \right)
		\Phi^2_\beta(\ra, \rb),  
\label{eq:IDMhat}
\ee
For two equally weighted spot frequencies, these are 
\be
\Itinfty(R, \beta)  = \IDMhat(R, \beta)  = 
		\displaystyle
		\left[
		\left( \frac{R^2}{R^2-1}  \right)
		F_\beta(R)
		\right]^2 ,
\label{eq:Itinfty_DM_hat_twofreq}
\ee
which is obtained by using the identify $\Phi(x, x) = F_\beta(x)$, where $F_\beta$, first defined in CSS16,
is given in Eq.~\eqref{appeq:Fbeta}.
These integrals are shown in Fig.~\ref{fig:Iintegrals} for three values of $\beta$ and 
for both the wideband and two-frequency cases. 

The  quantity   $J_{\rm DM} \propto \int d\zbar\, \cnsq(\zbar) \,\sigma_X^{\beta-2}(\zbar, \nuH)$  
integrates  $\cnsq$ over the  bundle of ray paths that reach the observer. Because
$\sigma_X$ also depends on $\cnsq$, the integral generally depends on the square of the scattering measure,
$J_{\rm DM} \propto \SM^2$, but it also depends on the detailed distribution of
$\cnsq(\zbar)$ in both factors comprising the integrand.  

We simplify the expression for $J_{\rm DM}$ by first expressing $\sigma_X$ in a form that isolates quantities that depend on the distribution of $\cnsq$ along the \LoS\, 
\be
\sigma_X(s, \nuH)  = \dso \sigthetad \left(\frac{\SMeff}{\SM} \right)^{1/{(\beta-2)}} h(s) ,
\ee
where 
$\dso$ is the source-observer distance (as used elsewhere), 
$\sigthetad$ is the \RMS\ scattering angle produced by a medium with plane wave incidence,
$\SMeff$ is the effective scattering measure that takes into account the geometry of the scattering medium (thin screen, thick slab, uniform, etc.), and 
$h(s)$ is a dimensionless quantity that defines the extent of the raypath bundle and is also geometry dependent.

The \RMS\ scattering angle is 
\be
\sigthetad 
= \frac{\lambda }{\sqrt{2} \pi}  
	\left[  \frac{f_\beta (\lambda \relec)^2 \SM }{2}\right]^{1/(\beta-2)}
= \frac{1}{\pi}  
	\left[  \frac{c^{\beta}  \relec^2 f_\beta   \SM }{2^{\beta/2} \nu^{\beta}}\right]^{1/(\beta-2)}
\ee
where $f_\beta$ is given in Eq.~\eqref{eq:fbetadef}. 
The effective scattering measure is given by
\be
\frac{\SMeff}{\SM} =  \frac{\displaystyle  \int ds\, \cnsq(s) (s / \dso)^{\beta-2} }  {\displaystyle  \int ds\, \cnsq(s) }
\ee
and has representative values,
\be
\frac{\SMeff}{\SM} = 
	\left \{
	\begin{array}{ll}
		1 							& \quad\quad \text{Plane wave incidence} \\
		\\
		(\dsl / \dso)^{\beta-2} 			& \quad\quad \text{Thin screen} \\
		\\
		(\beta-1)^{-1}					& \quad\quad \text{Uniform medium},
	\end{array}
	\right.
\ee
where we have used $\cnsq(s) = \SM \delta(s - \dsl)$ for a thin screen at a source-screen distance $\dsl$.

The ray bundle shape $h(s)$ generally requires a numerical solution.   For a thin screen it is
\be
h(s) =  
	\left\{
		\begin{array}{ll}
		\displaystyle
			\frac{s}{\dsl} \left(1 - \frac{s}{\dso} \right)		& \quad\quad s \le \dsl, \\
			\\
			\displaystyle
			1 - \frac{s}{\dso}		& \quad\quad s \ge \dsl, 
		\end{array}
	\right.
\ee
which has a value $h(\dsl) = \dlo / \dso$ at the screen itself. 
For a uniform medium it is 
\be
\displaystyle 
h(s) =  \frac{\dsl}{\dso}     \left( 1 - \frac{\dsl}{\dso}  \right). 
\ee

The ray bundle shape is integrated along the line of sight to yield another dimensionless quantity, 
\be
H_\beta = \frac{\displaystyle  \int ds\, \cnsq(s) [h(s)]^{\beta-2}}  {\displaystyle  \int ds\, \cnsq(s) },
\ee
which has values of $(\dlo / \dso)^{\beta-2}$ for a thin screen and 
$H_\beta = \int_0^1 ds\, [s (1-s)]^{\beta-2} = 0.0564$ for a uniform medium with $\beta=11/3$. 

Defining, as in CSS16,
\be
G_\beta^2 &=& H_\beta \SMeff / \SM
\\
Q_\beta^2 &=& 
	\frac{(\sqrt{2}\pi)^{4-\beta} \Gamma(2-\beta/2) f_\beta }{\beta-2},
\ee
where $Q_\beta$ and $G_\beta$ depend on the spectral index of the density spectrum while $G_\beta$ also
depends on the specific distribution of $\cnsq$ along the \LoS. 
We then have
\be
\JDM(\nuH, \beta) = 
(G_\beta Q_\beta)^2 c^\beta \relec^2 \nuH^{-\beta} \dso^{\!\!\!\!\beta-2} \, \SM^2
 \label{eq:JDM2}
\ee

The final expressions for the variances are given by Eq.~\eqref{appeq:Var_tinfty}--\eqref{appeq:Var_DMhat} using the above defined quantities.    Evaluating for a Kolmogorov index $\beta = 11/3$ with a nominal scattering measure
$\SM = 10^{-3.5}\, \SMunits = 4.53\times10^4\, {\rm cm^{-17/3}}$, we obtain
$Q_\beta^2 = 485$ and 
\be
\sigma_{\dtinfty} &=& [\Var(\dtinfty)]^{1/2}   = 156\,{\rm ns}\, \times G_\beta\, \nuH^{-23/6} \dso^{\!\!\!5/6}\, \SM_{-3.5} \Itinfty(R, \beta)
\\
\sigma_{\delta \DMhat} &=& [\Var(\delta \DMhat)]^{1/2}  = 3.76\times 10^{-5}\,\DMunits \times 
	G_\beta \nuH^{-11/6} \dso^{\!\!5/6} \, \SM_{-3.5} \IDMhat(R, \beta),
\ee
where $\SM_{-3.5} = \SM / (10^{-3.5}\,\SMunits)$.
We note that the dependence on the upper frequency $\nuH$ for the excess DM variation is
$\sigma_{\delta \DMhat} =  \propto \nuH^{-11/6}$ for a
Kolmogorov spectrum (inertial subrange) while the \ToA\ variation has a much stronger dependence, 
$\sigma_{\dtinfty} = \propto \nuH^{-23/6}$.   The actual \ToA\ error also depends on the frequency ratio $R$ and can be much smaller than the nominal values for a narrowband system, particularly if the upper frequency $\nuH$ is well above 1\,GHz.

\begin{figure}[htbp] 
   \centering
   \includegraphics[width=0.475\linewidth]{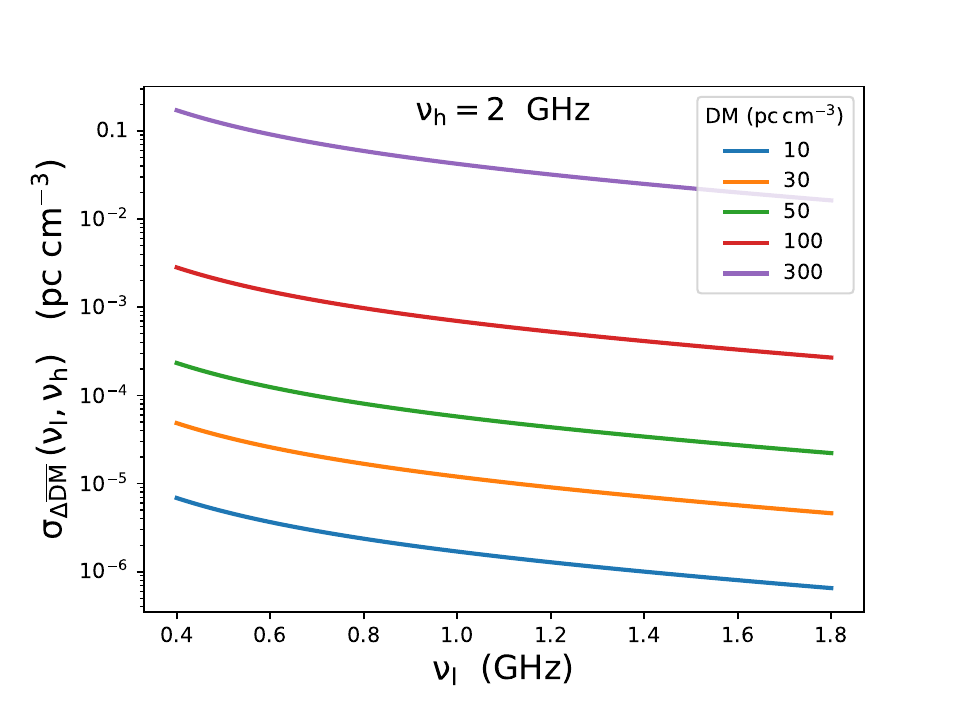}
    \includegraphics[width=0.475\linewidth]{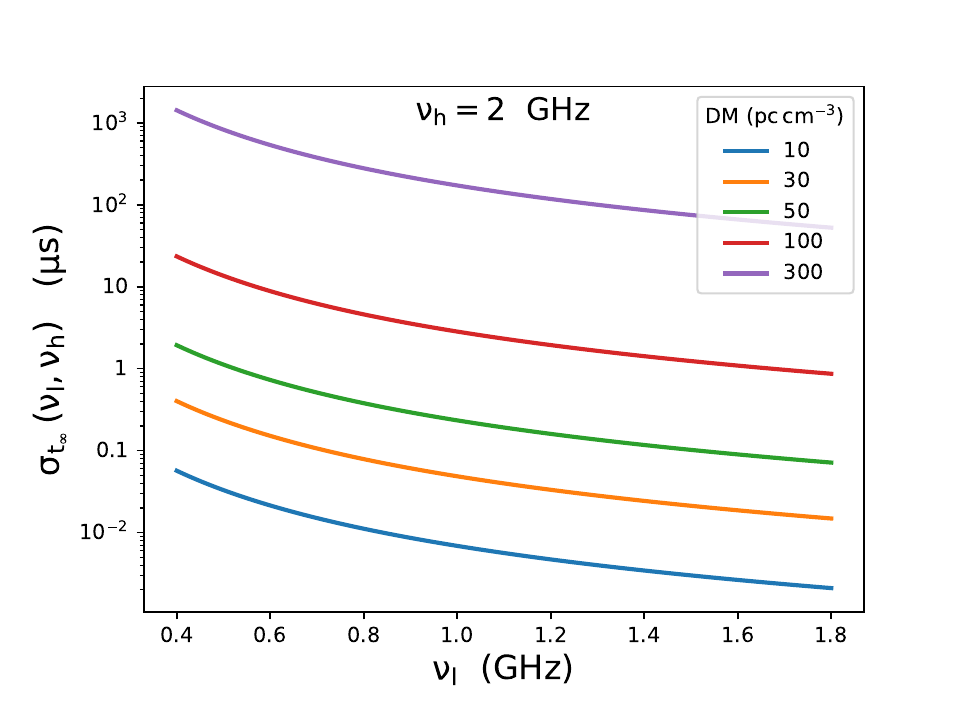} 
   \caption{Left: Plot of the \RMS\ excess DM variation due to chromatic multipath propagation for different values of 
   the total DM that are used to calculate the total DM.
   Right: Plot of the \RMS\  variation in $\tinfty$  due to chromatic multipath propagation for different values of 
   the total DM that are used to calculate the total DM.   
   }
   \label{fig:rmsDMhat}
   \label{fig:rmstinfty}
\end{figure}

\subsection{Derivation of the cross correlation $\GammaDDM(\nua, \nub, \nuc) $}
\label{app:derivationGamma}

The cross correlation  is expanded into the sum of four autocorrelation functions,
\be
\GammaDDM(\nua, \nub, \nuc)  
= 
\Cdmbar(\nua, \nub) + \Cdmbar(\nuc, \nuc) - \Cdmbar(\nub, \nuc) - \Cdmbar(\nua, \nuc),
\label{eq:GammaDDM}
\ee
where the cross correlation of $\delta\DMbar = \DMbar - \left\langle \DM \right\rangle$ between frequency pairs is
\be
C_{\delta \DMbar}(\nua, \nub) &=& \left\langle \delta\DMbar(\nua, \xvec) \delta\DMbar(\nub, \xvec) \right\rangle
\nonumber \\
    &=&
     \iint \! d\xvecp d\xvecpp
     \iint \! d\zp d\zpp\,
    \Anuxpz \Anuxppz
    A_{\nua}(\xvecp) A_{\nub}(\xvecpp) 
    \langle \delta n_e(\xvecp, \zp) \delta n_e (\xvecpp, \zpp)\rangle.
\ee

The $z$ integrals are from $0$ to $d$ and the $\xvec$ integrals are over an infinite plane\footnote{Extension over an infinite plane will not apply to bounded media, as discussed by \cite{2001ApJ...549..997C}.   A bounded medium will alter the frequency dependence of any multipath propagation.}.  
To proceed, a change of variables is made from $\xvecpr, \xvecpp$ to
$\xbarvec = (\xvecpr + \xvecpp)/2$ and $\dxvec = \xvecpr - \xvecpp$
and from $\zp, \zpp$ to $\zbar = (\zp+\zpp)/2$ and $\dz = \zp-\zpp$.
The resulting integral over $\xbar$ yields  the cross-correlation function 
for the averaging function, 
\be
C_A(\dxvec, \delta z, \zbar, \nua, \nub) =  \int d\xbarvec \, 
	A_{\nua}(\xbar+\dxvec/2, \zbar+\dz) 
	A_{\nub}(\xbar-\dxvec/2, \zbar-\dz),
\ee
which we assume  changes slowly in $\zbar$ compared to density variations,
and thus set $\delta z = 0$ in the argument and rewrite simply as
$C_A(\dxvec,  \zbar, \nua, \nub)$. 

\begin{figure}[h!] 
   \centering
   \includegraphics[width=0.6\linewidth]{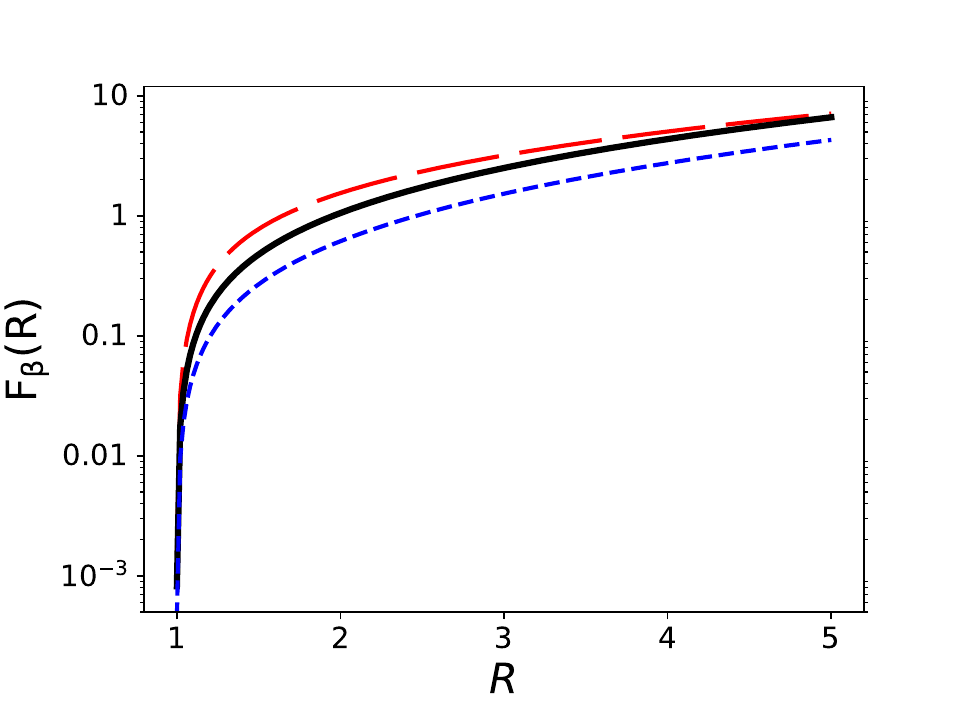} 
   \caption{Plot of $F_\beta(R)$ vs. frequency ratio $R = \nuH / \nuL$ for several values of $\beta$.   This function is applicable to the
   two-frequency case as presented in CSS16.}
   \label{fig:Fbeta}
\end{figure}

The density correlation function is written in terms of the wavenumber
spectrum,
\be
\langle \delta n_e(\xbarvec+\dxvec, \zbar+\dz) \delta n_e (\xbarvec-\dxvec, \zbar-\dz)\rangle
=
\int d\qvec\, e^{i\qperpvec\cdot\dxvec + q_z \dz} \Pne(\qvec, \zbar).
\ee
The integration over $\dz$ gives $\dunit \delta(q_z)$ and the integration 
over $d\qperpvec$ yields the Fourier transform\footnote{Note that the convention used for Fourier transforms  is $\widetilde g(q) = (2\pi)^{-1} \int dx\, e^{-iqx} g(x)$.  
Thus a factor $(2\pi)^{-1}$ is introduced for each dimension of a Fourier transform.}
 of  $C_A(\dxvec,  \zbar, \nua, \nub)$,

We adopt a Gaussian smoothing function with \RMS\ spatial scale $\sigma_X$ for the bundle
of rays comprising the scattering cone, 
\be
\Anuxz = \left[2\pi \sigma_X^2(\nu,z)\right]^{-1} e^{-x^2/2\sigma_X^2(\nu,z)},
\label{eq:smooth}
\ee
that yields a Fourier transform for the  cross-correlation function also Gaussian in form,
\be
\widetilde{C}_A(\qperpvec, \nua, \nub, \zbar) = 
(2\pi)^{-2} e^{(\qperp^2/2) [\sigma_X^2(\nua, \zbar) + \sigma_X^2(\nub, \zbar]}.
\ee
The two-frequency cross correlation becomes
\be
\Cdmbar(\nua, \nub) = 
	2\pi \int d\zbar \int d\qperpvec \, \Pne(\qperpvec, 0, \zbar)
	e^{-(\qperp^2/2) [\sigma^2_X(\nua, \zbar) + \sigma^2_X(\nub, \zbar)]}.
\ee
Using the power-law form for $\Pne(\qvec)$ 
for $2 < \beta < 4$
with an exponential rolloff  from the inner scale and using 
$d\qperpvec = 2\pi \qperp d\qperp$, 
we require an integral of the form
\be
\int_{y_0}^{\infty} dy\, y^{1-\beta} e^{-ay^2} 
= \int_0^{\infty} dy\, y^{1-\beta} e^{-ay^2} - \int_0^{y_0} dy\, y^{1-\beta} e^{-ay^2} .
\ee
We make use of the integral (GR 3.478.1),
\be
\int_0^{\infty} dy\, x^{v-1} e^{-u x^p} = p^{-1} u^{-v/p} \Gamma(v/p), 
\ee
which holds for $u, v, p > 0$.   Thus to apply it for $2 < \beta < 4$ we integrate by parts to obtain
\be
I 
= 	
	\int_{\qouter}^{\infty}  d\qperp\,  \qperp^{1-\beta} e^{-(\qperp /\qinner)^2}
= 
	\frac{1}{(\beta-2)}
	\left[  
	\qouter^{2-\beta} e^{-a \qouter^2} + \frac{2a \qouter^{4-\beta}}{4-\beta} 
	- a^{(\beta-2)/2} \Gamma(2-\beta/2)
	\right] ,
\ee
with $a = [\sigma^2_X(\nua, \zbar) + \sigma^2_X(\nub, \zbar)]/2 + 1 / \qinner^2$.
This is further simplified by taking into account that  the outer scale is much larger than  the width of the scattering cone $\sim \sigma_X$ at relevant radio  frequencies which, in turn, is much larger than the inner scale.    
This impies
 $a\qouter^2 \lll 1 \lll (\qouter/\qinner)^2$ and
\be
I  \approx 
	\frac{1}{(\beta-2)}
	\left[  
	\qouter^{2-\beta}  + \frac{2a \qouter^{4-\beta}}{4-\beta} 
	- a^{(\beta-2)/2} \Gamma(2-\beta/2))
	\right]. 
\ee
We then have
\be
\Cdmbar(\nul, \nuh) \approx 
	\frac{(2\pi)^2 }{\beta-2}
	\int d\zbar\, \cnsq(\zbar) 
	\left[  
	\qouter^{2-\beta}  + \frac{2a \qouter^{4-\beta}}{4-\beta} 
	- a^{(\beta-2)/2} \Gamma(2-\beta/2)
	\right]. 
\label{eq:CdDMbar}
\ee
The first term in square brackets is independent of frequency so its contribution  vanishes when substituted into Eq.~\eqref{eq:GammaDDM}.  The second term is linear in
the $\sigma^2_X$ terms at the two frequencies and also cancels.    Using the third term and substituting for the 
relevant values of $a$ we obtain
\be
\GammaDDM(\nua, \nub, \nuc)  
	= 
	\frac{(2\pi)^2 \Gamma(2-\beta/2) }{\beta-2}
	\Phi_\beta^2(\ra, \rb) 
	\int d\zbar\, \cnsq(\zbar) \,\sigma_X^{\beta-2}(\zbar, \nuc) \,
\ee
where the frequency ratios $\ra = \nuc / \nua$ and $\rb = \nuc / \nub$ are used in the dimensionless function, 
\be
\Phi_\beta^2(\ra, \rb) 
&=&
2^{-(\beta-2)/2} 
	[ \Vbeta(\ra) + \Vbeta(\rb) - \rb^{\beta} \Vbeta(\ra/\rb)  ] - 1
\ee
with
\be
\Vbeta(x) &\equiv& [1 + x^{2\beta/(\beta-2)}]^{(\beta-2)/2}.
\ee
We note that for equal values of the arguments,
\be
\Phi_\beta(r, r) = F_\beta(r) ,
\ee
where $F_\beta(r)$ is the function defined in CSS16 for the case where a pair of frequencies is used 
to calculate $\DMhat$ and $\tinftyhat$ rather than the wideband continuum range of frequencies considered here,
\be
F_\beta(r) &=&
	\left\{
	2^{(4 - \beta)/2} 
		\left[1 + r^{2\beta / (\beta-2)}   \right]^{(\beta-2)/2} - r^{\beta} -1 
	\right\}^{1/2} .
\label{appeq:Fbeta}
\ee
This quantity is shown in Fig.~\ref{fig:Fbeta} for three values of $\beta$.

\clearpage
\section{Simulation of propagation through thin plasma screens}
\label{app:sims}

\subsection{Diffraction and lensing from a thin screen}
\label{app:diffsims}

To create Fig.~\ref{fig:kdiplasmalens}, we evaluated the wavefield at an observer location $(\xvec,d)$ using a
 two dimensional Fresnel-Kirchhoff diffraction integral for a thin screen that imposes a phase  $\phi_{\rm s}(x)$ on an incident plane wave, 
\be
E(\xvec) = \int d\xvecp\, K(\xvec-\xvecp) e^{i\phis(\xvecp)}, 
\label{eq:kdi}
\ee
where $K(\xvec) = (2\pi i \rF^2  )^{-1} e^{i \vert\xvec-\xvecp\vert^2 / 2\rF^2}$ is the propagation kernel and the Fresnel scale is 
$\rF = \sqrt{\lambda d / 2\pi}$.   The screen phase includes a Gaussian lens 
$\phi_g(\xvec) \propto \exp(-\vert\xvec\vert^2 / a^2)$ combined with a stochastic phase  $\phi_{\rm K}(\xvec)$.
The latter is  drawn from a Kolmogorov medium with spectrum of the form of Eq.~\eqref{eq:Pne} with an inner scale 
$\linner = 2\pi / \qinner$ imposed as an exponential factor $\exp[-(q / \qinner)^2]$
and outer scale determined by the size of the spatial grid used to evaluate $E(\xvec)$.  A realization of $\phi_{\rm K}$ is calculated by using the average phase spectrum 
$S_{\phi}(\qperpvec) = 2\pi(\lambda\relec)^2 \SM\,\exp[-(q / \qinner)^2] \qperp^{-\beta}$ to shape
 complex white noise ${\cal N}(\qperpvec)$,
 \be
 \phi_{\rm K}(\qperpvec) \sim S^{1/2}_{\phi}(\qperpvec) {\cal N}(\qperpvec).
 \ee 
 The convolution in Eq.~\eqref{eq:kdi} is executed by taking the inverse FT of the product of FTs of the kernel and the screen phase factor. 

\subsection{Refractive distortions of scattered Images and PBFs}
\label{app:refsims}

We outline  simulations described in the main text
 that demonstrate the effects of stochastic refraction screens on observable quantities. 
For simplicity, we employ a circularly symmetric  scattered image shape in the absence of refraction. 
Non-Gaussian images like those from a Kolmogorov medium and images caused by asymmetric scattering could be used instead, though the essential features of the examples shown here would not be altered significantly.  
 We thus use an unrefracted scattered image seen by an observer  with \RMS\ angles $\sigma_x = \sigma_y  = \sigmao$.
 The corresponding \RMS\ diffraction angle is $\sigmad = (\dso / \dsl) \sigmao$.   For a source much further from the observer than the screen, $\dso / \dsl \to 1$ and $\sigmad \to \sigmao$. 
 
We calculate the scattering parameters of a thin screen by  specifying the DM for the \LOS\ and 
inferring the scattering strength  from the $\taud(\DM)$ relation for Galactic pulsars (the `hockey stick' relation in Eq.~\eqref{eq:taudm}).   After  scaling from 1\,GHz to the frequency of interest,  the diffraction and refraction scales $\ld$ and $\lr$ are calculated along with the value of the phase structure function across a Fresnel scale.  Using the electromagnetic wavenumber $k = 2\pi/\lambda$ and Fresnel scale
$\rF = \sqrt{\deff / k} = \sqrt{\lambda \deff / 2\pi}$, 
\be
\ld &=& \frac{1}{k \sigmad} 
	= \frac{1}{k} \left(\frac{\deff}{c \taud}  \right)^{1/2} 
	= \frac{1}{2\pi\nu} \left(\frac{c \deff}{\taud}  \right)^{1/2}
	= \frac{\rF}{(2\pi\nu\taud)^{1/2} } 
	\equiv \frac{\rF}{u}
\\
\lr & = & \dlo \sigmao 
	= \deff \sigmad 
	=  \frac{\rF^2}{\ld} 
	\equiv u \rF,
\ee
where another quantity $u$ is also used to relate the Fresnel scale to the diffraction and refraction scales.
For a Gaussian image, the scattering time and scintillation bandwidth $\dnud$ are related as
\be
2\pi \taud \dnud = 1,
\ee  
implying that
\be
u = \sqrt{2\pi\taud} =  \left(\frac{\nu}{\dnud}\right)^{1/2}.
\ee

By definition, $\Dphi(\ld) = 1\,$rad$^2$ and we use a scaling of the phase structure function with baseline $\bvec$ 
 for a medium with an isotropic  power-law wavenumber spectrum with index $\beta$ and assume that $b$ is intermediate between the inner and outer scales (the intertial subrange), 
 \be
 \Dphi(\bvec) = (b / \ld)^{\beta-2} \equiv \phiF^2 (b / \rF)^{\beta-2}, \quad \linner \ll b \ll \louter, 
\ee
where the  second equality defines the \RMS\ phase  between a pair of points in the screen separated by one Fresnel scale, 
\be
\phiF = (\rF / \ld)^{(\beta-2)/2} = u^{(\beta-2)/2} .
\ee

 Since we specify the unrefracted scattered image using the parameters given above, we do not need to simulate the fine spatial scales in the screen that would cause scattering.  Instead, we simulate only those scales $l \gtrsim \lr$ that cause refraction by drawing realizations that conform to the wavenumber spectrum with wavenumbers
 $q \lesssim 2\pi / \lr$.   
 
 A refraction screen is generated by shaping complex Gaussian noise in the frequency domain with a shape function related to the  target wavenumber spectrum as $[\Pne(q)]^{1/2}$ with the required wavenumber cutoffs
 (cf.\ Eq.~\eqref{eq:Pne}).   To exclude any scales that would diffract rather then refract,  the upper wavenumber cutoff is set to $q_{\rm max} = 2\pi / \lr$.   The largest scale corresponds to half the screen size.  The shaped complex noise is inverse Fourier transformed to give $\phi_{\rm r}(\xvec)$ which is then normalized to give a phase structure function equal to $\phiF^2$ across a length of one Fresnel scale.   Enhanced refraction can be implemented by increasing $\phiF$ from the value needed to account for the diffraction.
 This approach can be generalized to anisotropic scattering by using a different wavenumber spectrum.


\clearpage
\section{Inventory of timing effects from dispersion, scattering, and refraction}
\label{app:notation_and_toa_estimation_errors}

This section itemizes and aggregates contributions to arrival time estimates when 
a model template is applied to a measured profile.    
The error from additive noise is excluded here except in our final expressions.
We assume that the  profile and template both  include an intrinsic pulse shape   convolved with a pulse broadening function.   The time argument $t$ extends across one pulse period (one cycle of phase), $\nubar$ is the center frequency of a channelized signal with narrow bandwidth $\Dnuch$,
and $\tbar$ is the epoch of observation.   The epoch corresponds to the mean time of an observation  block of duration $T$ used to construct an average profile and estimate a \TOA.   Multiple profiles obtained in the same observing session (e.g. subaverages obtained over $\sim 1$\,hr) have different epochs in this framework.  However, profiles from different frequency channels spanning a total bandwidth $B$  obtained during the same time block are considered to  have identical epochs\footnote{Typically the difference in arrival times due to dispersion delays  across a wide band $B$  are much smaller than the integration time $T$.   The main utility of the epoch variable here  is to provide a time argument for slowly changing processes from the ISM and, less prevalent, intrinsic changes in profile shape.}.   In the following, it is assumed that sufficient averaging has been  done over time and frequency (i.e. time-frequency product $\gg 1$) that we can ignore self-noise (as defined in Sects.~\ref{sec:pulse_shapes}, \ref{sec:selfnoise}) and ignore cross terms between radiometer noise and the pulsar signal. 

The  intrinsic (emitted) pulse $\Ui$, pulse broadening function $\PBF$, and  scattered pulse $\Us$ are  defined according to:
$$
\begin{array}{lllcl}
\text{\bf Actual pulse:} \\
\text{Quantity}	&     & \text{Terms}	& \text{Arguments} 	&  \text{Description}	\\
\Ui    				&  = & \Uiave  + \dUi 						& \quad (t, \nubar, \tbar) 	&  \text{Intrinsic pulse} \\
\PBF 			&  = & \PBFave{} + \dPBF 				& \quad (t, \nubar, \tbar) 	&  \text{Pulse broadening function} \\
\Us	:			& = & \Ui * \PBF = \Usave{} + \dUs 			& \quad (t, \nubar, \tbar) 	&  \text{Scattered pulse} \\
	\quad\quad\Usave	& = & \Uiave  * \PBFave{} 					& \quad (t, \nubar, \tbar) 	&  \text{Ensemble average scattered pulse } \\
	\quad\quad\dUs		& = & \dUi * \PBFave{} + \Uiave * \dPBF + \dUi * \dPBF					
													& \quad (t, \nubar, \tbar) 	&  \text{Deviation from ensemble average scattered pulse } \\
{\Usave}_{\tbar}		& = &  \Usave{} + \langle \dUs\rangle_{\tbar}	& \quad (t, \nubar) 		&  \text{Scattered pulse averaged over epochs}, 
\end{array}.
$$
where we use the notation here, as elsewhere,  $\langle \cdots \rangle_{x}$ to denote an average of a quantity over $x$. 
\\

The model pulse includes analogous elements,
$$
\begin{array}{lllcl}
\text{\bf Model pulse:} \\
\text{Quantity}	&     & \text{Terms}	& \text{Arguments} 	& \text{Description}	\\
\Um	:			& = & \Uiave + \dUm, \quad 	& \quad (t, \nubar) 							&  \text{Model intrinsic pulse} \\
	\quad\quad \dUm 	& = &  \Um-\Uiave							&								& \text{Model - actual} \\
\PBFm :			& = & \PBFave{} + \dPBFm				& \quad (t, \nubar) 			&  \text{model Pulse broadening function} \\
	\quad\quad \dPBFm	& = & \PBFm-\PBFave{}  					&								& \text{Model - actual} \\
\Ums	 :		& = & \Um * \PBFm = \Usave + \dUms		& \quad (t, \nubar) 			&  \text{Model scattered pulse} \\
	\quad\quad \dUms     & = & \Ums - \Usave						& \quad (t, \nubar) 			& \text{Model -  actual average scattered pulse} \\
				& = &  \multicolumn{2}{l}{\dUm * \PBFave{} + \Uiave * \dPBFm + \dUm * \dPBFm} 		& \text{Deviation of model scattered pulse from}	 \\
				&    &   &																& \quad \text{ensemble average scattered pulse} \\
\end{array}
$$

The scattered pulse $\Us$ is fitted with the model by  cross correlating it with the model scattered pulse, $\Ums$, and finding the time lag of maximum correlation.    This is equivalent to least-squares fitting in the Fourier domain, 
as described in the main text.   The cross correlation function (CCF) includes four terms (where we use the same asterisk notation for both cross-correlation and convolution; context should help avoid any confusion between the two operations), 
$$
\begin{array}{lllcl}
{\text{\bf Template fit:}} & &  \Ums = \text{template} \quad \Us = \text{profile}  & \multicolumn{2}{l}{\text{\quad CCF} = \langle \Ums(t + \tau) \Us(t-t_0) \rangle_{t} \to \Ums * \Us} \\
\Ums * \Us 		& = & \multicolumn{3}{l}{\Usave * \Usave + \Usave * \dUs +  \dUms * \Usave + \dUms * \dUs} \\
\end{array}
$$

The scattered profile  includes terms that are stochastic due to  pulse jitter and from PBF variations (rapid variations  from the finite scintle effect and slow variations due to refraction). The template includes a term from  mismatch between its shape and that of the measured profile, which leads to a systematic \TOA\ error.   

The properties of fluctuating terms are:
$$
\begin{array}{lllll}
\text{\bf Quantity}	& \quad	& \text{\bf Cause}						& \quad \text{\bf Change rate}	& \text{\bf Correlation scales} \\
\dUi				& \quad 	& \text{Jitter}  							& \quad \text{F = Fast}  			& \text{High correlation  across}\ \nubar \\
\dPBF:			& \quad 	& \text{Finite scintle number} 				& \quad \text{F}			& \text{DISS} \\
				& \quad 	& \text{Refraction}						& \quad \text{S = Slow}			& \text{RISS} \\
\dUs				& \quad 	& \text{Jitter + finite scintles + refraction} 		& \quad \text{F + S} 		& \text{Jitter + DISS + RISS} \\
\dUm 			& \quad 	& \Um-\Uiave:  \ \text{shape mismatch} 	& \quad \text{None (S)} 		& \text{Systematic error} \\
\dPBFm 			& \quad 	& \PBFm-\PBFave{}:  \ \text{shape mismatch} 	& \quad \text{None (S)} 	& \text{Systematic error} \\
\dUms  			& \quad 	& \text{Combined}\ \dUm, \dPBFm:  \ \text{shape mismatches} \, 			
															& \quad \text{None (S)} 	& \text{Systematic error} \\
\end{array}		
$$
\\
\\

\centerline{\bf Terms in the CCF,  fluctuation properties, \TOA\ contribution} 

The CCF of template and profile is an average over time $t$, 
\be
\widehat{C}(\tau) = \langle \Ums(t+\tau)  * \Us(t-t_0) \rangle_{t} = \left( \Usave + \dUms) \right) * \left( \Usave + \dUs  \right),
\ee
where we drop the arguments in the rightmost equality.   Note that $\dUms$ and $\dUs$ each comprise three terms.  
The CCF expands into 16 terms  that  yield a total of 21 terms in the \TOA,  one equal to the  true \TOA\  (in the absence of additive noise) 
from the cross correlation of $\Usave$ in the template model with $\Usave$ in the profile, 
and the remaining 20 from  pulse stochasticity (jitter and scattering)  and shape mismatches between the template and profile. 
These 21 terms are itemized as,
{  
$$
\begin{array}{lllll}
\\
\text{\bf Term }			& \quad		& \text{\bf Cause}							& \quad \text{\bf Term in \TOA}	&  \text{\bf Comment} \\
\Usave * \Usave 		& \quad 		& \text{ACF of $\Usave$ with argument $\tau - t_0$} 	& \quad t_0(\nubar, \tbar) 		&  \text{True \TOA}  \\
\dUms * \Usave 		& \quad 		& \text{Mismatch of intrinsic profile}\ (\dUm)		& \quad \DtUUm			&  \text{Systematic error} \\
					& \quad 		& \text{Mismatch of PBFs}\,(\dPBFm) 			& \quad \DtPBFPBFm		&  \text{Systematic error} \\
					& \quad 		& \text{Combined mismatches}\, (\dUm, \dPBFm) 	& \quad \DtdUdPBF			&  \text{Systematic error} \\	
\Usave * \dUs			& \quad 		& \text{Jitter} 								& \quad \DtJ 				&  \quad \text{F} \\
					& \quad 		& \text{DISS finite scintles } 					& \quad \DtdPBF\equiv \Delta t_{\dPBF, \rm FSE} 			&  \quad \text{F}  \\
					& \quad 		& \text{Refraction} 							& \quad \DtdPBFbar \equiv \Delta t_{\dPBF, \rm Refraction}  			&  \quad \text{S} \\
\dUms * \dUs  = 			& \quad 		& \text{Combined model and actual terms}					\\
\quad (\dUm*\PBFave{}) * (\dUi * \PBFave) 	& \quad	& \text{$U$ mismatch + jitter}							& \quad \DtdUmdUi							& \quad F 	\\
\quad (\dUm*\PBFave{}) * (\Uiave * \dPBF) 	& \quad	& \text{$U$ mismatch}, \, \dPBF, \, \dPBFbar				& \quad \DtdUmdPBF 						& \quad F  	\\
									&		&												& \quad \DtdUmdPBFbar						& \quad S	\\
\quad (\dUm*\PBFave{}) * (\dUi * \dPBF) 		& \quad	& \text{$U$ mismatch, jitter},  \, \dPBF, \,\dPBFbar			& \quad \DtdUmddPBF, 						& \quad F   \\
									&		&												& \quad \DtdUmddPBFbar						& \quad F	\\
\quad (\Uiave *\dPBFm) * (\dUi * \PBFave) 	& \quad	& \text{$\PBF$ mismatch,  jitter}						& \quad \DtdPBFmdUi 						&   \quad F 	\\ 				
\quad (\Uiave *\dPBFm) * (\Uiave * \dPBF) 	& \quad	& \text{$\PBF$ mismatch}, \, \dPBF, \, \dPBFbar			& \quad \DtdPBFmdPBF   						&  \quad F 	\\
									&		&												& \quad \DtdPBFmdPBFbar 					&  \quad S	\\
\quad (\Uiave *\dPBFm) * (\dUi * \dPBF) 		& \quad	& \text{$\PBF$ mismatch + jitter}, \, \dPBF, \, \dPBFbar		& \quad \DtdPBFmdUidPBF					&  \quad F  \\

\quad (\dUm*\dPBFm * (\dUi * \PBFave) 		& \quad	& \text{$U, \PBF$ mismatches, jitter}						& \quad \DtdUmdUi 							&  \quad F 	\\
									
\quad (\dUm*\dPBFm * (\Uiave * \dPBF) 		& \quad	& \text{$U, \PBF$ mismatches}, \dPBF, \, \dPBFbar			& \quad \DtdUmdPBFmdPBF 					&  \quad F 	\\
									&		&												& \quad  \DtdUmdPBFmdPBFbar 				&  \quad S	\\
\quad (\dUm*\dPBFm * (\dUi * \dPBF) 		& \quad	& \text{$U, \PBF$ mismatches, jitter}, \, \dPBF, \, \dPBFbar	& \quad \DtdUmdPBFmdUidPBF 				&  \quad F \\
									&		&												& \quad \DtdUmdPBFmdUidPBFbar				&  \quad F	\\
\end{array}
$$
}

{\bf Comments:}

We have separated those effects that simply delay the emitted pulse  from those that introduce \TOA\ variations by  distortion of  the pulse.  Dispersion delays are included in the former along with several achromatic contributions while scattering and multiple imaging from lensing are included in the latter. 

In the first column $\dPBF$ refers to all temporal changes in the PBF (fast and slow) while in the middle two columns $\dPBF$ refers to rapid changes from the finite scintle effect
and $\dPBFbar$ corresponds to slow changes in the envelope shape caused by refraction. 

Systematic \TOA\ terms are denoted  as $t_X$   while random errors are denoted as $\Delta t_{X}$.   

Comments in the  last column include designation of a term as varying with epoch  fast (F) or slow (S).

The 21 terms in the third column  are described as follows:
\begin{enumerate}[label=\Roman*.]
\item
The true \TOA\ $t_0(\nubar, \tbar)$ comprises:
\begin{enumerate}
	\item
 	$\tinfty$, which includes all achromatic contributions (spin, orbit, astrometric, gravitational) 
	\item
	$t_{\DM}(\nubar, \tbar)$ = dispersion delays, including frequency dependent DMs
	\item 
	$\dtAOABary$, the observatory to solar system barycentric time offset due to  refraction (Sect.~\ref{sec:aoatoa}).
\end{enumerate}

\item 
The next three terms are  time offsets, $\DtUUm, \DtPBFPBFm$, and  $\DtdUdPBF$,  caused by differences  between model shapes $\Um$ and $\PBFm$ 
and the true ensemble average shapes,  $\Uiave$ and $\PBFave{}$. 

\item
Terms 5 through 7 are random errors from jitter and the finite scintle effect, which vary rapidly, and a slowly varying term resulting from changes in the PBF vs. epoch.   We denote rapid PBF changes as $\dPBF$ and slow changes as $\dPBFbar$, where the overline connotes the envelope of the PBF discussed earlier. 

\item
The remaining 14 \TOA\ contributions result from the  nine terms in  the cross correlation contribution $\dUms * \dUs$; we group into  fast (F) or  slow(S) terms as labeled in the last column.  We obtain a consolidated fast term
 $\DtdUmsdUsF$ that  varies rapidly due to jitter and the finite scintle effect (DISS time scales $\sim$\,minutes),
\be
\DtdUmsdUsF &=& \DtdUmdUi + \DtdUmdPBF + \DtdUmddPBF + \DtdPBFmdUi  +  \DtdPBFmdPBF  
\nonumber \\
		&+& \DtdPBFmdUidPBF + \DtdUmdUi  + \DtdUmdPBFmdPBF  + \DtdUmdPBFmdUidPBF ,
\ee
and $\DtdUmsdUsS$ varies slowly due to refraction (RISS time scales) and rarely from secular or episodic changes in intrinsic pulse shape (cf.\ J1713+0747), 
\be
\DtdUmsdUsS = \DtdUmdPBFbar + \DtdPBFmdPBFbar  + \DtdUmdPBFmdPBFbar.
\ee
\end{enumerate}

Fortunately many of these terms are similar in statistical character and can be merged for modeling and statistical inference.
We write the net \TOA\ from model fitting now also including   the matched filtering error from additive noise $\DtSNR$ (recall up to now in this appendix we have assumed noiseless signals), 
\be
\tnuhat 	&=& t_0 + \DtUUm + \DtPBFPBFm + \DtdUdPBF 
\nonumber \\
		&+& \left[\DtSNR + \DtJ + \DtdPBF + \DtdUmsdPBF  \right]_F
\nonumber \\
		&+& \left[\DtdPBFbar +  \DtdUmsdPBFbar  \right]_S .
\label{eq:timing_equation_long}
\ee

This can be consolidated further by combining the two terms involving $\dPBF$ and the two terms involving $\dPBFbar$: 
\be
\tnuhat 	&=& t_0 +  \Dtmismatch 
+  \left[\DtSNR + \DtJ + \DtdPBFall  \right]_F   +  \left[\DtdPBFbarall  \right]_S .
\label{eq:timing_equation_short}
\ee
The systematic offsets from template-profile mismatch are combined into the term $\Dtmismatch = \DtUUm + \DtPBFPBFm + \DtdUdPBF$.   
 If a single template at a reference frequency $\nuref$ is used, the `FD' correction  used by NANOGrav is associated with this term.   Since `FD' (meaning frequency dependent) applies to all terms, we refer to this particular term as the `chromatic mismatch'  term for this choice of template. 

The rapidly varying random errors are the usual S/N,  jitter, and FSE errors with the latter affected by template mismatch. 

The slowly varying error is from changes in the PBF with epoch, manifested as a change in apparent scattering time $\taud$, which is most likely caused by refraction that alters  the  PBF shape.  
Note again that refraction also alters the centroid of the scattered image and thus also the width and shape of the PBF (Sect.~\ref{sec:aniso_refract}).   The centroid shift yields a purely refractive time delay $\dtAOA$ given by Eq.~\eqref{eq:dtoa_thetar} that could be itemized separately in  Eq.~\eqref{eq:timing_equation_short}, but we subsume it here into the $\DtdPBFbarall$ term. 
 This term is not equal to $\taud$ for the case where a PBF is excluded from the template and it is not equal to $\delta \taud = \taud - {\taud}_{\rm template}$  when a fixed value 
${\taud}_{\rm template}$ for the scattering time is included in the template but the $\taud$ for the profile differs.

Mismatch is from the quantities $\dUm = \Um - \Uiave$ and $\dPBFm = \PBFm - \PBFave{}$ that appear in 
$$
\dUms * \Usave = \dUm * \PBFave{} + \Uiave * \dPBFm + \dUm * \PBFm. 
$$
The PBFs are normalized to unit area so the area of the difference  $\dPBFm$ is zero,   Suppose the actual, mean PBF is a heavy-tailed PL form while the model PBF is an exponential.   Their difference is positive at small lags $t \lesssim \taud$, negative for $t \gtrsim \taud$, and asymptotes to zero at $t \gg \taud$.




\subsection{With  epoch-dependent dispersion and multipath propagation}

Motion of the pulsar and observer introduce epoch dependence in all propagation effects, including  refraction and multipath propagation from scattering that also require complex chromatic terms in  the timing equation.  Pulse distortion from the convolution with the PBF induces short term and long term timing offsets.   Refraction, which varies on day to several month time scales, causes image wander and distortion of the PBF. 
Overall we obtain the full expression in Eq.~\eqref{eq:timing_equation_short} with $t_0(\nu)$ expanded, 
\be
\tnuhat 	&=& 
\tinfty + \frac{\KDM\DMbar(\nu, \tbar)}{\nu^2}    
+  \left[\DtSNR + \DtJ + \DtdPBFall  \right]_F 
+  \Dtmismatch(\nubar, \nuref, \tbar)
+  \left[\DtdPBFbarall  \right]_S  .
\label{eq:toa_with_dm_scattering1}
\ee
where $\DMbar(\nu, \tbar)$ is the DM applicable to frequency $\nu$ at epoch $\tbar$. 
Expanding $\DMbar(\nu, \tbar) = \langle \DMbar(\nu, \tbar)\rangle_{\nu} + \delta\DMbar(\nu, \tbar)$ as an average over the receiver's frequency range with a fluctuating part,
the \TOA\ equation becomes
\be
\tnuhat   =  
\tinfty &+& \frac{\KDM \langle \DMbar(\nu, \tbar) \rangle_{\nu}}{\nu^2}   +  \left[\DtSNR + \DtJ + \DtdPBFall  \right]_F
\nonumber \\
&+&  \Dtmismatch(\nubar, \nuref, \tbar)
+ \left[\DtdPBFbarall  \right]_S 
+ \frac{\KDM \delta\DMbar(\nu, \tbar) }{\nu^2}  .
\label{eq:toa_with_dm_scattering3}
\ee
In this last equation, the first line includes the targeted $\tinfty$ term along with the simplest DM term that scales strictly as $\nu^{-2}$  accompanied by the rapid white noise terms.    The second line includes the $\Dtmismatch$ term  conventionally handled through the `FD' formulation for chromatic mismatch along with the slow term from PBF envelope variations and the frequency-dependent DM terms.   These terms vary with frequency differently  from each other and from  the $\nu^{-2}$ scaling of the simplest DM term. 

\clearpage
\section{Temporal spectra for screen phase and refraction angle}
\label{sec:AOA_spectrum}
\label{app:AOA_spectrum}

Here we calculate  temporal spectra in terms of the electron-density  wavenumber spectrum  $\Pne(\qvec, q_z)$,
where the arguments are the  two-dimensional wavenumber $\qvec$   in the screen plane and an orthogonal wavenumber  $q_z$ in the mean direction of propagation. 
We use $S_a(f)$ to denote a temporal spectrum for a quantity $a$
and $P_a(\qvec)$ for its spatial spectrum, 
Our analysis applies to a screen with time-frozen fluctuations (the Taylor hypothesis)  as it is swept across the line of sight in the $x$ direction with velocity $v_x$. 

A general relation between temporal and wavenumber spectra is then
\be
S_a(f) = \frac{2\pi}{v_x} \int dq_y\, P_a(2\pi f / v_x, q_y), 
\ee
with spectra  related to the temporal and spatial Fourier transforms (FTs)  
$\widetilde a(f)$ and $\widetilde a(\qvec)$ as
\be
S_a(f) =  \frac{\displaystyle \langle \vert \widetilde a(f) \vert^2 \rangle} {\int dt} 
\quad {\rm and} \quad
P_a(\qvec) = \frac{\displaystyle (2\pi)^2   \langle\vert \widetilde a(\qvec) \vert^2 \rangle} {\int d\xvec},
\ee
where integrals in the denominators are over the same domains used to calculate the FTs.   Strictly speaking, the expressions are in the limit where these domains become infinite.

The phase is the integral 
$\phi(\xvec) = -\lambda \relec \int ds\, \nelec(\xvec, s) $
through a screen with thickness  much smaller than the source's distance but  much larger than the outer scale of density fluctuations. 
The phase spectra are
\be
\Phi(\qvec) &=& 2\pi (\lambda \relec)^2 \int  ds\, \Pne(\qvec, 0)
\\
\Sphi(f)       &=&  \frac{(2\pi \lambda \relec)^2}{v_x} \int ds\,  \int dq_y\, \Pne(2\pi f / v_x, q_y, 0).
\ee

The spectrum of the refraction angle  $\thetarvec(\xvec) = k^{-1}\grad_\xvec \phi(\xvec)$ (with $k = 2\pi/\lambda$)  is related to the phase spectrum using the Fourier derivative theorem, giving
\be
\Pthetar(\qperpvec) 
= k^{-2}\vert \qvec \vert^2  \Pphi(\qvec)
= \frac{(\lambda^2 \relec)^2}{2\pi} q^2 \int ds\, \Pne(\qvec, 0)
\ee
and a temporal spectrum
\be
\Sthetar(f) 
=
\frac{(\lambda^2 \relec)^2}{v_x} 
\int ds\,
\int dq_y\,
\left[ (2\pi f / v_x )^2  + q_y^2\right]
 \Pne(2\pi f / v_x, q_y).
 \label{eq:Sthr1}
\ee
The fluctuation spectrum of $\thetar^2$ results from  the self convolution of $\Sthetar$ under the assumption that $\thetar(t)$ is a Gaussian process,
\be
\Sthetarsq(f) = 2 \Sthetar(f) * \Sthetar(f).  
\label{eqn:Sthsq1}
\ee 
A useful expression follows by relating the fourth and second moments of $\thetar$, again for a Gaussian process,
$\Var(\thetar^2) = 2 \Var(\thetar)^2$.   Letting $\Sthetarsq(f) \equiv A_2 g(f)$ where $g(f)$ is the spectral shape,  this implies
\be
\Sthetarsq(f)   = \frac{2 [ \int df\, \Sthetar(f) ]^2} {\int df\, g(f)} g(f). 
\label{eq:Sthrsq2}
\ee

Evaluation  for electron densities having a  power-law spectrum 
that is isotropic in wavenumber, $\Pne(\qvec)  = \cnsq q^{-\beta} \exp(-(q / \qinner)^2)$ for $q\ge \qouter$
and $3 < \beta < 4$,  yields temporal spectra in terms of the scattering measure $\SM = \int ds\, \cnsq(s)$,
\be
\Sphi(f) = 
\frac{\sqrt{\pi}\ \ \  } {(2\pi)^{\beta-3}} 
\frac{\Gamma(\frac{\beta-1}{2})}  {\Gamma(\frac{\beta}{2})} 
\frac{(\lambda \relec)^2}{v_x} 
\SM
\left(\frac{f}{v_x}\right)^{-(\beta-1)} .
\ee
and
\be
\Sthetar(f) = 
\frac{\sqrt{\pi}\ \ \  } {(2\pi)^{\beta-3}} 
\frac{\Gamma(\frac{\beta-3}{2})}  {\Gamma(\frac{\beta-2}{2})} 
\frac{(\lambda^2 \relec)^2}{v_x} 
\SM
\left(\frac{f}{v_x}\right)^{-(\beta-3)} .
\label{eq:Sthr2}
\ee
These forms assume that the exponential factor $\exp(-(q / \qinner)^2) \simeq 1$ 
for wavenumbers that dominate the 
integral  in Eq.~\eqref{eq:Sthr1} at relevant frequencies $f$. 
The spectrum for $\thetar^2$ follows from Eq.\, \eqref{eq:Sthrsq2}   and   \eqref{eq:Sthr1},
\be
\Sthetarsq(f) \simeq 
\frac{2}{(4-\beta) (2\pi)^{2\beta-7}} 
\left[ \frac{\Gamma(\frac{\beta-3}{2})}  {\Gamma(\frac{\beta-2}{2})} 
\frac{(\lambda^2 \relec)^2}{v_x^{4-\beta} }
\SM
\right]^2 f^{-(2\beta-7)}.
\ee
This expression holds for $7/2 < \beta < 4$, including the Kolmogorov spectrum.

For a Kolmogorov spectrum ($\beta = 11/3$), these become
\be
\Sphi(f) = 
\frac{1 } {\sqrt{2}(2\pi)^{1/6}} 
\frac{\Gamma(4/3)}  {\Gamma(11/6)} 
\frac{(\lambda \relec)^2}{v_x} 
\SM
\left(\frac{f}{v_x}\right)^{-8/3} 
\simeq (4.65\times 10^3\, {\rm rad})^2 \,{\rm yr} \times \nu^{-2} v_{100}^{5/3} \, \SM_{-3.5} f^{-8/3},
\ee
\be
\Sthetar(f) = 
\frac{1 } {\sqrt{2}(2\pi)^{1/6}} 
\frac{\Gamma(1/3)}  {\Gamma(5/6)} 
\frac{(\lambda^2 \relec)^2}{v_x} 
\SM
\left(\frac{f}{v_x}\right)^{-2/3} 
\simeq  (0.0423 \, {\rm mas})^2 {\rm yr} \times \nu^{-4} v_{100}^{-1/3} \, \SM_{-3.5} f^{-2/3},
\label{eq:Sthr3}
\ee
and
\be
\Sthetarsq(f)  
=
\frac{6}{ (2\pi)^{1/3}} 
\left[ 
	\frac{\Gamma(1/3)}  {\Gamma(5/6)}
	\frac{(\lambda^2 \relec)^2}{v_x^{1/2} }
	\SM 
\right]^2
\left(\frac{f}{v_x}\right)^{-1/3} 
\simeq  {\rm (0.0787 \, mas)^4\,yr} \times
\nu^{-8} v_{100}^{-2/3} \, \SM_{-3.5}^2 f^{-1/3},
\ee
where the approximate equalities are  for $\nu$ in GHz,  $v_x$ in units of 100\,km\,s$^{-1}$,
a fiducial scattering measure  $\SM = 10^{-3.5}\,\SMunits$, and frequencies in cycles\,yr$^{-1}$. 
For  a source at finite distance the AOA is $\thetaAOA = (\dsl / \dso) \thetar$ and the spectra for $\thetar$
and $\theta_r^2$ should be multiplied by $(\dsl / \dso)^2$ and $(\dsl / \dso)^4$, respectively, where
$\dsl$ is the distance of the screen layer from the source and $\dso$ is the source-observer distance. 


The temporal spectrum for DM is related to the phase spectrum as
$\SDM(f) = (\lambda \relec)^{-2} \Sphi(f)$ or
\be
S_{\DM}(f) \simeq  (1.783\times 10^{-4} \, \DMunits)^2\,{\rm yr} \times  v_{100}^{5/3} \, \SM_{-3.5} f^{-8/3}.
\ee

\renewcommand{\hismtilde}{\tilde{\hism}}

\newcommand{\bbfield}{\varepsilon}
\newcommand{\bbfieldi}{\varepsilon_{\rm i}}

\newcommand{\bpass}{\tilde b}
\newcommand{\bt}{b}
\newcommand{\Gammaeps}{\Gamma_{ \widetilde{\varepsilon}}}
\newcommand{\Gammaepsp}{\Gammaeps^{^\prime}}
\newcommand{\Gammah}{\Gamma_{ \hismtilde}}
\newcommand{\Gammahp}{\Gammah^{^\prime}}
\newcommand{\Gammab}{r_{\tilde b}}

\newcommand{\Gammahmean}[1]{\left\langle\Gamma_{\hismtilde}{(#1)}\right\rangle}
\newcommand{\Gammahmeanp}[1]{\left\langle\Gamma^{^\prime}_{\hismtilde}{(#1)}\right\rangle}

\newcommand{\tbarmean}{\left\langle \tbar \right\rangle}

\newcommand{\tbarh}{{t}_h}
\newcommand{\tbarb}{{t}_b}
\newcommand{\tbarhmean}{\langle \tbarh \rangle}
\newcommand{\tbarbmean}{\langle \tbarb \rangle}

\newcommand{\Bnux}{B_{\nu, x}}
\newcommand{\Bnupxp}{B_{\nup, x^{\prime}}}
\newcommand{\Bnuzero}{B_{\nu, 0}}
\newcommand{\Bnupzero}{B_{\nup, 0}}
\newcommand{\tsp}{t_s^{^\prime}}

\newcommand{\tpbar}{t_{\PBF}}
\newcommand{\tpbarmean}{\langle \tpbar \rangle}
\newcommand{\tpbarmeansq}{\langle \tpbar^2 \rangle}

\clearpage
\section{Derivation of TOA errors from rapid changes in   pulse broadening functions}     
\label{app:FSE}

We analyze measured signals for a heterodyned system with a bandpass of bandwidth $B$ 
centered on frequency $\nu_0$ and shifted to baseband (zero frequency).  
After propagation through the ISM, the signal is
\be
\epsism (t) = \hism(t) * \bbfieldi(t),
\ee
where the asterisk denotes convolution and $\hism(t)$ is the impulse response of the ISM over the band
sampled by  the receiver system.   Generally $\hism$ would include dispersion delays, refraction, and multipath effects but here we restrict it to include only the delays associated with diffraction. 

A receiver system with  a bandpass function $\bpass(\nu)$ yields
\be
\varepsilon(t) =  \bt(t)  * \epsism(t)   = \bt(t) * \hism(t) * \bbfieldi(t), 
\ee
where $b(t)$ is the temporal response equal to the inverse Fourier transform of $\bpass$. 
In the following let  $\bbfieldi(t) = \delta(t)$ to get
$\varepsilon(t) = \bt(t) * \hism(t)$ as the impulse response from  scattering
combined with the receiver response.  The bandpass function $\bpass$ is a `low pass' function centered on 
zero frequency with total bandwidth $B$ (across both negative and positive frequencies).  
For convenience, we define 
 $\vert\bpass\vert^2$ to have  unit area.  Its  autocorrelation function (ACF) is then
\be
        r_{\bpass}(\delta\nu) =
        \int d\nu\, \bpass(\nu) \bpass^*(\nu+\delta\nu).
\ee
with normalization $r_{\bpass}(0) =1$ and where the superscript asterisk denotes a complex conjugate.   The area of $\vert\bpass(\nu)\vert^4$  is the reciprocal of 
the effective bandwidth, $\Beff$. 
A rectangular bandpass function with bandwidth $B$ has
$\Beff = B$ and $r_{\bpass}(\delta\nu) = 1 - \vert\delta\nu\vert/B$. 

We call $\hism$ the `field' pulse broadening function (PBF) to distinguish it from the  squared magnitude 
$\pism(t) = \vert \hism \vert^2$, which  is the standard  intensity PBF in the limit where the receiver response is much narrower than $h(t)$.

\subsection{Pulse Broadening Function}

Define $\Gammaeps(\delta\nu)$ as the spectral correlation function of scintillation variations.
When averaged over an ensemble, its half-width at half maximum is the scintillation bandwidth, $\dnud$. 

The field PDF $\hism(t)$ has  the form of an envelope function that multiplies a noise-like process.  
The envelope has  duration $\sim \tau_d$, the pulse broadening time, while the noise process varies
on inverse-bandwidth time scales.  The noise process is  a consequence of the particular paths  taken by the  radiation
due to  diffraction, which persist over time scales up to about  the scintillation time scale, $\Dtiss$.
It is therefore useful to define two time scales: one for variations on time scales of
order the inverse bandwidth (ns) and the other `epochal'  time over an observation of order minutes to
hours.  The time ``t'' in $h(t)$ refers to the fast time scale and the epochal time
dependence is implicit for now.  Here we consider TOA variations due to the finite number of scintles
within the receiver bandpass at a fixed epochal time.  Later we consider how integrations over epochal time reduce TOA variations.

The total impulse response,  including the finite receiver bandwidth,   
$\epsilon(t) =  h(t) * b(t)$, has a frequency-domain ACF
\be
\Gammaeps(\delta\nu) 
	= \int d\nu\, \bpass(\nu)\, \bpass^*(\nu+\delta\nu)\, \hismtilde(\nu) \hismtilde^*(\nu+\delta\nu)
	=  \int d\nu\, \bpass(\nu)\, \bpass^*(\nu+\delta\nu)\, \Gammah(\nu, \delta\nu).
\ee 
The  corresponding PBF is  
\be
\PBF(t) = \int dx\,  \Gammaeps\,(x) e^{2\pi i x t}
\label{eq:pbf1}
\ee
The ensemble averages of these quantities are  
\be
\left\langle\Gammaeps(\delta\nu)\right\rangle &=& r_{\bpass}(\delta\nu) \Gammahmean{\delta\nu}
\label{eq:Gammaeps}
 \\
\left\langle \PBF(t)\right\rangle  &=& \int dx\, \left\langle \Gammaeps\,(x)\right\rangle e^{2\pi i x t},
\label{eq:pbfmean1}
\ee
where we have  dropped the $\nu$-dependence of $\Gammahmean{\delta\nu}$ by assuming that 
$\hismtilde(\nu)$ has stationary statistics over the bandwidth 
(i.e. the frequency range is small enough that the scintillation bandwidth does not change significantly).
Both $\vert h(t) \vert^2$ and   $\PBF(t)$ have unit area and
   $\Gammah(0) =  \Gammab(0) = \Gammaeps(0) = 1$. 
For wide bandwidths, similar quantities can be defined for subbands. 

\subsection{Mean Delay}

The sum of a large number of pulses converges to  the  pulse shape intrinsic to the pulsar
convolved with the ensemble average PBF.     Because the PBF is causal,  the measured pulsar signal
is always delayed by scattering.  The TOA shift (of an impulse)  is the mean of the PBF, 
\be
\tpbar = \int dt\, t\, \PBF(t),
\label{eq:tpbar}
\ee
which has  an ensemble average 
\be
\tpbarmean = \int dt\, t \left\langle \PBF(t) \right\rangle.
\label{eq:tbarmean1}
\ee
Substituting for $\PBF(t)$ using Eq.~\eqref{eq:pbf1} and 
using the derivative theorem for Fourier transforms yields
\be
\tpbarmean 
= 
\int dt\, t  \int dx\, \left\langle \Gammaeps(x)\right\rangle e^{2\pi x t}
= 
\frac{1}{2\pi i}  \int dx\, \left\langle \Gammaeps(x)\right\rangle \delta^{\,\prime}(x)
= 
- \frac{1}{2\pi i} \left\langle \Gammaepsp(0) \right\rangle,
\label{eq:tbarmean2}
\ee
where $\Gammaepsp(x) \equiv d\Gammaeps(x)/dx$ and we used the derivative property of the delta function. 

The time delay comprises  two terms, 
\be
\tpbarmean = \tbarbmean + \tbarhmean = 
	-\frac{1}{2\pi i} \left[ r_{\bpass}^{^\prime}(0) + \Gammahp(0) \right],
\ee
where the  delay due to the causal bandpass filter is $\tbarbmean$  while $\tbarhmean$ is  from scattering in the ISM; both are calculated analogously to Eq.~\eqref{eq:tpbar} using $\vert b(t)\vert^2$ and $\vert h(t)\vert^2$ in place of $p(t)$, respectively. 
The range of cases encountered includes $\tbarhmean \ll \tbarbmean$ for low-DM pulsars observed at
high frequencies while $\tbarhmean \gg \tbarbmean$ for high-DMs and low frequencies.

A commonly-assumed  PBF is a one-sided exponential with time constant $\taud$,
\be
\PBF(t) = \taud^{-1} \exp(-t/\taud)\Theta(t),
\ee
where $\Theta(t)$ is the Heaviside function.
For this case, $\tpbarmean = \taud$ is obtained from either Eq.~\eqref{eq:tbarmean1}  or \eqref{eq:tbarmean2}.   For more realistic PBFs, such as  the heavy-tailed PBFs  associated with
a Kolmogorov  or other power-law media, $\tpbarmean$ is generally larger than $\taud$. 

\subsection{Delay Variance}
First we consider the case where the integration time is much less than the scintillation time, corresponding to the case of static geometry.   Later we  consider the opposite case where the scintillation time is much larger than the scintillation time. 

The stochasticity of $\PBF(t)$ causes mean square TOA fluctuations  
 
\be
\tpbarmeansq = \tpbarmean^2 + \sigma_{\tpbar}^2 = 
		\iint dt d\tp\, t \tp \left\langle\PBF(t) \PBF(\tp) \right\rangle
		= -\frac{1}{(2\pi)^2}
			\frac{d^2}{dx dx^{\prime}}
			\bigg[
				\left\langle
					\Gammaeps\,(x) \Gammaeps\,(x^{\prime})
				\right\rangle
			\bigg]_{x = x^{\prime} = 0}.
\label{eq:tbarmeansq}		
\ee
The expectation in  Eq.~\eqref{eq:tbarmeansq} is
\be
\left\langle  \Gammaeps\,(x) \Gammaeps\,(x^{\prime})   \right\rangle &=& 
\iint d\nu d\nup\, \bpass(\nu) \bpass^*(\nu + x) \bpass(\nup) \bpass^*(\nup + x^{\prime})
	\nonumber \\
&& \quad\quad
	\times
	\left\langle  
	\hismtilde(\nu) \hismtilde^*(\nu+x) \hismtilde(\nup) \hismtilde^*(\nup+x^{\prime})
	\right\rangle .
\ee
To an excellent approximation,  $\hismtilde(\nu)$  is a complex Gaussian process when scintillations are strong,  so the fourth moment can be expanded into products of second moments\footnote{Note that the fourth moment of a complex Gaussian process  expands into  two terms while a real process expands into  three.},
\be
\left\langle  
	\hismtilde(\nu) \hismtilde^*(\nu+x) \hismtilde(\nup) \hismtilde^*(\nup+x^{\prime})
\right\rangle 
= 
	\Gammahmean{x} \Gammahmean{x^{\prime}} + 
	\Gammahmean{\nup - \nu + x^{\prime}}\Gammahmean{\nu-\nup + x}.
\label{eq:fourth}
\ee
The first term in the expansion yields the square of the mean delay, $\tpbarmean^2$, while the second
term yields the variance of $\tpbar$, 
\be
 \sigma_{\tpbar}^2 = 
 			-\frac{1}{(2\pi)^2}
 			\iint d\nu d\nup\,
			\frac{d^2}{dx dx^{\prime}}			 
				\bigg[
					B_{\nu, x} B_{\nup, x^{\prime}}
					\Gammahmean{\nup - \nu + x^{\prime}} \Gammahmean{\nu-\nup + x}
				\bigg]_{x = x^{\prime} = 0}.			
\label{eq:tbarvariance1}	
\ee	
where we have defined 
$B_{\nu, x} \equiv \bpass(\nu) \bpass^*(\nu + x)$.  

The double derivative yields four combinations of individual derivatives of the $\Bnux$ and 
$\left\langle\Gamma_{\tilde h} \right\rangle$
factors evaluated at $x = x^{\prime} = 0$,
\be
\text{(I)}\quad &&
\Bnuzero^{^\prime} \Bnupzero^{^\prime}\Gammahmean{\nup - \nu}\Gammahmean{\nu-\nup}
	\equiv \Bnuzero^{^\prime} \Bnupzero^{^\prime} \left\vert \Gammahmean{\nu-\nup}\right\vert^2
\nonumber \\
\text{(II)}\quad &&
\Bnuzero \Bnupzero^{^\prime} \Gammahmean{\nup - \nu}\Gammahmeanp{\nu-\nup}
\nonumber \\
\text{(III)}\quad &&
\Bnuzero^{^\prime} \Bnupzero\Gammahmeanp{\nup - \nu}\Gammahmean{\nu-\nup}
\nonumber \\
\text{(IV)}\quad &&
-\Bnuzero \Bnupzero \Gammahmeanp{\nup - \nu}\Gammahmeanp{\nu-\nup}
	\equiv \Bnuzero \Bnupzero \left\vert \Gammahmeanp{\nu-\nup}\right\vert^2.
\nonumber 
\ee
Generally, all terms contribute to the delay variance.  However terms I - III are important only 
for the case where the scintillation bandwidth is comparable to or larger than the measurement bandwidth $\Beff$.
In the regime where $\dnud \ll \Beff$, the four terms contribute to the variance respectively as 
$\Beff^{-2} (\dnud/\Beff)$, 
$\Beff^{-2}$,
$\Beff^{-2}$,
and
$\Beff^{-2} (\Beff/\dnud)$.
In this regime, therefore, term IV will exceed the next largest terms by a factor proportional to the number
of scintles across the band.  

\subsubsection{Narrow-scintle Regime} 

For $\dnud \ll \Beff$, the delay variance becomes
\be
 \sigma_{\tpbar}^2 \approx  \frac{1}{(2\pi)^2}
 			\iint d\nu d\nup\,
				\Bnuzero \Bnupzero \left\vert \Gammahmeanp{\nu-\nup}\right\vert^2.			
\label{eq:tbarvariance2}	
\ee	

We simplify Eq.~\eqref{eq:tbarvariance2} by changing variables
from $\nu, \nup$ to $\nubar = (\nu+\nubar)/2$ and $\delta\nu = \nu-\nup$ and by defining
$B_\nu = \vert \bpass(\nu)\vert^2$ (which equals $\Bnuzero$ used before but simplifies 
the notation) and  its frequency-domain correlation function,
\be
\widetilde R_B(\delta\nu) = \int d\nu\, B_\nu B_{\nu+\delta\nu}.
\ee
The inverse Fourier transform  of $\widetilde R_B$ is  $R_B(t)$, which has an   
 area identically equal to $\Beff$.  This yields
 \be
 \sigma_{\tpbar}^2 \approx  -\frac{1}{(2\pi)^2}
 			 \int d\delta\nu\,
			\widetilde R_B(\delta\nu)
			 \left\vert \Gammahmeanp{\delta\nu}\right\vert^2.			
\label{eq:tbarvariance3}	
\ee	
By using the relation between $\pism(t)$ and $\Gammahmean{\delta\nu}$
we obtain  
\be
\sigma_{\tpbar}^2 = 
	\iint dt d\tp\, t \tp 
		\left\langle\pism(t)\right\rangle \left\langle\pism(\tp) \right\rangle 
		R_B(t-\tp). 
\label{eq:tbarvariance4}
\ee
In the limit of narrow scintles where $\dnud \ll B$, the bandpass correlation  $R_B(t-\tp)$ tends to a delta function $\delta(t-\tp)$ with  area 
$\Beff^{-1}$ and the delay variance simplifies to 
\be
 \sigma_{\tpbar}^2 \approx  
			 \Beff^{-1} \int dt\, t^2 \left\langle \pism(t) \right\rangle^2	.
\label{eq:tbarvariance5}	
\ee	

{\it Exponential PBF:} For  
$\left\langle \pism(t) \right\rangle  = \taud^{-1} \exp(-t/\taud) \Theta(t)$, 
the time integral yields $\taud/4$ and $\sigma_{\tpbar} \approx \sqrt{\taud / \Beff} / 2$.    Using the relation
$2\pi\taud\dnud = 1$  for this PBF\footnote{Generally $2\pi\taud\dnud = C_1$ with $C_1$ ranging from 1 to 2, depending on the scattering medium but the constant is unity for a thin screen that produces an exponential PBF.}, we obtain
\be
\sigma_{\tpbar} =  \frac{\taud}{\sqrt{2 \Beff / \pi\dnud}}
\ee
which can be expressed in terms of the number of scintles within the band\footnote{The number of scintles is defined   so that the rms intensity over the band is
$\sigma_I  / \langle I \rangle = \nissnu^{-1/2}$.} 
\be
\nissnu= \frac{\Beff}{\pi\dnud},
\ee
to yield
\be
\sigma_{\tpbar} =  \frac{\taud}{\sqrt{2 \nissnu}}.
\label{eq:rms1}
\ee

\subsubsection{Integration Times Longer than the Scintillation Time}

For integration times $T$ smaller than the diffractive scintillation time scale ($\Dtiss$)
the PBF is static because  multipath scattering has a  steady (but frequency dependent) configuration of propagation paths.  
Observations of nearby pulsars at high frequencies often involve this situation for the entirety of a
1/2 to one hour observation.   However for other cases $T$ can be  much longer than the scintillation
time scale, causing the pattern of scintles within the receiver bandpass to change.   To distinguish between
short time scales as used above --- which range from the inverse bandwidth to some multiple of the pulse
broadening time --- from longer times, we  introduce the time $t_s\in[0,T]$   that can span many scintillation time scales.   The correlation function of $\hismtilde$ at time $t_s$ becomes
\be
\Gammah(\delta\nu,  \delta t_s) = \hismtilde(\nu, t_s) \hismtilde^*(\nu, t_s + \delta t_s)
\ee
and the time-integrated PBF  is
\be
\PBF(t) = T^{-1} \int_0^T dt_s\, \
\bpass(\nu)\, \bpass^*(\nu+\delta\nu)\, \Gammah(\delta\nu; 0).
\ee
The ensemble average of $\PBF(t)$ is independent of $T$ in this treatment though in reality 
complications arise from  epoch to epoch variations due to refraction in the ISM.   The delay variance is calculated as before and the fourth moment of $\hismtilde$ becomes (cf.\ Eq.~\eqref{eq:fourth}),
\be
	&& \left\langle  
	\hismtilde(\nu, t_s) \hismtilde^*(\nu+x, t_s) \hismtilde(\nup, \tsp) \hismtilde^*(\nup+x^{\prime}, \tsp)
	\right\rangle = 
	\nonumber \\
	&& 
	\quad\quad\quad
		\Gammahmean{x, 0} \Gammahmean{x^{\prime}, 0} + 
		\Gammahmean{\nup - \nu + x, t_s - \tsp} \Gammahmean{\nu-\nup + x^{\prime}, t_s - \tsp}.
\label{eq:fourth2}
\ee
The first term in the expansion corresponds to  $\tpbarmean^2$ and the second term yields the variance. 
For scintles that are narrow compared to the bandwidth,  we obtain the extension of 
Eq.~\eqref{eq:tbarvariance3}
\be
 \sigma_{\tpbar}^2 &\approx&  \frac{1}{(2\pi)^2 T}
			\int d\delta t_s
			\left(1 - \frac{\vert \delta t_s \vert}{T} \right)
 			\iint d\nu d\nup\,
				\Bnuzero \Bnupzero \left\vert \Gammahmeanp{\nu-\nup, \delta t_s}\right\vert^2,			
\label{eq:tbarvariance6}	
\ee	
where the derivative of the ACF  is with respect to the frequency lag, $\nu-\nup$.
A change of variables gives
 \be
 \sigma_{\tpbar}^2 \approx  \frac{1}{(2\pi)^2T}
 			\int d\delta t_s
			\left(1 - \frac{\vert \delta t_s \vert}{T} \right)
 			 \int d\delta\nu\,
			\widetilde R_B(\delta\nu)
			 \left\vert \Gammahmeanp{\delta\nu, \delta t_s}\right\vert^2.			
\label{eq:tbarvariance7}	
\ee
When scintles are also short in duration compared to the integration time $T$, 
we obtain
\be
 \sigma_{\tpbar}^2 \approx  \frac{1}{(2\pi)^2 T}
			\int d\delta t_s
			\int d\delta\nu\,
			\widetilde R_B(\delta\nu)
			 \left\vert \Gammahmeanp{\delta\nu, \delta t_s}\right\vert^2.			
\label{eq:tbarvariance8}	
\ee	
 
Generally, $\Gammahmeanp{\delta\nu, \delta t_s}$ is not factorable into time-lag and frequency-lag
factors so there is no general analog of Eq.~\eqref{eq:tbarvariance4}, \eqref{eq:tbarvariance5}.
However, the rms $\tpbar$ will still scale similarly to Eq.~\eqref{eq:rms1} as $\taud / \sqrt{2 N_s}$ where the
number of scintles in the large $\Beff T$ limit is 
$N_s \propto \Beff T / \dnud \dtd$. 


\newcommand{\assessdir}{/Users/cordes/Documents/Papers/InProgress/New_Fundamental_Noise_Processes_for_PTAs/Code/ism_calcs}

\clearpage
\section{Dynamic spectra computation and estimation errors}
\label{app:ds_errors_prs_cs}

The dynamic spectrum (DS) is considered for a wide total bandwidth and time span encompassing many scintillation time scales.  It is simply the power spectrum calculated from Fourier transforms, but the lengths of those FTs relative to the pulse period offer alternative specific methods. 
The spectrum comprises frequency structure from both the emitted signal (amplitude modulated noise) and from  multipath propagation through the ISM.  The primary utility of the DS for pulsar timing  is to provide
a good estimate of the scintillation bandwidth.  Fortunately, averaging spectra over multiple pulses diminishes emission structure so that scintillation features dominate.   This contrasts with  the average pulse profile that includes the intrinsic pulse shape  along with scattering, regardless of how many pulses are averaged. 

Good estimation of DISS parameters requires a DS containing many  scintles   resolved in frequency and time.    Larger scattering yields smaller scintles, so sampling requirements are more stringent for larger \DM\  pulsars observed at lower frequencies.   For the largest DMs and lowest frequencies,  the scintles are unresolvable and only pulse-shape fitting will yield an estimate for $\taud$. 

Denoting  the lowest and highest frequencies of the DS as $\nul$ and $\nuh$, respectively,  
voltage data are obtained over a total bandwidth $B = \nuh-\nul$ and time span $T$
with resolutions $\Dnu$ and $\Dt$.  
To satisfy the estimation conditions , we require $\Dnu \ll \dnud \ll B$ and $\Dt \ll \dtd \ll T$,
 where the DISS bandwidth $\dnud$ and DISS time scale
$\dtd$ are defined in the standard way\footnote{$\dnud$ Is the  HWHM of the intensity correlation function vs. frequency lag and $\dtd$ is the HW at $1/e$ time lag.}.  
%
Two approaches for obtaining dynamic spectra are as follows.   

\subsection {Phase-resolved spectroscopy (PRS)}

If the DISS bandwidth is large enough, a running spectrum across pulse phase is obtained with FFT lengths $\TFT \lesssim \WA$, where $\WA$ is the pulse width (i.e. the width of the squared amplitude modulation in the AMN model.).   Multiple FFTs done across the pulse width (if $\TFT \ll \WA$) and over multiple periods are then  combined (as squared magnitudes) to form a DS with frequency resolution $\Delta\nu = \TFT^{-1}$ and time resolution $\Delta t = \Np\times P$ by  averaging spectra in phase bins over $\Np$ periods, corresponding to  1 to 10\,s, typically.

Nyquist sampling of complex voltage data requires  sample intervals $\dtNy = 1/B$.   
Resolving scintles requires $\Delta \nu \lesssim \dnud$ and $\Delta t \lesssim \dtd$.   Using the maximum FFT length with the uncertainty relation Eq.~\eqref{eq:uncertainty1}  gives $1 / \WA \lesssim 1/\TFT = 1 / (\Np P)  \lesssim C_1 / 2\pi\taud$, or
\be
\frac{2\pi\taud}{C_1} \lesssim \TFT \lesssim \WA \lesssim P \ll \dtd,
\label{eq:condition1}
\ee
where we consider the DISS time scale $\dtd$  to be the largest. 
For a thin screen, the scintillation time scale is
\be
\dtd (\nu) =    
    \frac{a_\beta}{2\pi\nu\veff} \left( \frac{c\deff}{\taud} \right)^{1/2} 
    \simeq 
    143\,{\rm s} \times \frac{\nu^{6/5}}{v_{100}} \left(\frac{\deff}{\tau_1} \right)^{1/2} ,
\label{eq:dtd}
\ee
where $a_\beta \equiv 1 / 2^{(4-\beta)/2(\beta-2)} \simeq  0.93$ for $\beta = 11/3$ and
where $\nu$ is in GHz, $v_{100} = \veff / (100\, \rm km\,s^{-1})$,
$\deff = s(1-s) d$ is the effective distance in kpc for a screen at fractional distance $s$ from the pulsar, 
and $\tau_1$ is the scattering time in $\mu s$ at 1\,GHz.

The constraint on the scattering  time,
\be
\taud \lesssim \frac{\WA}{2\pi C_1} \simeq 0.16\,{\rm ms}\times \frac{\WA({\rm ms})}{C_1},
\label{eq:tau_constraint1}
\ee
is satisfied by many long period pulsars with $\WA \sim 10$ to 100\,ms except at low frequencies
where $\taud(\nu) \propto \nu^{-4.4}$ can exceed the requirement. 
 
PRS clearly fails if scintles are unresolved using the maximum FFT length indicated in  Eq.~\eqref{eq:condition1}.  The MSP B1937+21 is a textbook example: with $\WA \simeq 45\,\mu s$, the requirement
$\taud \lesssim 7\,\mu s$ is violated by   the measured $\taud \simeq 25\,\mu s$ at 0.43\,GHz.

\subsubsection{Degrees of freedom in the DS}

In a time series comprising $N$ complex samples, there are $\ndof = 2N$ degrees of freedom (dof). 
For a total time $\Ttotal$ and Nyquist sampling over a bandwidth $B$, this is 
$\ndof = 2 B \Ttotal$. 

A FT of length $\NFT$ spanning a time $\TFT$ yields a  spectrum with frequency resolution 
$\Delta\nu = \TFT^{-1}$ and $2$ dof per frequency bin. 

When $\Nb$ spectra are averaged over a time $\Tstep = \Nb\TFT$ to give a single time step in the DS, there are $2\Nb$ dof/bin.

In a fixed total time $\Ttotal$ there are $\Nsteps = \Ttotal / \Tstep$ total time steps in the DS, yielding 
$2\NFT\Nb\Nsteps$ dof/bin.   With $\NFT$ frequency bins, the total number of dof in the DS is
\be
\ndof 
&=& 2\Nb\Nsteps \times \NFT 
\nonumber \\
&=& \frac{2 \Nb \NFT \Ttotal}{\Nb\TFT} 
=  \frac{2 \NFT\Ttotal}{\NFT \dtNy} 
\nonumber \\
&=& 2 B \Ttotal,
\ee
in agreement with $\Ndof$ in the original time series. 

For PRS, spectra are calculated only for an on-pulse window with duty cycle $\sim \WA/P$.   The number of degrees of freedom for PRS is therefore
\be
\ndof^{\rm (PRS)}  = 2 B \Ttotal \left(\frac{\WA}{P} \right) .
\label{eq:dofprs}
\ee

\subsubsection{Error budget of DS}

In building a DS from a voltage time series $\varepsilon(t)$, the spectral estimate from a single FT is
\be
S(\nu) = \NFT^{-1} \vert \varepsilon(\nu) \vert^2. 
\ee
For a stochastic signal,    fluctuations about the spectral mean are 100\% because
$\sigma_S / \langle S \rangle = \sqrt{2 / \Ndofbin}$ and $\Ndofbin = 2$.    We express this as
\be
S(\nu) = \langle S \rangle \times [1  + {\cal E}(\sigma^2_S) ], 
\ee
where $\sigma_S = 1$ and the second term denotes random values drawn from an exponential PDF,
${\cal E}(\sigma^2_S) = \sigma^{-2}_S \exp(-S / \sigma^{-2}_S) \Theta(S)$
where $\Theta(S)$ is the Heaviside (unit step) function.
The  expression implies that the spectrum  is given by the mean plus an  error  of similar amplitude drawn from an exponential distribution.  This form leaves unspecified the correlation frequency of the error: for AMN spectral noise this is $\delta\nu_{\rm corr} \sim \WA^{-1}$  while for radiometer noise it is 
$\delta\nu_{\rm corr} \sim \TFT^{-1}$.   For diffractive scintillations $\delta\nu_{\rm corr} \equiv \dnud$, 
the scintillation bandwidth. 

With PRS, there are $\sim \WA / \TFT$ spectra averaged across the pulse  in each period and
$\Np$ periods worth of such spectra are averaged to get a single time step in the DS.  This gives
$\Nb = \Np \WA / \TFT = \Np / \epsW$, where  $\TFT \equiv \epsW \WA$ and   $\Ndofbin= 2\Nb$ per bin. 

We then write, after invoking the central limit theorem ($\Ndofbin \gg 1$) and denoting a
zero mean Gaussian distribution with variance  $\sigma^2$ as  ${\cal N}(0, \sigma^2)$,  
\be
\langle S(\nu)\rangle_{\Nb} = \langle S \rangle \times \left[1  + {\cal N}(0, \Nb^{-1}) \right] .
\ee

For the scintillated AMN signal with additive noise, we write $I(\nu) =  S(\nu) + N(\nu)$ and express the
block-averaged spectrum as 
\be
\langle I(\nu)\rangle_{\Nb} 
=  
H(\nu) \Sphiave \left(\frac{P}{\WA}\right)  \times \left[1  + {\cal N}(0, \Nb^{-1}) \right]
+
\sigma_N\times  {\cal N}(0, \Nb^{-1}), 
\ee
where  the DISS modulation $H(\nu)$ is assumed constant over the block average. 
It has unit mean and unit variance (in strong scattering with no quenching from time-frequency averaging,
as here). 
The \rms\ noise variance in a single FT is for  radiometer noise with unit time-bandwidth product, 
$\sigma_N = \Ssys$. 
Adjusting terms,  we rewrite the spectrum as
\be
\langle I(\nu)\rangle_{\Nb} 
=  
\Sphiave \left(\frac{P}{\WA}\right) 
\left\{
    H(\nu)  \times \left[1  + {\cal N}(0, \Nb^{-1}) \right]
+
   \frac{\Ssys}{\Sphiave} \left(\frac{\WA}{P} \right) \times  {\cal N}(0, \Nb^{-1})
\right \}
\ee
The signal to noise factor for a single spectrum (i.e. based on one FT) is
\be
\SNRone 
= \frac{\Sphiave} {\Ssys} \frac{P}{\WA}
\simeq 0.011 \times \left( \frac{\Sphiave / 1\, \rm mJy} {\Ssys / 3\, {\rm Jy}}\right) \left(\frac{0.03}{\WA/P}\right),
\ee
where a nominal 1\,mJy period averaged flux density and 3\% duty cycle are used.   The value of $\Ssys = 3$\,Jy
is appropriate for  Arecibo/DSA-2000/FAST/SKA-mid class telescopes. 
In most cases $\SNRone \ll 1$ applies and the errors are radiometer noise dominated.   
Exceptions are the Vela pulsar with $\SNRone \simeq 90$ and 20 at 0.4 and 1.4 GHz respectively.
Among MSPs, J0437-4715 has $\SNRone \simeq 8$ and 2  while 
J1939+2134 (B1937+21)  has $\SNRone \simeq 3.1$ and 0.2 at these frequencies.

The spectrum at a single time step in the DS comprises
a block average  of $\Nb$ individual spectra, which we express  as 
\be
\langle I(\nu)\rangle_{\Nb} 
=  
\Sphiave \left(\frac{P}{\WA}\right) 
\left[
    H(\nu)  +  {\cal N}(0, \rmsDSN^2)
\right ], 
\label{eq:blockspectrum}
\ee
where the dimensionless variance of the Gaussian distribution is
\be
\rmsDSN^2 
	= \left[  \frac{ 1 + \SNRone^{-2}} {\Nb}  \right].
\ee
Block averaging is critical to ensuring adequate sensitivity to scintillations. 

\subsubsection{Errors in estimating $\dnud$ from the spectral ACF}
\label{sec:acfdnuderr}

The utility of the DS in the pulsar timing context is for estimating the scintillation bandwidth,  then the scattering time, and finally a correction to the raw \TOA.   An ACF of  a single spectrum  has a width with an associated error due to ACF fluctuations.
For a single instance of $\langle I(\nu)\rangle_{\Nb}$ we do not include fluctuations in $H(\nu)$ because we wish to know the scattering time {\it for that particular realization of $H(\nu)$}.  In another context, we might want an epoch averaged estimate, which would calculate the average ACF from the entire DS (i.e. the average of individual ACFs). 

The ACF $R_{I_{\rm b}}(\delta\nu)$ of the spectrum  in Eq.~\eqref{eq:blockspectrum} has  a fractional error from radiometer noise near  zero lag $\delta\nu = 0^+$ (see section \ref{app:acfstats}), 
\be
\varepsilon^{(N)}_{\rm ACF} 
\equiv
\frac{ [\Var  R_{I_b}(0+) ]^{1/2}} {\left \langle  R_{I_b}(0+) \right\rangle}
\simeq   \frac{\displaystyle  \left[  \rmsDSN^4 + 4 \rmsDSN^2 R_S(0^+) \right] ^{1/2}}{\sqrt{\NFT}\times R_S(0^+)}.
\ee
%

Substituting other quantities yields
\be
\varepsilon^{(N)}_{\rm ACF} 
&=& 
\frac{2}{\sqrt{\NFT \Nb} }
\left[ 1 + \SNRone^{-2}\right]^{1/2} 
\left\{ 1 + \Nb^{-1} [1 + \SNRone^{-2}]\right\}^{1/2} 
\nonumber \\ 
&\overset{\SNRone \ll 1}{\simeq}&
\frac{2}{\SNRone\sqrt{\NFT \Nb } } 
\left\{ 1 + \Nb^{-1}  \SNRone^{-2}\right\}^{1/2} 
\nonumber \\ 
&\overset{\Nb^{1/2}\SNRone \gg 1}{\simeq}&
\frac{2} {\SNRone\sqrt{\NFT\Nb}}  
\nonumber \\ 
%
&\simeq& 
\frac{2}{\SNRone} 
\left( \frac{ P} {\Ttotal B \WA}  \right)^{1/2}
\nonumber \\ 
&\simeq& 
2
\left(\frac{\Ssys}{\Sphiave}  \right)
\left( \frac{ \WA} {\Ttotal B P}  \right)^{1/2}
\label{eq:epsacf}
\ee
In the second to last step we assume the DS comprises $\Nb = (\WA/P)  (\Ttotal / \TFT)  = (\WA/P) (\Ttotal  B / \NFT) $ total blocks where we include a duty cycle factor $\WA/P$ and we relate the sampling time to bandwidth, $\delta t = 1/B$, which gives $\TFT = \NFT / B$.

Note that the last result can be obtained by considering the number of noise degrees of freedom included in
the ACF for PRS (Eq.~\eqref{eq:dofprs})
\be
\varepsilon^{(N)}_{\rm ACF} 
\simeq 
\frac{2}{\SNRone} \left(\frac{2}{\ndof^{\rm (PRS)}} \right)^{1/2}.
\ee

The resulting fractional error on  the scintillation bandwidth $\dnud$ (defined as the HWHM of the ACF)  due to radiometer noise only is 
\be
\varepsilon^{(N)}_{\rm \dnud} = 
\Kdnud \varepsilon^{(N)}_{\rm ACF}
\left(\frac{\Delta\nu}{\dnud} \right)^{1/2} ,
\ee
where $\Delta\nu$ is the spectral frequency resolution and the constant $\Kdnud \sim 1$ is dependent on the shape of the ACF\footnote{E.g. $\Kdnud \sim 1.16$ for a Gaussian shape \citep[][]{1992PASP..104.1104L}.    For an exponential PBF the ACF  is  Lorentzian and $\Kdnud \sim 1.58$.}, which depends on the shape of the PBF.   In practice we know the  PBF shape  only approximately because it is stochastic on both short and long time scales.  
Substituting for $\varepsilon^{(N)}_{\rm ACF}$ using the last expression in Eq.~\eqref{eq:epsacf} gives
 \be
\sigma^{(N)}_{\dnud} 
\simeq 
2\Kdnud 
		\left(\frac{\Ssys}{\Sphiave}  \right)
		\left ( 
			\frac{\WA \Delta\nu \dnud } {\Ttotal B P} 
		\right)^{1/2} .	
\ee

%

\subsection{Cyclic spectroscopy (CS)}     

The primary  utility of the cyclic spectrum (CS) is that it  can provide high-frequency resolution dynamic spectra, as noted.   Second is its potential for estimating the  full impulse response $h(t)$ contemporaneously with arrival time data, which could improve efforts to correct  \TOAs\ for  multipath propagation delays \citep[e.g.][]{2011MNRAS.416.2821D, 2013ApJ...779...99W, Palliyaguru_2015,2021ApJ...913...98D, 2023ApJ...944..191T}.   

The CS is defined as 
\be
\GammaFTE(\delta\nu, \nu) =  \EfieldFT(\nu+\delta\nu/2)  \EfieldFT^{\,*}(\nu-\delta\nu/2)
\ee
and is based on a single FT; averaging over multiple FTs is done later.
Inserting the scattered  baseband field  defined earlier, the CS is
\be
\GammaFTE(\delta\nu, \nu)  
= \Gamma_{\htilde}(\delta\nu, \nu)    \iint dt_1 dt_2\,  a(t_1) a(t_2)  m(t_1) m^*(t_2) e^{2\pi i [\nu(t_1-t_2) + \delta\nu (t_1+t_2)/2]} ,
\label{eq:CSfull}
\ee
where 
\be
\Gamma_{\htilde}(\delta\nu, \nu) =  \htilde(\nu+\delta\nu/2)  \htilde^{\,*}(\nu-\delta\nu/2)
\ee
is the CS of the field PBF.  

Resolving  small scintle bandwidths $\dnud$  simply requires longer FTs than those used with phase resolved spectroscopy.  The CS  employs  FTs spanning multiple periods,
$\TFT \equiv \Np P$, but smaller than the scintillation time $\dtd$. 
In fact they need to be much smaller than $\dtd$ in order to reduce errors by summing multiple blocks. 

 For a single FT   the full CS that includes radiometer noise and all cross terms  is
\be
\GammaFTE(\Gargs) 
&= & \EfieldFT(\nup)  \EfieldFT^{\,*}(\num)
= \Gamma_{\htilde}(\Gargs)  \Gamma_{\EfieldiFT}(\Gargs)  + \GammaFTn(\Gargs)
+ [\htilde(\nuplus) \EfieldFT_{\rm i}(\nuplus) \ntilde_{\numinus}^* 
+  \htilde(\numinus) \EfieldFT_{\rm i}(\numinus)^* \ntilde(\nuplus)]
\ee
where $\nu_{\pm} \equiv \nu\pm\delta\nu/2$ and cross terms in square brackets between $\widetilde n$ and $\EfieldiFT$ average to zero.  However, they contribute significantly to the variance of  CS estimates, as shown below. 
For zero lag,  
$\GammaFTE(\delta\nu=0, \nu) 
= \vert \EfieldFT(\nu)\vert^2 
= H(\nu)  \vert \EfieldiFT(\nu)\vert^2 + \vert \widetilde n(\nu) \vert^2$ 
is proportional to the standard FT spectral estimate for $\Efield$ (when the cross terms are averaged out). 

At minimum, the scintillation spectrum $H(\nu) \equiv \vert \htilde(\nu)\vert^2$ 
is the quantity of interest to be  extracted from the CS. 
Fluctuations in $H(\nu)$ and $\Gamma_{\EfieldiFT}$  from a single FT are 100\% about the mean.  The exponential statistics of these quantities originates from the stochasticity of the signal, which yields  complex Gaussian statistics for the FT by virtue of the central limit theorem.   They are reduced by smoothing 
the spectrum  or by  averaging spectra obtained from multiple FTs.     Assuming the FT length $\TFT = \Np P$ is chosen to yield the required frequency resolution, spectral smoothing is not an option, so averaging over multiple blocks is mandatory\footnote{In this regard,  the example in \citet[][]{2011MNRAS.416.2821D} for B1937+21 at 430 MHz is informative.   Long FTs with $\Np=87$ were used for coherent dedispersion of the 30\,ms dispersion time across the  4\,MHz band.   This yields  frequency resolution.
$\Delta\nu \sim 1 / \Np P \simeq 7.4$\,Hz, far smaller than scintle widths $\dnud \sim 1 / 2\pi\taud \sim 25$\,kHz
for $\taud \sim 40 \, \mu s$.   The CS was then degraded by a factor of 87 to yield a resolution
$\Delta\nu \sim 0.642$\,kHz which is the resolution corresponding to $1/P$.  Paradoxically, if FT lengths of 
$\Np = 1$ were used for the CS computation (after initial coherent dedispersion), the resolution would be limited by the pulse width to $\Delta\nu \sim (30\,\mu s)^{-1} \simeq  33$\,kHz, insufficent for resolving  scintles.   Thus the CS using $\Np \gg 1$ exploits the rapid sampling of voltage data, $\Delta t = 0.25\,\mu s$ to provide a large Nyquist frequency and also provides the needed frequency resolution.}.

Ultimately the CS involves the FT $\amnAmpFT(\delta\nu)$ of the periodic  modulation function $A(t) \equiv a^2(t)$, which is defined   before propagation and so does not include scattering. 
A strictly periodic train of  $\Np\gg 1$ identical pulse shapes $A_0(t)$ with $A_0(0) = 1$ multiplied by an   average amplitude $\Spknu$  has an FT,
\be
\amnAmpFT(\delta\nu) 
\simeq \Np \WA \Spknu
\amnAmpFT_{\rm n}(\delta\nu) \times \vert \sincd(\delta\nu P, \Np)\vert ,
\ee
where $\amnAmpFT_{\rm n}(\delta\nu) = \amnAmpFT_0(\delta\nu) / \amnAmpFT_0(0)$ is  normalized to unit maximum and its effective width is $\WA \equiv \amnAmpFT_0(0)$. The Dirichlet sinc function, 
$\sincd(x, N) = \sin(N\pi x) / N\sin(\pi x)$,  yields peaks at discrete frequencies $\dnuk = k / P$ with $k$ extending to
$k_{\rm max }\sim P / \WA$, the reciprocal of the pulse duty cycle.     CS can also be used to extract the full profile
(i.e. across all of pulse phase) by using a larger number of harmonics than $\kmax$,  integrating over $\nu$, 
and inverse transforming $\amnAmpFT_{\rm n}$.

The  CS for a single FT spanning a time much greater than the inverse bandwidth but much shorter than
the scintillation time $\dtd$ is given by an ensemble average over noise terms while $\Gamma_{\htilde}$ is fixed. This gives
\be
\left\langle \GammaFTE(\Gargsk) \right\rangle
= (\TFT / B)\, \Sphiave\, \bandpass(\nu) 
\left[ \amnAmpFT_{\rm n}(\dnuk)  \Gamma_{\htilde}(\Gargsk) + \frac{\sinc\, \dnuk T}{\SNRphi} \right],
\label{eq:Gepsmean}
\ee
where 
the bandpass factor is $\bandpass(\nu) = \widetilde\Delta(\nu) / \widetilde\Delta(0) =  B \widetilde\Delta(\nu)$.
Also, 
$\Sphiave \simeq (\WA/P)\Spknu$ is the period averaged flux density (as usually reported for pulsars),
$\bandpass(\nu)$ is the receiver bandpass function with width $B$,
and $\sinc\, x \equiv (\sin \pi x) / \pi x$.   The signal to noise ratio is 
$\SNRphi \equiv \Sphiave / \Ssys$ where $\Ssys = \Tsys / G$ is the system effective flux density (SEFD) for a telescope with gain $G$.    The first term in square brackets is from the pulsar signal while the second term is from additive noise reduced by the $\SNRphi^{-1}$ factor.   Though $\SNRphi \ll 1$ for most pulsars,  the second term is important only for $k=0$. 

The inverse transform of $\GammaFTE(\Gargsk)$ with respect to $\dnuk$ yields the phase resolved CS
or `profile spectrum'  with a time argument replacing the $\delta\nu$ argument, 
\be
\left\langle \GammaFTE(t, \nu) \right\rangle
\propto  \Sphiave\, \bandpass(\nu)  \amnAmp_{\rm n}(t) *  \Gamma_{\htilde}(t, \nu) ,
\ee
where a constant term from the $\sinc$ function has been excluded. 

\subsubsection{CS estimation errors}

The 100\% errors in the CS from a single FT (even with infinite S/N)  are a consequence of the  complex Gaussian statistics of the FT.    To show this we calculate the variance of $\GammaFTE(\Gargsk)$,
\be
\Var\left[\GammaFTE(\Gargsk) \right]
= 
\left[ (\TFT / B)\, \Sphiave \right]^2  \bandpass(\nuplus) \bandpass(\numinus) 
\left\{
\left\vert  \Gamma_{\htilde}(\Gargsk)\right\vert^2 + \SNRphi^{-2} + \SNRphi^{-1} [H(\nuplus) + H(\numinus) ]
\right\},
\label{eq:Gepsvar}
\ee
There is only weak dependence of the variance on $\dnuk$ and no explicit dependence on the shape $\Atilde_{\rm n}(\dnuk)$ in the total variance.  
{\it However}, the dependence on $\Atilde_{\rm n}(\dnuk)$ is hidden in the partition of variance into the real and imaginary parts of  $\GammaFTE$,  which evolves with the variation of  $\Atilde(\dnuk)$ vs. $\dnuk$.

\begin{figure}[htbp] 
   \centering
   \includegraphics[width=0.475\linewidth]{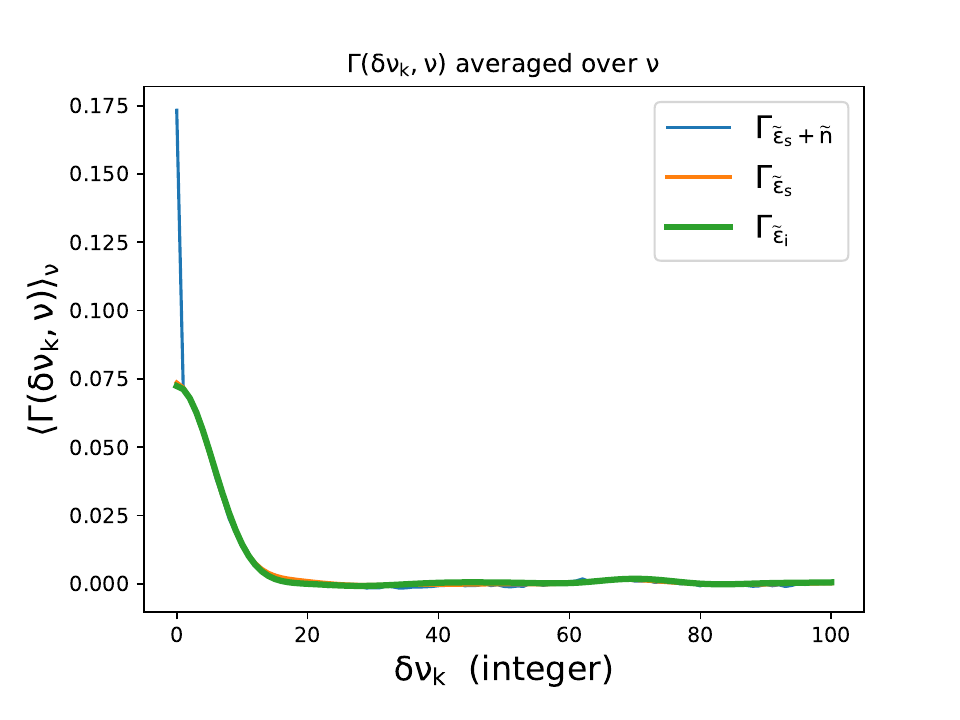} 
    \includegraphics[width=0.475\linewidth]{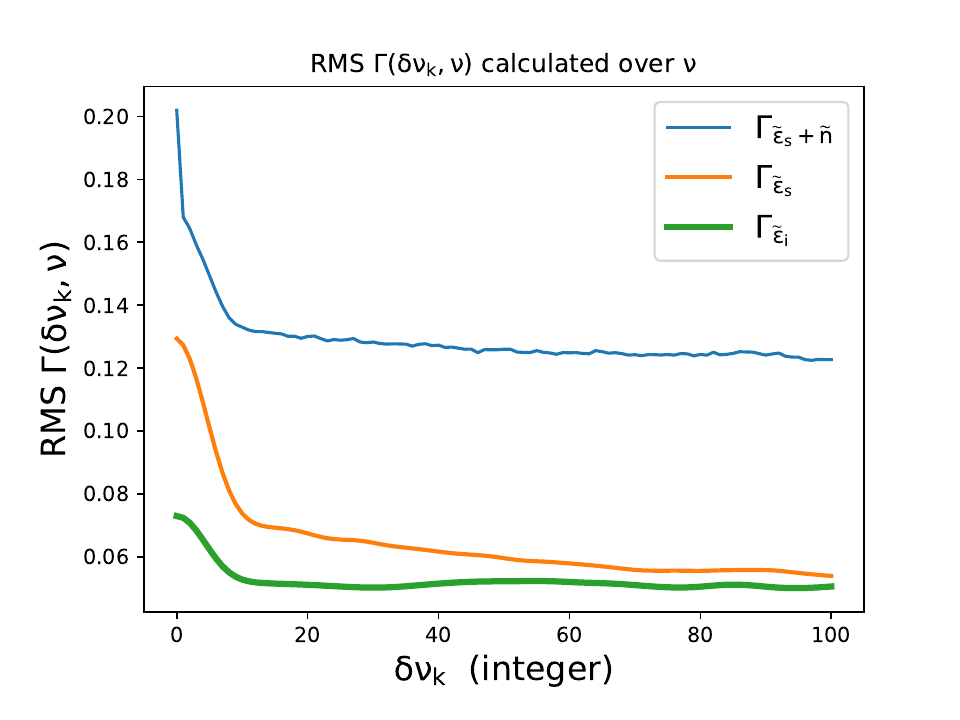} 
   \caption{Average and \RMS\ cyclic spectra vs. frequency lag,  
   Left:  Averages over frequency $\nu$ for the cyclic spectra of 
    the intrinsic field $\EfieldFT_{\rm i}$, 
   the scattered field $\EfieldFT_{\rm s}$ that includes scintillations, 
   and the total field that includes radiometer noise.  Radiometer noise decorrelates in one sample, producing 
   the spike in the CS. 
   Right: \RMS\ values of the cyclic spectra which combine  contributions from the real and imaginary parts.  
   }
   \label{fig:CSave_rms}
\end{figure}

Fig.~\ref{fig:CSave_rms} shows the mean and \RMS\  variation of  $\GammaFTE(\dnuk,\nu)$ vs $\nu$ for 
fixed $\dnuk$.  

\begin{figure}[htbp] 
   \centering
   \includegraphics[width=0.6\linewidth]{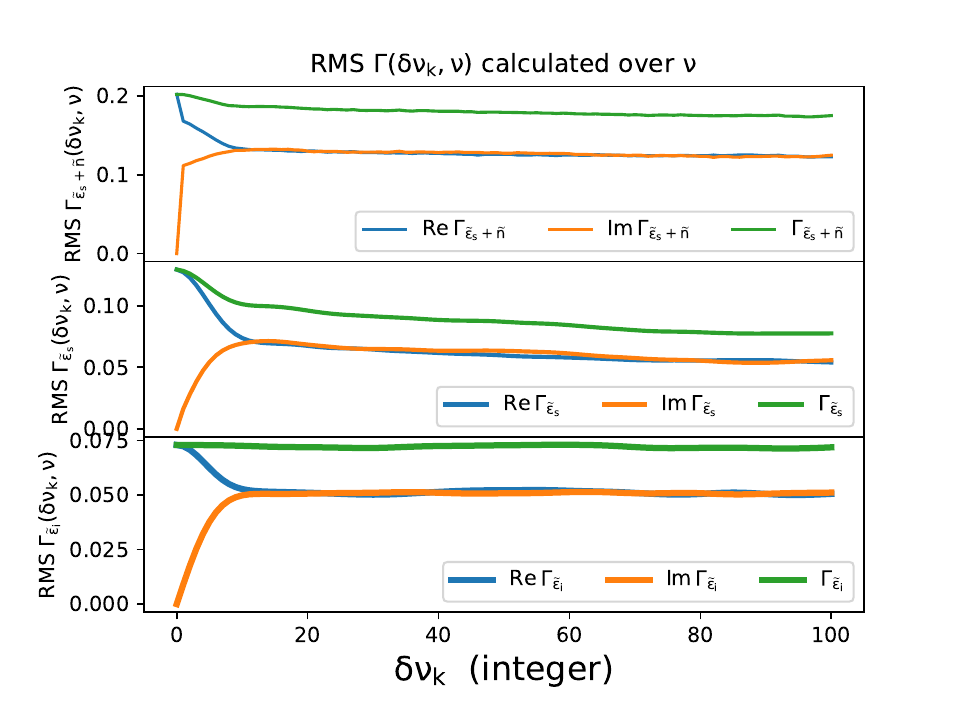}
   \caption{RMS of cyclic spectra vs $\dnuk$ showing how fluctuations are distributed between the real and 
   imaginary parts.  In each case, the fluctuations are all in the real part for $\delta\nu_k = 0$. 
    }
   \label{fig:re_im_total_cs_rms}
\end{figure}

Partition of variance in the real and imaginary parts is shown in Fig.~\ref{fig:re_im_total_cs_rms}, which shows 
the \RMS\ $\GammaFTE$ for the total (scattered + noise) signal (top panel), scintillated intrinsic signal (middle), and intrinsic signal (bottom).  The variance  is entirely in the real part at $\dnuk = 0$, but equipartitions into the real and imaginary parts for 
$\dnuk \gtrsim \WA^{-1}$.   The asymptotic variances for $\GammaFTE$ are constant in $\dnuk$ while the
other correlations, $\Gamma_{\widetilde \epsilon + \widetilde n}$ and $\Gamma_{\widetilde \varepsilon_{\rm s}}$ decline slowly due to the scintillation modulation. 

{\it Fluctuations near the origin:}

\begin{quote}
 For small $0 < \dnuk \ll \WA^{-1} \ll \dnud \ll B$ most of the variance is in the real part of $\GammaFTE$ and
 we have $ \Gamma_{\htilde}(\Gargsk)  \sim H(\nuplus) \sim H(\numinus) \simeq H(\nu)$,
 
 {\it Single block}:
The variance is
\be
\Var\left[\GammaFTE(\Gargsk) \right]
&\simeq&
\left[ (\TFT / B)\, \Sphiave  \bandpass(\nu) \right]^2 
\SNRphi^{-2}
\left[   1 + H(\nu) \SNRphi \right]^2
\nonumber \\
&\simeq& 
\left\langle \GammaFTE(0, \nu) \right\rangle^2 
\left\{1 + [H(\nu) \SNRphi]^{-1}  \right\}^2
\label{eq:Gepsvar2}
\ee
where the variance is written in terms of the mean CS in the last step.
For large $\SNRphi$, we see that the CS has 100\% errors.

We write the unaveraged CS in terms of the average CS and a statistical part, 
\be
\GammaFTE(\Gargsk)  
\equiv 
\left\langle \GammaFTE(\Gargsk) \right\rangle
+ 
\delta \GammaFTE(\Gargsk) ,
\ee
where  for $\SNRphi \gg 1$,  $\delta \GammaFTE(\Gargsk)$ is drawn from a one-sided exponential PDF
(see Table)
\be
\delta \GammaFTE(\Gargsk)  \sim {\cal E}_1(\mu_{\GammaFTE}),
\quad\quad
\text{where} \ \ 
\mu_{\GammaFTE} = \sqrt{\Var(\left[\GammaFTE(\Gargsk) \right])}  .
\ee
For smaller $\SNRphi$,  the distribution is a combination of  exponential PDFs from the signal and from radiometer noise.

{\it Block averaging:}
The only way to reduce CS estimation errors is by averaging over multiple data blocks (assuming the frequency resolution is fixed).
This reduces the variance by a factor $\Nb^{-1}$ for a PDF that converges to a Gaussian form.  In this case,
\be
\delta \GammaFTE(\Gargsk)  \sim {\cal N}(0, \sigma^2_{\GammaFTE}),
\quad\quad
\text{where}\ \ 
\sigma^2_{\GammaFTE} 
=
\Var(\left[\GammaFTE(\Gargsk) \right]) 
\simeq 
\Nb^{-1} \left\langle \GammaFTE(\Gargsk) \right\rangle^2 .
\ee
\end{quote}

{\it Fluctuations away from the origin:}   
For $\dnuk \gtrsim \WA^{-1}$ where $\Atilde(\dnuk)$ is negligible,
the real and imaginary parts of $\GammaFTE$ have equal variance. 
\begin{quote}
{\it Single block:} The real and imaginary parts are each drawn independently 
 from a two-sided exponential distribution, which we write as a  a complex, 
two sided exponential distribution (see Table),
\be
\delta \GammaFTE(\dnuk \gg \WA^{-1}, \nu)
\sim
{\cal E}_2^{\rm(c)} (0, \sigma^2_{\GammaFTE} ).
\ee

{\it Block averaging:}  The CLT implies convergence to a complex Gaussian PDF such that
\be
\delta \GammaFTE(\Gargsk)  \sim {\cal N}^{\rm (c)} (0, \sigma^2_{\GammaFTE}),
\quad\quad
\text{where}\ \ 
\sigma^2_{\GammaFTE} 
\simeq 
\Nb^{-1} \left\langle \GammaFTE(\Gargsk) \right\rangle^2 .
\ee
\end{quote}

{\it Implications for CS utility:}  If the CS is used solely to produce a dynamic spectrum,   the sensitivity analysis for 
estimation of $\dnud$ using  PRS also applies here.     The signal to noise ratio of the scintillated pulsar signal in the CS need not be large in that case because determination of $\dnud$ is often dependent only on the total number of scintles in the DS.  

However, if the CS is used to determine the field PBF $\htilde(\nu)$ itself using methods like those discussed in 
 \citet[][]{2013ApJ...779...99W, 2021ApJ...913...98D}, the requirements are much more stringent.   In this case, 
 the statistical errors must be small enough to allow solution for $\htilde(\nu)$ from data spanning less than 
 $\dtd$ rather than utilizing the full DS extending over a large multiple of $\dtd$.    If the CS errors are scintle  (not noise) dominated,  the single block errors are 100\% and the $\Nb$-block errors are $\simeq \Nb^{-1/2}$.   If we require that $H(\nu)$ be determined to a precision  better than $\epsH \ll 1$, then we require  
 $\Nb > \epsH^{-2}$ while also requiring that the blocks extend over no more than a fraction $\epsdtd$ of the scintillation time,  $\Nb \Np P = \epsdtd \dtd$.    The combined constraints  on $\Nb$ are 
 \be
 \epsH^{-2} \lesssim \Nb \lesssim \epsdtd \dtd / \Np P. 
 \ee
 For 10\% error on $H$ and ten samples per scintle ($\epsdnud = 0.1$), we have 
 \be
 100 \times (0.1 / \epsH)^2  
 \lesssim \Nb \lesssim 
 10^4 \times (\epsdtd / 0.3) (\dtd / 100\, {\rm s}) (3\, {\rm ms} / P) \Np^{-1}.
 \ee 
To resolve scintles, we must have $\Np \gtrsim (P\epsdnud \dnud)^{-1} = 2\pi\taud / \epsdnud P$. If $\Np \gg 1$,
the number of blocks is more constrained and may be insufficent for heavily scattered millisecond pulsars. 
 
\subsubsection{Role of Pulse Jitter}

Departure from strict periodicity results from time jitter  of  pulse centroids. 
 Gaussian distributed  jitter $\DtJ$ of each pulse with \RMS\ $\rmsJ$ reduces $\amnAmpFT$ by a form factor,
\be
\etaJ(\delta\nu)= e^{-2(\pi \delta\nu \rmsJ)^2} .  
\ee
If we restrict the reduction in all $\kmax=P/\WA$ harmonics to no more than $1/x$ 
the  \RMS\ jitter is bounded by
\be
\rmsJ < \frac{\WA}{\pi} \left( \frac{\ln x}{2} \right)^{1/2}.
\ee
For $x=2$ ($<50$\% reduction),   $\rmsJ  < 0.19 \WA$,
roughly the amount of single-pulse jitter seen in most pulsars
\citep[e.g.][]{2012ApJ...761...64S, 2016ApJ...819..155L}.   Larger \RMS\ jitter reduces the number of harmonics that otherwise could have been detectable.    This reduction is unavoidable because  measured average pulses include jitter.  In effect, the pulse shape $A(t)$ used above should be viewed as including phase jitter. 

\subsubsection{Constraints on CS parameters}

The viability of CS for obtaining dynamic spectra depends on sampling constraints and on  estimation errors.  The conditions needed are therefore pulsar and frequency dependent:
\begin{enumerate}
\item CS is unnecessary if the scintillation bandwidth exceeds half the receiver bandwidth,
$\dnud \ge B/2$.    Using $B = \fB \nu$ and $\dnud = C_1 / 2\pi \taud$ along with
$\taud = \tau_1 \nu^{-\Xtau}$ with $\Xtau = 22/5$.  CS cannot be applied for  frequencies above
\be
\nu_{\rm no\, CS} = \left(\frac{1000\pi\fB \tau_1 }{C_1} \right)^{1 / (\Xtau-1)}
\simeq 10.7\,{\rm GHz} \times \left(\frac{\fB \tau_1 }{C_1} \right)^{5/17}.
\label{eq:nucond1}
\ee
\item 
Resolving scintles requires $\TFT^{-1} = (\Np P)^{-1}  \ll  \dnud = C_1 / 2\pi\taud$ which we write as 
$\TFT^{-1} \equiv \epsdnud \dnud \ll \dnud$ with  $\epsdnud \ll 1$. For ten samples per $\dnud$ 
(i.e. $\epsdnud = 0.1$)
\be
\Np \gg \frac{2\pi \taud}{C_1 P} \quad {\rm or} \quad 
\Np  = \frac{2\pi\taud}{\epsdnud C_1 P} 
\simeq \frac{0.063 \, \nu^{-22/5} \tau_1}{(\epsdnud/0.1)C_1 P({\rm ms}) }
\ee 
where $\tau_1$ is the scattering time at 1\,GHz in $\mu s$ units. 
This expression suggests that CS is indicated only for large scattering times $\taud$ and low frequencies.   However, even for 
moderate scattering times,  CS is needed because phase resolved spectroscopy is limited to a resolution
$\Delta\nu \simeq \WA^{-1}$ that is sufficient only for 
$\WA \gtrsim 63\,\mu s \,\taud(\mu s) /  C_1 \epsdnud/0.1)$.   MSPs with narrow pulse widths, such as
B1937+2214 mentioned above, do not satisfy this condition and require $\Np > 1$ at 0.4\,GHz. 

This condition can be written as $\nu \ge \nu_{\dnud}$ with
\be
\nu_{\dnud} 
=  \left( \frac{ 2\pi \tau_1}{\epsdnud C_1 \Np P}  \right)^{1/\Xtau}
\simeq  0.53\, {\rm GHz}  \times  \left[\frac{  \tau_1}{(\epsdnud/0.1)C_1 \Np \,P({\rm ms}) }\right]^{5/22}.
\label{eq:nucond2}
\ee
\item Reducing  CS variance 
 by maximizing the  number of blocks $\Nb$  is in tension with the use of $\Np \gg 1$ FTs  to resolve scintles (condition 2). Their product $\Np\Nb$  is constrained by $\Nb \TFT = \Np\Nb P \lesssim  \dtd$,
where $\dtd(\nu) \propto\nu^{6/5}$ is the scintillation time (Eq.~\eqref{eq:dtd}).    If an error $\epstol = 1 / \sqrt{\Nb}$ is specified, the FT length is limited by
\be
\Np 	\lesssim \frac{\epstol^2 \dtd}{P}  
	\simeq 1.43\times 10^3\, 
	\left(
		\frac{\epstol}{0.1} 
	\right)^2
	\left(
	\frac{\nu^{6/5}}{v_{100} P({\rm ms})}
	\right) 
	 \left(\frac{\deff}{\tau_1} \right)^{1/2},
\ee
where, as before, $\nu$ is in GHz, $v_{100}$ is the effective transverse velocity in units of 100\, $\kms$,
and the effective distance $\deff$ is in kpc. 
This limit on $\Np$ becomes small for longer periods, lower frequencies, and larger scattering times. 

Expressed as a frequency limit $\nu \gtrsim \nu_{\dtd}$ and using the scintillation time expressed  in terms of
its value at 1\,GHz,  $\dtd(\nu) = {\dtd}_1 \nu^{2/(\beta-2} \to {\dtd}_1 \nu^{6/5}$, 
\be
\nu_{\dtd} = \left( \frac{\Np\Nb P}{{\dtd}_1} \right)^{(\beta-2)/2}
\simeq 0.016\,{\rm GHz} 
	\left(\frac{\Np \Nb v_{100} P({\rm ms})}{1000} \right)^{5/6}
	\left(\frac{\tau_1}{\deff} \right)^{5/12}.
\label{eq:nucond3}
\ee

\item 
The last  constraint derives from the net fractional errors on $\GammaEFT$  and $\widehat H(\nu)$ from  
S/N considerations.  Large values of
$\SNRphi$ are uncommon.   The $\sim 5$\,Jy flux density of the Vela pulsar at 0.4\,GHz exceeds $\Ssys$ only for the largest telescopes with effective diameters $\gtrsim 300$\,m.   For most other cases,  $\SNRphi \ll 1$. Restricting the fractional error to $\epsH \ll 1$   implies 
$\epsH \sqrt{\Nb\kmax}  \SNRphi\ge 1$ or 
\be
\SNRphi \ge \left( \epsH \sqrt{\Nb\kmax} \right)^{-1}
\simeq 0.058 \times 
\left( \frac{0.1}{\epsH} \right)
\left( \frac{10^3}{\Nb} \frac{30}{\kmax} \right)^{1/2}.
\ee
The MSP J1939+2134 (B1937+21) satisfies this condition in Arecibo data with
$\Ssys \simeq 4$\, Jy and   $\Sphi(0.43\,{\rm GHz}) =  240$\,mJy  while using $\Nb = 7.7\times10^4$ blocks across  120\,s and $\kmax \simeq  20$ \citep[][]{2011MNRAS.416.2821D}.   This yields $\SNRphi\simeq 0.06 \gg 0.008$.
\end{enumerate}

Conditions 1 to 3 can be rewritten as constraints on the period $P$,  dispersion measure \DM, and frequency
$\nu$ evaluating $\tau_1$ and the effective distance $\deff$ in terms of DM.  We evaluate the constraints by
calculating $\tau_1$ from the empirical $\taud(\DM)$ relation.  That relation shows substantial scatter 
($\sigma_{\log_{10}\tau} \simeq 0.76$) but these amount to only a factor of 1/2 to 2 spread in the constraints 
given the indices of $\tau_1$ in Eq.~\eqref{eq:nucond1}, \eqref{eq:nucond2}, and \eqref{eq:nucond3}.

\subsection{ACF Statistics}
\label{app:acfstats}

From a given spectrum we want to determine the autocorrelation width of frequency structure.   
Here we wish to know the error on the signal part of the ACF caused by additive white noise.
Let $I = S + N$ where $S$ corresponds to the DISS spectrum and $N$ is additive noise.   Assume $N$ has Gaussian statistics (via the CLT applied to a spectrum produced with a sufficient degree of averaging) and let it have zero mean and correlation function $\langle N_j N_k \rangle = \sigma_N^2 \delta_{jk}$ using
the Kronecker delta $\delta_{jk}$.   Assume that $S$ is correlated over a relatively large number of samples. 

The ACF of a spectrum with $N$ samples is
\be
R_I(\delta\nu) = N^{-1} \sum_{j=0}^{N-\delta\nu} I_j I_{j+\delta\nu} 
\ee
where $I_j I_{j+\delta\nu} = S_j S_{j+\delta\nu} + N_j N_{j+\delta\nu} + S_j N_{j+\delta\nu} + S_{j+\delta\nu} N_j$.

The ensemble mean ACF receives no contributions from the $S\times N$ cross terms, 
\be
\left\langle R_I(\delta\nu)\right\rangle  
= N^{-1} \sum_{j=0}^{N-\delta\nu} S_j S_{j+\delta\nu}  + \sigma_N^2 \delta_{\delta\nu, 0}
\equiv R_S(\delta\nu) + \sigma_N^2 \delta_{\delta\nu 0}.
\ee

The mean-square ACF at a lag $\delta\tau = 0+$  avoids the noise spike at zero lag but is contributed to by cross terms
between $S$ and $N$, 
\be
\left\langle R_I^2(0+)\right\rangle  
\simeq \left\langle R_I(\delta\nu)\right\rangle^2  
	+  N^{-1} \left[  \sigma_N^4 + 4 \sigma_N^2 R_S(0) \right]. 
\ee
The second term is the variance of the ACF that implies a fractional ACF error,
\be
\frac{ [\Var  R_I(0+) ]^{1/2}} {\left \langle  R_I(0+) \right\rangle}
	\simeq   \frac{\displaystyle  \left[  \sigma_N^4 + 4 \sigma_N^2 R_S(0) \right] ^{1/2}}{\sqrt{N}\times R_S(0)}.
\ee
The first term dominates for low S/N.  For large S/N, i.e.  $R^{1/2}_S(0) / \sigma_N \gg 1$, the fractional error is
\be
\frac{ [\Var  R_I(0+) ]^{1/2}} {\left \langle  R_I(0+) \right\rangle}
	\simeq  \frac{\displaystyle   2 \sigma_N}{\sqrt{N \times R_S(0)}}.
\ee
Errors in the ACF translate into an error  on the estimate of  its characteristic width,  $\widehat W_S$. 
Let $W_S$ equal the true width  in samples.   The fractional error on the estimate is
\be
\frac{\sigma_{\widehat W_S}} {W_S} \simeq 
	\frac{1}{\sqrt{W_S}} 
	\frac{ [\Var  R_I(0+) ]^{1/2}} {\left \langle  R_I(0+) \right\rangle} . 
\ee
Using these expressions, a specified error on $W_S$ can be back propagated into a requirement on the 
S/N of the data.

%

%

\clearpage
\section{Hilbert transform solutions for non-minimum delay functions}
\label{sec:ht_nonmin}

For a function that does not have the minimum-delay property, the minimum-delay phase still has a role.  Let $\tilde h_{\rm HT}(\nu)$ be the FT obtained using
the HT to calculate the minimum-delay solution.    The true FT for more general functions can be written as
\citep[e.g.][Chapter 5]{alma991011802429705251} 
 \be
\tilde h(\nu) = \tilde h_{\rm HT}(\nu) Z(\nu), 
\ee 
where $Z(\nu)$ is an  ``all-pass'' filter with unit amplitude that
comprises some number
of zeros of $\exp(-2\pi i\nu)$  in the complex plane and that changes only
the phase.
The all-pass filter can be written as the product of $Q$ individual factors
involving the zeros $s_j^{-1}$ where $\exp(2\pi i \nu) = 0$,
\be
Z(\nu) = \prod_{j=0}^{Q-1} 
	\left(
	\frac{e^{-2\pi i \nu} - s_j}{1- s_j^*e^{-2\pi i \nu}}
	\right),
\ee
and where each factor and thus $Z$ has unity magnitude. 
The numerical task is to determine the values for $s_j$ ($2Q$ unknowns)
and the value of $Q$ itself.  Determination is aided by the fact that the
zeros $s_j^{-1}$ are outside the unit circle and by the possibility that $Q$ is
much smaller than the overall length of a time series.   

\clearpage
\section{Quality Measure Evaluation Code}
\label{app:QMcode}

Quality measures can be evaluated for an individual pulsar using {\tt PTQ.py}, available at GitHub
(J. M. Cordes, PTQ, (2025), GitHub respository, \url{https://github.com/jmcordes/PTQ)}, which includes a description and several convenience functions.

{\tt
\begin{verbatim}
usage: PTQ.py [-h] [-psr PSR] [-p0 P0] [-p1 P1] [-d D] [-vtrans VTRANS] [-wc WC] [-nc NC] 
            [-s S1400] [-tau TAU] [-nu_tau NU_TAU] [-B BW] [-Tint TINT] [-Ssys SSYS] 
            [-npol NPOL] [-Tyr TYR] [-bKol BKOL] [-Fj FJ] [-nu_eval NU_EVAL] [-v]

Calculates quality measures using input data for an MSP. Outputs metrics that compare 
the pulsar to other MSPs. JMC 2023 Jun - 2024 Nov 5, 2025 May 19-22 Input information needed: 
psrname P, Pdot period, period derivative sec and s/s D distance kpc vtrans transverse pulsar 
speed km/s tau pulsar broadening time microsec nu_tau RF for pulsar broadening time GHz Wc
Component width microsec nc Number of components dimensionless Alternative to Wc, nc 
(not yet implemented): template file from which Weff is determined (equivalent to Wc, nc) 
Metrics file: 'QMcalcs_metrics_20241105.npz'

optional arguments:
  -h, --help            show this help message and exit
  -psr PSR, --psr PSR   Pulsar name (for labeling) (default: J1909-3744)
  -p0 P0, --p0 P0       pulse period (sec) (default: 0.002947108069160717)
  -p1 P1, --p1 P1       period derivative (s/s) (default: 1.402541e-20)
  -d D, --d D           pulsar distance (kpc) (default: 1.14)
  -vtrans VTRANS, --vtrans VTRANS
                        Transverse pulsar speed (km/s) (default: 200.19)
  -wc WC, --wc WC       Profile component width (microsec)) (default: 45.6489)
  -nc NC, --nc NC       Number of profile components (default: 1)
  -s S1400, --s1400 S1400
                        Period averaged flux density (mJy)) (default: 1.8)
  -tau TAU, --tau TAU   Pulse broadening time (microsec) (default: 0.033263530451980036)
  -nu_tau NU_TAU, --nu_tau NU_TAU
                        Radio frequency of pulse broadening time (GHz) (default: 1)
  -B BW, --BW BW        Bandwidth (GHz) (default: 0.1)
  -Tint TINT, --Tint TINT
                        Total integration time per epoch (sec) (default: 1000.0)
  -Ssys SSYS, --Ssys SSYS
                        SEFD (Jy (default: 3)
  -npol NPOL, --npol NPOL
                        Number of polarization channels (int (default: 2)
  -Tyr TYR, --Tyr TYR   Time range for calculating spin noise (yr) (default: 1)
  -bKol BKOL, --bKol BKOL
                        Wavenumber spectrum spectral index (dimensionless) 
                        (default: 3.6666666666666665)
  -Fj FJ, --FJ FJ       Jitter parameter (dimensionless) (default: 0.3333333333333333)
  -nu_eval NU_EVAL, --nu_eval NU_EVAL
                        RF for evaluating pulsar (GHz) (default: 1.4)
  -v, --V               Verbose to get detailed output (default: False)
\end{verbatim}
}

  
\clearpage

\phantomsection
\addcontentsline{toc}{section}{References}
\bibliographystyle{aasjournalv7.bst}
\bibliography{FNP2026Oct9}

\end{document}